\pdfoutput=1

\documentclass[11pt,letterpaper,hyperref,dvipsnames]{cernyrep}
\usepackage[utf8]{inputenc}

\hypersetup{
  colorlinks=true,
  linkcolor=red,
  urlcolor=blue,
  citecolor=darkgreen,
  pdfauthor={LH19 participants},
  pdftitle={Les Houches 2019 Proceedings}
}

\usepackage{cite}
\usepackage{amsmath,amssymb,graphicx,xspace,subfigure}
\usepackage{mathtools}
\usepackage{wasysym}
\usepackage{todonotes}
\usepackage{bigints}
\usepackage{wrapfig}

\usepackage[T1]{fontenc}
\usepackage{lmodern}
\usepackage{rotating}
\usepackage{xcolor}
\usepackage{booktabs}
\usepackage{arydshln}
\usepackage{lscape}
\usepackage{floatpag}
\floatpagestyle{plain}
\usepackage{axodraw}
\usepackage{lineno}
\usepackage{graphpap}

\usepackage{import}
\usepackage{enumitem}
\usepackage{caption}
\usepackage{xspace}
\usepackage{adjustbox}
\usepackage{tikz}
\usetikzlibrary{positioning,arrows}
\usetikzlibrary{decorations.pathmorphing}
\usetikzlibrary{decorations.markings}

\newcommand{\eV}{\ensuremath{\text{e\kern-0.15ex{}V}}\xspace}

\newcommand{\GeV}{\ensuremath{\text{G\eV}}\xspace}
\newcommand{\TeV}{\ensuremath{\text{T\eV}}\xspace}

\definecolor{darkred}{rgb}{.8, 0.1, 0.1}
\definecolor{darkyellow}{rgb}{0.45, 0.45, 0.}
\definecolor{violet}{rgb}{1.0, 0.0, 1.0}
\definecolor{darkgreen}{rgb}{0.15, .8, 0.15}

\makeatletter
\renewcommand{\thechapter}{\@Roman\c@chapter}
\makeatother

\begin{document}

{\centering{\LARGE{\bf{Les Houches 2019: Physics at TeV Colliders\\ Standard Model Working Group Report \par }}}}

\pagenumbering{roman}

\vspace{0.7cm}
\leftline{\bf Conveners}


\noindent \emph{Higgs physics: SM issues} \\
         D.~de~Florian (Theory),\,
         M.~Doneg\`a (CMS), \,
         M.~D\"{u}hrssen-Debling (ATLAS), \,
         S.~Jones (Theory) \\
\vspace{0.1cm}

\noindent \emph{SM: Loops and Multilegs}  \\	
         J.~Bendavid (CMS), \,
         A.~Huss (Theory), \,
         J.~Huston (ATLAS), \,
         S.~Kallweit (Theory), \,
         D.~Ma\^{\i}tre (Theory), \,
         S.~Marzani (Jets contact),\,
         B.~Nachman (Jets contact, ATLAS) \\
\vspace{0.1cm}

\noindent \emph{Tools and Monte Carlos} \\	
         V.~Ciulli (CMS), \,
         S.~Prestel (Theory), \,
         E.~Re (Theory) \\
\vspace{1.0cm}

{\leftline{\bf{Abstract}}}

\vspace{0.5cm}
This Report summarizes the proceedings of the 2019 Les Houches workshop
on Physics at TeV Colliders. Session 1 dealt with (I) new developments for high precision Standard Model calculations, (II) the sensitivity of parton distribution functions to the experimental inputs, (III) new developments in jet substructure techniques and a detailed examination of gluon fragmentation at the LHC, (IV) issues in the theoretical description of the production of Standard Model Higgs bosons and how to relate experimental measurements, and (V) Monte Carlo event generator studies relating to PDF evolution and comparisons of important processes at the LHC.


\vspace{0.5cm}

{\leftline{\bf{Acknowledgements}}}

\vspace{0.5cm}

We would like to thank the organizers (N. Berger, F. Boudjema,
C. Delaunay, M. Delmastro, B. Fuks, S. Gascon, M. H. Genest, P. Gras,
J. P. Guillet, B. Herrmann, S. Kraml, N. Makovec, G. Moreau, E. Re)
and the Les Houches staff for the stimulating environment always
present at Les Houches. We thank the Formation Permanente du CNRS, the
IDEX Universit\'e Grenoble Alpes, the Universit\'e Savoie Mont Blanc,
LAPP and LAPTh for support.

\newpage
{\centering{\bf{Authors\par}}}
\begin{flushleft}
  S.~Amoroso$^{1}$,
  P.~Azzurri$^{2}$,
  J.~Bendavid$^{3}$,
  E.~Bothmann$^{4}$,
  D.~Britzger$^{5}$,
  H.~Brooks$^{6}$,
  A.~Buckley$^{7}$,
  M.~Calvetti$^{2,8}$,
  X.~Chen$^{9}$,
  M.~Chiesa$^{10}$,
  L.~Cieri$^{11}$,
  V.~Ciulli$^{11,12}$,
  J.~Cruz-Martinez$^{13}$,
  A.~Cueto$^{14}$,
  A.~Denner$^{15}$,
  S.~Dittmaier$^{16}$,
  M.~Doneg\`a$^{17}$,
  M.~D\"{u}hrssen-Debling$^{3}$,
  I.~Fabre$^{9,18}$,
  S.~Ferrario-Ravasio$^{19}$,
  D.~de~Florian$^{18}$,
  S.~Forte$^{13}$,
  P.~Francavilla$^{2,8}$,
  T.~Gehrmann$^{9}$,
  A.~Gehrmann-De Ridder$^{9,20}$,
  L.~Gellersen$^{21}$,
  E.~W.~N.~Glover$^{19}$,
  P.~Gras$^{22}$,
  C.~Gwenlan$^{23}$,
  Y.~Haddad$^{24}$,
  G.~Heinrich$^{5}$,
  J.~Hessler$^{5}$,
  T.~J.~Hobbs$^{25,26}$,
  M.~H\"{o}fer$^{9}$,
  A.~Huss$^{19,27}$,
  J.~Huston$^{28}$,
  T.~Je\v{z}o$^{29}$,
  S.~P.~Jones$^{27}$,
  S.~Kallweit$^{30}$,
  M.~Klasen$^{31}$,
  G.~Knippen$^{16}$,
  A.~Larkoski$^{32}$,
  M.~LeBlanc$^{33}$,
  P.~Loch$^{33}$,
  K.~Long$^{3,34}$,
  D.~Ma\^{\i}tre$^{19}$,
  S.~Marzani$^{35}$,
  J.~Mazzitelli$^{5}$,
  J.~A.~Mcfayden$^{3}$,
  E.~Metodiev$^{36}$,
  J.~K.~L.~Michel$^{37}$,
  M.~Moreno Ll\'{a}cer$^{38}$,
  B.~Nachman$^{39}$,
  P.~Nadolsky$^{25}$,
  D.~Napoletano$^{30}$,
  E.~R.~Nocera$^{40}$,
  C.~Oleari$^{30}$,
  C.~Pandini$^{41}$,
  M.~Pellen$^{42}$,
  S.~Pigazzini$^{17}$,
  J.~Pires$^{43}$,
  S.~Pl\"atzer$^{44}$,
  S.~Prestel$^{21}$,
  K.~Rabbertz$^{45}$,
  E.~Re$^{10}$,
  P.~Richardson$^{19,27}$,
  F.~Ringer$^{46,47}$,
  J.~Rojo$^{40,48}$,
  J.~Roloff$^{49}$,
  R.~R\"ontsch$^{27}$,
  M.~Sch\"onherr$^{19}$,
  C.~Schwan$^{13}$,
  F.~Siegert$^{50}$,
  D.~Soper$^{51}$,
  G.~Soyez$^{52}$,
  M.~Spira$^{53}$,
  M.~R.~Sutton$^{54}$,
  F.~J.~Tackmann$^{37}$,
  V.~Theeuwes$^{4}$,
  S.~L.~Villani$^{4}$,
  J.~Whitehead$^{19}$,
  H.~T.~Yang$^{39}$,
  J.~Zhou$^{55}$
\end{flushleft}
\begin{itemize}
\item[$^{1}$] Deutsches Elektronen-Synchrotron (DESY), Hamburg, Germany
\item[$^{2}$] INFN, Sezione di Pisa, Pisa, Italy
\item[$^{3}$] Experimental Physics Department, CERN, Geneva, Switzerland
\item[$^{4}$] Institut f\"ur Theoretische Physik, Georg-August-Universit\"at G\"ottingen, G\"ottingen, Germany
\item[$^{5}$] Max-Planck-Institut f\"ur Physik, M\"unchen, Germany
\item[$^{6}$] School of Physics and Astronomy, Monash University, Clayton, VIC, Australia
\item[$^{7}$] School of Physics and Astronomy (SUPA), University of Glasgow, Glasgow, Scotland, UK
\item[$^{8}$] Dipartimento di Fisica E. Fermi, University of Pisa, Pisa, Italy
\item[$^{9}$] Physik-Institut, Universit\"at Z\"urich, Z\"urich, Switzerland
\item[$^{10}$] LAPTh, Universit\'e Grenoble Alpes, Universit\'e Savoie Mont Blanc, CNRS, Annecy-le-Vieux, France
\item[$^{11}$] INFN, Sezione di Firenze, Firenze, Italy
\item[$^{12}$] Dipartimento di Fisica e Astronomia, Universit\`a di Firenze, Firenze, Italy
\item[$^{13}$] Tif Lab, Dipartimento di Fisica, Universit\`a di Milano and INFN, Sezione di Milano, Milano, Italy
\item[$^{14}$] Laboratoire d'Annecy de Physique des Particules (LAPP), Annecy-le-Vieux, France
\item[$^{15}$] Universit\"at W\"urzburg, Institut f\"ur Theoretische Physik und Astrophysik, W\"urzburg, Germany
\item[$^{16}$] Albert-Ludwigs-Universit\"at Freiburg, Physikalisches Institut, Freiburg, Germany
\item[$^{17}$] ETH, Z{\"u}rich, Switzerland
\item[$^{18}$] International Center for Advanced Studies (ICAS) and ICIFI, ECyT-UNSAM, Buenos Aires, Argentina
\item[$^{19}$] Institute for Particle Physics Phenomenology, Durham University, Durham, UK
\item[$^{20}$] Institute for Theoretical Physics, ETH, Z\"urich, Switzerland
\item[$^{21}$] Department of Astronomy and Theoretical Physics, Lund University, Lund, Sweden
\item[$^{22}$] IRFU, CEA, Universit\'e Paris-Saclay, Gif-sur-Yvette, France
\item[$^{23}$] Department of Physics, The University of Oxford, Oxford, UK
\item[$^{24}$] Northeastern University, Boston, MA, U.S.A.
\item[$^{25}$] Department of Physics, Southern Methodist University, Dallas, TX, U.S.A.
\item[$^{26}$] Jefferson Lab, EIC Center, Newport News, VA, U.S.A.
\item[$^{27}$] Theoretical Physics Department, CERN, Geneva, Switzerland
\item[$^{28}$] Department of Physics and Astronomy, Michigan State University, East Lansing, MI, U.S.A.
\item[$^{29}$] Institute for Theoretical Physics, KIT, Karlsruhe, Germany
\item[$^{30}$] Dipartimento di Fisica, Universit\`{a} degli Studi di Milano-Bicocca and INFN, Sezione di Milano-Bicocca, Milano, Italy
\item[$^{31}$] Institut  f\"ur  Theoretische  Physik,  Westf\"alische  Wilhelms-Universit\"at  M\"unster, M\"unster, Germany
\item[$^{32}$] Physics Department, Reed College, Portland, OR, U.S.A.
\item[$^{33}$] Department of Physics, University of Arizona, Tucson, AZ, U.S.A.
\item[$^{34}$] University of Wisconsin - Madison, Madison, WI, U.S.A.
\item[$^{35}$] Dipartimento di Fisica, Universit\`a di Genova and INFN, Sezione di Genova, Genova, Italy
\item[$^{36}$] Center for Theoretical Physics, Massachusetts Institute of Technology, Cambridge, MA, U.S.A.
\item[$^{37}$] Theory Group, Deutsches Elektronen-Synchrotron (DESY), Hamburg, Germany
\item[$^{38}$] Instituto de F\'{i}sica Corpuscular (IFIC), Centro Mixto Universidad de Valencia - CSIC, Valencia, Spain
\item[$^{39}$] Physics Division, Lawrence Berkeley National Laboratory and University of California, Berkeley, CA, U.S.A.
\item[$^{40}$] Nikhef, Amsterdam, The Netherlands
\item[$^{41}$] University of Geneva, D\`{e}partement de physique nucl\`{e}aire et corpusculaire, Switzerland
\item[$^{42}$] University of Cambridge, Cavendish Laboratory, Cambridge, UK
\item[$^{43}$] LIP, Lisboa, Portugal
\item[$^{44}$] Particle Physics, Faculty of Physics, University of Vienna, and Erwin Schr{\"o}dinger International Institute for Mathematical Physics (ESI), Vienna, Austria
\item[$^{45}$] Institut f\"ur Experimentelle Teilchenphysik (ETP), Karlsruhe Institute of Technology (KIT), Karlsruhe, Germany
\item[$^{46}$] Department of Physics, University of California, Berkeley, CA, U.S.A.
\item[$^{47}$] Nuclear Science Division, Lawrence Berkeley National Laboratory, Berkeley, CA, U.S.A.
\item[$^{48}$] Department of Physics and Astronomy, VU Amsterdam, Amsterdam, The Netherlands
\item[$^{49}$] Physics Department, Brookhaven National Laboratory, Upton, NY, U.S.A.
\item[$^{50}$] TU Dresden, Institut f\"ur Kern- und Teilchenphysik, Dresden, Germany
\item[$^{51}$] Institute for Fundamental Science, University of Oregon, Eugene, OR, U.S.A.
\item[$^{52}$] Universit\'e Paris-Saclay, CNRS, CEA, Institut de physique th\'eorique, Gif-sur-Yvette, France
\item[$^{53}$] Paul Scherrer Institut (PSI), Villigen, Switzerland
\item[$^{54}$] Department of Physics and Astronomy, The University of Sussex, Brighton, UK
\item[$^{55}$] Amherst Center for Fundamental Interactions, Physics Department, University of Massachusetts Amherst, Amherst, MA, U.S.A.
\end{itemize}

\newpage
{\parskip=1.5ex \tableofcontents}

\newpage
\pagenumbering{arabic}
\setcounter{footnote}{0}



\section{Introduction}

These past two years have seen the integrated luminosities at 13~TeV for ATLAS and CMS increase to approximately $140~\mathrm{fb}^{-1}$ each. At the same time, there have been reductions in many systematic errors (modulo the impact of increased pileup), brought on by a better understanding of the detectors and of the reconstruction algorithms. Thus, there is a continuing pressure for improvements on the corresponding theoretical predictions.
These predictions include those defined at fixed-order, those resumming large logarithms due to kinematic thresholds and boundaries, and those involving parton showering, and subsequent hadronization. The latter allows for a direct comparison to data at the hadron level. All
levels of theoretical predictions are needed for a full exploration of LHC physics. We
continue in these proceedings to discuss advances in theoretical predictions, while also examining their connections, their limitations, and their prospects for improvement.

Calculations for many important $2\to2$ NNLO processes at the LHC are now available and are reviewed in the wishlist update. Progress has also been made in calculations involving elliptic integrals, such as $H+\ge1$ jet at two loops, with finite top mass corrections~\cite{Bonciani:2019jyb,Francesco:2019yqt,Frellesvig:2019byn}. This process has been known for over 2 years, but using numerical techniques~\cite{Jones:2018hbb}.   Considerable progress  on  the calculation of $2\rightarrow3$ processes at NNLO has also been made and is also discussed. The first $2\rightarrow3$ process, 3 photon production, has been calculated at NNLO~\cite{Chawdhry:2019bji}. Dissemination of complex NNLO results continues to be a problem. Often NNLO/NLO point-by-point K-factors are provided by the authors of the NNLO calculation. These proceedings provide an update on the use of the NNLO grid tables that will allow the flexible use of these calculations.  Another technique, also discussed in these proceedings,  is the use of ROOT ntuples in which enough information is stored to produce new  event weights while applying different scales, PDFs, kinematic cuts, etc. There needs to be  a balance between speed and disk storage requirements in order for this technique to be practical. Both of these techniques cannot get by using  the same ``brute force'' methods that were successful at NLO.

There has also been great progress on N${}^3$LO predictions, with the calculation of the differential Higgs boson rapidity cross section~\cite{Dulat:2018bfe} and the calculation of the Drell-Yan production cross section~\cite{Duhr:2020seh}, at this order. The latter calculation shows N${}^3$LO corrections that are surprisingly large, outside the uncertainty bands at NNLO. This is attributed to accidental cancellations at NNLO that resulted in an artificially reduced uncertainty band. Although scale uncertainty bands at a given order are one of the few tools we have to (i) estimate the uncertainty of the cross section at that order, and (ii) estimate the possible size of higher order corrections, this is another lesson of the dangers that this can lead to.

Part of the impressive progress in NNLO results is due to the development of subtraction methods to treat infrared divergences at this order.
An account on the current status of the different methods and their recent progress is given in the wishlist update.
A different approach aims to avoid the occurrence of (dimensionally regulated) poles which need to be isolated and cancelled between real and virtual parts by combining all the contributions at integrand level and then performing numerical integration in $D=4$ dimensions.
Such four-dimensional frameworks are promising as they avoid some technicalities related to calculations in general $D$ dimensions.
However these methods face hurdles of different type, related to the purely numerical approach.
%

The LHC is accessing kinematic regions and processes where electroweak corrections have significant impact, with a great deal of progress in recent years being observed on the automation of electroweak NLO corrections. A comparison of two predictions for off-shell WWW production, including NLO QCD and EW corrections, was carried out
in the context of the  Les Houches workshop. This process is interesting as it is sensitive to quartic gauge boson self-interactions and to
off-shell Higgs boson exchange. ATLAS has recently reported 4.1 sigma evidence for this final state.

Two recent independent calculations, with leptonic decaying W bosons, including all off-shell effects, did not entirely agree.
Detailed cross-checks using the same parameter/PDF setup resulted in good agreement being observed between the two calculations,
after a revision in one of them after the original publication. This comparison is included in these proceedings.


A key ingredient for any theoretical prediction at hadron colliders are parton distribution functions
(PDFs).
In the last two years, a great deal of LHC data (still mostly at 7 and 8 TeV) has been incorporated into global PDF fits,
supplementing, but not overwhelming, existing data from HERA, the Tevatron and from fixed target experiments. The LHC data
provide PDF information for, by definition, the kinematics encountered at the LHC. The availability of multiple processes allows
for cross-checks of the PDF constraints, as does the information from the two separate experiments.

There can be tensions between data taken by the two experiments, or even data taken within the same experiment that can reduce the
decrease in PDF uncertainty that might be expected given the precision of the data. 2020 marks the start of a benchmarking process by the PDF4LHC group that
will eventually lead to the creation of a PDF4LHC20 set of PDFs. There are two contributions relevant to this study in these proceedings, one
on a detailed study of top pair production and its impact on PDF fits, and one on the use of the $L_2$ sensitivity variable, to better
understand the level of data constraints on the PDFs and the size of the various tensions between data sets.
Understanding the source of these tensions will become even more important as the more copious 13 TeV data are included in the fits.




Studies related to gluon jets have played a key role in particle and nuclear physics since their discovery at PETRA exactly four decades prior to the 2019 Les Houches workshop.  We have used jet substructure techniques to investigate gluon fragmentation at the LHC, covering nearly four decades in energy scales.  Low energy scales involving gluon (sub)jets are studied from the point of view of hadronization and Monte Carlo tuning.  We find that small values of the groomed jet mass provide a sensitive handle for studying non-perturbative effects.  Higher-order effects in parton shower programs are investigated using deep learning, where subtle QCD corrections are identified using state-of-the-art neural networks.  One of the main investigations in the context of jet substructure was a study about the usefulness of a gluon jet differential cross section measurement for parton distribution functions.  We find that with a careful choice of observables and advances in jet substructure calculations, it may be possible to use jet substructure to constrain the high-$x$ gluon PDF.

Many of the interesting final states measured at the LHC involve one or more photons, for example $H\rightarrow \gamma \gamma$. An experimental measurement of photons requires the imposition of an isolation cut, in order to reduce the background from jets. Such isolation cuts also
reduce photon fragmentation contributions. The imposition of an isolation cut using an angular energy profile around the photon direction, a la Frixione, removes all of the fragmentation contribution, greatly simplifying the calculation. However, Frixione isolation is not well-adapted to
the LHC environment, where there is not only additional energy from the underlying event, but also energy deposited by pileup interactions. The
experimental preference is to require the energy in a cone (typically of radius 0.4 about the photon) to be less than a given amount, with no
requirement on its profile. To make matters perhaps more complicated, the pileup energy is subtracted before any isolation cut is implemented,
often leaving a net negative energy in the isolation cone. This mis-match of experimental and theoretical definitions of photon isolation can be
an additional uncertainty that becomes important as both theoretical and experimental precisions improve. Benchmark comparisons have been
performed as part of the Les Houches activities and are included as a contribution to these proceedings.



One of the pillars of the LHC program is the detailed study of the Higgs
boson. Understanding if the discovered particle has the properties as
predicted by the Standard Model, or if deviations from the Standard
Model predictions point to beyond-the-Standard-Model effects in the
Higgs sector, requires increased precision both in theoretical
predictions as well as experimental measurements.
The greater integrated luminosity for the LHC has allowed a better probe of high $p_T$ Higgs boson
production, using the boosted $H\rightarrow b\bar{b}$ mode as well as the $H\rightarrow\gamma\gamma$, $H\rightarrow WW$ and $H\rightarrow\tau\tau$ modes.
Any deviations at high $p_T$ can be cross-checked using these modes (and the two experiments). As mentioned above, the finite
top mass correction for the gluon-gluon process at NLO has been known for over two years. One topic explored in these proceedings
relates to the uncertainty caused by the top mass scheme (i.e. $\overline{\rm MS}$ or on-shell) used in the NLO finite top mass calculation.
The variation in the production cross section from the two schemes create a non-negligible addition to the theoretical uncertainty at
high $p_T$. The top mass scheme ambiguity affects other Higgs boson processes as well, such as off-shell Higgs boson production through
gluon-gluon fusion, off-shell Higgs boson decays into diphotons and Higgs boson pair production. This contribution also discusses the impact
of the top mass scheme on these processes.

In order to facilitate the Higgs high $p_T$ measurements, the studies of the Simplified Template Cross Section (STXS) framework were continued from LH15 and LH17 with a special focus on the definition of $p_T$ and jet bins for the $t\bar{t} H$ and $gg H$ production processes as well as bins that are sensitive to azimuthal angle correlations of the jets in the VBF production process.

A powerful and model-independent approach to parametrize Beyond the Standard Model effects that manifests at higher scales consists in considering the low energy effective field theory (EFT) that remains after integrating out the heavy fields of new physics. There are various implementations for the EFT approach in the Higgs sector which might differ in the basis used for the higher dimensional operators and the inclusion of loop induced effects in their approach. Here we  compare two different tools for the $pp \rightarrow t\bar{t} H$ and $pp \rightarrow Z H$  production processes as a way to clarify the situation for precise phenomenological analysis.

Another key process  is Higgs pair production, which provides the first  direct way to test the Higgs trilinear coupling. The dominant production mode of Higgs boson pairs at hadron colliders  is gluon fusion mediated by a top-quark loop, which considerably increases the difficulty in higher order calculations. Only full NLO accuracy has been reached for this observable, while N$^3$LO corrections were computed in the infinite top mass limit. In LH19 we combine the full NLO calculation with the approximated NNLO result applying a reweighting technique in order to provide NNLO improved results including EFT effects.

As a followup to LH17, where we carried out a comprehensive comparison of fixed order and ME+PS predictions for
gluon-gluon fusion Higgs boson production, dijet production and Z+jet production as a function of jet radius R, in LH19
we carry out a similar study for VBF Higgs boson production. The 2017 study resulted in a separate publication~\cite{Bellm:2019yyh}, as will
the 2019 study. In these proceedings, we will only be able to show preliminary results, and only for fixed order.

General-Purpose event generators remain a pillar of high-energy
physics phenomenology. As such, the quest for quantifying their
uncertainties continues. Parton shower Monte-Carlo (PSMC) programs are
an important aspect of event generation, not least because it is
likely that systematic uncertainties due to choices in their
construction can eventually be assessed rigorously, since their
construction is after all rooted in perturbative QCD factorization and
resummation. Thus, one major avenue of MC developments has been to
include as much perturbative information from fixed-order calculations
into event generators as possible, e.g.\ by including NLO matrix
element information for the production of additional jets in a process
using matching/merging techniques. Such calculations nominally
provide a better description of hard jets, but also come at the cost
of introducing some choices in the matching/merging procedure. Various
discussions at Les Houches were centered around understanding such
choices and their interplay with choices made in the PSMC algorithms.
Soft and collinear radiation is then still generated by the PSMC
programs, which currently operate at a leading-logarithmic accuracy.
We continued the quest towards a better understanding of PSMCs started
in Les Houches 2015 and continued in 2017, through phenomenological
studies as well as more theoretically-oriented ones.
We would like to note a useful development that was posted on the archive
as the proceedings were being finalized, related to the improvement of
parton showers beyond leading logarithmic accuracy~\cite{Dasgupta:2020fwr}.
This result, together with insights from earlier developments~\cite{Catani:1990rr}
should help the progress towards the ultimate goal of parton showers that can
achieve NLL accuracy for arbitrary observables.

One aspect that we assessed through a phenomenological study is the MC
modeling of the $gg \rightarrow ZH$ process, which is crucial for the
future experimental analyses targeting $VH$ final states, due to the
fact that NLO QCD corrections to this process with exact top mass
effects are not yet known. Here we focused on the comparison between
different LO+PS tools with respect to the improved merged MEPS
0,1-jets prediction available in {\tt Sherpa}. We confirmed that in
the high transverse momentum regime the inclusion of $2\rightarrow 3$
matrix elements is important, and we also show for the first time the
effects due to parton-shower and matching variations for this process.

At previous Les Houches workshops, studies of PSMCs were mostly
focused on the impact and definition of renormalization scale
uncertainties, or on uncertainties on modeling final-states in
lepton-lepton collisions. An uncertainty assessment at LHC is
complicated by the presence of parton distribution functions, which
lead to uncertainties due to their parametrization, as well as from
factorization scale variations. Before the latter can be addressed, it
is important to determine how well the initial-state evolution
produced by PSMCs recovers known DGLAP results. We start this
discussion here by performing a simple check of the self-consistency
of backward evolution in PSMCs: Certain products of PDFs and PSMC
no-emission probabilities should lead to Bjorken-$x$ independent
result; this fact can be explicitly checked. This highlights visible
deformations of the PDF evolution when using PSMCs alone for long
evolution spans without branching, and hints at the necessity of NLO
parton showers to be able to handle NLO PDF sets correctly. The
discussions in Les Houches have further triggered separate studies by
one group on the same topic~\cite{Nagy:2020gjv}.

Another aspect that was dicussed in Les Houches was the long-standing
problem of finding a procedure to quantify the theoretical uncertainty
due to PSMCs. The original goal that was considered to be relevant
both from the theoretical and experimental side was to perform a study
similar to the one completed in LH15, but using NLO+PS-accurate tools
and frameworks recently developed to perform ``PS-reweighting''
efficiently. The aim was to establish if, for the main ``variations''
of perturbative nature available in different PS algorithms, the
results obtained with different generators are mutually compatible or
not. Although we didn't manage to perform such an ambitious study, in
one contribution we have looked at several sources of perturbative and
algorithmical uncertainties in a NLO+PS simulation of top-pair
production in hadronic collisions. We have found that, in most cases,
each considered variation has the expected impact on differential
distributions, and pointed out some aspects that would deserve more
detailed studies.

Several other topics related to improving the efficiency of event
generators were discussed at Les Houches. In particular, with
increased need for precision calculations but flat/reduced computing
resources, it becomes pressing to try everything to avoid widely
varying event weights or negatively weighted contributions. This is
especially relevant when put into the context of the large resource
demand from detector simulation -- ``wasting'' resources by processing
large event samples with a small total statistical power should be
avoided. The latter situation can arise if samples have an appreciable
fraction of negatively weighted events, as is often the case for
bleeding-edge precision calculations. A general solution is of course
extremely challenging, as potential strategies may depend very much on
the details of the MC producing the event sample. Thus, new MC
developments are first necessary. One such development effort,
tentatively coined ``Posterior importance sampling of MC events'' has
been kick-started in Les Houches.
Here, the idea is that during event generation, all events \emph{and} the MC
weight distribution are kept. Then, only a \emph{statistically equivalent
subset} of events, chosen using the multi-dimensional weight
distributions, are passed on to detector simulation.
This study will appear as a separate
publication; the authors gratefully acknowledge the
stimulating atmosphere at Les Houches.

Finally, another topic which does not appear in these proceedings, but on which
there were several dedicated discussions and preliminary studies
during the workshop, is vetoing the central hadronic activity in
Vector Boson Fusion and Vector Boson Scattering final states. A lot of
emphasis was put in particular on the results for the veto efficiency
reported by CMS in the measurement of electroweak production of a W
associated with two jets~\cite{Sirunyan:2019dyi}, where data seems to
prefer one particular PSMC. It was pointed out that it would be
envisageable to further exploit these results by comparing them to
modified versions of PSMCs to pin down the origin of the observed
differences. A limitation of this measurement however comes from the
usage of a Boosted Decision Tree to define the signal region, which
prevents performing the same selection on stable-particles in generated
events, prior to the simulation of their interaction with the
detector. In the past months a lot of work has been done by several
CMS members to tackle this issue, without reaching yet conclusive
results, but with some hope that this can be done in the near
future. In any case, whether this is achieved or not, the
experimentalists are thankful for the Les Houches discussions with
their fellow theorists, which renovated the interest in these results
and brought up the issue of unfolding measurements based on machine
learning techniques.


\newpage

\chapter{NLO automation and (N)NLO techniques}
\label{cha:nnlo}

\newcommand{\NLLgen}[1]{N${}^{#1}$LL\xspace}

\newcommand{\NLL}[1]{N${}^{#1}$LL\xspace}
\newcommand{\NLLone}{NLL\xspace}

\newcommand{\LL}{LL\xspace}
\newcommand{\NNLL}{NNLL\xspace}
\newcommand{\NNLLp}{NNLL'\xspace}
\newcommand{\NNNLL}{\NLLgen3\xspace}

\newcommand{\NLO}[1]{N${}^{#1}$LO\xspace}
\newcommand{\NLOone}{NLO\xspace}
\newcommand{\NLOgen}{NLO\xspace}
\newcommand{\NNLOgen}{NNLO\xspace}
\newcommand{\NNNLOgen}{NNNLO\xspace}

\newcommand{\NLOH}[1]{N${}^{#1}$LO${}_{\rm HTL}$\xspace}
\newcommand{\NLOHone}{NLO${}_{\rm HTL}$\xspace}
\newcommand{\NLOHTL}{NLO${}_{\rm HTL}$\xspace}
\newcommand{\NNLOHTL}{NNLO${}_{\rm HTL}$\xspace}
\newcommand{\NNNLOHTL}{\NLOH3}

\newcommand{\NLOQ}[1]{N${}^{#1}$LO${}_{\rm QCD}$\xspace}
\newcommand{\LOQ}{LO${}_{\rm QCD}$\xspace}
\newcommand{\NLOQone}{NLO${}_{\rm QCD}$\xspace}
\newcommand{\LOQCD}{LO${}_{\rm QCD}$\xspace}
\newcommand{\NLOQCD}{NLO${}_{\rm QCD}$\xspace}
\newcommand{\NNLOQCD}{NNLO${}_{\rm QCD}$\xspace}
\newcommand{\NNNLOQCD}{\NLOQ3}

\newcommand{\NLOE}[1]{N${}^{#1}$LO${}_{\rm EW}$\xspace}
\newcommand{\NLOEone}{NLO${}_{\rm EW}$\xspace}
\newcommand{\LOEW}{LO${}_{\rm EW}$\xspace}
\newcommand{\NLOEW}{NLO${}_{\rm EW}$\xspace}
\newcommand{\NNLOEW}{NNLO${}_{\rm EW}$\xspace}

\newcommand{\NLOD}[1]{N${}^{#1}$LO${}_{\rm QED}$\xspace}
\newcommand{\NLODone}{NLO${}_{\rm QED}$\xspace}
\newcommand{\NLOQED}{NLO${}_{\rm QED}$\xspace}
\newcommand{\NNLOQED}{NNLO${}_{\rm QED}$\xspace}

\newcommand{\NLOSM}{NLO${}_{\rm SM}$\xspace}

\newcommand{\NLOQE}[2]{N${}^{(#1,#2)}$LO${}_{{\rm QCD}\otimes{\rm EW}}$\xspace}

\newcommand{\NLOHE}[2]{N${}^{(#1,#2)}$LO${}^{\rm (HTL)}_{{\rm QCD}\otimes{\rm EW}}$\xspace}

\newcommand{\NLOmixQED}[2]{N${}^{(#1,#2)}$LO${}_{{\rm QCD}\otimes{\rm QED}}$\xspace}

\newcommand{\NLOQmtsix}[1]{N${}^{#1}$LO${}_{\rm QCD}^{(1/{m_t^8})}$\xspace}
\newcommand{\NLOQzzero}[1]{N${}^{#1}$LO${}_{\rm QCD}^{(z\to0)}$\xspace}
\newcommand{\NLOQVBF}[1]{N${}^{#1}$LO${}_{\rm QCD}^{(\rm VBF)}$\xspace}
\newcommand{\NLOQoneVBF}{NLO${}_{\rm QCD}^{(\rm VBF)}$\xspace}
\newcommand{\NLOQCDVBF}{NLO${}_{\rm QCD}^{(\rm VBF)}$\xspace}
\newcommand{\NNLOQCDVBF}{NNLO${}_{\rm QCD}^{(\rm VBF)}$\xspace}
\newcommand{\NLOQoneDIS}{NLO${}_{\rm QCD}^{(\rm DIS)}$\xspace}
\newcommand{\NLOQDIS}[1]{N${}^{#1}$LO${}_{\rm QCD}^{(\rm DIS)}$\xspace}
\newcommand{\NLOEoneVBF}{NLO${}_{\rm EW}^{(\rm VBF)}$\xspace}
\newcommand{\NLOEWVBF}{NLO${}_{\rm EW}^{(\rm VBF)}$\xspace}

\newcommand{\NLOQoneVBFstar}{NLO${}_{\rm QCD}^{(\rm VBF^{*})}$\xspace}
\newcommand{\NLOQVBFstar}[1]{N${}^{#1}$LO${}_{\rm QCD}^{(\rm VBF^{*})}$\xspace}
\newcommand{\NLOQCDVBFstar}{NLO${}_{\rm QCD}^{(\rm VBF^{*})}$\xspace}
\newcommand{\NNLOQCDVBFstar}{NNLO${}_{\rm QCD}^{(\rm VBF^{*})}$\xspace}
\newcommand{\NNNLOQCDVBFstar}{\NLOQVBFstar3}

\newcommand{\NLOEoneVBFstar}{NLO${}_{\rm EW}^{(\rm VBF^{*})}$\xspace}
\newcommand{\NLOggHVtb}[1]{N${}^{#1}$LO${}_{gg\to HZ}^{(t,b)}$\xspace}

\newcommand{\xs}{$\sigma$}
\newcommand{\tb}{\bar{t}}
\newcommand{\bb}{\bar{b}}
\newcommand{\qb}{\bar{q}}

\newcommand{\wodecay}{(w/o decay)}
\newcommand{\wdecay}{}
\newcommand{\wodecays}{(w/o decays)}
\newcommand{\wdecays}{}
\newcommand{\wleptdecays}{}

\newcommand{\MadgraphaMCatNLO}{\textsc{Madgraph5}\_a\textsc{MC@NLO}\xspace}
\newcommand{\OpenLoops}{O\protect\scalebox{0.8}{PENLOOPS}\xspace}
\newcommand{\Recola}{R\protect\scalebox{0.8}{ECOLA}\xspace}
\newcommand{\GoSam}{G\protect\scalebox{0.8}{O}S\protect\scalebox{0.8}{AM}\xspace}
\newcommand{\MadLoop}{M\protect\scalebox{0.8}{AD}L\protect\scalebox{0.8}{OOP}\xspace}
\newcommand{\Powheg}{P\protect\scalebox{0.8}{OWHEG}\xspace}
\newcommand{\Powhegboxres}{P\protect\scalebox{0.8}{OWHEG-BOX-RES}\xspace}
\newcommand{\PowhegboxVtwo}{P\protect\scalebox{0.8}{OWHEG-BOX-V2}\xspace}
\newcommand{\Herwig}{H\protect\scalebox{0.8}{ERWIG}\xspace}
\newcommand{\Matrix}{M\protect\scalebox{0.8}{ATRIX}\xspace}
\newcommand{\Munich}{M\protect\scalebox{0.8}{UNICH}\xspace}
\newcommand{\Geneva}{G\protect\scalebox{0.8}{ENEVA}\xspace}
\newcommand{\Sherpa}{S\protect\scalebox{0.8}{HERPA}\xspace}
\newcommand{\NNLOjet}{NNLO\protect\scalebox{0.8}{JET}\xspace}
\newcommand{\MiNLO}{M\protect\scalebox{0.8}{iNLO}\xspace}
\newcommand{\NLOX}{NLOX\xspace}

\section{Update on the precision Standard Model wish list~\protect\footnote{
  A.~Huss,
  J.~Huston,
  S.~Jones,
  S.~Kallweit}{}}
\label{sec:SM_wishlist}

Identifying key observables and processes that require improved theoretical input has been
a key part of the Les Houches programme. In this contribution we briefly summarise progress since the previous
report in 2017 and explore the possibilities for further advancements.
We also provide an estimate of the experimental uncertainties for a few key processes. A summary of this sort is perhaps unique in the field and serves a useful purpose for both practitioners in the field and for other interested readers. Given the amount of work that has been, and is being, done, this summary will no doubt be incomplete, and we apologize for any omissions.%
\footnote{The Les Houches Disclaimer}

\subsection{Introduction}

While the years before the Les Houches 2017 report~\cite{Bendavid:2018nar} had been marked
by significant progress in the production of \NNLOgen results in an almost industrial manner
with most useful $2\rightarrow2$ processes having been calculated,
the last two years have seen kind of a saturation due to the unavailability of 2-loop amplitudes
beyond $2\rightarrow2$ scattering.
However, a remarkable progress was achieved in this direction by several groups and approaches,
cumulating
in a first $2\rightarrow3$ calculation of a hadron collider process (triphoton production) that
has been completed very recently~\cite{Chawdhry:2019bji}.
Closely related is the huge progress in the calculation of 2-loop amplitudes for 3-jet production,
but also 2-loop amplitudes for $2\to2$ processes with internal masses have seen impressive
developments.

However, not only the amplitude community has seen impressive development. There have also
been significant steps forward on the side of subtraction schemes, and there are in the meanwhile
several subtraction and slicing methods available to deal (in principle) with higher-multiplicity
processes at \NNLOgen (see below).

On the parton shower side,
NLO QCD matched results and matrix element improved multi-jet merging techniques have become a standard
level of theoretical precision. The automation of full SM corrections including NLO
electroweak predictions has also seen major improvements.

Another challenge is to make the \NNLOgen $2\to2$ predictions or complex NLO predictions publicly
available to experimental analyses, and there has been major progress to achieve this goal.
{\sc Root nTuples} have been a useful tool for complicated final states at NLO and
allow for very flexible re-weighting and analysis. The cost for this is
the large disk space required to store the event information.
A feasibility study about using {\sc Root nTuples} for the Drell--Yan process at
\NNLOgen is described in Sec.~\ref{cha:nnlo}.\ref{sec:SM_nnlontuples}.

An extension of APPLgrid~\cite{Carli:2010rw} and fastNLO~\cite{Kluge:2006xs} offers a simpler, but less flexible method to
distribute higher order predictions. The latter option is likely to be used heavily in
precision PDF fits, and new developments in the APPLfast project are described in Sec.~\ref{cha:nnlo}.\ref{sec:SM_applfastLHCgrids}.

\subsection{Developments in theoretical methods}

Precision predictions require a long chain of various tools and methods, all
of which demand highly technical computations.

Computational methods for the amplitude level ingredients have seen substantial
progress in the last few years.  Scattering amplitudes at $L$ loops are
generally decomposed into a basis of integrals together with rational coefficients,
\begin{equation}
  A^{(L)}_{2\to n} = \sum_i (\text{coefficients})_i (\text{integrals})_i\,.
\end{equation}
One must then remove infrared singularities to obtain a finite cross section,
\begin{equation}
  d\sigma_{2\to n} \text{\NLO{k}} = {\rm IR}_k(A^{k}_{2\to n}, A^{k-1}_{2\to n+1},\cdots, A^{0}_{2\to n+k})\,,
\end{equation}
where the function ${\rm IR}_k$ represents an infrared subtraction technique.
Ultraviolet renormalisation must also be performed but in a (semi-)analytic approach
presents no technical difficulties.
There are also fully numerical approaches, aiming to calculate higher order corrections without the
separation into individually divergent components, such that 4-dimensional methods can be applied.

\subsubsection{Loop integrals}

Most of the new analytic results for two-loop integrals and beyond have been calculated employing the differential equations technique~\cite{Kotikov:1990kg,Gehrmann:1999as},
which got a significant boost through Henn's canonical form~\cite{Henn:2013pwa}.
In the last few years, several tools to find a canonical basis automatically have been developed: {\sc epsilon}~\cite{Prausa:2017ltv}, {\sc Fuchsia}~\cite{Gituliar:2017vzm}, {\sc Canonica}~\cite{Meyer:2017joq}, and {\sc DlogBasis}~\cite{Henn:2020lye}.
Important new developments concerning the differential equations technique to calculate multi-loop integrals can be found in  Refs.~\cite{Primo:2016ebd,Frellesvig:2017aai,Zeng:2017ipr,Bosma:2017ens,Primo:2017ipr,Harley:2017qut,Lee:2017qql,Bosma:2017hrk,Francesco:2019yqt,Dlapa:2020cwj}.
For a review on the method of differential equations we refer to Ref.~\cite{Argeri:2007up,Henn:2014qga}.

Major progress has been made in the calculation of two-loop master integrals with massive propagators.
For example, analytic results are now known for all integrals entering Higgs+jet~\cite{Bonciani:2016qxi,Primo:2016ebd,Bonciani:2019jyb,Francesco:2019yqt,Frellesvig:2019byn}.
Significant progress has been made analytically computing $gg\to \gamma\gamma$~\cite{Becchetti:2017abb} and $gg \to ZZ$~\cite{Grober:2019kuf,Davies:2020lpf} via massive top quark loops,
$HH$~\cite{Grober:2017uho,Davies:2018ood,Bonciani:2018omm,Xu:2018eos,Davies:2018qvx,Mishima:2018olh},
the mixed QCD--EW corrections to the Drell--Yan process~\cite{Bonciani:2016ypc,vonManteuffel:2017myy,Heller:2019gkq,Frellesvig:2020vii},
electron-muon scattering (with $m_e=0,m_{\mu}\not =0$)~\cite{Mastrolia:2017pfy,DiVita:2018nnh,Lee:2019lno},
and three-loop corrections to the heavy flavour Wilson coefficients in DIS with two different masses~\cite{Ablinger:2017err,Ablinger:2017xml,Ablinger:2018brx,Ablinger:2019gpu}.
The integrals entering top quark pair production at \NNLOgen~\cite{Czakon:2013goa}
were calculated numerically some time ago~\cite{Baernreuther:2013caa}.
Complete analytic results for integrals entering $q\bar{q} \to t \bar{t}$ have now been obtained~\cite{%
Bonciani:2013ywa,
Abelof:2015lna,
Adams:2018bsn,
Adams:2018kez,
Chen:2019zoy,
Becchetti:2019tjy,
DiVita:2019lpl%
}.

A major complication appearing for integrals beyond 1-loop (especially those with massive propagators) is related to the fact that
the basis for an analytic representation of such integrals may go beyond the function class of generalized polylogarithms (GPLs),
i.e.  integrals of elliptic type occur. The latter have been subject of intense studies recently, see \eg~Refs.~\cite{%
MullerStach:2011ru,
Remiddi:2013joa,
Adams:2013kgc,
Bloch:2013tra,
Adams:2014vja,
Adams:2015gva,
Adams:2015ydq,
Bloch:2016izu,
Remiddi:2016gno,
Broedel:2017siw,
Broedel:2017kkb,
vonManteuffel:2017hms,
Bogner:2017vim,
Broedel:2018iwv,
Broedel:2018qkq,
Blumlein:2018jgc,
Broedel:2019hyg,
Broedel:2019kmn,
Bogner:2019lfa%
}.

For integrals which do not leave the class of GPLs, improvements in the understanding of the basis of multiple polylogarithms through symbol calculus
and Hopf algebras (see \eg~\cite{Duhr:2014woa,Abreu:2017enx}) has led to a high degree of automation for these integral computations.
This is a necessary step in order to apply such techniques to phenomenologically relevant cases, most
notably \eg~of $pp\to H$~\cite{Dulat:2017prg,Mistlberger:2018etf,Dulat:2018rbf} and Drell--Yan~\cite{Duhr:2020seh} at \NLO3.
A new public package for the manipulation of multiple polylogarithms, {\sc PolyLogTools}, was presented in Ref.~\cite{Duhr:2019tlz}.

At the multi-loop front, remarkable recent achievements include the complete four-loop (and part of the five-loop) contributions to the cusp anomalous dimension and the progress towards N$^3$LO splitting functions~\cite{%
vonManteuffel:2015gxa,
Davies:2016jie,
vonManteuffel:2016xki,
Lee:2016ixa,
Ruijl:2017eht,
Lee:2017mip,
Moch:2017uml,
Moch:2018wjh,
Herzog:2018kwj,
Henn:2019rmi,
Bruser:2019auj,
vonManteuffel:2019wbj,
vonManteuffel:2019gpr,
Catani:2019rvy,
Henn:2019swt,
vonManteuffel:2020vjv%
}.
In recent years, the five-loop QCD beta-function~\cite{Baikov:2016tgj,Herzog:2017ohr,Luthe:2017ttg,Chetyrkin:2017bjc} and Higgs decays to hadrons and the R-ratio at N$^{4}$LO~\cite{Herzog:2017dtz} have also been calculated.

There have also been developments in the direct evaluation of Feynman integrals with fewer scales, but higher loops.
The {\sc HyperInt}~\cite{Panzer:2014caa} and {\sc mpl}~\cite{Bogner:2015nda} packages
have focused mainly on zero and one scale integrals with a high number of loops,
but the algorithms employed have potential applications to a wider class of integrals.
Another newly developed tool is {\sc Dream}~\cite{Lee:2017ftw}, a program for the computation of multiloop integrals within the {\sc dra} (Dimensional Recurrence \& Analyticity) method.
A method to systematically approximate multi-scale integrals based on Taylor expanding in Feynman parameter space was presented in Ref.~\cite{Borowka:2018dsa}.

In order to facilitate the search for analytic results for multi-loop integrals in the literature, a database
Loopedia~\cite{Bogner:2017xhp} has been created. At \url{https://loopedia.mpp.mpg.de} results for integrals can be searched for by topology.
The webpage also allows to upload results for newly calculated integrals and literature information.

Direct numerical evaluation remains a powerful technique.
It is an especially promising strategy for tackling multi-loop integrals with a rather large number of kinematic scales.
Since the last workshop, the first calculation of $gg \to \gamma \gamma$ with full top quark mass dependence was completed~\cite{Maltoni:2018zvp,Mandal:2018cdj}, using the numerical solution of the differential equations~\cite{Caffo:1998du,Boughezal:2007ny,Czakon:2007qi,Czakon:2008zk}.
The sector decomposition algorithm~\cite{Binoth:2000ps} has seen a number of optimisations, implemented into the publicly available
updates of the codes {\sc (py)SecDec}~\cite{Borowka:2015mxa,Borowka:2017idc,Borowka:2018goh} and {\sc Fiesta}~\cite{Smirnov:2015mct}.
By numerically integrating in Feynman parameter space, computations of $gg \to \gamma \gamma$~\cite{Chen:2019fla}, $pp\to HH$~\cite{Borowka:2016ehy,Borowka:2016ypz,Baglio:2018lrj}
and $pp\to H+$\,jet~\cite{Jones:2018hbb} at NLO including the full top quark mass dependence have been completed.
Within the framework of Loop--Tree Duality, significant progress has been made in the numerical evaluation of loop integrals directly in momentum space ~\cite{Capatti:2019edf,Capatti:2019ypt,Runkel:2019zbm,Runkel:2019yrs}.
This work benefits directly from recent advances made in methods to remove infrared divergences from two-loop amplitudes~\cite{Anastasiou:2018rib}.

\subsubsection{Loop integral reduction}

Many recent calculations rely on the use of integration-by-parts (IBP) reduction identities~\cite{Tkachov:1981wb,Chetyrkin:1981qh,Laporta:2001dd} (see \eg~\cite{Zhang:2016kfo,Grozin:2011mt} for a review).
Several efficient codes exist to facilitate their use, including:
{\sc Air}~\cite{Anastasiou:2004vj}, 
{\sc Fire}~\cite{Smirnov:2008iw,Smirnov:2013dia,Smirnov:2014hma,Smirnov:2019qkx}, 
{\sc LiteRed}~\cite{Lee:2012cn,Lee:2013mka},
{\sc Reduze}~\cite{Studerus:2009ye,vonManteuffel:2012np}
and
{\sc Kira}~\cite{Maierhoefer:2017hyi,Maierhofer:2018gpa},
many of which have been significantly developed since the last workshop.
For example, new ideas have been explored which can greatly improve multivariate functional reconstruction in the context of amplitude reduction~\cite{vonManteuffel:2014ixa,Peraro:2016wsq}.
Public tools implementing these techniques include {\sc FireFly}~\cite{Klappert:2019emp} and {\sc FiniteFlow}~\cite{Peraro:2019svx}.
Specialist programs, such as {\sc Forcer}, for the reduction of four-loop massless propagator diagrams~\cite{Ruijl:2017cxj}, have also played a key role especially at high-loop order.

The use of IBP identities usually requires solving a system of linear equations, this involves algebraic manipulations which can become computationally demanding.
Recently, a promising alternative approach to finding relations between Feynman integrals using intersection numbers has been explored~\cite{Weinzierl:2020xyy,Mizera:2019vvs,Abreu:2019wzk,Frellesvig:2019uqt,Frellesvig:2019kgj,Mastrolia:2018uzb,Mizera:2017rqa}.

\subsubsection{Generalised unitarity, integrand reduction and amplitudes}
\label{sec:SM_wishlist:loop}

Extending the current multi-loop methods to higher multiplicity still represents
a serious challenge. The increased complexity in the kinematics, and large amount
of gauge redundancy in the traditional Feynman diagram approach, at one-loop has been solved numerically through on-shell
and recursive off-shell methods. This breakthrough has led to the development
of the now commonly used automated one-loop codes~\cite{Berger:2008sj,Bevilacqua:2011xh,Cascioli:2011va,Buccioni:2019sur,Badger:2012pg,Cullen:2011ac,Cullen:2014yla,Alwall:2014hca,Frederix:2018nkq,Actis:2012qn,Actis:2016mpe,Denner:2017wsf}.

In the context of one-loop amplitudes, a method to extract analytic amplitudes from high-precision floating point numerical evaluations has been presented in Ref.~\cite{DeLaurentis:2019phz}.
The use of neural networks to efficiently evaluate high multiplicity amplitudes was recently studied in Ref.~\cite{Badger:2020uow}.

The $D$-dimensional generalised unitarity cuts algorithm~\cite{Bern:1994zx,Bern:1994cg,Britto:2004nc,Giele:2008ve,Forde:2007mi} has
been extended to multi-loop integrands using integrand reduction~\cite{Ossola:2006us,Ellis:2008ir}\footnote{We do not attempt a complete review of integrand reduction here. Further information can be found in the review article \cite{Ellis:2011cr} and references therein.}
and elements of computational algebraic geometry~\cite{Mastrolia:2011pr,Badger:2012dp,Zhang:2012ce,Kleiss:2012yv,Feng:2012bm,Mastrolia:2012an,Mastrolia:2013kca,Badger:2013gxa,Mastrolia:2016dhn}.
In contrast to the one-loop case, the basis of integrals obtained through this
method is not currently known analytically and is much larger than the set of
basis functions defined by standard integration-by-parts identities.
The maximal unitarity method~\cite{Kosower:2011ty}, which incorporates IBP identities, has been applied to a
variety of two-loop examples in four dimensions~\cite{Larsen:2012sx,CaronHuot:2012ab,Johansson:2012zv,Johansson:2013sda,Johansson:2015ava}.
Efficient algorithms to generate unitarity compatible IBP identities are a key ingredient
in both approaches and have been the focus of on-going investigations~\cite{Gluza:2010ws,Schabinger:2011dz,Ita:2015tya,Larsen:2015ped,Hirschi:2016mdz,Mastrolia:2016dhn}.
Automated tools for IBP reductions based on algebraic geometry have also been developed, see \eg~{\sc Cristal} and {\sc Azurite}~\cite{Georgoudis:2016wff,Larsen:2017aqb,Georgoudis:2017iza} and Ref.~\cite{Bendle:2019csk}.

Tremendous progress has been made in the computation of 2-loop 5-point amplitudes, as for instance needed for the 3-jet process at \NNLOQCD.
In the case of the 2-loop 5-gluon amplitudes, computations based on numerical unitarity~\cite{Badger:2015lda,Abreu:2017hqn,Badger:2017jhb,Boehm:2017wjc,Badger:2018enw,Badger:2018gip,Chicherin:2018yne,Abreu:2018jgq} resulted in the analytic expression at leading-colour for all helicity configurations~\cite{Abreu:2018zmy} and the full-colour result in the case of the all-plus helicity configuration~\cite{Badger:2019djh}.
The result for 2-loop amplitudes relevant for the scattering of 5 massless partons were obtained at leading colour in Ref.~\cite{Abreu:2019odu}.
The relevant master integrals are known and have been computed in Refs.~\cite{Chicherin:2018old,Chicherin:2018mue,Gehrmann:2018yef}.
Progress has also been made in evaluating 2-loop 5-point amplitudes with one off-shell leg~\cite{Hartanto:2019uvl}, and in computing the associated integrals~\cite{Papadopoulos:2019iam}.
IBP reductions relevant for massless 2-loop 5-point amplitudes have been completed in Ref.~\cite{Chawdhry:2018awn,Guan:2019bcx}.
Very recently, the full-colour 2-loop 6-gluon all-plus helicity amplitude was obtained~\cite{Dalgleish:2020mof}.

An alternative, perhaps more traditional, method for computing multi-loop scattering amplitudes consists of identifying Lorentz-invariant form factors which can be extracted from Feynman diagrams
using projector operators.
Although this strategy has been very successful for computing amplitudes with up to four external particles, the complexity of directly deriving suitable projection operators can become prohibitive at higher multiplicity.
Recent work has investigated how projection operators for physical helicity amplitudes can be much more efficiently derived~\cite{Chen:2019wyb,Peraro:2019cjj}.

\subsubsection{Infrared subtraction methods for differential cross sections}

The construction of fully differential \NNLOQCD cross sections for $2\to2$
processes has been a major theoretical challenge over the past years.
This programme has seen remarkable progress with many different approaches now
applied to LHC processes. We give a brief characterisation of the main methods below,
as well as some of their LHC applications.
The next challenges include progress towards $2\to3$ processes at \NNLOQCD as well as extensions towards fully differential \NNNLOQCD predictions.

\begin{itemize}
\item Antenna subtraction~\cite{GehrmannDeRidder:2005cm,Currie:2013vh}:\\
  Analytically integrated counter-terms, applicable to hadronic initial and final states. Almost
  completely local, requires averaging over azimuthal angles. Applied
  to $e^+e^-\to 3j$~\cite{Ridder:2014wza,Gehrmann:2017xfb}, (di-)jets in DIS~\cite{Currie:2017tpe,Niehues:2018was}, $pp\to j+X$~\cite{Currie:2016bfm}, $pp\to 2j$~\cite{Currie:2017eqf}, $pp\to \gamma+j/X$~\cite{Chen:2019zmr},
  $pp\to Z+j$~\cite{Ridder:2015dxa,Ridder:2016nkl}, $pp\to W+j$~\cite{Gehrmann-DeRidder:2017mvr}, $pp\to H+j$~\cite{Chen:2016zka}, $pp\to VH$~\cite{Gauld:2019yng}, and Higgs production in VBF~\cite{Cruz-Martinez:2018rod}.

\item Sector Improved Residue Subtraction~\cite{Czakon:2010td,Czakon:2011ve,Boughezal:2011jf}:\\
  Fully local counter-terms, based on a sector decomposition~\cite{Binoth:2000ps} approach
  for IR divergent real radiation~\cite{Heinrich:2002rc,Anastasiou:2003gr,Binoth:2004jv} and an extension of the FKS approach at NLO \cite{Frixione:1995ms,Frederix:2009yq}.
  Numerically integrated counter-terms, capable of treating hadronic initial and final states.
  Improvements through a four-dimensional formulation~\cite{Czakon:2014oma}.
  Applied to
  top-quark processes~\cite{Czakon:2013goa,Czakon:2014xsa,Czakon:2015owf,Czakon:2016ckf,Brucherseifer:2013iv,Brucherseifer:2014ama}, to $pp\to H+j$~\cite{Boughezal:2015dra,Caola:2015wna}, inclusive jet production~\cite{Czakon:2019tmo}, and $pp\to3\gamma$~\cite{Chawdhry:2019bji}.

\item $q_T$ \cite{Catani:2007vq}:\\
Phase-space slicing approach for colourless final states, applied to
$H$~\cite{Catani:2007vq,Grazzini:2008tf},
$V$~\cite{Catani:2009sm,Catani:2010en}
and $VV'$ production processes~\cite{Catani:2011qz,Grazzini:2013bna,Gehrmann:2014fva,Cascioli:2014yka,Grazzini:2015nwa,Grazzini:2015hta,Grazzini:2016swo,Grazzini:2016ctr,Grazzini:2017ckn,Catani:2018krb,Kallweit:2018nyv}.
All those processes available in \Matrix~\cite{Grazzini:2017mhc}.
Also applied to obtain \NNLOQCD differential results for $VH$~\cite{Ferrera:2011bk,Ferrera:2013yga,Ferrera:2014lca}
and $HH$~\cite{deFlorian:2016uhr,Grazzini:2018bsd},
as well as for $WHH$~\cite{Li:2016nrr}
and $ZHH$~\cite{Li:2017lbf}.
An extension for $t\tb$ final states has been
proposed \cite{Bonciani:2015sha}, and was realized in Ref.~\cite{Catani:2019iny,Catani:2019hip}.
The \NNLOQCD soft function for $t\tb$ production was independently calculated in
Ref.~\cite{Angeles-Martinez:2018mqh}.
Based on these developments, an extension of the $q_T$ subtraction method towards electroweak
corrections for massive lepton pairs was discussed in Ref.~\cite{Buonocore:2019puv}.
Much progress has also been made in an extension to \NNNLOQCD~\cite{Luo:2019szz} with a first application in Higgs production~\cite{Cieri:2018oms}.

\item $N$-jettiness \cite{Boughezal:2015eha,Boughezal:2015dva,Gaunt:2015pea}:\\
Extension of the $q_T$ method to final states including a jet,
matching to soft-collinear effective theory (SCET)
below the $N$-jettiness cut-off parameter. Applied to $2\to 2$
processes containing vector bosons or a boson plus one jet in the final
state~\cite{Boughezal:2015dva,Boughezal:2015aha,Boughezal:2015ded,Boughezal:2016isb,Boughezal:2016yfp,Boughezal:2016dtm,Campbell:2016jau,Campbell:2016lzl,Campbell:2017aul};
for colourless final states see also MCFM version\,8~\cite{Boughezal:2016wmq}
and version\,9~\cite{Campbell:2019dru}.
Similar techniques also applied to top decay~\cite{Gao:2012ja} and $t$-channel single top production~\cite{Berger:2016oht}.
Important steps towards applicability for \NNNLOQCD calculations have been presented
in Refs.~\cite{Melnikov:2018jxb,Melnikov:2019pdm,Behring:2019quf,Billis:2019vxg}.

\item ColorFull \cite{DelDuca:2015zqa}:\\
Fully local counter-terms extending the Catani--Seymour dipole method~\cite{Catani:1996vz}.
Analytically integrated for infrared poles, numerical integration for finite parts.
Currently developed for hadronic final states such as $H\to b\bb$
\cite{DelDuca:2015zqa} and $e^+e^-\to$ 3 jets~\cite{DelDuca:2016csb,DelDuca:2016ily,Tulipant:2017ybb}.

\item Nested Soft-Collinear Subtraction~\cite{Caola:2017dug}:\\
Initial proposal featured fully local subtraction terms, partially numerical cancellation of IR poles, and allowed matrix elements to be evaluated in four dimensions. Subsequently, full analytic results for subtraction counterterms have been calculated~\cite{Caola:2018pxp,Delto:2019asp}, allowing full analytic cancellation of IR poles. All required building blocks required for the application to computations of \NNLOQCD corrections to arbitrary processes at hadron colliders have been subsequently achieved~\cite{Caola:2019nzf, Caola:2019pfz,Asteriadis:2019dte}.

\item Analytic local sector subtraction~\cite{Magnea:2018hab, Magnea:2018ebr}:\\
Local subtraction, aiming at the minimal counterterm structure arising from a sector partition of the radiation phase space. Analytic integration of the counterterms. Proof of
principle example from $e^+e^-\to 2$\,jets~\cite{Magnea:2018hab}.

\item Projection to Born~\cite{Cacciari:2015jma}:\\
Range of applicability limited (it requires the knowledge of inclusive corrections), however, generalisation to higher orders more straightforward once individual ingredients available.
Applied to VBF Higgs~\cite{Cacciari:2015jma} and Higgs-pair~\cite{Dreyer:2018rfu} production, and $t$-channel single top production~\cite{Berger:2016oht} at \NNLOQCD.
Fully differential predictions at N${}^3$LO using this method were obtained for jet production in DIS~\cite{Currie:2018fgr,Gehrmann:2018odt} and $H\to b\bar{b}$~\cite{Mondini:2019gid}.
\end{itemize}


\subsection{The precision wish list}
\label{sec:SM_wishlist:precision_wish_list}
We break the list of precision observables into
four sections: Higgs,  jets, vector bosons and top quarks.

Corrections are defined with respect to the leading order, and we organise
the perturbative expansion into QCD corrections, electroweak (EW) corrections and
mixed QCD$\otimes$EW,
\begin{equation}
  d\sigma_X = d\sigma_X^{\rm LO} \left(1 +
      \sum_{k=1} \alpha_s^k d\sigma_X^{\delta \text{\NLOQ{k}}}
    + \sum_{k=1} \alpha^k d\sigma_X^{\delta \text{\NLOE{k}}}
    + \sum_{k,l=1} \alpha_s^k \alpha^l d\sigma_X^{\delta \text{\NLOQE{k}{l}}}
    \right).
  \label{eq:SM_wishlist:dsigmapertexp}
\end{equation}
We explicitly separate the mixed QCD and EW corrections to distinguish between additive predictions
QCD+EW and mixed predictions QCD$\otimes$EW. The definition above only applies in the case where the leading order
process contains a unique power in each coupling constant. For example, in the case of $q\qb\to q\qb Z$ two
leading order processes exist: via gluon exchange of $\mathcal{O}(\alpha_s^2\alpha)$, via electroweak boson
exchange of $\mathcal{O}(\alpha^3)$ and the interference $\mathcal{O}(\alpha_s\alpha^2)$.
In these cases it is customary to classify the Born process with highest power in
$\alpha_s$ (and typically the largest cross section) as the leading order, and
label the others as subleading Born processes. The above classification is then
understood with respect to the leading Born process, unless otherwise stated.
We will also use the notation \NLOSM for an NLO calculation that includes the complete
Standard Model, \ie QCD and EW, corrections to the full set of LO processes.

In the following we attempt to give a current snapshot of the
available  calculations of higher (fixed) order corrections in both QCD and EW theory.
The main goal is to summarise the state of the art for computations at the time of the  2017
wish list (labelled as LH17 status) with the following paragraphs discussing the advances since LH17.  We also identify
processes with a large mismatch between the (expected for the HL-LHC)
experimental precision and the current theoretical uncertainties.%
\footnote{Unfortunately, time has allowed a discussion of
experimental uncertainties only for the Higgs sector and the inclusive $W, Z$ and $t\tb$ processes. This will be rectified in future updates. Extrapolating to a data
sample of 3000~fb$^{-1}$ can be problematic. Assuming a center-of-mass energy of 14 TeV for most of the running leads to a decrease in statistical errors
by a factor of 10. We make the assumption that the systematic errors stay the same; this may be optimistic given the environment
in the high luminosity LHC, so take this with a grain of salt. In almost all cases, the systematic errors will dominate over
the statistical ones for this large data sample. We assume a luminosity uncertainty of 2\%, as current.}
We are aware that there are obvious difficulties in compiling such lists, which make it difficult to address every possible relevant computation.
Specific approximations and/or extensions beyond fixed order are often necessary when comparing theory to data.

Following the 2017 wishlist we clarify that it is desirable to have a prediction that combines all the known corrections.
For example \NNLOQCD\!+\,\NLOEW refers to a single code that produced differential predictions including
$\mathcal{O}(\alpha_s^2)$ and $\mathcal{O}(\alpha)$ corrections. In most cases this is a non-trivial task and when considered
in combination with decays can lead to a large number of different sub-processes.

\paragraph*{Electroweak corrections}
Complete higher order corrections in the SM are technically more involved than
the better known corrections in QCD. An exhaustive review on electroweak
corrections within the Standard Model has been presented very recently in
Ref.~\cite{Denner:2019vbn}.

As a basic rule of thumb $\alpha_s^2 \sim \alpha$, and consequently
corrections at \NNLOQCD and \NLOEW are typically desirable together.
Moreover, for energy scales that are large compared to
the $W$-boson mass, EW corrections are enhanced by large --- so-called Sudakov --- logarithms.
There had been vast progress in a complete automation of
\NLOEW corrections within one-loop programs such as
\OpenLoops~\cite{Cascioli:2011va,Buccioni:2019sur}, \GoSam~\cite{Cullen:2011ac,Cullen:2014yla}, \Recola~\cite{Actis:2012qn,Actis:2016mpe,Denner:2017wsf}, \MadLoop~\cite{Alwall:2014hca,Frederix:2018nkq} and \NLOX~\cite{Honeywell:2018fcl},
which lead to a plethora of \NLOEW computations for final states of
previously unthinkable complexity within the recent years. Various examples are given in
this section of the report.
A detailed tuned comparison of these amplitude generators
for the production of the $4\ell$ and $2\ell2\nu$ final states (off-shell $ZZ$ and $WW$ production)
was presented in the previous Les Houches report\cite{Bendavid:2018nar} at the level of amplitudes,
and in combination with Monte Carlo integration frameworks that are capable of dealing with \NLOEW
corrections at the level of integrated and differential cross sections.
\GoSam~\cite{Cullen:2011ac,Cullen:2014yla} and
\Recola~\cite{Actis:2012qn,Actis:2016mpe,Denner:2017wsf} were already public at the stage of that study.
\MadLoop for electroweak corrections was released shortly after as part of the
\MadgraphaMCatNLO framework~\cite{Frederix:2018nkq}. The one-loop amplitude provider \NLOX
was made public a bit later~\cite{Honeywell:2018fcl}, and was first applied in a phenomenological
study in Ref.~\cite{Figueroa:2018chn}. Finally, the electroweak features of \OpenLoops were released
as part of a new version of the program~\cite{Buccioni:2019sur} that relies on its own on-the-fly
reduction~\cite{Buccioni:2017yxi}.

\paragraph*{Heavy top effective Higgs interactions and finite mass effects}

Many calculations of SM processes involving Higgs bosons use the effective
gluon--Higgs couplings that arise in the $m_t\to\infty$ limit,
often referred to as ``Higgs Effective Field Theory''. To avoid conflicts with the original
usage of the abbreviation HEFT in BSM contexts, we refrain from employing it here, but
rather refer to the heavy-top limit as HTL. In the HTL the Higgs bosons couple directly to gluons via the effective Lagrangian
\begin{equation}
\mathcal{L}_{\rm eff} = - \frac{1}{4} G^a_{\mu \nu} G_a^{\mu \nu} \left(C_H \frac{H}{v} - C_{HH} \frac{H^2}{2 v^2} + C_{HHH} \frac{H^3}{3 v^3} + \ldots\right)\,.
\end{equation}
The matching coefficients $C_H$, $C_{HH}$ and $C_{HHH}$ can be expanded in powers of $\alpha_S$ and
are known up to fourth order~\cite{Chetyrkin:1997iv,Chetyrkin:2005ia,Kramer:1996iq,Schroder:2005hy,Djouadi:1991tka,Grigo:2014jma,Spira:2016zna}.

At high energy hadron colliders, gluon fusion is the most dominant production process for Higgs bosons.
However, at high momentum transfers, where the top quark loops are resolved, the
approximation will break down.

For the data collected during Run II, and even more so at the HL/HE LHC,  it is certainly true that they
probe regions where the HTL approximation becomes invalid and finite mass effects are important.
Calculating the complete top mass dependence of such loop-induced processes at NLO is difficult
since it involves two-loop integrals with several mass scales.
While the analytic calculations of such integrals
have seen much progress in the last two years, as reported here, the phenomenological results available so far for this class of processes
mostly rely on either numerical methods or approximations.
We list processes in the wishlist as \NLOH{k}$\!\otimes\,$\NLOQ{l} when re-weighting including the full top mass dependence up to
order $l$ has been performed.

Some parts of the finite-mass effects can be accounted for using the so-called ``FT${}_\text{approx}$'' approximation that uses the virtual amplitudes within the HTL, while retaining the exact $m_t$ dependence in the real-emission diagrams~\cite{Frederix:2014hta, Maltoni:2014eza}.

\paragraph*{Resummation}

We do not attempt a complete classification of all possible resummation
procedures that have been considered or applied to the processes in the list.
In many cases precision measurements will require additional treatment beyond
fixed order, and since resummed predictions always match onto fixed order
outside the divergent region it would be desirable for most predictions to be available
this way. Since this is not feasible, some specific cases are highlighted in
addition to the fixed order.

There are several important kinematic regions where perturbative predictions
are expected to break down. Totally inclusive cross sections often have large
contributions from soft-gluon emission in which higher order logarithms can be
computed analytically. The $q_T$ and $N$-jettiness subtraction methods naturally
match onto resummations of soft/collinear gluons, in the latter case through
soft-collinear effective theory. A study using the $q_T$ method has been
applied in the case of $pp\to ZZ$ and $pp\to W^+ W^-$ \cite{Grazzini:2015wpa} where further details can be found.
$0$-jettiness resummations within SCET have also been
considered for Higgs boson production~\cite{Alioli:2015toa}, recently also extending to next-to-leading-logarithmic power corrections~\cite{Moult:2016fqy,Moult:2017jsg,Beneke:2017ztn,Boughezal:2018mvf}, the importance of which had been pointed out in Refs.~\cite{Kramer:1996iq,Catani:2003zt}.

Observables with additional restrictions on jet transverse momenta can also
introduce large logarithms, and jet veto resummations have been studied extensively in the case of
$pp\to H$ and $pp\to H+j$~\cite{Boughezal:2013oha,Boughezal:2014qsa,Banfi:2015pju}.
More in general, the logarithmic structure of Higgs production in gluon fusion has been recently investigated
in details, see \eg~\cite{Monni:2016ktx,Ebert:2016gcn,Caola:2016upw,Ebert:2017uel,Bizon:2017rah,Spira:1995rr,Melnikov:2016emg,Liu:2017vkm}.

There has been recent progress in resummation for double-differential observables involving a jet algorithm, specifically for the case of the transverse
momentum of the Higgs boson in the presence of a jet veto~\cite{Monni:2019yyr}. This technique combines the resummation of logarithms previously examined only
individually at \NNLL, for the Higgs boson transverse momentum, and for the presence of a jet veto in the final state.
An experimental comparison to the prediction is currently underway in the ATLAS experiment, which will allow for a more precise understanding of the QCD physics involving the production of a Higgs boson. This formalism will also allow for the calculation of any joint observable involving
the Higgs boson and the leading jet.

With increasing precision of both experimental data and fixed order
calculations other regions may also begin to play a role. A method for the
resummation of logarithms from small jet radii has been developed
\eg~in Refs.~\cite{Dasgupta:2014yra,Banfi:2015pju,Dasgupta:2016bnd,Kolodrubetz:2016dzb,Liu:2017pbb,Liu:2018ktv}.
A clear understanding of these effects is important as the most popular jet radius
for physics analyses at the LHC is 0.4, a size for which resummation may start to
become noticeable. These logarithms are implicitly resummed in parton shower Monte Carlos.

The cusp anomalous dimensions for quarks and gluons determine the leading infra-red singularities of massless
scattering amplitudes and are crucial for resummation calculations. The first complete quark and gluon cusp anomalous
dimensions, calculated from first principles,  in four-loop massless QCD was given in Ref.~\cite{vonManteuffel:2020vjv},
following earlier numerical approximations  ~\cite{Moch:2017uml,Moch:2018wjh}  and analytical results \cite{Henn:2019swt} employing conjectural input.

These represent only a tiny fraction of the currently available tools and predictions
with resummed logarithms. For a review the interested reader may refer to~\cite{Luisoni:2015xha}
and references therein.

\paragraph*{Parton showering}

As is the case of resummation, we refrain from listing all improvements
necessary for parton-shower Monte-Carlo (PSMC) programs. However, since PSMCs
have turned into crucial components of simulations that combine precision
fixed-order calculations with event generators, it is extremely desirable
to define, implement and assess the effects of PSMCs at higher perturbative
order. This includes their relation to (semi-)analytic resummation,
observable-independent definitions of NLO parton showers, and the relation
of PSMCs to higher-order evolution of parton distributions or
fragmentation functions. Furthermore, electro-weak effects should
systematically be included in PSMCs and in their matching to fixed-order
calculations.

\paragraph*{Decay sub-processes}

The description of decay sub-processes is incomplete though we do list a few
notable cases. Ideally all on-shell (factorised) decays
in a narrow-width approximation~(NWA) would be available up
to the order of the core process. In some cases this is potentially an
insufficient approximation and full off-shell decays including background
interference would be desirable, but are often prohibitive. The $t\tb$ final state
is an obvious example where the off-shell decay to $WWb\bb$ at \NNLOQCD is beyond the scope
of current theoretical methods.

Decays in the context of electroweak corrections are usually much more
complicated. Full off-shell effects at NLO are expected to be small, but higher-order
corrections within factorisable contributions to the decay can be important.
However, with the great progress of automated tools, NLO calculations not only in
QCD, but also in electroweak theory for $2\to6$ processes and beyond have
become feasible.

\subsection{Higgs boson associated processes}
An overview of the status of Higgs boson associated processes is given in Table~\ref{tab:SM_wishlist:wlH}.

\begin{table}
  \renewcommand{\arraystretch}{1.5}
\setlength{\tabcolsep}{5pt}
  \begin{center}
  \begin{tabular}{lll}
    \hline
    \multicolumn{1}{c}{process} & \multicolumn{1}{c}{known} &
    \multicolumn{1}{c}{desired} \\
    \hline
    $pp\to H$ &
    \begin{tabular}{l}
      \NNNLOHTL (incl.) \\
      \NLOHE11 \\ 
      \NNLOHTL$\!\otimes\,$\NLOQCD
    \end{tabular} &
    \begin{tabular}{l}
      \NNNLOHTL (partial results available) \\
      \NNLOQCD
    \end{tabular} \\
    \hline
    $pp\to H+j$ &
    \begin{tabular}{l}
      \NNLOHTL \\
      \NLOQCD
    \end{tabular} &
    \begin{tabular}{l}
      \NNLOHTL$\!\otimes\,$\NLOQCD\!+\,\NLOEW
    \end{tabular} \\
    \hline
    $pp\to H+2j$ &
    \begin{tabular}{l}
      \NLOHone$\!\otimes\,$\LOQCD \\
      \NNNLOQCDVBFstar (incl.) \\
      \NNLOQCDVBFstar \\
      \NLOEWVBF
    \end{tabular} &
    \begin{tabular}{l}
      \NNLOHTL$\!\otimes\,$\NLOQCD\!+\,\NLOEW\\
      \NNLOQCDVBF\!+\,\NLOEWVBF
    \end{tabular} \\
    \hline
    $pp\to H+3j$ &
    \begin{tabular}{l}
      \NLOHone \\
      \NLOQCDVBF
    \end{tabular} &
    \begin{tabular}{l}
      \NLOQCD\!+\,\NLOEW \\
    \end{tabular} \\
    \hline
    $pp\to H+V$ &
    \begin{tabular}{l}
      \NNLOQCD\!+\,\NLOEW \\
    \end{tabular} &
    \begin{tabular}{cl}
      \NLOggHVtb{} \\
    \end{tabular} \\
    \hline
    $pp\to HH$ &
    \begin{tabular}{l}
      \NNNLOHTL$\!\otimes\,$\NLOQCD \\
    \end{tabular} &
    \begin{tabular}{cl}
      \NLOEW \\
    \end{tabular} \\
    \hline
    $pp\to H+t\tb$ &
    \begin{tabular}{l}
      \NLOQCD\!+\,\NLOEW\\
    \end{tabular} &
    \begin{tabular}{l}
     \NNLOQCD
    \end{tabular}  \\
    \hline
    $pp\to H+t/\tb$ &
    \begin{tabular}{l}
      \NLOQCD\\
    \end{tabular} &
    \begin{tabular}{l}
      \NLOQCD\!+\,\NLOEW
    \end{tabular} \\
    \hline
  \end{tabular}
  \caption{Precision wish list: Higgs boson final states. \NLOQVBFstar{x} means a
   calculation using the structure function approximation.}
  \label{tab:SM_wishlist:wlH}
  \end{center}
\renewcommand{\arraystretch}{1.0}
\end{table}

\begin{itemize}[leftmargin=2cm]

\item[$H$:] \textit{LH17 status:}
\NNLOHTL results known for almost two decades~\cite{Harlander:2002wh,Anastasiou:2002yz,Ravindran:2003um,Catani:2007vq,Grazzini:2008tf};
supplemented by an expansion in $1/m_t^n$~\cite{Harlander:2009my}, and matched to a calculation in the high energy limit~\cite{Pak:2011hs};
first steps towards differential results at \NNNLOHTL presented in Ref.~\cite{Dulat:2017brz},
and results beyond threshold approximation in Refs.~\cite{Dulat:2017prg,Mistlberger:2018etf,Dulat:2018rbf};
\NLOHE11 corrections at order $\alpha\alpha_s^2$ calculated in the soft gluon
approximation~\cite{Bonetti:2017ovy,Bonetti:2018ukf}; 
comprehensive phenomenological study presented in~\cite{Anastasiou:2016cez},
and available in the program {\sc iHixs}~\cite{Dulat:2018rbf};
\NNLOgen\!+\,PS computations~\cite{Hamilton:2013fea,Hoche:2014dla} extended to include
finite top and bottom mass corrections at \NLOgen~\cite{Hamilton:2015nsa}.

The rapidity spectrum for Higgs production in gluon fusion has been calcuated to
\NNNLOHTL~\cite{Dulat:2018bfe,Cieri:2018oms} accuracy.
The \NNNLOHTL corrections lead to a mild enhancement compared to the \NNLOHTL results,
and significantly reduce the scale dependence throughout the entire rapidity range.

The transverse momentum spectrum of the Higgs boson has been studied at \NNLOgen\!+\,\NNNLL both
at the inclusive level~\cite{Chen:2018pzu} and in the $H\rightarrow\gamma\gamma$ channel
with fiducial cuts~\cite{Bizon:2018foh}.
In both cases resummation reduces the theoretical uncertainties and stabilises the result
for $p_T < 40\ \mathrm{GeV}$. The \NNNLL corrections were found to be moderate in size,
growing to 5\% at very small $p_T$.
However, the perturbative uncertainty was reduced significantly below $10\ \mathrm{GeV}$
with respect to the \NNLOgen\!+\,\NNLL case.

Significant progress has been made in the ongoing effort to include also quark mass effects.
The 3-loop virtual corrections to Higgs production, including the effect of one massive quark,
were first obtained by combining large-$m_t$ and threshold expansions using a conformal mapping
and a Pad{\'e} approximation~\cite{Davies:2019nhm} and subsequently via the numerical solution
of the differential equations~\cite{Czakon:2020vql}.
For the subset of three-loop diagrams which contain a closed light-quark loop,
analytic results are available~\cite{Harlander:2019ioe}.
At 4-loops virtual corrections have also been computed in a large-$m_t$ expansion~\cite{Davies:2019wmk}.

For Higgs bosons with intermediate transverse momenta in the range $m_b < p_T < m_t$
the effect of bottom quarks can also be important.
Such effects have been studied at \NLOgen\!+\,\NNLL~\cite{Caola:2018zye}.
It was found that the uncertainty on the top--bottom interference contribution
is around 20\% and that ambiguities related to the resummation procedure are
of the same order as the fixed-order uncertainties.

Mixed QCD--EW corrections to Higgs production via gluon fusion have recently been computed
in the limit of a small mass of the electroweak gauge bosons~\cite{Anastasiou:2018adr}.
This work provides an important check of earlier results at order $\alpha\alpha_s^2$ which
were obtained using the soft gluon approximation~\cite{Bonetti:2017ovy,Bonetti:2018ukf}.

The experimental uncertainty on the total Higgs boson cross section is currently
of the order of 8\%~\cite{ATLAS:2019mju}
based on a data sample of 139~fb$^{-1}$,
and is expected to reduce to the order of 3\% or less with a data sample
of 3000~fb$^{-1}$~\cite{Campbell:2286381}.
To achieve the desired theoretical uncertainty, it may be necessary to calculate the
finite-mass effects to \NNLOQCD, combined with fully differential \NNNLOHTL corrections.

\item[$H+j$:] \textit{LH17 status:}
Known to \NNLOHTL in the infinite top mass limit~\cite{Chen:2014gva,Chen:2016zka,Boughezal:2015dra,Boughezal:2015aha,Caola:2015wna};
later calculated at \NLOQCD with full top-quark mass dependence~\cite{Jones:2018hbb},
based on numerical methods~\cite{Borowka:2015mxa,Borowka:2017idc},
revealing a fairly constant (NLO/LO) K-factor over the Higgs $p_T$ range
above the top quark threshold when using the scale $H_T/2$ and
a roughly 9\% (6\%) larger full result than in HTL (FT$_{\rm{approx}}$~\cite{Buschmann:2014sia,Hamilton:2015nsa,Frederix:2016cnl,Neumann:2016dny});
also top--bottom interference effects calculated~\cite{Melnikov:2016qoc,Lindert:2017pky},
as well as the mass effects in the large transverse momentum
expansion~\cite{Lindert:2018iug,Neumann:2018bsx};
Higgs $p_T$ spectrum with finite quark mass effects calculated beyond the LO
using high-energy resummation techniques at \LL accuracy~\cite{Caola:2016upw};
parton shower predictions including finite mass effects available in various
approximations~\cite{Frederix:2016cnl,Neumann:2016dny,Hamilton:2015nsa,Buschmann:2014sia}.

Concerning the above-mentioned \NNLOHTL calculations, an investigation of the implementation
and stability of the $N$-jettiness result has recently helped to resolve a long standing small
discrepancy between the results of the different groups~\cite{Campbell:2019gmd}.

Fiducial cross sections for the four-lepton decay mode in Higgs-plus-jet production were computed
up to \NNLOHTL and reweighted by the \LOQCD result
to include the top-quark mass dependence~\cite{Chen:2019wxf}.
It was found that the acceptance factors used to infer simplified template cross sections
are perturbatively stable for ATLAS measurements but less so for CMS,
the difference in stability was understood to be due to the different lepton isolation
prescriptions used by the experiments.

The Higgs transverse momentum spectrum has been studied at \NLOgen\!+\,\NNLL in the case a jet veto,
$p_t^j \ll p_t^{j,v}$, is applied~\cite{Monni:2019yyr}.
In the region $p_t^H, p_t^{j,v} \ll m_H$, logarithms involving either $p_t^H$ or $p_t^{j,v}$ can become large,
and a joint resummation of both classes of logarithms is applied.
This work contains the first resummation for a double-differential observable
involving a jet algorithm in hadronic collisions.

The current experimental uncertainty on the Higgs + $\ge$ 1 jet differential cross section is of the order of 10--15\%, dominated by
the statistical error, for example the fit statistical errors for the case of the combined $H\rightarrow \gamma \gamma$ and $H\rightarrow 4\ell$ analyses~\cite{ATLAS:2019mju}.
With a sample of
3000 fb$^{-1}$, the statistical error will nominally decrease by about a factor of 5, resulting in a statistical error of
the order of 2.5\%.  If the remaining systematic errors
(dominated for the diphoton analysis by the spurious signal systematic error) remain the same,
the resultant systematic error would be of the order of 9\%, leading to a
total error of approximately 9.5\%.  This is similar enough to the current theoretical uncertainty that it may motivate
improvements on the $H+j$ cross section calculation. Of course, any improvements in the systematic errors would reduce the experimental uncertainty further.
Improvements in the theory could entail a combination of the \NNLOHTL results with the
full \NLOQCD results,
similar to the reweighting procedure that has been done one perturbative order lower.

\item[$H+\geq 2j$:] \textit{LH17 status:}
VBF production of a Higgs boson known at \NNNLOHTL accuracy
for the total cross section~\cite{Dreyer:2016oyx} and to \NNLOHTL accuracy
differentially~\cite{Cacciari:2015jma,Cruz-Martinez:2018rod}
in the ``DIS'' approximation~\cite{Han:1992hr};
in the VBF channel, full \NLOQCD corrections for $H + 3j$ in the VBF channel
available~\cite{Campanario:2013fsa,Campanario:2018ppz};
a phenomenological study of $H + \le 3j$ in the gluon fusion channel performed in
Ref.~\cite{Greiner:2016awe} and an assessment of the mass dependence of the various
jet multiplicities in Ref.~\cite{Greiner:2015jha};
\NLOEW corrections to stable Higgs boson production in VBF calculated~\cite{Ciccolini:2007jr} and
available in {\sc Hawk}~\cite{Denner:2014cla}.

The ``DIS'' approximation used to study VBF at \NNNLOHTL and \NNLOHTL neglects interactions
between incoming QCD partons, retaining only QCD effects confined to a single fermion line.
Non-factorizable QCD effects beyond this approximation have recently been studied using the
eikonal approximation~\cite{Liu:2019tuy}, corrections were found
to be $\sim 0.5\%$ growing to $1\%$ in certain kinematic regions.
Mass effects in $H+2j$ at large energy have also recently been studied~\cite{Andersen:2018kjg},
within the ``High Energy Jets''
framework~\cite{Andersen:2009nu,Andersen:2009he,Andersen:2011hs,Andersen:2017kfc,Andersen:2018tnm}.

The current experimental error on the $H+\geq 2j$ cross section is on the order of
20\%~\cite{ATLAS:2019jst}, again dominated by statistical errors,
and again for the diphoton final state, by the fit statistical error. With the same assumptions
as above, for 3000 fb$^{-1}$, the statistical error will reduce to the order of 3.5\%.
If the systematic errors remain the same, at approximately 12\% (in this case the largest
systematic error is from the jet energy scale uncertainty and the jet energy resolution uncertainty),
a total uncertainty of approximately 12.5\% would result, less than the
current theoretical uncertainty.
To achieve a theoretical uncertainty less than this value would require the calculation
of $H+\geq 2j$ to \NNLOHTL$\!\otimes\,$\NLOQCD in the gluon fusion production mode.

\item[$VH$:] \textit{LH17 status:}
inclusive \NNLOQCD corrections available for some time~\cite{Brein:2003wg,Brein:2011vx}, available
in {\sc VH@NNLO}~\cite{Brein:2012ne,Brein:2003wg,Brein:2011vx};
total inclusive cross considered in the threshold limit at \NNNLOQCD~\cite{Kumar:2014uwa}; 
differential predictions at \NNLOQCD calculated in $q_T$ subtraction for $WH$~\cite{Ferrera:2011bk}
and $ZH$~\cite{Ferrera:2014lca}, later extended to include \NNLOQCD $H\to b\bb$ decays~\cite{Ferrera:2017zex};
\NNLOQCD with $H\to b\bar{b}$ decays at \NLOQCD calculated in N-jettiness subtraction in
MCFM~\cite{Campbell:2016jau}, and with \NNLOQCD decays in
nested soft-collinear subtraction~\cite{Caola:2017xuq};
soft-gluon resummation effects found to be small compared to \NNLOQCD result~\cite{Dawson:2012gs};
\NNLOQCD for $WH$ production matched to parton shower by the MiNLO procedure in \Powheg~\cite{Astill:2016hpa};
\NLOEW corrections calculated~\cite{Ciccolini:2003jy,Denner:2011id,Obul:2018psx,Granata:2017iod},
also including parton shower effects~\cite{Granata:2017iod};
loop-induced $gg\to ZH$ known at \NLOQCD~\cite{Altenkamp:2012sx} by reweighting the full
LO cross section with a $K$-factor in the limit $m_t\to \infty$ and with $m_b = 0$;
threshold resummation for $gg\to ZH$ calculated in Ref.~\cite{Harlander:2014wda};
$m_t$ effects at \NLOQCD considered in the framework of an $1/m_t$ expansion~\cite{Hasselhuhn:2016rqt};
\NLOQCD with dimension-six SMEFT operators investigated~\cite{Degrande:2016dqg}, matched to a parton
shower in the \MadgraphaMCatNLO framework;
Higgs pseudo-observables investigated at \NLOQCD~\cite{Greljo:2017spw}.

At \NNLOQCD the process $pp \rightarrow VH + X \rightarrow l\bar{l} b\bar{b} + X$ was studied
fully differentially using the antenna subtraction formalism~\cite{Gauld:2019yng}.
Independent variation of the production and decay scales was found to yield percent-level
uncertainties.

The process $b\bb\to ZH$ in the 5FS, but with a non-vanishing
bottom-quark Yukawa coupling was investigated in the soft-virtual approximation at
\NNLOQCD~\cite{Ahmed:2019udm}.

$ZH$ production has also been studied~\cite{Astill:2018ivh} in a \NNLOgen\!+\,PS approach
based on the MiNLO procedure and implemented in the \Powhegboxres framework,
with the Higgs boson decay to bottom quarks treated at \NLOQCD.
It was found that the loop-induced $gg \rightarrow ZH$ channel, which enters
at $\mathcal{O}(\alpha_s^2)$, is relevant at the level of the total cross section and
can lead to substantial distortions in kinematic distributions.

Higgs boson production in association with a vector boson at \NNLOQCD 
was supplemented with \NNLLp resummation in the 0-jettiness variable and matched
to a parton shower within the \Geneva Monte Carlo framework~\cite{Alioli:2019qzz}.

Published results for the $VH$ cross section are available for data samples up to 80 $fb^{-1}$,
with uncertainties on the order of 20\%, equally divided between statistical and
systematic errors~\cite{Aaboud:2018zhk}. For 3000 $fb^{-1}$, the statistical error will reduce
to 2-3\%, resulting in a measurement that is systematically limited, unless there are significant
improvements to the systematic errors.
The general $VH$ process has been calculated to \NNLOQCD,
leading to a small scale uncertainty. However, the $gg\rightarrow ZH$ sub-process is
still only known at LO with the exact top-mass dependence and at \NNLOHTL with
reweighting~\cite{Altenkamp:2012sx}. The best understanding of the $ZH$ process requires a
calculation of the $gg\rightarrow ZH$ sub-process beyond that.

\item[$HH$:] \textit{LH17 status:}
\NNLOHTL corrections known inclusively~\cite{deFlorian:2013jea}
and differentially~\cite{deFlorian:2016uhr};
threshold resummation performed at \NLOgen\!+\,\NNLL~\cite{Shao:2013bz}
and \NNLOgen\!+\,\NNLL~\cite{deFlorian:2015moa};
power corrections in $1/m_t$ computed for \NLOHTL and \NNLOHTL cross
sections~\cite{Grigo:2013rya,Grigo:2015dia}; complete $m_t$ dependence included at \NLOQCD using
numerical methods~\cite{Borowka:2016ehy,Borowka:2016ypz}; matched to
parton showers~\cite{Heinrich:2017kxx,Jones:2017giv} and publicly available in \PowhegboxVtwo;
ansatz with Pad{\'e} approximants based on the large-$m_t$ expansion and analytic
results near the top threshold presented~\cite{Grober:2017uho}, reproducing the full result very well;
planar two-loop integrals entering $gg\to HH$ computed in the
high-energy limit~\cite{Davies:2018ood};
finite $m_t$ effects incorporated in \NNLOHTL calculation by suitable reweighting, combined with
full-$m_t$ double-real corrections~\cite{Grazzini:2018bsd};
also studied for $b\bb\to HH$ at \NNLOQCD~\cite{H:2018hqz}

The \NNNLOHTL corrections were recently computed in the infinite top mass limit~\cite{Chen:2019lzz,Banerjee:2018lfq} and have been reweighted by the \NLOQCD result (\ie including finite top-quark mass effects)~\cite{Chen:2019fhs}.
The new results reduce the perturbative scale uncertainty to the level of a few percent,
such that the remaining uncertainty is dominated by the missing higher-order $m_t$ corrections.
Additional sources of uncertainty include the top-quark mass scheme dependence,
electroweak corrections and parametric uncertainties on the input parameters and PDFs.

The sensitivity of Higgs boson pair production to the quartic self-coupling (which enters via EW
corrections) was studied in Refs.~\cite{Liu:2018peg,Bizon:2018syu,Borowka:2018pxx}.

At \NLOQCD results including the full top-quark mass dependence were computed
numerically~\cite{Baglio:2018lrj}, providing an important cross-check of earlier
results~\cite{Borowka:2016ehy,Borowka:2016ypz}.
The new calculation treated the top-quark mass in both the $\overline{\mathrm{MS}}$ scheme and
the on-shell scheme, demonstrating that mass-scheme and scale uncertainties can be as large
as other perturbative scale uncertainties.
A dedicated study of the top-quark scheme dependence was performed in
Sect.~\ref{cha:higgs}.\ref{sec:higgs_msu} of this report.
The exact numerical results at \NLOQCD have also been supplemented by results obtained
in a high-energy expansion~\cite{Davies:2019dfy,Davies:2018qvx}.

Considerable progress has been made in the direction of studying mass effects beyond \NLOQCD.
At \NNLOQCD real-virtual corrections to Higgs boson pair production involving three closed
top-quark loops have been computed in a large-$m_t$ expansion~\cite{Davies:2019xzc}.
The 3-loop form factors (\NNLOQCD virtual corrections) have also been computed in an expansion
around the large top mass limit~\cite{Grigo:2015dia,Davies:2019djw}.

Fully differential results for VBF $HH$ production are now available at
\NNLOHTL~\cite{Dreyer:2018rfu} and at \NNNLOHTL for the inclusive cross section~\cite{Dreyer:2018qbw}.
The \NNLOHTL corrections were found to be at the level of $3-4\%$ after typical VBF cuts;
while the \NNNLOHTL corrections were found to be negligible at the central scale choice,
they reduce the remaining scale uncertainty by a factor of four.

The experimental limits on $HH$ production are currently at the level of 10--12 times the
SM cross section~\cite{Aad:2019uzh,Sirunyan:2018two} based on a data sample of 38 fb$^{-1}$.
The most stringent limits on the Higgs boson cubic self-coupling,
$-2.3 < \lambda_{hhh}/\lambda_{hhh,\mathrm{SM}} < 10.3$, come from combining
single Higgs boson production data with limits from $HH$ production~\cite{ATLAS:2019pbo}.
With a data sample of 3000 fb$^{-1}$ it is projected that a limit of
$0.5 < \lambda_{hhh}/\lambda_{hhh,\mathrm{SM}} < 1.5$ can be achieved at the $68\%$ CL
for ATLAS and CMS combined~\cite{Cepeda:2019klc}.

\item[$HHH$:] \textit{LH17 status:}
two-loop results in soft-virtual approximation known~\cite{deFlorian:2016sit}.

Triple Higgs boson production has recently been calculated at \NNLOHTL~\cite{deFlorian:2019app}
extending on previous work~\cite{deFlorian:2016sit}.
Finite quark mass effects are included by reweighting with the full Born result.
The remaining uncertainty was found to be dominated by the currently unknown
higher-order finite top-quark mass effects.

\item[$t\bar{t}H$:] \textit{LH17 status:}
\NLOQCD corrections for on-shell $t\bar{t}H$ production known for many
years~\cite{Beenakker:2001rj,Reina:2001sf,Beenakker:2002nc,Dawson:2003zu};
\NLOEW~corrections studied within the \MadgraphaMCatNLO
framework~\cite{Frixione:2014qaa,Frixione:2015zaa};
combined \NLOQCD and \NLOEW corrections with NWA top-quark decays calculated~\cite{Yu:2014cka};
\NLOQCD results merged to parton showers~\cite{Garzelli:2011vp,Hartanto:2015uka};
\NLOgen\!+\,\NNLL resummation performed in
Refs.~\cite{Kulesza:2015vda,Broggio:2015lya,Broggio:2016lfj,Kulesza:2017ukk};
\NLOQCD results in the Standard Model Effective Field Theory calculated~\cite{Maltoni:2016yxb};
\\
corrections to $t\bar{t}H$ including top quark decays and full off-shell effects
computed at \NLOQCD~\cite{Denner:2015yca},
and later even combined with \NLOEW~\cite{Denner:2016wet},
involving up to 9-point functions in the virtual amplitudes.

The cross section for $t\bar{t}H$ has been measured with a data sample of 80 $fb^{-1}$,
with a total uncertainty on the order of 20\%, equally divided between statistical and
systematic errors~\cite{Aaboud:2018urx}.
Again the statistical error will shrink to the order of 2--3\% for 3000 $fb^{-1}$,
leaving a systematics-dominated measurement. Given that this calculation is currently known
only at \NLOQCD, with a corresponding scale uncertainty of the order of 10--15\%,
this warrants a calculation of the process to \NNLOQCD.

\item[$tH$:] \textit{LH17 status:}
\NLOQCD corrections to $tH$ associated production known~\cite{Campbell:2013yla,Demartin:2015uha}.

\item[$b\bar{b}H$:] (including $H$ production in bottom quark fusion treated in 5FS)

\textit{LH17 status:}
\NNLOQCD predictions in bottom quark fusion in the 5FS known for a long time,
inclusively~\cite{Harlander:2003ai} and later differentially~\cite{Harlander:2011fx,Buehler:2012cu};
resummed calculation at \NNLOgen\!+\,\NNLL available~\cite{Harlander:2014hya};
three-loop $Hb\bar{b}$ form factor known~\cite{Gehrmann:2014vha};
\NNNLOQCD in threshold approximation~\cite{Ahmed:2014cha,Ahmed:2014era} calculated;\\
\NLOQCD corrections in the 4FS known since long ago~\cite{Dittmaier:2003ej,Dawson:2003kb};
\NLOQCD matched to parton shower, also comparing with 5FS~\cite{Wiesemann:2014ioa};
various methods proposed to combine 4FS and 5FS
predictions~\cite{Harlander:2011aa,Bonvini:2015pxa,Forte:2015hba,Bonvini:2016fgf,Forte:2016sja};
\NLOEW corrections calculated\cite{Zhang:2017mdz}.

More recently, the complete inclusive \NNNLOQCD calculation in bottom quark fusion, treating
the bottom quark as massless while retaining a non-vanishing Yukawa coupling to the $H$,
was presented in Ref.~\cite{Duhr:2019kwi}. A reduced dependence on renormalisation and
factorisation scales and a convergence of the series is found for judicious scale choices.
Compared to the cross section obtained from the so-called Santander-matching~\cite{Harlander:2011aa}
of 4FS and 5FS results, a slightly higher, though consistent cross section is predicted.

Based upon these results in the 5FS a resummed calculation up to \NNNLOgen\!+\,\NNNLL was presented in
Ref.~\cite{H:2019dcl}, and \NLOmixQED{1}{1} as well as \NNLOQED predictions were derived in
Ref.~\cite{H:2019nsw}.

The $b\bar{b}H$ final state has been studied at \NLOQCD
(including the formally \NNLOHTL $y_t^2$ contributions) using the 4FS~\cite{Deutschmann:2018avk}.

\end{itemize}

\subsection{Jet final states}
An overview of the status of jet final states is given in Table~\ref{tab:SM_wishlist:wljets}.

\begin{table}
  \renewcommand{\arraystretch}{1.5}
\setlength{\tabcolsep}{5pt}
  \centering
  \begin{tabular}{lll}
    \hline
    \multicolumn{1}{c}{process} & \multicolumn{1}{c}{known} & \multicolumn{1}{c}{desired} \\
    \hline
    $pp\to 2$\,jets &
    \begin{tabular}{l}
      \NNLOQCD \\
      \NLOQCD\!+\,\NLOEW
    \end{tabular} &
    \begin{tabular}{cl}
      \\
    \end{tabular} \\
    \hline
    $pp\to 3$\,jets &
    \begin{tabular}{l}
      \NLOQCD\!+\,\NLOEW
    \end{tabular} &
    \begin{tabular}{l}
      \NNLOQCD
    \end{tabular} \\
    \hline
  \end{tabular}
  \caption{Precision wish list: jet final states.}
  \label{tab:SM_wishlist:wljets}
  \renewcommand{\arraystretch}{1.0}
\end{table}

\begin{itemize}[leftmargin=2cm]

\item[j+X:] \textit{LH17 status:}
Differential \NNLOQCD corrections calculated in the \NNLOjet framework~\cite{Currie:2016bfm} with a detailed study of scale choices performed in Ref.~\cite{Currie:2017ctp}.

Single-jet inclusive rates with exact colour at $\mathcal{O}({\alpha}_s^4)$ were recently completed
in the sector-improved residue subtraction formalism~\cite{Czakon:2019tmo}.
This full calculation confirmed that the approximation applied in the previous one, \ie
leading-colour approximation in the case of channels involving quarks and exact calculation
in colour only in the pure-gluon channel, is perfectly justified for phenomenological
applications.

\item[2j:] \textit{LH17 status:}
\NNLOQCD corrections calculated in the \NNLOjet framework~\cite{Currie:2017eqf}; complete NLO QCD+EW corrections available~\cite{Frederix:2016ost}.

Building upon the \NNLOjet framework, which implements the antenna subtraction formalism,
in Ref.~\cite{Gehrmann-DeRidder:2019ibf} a first dedicated \NNLOQCD study of triple-differential 2-jet
cross sections has been performed.

\item[$\geq$3j:] \textit{LH17 status:} \NLOQCD corrections for 3-jet~\cite{Nagy:2003tz}, 4-jet~\cite{Bern:2011ep,Badger:2012pf} and 5-jet~\cite{Badger:2013yda} known.

\NLOEW corrections for 3-jet production were first reported inclusively, calculated in the
automated framework \MadgraphaMCatNLO~\cite{Frederix:2018nkq}. A full \NLOSM calculation for 3-jet production was performed using \Sherpa and amplitudes from \Recola in Ref.~\cite{Reyer:2019obz}.

Complete \NNLOQCD corrections could not be achieved to date, but huge progress was made on
the calculation of 5-point two-loop amplitudes for that process as summarised in Sect.~\ref{sec:SM_wishlist:loop}.

\end{itemize}

\subsection{Vector boson associated processes}

The numerous decay channels for vector bosons and the possible inclusion
of full off-shell corrections versus factorised decays in the narrow width approximation
make vector boson processes complicated to classify.
A full range of decays in the narrow width approximation would be a desirable minimum precision.
For leptonic decays, this goal is met for essentially all processes in the list.
In terms of QCD corrections, full off-shell decays don't mean a significant complication of
the respective QCD calculations and are available almost everywhere.
In the case of EW corrections, on the other hand, leptonic decays increase the complexity of
the calculation.
In the meantime, they have become available for many high-multiplicity processes (up to six final-state particles and beyond).
Hadronic decays are even harder to classify because they are formally part of subleading Born
contributions to processes involving jets and possibly further leptonically decaying vector bosons.
Including higher-order corrections in a consistent way here will usually require full
SM corrections to the complete tower of Born processes,
as briefly discussed in Sec.~\ref{sec:SM_wishlist:precision_wish_list}.
An overview of the status of vector boson associated processes is given in Table~\ref{tab:SM_wishlist:wlV},
where leptonic decays are understood if not stated otherwise.
Also $\gamma$-induced processes become increasingly important in cases where EW
corrections are highly relevant. While often included only at their leading order, first
computations involving also full EW corrections to $\gamma$-induced channels were recently achieved.

\begin{table}
  \renewcommand{\arraystretch}{1.5}
\setlength{\tabcolsep}{5pt}
  \centering
  \begin{tabular}{lll}
    \hline
    \multicolumn{1}{c}{process} & \multicolumn{1}{c}{known} & \multicolumn{1}{c}{desired} \\
    \hline
    $pp\to V$ &
    \begin{tabular}{l}
      \NLOQzzero3 (incl.) \\
      \NNNLOQCD (incl., $\gamma^*$) \\
      \NNLOQCD \\
      \NLOEW
    \end{tabular} &
    \begin{tabular}{l}
      \NNNLOQCD\!+\,\NLOE2\!+\,\NLOQE11 \wdecay{}
    \end{tabular} \\
    \hline
    $pp\to VV'$ &
    \begin{tabular}{l}
      \NNLOQCD\!+\,\NLOEW{ }\wleptdecays{} \\
      \!+\,\NLOQCD{ }($gg$ channel) \wleptdecays{} \\
    \end{tabular} &
    \begin{tabular}{l}
      \NLOQCD{ }($gg$ channel, w/ massive loops) \\
    \end{tabular} \\
    \hline
    $pp\to V+j$ &
    \begin{tabular}{l}
      \NNLOQCD\!+\,\NLOEW{ }\wleptdecays{} \\
    \end{tabular} &
    \begin{tabular}{l}
      hadronic decays
    \end{tabular} \\
    \hline
    $pp\to V+2j$ &
    \begin{tabular}{l}
      \NLOQCD\!+\,\NLOEW{ }\wleptdecays{} \\
      \NLOEW{ }\wleptdecays{}
    \end{tabular} &
    \begin{tabular}{l}
      \NNLOQCD \wdecays{} \\
    \end{tabular}\\
    \hline
    $pp\to V+b\bar{b}$ &
    \begin{tabular}{l}
      \NLOQCD{ }\wleptdecays{} \\
    \end{tabular} &
    \begin{tabular}{l}
      \NNLOQCD \!+\,\NLOEW{ }\wdecays{} \\
    \end{tabular} \\
    \hline
    $pp\to VV'+1j$ &
    \begin{tabular}{l}
      \NLOQCD{ }\wdecays{} \\
      \NLOEW{ }\wodecays{}
    \end{tabular} &
    \begin{tabular}{l}
      \NLOQCD\!+\,\NLOEW{ }\wdecays{} \\
    \end{tabular} \\
    \hline
    $pp\to VV'+2j$ &
    \begin{tabular}{l}
      \NLOQCD \wleptdecays{} \\
    \end{tabular} &
    \begin{tabular}{l}
      \NLOQCD\!+\,\NLOEW{ }\wdecays{} \\
    \end{tabular} \\
    \cline{2-2}
    $pp\to W^+W^++2j$ &
    \begin{tabular}{l}
      \NLOQCD\!+\,\NLOEW{ }\wleptdecays{} \\
    \end{tabular} &
    \begin{tabular}{l}
      \\
    \end{tabular} \\
    \cline{2-2}
    $pp\to W^+Z+2j$ &
    \begin{tabular}{l}
      \NLOQCD\!+\,\NLOEW{ }\wleptdecays{} \\
    \end{tabular} &
    \begin{tabular}{l}
      \\
    \end{tabular} \\
    \hline
   $pp\to VV'V''$ &
    \begin{tabular}{l}
      \NLOQCD \\
      \NLOEW{ }\wodecays{}
    \end{tabular} &
    \begin{tabular}{l}
      \NLOQCD\!+\,\NLOEW \wdecays{} \\
    \end{tabular} \\
    \cline{2-2}
   $pp\to W^\pm W^+W^-$ &
    \begin{tabular}{l}
      \NLOQCD + \NLOEW{ }\wdecays{}
    \end{tabular} &
    \begin{tabular}{l}
      \\
    \end{tabular} \\
    \hline
    $pp\to \gamma\gamma$ &
    \begin{tabular}{l}
      \NNLOQCD\!+\,\NLOEW
    \end{tabular} &
    \begin{tabular}{l}
      \\
    \end{tabular} \\
    \hline
    $pp\to \gamma+j$ &
    \begin{tabular}{l}
      \NNLOQCD\!+\,\NLOEW
    \end{tabular} &
    \begin{tabular}{l}
      \\
    \end{tabular} \\
    \hline
    $pp\to \gamma\gamma+j$ &
    \begin{tabular}{l}
      \NLOQCD \\
      \NLOEW
    \end{tabular} &
    \begin{tabular}{cl}
      \NNLOQCD\!+\,\NLOEW
    \end{tabular} \\
    \hline
    $pp\to \gamma\gamma\gamma$ &
    \begin{tabular}{l}
      \NNLOQCD
    \end{tabular} &
    \begin{tabular}{cl}
      \\
    \end{tabular} \\
    \hline
  \end{tabular}
  \caption{Precision wish list: vector boson final
    states. $V=W,Z$ and $V',V''=W,Z,\gamma$.
    Full leptonic decays are understood if not stated otherwise.}
  \label{tab:SM_wishlist:wlV}
  \renewcommand{\arraystretch}{1.0}
\end{table}

\begin{itemize}[leftmargin=2cm]
\item[$V$:] \textit{LH17 status:}
Fixed-order \NNLOQCD and \NLOEW corrections to the Drell--Yan process known for many years, see \eg~Ref.~\cite{Alioli:2016fum} and references therein;
inclusive cross sections and rapidity distributions in the
threshold limit at \NNNLOQCD extracted from the $pp\to H$ results at this
order~\cite{Ahmed:2014cla,Ahmed:2014uya};
dominant factorizable corrections at $\mathcal{O}(\alpha_s \alpha)$ (\NLOQE11) known
differentially~\cite{Dittmaier:2015rxo} for the off-shell process including the leptonic decay;
\NNLOQCD computations matched to parton shower available using the
MiNLO method~\cite{Karlberg:2014qua}, SCET resummation \cite{Alioli:2015toa}
and the UN${}^2$LOPS technique \cite{Hoeche:2014aia};

Most recently a novel method to match parton showers to \NNLOQCD, the MINNLO${}_\text{PS}$ method,
was proposed and applied also to Drell--Yan production~\cite{Monni:2019whf}.

The calculation of the inclusive cross section at \NNNLOQCD (for an off-shell photon) was recently completed~\cite{Duhr:2020seh}, exhibiting rather sizeable corrections that are traced back to accidental cancellations occurring at \NNLOQCD.
Completing the inclusive as well as the fully differential \NNNLOQCD computation
for Z- and W-boson exchange is an important step for phenomenological studies.

Very recently, the total cross section for the $q\bar{q}$ channel at \NLOQE11 was computed for on-shell Z bosons~\cite{Bonciani:2019nuy}.
Corrections at this order, but considering the photonic part of the EW corrections only, \ie
\NLOmixQED11,
were completed for the on-shell production of the Z both for the inclusive cross section~\cite{deFlorian:2018wcj},
and differentially using the nested soft-collinear subtraction formalism~\cite{Delto:2019ewv}.

The inclusive production cross section for $W$ and $Z$ bosons has been measured at the LHC using the leptonic decays of the vector bosons.
The precision in those measurements already reached the barrier of the luminosity uncertainty $\sim 2\%$, which is not easy to further improve.
For example, the most precise measurement of the $W$ and $Z$ bosons integrated fiducial cross sections
is for the $\sqrt{s} = 7$~\TeV sample having
$\Delta\sigma_W/\sigma_W = 1.87\%$ and $\Delta\sigma_Z/\sigma_Z = 1.82\%$ uncertainty, with the luminosity uncertainty ($\sim1.8\%$)
accounting for most of it~\cite{Aaboud:2016btc}.

While the inclusive integrated cross sections have been already measured and compared fairly well with the present theoretical predictions, this is not the case for
differential distributions.
A key observable, both for precision studies as well as for new physics searches, is the transverse momentum of the vector bosons, as well as the $\phi^\ast$ variable which is
also very much related with the momentum of the vector boson, without being affected by the leptons' energy scale uncertainties.
For neutral Drell--Yan, those have been measured both at $8$ and $13$~\TeV with precision that is $<1\%$ for $0<p_\mathrm{T}<20$~\GeV~\cite{Aad:2015auj,Sirunyan:2017igm,Sirunyan:2019bzr,Aad:2019wmn}.
These spectra are known at \NNLOQCD\!+\,\NNNLL accuracy~\cite{Bizon:2018foh}, which exhibit a substantial reduction in scale uncertainties and a good perturbative convergence.
Special runs with very low pileup have been taken from the LHC, with the experiments targeting to measure with $<1\%$ accuracy in very fine grained bins
the low $p_\mathrm{T}<20$~\GeV part of the distribution.
Data with high $p_\mathrm{T}$ vector bosons could be used to study the strong coupling constant at \NNLOQCD accuracy.

\item[$V/\gamma+j$:] \textit{LH17 status:}
both $Z+j$~\cite{Ridder:2015dxa,Boughezal:2015ded,Boughezal:2016isb,Boughezal:2016yfp,Gehrmann-DeRidder:2017mvr} and
$W+j$~\cite{Boughezal:2015dva,Boughezal:2016dtm,Boughezal:2016yfp,Gehrmann-DeRidder:2017mvr}
completed through \NNLOQCD including leptonic decays,
via antenna subtraction and $N$-jettiness slicing;
also $\gamma+j$ available through \NNLOQCD from calculations using the $N$-jettiness slicing~\cite{Campbell:2016lzl};
all processes of this class, and in particular their ratios, investigated in great
detail in Ref.~\cite{Lindert:2017olm}, combining \NNLOQCD predictions with full NLO EW and
leading \NNLOEW effects in the Sudakov approximation, including also approximations for leading
\NLOQE11 effects, devoting particular attention to error estimates and
correlations between the processes.

More recently, an independent calculation of the $\gamma+$jet calculation at \NNLOQCD was completed using the antenna subtraction method and employing the hybrid isolation prescription~\cite{Chen:2019zmr}.
A study of the impact of different photon isolation criteria for different processes, in cluding $\gamma+$jet, was performed in Sect.~\ref{cha:nnlo}.\ref{sec:SM_phiso} of this report.

\item[$V+\geq2j$:] \textit{LH17 status:}
\NLOQCD computations known for $V+2j$ final states
in QCD~\cite{Campbell:2002tg,Campbell:2003hd}
and EW~\cite{Oleari:2003tc}
production modes,
for $V+3j$~\cite{Ellis:2009zw,Berger:2009zg,KeithEllis:2009bu,Berger:2009ep,Melnikov:2009wh,Berger:2010vm},
for $V+4j$~\cite{Berger:2010zx,Ita:2011wn} and for $W+5j$~\cite{Bern:2013gka};
\NLOEW corrections known~\cite{Denner:2014ina} including merging and showering~\cite{Kallweit:2014xda,Kallweit:2015dum}.

Advances in event-generation methods achieved multi-jet merged predictions for this process with up to 9 jets at LO~\cite{Hoeche:2019rti}.
First progress in the computation of two-loop amplitudes for the $W+2$ jet process were reported
in Ref.~\cite{Hartanto:2019uvl}, which are an important ingredient for predictions at \NNLOQCD.

\item[$V+b\bb$:] \textit{LH17 status:} Known at \NLOQCD for a long time~\cite{FebresCordero:2006nvf,Campbell:2008hh,Cordero:2009kv,Badger:2010mg}, and matched
to parton showers~\cite{Frederix:2011qg,Oleari:2011ey,Krauss:2016orf,Bagnaschi:2018dnh};
\NLOQCD for $Wb\bb j$ calculated with parton shower matching~\cite{Luisoni:2015mpa};
$Wb\bb$ with up to three jets computed at \NLOQCD in Ref.~\cite{Anger:2017glm}

In Ref.~\cite{Hoche:2019ncc} a novel technique was proposed for the combination of
multi-jet merged simulations in the five-flavour scheme with calculations for the
production of b-quark associated final states in the four-flavour scheme.
This multi-jet merging in a variable flavour number scheme was applied to
$Z+b\bb$ production at the LHC~\cite{Hoche:2019ncc}.

\item[$VV'$:] \textit{LH17 status:} \NNLOQCD publicly available for all vector-boson
pair production processes with full leptonic decays, namely
$WW$~\cite{Gehrmann:2014fva,Grazzini:2016ctr},
$ZZ$~\cite{Cascioli:2014yka,Grazzini:2015hta},
$WZ$~\cite{Grazzini:2016swo,Grazzini:2017ckn},
$Z\gamma$~\cite{Grazzini:2013bna,Grazzini:2015nwa},
$W\gamma$~\cite{Grazzini:2015nwa}, using the $q_T$ subtraction method within
the \Matrix framework~\cite{Grazzini:2017mhc},
exploiting the {\sc VVamp} implemenation~\cite{Gehrmann:2015ora} of the
2-loop amplitudes~\cite{Caola:2014iua,Gehrmann:2015ora};
\NNLOQCD results calculated for $Z\gamma$~\cite{Campbell:2017aul} and
$ZZ$~\cite{Heinrich:2017bvg} in the $N$-jettiness method;
\NLOQCD corrections to the loop-induced $gg$ channels
computed for $ZZ$~\cite{Caola:2015psa} and
$WW$~\cite{Caola:2015rqy} involving full off-shell leptonic dacays,
based on the two-loop amplitudes of Refs.~\cite{Caola:2015ila,vonManteuffel:2015msa};
also interference effects
with off-shell Higgs contributions studied~\cite{Caola:2016trd,Campbell:2016ivq};
NLO EW corrections known for
all vector-boson pair production processes including full leptonic
decays~\cite{Denner:2014bna,Denner:2015fca,Biedermann:2016yvs,Biedermann:2016guo, Biedermann:2016lvg, Biedermann:2017oae}, extensively validated between several automated tools in Ref.~\cite{Bendavid:2018nar};
combination of \NLOQCD and \NLOEW corrections, including $\gamma$-induced channels, discussed
for all $2\ell2\nu$ final states in Ref.~\cite{Kallweit:2017khh}.

All relevant leptonic $ZZ$ signatures were discussed at \NNLOQCD in Ref.~\cite{Kallweit:2018nyv} within the \Matrix framework, for the first time including the same-flavour $2\ell2\nu$ channels that mix the double-resonant processes $ZZ$ and $WW$. The finding of Ref.~\cite{Kallweit:2017khh}
that interference effects between the two resonant core processes are tiny, is confirmed at \NNLOQCD.

The combination of \NNLOQCD and \NLOEW corrections to all massive diboson processes has been discussed
in \cite{Kallweit:2019zez}, with a focus on the high-energy tails of distributions where both
types of corrections typically become significant. The calculation is carried out in
in the \Matrix+\OpenLoops framework, using amplitudes from \OpenLoops~\cite{Buccioni:2019sur}.

The presumably leading \NNNLOQCD corrections, namely the NLO QCD for the loop-induced gluon--gluon
channel, have been calculated within this framework as well for $ZZ$~\cite{Grazzini:2018owa}
and $WW$~\cite{Grazzini:2020stb} production with full leptonic decays,
including for the first time also the loop-induced quark--gluon channels.

Both achievements, \ie \NNLOQCD\!+\,\NLOEW\!+\,\NLOQCD($gg$ channel)\wleptdecays{}, were announced to
become public in an upcoming new release of \Matrix~\cite{Kallweit:2019zez,Grazzini:2020stb}.

In Ref.~\cite{Chiesa:2018lcs} anomalous triple-gauge-boson interactions were studied for all massive vector-boson pair production processes at NLO QCD, on top of combined \NLOQCD\!+\,\NLOEW predictions. This calculation was carried out using amplitudes from \Recola\,2~\cite{Denner:2017wsf}.

\NNLOQCD corrections to off-shell $WW$ production were for the first time matched to a parton shower
in Ref.~\cite{Re:2018vac}. The calculation combines the fully differential \NNLOQCD corrections
available in \Matrix~\cite{Grazzini:2017mhc} and the MiNLO computation for $WWj$ production
(WWJ-MiNLO) of \Ref.~\cite{Hamilton:2016bfu}.

\item[$VV'+j$:] \textit{LH17 status:}
\NLOQCD corrections known for many years~\cite{Dittmaier:2007th,Campbell:2007ev,Bern:2008ef,Dittmaier:2009un,Binoth:2009wk,Binoth:2010nha,Campanario:2010hp,Campanario:2009um,Campbell:2012ft,Campbell:2015hya};
\NLOEW corrections available for some on-shell processes, with subsequent leptonic decays treated
in NWA~\cite{Li:2015ura,Yong:2016njr}; full \NLOEW corrections
including decays in reach of the automated tools.

\item[$VV'+\geq2j$:] \textit{LH17 status:}
\NLOQCD corrections known for
the EW~\cite{Jager:2006zc,Jager:2006cp,Bozzi:2007ur,Jager:2009xx,Denner:2012dz,Campanario:2013eta,Campanario:2017ffz}
and QCD~\cite{Melia:2010bm,Melia:2011dw,Greiner:2012im,Campanario:2013qba,Campanario:2013gea,Campanario:2014ioa,Campanario:2014dpa,Campanario:2014wga}
production modes;
\NLOQCD calculated for $WW+3j$~\cite{Cordero:2015hem};
full \NLOSM corrections (\NLOQCD, \NLOEW and mixed \NLOgen) available for $W^+W^++2j$ production
with full leptonic decays~\cite{Biedermann:2016yds,Biedermann:2017bss}.

For same-sign $WW$ scattering, an event generator based on the Monte Carlo program \Powheg
in combination with the matrix-element generator \Recola was presented in Ref.~\cite{Chiesa:2019ulk}.

A first calculation of complete \NLOQCD\!+\,\NLOEW predictions for $WZ$ scattering with full leptonic
decays was presented in Ref.~\cite{Denner:2019tmn}, based on the automated matrix element generators
\OpenLoops and \Recola and in-house Monte Carlo integrators.
Again sizeable negative EW corrections are found,
as previously for like-sign $WW$ scattering, confirming the expectation that
large EW corrections are an intrinsic feature of VBS processes
(and the corresponding event selections) at the LHC.

Furthermore, a strategy how to measure VBS at LHCb was proposed in Ref.~\cite{Pellen:2019ywl}.
An extensive study of the same-sign WW process was performed in Ref.~\cite{Ballestrero:2018anz} with a systematic comparison of approximations both in the fiducial and a more inclusive phase-space setup as well as an investigation into the impact that the details of the matching to parton showers entail.

\item[$VV'V''$:] \textit{LH17 status:}
\NLOQCD corrections known for many years~\cite{Hankele:2007sb,Binoth:2008kt,Campanario:2008yg,Bozzi:2009ig,Bozzi:2010sj,Bozzi:2011wwa,Bozzi:2011en,Campbell:2012ft},
also in case of $W\gamma\gamma j$~\cite{Campanario:2011ud};
\NLOEW corrections available for the
on-shell processes involving
three~\cite{Nhung:2013jta,Yong-Bai:2015xna,Yong-Bai:2016sal,Hong:2016aek,Dittmaier:2017bnh}
and two~\cite{Chong:2014rea,Yong:2017nag} massive vector bosons, some with leptonic
decays in NWA; $V\gamma\gamma$ processes with full leptonic decays calculated
at \NLOQCD and \NLOEW accuracy~\cite{Greiner:2017mft}.

A first off-shell \NLOEW calculation for $WWW$ production was presented
in \cite{Schonherr:2018jva},
and in Ref.~\cite{Dittmaier:2019twg} an independent calculation of \NLOQCD and \NLOEW
corrections to that process was performed, both involving full leptonic decays with all
off-shell effects, spin correlations and interferences.
More details on both calculations are discussed in Sect.~\ref{cha:nnlo}.\ref{sec:SM_WWW}
of this report.

\item[$\gamma\gamma$:] \textit{LH17 status:}
\NNLOQCD results for $\gamma\gamma$ production calculated by using
$q_T$ subtraction~\cite{Cieri:2015rqa,Catani:2018krb}, and by using
$N$-jettiness subtraction in the MCFM framework~\cite{Campbell:2016yrh};
\NNLOQCD also available within the public \Matrix program~\cite{Grazzini:2017mhc};
$q_T$ resummation computed at \NNLL~\cite{Cieri:2015rqa};
\NLOEW corrections available for $\gamma\gamma$~\cite{Bierweiler:2013dja,Chiesa:2017gqx}.

While the massless \NLOQCD corrections to the loop-induced $gg$ channel had been known
for decades, top-quark mass effects to this contribution, which become particularly
important at the $t\tb$ threshold and above, were first computed
in Ref.~\cite{Maltoni:2018zvp}. In Ref.~\cite{Chen:2019fla} an independent calculation
of these \NLOQCD corrections was presented, where NRQCD methods have been applied
to resum the bound-state effects to obtain a more reliable description of the
threshold region.

This process remains an important ingredient in Higgs measurements at Run II.
Prospects for \NNNLOQCD corrections remain
closely connected with differential Higgs and Drell--Yan production at \NNNLOQCD.

\item[$\gamma\gamma+j$:] \textit{LH17 status:}
\NLOQCD corrections calculated long ago~\cite{DelDuca:2003uz,Gehrmann:2013aga}, later also for
$\gamma\gamma+2j$~\cite{Gehrmann:2013bga,Badger:2013ava,Bern:2014vza} and
$\gamma\gamma+3j$~\cite{Badger:2013ava};
photon isolation effects studied at \NLOQCD~\cite{Gehrmann:2013aga};
\NLOEW corrections available for $\gamma\gamma j(j)$~\cite{Chiesa:2017gqx};

Very recently, a first \NLOQCD calculation for the EW production mode for $\gamma\gamma+2j$
was presented in Ref.~\cite{Campanario:2020xaf}, studying also anomalous gauge coupling
effects via bosonic dimension-6 and 8 operators.

At high transverse momentum it may also be interesting to have \NNLOQCD
predictions for $\gamma\gamma+j$. Given that the two-loop amplitudes are of comparable complexity
as those of $\gamma\gamma\gamma$ production, and that subtraction methods to address
this process are available, the \NNLOQCD calculation may be considered in reach in the nearer
future.

\item[$\gamma\gamma\gamma$:] \textit{LH17 status:}
\NLOQCD corrections calculated in Ref.~\cite{Bozzi:2011en} and later in MCFM~\cite{Campbell:2014yka}.

Very recently, as the first \NNLOQCD calculation for a $2\to3$ process
in hadronic collisions, three-photon production has been computed in the
sector-improved residue subtraction formalism~\cite{Chawdhry:2019bji}.
The involved two-loop amplitudes 
apply a leading-colour approximation, but the impact of the neglected contributions
is estimated to be phenomenologically irrelevant.

\end{itemize}

\subsection{Top quark associated processes}
An overview of the status of top quark associated processes is given in Table~\ref{tab:SM_wishlist:wlTJ}
\begin{table}
  \renewcommand{\arraystretch}{1.5}
\setlength{\tabcolsep}{5pt}
  \centering
  \begin{tabular}{lll}
    \hline
    \multicolumn{1}{c}{process} & \multicolumn{1}{c}{known} &
    \multicolumn{1}{c}{desired} \\
    \hline
    $pp\to t\tb$ &
    \begin{tabular}{l}
      \NNLOQCD\!+\,\NLOEW \\
      \NLOQCD{ }(w/ decays, off-shell effects)\\
      \NLOEW{ }(w/ decays, off-shell effects)\\
    \end{tabular} &
    \begin{tabular}{l}
      \NNLOQCD{ }(w/ decays)
    \end{tabular} \\
    \hline
    $pp\to t\tb+j$ &
    \begin{tabular}{l}
      \NLOQCD{ }(w/ decays) \\
      \NLOEW
    \end{tabular} &
    \begin{tabular}{l}
      \NNLOQCD\!+\,\NLOEW{ }(w/ decays)
    \end{tabular} \\
    \hline
    $pp\to t\tb+2j$ &
    \begin{tabular}{l}
      \NLOQCD{ }(w/ decays)
    \end{tabular} &
    \begin{tabular}{l}
      \NLOQCD\!+\,\NLOEW{ }(w/ decays)
    \end{tabular} \\
    \hline
    $pp\to t\tb+Z$ &
    \begin{tabular}{l}
      \NLOQCD\!+\,\NLOEW{ }(w/ decays)
    \end{tabular} &
    \begin{tabular}{l}
      \NNLOQCD\!+\,\NLOEW{ }(w/ decays)
    \end{tabular} \\
    \hline
    $pp\to t\tb+W$ &
    \begin{tabular}{l}
      \NLOQCD \\
      \NLOEW
    \end{tabular} &
    \begin{tabular}{l}
      \NNLOQCD\!+\,\NLOEW{ }(w/ decays)
    \end{tabular} \\
    \hline
    $pp\to t/\tb$ &
    \begin{tabular}{l}
      \NNLOQCD{*}(w/ decays)
    \end{tabular} &
    \begin{tabular}{l}
      \NNLOQCD\!+\,\NLOEW{ }(w/ decays)
    \end{tabular} \\
    \hline
  \end{tabular}
  \caption{Precision wish list: top quark  final states. \NNLOQCD$^{*}$ means a
   calculation using the structure function approximation.}
  \label{tab:SM_wishlist:wlTJ}
  \renewcommand{\arraystretch}{1.0}
\end{table}

\begin{itemize}[leftmargin=2cm]

\item[$t\tb$:] \textit{LH17 status:}
Fully differential \NNLOQCD computed for on-shell top-quark pair production~\cite{Czakon:2015owf,Czakon:2016ckf,Czakon:2016dgf}, also available as {\tt fastNLO} tables~\cite{Czakon:2017dip};
polarized two-loop amplitudes known~\cite{Chen:2017jvi};
combination of \NNLOQCD and \NLOEW corrections performed~\cite{Czakon:2017wor};
also multi-jet merged predictions with \NLOEW corrections available~\cite{Gutschow:2018tuk};
resummation effects up to \NNLL computed~\cite{Beneke:2011mq,Cacciari:2011hy,Ferroglia:2013awa,Broggio:2014yca,Kidonakis:2015dla,Pecjak:2016nee};
top quark decays known at \NNLOQCD~\cite{Gao:2012ja,Brucherseifer:2013iv};
$W^+W^- b\bar{b}$ production with full off-shell effects calculated
at \NLOQCD~\cite{Denner:2010jp,Denner:2012yc,Bevilacqua:2010qb,Heinrich:2013qaa}
including leptonic $W$ decays, and in the lepton plus jets channel~\cite{Denner:2017kzu};
full \NLOEW corrections for leptonic final state available~\cite{Denner:2016jyo};
calculations with massive bottom quarks available at
\NLOQCD~\cite{Frederix:2013gra,Cascioli:2013wga};\\
\NLOQCD predictions in NWA matched to parton shower~\cite{Campbell:2014kua}, and multi-jet merged
for up to 2 jets in \Sherpa~\cite{Hoeche:2014qda} and \Herwig\,7.1~\cite{Bellm:2017idv};
$W^+W^- b\bar{b}$ at \NLOQCD first matched to a parton shower in the \Powheg
framework~\cite{Garzelli:2014dka}; improved resonance treatment, called ``resonance aware matching'',
done in \Powhegboxres~\cite{Jezo:2015aia,Jezo:2016ujg}; alternative approach
in the \Powheg \NLOgen\!+\,PS framework presented in Ref.~\cite{Buonocore:2017lry};\\
Various aspects of the definition and extraction of the top quark mass studied in Refs.~\cite{Beneke:2016cbu,Butenschoen:2016lpz,Kawabata:2016aya,Hoang:2017btd,Hoang:2017kmk,Heinrich:2017bqp,Bevilacqua:2017ipv,Corcella:2017rpt,Ravasio:2018lzi}.

The first resummed calculation for a final state with non-trivial colour structure
at \NNLOgen\!+\,\NNLL was performed for (boosted) top-quark pair production in Ref.~\cite{Czakon:2018nun}.
This computation combined state-of-the-art \NNLOQCD predictions with double resummation of
threshold logarithms arising from soft-gluon emissions and of small-mass logarithms.

A new \NNLOQCD calculation of on-shell $t\bar{t}$ has been performed within the
\Matrix framework using the extension of the $q_T$ subtraction method to massive coloured
final states, inclusively in Ref.~\cite{Catani:2019iny}, and comparing to
CMS data~\cite{Sirunyan:2018wem} for single- and double-differential distributions
in Ref.~\cite{Catani:2019hip}. The \NNLOQCD soft function for $t\tb$ production
was also calculated independently in Ref.~\cite{Angeles-Martinez:2018mqh}.

The impact of double-differential top distributions from CMS on
parton distribution functions was studied in Ref.~\cite{Czakon:2019yrx}.
In Ref.~\cite{Behring:2019iiv}, for the first time, the complete set of \NNLOQCD corrections to
top-pair production and decay at hadron colliders was calculated in the NWA for both intermediate
top quarks and $W$ bosons.

In terms of experimental precision, the inclusive $t\tb$ production cross section has been measured by ATLAS and CMS Collaborations
at $\sqrt{s} = 7, 8$ and $13$~\TeV.
\begin{table}[t]
\begin{center}
\begin{tabular}{cccc}
$\sqrt{s}$ & ATLAS & CMS & NNLO\!+\,\NNLL \\\hline
7 TeV  & 3.9\% & 3.6\% & 4.4\%     \\
8 TeV  & 3.6\% & 3.7\% & 4.1\%   \\
13 TeV & 4.4\% & 5.3\% & 5.5\%   \\
\end{tabular}
\caption{
Experimental uncertainty $\Delta\sigma_{t\tb}/\sigma_{t\tb}$ on the inclusive $t\tb$ production cross section measurements, in the electron-muon channel at the
LHC~\cite{Aad:2014kva, Aaboud:2016pbd, Khachatryan:2016mqs, Khachatryan:2016kzg} compared to the precision of the \NNLOgen\!+\,\NNLL calculation~\cite{Czakon:2013goa,Czakon:2011xx}.}
\label{tab:SM_wishlist:expTTbar}
\end{center}
\end{table}
The measurements' uncertainty is a bit smaller than the corresponding theoretical calculations (Table~\ref{tab:SM_wishlist:expTTbar}).
Significant part of the theory uncertainty  stems from PDFs and $\alpha_\mathrm{s}$. For example, in the $13$~\TeV calculation
$\sim 4.2$\% comes from PDFs and $\alpha_\mathrm{s}$, while the scale uncertainty is about $3.5$\%.
In terms of the total production cross section, the measurements agree with the theoretical predictions within the quoted uncertainties.
However, a long standing problem related to the discrepancy observed in the transverse momentum distribution of the top-quarks
(see \eg~\cite{Khachatryan:2015oqa, Aad:2015mbv}), still misses from a complete resolution though higher
order effects~\cite{Czakon:2016ckf} seem to alleviate at least partially the effect.
Understanding the origin of this discrepancy is important for the LHC physics programme
since it affects directly (or indirectly) many physics analyses for
which the $t\tb$ is a dominant source of background.

\item[$t\tb\,j$:] \textit{LH17 status:}
\NLOQCD corrections calculated for on-shell top quarks~\cite{Dittmaier:2007wz,Melnikov:2010iu,Melnikov:2011qx},
also matched to parton showers~\cite{Kardos:2011qa,Alioli:2011as}; full off-shell decays
included at \NLOQCD~\cite{Bevilacqua:2015qha,Bevilacqua:2016jfk}; studies for top-quark mass extraction
based on $t\tb j$ production performed~\cite{Fuster:2017rev,Bevilacqua:2017ipv};
\NLOEW corrections known~\cite{Gutschow:2018tuk}.

\item[$t\tb+\geq2j$:]\textit{LH17 status:}
\NLOQCD corrections to $t\tb jj$  known for many years~\cite{Bevilacqua:2010ve,Bevilacqua:2011aa};
$t\tb jjj$ at \NLOQCD calculated~\cite{Hoche:2016elu} using \Sherpa{}+\OpenLoops.

\item[$t\tb+b\bb$:]\textit{LH17 status:}
\NLOQCD corrections to $t\tb b\bb$ with massless bottom quarks known
for a long time~\cite{Bredenstein:2009aj,Bevilacqua:2009zn,Bredenstein:2010rs};
\NLOQCD with massive bottom quarks and matching to a parton shower investigated~\cite{Cascioli:2013era,Jezo:2018yaf}.

Recently, a first \NLOQCD study for $t\bar{t}b\bar{b}$ production in association with a
light jet~\cite{Buccioni:2019plc} was performed using \OpenLoops in combination with \Sherpa and \Munich,
in order to address the large uncertainties associated with the modelling of extra QCD radiation
in $t\bar{t}b\bar{b}$ events and thus to validate this modelling in $t\bar{t}b\bar{b}$ generators.

\item[$t\tb V$:] \textit{LH17 status:}
\NLOQCD corrections to $t\tb Z$ including NWA decays considered~\cite{Rontsch:2014cca,Rontsch:2015una};  \NLOQCD corrections to $t\tb\gamma\gamma$ production matched to parton shower, focussing
on the top-quark polarisation observables~\cite{vanDeurzen:2015cga};
\NLOEW and \NLOQCD corrections to $t\tb Z/W/H$ computed
within \MadgraphaMCatNLO~\cite{Frixione:2015zaa};
dedicated studies on complete \NLOSM corrections for $t\tb W$ and $t\tb t\tb$ production~\cite{Frederix:2017wme};
resummed calculations up to \NNLL to $t\tb W$~\cite{Broggio:2016zgg}
and $t\tb Z$~\cite{Broggio:2017kzi} production.

Very recently, a comprehensive study of top-quark pair hadro-production in association with a
heavy boson was performed in order to provide the most complete predictions to
date~\cite{Broggio:2019ewu}.
Here the full \NLOSM corrections for the processes $t\tb Z/W/H$ were combined with
soft gluon emission corrections resummed to \NNLL accuracy.

Studies in this process class will help to improve the constraints on anomalous EW couplings in the top quark sector during Run II.

\item[$t$/$\tb$:] \textit{LH17 status:}
Fully differential \NNLOQCD corrections for the dominant $t$-channel production process completed
in the structure function approximation, for stable top quarks~\cite{Brucherseifer:2014ama} and
later including top-quark decays to \NNLOQCD accuracy in the NWA~\cite{Berger:2016oht,Berger:2017zof};
\NLOQCD corrections to $t$-channel electroweak $W+bj$ production available
within MG5\_aMC@NLO~\cite{Papanastasiou:2013dta}, beforehand in MC@NLO~\cite{Frixione:2005vw};
\NLOQCD corrections to single-top production in the $t, s$ and $tW$ channels also
available in \Sherpa~\cite{Bothmann:2017jfv}
and in \Powheg~\cite{Alioli:2009je,Re:2010bp};
top-quark mass determination from single-top hadro-production performed
in Ref.~\cite{Alekhin:2016jjz}, and in~\cite{Martini:2017ydu} using the Matrix Element Method
at \NLOQCD.

A calculation of $t$-channel single-top plus jet production matched to a parton shower was
completed in the \MiNLO method~\cite{Carrazza:2018mix}.
The impact of EW corrections and a QCD parton shower on the $t$-channel signature was discussed
in Ref.~\cite{Frederix:2019ubd}.
Soft-gluon resummation at \NLLone for single-top production was investigated in the $t$-channel~\cite{Cao:2019uor} and the $s$-channel modes~\cite{Sun:2018byn}.
A first calculation of single top-quark production in the $s$-channel and decay at \NNLOQCD,
neglecting the colour correlation between the light and heavy quark lines and applying the NWA was
achieved in Ref.~\cite{Liu:2018gxa}.
A \NLOQCD calculation for single top-quark production in association with two jets was recently performed in Ref.~\cite{Molbitz:2019uqz}.

\end{itemize}

\subsection*{Acknowledgements}
We thank all of our colleagues who provided us with valuable input to update the wishlist.
S.\,K.~ is supported by the ERC Starting Grant REINVENT-714788.

\let\NLO\undefined
\let\NLL\undefined
\let\NLOH\undefined
\let\NLOQ\undefined
\let\NLOE\undefined
\let\NLOHone\undefined
\let\NLOQone\undefined
\let\NLOEone\undefined
\let\NLOQE\undefined
\let\LOQ\undefined
\let\NLOQonetb\undefined
\let\NLOQtb\undefined
\let\NLOQmtsix\undefined
\let\NLOQzzero\undefined
\let\NLOQoneVBF\undefined
\let\NLOQVBF\undefined
\let\NLOQoneDIS\undefined
\let\NLOQDIS\undefined
\let\NLOEoneVBF\undefined
\let\NLOQoneVBFstar\undefined
\let\NLOQVBFstar\undefined
\let\NLOEoneVBFstar\undefined
\let\NLOggHVtb\undefined

\let\xs\undefined
\let\tb\undefined
\let\bb\undefined
\let\qb\undefined
\let\VdkL\undefined
\let\VdkQ\undefined
\let\VdkLNWA\undefined
\let\VdkQNWA\undefined

\let\wodecay\undefined
\let\wdecay\undefined
\let\wodecays\undefined
\let\wdecays\undefined
\let\wleptdecays\undefined

\let\VdkALLNWA\undefined
\let\VdkALL\undefined
\let\tdk\undefined
\let\tdkNWA\undefined
\let\TVdkALLNWA\undefined

\let\MadgraphaMCatNLO\undefined
\let\Herwig\undefined
\let\Powheg\undefined
\let\Powhegboxres\undefined
\let\PowhegboxVtwo\undefined
\let\GoSam\undefined
\let\Recola\undefined
\let\OpenLoops\undefined
\let\MadLoop\undefined
\let\Matrix\undefined
\let\Munich\undefined
\let\Geneva\undefined
\let\Sherpa\undefined
\let\NNLOjet\undefined
\let\MiNLO\undefined
\let\NLOX\undefined

\newcommand{\Herwig}{H\protect\scalebox{0.8}{ERWIG}\xspace}
\newcommand{\Pythia}{P\protect\scalebox{0.8}{YTHIA}\xspace}
\newcommand{\Sherpa}{S\protect\scalebox{0.8}{HERPA}\xspace}
\newcommand{\Rivet}{R\protect\scalebox{0.8}{IVET}\xspace}
\newcommand{\Professor}{P\protect\scalebox{0.8}{ROFESSOR}\xspace}
\newcommand{\eps}{\varepsilon}
\newcommand{\mc}[1]{\mathcal{#1}}
\newcommand{\mr}[1]{\mathrm{#1}}
\newcommand{\mb}[1]{\mathbb{#1}}
\newcommand{\tm}[1]{\scalebox{0.95}{$#1$}}

\section{NNLO nTuples for Drell-Yan~\protect\footnote{
  D.~Ma\^{\i}tre,
  R.~R\"ontsch}{}}

\label{sec:SM_nnlontuples}
In this contribution we investigate the possibility of using nTuples for the Drell-Yan process at NNLO.  

\subsection{Introduction}

NNLO calculations are extremely computationally intensive, typically involving the use of a complex code base that requires a significant amount of insider knowledge. In contrast, the output of the calculation is rather simple: a set of phase space momenta and associated weights, which are usually presented through differential cross sections. In order to leverage the CPU cost of producing an NNLO calculation and to facilitate its usage, one can store the simple output of the NNLO simulation in so-called nTuple files. This approach has been used extensively at NLO \cite{Bern:2013zja}. In a past Les Houches workshop the approach of using nTuple files for an NNLO process has been investigated \cite{Bendavid:2018nar} using {\tt NNLOjet} \cite{Currie:2017eqf}.

In this contribution we investigate the Drell-Yan process using a private code written by F.~Caola, K.~Melnikov, and R.~R\"ontsch, which we refer to as {\tt CaMeRo-DY}.
This program implements the so-called nested soft-collinear scheme (NSS) for the subtraction of infrared singularities in colour-singlet hadroproduction at NNLO~\cite{Caola:2017dug,Caola:2018pxp,Delto:2019asp,Caola:2019nzf}.  Since the way real subtraction is organised in this program is different from  {\tt NNLOjet}, it is interesting to see whether it leads to a similar, smaller or larger storage cost.
 Using the same approach and software with two very different NNLO programs also demonstrates the versatility of the strategy and its robustness.     

\subsection{Information extraction}
\label{sec:SM_nnlontuples:section1}

In this section we explain how the information required for the nTuple files is extracted. The general strategy is described in Ref.~\cite{Maitre:2018gua}, so we limit our discussion here to some aspects that were particularly relevant for this process and the program we considered in this study. 

For a fixed scale one can decompose the weight into several components, according to the pdf function that enters it:

\begin{eqnarray}
w &=& \sum\limits_{i \leq j}^{n_s} pdf_i pdf_j w_{ij},
\end{eqnarray}
where the index $i$ in $pdf_i$ enumerates all possible combinations of argument $x$ and parton flavour of the pdf. For example, if we have 
\begin{eqnarray}
w = c_{u\bar u} f_u(x_1) f_{\bar u}(x_2) + c_{gg} f_g(x_1) f_g(x_2) + c_{ug}f_u(x_1) f_g(x_2),
\end{eqnarray}
we could set 
\begin{eqnarray}
pdf_0&=& f_u(x_1),\\
pdf_1&=& f_{\bar u}(x_2),\\
pdf_2&=& f_g(x_1),\\
pdf_3&=& f_g(x_2),
\end{eqnarray}  
and we would have 
\begin{eqnarray}\label{eq:SM_nnlontuples:example}
w_{01} = c_{u\bar u}\,,w_{23}=c_{gg}\,,w_{03}=c_{ug} ,
\end{eqnarray}
and all other $w_{ij}$ vanish.

We modify the LHAPDF library so that it keeps track of the number of times the pdf evaluation has been called; that is, each call is aware of its placement in the succession of calls to the library. We call this (zero-based) index $n_e$. We need to specify the maximum number of pdf calls for a single weight, $n_s$. We can also specify for each thread two "stride numbers" $s_A$ and $s_B$ and the pdf call will return 0 unless either 
\begin{equation}
n_e \mod n_s = s_A\qquad\mbox{or} \qquad n_e \mod n_s = s_B.
\end{equation}
We can illustrate what happens in the Table~\ref{tab:SM_nnlontuples:nnlontuples.example} . We assume the calls for the pdf evaluations are issued in the order they are written in Eq.~(\ref{eq:SM_nnlontuples:example}). With $n_s=4$ we get six combinations of pdfs, so we need six individual threads. Each thread is represented by a column in Table~\ref{tab:SM_nnlontuples:nnlontuples.example}. The two values for $s_A$ and $s_B$ are displayed at the top of the table and the values returned by our modified LHAPDF library are shown for each pdf call. The value of the weight that is calculated by the NNLO code is given at the bottom of the table.
For the first evaluation of the expression in Eq.~\ref{eq:SM_nnlontuples:example}, the values of $n_e$ will be $n_e=0,1,2,3$ for $pdf_{0,1,2,3}$ respectively. The value returned e.g. for $pdf_1$ will be 1 if any of $s_A$ or $s_B$ are equal to $1$ (modulo $n_s=4$).
\begin{table}[t]
\begin{center}
\begin{tabular}{c|cccccc}
$s_A$   &  0 & 0 & 0 & 1 & 1 & 2 \\
$s_B$   &  1 & 2 & 3 & 2 & 3 & 3 \\\hline
$pdf_0$ &  1 & 1 & 1 & 0 & 0 & 0   \\
$pdf_1$ &  1 & 0 & 0 & 1 & 1 & 0 \\
$pdf_2$ &  0 & 1 & 0 & 1 & 0 & 1 \\
$pdf_3$ &  0 & 0 & 1 & 0 & 0 & 1 \\\hline
$w$     &  $c_{u\bar u} $ & 0 & $c_{ug}$ & 0 & 0 & $c_{gg}$ 
\end{tabular}
\end{center}
\caption{Six threads for $n_s=4$.}\label{tab:SM_nnlontuples:nnlontuples.example}
\end{table}
Evaluating the same expression again for a new phase space point would give the same picture. After an exploratory run one can realise that only three out of the six threads contribute any information and one can only run these threee threads.
 
One might wonder what happens when $n_s$ is not set correctly (for example in an exploratory step). If it is set too high there is no problem beyond wasting resources. For the first phase space point the pdfs are evaluated with $n_e=0,1,2,3$, as illustrated in Table~\ref{tab:SM_nnlontuples:nnlontuples.example_ns5a}.
\begin{table}[t]
\begin{center}
\begin{tabular}{c|cccccccccc}
$s_A$   &  0 & 0 & 0 & 0 & 1 & 1 & 1 & 2 & 2 & 3 \\
$s_B$   &  1 & 2 & 3 & 4 & 2 & 3 & 4 & 3 & 4 & 4  \\\hline
$pdf_0$ &  1 & 1 & 1 & 1 & 0 & 0 & 0 & 0 & 0 & 0   \\
$pdf_1$ &  1 & 0 & 0 & 0 & 1 & 1 & 1 & 0 & 0 & 0 \\
$pdf_2$ &  0 & 1 & 0 & 0 & 1 & 0 & 0 & 1 & 1 & 0 \\
$pdf_3$ &  0 & 0 & 1 & 0 & 0 & 1 & 0 & 1 & 0 & 1 \\\hline
$w$     &  $c_{u\bar u} $ & 0 & $c_{ug}$  & 0 & 0 & 0 & 0 & $c_{gg}$ & 0& 0 
\end{tabular}
\end{center}
\caption{Ten threads for $n_s=5$, first evaluation.}\label{tab:SM_nnlontuples:nnlontuples.example_ns5a}
\end{table}
All components are identified separately.
For the second phase space point the pdfs are evaluated with $n_e=4,5,6,7$ (see Table~\ref{tab:SM_nnlontuples:nnlontuples.example_ns5b}).
\begin{table}[t]
\begin{center}
\begin{tabular}{c|cccccccccc}
$s_A$   &  0 & 0 & 0 & 0 & 1 & 1 & 1 & 2 & 2 & 3 \\
$s_B$   &  1 & 2 & 3 & 4 & 2 & 3 & 4 & 3 & 4 & 4  \\\hline
$pdf_0$ &  0 & 0 & 0 & 1 & 0 & 0 & 1 & 0 & 1 & 1   \\
$pdf_1$ &  1 & 1 & 1 & 1 & 0 & 0 & 0 & 0 & 0 & 0 \\
$pdf_2$ &  1 & 0 & 0 & 0 & 1 & 1 & 1 & 0 & 0 & 0 \\
$pdf_3$ &  0 & 1 & 0 & 0 & 1 & 0 & 0 & 1 & 1 & 0 \\\hline
$w$     &  0 & 0 & 0 & $c_{u\bar u} $ & $c_{gg}$ & 0 & 0 & 0 & $c_{ug}$ & 0 
\end{tabular}
\end{center}
\caption{Ten threads for $n_s=5$, second evaluation.}\label{tab:SM_nnlontuples:nnlontuples.example_ns5b}
\end{table}
The individual components are still isolated, albeit in different threads.

The situation is worse if we choose $n_s$ too small. This is illustrated in Table~\ref{tab:SM_nnlontuples:nnlontuples.example_ns3} where $n_s=3$ while we need at least $n_s=4$.
\begin{table}[t]
\begin{center}
\begin{tabular}{c|cccccc}
$s_A$   &  0 & 0 & 1  \\
$s_B$   &  1 & 2 & 2   \\\hline
$pdf_0$ &  1 & 1 & 0   \\
$pdf_1$ &  1 & 0 & 1  \\
$pdf_2$ &  0 & 1 & 1  \\
$pdf_3$ &  1 & 1 & 0  \\\hline
$w$     &  $c_{u\bar u}+c_{ug} $ & $c_{gg}$  &  0 
\end{tabular}
\end{center}
\caption{Six threads for $n_s=3$.}\label{tab:SM_nnlontuples:nnlontuples.example_ns3}
\end{table}
The individual components are not isolated.

\subsection{Using multiple values per LHAPF calls}

LHAPDF gives the option to get the pdf values for all 13 parton flavours in one call. This makes it difficult to know which components are actually used to calculate the weight. For example consider the two options
\begin{eqnarray}
w_1 &=& f_u(x_1)f_g(x_2) c, \\
w_2 &=& \left(f_u(x_1)+ f_c(x_1) \right) f_g(x_2).
\end{eqnarray}
They would be difficult to discriminate if all values of $f_\star(x_1)$ are used from the same call to LHAPDF. The solution is to run two threads, one where the result of the pdf call is replaced by one, and one where the pdf result is returned as calculated by LHAPDF:
    
\begin{eqnarray}
w_1^{(one)}  &=& c, \\
w_2^{(one)}  &=& \left(1+ 1 \right) f_g(x_2), \\
w_1^{(true)} &=& f_u(x_1)f_g(x_2) c, \\
w_2^{(true)} &=& \left(f_u(x_1)+ f_c(x_1) \right) f_g(x_2).
\end{eqnarray}
One can then compare the ratio $r$ of the weight from both threads and if it matches
\begin{equation}
r=\frac{w_1^{(true)}}{w_1^{(one)}} = f_u(x_1)f_g(x_2),
\end{equation}
then we can conclude that the weight calculated by the NNLO program must be of the form $w_1$.
On the other hand, if it matches 
\begin{equation}
r=\frac{w_2^{(true)}}{w_2^{(one)}} = \frac{1}{2} \left(f_u(x_1)+ f_c(x_1) \right) f_g(x_2),
\end{equation}
then we can conclude that the weight is of the form $w_2$. In extracting nTuples from a NNLO program it is necessary to compare the ratio $r$ for each phase space point with a list of different combinations of pdf values that can arise. This list can differ from process to process or from program to program. 

If the ratio does not correspond to any known combination, one can find out what the combination is using another feature of our modified LHAPDF library. We can instruct the library to set all pdf results to zero except for one flavour. This allows one to collect information on each of the $13\times 13$ combinations
\begin{equation}
\sum\limits_{i,j=q,\bar q ,g} c_{ij} f_i(x_1)f_j(x_2)\;.
\end{equation} 
This is normally only necessary in an exploratory step to find out which combinations of pdf is used by the program and needs to be added to the list of potential combinations described above. 

The coefficients $c_{ij}$ above can be further split into coefficients of the logarithm of the renormalisation scale, and terms that are independent of the renormalisation scale. The procedure is similar to the one for separating the $c_{ij}$ coefficients themselves: we run copies of the program with different scale setting and compare the weights to solve for the coefficients. This strategy has been detailed in ref.~\cite{Maitre:2018gua} and we will not repeat it here.

\subsection{NNLO Drell-Yan}

We used the strategy outlined above to extract nTuples for the {\tt CaMeRo-DY} program. To obtain a complete prediction we need to run several different parts. Each part can have a different number of contributions and different number $n_s$. The program execution is separated in ten separate parts: 
\begin{itemize}
\item the LO contribution, labelled {\tt LO};
\item three NLO contributions, corresponding to the real corrections {\tt R}, the virtual corrections {\tt V}, and the subtraction counterterms {\tt S};
\item five NNLO contributions, corresponding to the double-real {\tt RR}, real-virtual {\tt RV}, and double virtual {\tt VV} corrections, as well as subtraction counterterms {\tt sub12}, {\tt subv}, and {\tt sub124}.
\end{itemize}
The contributions with real radiation are all regulated against IR singularities according to the NSS, while all virtual corrections have their IR poles subtracted according to the well-known formula of Catani~\cite{Catani:1998bh}. Thus the ten contributions listed above are all finite. We refer the reader to Refs.~\cite{Caola:2017dug,Caola:2019nzf} for a detailed description of the NSS.

We summarize our findings in Table \ref{tab:SM_nnlontuples:nnlontuples.nthreads}.
We list the ten parts detailed above, together with their order in $\alpha_s$, the number $n_s$, and the total number of threads used for each part. The number of threads is the number we used for the proof-of-principle study, and this number could be reduced for production runs. We also show the storage needed  per event for each part. This has been calculated by generating 5000 events for each part and storing the information in an NNLO nTuple file. In the second-to-last column we give an order of magnitude estimate of the number of events required to generate differential cross sections. These numbers will vary from application to application but here we are  only interested in orders of magnitude. We note that if one is interested in cross sections -- either total or fiducial -- then these numbers can be 2-3 orders of magnitude smaller. 
The last column of Table~\ref{tab:SM_nnlontuples:nnlontuples.nthreads} shows the storage needed for the estimated number of events. We can see that the total storage required is approximately 6 TB.
\begin{table}[t]
\begin{center}
\begin{tabular}{c|cccccc|}
  part & $\alpha_s$ order & $n_s$ & threads number & kB/event & events  & storage  \\
  & & & & & needed$/10^{9}$& needed [TB]\\\hline
{\tt LO    } & 0  &      8 &   1206 &                 0.10  &         0.1 &        0.0  \\
{\tt V     } & 1  &      2 &     15 &                 0.09  &         0.1 &        0.0  \\
{\tt R     } & 1  &      2 &     39 &                 0.33  &         0.5 &        0.2  \\
{\tt S     } & 1  &      6 &    185 &                 0.28  &         0.1 &        0.0  \\
{\tt VV    } & 2  &      8 &    246 &                 0.12  &         0.1 &        0.0  \\
{\tt RV    } & 2  &      8 &    246 &                 0.36  &         1.0 &        0.3  \\
{\tt sub12 } & 2  &      8 &   2834 &                 0.95  &         0.5 &        0.4  \\
{\tt sub124} & 2  &      8 &    846 &                 2.04  &         2.0 &        3.8  \\
{\tt subv  } & 2  &      8 &    246 &                 0.18  &         0.1 &        0.0  \\
{\tt RR    } & 2  &      6 &     86 &                 0.39  &         5.0 &        1.8  \\
\hline {\tt total } &   &  &  &   &   &        6.6  \\
\hline

\end{tabular}
\end{center}
\caption{$\alpha_S$ power, $n_s$, number of threads, size per event, estimated number of events needed and estimated storage needed for each calculation part.}\label{tab:SM_nnlontuples:nnlontuples.nthreads}
\end{table}

\subsection*{Conclusions}
In this contribution we investigated the prospect of using nTuples to store the phase space configurations and weights produced by {\tt CaMeRo-DY} for NNLO Drell-Yan production. This study shows that the technique outlined in Ref.~\cite{Maitre:2018gua} can be used with different NNLO programs with very different structures. We estimated the number of events that would be necessary to generate useful NNLO nTuples for the Drell-Yan process and found that around 10 TB should be sufficient. While not a small amount of storage, it is in the same ball park as the storage needed for high multiplicity NLO processes. It would be interesting to compare the storage need of {\tt CaMeRo-DY} with those of other NNLO programs for the same process, in order to assess which subraction scheme is most promising when the focus is on minimising the nTuple storage. We leave this as a topic for a future study.

\let\Herwig\undefined
\let\Pythia\undefined
\let\Sherpa\undefined
\let\Rivet\undefined
\let\Professor\undefined
\let\eps\undefined
\let\mc\undefined
\let\mr\undefined
\let\mb\undefined
\let\tm\undefined


\section{Towards APPLfast interpolation grids at NNLO in QCD for LHC observables~\protect\footnote{
  D.~Britzger,
  A.~Gehrmann-De Ridder,
  T.~Gehrmann,
  E.W.N.~Glover,
  C.~Gwenlan,
  J.~Hessler,
  A.~Huss,
  J.~Pires,
  K.~Rabbertz,
  M.R.~Sutton
  }{}}

\label{sec:SM_applfastLHCgrids}

The current status of the production of fast interpolation grids for
the NNLO QCD calculation of the differential cross section for both
ATLAS and CMS jet measurements is discussed. Fast interpolation grids
in both fastNLO and APPLgrid format for the NNLO jet cross section at
HERA have previously been made available as part of the APPLfast
project~\cite{Britzger:2019kkb}.  Here, the generation of cross
sections for the LHC is discussed.

\subsection{Motivation}
\label{sec:SM_applfastLHCgrids:motivation}

The exceptional performance of the LHC during Run~I and~II enabled the
experiments to collect large datasets of proton-proton collision
events. Moreover, the use of advanced experimental techniques has
resulted in an improved and solid understanding of the experimental systematic
uncertainties. As an example, the latest measurements of inclusive jet
production from the CMS~\cite{Khachatryan:2016mlc} and
ATLAS~\cite{Aaboud:2017dvo,Aaboud:2017wsi} collaborations report, over a wide range
in jet-$p_\mathrm{T}$, a systematic uncertainty at the 5\% level and a
statistical uncertainty at the subpercent level.

To exploit the full potential of such high-precision experimental
data, they must be compared with theoretical predictions of similar or
better accuracy. Searches for new phenomena at the LHC so far have
been unsuccessful. However, history has shown that evidence of new
phenomena is often observed as small deviations from highly precise
predictions. Therefore, it is mandatory to reduce the theoretical
uncertainties of the predictions for LHC processes to the percent
level. Such effort demands, in particular, improvements in perturbative
calculations by advancing to next-to-next-to-leading order (NNLO) or
higher accuracy. The perturbative calculations for the inclusive
jet~\cite{Currie:2016bfm,Czakon:2019tmo} and
dijet~\cite{Currie:2017eqf,Gehrmann-DeRidder:2019ibf} cross sections
have recently been completed. First phenomenological studies using
these new results promise a significant reduction in the theoretical
uncertainty, in particular, residual scale dependencies as estimates
for missing higher orders are reduced. Further uncertainties
originating from the parton distribution functions (PDFs) of the
proton or the knowledge of the strong coupling constant $\alpha_s$ are
not considered here, but can potentially also be reduced once the
predictions at NNLO are available in a format suitable for PDF and
$\alpha_s$ fits.

Computations at NNLO presently require computing times of the order of
${\cal O}(10^5-10^6)$ CPU hours to perform the numerical integration
over the kinematics of the final state particles, necessary for the
cancellation of infrared and collinear singularities. At the end of
such a computation, the results are obtained in the form of binned
histograms and derived for a specific set of input parameters. As a
result, the full computation needs to be performed from scratch to generate a new prediction
if one desires a change to the input PDF or $\alpha_s$ value. 
For this reason, fast grid techniques~\cite{Adloff:2000tq,
  Carli:2005ji, Carli:2010rw, Kluge:2006xs, Britzger:2012bs} were
developed which allow the storage of the weights of the higher-order
calculation on an interpolation grid. In this way, the convolution of
the weights with the PDFs or $\alpha_s$ can be performed \textit{a
  posteriori}, such that the time consuming QCD computation needs to
be performed only once. This approach allows for an extremely fast
reevaluation of fixed observables with different PDF sets,
and~$\alpha_s$ values, enabling their systematic determination from
data available from collider experiments.

These techniques were first used for jet production at NLO in $ep$
collisions at HERA~\cite{Adloff:2000tq,Chekanov:2005nn}, and were then
extended to jet production at NLO at the LHC with the
APPLgrid~\cite{Carli:2005ji,Carli:2010rw} and the fastNLO
projects~\cite{Kluge:2006xs,Britzger:2012bs}. More recently, NNLO
grids for inclusive jet production at HERA~\cite{Andreev:2017vxu}
became publicly available in~\cite{Britzger:2019kkb}. NNLO grids for
top quark pair differential distributions at the LHC have been
produced in~\cite{Czakon:2017dip}.

This contribution reviews the status of the APPLfast project which
aims for a combined interface between the NNLO calculations performed
with the NNLOJET parton-level generator~\cite{Gehrmann:2018szu}, and
the fast grid technology in both fastNLO and APPLgrid formats. The
resulting APPLfast interface allows for the production of full NNLO
grids for any of the processes implemented within the NNLOJET
program. For this article, the focus is on dijet measurements from 
both the ATLAS and CMS collaborations.

\subsection{Interpolation grids for dijet cross sections}
\label{sec:SM_applfastLHCgrids:dijets}
In this section, the production of the fast interpolation grids is
briefly described. Some closure tests to determine how accurately 
the fast convolution using the grid technology reproduces the native 
NNLO predictions are also presented.  As an exemplary process, dijet
production as measured by the ATLAS collaboration at
$\sqrt{s}=7\,\text{TeV}$ and by CMS at $\sqrt{s}=8\,\text{TeV}$ is
considered.

\subsubsection{ATLAS and CMS dijet cross sections}
ATLAS has measured the double-differential dijet cross section at
$\sqrt{s}=7\,\text{TeV}$ as a function of the dijet mass $m_{12}$ and
half the rapidity separation
$y^* = \left|y_1 - y_2\right|/2$~\cite{Aad:2013tea}.  Jets are
defined using the anti-$k_t$ jet algorithm with a distance measure of
\mbox{$R=0.4$} with a range in $m_{12}$ spanning from 260.0\,GeV to
5040\,GeV. The NNLO predictions have been calculated for all data
points, but are presented only for the interval $1.0<y^*<1.5$ in
the following.  This interval is representative of the sample as a
whole, including both low- and high-mass $m_{12}$ regions,
$510<m_{12}<4640$\,GeV.  Due to the choice of $y^*$ interval, the
calculations receive relevant contributions from both low-$x$ and
high-$x$ regions of the PDFs, where $x$ is the parton fractional
momentum of the proton. This regime is particular challenging for grid
interpolation techniques since an accurate reproduction of the cross
section requires the interpolation to reproduce the behaviour of the
parton distributions at both low-$x$ and high-$x$ simultaneously. The
calculations for ATLAS dijets have been performed using the NNPDF3.1
PDF set, with the chosen scales $\mu_R=\mu_F=m_{12}$.

For CMS, the NNLO cross sections are calculated for dijet production
at $\sqrt{s}=8\,\text{TeV}$ triple-differentially as a function of the
average transverse momentum of the leading two jets $p_T^\text{avg}$,
$y^*$, and the longitudinal boost of the dijet system
$y_\text{boost} = \left|y_1 + y_2\right|/2$~\cite{Sirunyan:2017skj}.
Jets are defined using the anti-$k_t$ jet algorithm with a distance
measure of \mbox{$R=0.7$}. The following discussion focuses on the
\emph{central} rapidity interval $y^*<1.0$ and
$y_\text{boost}<1.0$, since this spans the largest range in
$p_T^\text{avg}$: $133<p_T^\text{avg}<1784$\,GeV. The calculation for
the CMS cross section employs the CT14nnlo PDF set and
two scales for $\mu_R$ and $\mu_F$, either
$\mu=p_T^\text{jet1}\exp({0.3\cdot y^*})$ as used by CMS or
$\mu=m_{12}$ as recommended in~\cite{Currie:2017eqf}.  Here, some care
must be taken when performing closure tests, since neither scale
choice coincides with the measured observable variable and therefore the
scale interpolation might require more support nodes per observable
bin (for technical details see e.g.\ Ref.~\cite{Britzger:2019kkb}).
Moreover, for purely technical reasons, the reference cross
section from NNLOJET for closure tests in this case is derived with a
mixed scale setting of $\mu_R=p_T^\text{jet1}\exp({0.3\cdot y^*})$ and
$\mu_F=m_{12}$. For comparison to the CMS data, of course, either of
the two central scale choices should be used for $\mu_R$ and $\mu_F$.

All presented interpolated cross sections are individually optimised
for numerical accuracy. Altogether, each of the dijet cross sections
is based on several 100,000 hours of single-core CPU compute time.

\subsubsection{Validation and closure tests}
\begin{figure}[t]
  \centering
  \begin{tabular}{ccc}
  \includegraphics[width=0.32\textwidth]{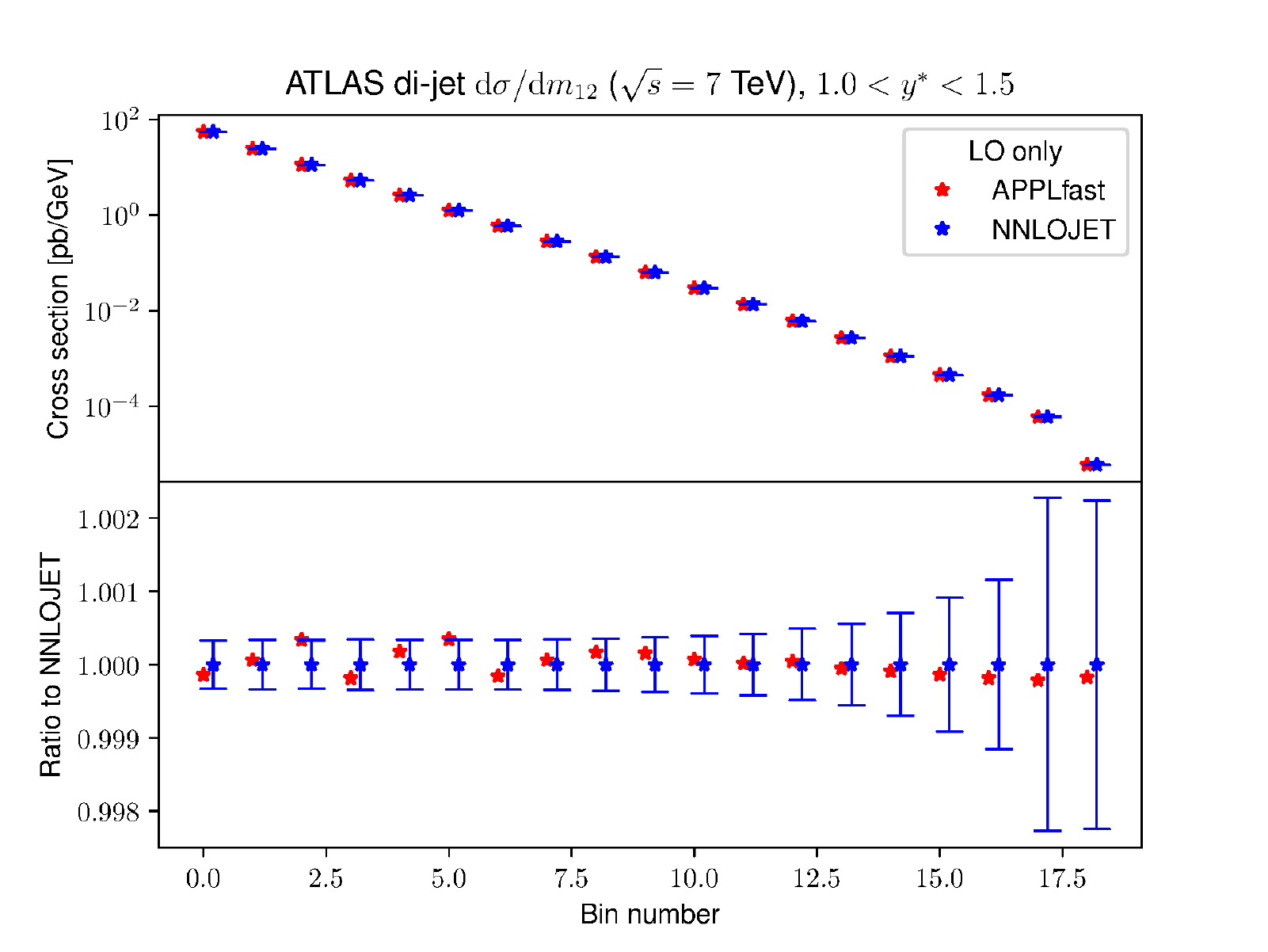} &
  \includegraphics[width=0.32\textwidth]{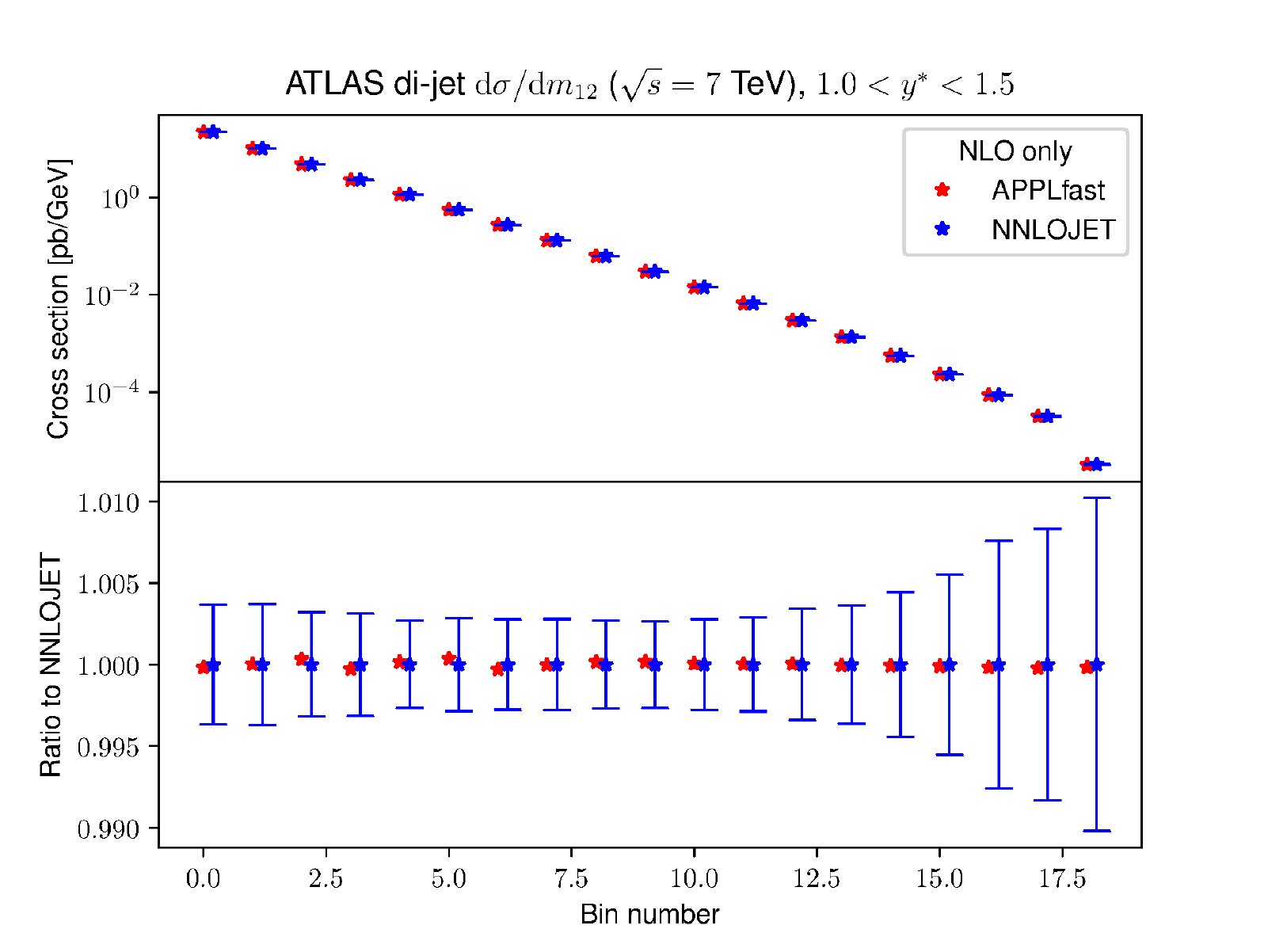} &
  \includegraphics[width=0.32\textwidth]{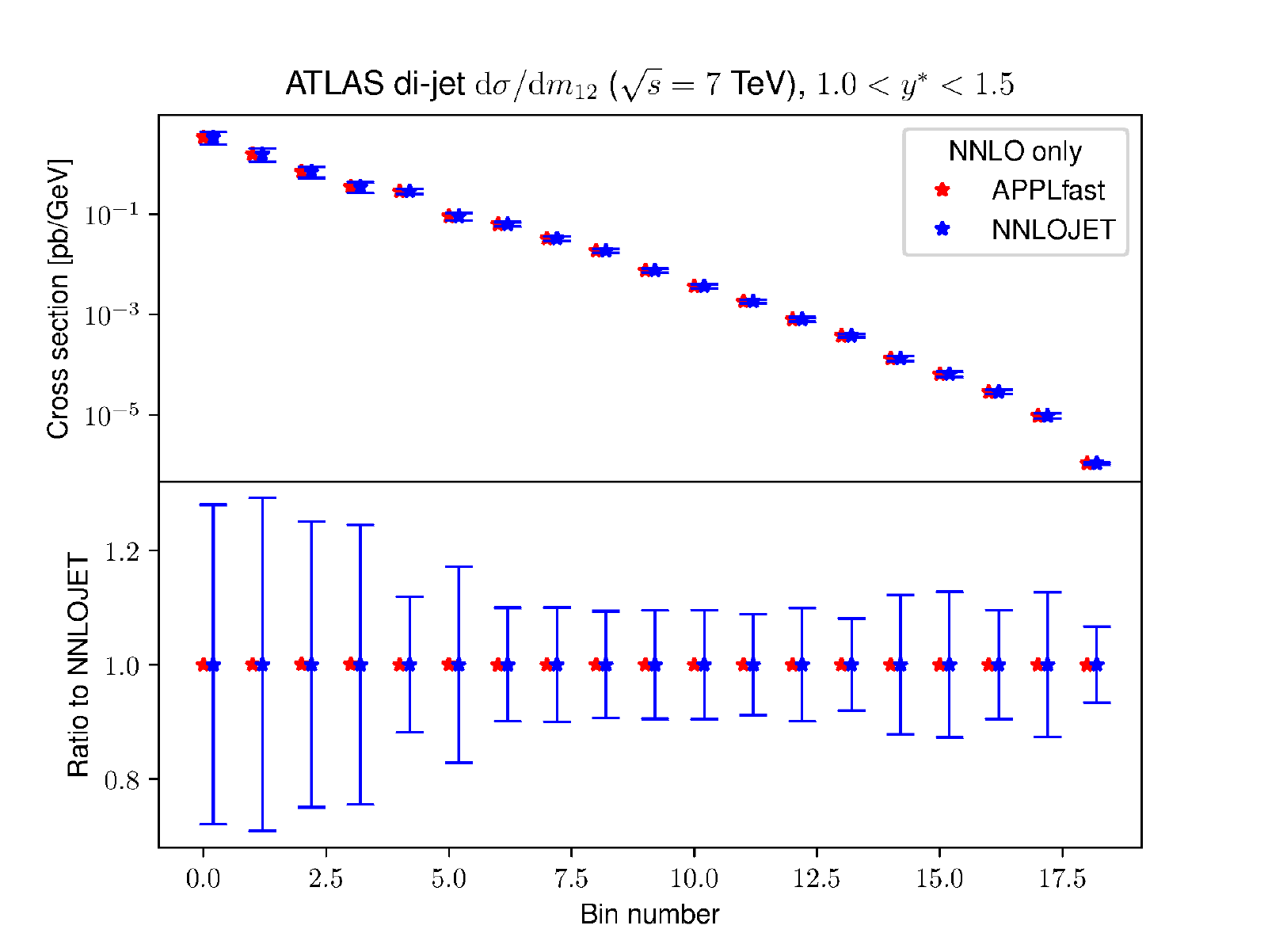}
  \end{tabular}
  \caption{Ratio of APPLfast to the original NNLOJET cross sections
    for the LO, NLO\_only, and NNLO\_only contribution in the $y^*$
    bin $1.0<y^*<1.5$ of the ATLAS dijet measurement
    $\text{d}\sigma^2/\text{d}m_{12}\text{d}y^*$ at
    $\sqrt{s}=7\,\text{TeV}$. The $x$-axis denotes the bin number of
    the respective $m_{12}$ bin ranging from
    $510<m_{12}<4640\,\text{GeV}$. Deviations from unity of the red
    symbols in the ratio are a measure of the interpolation bias. The
    vertical error bars represent the statistical uncertainty of the
    calculations, which are identical for NNLOJET and the respective
    APPLfast result.}
  \label{fig:SM_applfastLHCgrids:ATLASdijetClosureContr}
\end{figure}
Figure~\ref{fig:SM_applfastLHCgrids:ATLASdijetClosureContr} illustrates the closure of the
fast convolution using the interpolation grid for the calculation of
the contributions at each order for the ATLAS dijet calculations. The
lower panel shows the ratio of the cross sections from the
interpolation grid to the original result from NNLOJET for the LO, NLO
only and NNLO only contributions. For this comparison the
same PDF must be used in both cases.

Ideally, the grid technique should reproduce exactly the full
calculation in each bin and at each order, since both are based on
identical sets of parton-level events and weights as given by
NNLOJET\@. Since some approximation is involved, small deviations are
unavoidable. Most importantly, the approximation respectively
interpolation bias must be kept under control and should be smaller
than the common statistical uncertainty of the numerical integrations
in NNLOJET\@. Here, the target is for biases from the fast convolution 
to be at, or below the per mille level.

The left plot of Fig.~\ref{fig:SM_applfastLHCgrids:ATLASdijetClosureContr} shows that the 
cross section reproduced from the fast grid convolution at LO is in 
agreement with the native NNLOJET cross section in all bins to an 
accuracy much  better than $1\,\permil$. 
The statistical precision varies between 0.5 and
$2\,\permil$. %
At NLO, as visible from the middle plot, the statistical uncertainty
is somewhat larger with values between $2.5\,\permil$ and
1\%. Interpolation biases for the NNLO only contribution remain 
smaller then  $0.1\permil$. %
For the NNLO only contribution shown in the right plot of
Fig.~\ref{fig:SM_applfastLHCgrids:ATLASdijetClosureContr} the statistical uncertainty
increases to between 10 to 20\%. This may be sufficient precision 
for the full calculation for some use cases, as the NNLO 
contribution is a comparatively small component of the total 
cross section from the NNLO factor of $\alpha_s^2$. 
Whether to invest significant additional 
CPU resources to further improve the statistical precision
would need to be determined on a case by case basis, 
taking in to consideration the required uncertainty on the 
total cross section at NNLO, together with any additional 
uncertainties that might need to be considered, 
coupled with the intended use case.

\begin{figure}[t]
  \centering
  \begin{tabular}{cc}
  \includegraphics[width=0.48\textwidth]{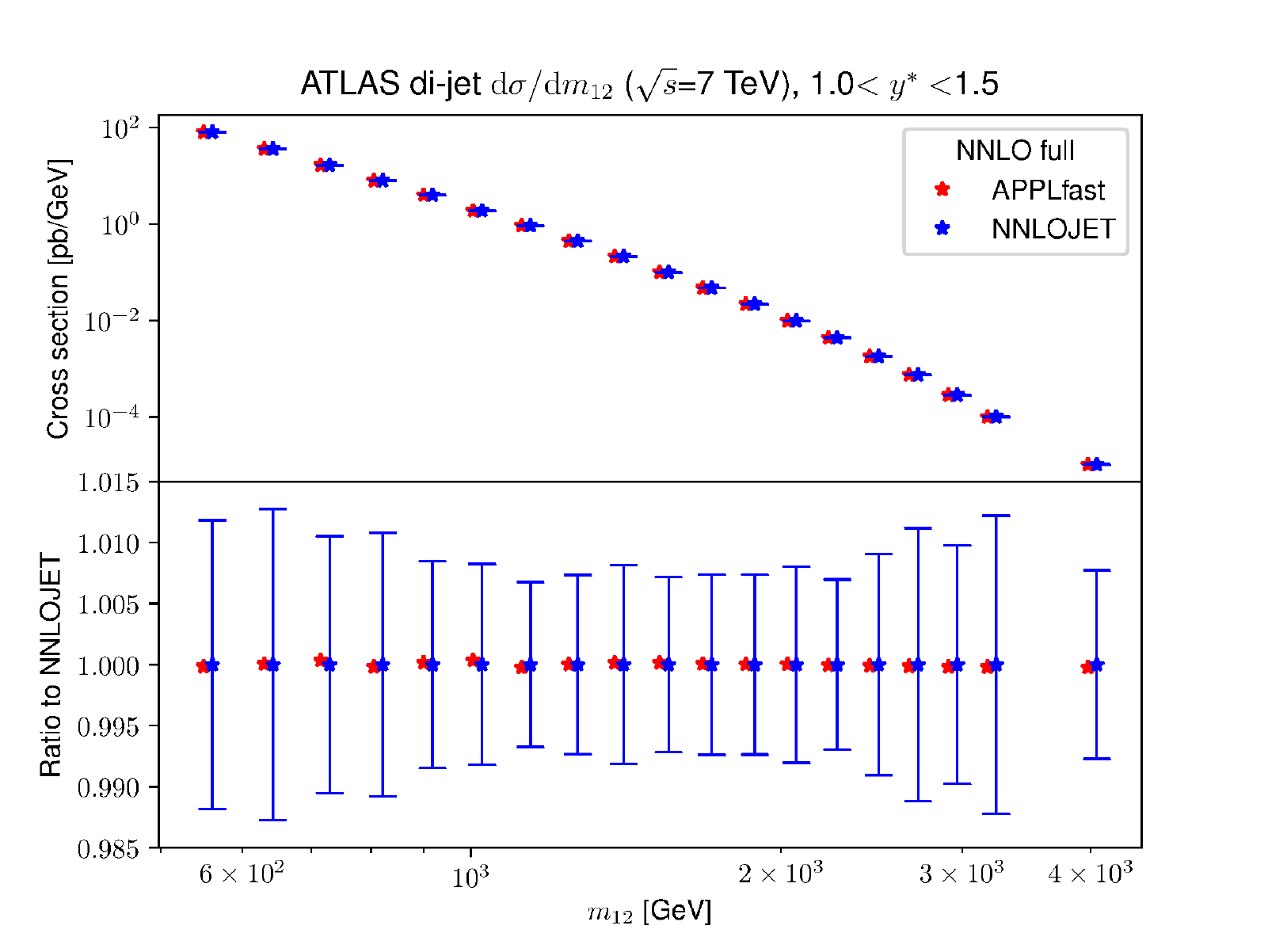} &
  \includegraphics[width=0.48\textwidth]{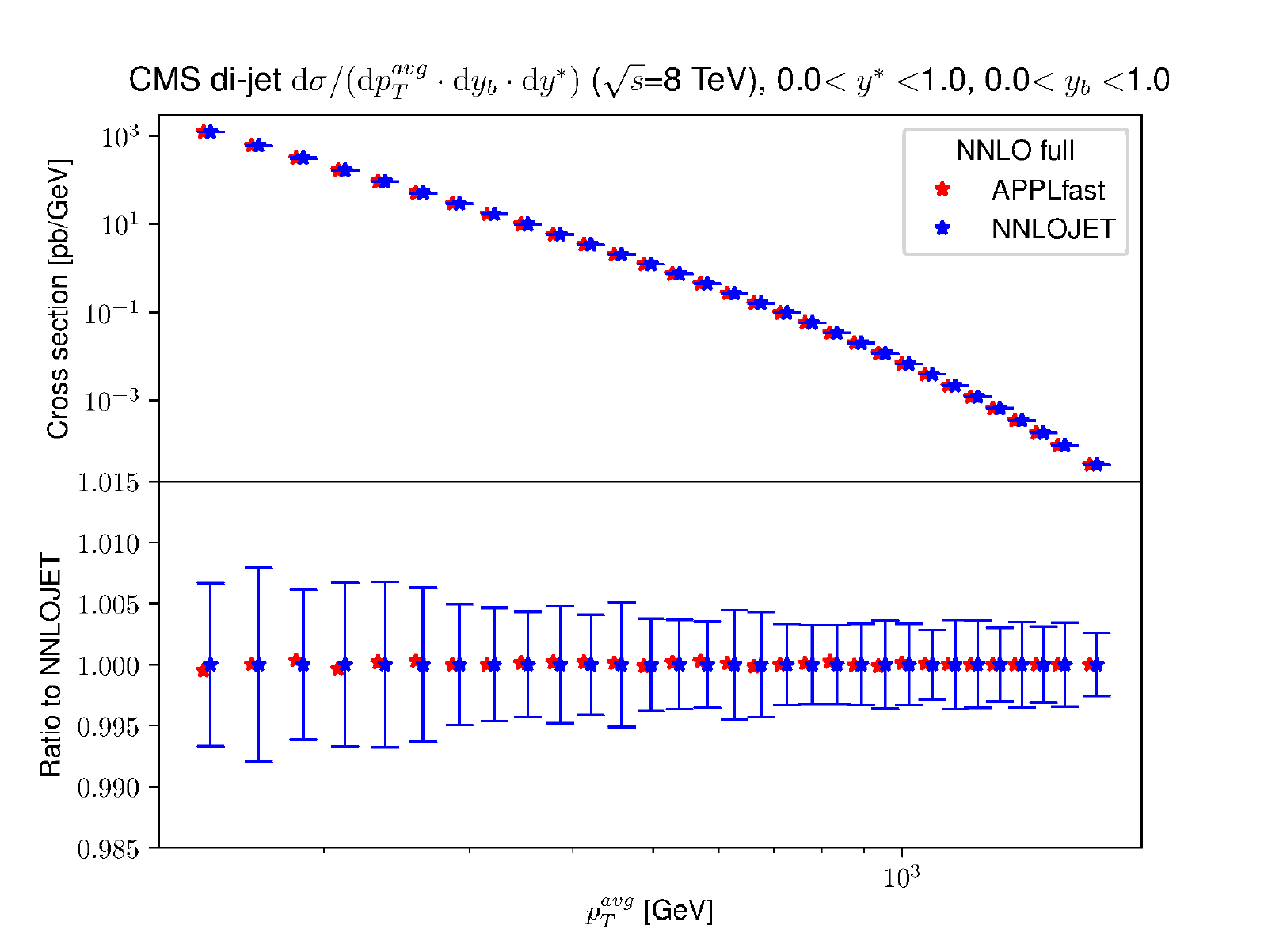}
  \end{tabular}
  \caption{Cross sections and ratio of APPLfast to the original
    NNLOJET cross sections for all orders summed up in the $y^*$ bin
    $1.0<y^*<1.5$ of the ATLAS dijet measurement
    $\text{d}^2\sigma/\text{d}m_{12}\text{d}y^*$ at
    $\sqrt{s}=7\,\text{TeV}$, and in the $y^*, y_\text{boost}$ bin
    $0.0<y^*<1.0$ and $0.0<y_\text{boost}<1.0$ of the CMS dijet
    measurement
    $\text{d}^3\sigma/\text{d}p_T^\text{avg}\text{d}y^*\text{d}y_\text{boost}$
    at $\sqrt{s}=8\,\text{TeV}$ (right). Deviations from unity of the
    red symbols in the ratio are a measure of the interpolation
    bias. The vertical error bars represent the statistical
    uncertainty of the calculations, which are identical for NNLOJET
    and the respective APPLfast result.}
    \label{fig:SM_applfastLHCgrids:closure1}
\end{figure}
\begin{figure}[t]
  \centering
  \begin{tabular}{cc}
  \includegraphics[width=0.48\textwidth]{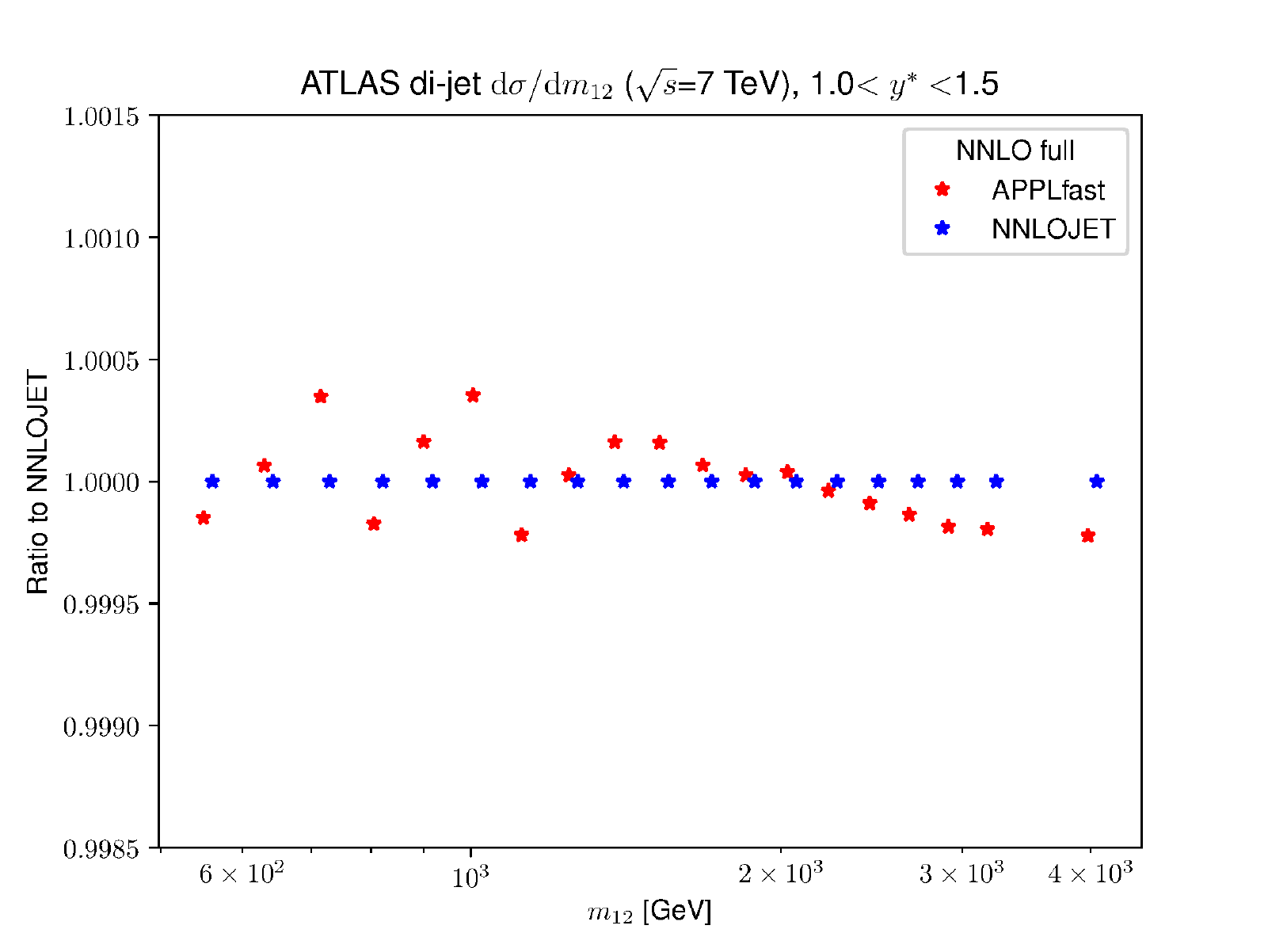} &
  \includegraphics[width=0.48\textwidth]{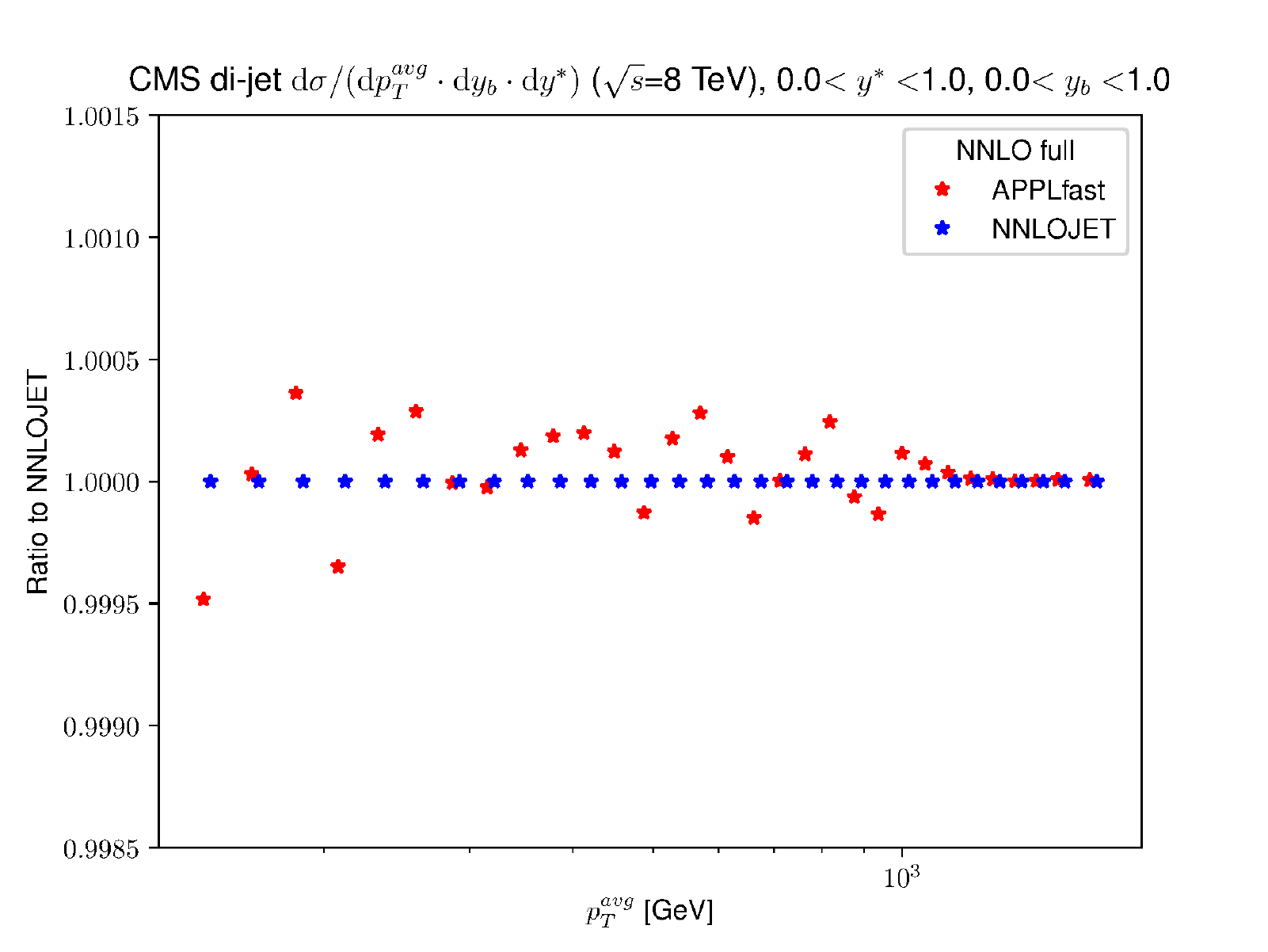}
  \end{tabular}
  \caption{Same ratio as Figure~\ref{fig:SM_applfastLHCgrids:closure1} but without
    statistical uncertainties. Interpolation biases visible as
    deviations from unity are at a negligible level of $0.5\,\permil$
    or smaller.}
    \label{fig:SM_applfastLHCgrids:closure2}
\end{figure}

In Figure~\ref{fig:SM_applfastLHCgrids:closure1} the closure of the full NNLO dijet cross
section for both the ATLAS and the CMS dijet predictions are shown.
It is observed that the interpolation grids reproduce the original
NNLOJET cross sections at full NNLO to better than
$\approx 0.5\,\permil$ over the complete range of the ATLAS and the
CMS dijet cross sections.
In order to study in greater detail the closure of the full NNLO
interpolation grids, Fig.~\ref{fig:SM_applfastLHCgrids:closure2} presents the ratio of the
interpolated result using APPLfast to the NNLOJET cross section on a
magnified scale without statistical uncertainties.
From this figure it is observed that the grid technique
does not introduce any visible systematic bias.
The largest uncertainty from  the NNLO calculations, at present
arises from the numerical integration, in this case particularly from the
limited statistical precision of the double-real contributions that
is very time consuming to evaluate.
Note that the grid technique accurately reproduces the statistical fluctuations
of the original NNLOJET calculation.
In summary, a combined total statistical precision of approximately 0.5--1.0\,\% has been achieved.

\subsubsection{Fast interpolation grids for NNLO dijet production}
\begin{figure}[t]
  \centering
  \begin{tabular}{cc}
  \includegraphics[width=0.48\textwidth]{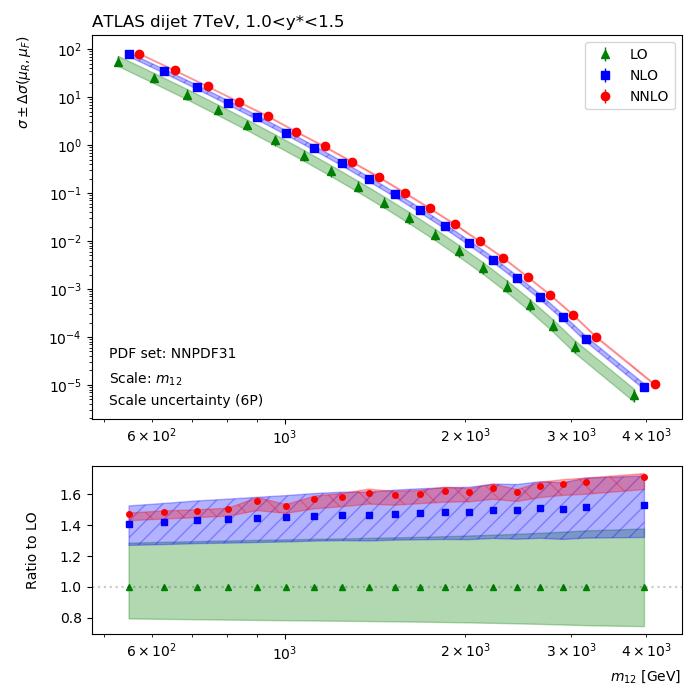} &
  \includegraphics[width=0.48\textwidth]{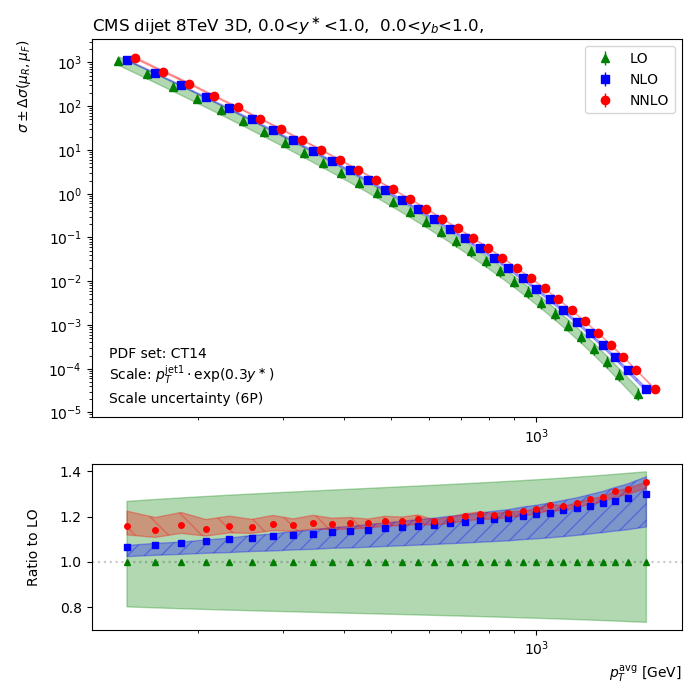}
  \end{tabular}
  \caption{NNLO predictions for dijet production obtained with fast
    interpolation grids. Left: LO, NLO, and NNLO dijet cross sections
    for an ATLAS measurement at $\sqrt{s}=7\,\text{TeV}$ as a function
    of $m_{12}$ in the interval $1.0<|y^*|<1.5$. Right: Dijet cross
    section for a CMS measurement at $\sqrt{s}=8\,\text{TeV}$ as a
    function of $p_{T}^\text{avg}$ in the interval
    $y_\text{boost}<1.0$ and $y^*<1.0$. The green, blue, and red bands
    indicate the LO, NLO, and NNLO scale uncertainties.}
    \label{fig:SM_applfastLHCgrids:crosssections}
\end{figure}
Figure~\ref{fig:SM_applfastLHCgrids:crosssections} illustrates the use of the
interpolation grids for the prediction of the NNLO dijet cross section
and the evaluation of the full scale uncertainties. While a full
computation of jet cross sections at NNLO involves very complex
multidimensional numerical integrations and therefore requires of the
order of $\mathcal{O}(100,000)$ hours of CPU time, the re-evaluation
with arbitrary scale variations or for other PDF sets is very fast.

\subsection{Summary and Outlook}
\label{sec:SM_applfastLHCgrids:outlook}
The rapidly increasing precision of Standard Model measurements at the LHC
demands improved precision in theoretical calculations.  
The NNLOJET code provides such improvements with calculations at NNLO
accuracy and beyond, for a wide range of physics processes. 
For the full exploitation of these predictions, for instance in fits 
of the proton parton distributions or even simply for the robust evaluation 
of PDF uncertainties, fast interpolation techniques as 
implemented in APPLgrid and fastNLO have proven to be essential.

In this article the current status of the production of interpolation
grids using the NNLOJET parton level calculations has been reported. For the first
time interpolation grids for dijet cross sections as measured by ATLAS
and CMS have been calculated.

In this case, the fast convolution using the interpolation tables can 
reproduce the native NNLO calculation to well within  0.5~per mille in 
all bins, significantly smaller than the current numerical accuracy 
of the published data, and significantly better than the statistical
uncertainty from the numerical integration of the NNLO contribution. 

The production of these interpolation grids at full NNLO with reasonable 
statistical precision together with their validation as 
detailed in this report, represents a significant milestone for the 
production of interpolation grids at NNLO.
The near future promises the production of interpolation grids at 
NNLO using NNLOJET for a variety of other processes, including 
inclusive jet production,  boson+jet production, inclusive $W$/$Z$
production, etc., at which point the detailed 
extraction of the proton PDF or determination of $\alpha_s$, both 
at full NNLO for a full portfolio of LHC processes will be possible. 

It is anticipated that the inclusion of multiple LHC measurements in PDF and
$\alpha_s$ fits with correspondingly precise 
theoretical predictions at NNLO, will the bring the PDF+$\alpha_s$ precision 
down to the percent level.



\newcommand{\Herwig}{H\protect\scalebox{0.8}{ERWIG}\xspace}
\newcommand{\Pythia}{P\protect\scalebox{0.8}{YTHIA}\xspace}
\newcommand{\Sherpa}{S\protect\scalebox{0.8}{HERPA}\xspace}
\newcommand{\Rivet}{R\protect\scalebox{0.8}{IVET}\xspace}
\newcommand{\Professor}{P\protect\scalebox{0.8}{ROFESSOR}\xspace}
\newcommand{\NNLOJET}{NNLO\protect\scalebox{0.8}{JET}\xspace}
\newcommand{\DIPHOX}{DIPHOX\xspace}
\newcommand{\tgNNLO}{2$\gamma$NNLO\xspace}
\newcommand{\JETPHOX}{JetPHOX\xspace}
\newcommand{\openloops}{OpenLoops\xspace}
\newcommand{\eps}{\varepsilon}
\newcommand{\mc}[1]{\mathcal{#1}}
\newcommand{\mr}[1]{\mathrm{#1}}
\newcommand{\mb}[1]{\mathbb{#1}}
\newcommand{\tm}[1]{\scalebox{0.95}{$#1$}}
\newcommand{\rd}{\ensuremath{\mathrm{d}}}

\section{Photon isolation studies~\protect\footnote{
  X.~Chen,
  M.~Chiesa,
  L.~Cieri,
  A.~Cueto,
  D.~de~Florian,
  A.~Denner,
  S.~Dittmaier,
  T.~Gehrmann,
  N.~Glover,
  M.~H\"{o}fer,
  A.~Huss,
  T.~Je\v{z}o,
  M.~Klasen,
  M.~Pellen,
  C.~Schwan,
  F.~Siegert,
  J.~Whitehead,
  J.~Zhou
}{}}

\label{sec:SM_phiso}

We study the impact of different isolation criteria for processes involving isolated photons.
In particular, we perform a detailed investigation of the impact of the parameters that characterise the isolation profile in the hybrid prescription compared with the standard and the smooth isolation criteria.
Moreover, we briefly describe the {\it photon-to-jet conversion function} to
treat the non-perturbative contribution to jet production via
$\gamma^*\to q\bar q$ splitting in the low-virtuality region of the photon.

\subsection{Isolation criteria}
\label{sec:SM_phiso:crit}

In the following, we briefly review the three isolation criteria that are used in this study.
The different algorithms can be characterised by the respective profile function that determines the maximal allowed amount of hadronic (partonic) transverse energy as a function of the angular separation $\Delta R = \sqrt{(\Delta y)^2 + (\Delta \varphi)^2}$ from the photon.
\begin{figure}[t]
  \begin{minipage}{.3\linewidth}
  \centering
  \includegraphics[width=\linewidth]{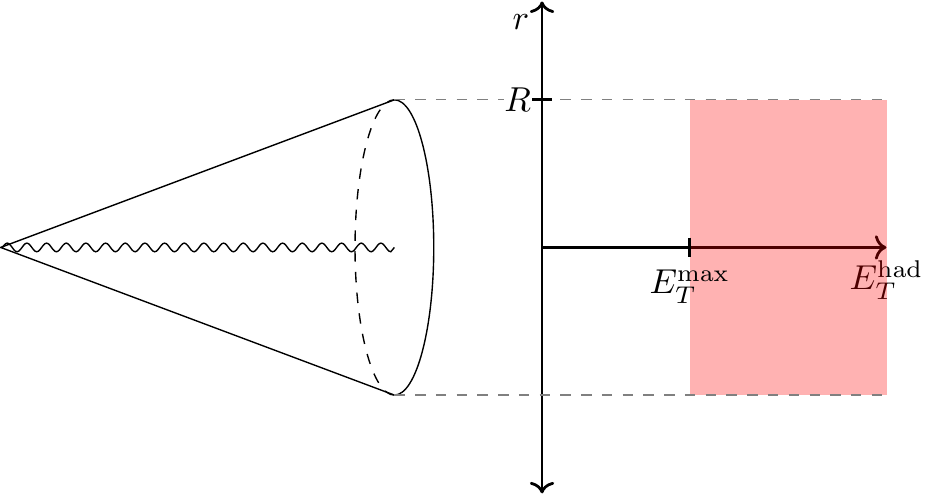}\\
  (a)~fixed-cone isolation
  \end{minipage}
  \hfill
  \begin{minipage}{.3\linewidth}
  \centering
  \includegraphics[width=\linewidth]{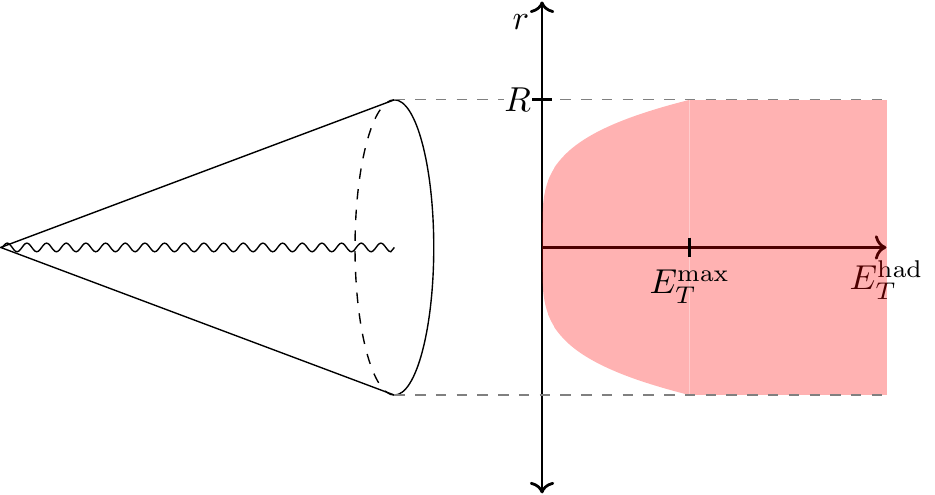}\\
  (b)~smooth-cone isolation
  \end{minipage}
  \hfill
  \begin{minipage}{.3\linewidth}
  \centering
  \includegraphics[width=\linewidth]{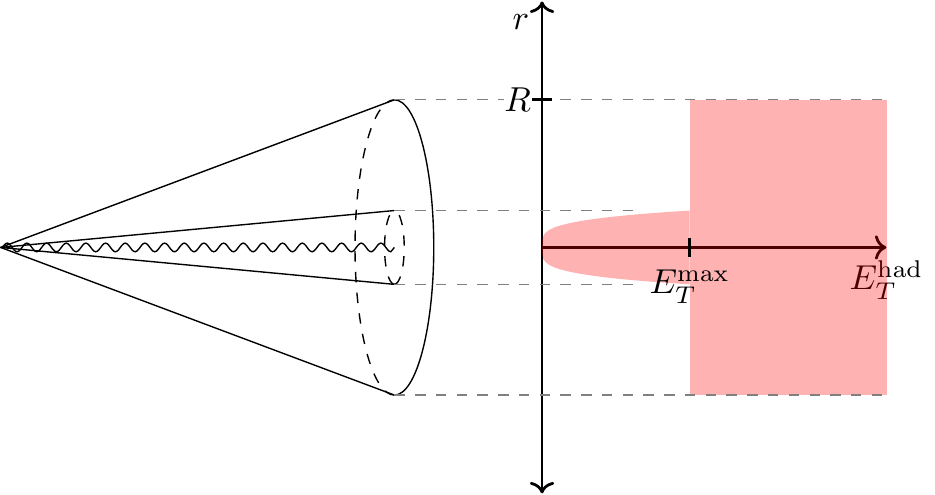}\\
  (c)~hybrid isolation
  \end{minipage}
  \medskip
  \caption{Profile functions for the constraint on the total hadronic (partonic) transverse energy around a photon.}
  \label{fig:SM_phiso:profiles}
\end{figure}
Figure~\ref{fig:SM_phiso:profiles} sketches the profile functions of the three isolation criteria which will be discussed in turn in the subsections that follow.
Note that this does not represent an exhaustive list by any means and arbitrary profiles can be introduced, e.g.\ variants that respect the smoothness of higher derivatives such as the smooth(er) step function.
We here focus on the hybrid isolation as a proxy for an isolation criterion that can be classified as being ``in between'' the fixed-cone and the smooth prescription (in the form sketched in Fig.~\ref{fig:SM_phiso:profiles}~(b) and initially proposed in Ref.~\cite{Frixione:1998jh}).

\subsubsection{Fixed-cone isolation (standard cone)}
\label{sec:SM_phiso:crit:fixed}

The commonly employed isolation criterion in the experimental measurements is the \emph{fixed-cone isolation}, which is defined as
\begin{align}
  E_\mathrm{T}^\text{had}(R)
  & \leq
  E_\mathrm{T}^\text{max} ,
\end{align}
where $E_\mathrm{T}^\text{had}(r) = \sum_i E_{\mathrm{T},i}^\text{had}\;\Theta(r-\Delta R_{\gamma i})$ denotes the total hadronic (partonic) transverse energy in a cone of size $r$ around the photon.
The maximum transverse energy can, in general, depend on the photon transverse momentum and is chosen as
\begin{align}
  E_\mathrm{T}^\text{max}
  &=
  \epsilon \; p_\mathrm{T}^\gamma + E_\mathrm{T}^\text{thresh}  ,
\end{align}
where a linear dependence is assumed with an additional constant offset.
Theory predictions employing this isolation criterion require the inclusion of the non-perturbative \emph{fragmentation functions}.
So far, the fragmentation component is known up to NLO and can be accompanied by rather sizeable uncertainties that arise from the challenges in constraining the parton fragmentation functions of the photon.

\subsubsection{Smooth-cone isolation (Frixione)}
\label{sec:SM_phiso:crit:smooth}

In order to avoid the fragmentation component altogether, the \emph{smooth-cone isolation}~\cite{Frixione:1998jh} proceeds by introducing a profile function $\chi(r;R)$ as follows:
\begin{align}
  E_\mathrm{T}^\text{had}(r)
  & \leq
  E_\mathrm{T}^\text{max} \; \chi(r;R) ,
  &
  &\forall r \leq R ,
\end{align}
with $E_\mathrm{T}^\text{had}(r)$ defined above.
Requiring $\chi(r;R)\to0$ for $r\to0$, rejects the fragmentation part while a smooth limit further avoids spoiling the soft region necessary for the proper cancellation of infrared singularities.
As such, the smooth-cone isolation can be applied at any perturbative order.
The function $\chi(r;R)$ in this study is chosen as%
\footnote{%
  Note that this function differs from the more commonly used choice $\chi(r;R)=\left(\frac{1-\cos(r)}{1-\cos(R)}\right)^n$.
  The difference between the two is however small and $\mathcal{O}(1\%)$ for the parameters considered in this study.%
}
\begin{align}
  \chi(r;R)
  & =
  \left( \frac{r}{R} \right)^{2n}
  \;.
\end{align}
This isolation prescription is thus determined by the cone size $R$ and the parameter $n$ that alters the profile, as well as the details of how $E_\mathrm{T}^\text{max}$ is chosen.
\begin{figure}[t]
  \centering
  \includegraphics[width=.45\linewidth]{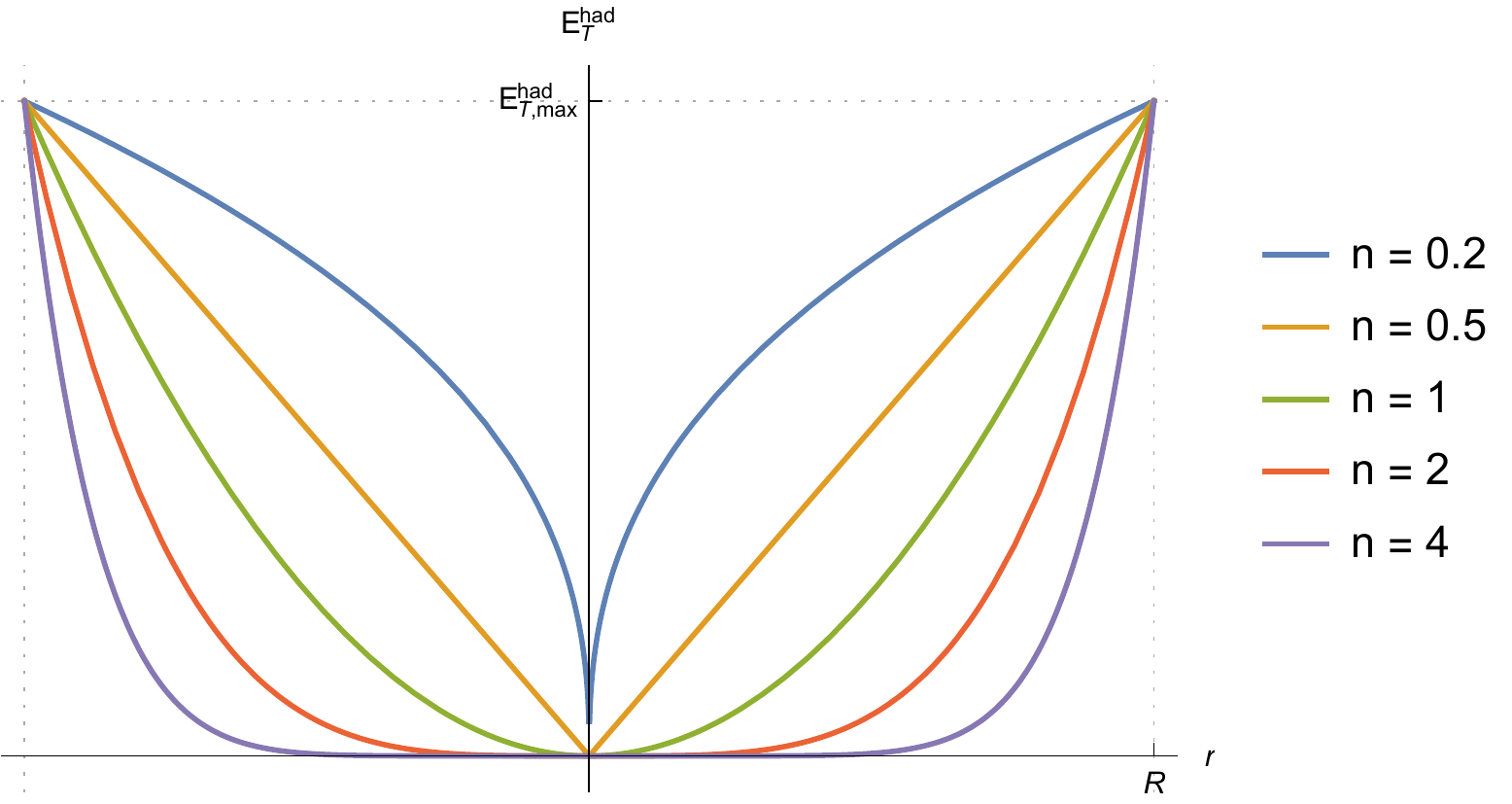}\\
  \caption{The dependence of the smooth profile function on the parameter $n$.}
  \label{fig:SM_phiso:profile_n}
\end{figure}
The change in the profile by varying the parameter $n$ is shown in Fig.~\ref{fig:SM_phiso:profile_n}.
We observe a smaller variation for $n\geq1$ with vanishing slopes at $r=0$, while a transition to $n<1$ induces bigger changes.
In particular, the slope of the profile at $r=0$ starts to become divergent for $n<0.5$, potentially inducing a stronger sensitivity to infrared emissions.
Clearly, the introduction of $\chi(r;R)$ substantially alters the isolation compared to the measurements that commonly employ the fixed isolation.
The comparison between the two latter has been the subject of past Les Houches proceedings~\cite{AlcarazMaestre:2012vp,Andersen:2014efa,Badger:2016bpw}, and detailed studies~\cite{Catani:2018krb,Cieri:2015wwa,Hall:2018jub}.

\subsubsection{Hybrid isolation}
\label{sec:SM_phiso:crit:hybrid}

More recently, a criterion dubbed as the \emph{hybrid isolation} has been put forward~\cite{Siegert:2016bre,Chen:2019zmr} that combines the fixed- and smooth-cone isolation with the aim to reduce the mismatch between the two.
More specifically, photons are required to pass \emph{both} criteria at the same time, where the smooth cone is embedded within a fixed cone with a smaller cone size ($R_d<R_\text{fixed}\equiv R$).
In small alteration of the procedure as used in Ref.~\cite{Chen:2019zmr}, we enforce $E_{\mathrm{T},\text{smooth}}^\text{max} = E_{\mathrm{T},\text{fixed}}^\text{max} \equiv E_{\mathrm{T}}^\text{max}$ in order to avoid the possibility of an effective reduction of the inner cone or discontinuities in the isolation profile, which are potential sources of instabilities~\cite{Catani:1997xc,Catani:2018krb} in fixed order perturbative calculations.
In our studies, the parameters of the fixed cone ($E_{\mathrm{T}}^\text{max}$ and $R$) will always be chosen to match the experimental analysis.
This leaves the inner smooth cone $R_d$ and $n$ as free parameters of the hybrid isolation prescription, for which we choose the nominal values
\begin{align}
  R_d &= 0.1 ,&
  n   &= 1 .
\end{align}
Note that choosing the inner cone to be larger or equal to the outer one, $R_d\geq R$, the hybrid prescription reduces to the smooth-cone criterion.

\subsubsection{Comparison of the prescriptions}
\label{sec:SM_phiso:crit:comp}

Comparing the isolation requirements of the standard, smooth, and hybrid isolation criteria with identical values of the cone size $R$ and the maximum energy $E_\mathrm{T}^\text{max}$ (and the same $n$ in the smooth and hybrid), we see that smooth cone and hybrid isolations are more restrictive than standard
cone isolation.
Therefore, the following physical constraint applies:
\begin{equation}
  \label{eq:relationisol}
  \rd\sigma_{\rm smooth}(R; E_\mathrm{T}^\text{max},n) <
  \rd\sigma_{\rm hybrid}(R; E_\mathrm{T}^\text{max},n,R_{d}<R) <
  \rd\sigma_{\rm standard}(R; E_\mathrm{T}^\text{max})
  \;\;,
\end{equation}
which is valid for any realistic implementation of the standard, smooth, and hybrid isolation prescriptions (see for instance Fig.~\ref{fig:SM_phiso:wgamma}).
In Eq.~\eqref{eq:relationisol}, $\rd\sigma$ generically denotes total cross sections and differential cross sections with respect to photon kinematical variables, and the subscripts ``smooth'', ``hybrid'' and ``standard'' refer to smooth, hybrid and standard isolation, respectively.

Considering independent variations of $n$ (variations of $R_d$ at fixed $n$) for the smooth (hybrid) isolation prescriptions respectively, the corresponding variation on the cross section increases by decreasing $n$ ($R_d$) at fixed $E_\mathrm{T}^\text{max}$ ($E_\mathrm{T}^\text{max}$ and $n$) and eventually the cross section diverges in the limit $n \to 0$ ($R_d \to 0$) as $1/n$ ($\log(R_d)$).
Since the cross section becomes arbitrarily large by decreasing $n$ ($R_d$ at fixed $n$),
it is obvious that at sufficiently small values of $n$ ($R_d$ at fixed $n$) the physical requirement~\eqref{eq:relationisol} is unavoidably violated.

\subsection{Study of isolation prescriptions}
\label{sec:SM_phiso:res}

\subsubsection{General Setup}
\label{sec:SM_phiso:res:setup}

If not explicitly stated otherwise, all predictions presented in this study employ the settings described in the following.
For hard and resolved photons, the appropriate scale of the emission is $Q^2=0$ and as a consequence the associated electromagnetic coupling is chosen to be $\alpha=\alpha_0$ ($\sim1/137$).
Theoretical uncertainties due to missing higher-order corrections are estimated using the standard 7-point variation around the central scale choice $\mu_R=\mu_F\equiv\mu_0$, where renormalization and factorization scales are independently varied up and down by a factor of two giving rise to the following combinations
\begin{align}
  (\mu_R,\mu_F)
  & =
  \left\{
    (1,1) ,\,
    (1,2) ,\,
    (2,1) ,\,
    (2,2) ,\,
    (1,\tfrac{1}{2}) ,\,
    (\tfrac{1}{2},1) ,\,
    (\tfrac{1}{2},\tfrac{1}{2})
  \right\} \times \mu_0
\end{align}
where the two extreme variations $\mu_R/\mu_F=4,\tfrac{1}{4}$ are excluded.
The choice of the central scale $\mu_0$ differs for each process and will be given in the respective section describing the process.
For the parton distribution functions, we use as default the \texttt{NNPDF3.1} sets.

\subsubsection{Di-photon production}
\label{sec:SM_phiso:res:2G}

\paragraph{Differential predictions at 8~\TeV}

For the di-photon production process, we follow the $8~\TeV$ ATLAS measurement~\cite{Aaboud:2017vol} and apply the fiducial cuts
\begin{align}
  E_\mathrm{T}^{\gamma_1} &> 40~\GeV  , &
  E_\mathrm{T}^{\gamma_2} &> 30~\GeV  , &
  \Delta R_{\gamma_1\gamma_2} &> 0.4  , \\
  |\eta^\gamma| &< 2.37 \quad
  \mathrlap{\text{(excluding $1.37 < |\eta^\gamma| < 1.56$) ,}}
\end{align}
with $\gamma_{1(2)}$ denoting the (sub-)leading photon.
The fixed-cone isolation used in the measurement is defined via
\begin{align}
  R &= 0.4, &
  E_\mathrm{T}^\text{max} &= 11~\GeV .
\end{align}
For the central scale in the theory predictions we choose the di-photon invariant mass $\mu_0 = m_{\gamma\gamma}$ and the following numerical programs are used in the study at 8~\TeV:
\begin{itemize}
  \item \textit{\DIPHOX}~\cite{Binoth:1999qq} (which contains the fragmentation contribution up to NLO) will be used along the standard cone isolation prescription.
  The uncertainty bands of the \DIPHOX predictions are obtained considering independent scale variations of $\mu_R$ and $\mu_F$ within the ranges
  $0.5 \leq \mu_R/\mu_0 \leq 2$ and $0.5 \leq \mu_F/\mu_0 \leq 2$
  around the central value  $\mu_0$. Practically, we obtain the results for nine
  scale configurations (we independently combine $\mu_R/\mu_0= \{ 0.5,1,2\}$
  and $\mu_F/\mu_0= \{ 0.5,1,2\}$) and we evaluate scale uncertainties by
  considering the maximum value and minimum value among these results.
  The fragmentation scale, $\mu_{\rm frag}$, is varied independently considering $\mu_{\rm frag}/\mu_0= \{ 0.5,1,2\}$. We have checked \cite{Catani:2018krb} that, for most
  of the computed quantities (including total cross
  sections), the maximum and minimum values correspond to the scale configurations
  $\{\mu_R=\mu_0/2$, $\mu_F=\mu_{\rm frag}=2\mu_0\}$ and $\{\mu_R=2\mu_0$ , $\mu_F=\mu_{\rm frag}=\mu_0/2\}$,
  respectively.
  Since the precedent two configurations are not present in the default 7-point scale configuration, the \DIPHOX prediction exhibits a larger dependence on the scale variation.
  \item \textit{\NNLOJET} for the fixed-order predictions up to NNLO that employ the smooth and hybrid isolation criteria.
  \item \Sherpa includes the direct and part of the fragmentation component by means of multileg matrix elements in a MEPS@NLO merging prescription and by setting the merging scale dynamically in the scheme of~\cite{Siegert:2016bre}.
    This includes $pp\to \gamma\gamma+0,1$j@NLO$+2,3$j@LO matrix elements, which are matched and merged with the \Sherpa parton shower~\cite{Schumann:2007mg,Hoeche:2011fd,Hoeche:2012yf}.
    In addition, the loop-induced $gg\to \gamma\gamma$ box process is included in these samples at LO accuracy.
    A hybrid isolation is used, i.e.\ the smooth-cone isolation with $\delta=0.1$, $\epsilon=0.1$ and $n=2$ is applied at parton level.
    The virtual QCD correction for matrix elements at NLO accuracy are provided by the \openloops library~\cite{Cascioli:2011va,Denner:2016kdg} and the NNPDF3.0 NNLO set~\cite{Ball:2014uwa} was used for PDFs.
\end{itemize}

For our phenomenological study, we investigate the invariant mass distribution of the two photons as shown in Fig.~\ref{fig:SM_phiso:2G:mgg}.
\begin{figure}[p]
  \begin{minipage}[b]{\linewidth}
  \centering
  \includegraphics[width=.5\linewidth]{{{figures/GG.mgg}}}%
  \includegraphics[width=.5\linewidth]{{{figures/GG.mgg.mc}}}%
  \\
  (a)~absolute predictions and comparison to data
  \end{minipage}
  \\[1.5em]
  \begin{minipage}[b]{.5\linewidth}
  \centering
  \includegraphics[width=\linewidth]{{{figures/GG.panels.mgg.NLO}}}\\
  (b)~hybrid isolation at NLO
  \end{minipage}
  \hfill
  \begin{minipage}[b]{.5\linewidth}
  \centering
  \includegraphics[width=\linewidth]{{{figures/GG.panels.mgg.NNLO}}}\\
  (c)~hybrid isolation at NNLO
  \end{minipage}
  \medskip
  \caption{
    The invariant mass distribution of the two photons in di-gamma production at the LHC for $\sqrt{s}=8~\TeV$.
    Comparison against the experimental data~(a) by the ATLAS measurement~\cite{Aaboud:2017vol}, and study of the dependence on the parameters of the hybrid isolation prescription at NLO~(b) and NNLO~(c).
  }
  \label{fig:SM_phiso:2G:mgg}
\end{figure}
We can immediately identify two distinct phase space regions in this distribution:
the low invariant-mass region $m_{\gamma\gamma} < 2 E_\mathrm{T}^{\gamma_1} $ populated only with events beyond Born kinematics and the complementary region ($m_{\gamma\gamma} > 2 E_\mathrm{T}^{\gamma_1} $) that is already non-vanishing at LO.
In Fig.~\ref{fig:SM_phiso:2G:mgg}~(a), we contrast the experimental data with various theory predictions:
NLO predictions using the \DIPHOX~\cite{Binoth:1999qq} program with a fixed-cone isolation as used by the experiment~(orange), fixed-order predictions up to NNLO obtained from the \NNLOJET program with the hybrid prescription~(LO: grey, NLO: green, NNLO: blue), as well as the NNLO prediction using the smooth-cone isolation~(red).
Higher-order QCD corrections are very sizeable for this process with the NLO $K$-factor ranging between $2$--$3$ and NNLO corrections at the level of $30\%$.
The inclusion of these corrections is essential in describing the data and the scale-uncertainty bands turn out to be an unreliable estimate of missing higher-order corrections, thus signalling still potentially sizeable corrections coming from the yet unknown N${}^\text{3}$LO corrections.

A direct comparison at NLO between the fixed-cone (\DIPHOX) and the hybrid prescription (\NNLOJET) is shown in the top panel of Fig.~\ref{fig:SM_phiso:2G:mgg}~(b).
The uncertainty bands of the \DIPHOX predictions are larger compared to the \NNLOJET results mainly due to the independent variation of the renormalisation and factorisation scales as described above.
We further note that the \DIPHOX results were obtained using the \texttt{CT10} PDF set.
Nonetheless, in the region above $m_{\gamma\gamma}>70~\GeV$ the two predictions almost coincide with the hybrid result being fully contained within the \DIPHOX uncertainty estimate band.
Sizeable differences are only visible in the first two bins, which however correspond to a phase-space region that is not populated at Born level.
As such, the perturbative accuracy degrades to only a LO prediction, which in the case of the hybrid prescription entails that there is no dependence on the isolation prescription.
At NLO, the direct part in \DIPHOX and the real corrections in fixed order theoretical tools (considering the smooth or hybrid prescriptions) are not affected by the isolation procedure, since the two forward photons are always far away from the only QCD parton contained in the final state (statement which is only true at NLO, \textit{i.e} with only one QCD parton). In summary, the sizeable NLO differences between standard and hybrid (or smooth) isolation results that are observed in the low-mass region (Fig.~\ref{fig:SM_phiso:2G:mgg}~(a)) are more an artifact of the NLO calculation than a physical
effect due to the two different isolation criteria \cite{Catani:2018krb}.
The discrepancy between standard and hybrid results in these two first bins therefore arise entirely from the fragmentation component included in the \DIPHOX calculation that employs the fixed-cone isolation.

\begin{figure}[t]
  \centering
  \includegraphics[width=.65\linewidth]{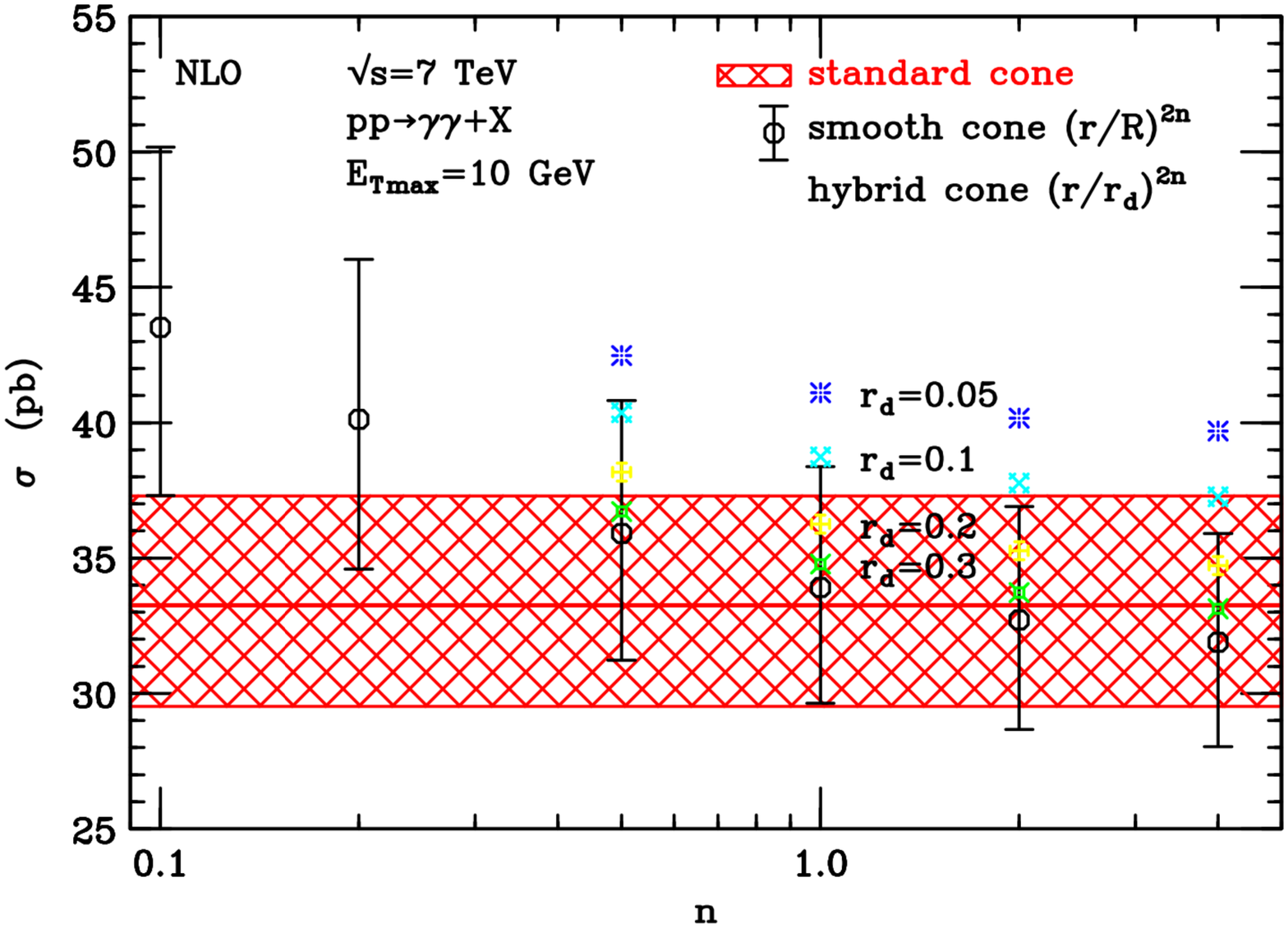}\\
  \caption{Value of the NLO total cross section,
including scale variation dependence, for the standard (red line and band)
and smooth (black error bars) isolation criteria.
The photon kinematical
cuts are described in the text.
The results are obtained for two different values of $E_\mathrm{T}^\text{max} = 10$~GeV.
In the case of smooth cone isolation,  different values of the
power $n$ ($n=0.1, 0.2, 0.5, 1, 2, 4$)
in the isolation function
$\chi(r;R)=\left( r/R \right)^{2n}$ are considered. In the case of the hybrid isolation, for fixed values of $n$ ($n=0.5, 1, 2, 4$) we vary the inner radius $R_{d}$ ($R_{d}=0.05, 0.1, 0.2, 0.3$).}
  \label{fig:SM_phiso:std_hybrid_smooth}
\end{figure}

\begin{figure}[t]
  \centering
  \includegraphics[width=.75\linewidth]{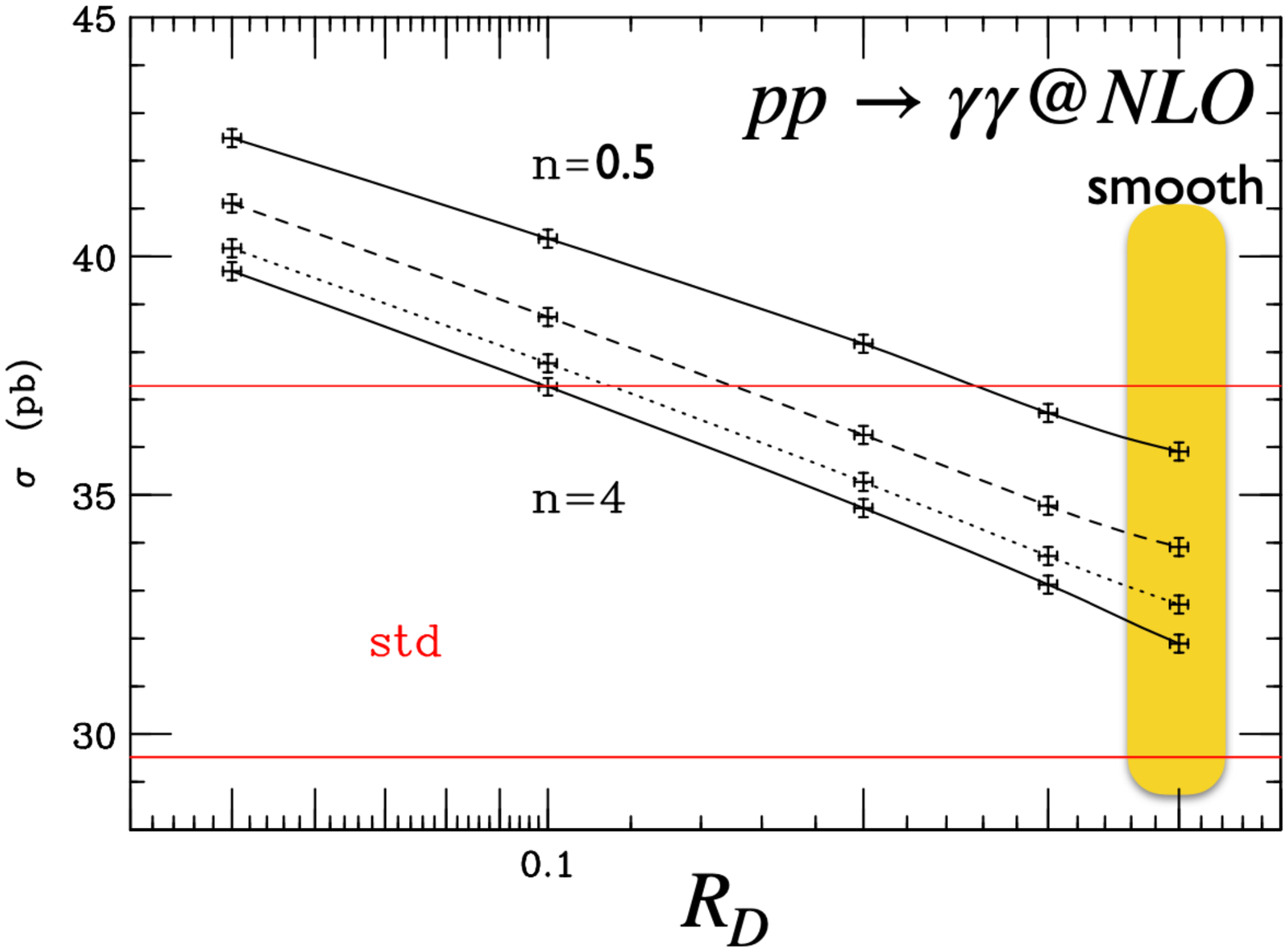}\\
  \caption{Considering the same setup of Fig.~\ref{fig:SM_phiso:std_hybrid_smooth} we show the logarithmic behaviour of the hybrid cross section in function of $R_{d}$.}
  \label{fig:SM_phiso:Rd_variation}
\end{figure}

The lower two panels in Fig.~\ref{fig:SM_phiso:2G:mgg}~(b) illustrate the dependence of the hybrid prescription at NLO w.r.t.\ the variation of the parameters $R_d$ and $n$.
Here, one parameter is always kept fixed while the other is varied by factors of $\{\tfrac{1}{5},\tfrac{1}{2},1,2,4\}$.
As mentioned above, no dependence on the isolation parameters is seen in the first two bins where the prediction effectively degrades to a LO one.
Above $m_{\gamma\gamma}>70~\GeV$, where the predictions are genuinely NLO, we observe that decreasing $R_d$ or $n$, effectively reduces the phase-space region that is rejected by the isolation criterion and thus increases the cross section.
This effect, however, becomes less and less pronounced as $m_{\gamma\gamma}$ increases, with almost no visible change above $700~\GeV$.
In this tail region, the photons are highly energetic (therefore less likely to be accompanied by low-energetic QCD radiation) and the smaller gluon luminosity further suppresses the dominant $qg \rightarrow q\gamma\gamma$ channel.
It is interesting to note that the variation of the inner cone size results in equidistant steps, confirming the expectation of a logarithmic dependence on $R_d$  (as can also be seen in Fig.~\ref{fig:SM_phiso:Rd_variation}).
The variation of $n$, on the other hand, shows much larger deviations for $n<1$ than for $n\geq1$.
This can be understood by the change in associated profile function as shown in Fig.~\ref{fig:SM_phiso:profile_n}, in particular, between $n\geq1$ and $n<0.5.$
A similar observation can be made in in Fig.~\ref{fig:SM_phiso:std_hybrid_smooth} for the fiducial cross section..

The corresponding results at NNLO are displayed in Fig.~\ref{fig:SM_phiso:2G:mgg}~(c).
The top panel illustrates the improved agreement with the experimental data by the inclusion of NNLO corrections; the variation in $R_d$ and $n$ follow the same pattern as at NLO with a similar relative impact on the cross sections.
The prediction using the smooth-cone isolation is shown as the red dotted curve and is equivalent to the $R_d=0.4$ setting in the hybrid prescription.
It results in a reduction of the cross section by about $10$--$20\%$ compared to the nominal setting.
Although the unphysical nature of the parameters $R_d$ and $n$ (and the divergent behaviour in the vanishing limit) make it difficult to define a range of variation for them,
it is worth noting that both at NLO and NNLO, a modification of either $R_d$ or $n$ up (down) by a factor of two (a half) is well contained within the respective scale uncertainty bands.

\paragraph{Fiducial cross sections at 7~\TeV}

The second study we present regarding diphoton production is following the kinematical cuts of the  $7~\TeV$ ATLAS measurement~\cite{Aad:2012tba}. In our theoretical study of standard, hybrid and smooth isolation
we apply the following kinematical cuts on photon transverse momenta and
rapidities: $E_\mathrm{T}^{\gamma_1} \geq 25$~GeV,
$E_\mathrm{T}^{\gamma_2}\geq 22$~GeV and the rapidity of both photons
is limited in the range $|\eta^\gamma|<2.37$.
The minimum angular distance
between the two photons is $R_{\gamma \gamma}^{\rm min}=0.4$. The isolation parameters for this setup are $E_\mathrm{T}^\text{max} = 10$~GeV and $R=0.4$. We use the
MMHT 2014 sets \cite{Harland-Lang:2014zoa} of parton distribution functions. In the case of the smooth and hybrid prescriptions the isolation parameters $n$ and $R_{d}$ are varied in order to asses their impact on the total cross section at NLO.
The predictions are obtained using the following numerical tools:
\begin{itemize}
  \item As before, the predictions with the standard cone isolation prescription are obtained from the \textit{\DIPHOX}~\cite{Binoth:1999qq} program.
  \item \textit{\tgNNLO}~\cite{Catani:2011qz} is used for the predictions with the smooth and hybrid isolation criteria.
\end{itemize}

These are basically the kinematical cuts used in the ATLAS Collaboration study of
Ref.~\cite{Aad:2012tba}.
The analysis of Ref.~\cite{Aad:2012tba}
is restricted to a smaller rapidity region since it excludes the rapidity
interval $1.37 < |\eta^\gamma|< 1.52$, which is outside the acceptance of the
electromagnetic calorimeter. For the sake of simplicity, in this subsection we do
not consider such additional rapidity restriction.

In Fig.~\ref{fig:SM_phiso:std_hybrid_smooth} we compare the standard cone NLO cross section (obtained using the numerical program \DIPHOX) with the corresponding results obtained with the hybrid and smooth cone isolation at NLO (using the \tgNNLO code). Since the smooth prediction with $n=1$ is at the same level of the standard result,
the use of the hybrid cone only can enlarge the NLO cross section obtained with the standard
and smooth isolation criteria. It is true that even for $R_{d}=0.05$ the results are
perturbatively still in accord with the standard and the smooth cone results (considering
the usual 9-point or 7-point scale variation), but the hierarchy presented
in Eq.~\eqref{eq:relationisol} starts to be violated. One can certainly argue
that it is very difficult to reliable estimate the fragmentation uncertainties
in the standard cone result, and this fact
can only enlarge the red band in Fig.~\ref{fig:SM_phiso:std_hybrid_smooth}, but has to
be aware of two independent facts: i) the NLO cross section using {\it both} prescriptions
diverges in the limit $n \rightarrow 0$ as $1/n$, but that divergence becomes visible in Fig.~\ref{fig:SM_phiso:std_hybrid_smooth}
only at  small $n$ that is lower than the typical values used for phenomenology
\footnote{It is worth noticing that this behaviour can be actually improved by choosing another
$\chi(r;R)$ function (with only one parameter: $n$) with a softer $n\rightarrow 0$ limit}; ii) the NLO cross section using
the hybrid cone diverges logarithmically in the $R_{d} \rightarrow 0$ limit.
That can be better observed in Fig.~\ref{fig:SM_phiso:Rd_variation} where we plot the same information as
in Fig.~\ref{fig:SM_phiso:std_hybrid_smooth} but plotting the
behaviour of the cross section as the parameter $R_{d}$ varies. There, the
logarithmic dependence of the cross section in terms of the
parameter $R_{d}$ is  visible
at values used for practical phenomenological implementations, starting to violate the physical constraint in Eq.~\eqref{eq:relationisol}  below $R_{d}=0.1$.


\subsubsection{Photon + jet production}
\label{sec:SM_phiso:res:GJ}

For the second process, we study the isolation criteria in the photon + jet process.
We adopt the same setup as in the $13~\TeV$ ATLAS measurement of Ref.~\cite{Aaboud:2017kff} using the fiducial cuts
\begin{align}
  E_\mathrm{T}^{\gamma} &> 125~\GeV  , &
  |\eta^\gamma| &< 2.37 \quad
  {\text{(excluding $1.37 < |\eta^\gamma| < 1.56$) ,}}
  \\
  p_\mathrm{T}^{j} &> 100~\GeV  , &
  |y^j| &< 2.37 ,
  \qquad
  \Delta R_{j\gamma} > 0.8 , \label{phiso:cut}
\end{align}
where jets are reconstructed using the anti-k${}_\mathrm{T}$ algorithm with $R_j=0.4$.
The parameters of the fixed-cone isolation are set to
\begin{align}
  R &= 0.4, &
  E_\mathrm{T}^\text{max}(E_\mathrm{T}^{\gamma}) &= 0.0042 \times E_\mathrm{T}^{\gamma} + 10~\GeV ,
\end{align}
and we choose the central scale $\mu_0 = p_\mathrm{T}^{\gamma}$.
In the phenomenological study of this process, we make use of the following code:
\begin{itemize}
  \item \textit{\JETPHOX}~\cite{Catani:2002ny} (which contains the fragmentation contribution up to NLO) will be used along the standard cone isolation prescription.
  \item \textit{\NNLOJET}: The fixed-order predictions up to NNLO using the smooth-cone and hybrid isolation criteria are obtained from the calculation of Ref.~\cite{Chen:2019zmr}.
  \item \textit{\Sherpa}:
  Samples generated with the Sherpa 2.2 Monte Carlo generator~\cite{Bothmann:2019yzt}.
  In this setup, NLO-accurate matrix elements for up to 2 jets, and LO-accurate matrix elements for up to 4 jets are calculated with the Comix~\cite{Gleisberg:2008fv} and OpenLoops~\cite{Cascioli:2011va,Denner:2016kdg} libraries.
  They are matched with the Sherpa parton shower using the MEPS@NLO~\cite{Hoeche:2012yf} prescription with a dynamic merging cut. Photons are required to be isolated according to a smooth-cone isolation criterion ($R=0.1$, $\epsilon=0.1$, $n=2$)~\footnote{Sherpa requirement on photon isolation follows $E_{\mathrm T}^{{\mathrm iso}}=\epsilon E_{\mathrm T}^{\gamma} \left(\frac{1-\cos(r)}{1-\cos(R)}\right)^n$}. Samples are generated using the \texttt{NNPDF3.0nnlo} PDF set,  along with the dedicated set of tuned parton-shower parameters developed by the Sherpa authors.
  The renormalisation and factorisation scales for the photon-plus-jet core process are set to the transverse momentum of the photon. QCD scale uncertainties are evaluated using 7-point variations of the renormalization and factorization scale.
\item \textit{POWHEG\,BOX\,V2/directphoton}~\cite{Jezo:2016ypn, Alioli:2010xd} will be used with all three isolation criteria.
  It implements direct photon production at NLO QCD supplemented by dijet production at LO matched to a parton shower.
  In the event samples, originally produced for the study of Ref.~\cite{Klasen:2017dsy}, the QED radiation was enhanced using the {\tt enhancedradfac} mechanism with its value set to 50.
  The samples were matched to Pythia~8.244~\cite{Sjostrand:2014zea} in which, other than switching off MPI and hadronization, default settings were used.
  Both the QED and QCD showers starting scales are set to the value of {\tt SCALUP}~\cite{Boos:2001cv}, QCD emissions are vetoed according to the $p_T$ value calculated by Pythia according POWHEG ISR $p_T$ and FSR $d_{ij}$ definitions while QED ones are not vetoed at all.
  Note that the photon-jet separation requirement in Eq.~\eqref{phiso:cut} is here implemented as an event selection cut rather than a jet selection cut as compared to the measurement and the other predictions.
\end{itemize}

\begin{figure}[p]
  \begin{minipage}[b]{\linewidth}
  \centering
  \includegraphics[width=.5\linewidth]{{{figures/GJ.pt_gam}}}%
  \includegraphics[width=.5\linewidth]{{{figures/GJ.pt_gam.mc}}}%
  \\
  (a)~absolute predictions and comparison to data
  \end{minipage}
  \\[1.5em]
  \begin{minipage}[b]{.5\linewidth}
  \centering
  \includegraphics[width=\linewidth]{{{figures/GJ.panels.pt_gam.NLO}}}\\
  (b)~hybrid isolation at NLO
  \end{minipage}
  \hfill
  \begin{minipage}[b]{.5\linewidth}
  \centering
  \includegraphics[width=\linewidth]{{{figures/GJ.panels.pt_gam.NNLO}}}\\
  (c)~hybrid isolation at NNLO
  \end{minipage}
  \medskip
  \caption{
    The transverse momentum distribution of the photon in photon+jet production at the LHC for $\sqrt{s}=13~\TeV$.
    Comparison against the experimental data~(a) by the ATLAS measurement~\cite{Aaboud:2017kff}, and study of the dependence on the parameters of the hybrid isolation prescription at NLO~(b) and NNLO~(c).
  }
  \label{fig:SM_phiso:GJ:pt_gam}
\end{figure}
Figure~\ref{fig:SM_phiso:GJ:pt_gam} displays the results for the transverse-momentum distribution of the photon in $\gamma+\text{jet}$ production following the same structure as in Fig.~\ref{fig:SM_phiso:2G:mgg} for di-gamma production.
The NLO prediction with a fixed-cone isolation is now obtained using the \JETPHOX~\cite{Catani:2002ny} program.
The comparison of the different predictions with the data shown in Fig.~\ref{fig:SM_phiso:GJ:pt_gam}~(a) reveals NLO corrections at the level of $\sim40\%$ and NNLO corrections of about $5\%$.
The latter move the central prediction on top of the data and further result in a substantial reduction of the scale uncertainty bands.
The NNLO predictions are further fully contained in the uncertainty estimate of the previous order, thus signalling a good perturbative convergence and a reliable estimate of missing higher orders though scale variations.

The comparison of the fixed-cone isolation (\JETPHOX) against the hybrid prescription (\NNLOJET) at NLO is shown in the top panel of Fig.~\ref{fig:SM_phiso:GJ:pt_gam}~(b).
We can observe that the prediction using the hybrid isolation lies slightly above the \JETPHOX results, however, this is far from alarming considering that the central predictions mutually lie within the error estimate of one another.
Investigating the dependence on the parameters $R_d$ and $n$ further supports this conclusion with all variations staying well below $\pm5\%$, which is fully covered by the NLO scale uncertainty band ($\sim\pm10\%$).
The qualitative features are similar to the case of the di-photon process discussed in Sect.~\ref{sec:SM_phiso:res:2G}.

The analogous study at NNLO is shown in Fig.~\ref{fig:SM_phiso:GJ:pt_gam}~(c), where we can appreciate the dramatic reduction of the residual scale uncertainty to about $\pm2\%$.
It is interesting to note that together with the scale uncertainties, also the relative impact from the variation of $R_d$ and $n$ is slightly reduced by going from NLO to NNLO.
While the most extreme variations in $R_d$ lie marginally outside of the scale uncertainty bands at around $p_\mathrm{T}^\gamma\sim180~\GeV$, the variation up (down) by a factor of two (a half) is still well captured by the uncertainty estimate, as was already seen in the case of di-photon production previously.

\subsubsection{W + photon production}
\label{sec:SM_phiso:res:GW}

An algorithm for the fully exclusive simulation of processes involving isolated
photons at NLO QCD matched to parton shower (PS) accuracy in the POWHEG framework was
developed in Ref.~\cite{Barze:2014zba} and applied to the process $pp\to e\nu\gamma$
({\sc Wgamma} package of {\sc Powheg-Box-V2}).
The strategy of Ref.~\cite{Barze:2014zba} follows the one of Ref.~\cite{DErrico:2011cgc}
for di-photon production, and consists in the calculation of both the $W\gamma$ and the
$Wj$ underlying Born (UB) production processes: this way the QED singularities coming
from the $q\to q\gamma$ splitting are cancelled by the usual QED subtraction terms
applied to the $Wj$ UB process and it is thus possible to generate the events in the
full phase space without imposing generation cuts or using the fragmentation functions.
Since the $Wj$ UB process is divergent in the limit of vanishing jet transverse momentum,
the MiNLO procedure~\cite{Hamilton:2012np,Hamilton:2012rf} is employed in order to
generate the events at NLO QCD+PS accuracy without generation cuts on the jet $p_T$.
The resulting events fall into three categories: events with $W\gamma$ UB where the
photon is harder than the other partons, events with $Wj$ UB where the hardest parton
is coloured and the $\gamma$ is the next-to-hardest parton, and events with two
coloured partons in the final state where the photons can be generated either by
the QED PS or from the hadronization (through the decay of unstable hadrons).
In the approach of Ref.~\cite{Barze:2014zba}, the perturbative part of the fragmentation
is provided by the {\sc Powheg} QED radiation on the $Wj$ UB process or by the QED PS, while the
non-perturbative part of the fragmentation is approximated by the hadronization
algorithm implemented in the shower Monte Carlo program used to process the {\sc Powheg} events.

\begin{figure}[t]
  \begin{center}
    \includegraphics[scale=1]{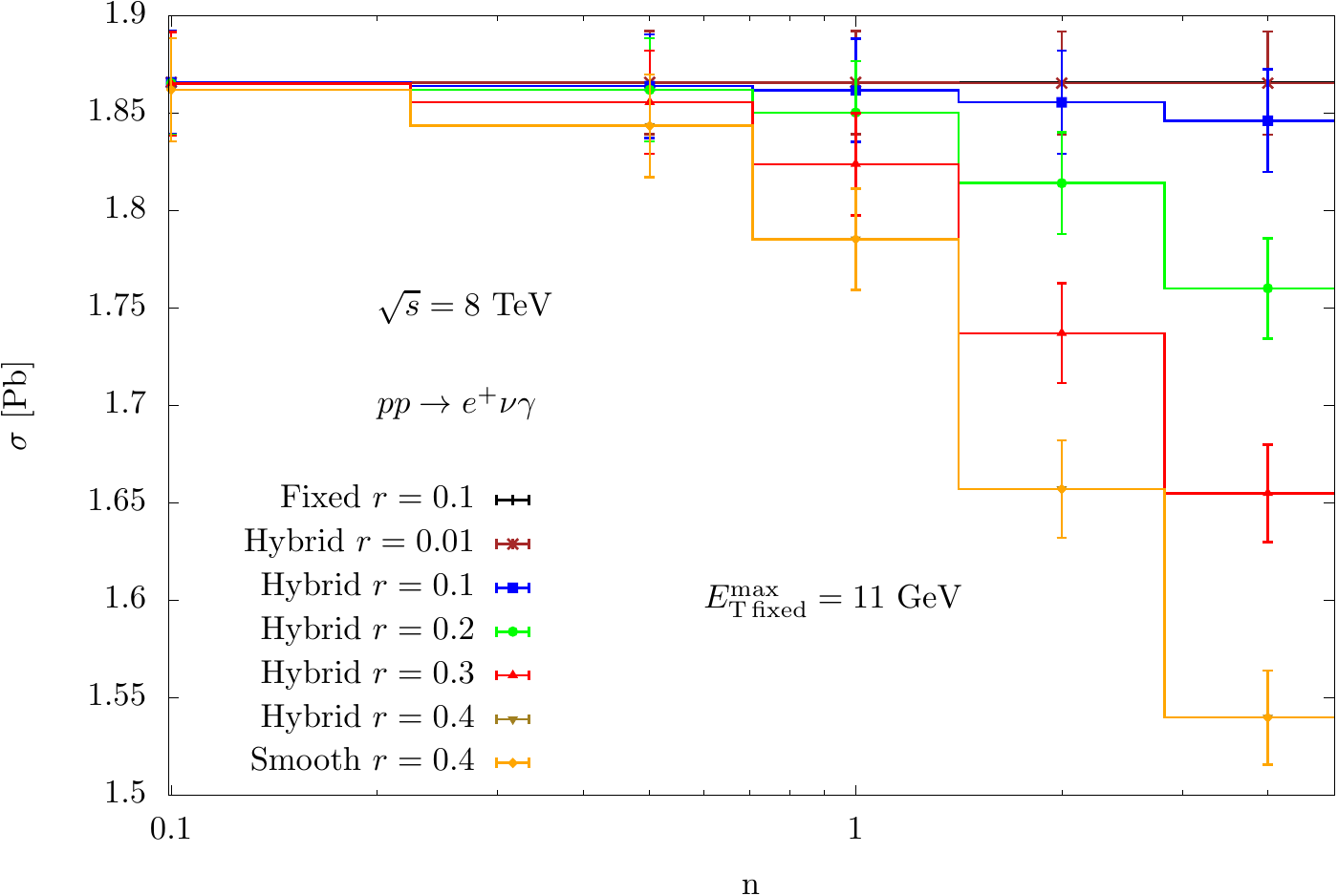}
    \caption{Integrated cross section for the process $pp\to e^+\nu\gamma$ at 8~TeV as a
      function of the isolation parameters $(n,r)$ under the event selection of Eq.~\eqref{eq:SM_phiso:wgamma_cuts}.
      The lines corresponding to the hybrid isolation with $r=R=0.4$ and to the Frixione isolation overlap.
      The results obtained in the hybrid isolation prescription tend to the ones computed with the Fixed cone
      isolation when the isolation parameter $r$ becomes small.     }
    \label{fig:SM_phiso:wgamma}
  \end{center}
\end{figure}

Figure~\ref{fig:SM_phiso:wgamma} summarizes the predictions of the {\sc Wgamma} package of {\sc Powheg-Box-V2}
for the integrated cross sections at NLO QCD+PS accuracy for different isolation strategies (fixed,
smooth, and hybrid isolation) and for different values of the isolation parameters $(n,r)$ at 8~TeV.
The calculation is performed in the $G_\mu$ scheme supplemented by the complex mass
scheme~\cite{Denner:1999gp,Denner:2005fg,Denner:2006ic,Denner:2019vbn} with the following input parameters:
\begin{eqnarray}
  & M_W^{\rm OS}= 80.398  \; {\rm GeV},\qquad & \Gamma_W^{\rm OS}=2.141 \; {\rm GeV}, \qquad |V_{\rm ud}|=|V_{\rm cs}|=0.975\nonumber \\
  & M_Z^{\rm OS} = 91.1876 \; {\rm GeV},\qquad & \Gamma_Z^{\rm OS}=2.4952 \; {\rm GeV}, \qquad |V_{\rm cd}|=|V_{\rm us}|=0.222 ,
  \label{eq:SM_phiso:wgamma_pars}
\end{eqnarray}
where the on-shell values of the masses and widths are converted internally to the corresponding pole values.
The {\tt NNPDF31\_nlo\_as\_0118\_luxqed} PDF set~\cite{Manohar:2016nzj,Manohar:2017eqh,Ball:2017nwa}
is used for the calculation via the {\sc Lhapdf6} interface~\cite{Buckley:2014ana}. For the PS evolution
of the events and the hadronization we use {\sc Pythia} version 8.235~\cite{Sjostrand:2006za,Sjostrand:2007gs,Sjostrand:2014zea}.
We consider the event selection
\begin{eqnarray}
  & & p_T^{\gamma} > 15 \; {\rm GeV},\qquad \Delta R_{l\gamma}> 0.7,\qquad p_T^{\nu} > 35 \; {\rm GeV},\nonumber \\
  & & p_T^{l} > 25 \; {\rm GeV},\qquad |\eta_\gamma|< 2.47, \qquad |\eta_l|< 2.37,\qquad M_T (l\nu) > 40 \; {\rm GeV},
  \label{eq:SM_phiso:wgamma_cuts}
\end{eqnarray}
where $M_T (l\nu)=\sqrt{2p_{T}^ep_{T}^\nu(1-\cos \theta_{e\nu})}$ is the transverse mass of the lepton-neutrino pair.

Figure~\ref{fig:SM_phiso:wgamma} shows that predictions obtained in the smooth isolation prescription are
systematically lower than the ones computed with the fixed cone approach (as expected), and the
difference increases with the value of the $n$ parameter. The predictions in the hybrid isolation
scheme fall between the ones in the Frixione and in the fixed isolation schemes, and they converge
to the latter when the $r$ parameter becomes small as stated by Eq.~\eqref{eq:relationisol}.

\subsection{Photon-to-jet conversion}
\label{sec:SM_phiso:GtoJ_conv}

\begin{figure}[t]
  \begin{minipage}{.5\linewidth}
  \centering
  \includegraphics{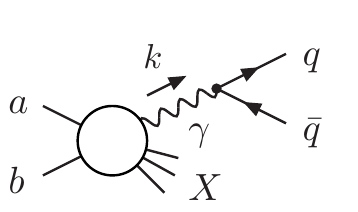}\\
  (a)~Generic diagram
  \end{minipage}
  \begin{minipage}{.5\linewidth}
  \centering
  \includegraphics[page=4,scale=0.8]{figures/diagrams}\\
  (b)~Example diagram for lepton pair+jet production
  \end{minipage}
\caption{Diagrams illustrating jet production via $\gamma^*\to q\bar q$
splitting of a photon with low virtuality~$k^2$.}
\label{fig:gamma2ffdiag}
\end{figure}

Jets or lower-multiplicity hadronic final states
may also be initiated by the electroweak mechanism of
photon-to-quark splittings $\gamma^*\to q\bar q$, see Fig.~\ref{fig:gamma2ffdiag}.
The perturbative treatment of this splitting leads to mass singularities
in cross-section predictions that would be cancelled against loop
corrections to the underlying hard process with a photon in the
final state instead of the $q\bar q$~pair.
In practice, this issue potentially occurs for electroweak corrections
to processes that involve explicit jets in the definition of their
experimental signature.
An example is lepton pair+jet production, in which at next-to-leading order
in $\alpha$ a quark pair can be produced from a virtual photon, as shown in
Fig.~\ref{fig:gamma2ffdiag}~(b).
Assuming that hadronic
activity can be experimentally distinguished from a hard photon,
a procedure is required to treat the effect of the low-virtuality $q\bar
q$~pair in a non-perturbative way.
To this end, in Ref.~\cite{Denner:2019zfp} the concept of a
{\it photon-to-quark conversion function} was introduced
similar to the concept of fragmentation functions for
identified-particle production.
This concept is briefly summarized in the following.

Technically, the phase-space integral over squared amplitudes that involve some
$\gamma^* \to q\bar q$ splitting process contains a mass singularity
for light quarks~$q$, originating from the collinear region, which is
characterized by some low virtuality $k^2$ of the photon.
The structure of this
singularity is universal in the sense that the squared
matrix elements factorize into a universal radiator function
and the square of the hard matrix element of the underlying
process with a real photon instead of the $q\bar q$ pair.
Note, however, that the physical final state is still a jet, or at least
some hadronic activity, emerging from the photon initiating the
splitting.
Perturbatively, the mass-singular cross-section contribution
can be calculated in a straightforward way, \eg via
two-cutoff slicing or dipole subtraction, as described in
Refs.~\cite{Dittmaier:2008md,Denner:2019vbn}.
The singular contributions show up as $1/\epsilon$ poles in
$D=4-2\epsilon$ dimensions or as logarithms $\ln m_q$ if
small quark masses $m_q$ are used as regulators.
Either way, the resulting singular contribution
is not yet described in a physically meaningful way, since the
splitting contains non-perturbative contributions,
which have to be taken from experiment.

The non-perturbative cross-section contribution can be combined
with the perturbative part by means of a {\it photon-to-quark conversion
function}
$D^{\mathrm{bare}}_{\gamma\to\mathrm{jet}}$
similar to the concept of fragmentation functions for
identified-particle production~\cite{Denner:2019zfp},
\begin{align}
\mathrm{d}\sigma^{\mathrm{conv}}_{ab\to \mathrm{jet}+X} &{}=
\mathrm{d}\sigma^{\mathrm{LO}}_{ab\to \gamma X} \, \int_0^1\mathrm{d} z\,
D^{\mathrm{bare}}_{\gamma\to\mathrm{jet}}(z).
\end{align}
Here $D^{\mathrm{bare}}_{\gamma\to\mathrm{jet}}(z)$ is the {\it bare}
$\gamma\to\mathrm{jet}$
conversion function, which depends on the variable $z$ describing
the fraction of the photon momentum~$k$ transferred to one of the
jets \mbox{($p_{\mathrm{jet}}=zk$)}. The bare conversion function
contains singular contributions so that the sum of the
conversion part $\mathrm{d}\sigma^{\mathrm{conv}}$ and the remaining
perturbative cross-section contribution is non-singular.
The extraction of the singular contribution from
$D^{\mathrm{bare}}_{\gamma\to\mathrm{jet}}(z)$ at some factorization scale
$\mu_{\mathrm{F}}$ requires a factorization scheme, for which
the $\overline{\mathrm{MS}}$ scheme is usually taken.
In dimensional regularization (DR), $D^{\mathrm{bare}}_{\gamma\to\mathrm{jet}}(z)$
is decomposed into a singular and a phenomenological part
$D_{\gamma\to\mathrm{jet}}(z,\mu_{\mathrm{F}}^2)$ as follows,
\begin{align}
\label{eq:Dbar_dreg}
D^{\mathrm{bare}}_{\gamma\to\mathrm{jet}}(z)\Big|_{\mathrm{DR}}
&{}=
\sum_q \frac{3Q_q^2\alpha}{2\pi} \,
\frac{(4\pi)^\epsilon}{\Gamma(1-\epsilon)}
\left(\frac{\mu^2}{\mu_{\mathrm{F}}^2}\right)^\epsilon
\frac{1}{\epsilon}
P_{f\gamma}(z)
+ D_{\gamma\to\mathrm{jet}}(z,\mu_{\mathrm{F}}^2),
\end{align}
where $\mu$ is the arbitrary reference scale of DR and
$P_{f\gamma}(z) = (1-z)^2 +z^2$ the $\gamma\to f\bar f$ splitting function.
In mass regularization (MR), $D^{\mathrm{bare}}_{\gamma\to\mathrm{jet}}(z)$ reads
\begin{align}
D^{\mathrm{bare}}_{\gamma\to\mathrm{jet}}(z)\Big|_{\mathrm{MR}}
&{}=
\sum_q \frac{3Q_q^2\alpha}{2\pi} \,
\ln\left(\frac{m_q^2}{\mu_{\mathrm{F}}^2}\right)
P_{f\gamma}(z)
+ D_{\gamma\to\mathrm{jet}}(z,\mu_{\mathrm{F}}^2),
\label{eq:Dbar_mreg}
\end{align}
where the finite non-perturbative part
$D_{\gamma\to\mathrm{jet}}(z,\mu_{\mathrm{F}}^2)$ is the same in the two
versions.

The non-perturbative contributions to
$D_{\gamma\to\mathrm{jet}}(z,\mu_{\mathrm{F}}^2)$ have to be extracted
from experimental data.  Ideally, this information would come from an
accurate differential measurement of a jet production cross section
(with low jet invariant mass) and of its corresponding prompt-photon
counterpart, \ie experimental information that is not available at
present.

In Ref.~\cite{Denner:2019zfp} it was shown that at least the inclusive
$z$-integral over $D_{\gamma\to\mathrm{jet}}(z,\mu_{\mathrm{F}}^2)$ can
be obtained from a dispersion integral for the $R$~ratio of the cross
sections for $\Pe^+\Pe^-\to\mathrm{hadrons}/\mu^+\mu^-$.  This
dispersion integral, in turn, can be tied to the quantity
$\Delta\alpha^{(5)}_{\mathrm{had}}(M_{\mathrm{Z}}^2)$,
which is fitted to experimental data
(see Refs.~\cite{Eidelman:1995ny,Keshavarzi:2018mgv} and references therein).
Based on this feature, it is possible to
predict the following form of $D_{\gamma\to
  \mathrm{jet}}(z,\mu_{\mathrm{F}}^2)$,
\begin{align}
D_{\gamma\to \mathrm{jet}}(z,\mu_{\mathrm{F}}^2) = \Delta\alpha^{(5)}_{\mathrm{had}}(M_{\mathrm{Z}}^2)
+ \sum_q \frac{3Q_q^2\alpha}{2\pi} \,
\left[\ln\left(\frac{\mu_{\mathrm{F}}^2}{M_{\mathrm{Z}}^2}\right)+\frac{5}{3}\right]\, P_{f\gamma}(z),
\end{align}
which is valid up to $z$-dependent terms that integrate to zero.
Here the sum over $q$ runs over all quarks but the top quark.
Since mostly the inclusive integral over $z$ is needed in predictions
for cross sections, and since the impact of $D_{\gamma\to \mathrm{jet}}(z)$
is quite small in general, this result should be sufficient for all
phenomenological purposes.

This conclusion is supported by the explicit examples in which the
photon-to-jet conversion function has been applied yet,
which comprise Z+jet production~\cite{Denner:2019zfp} and
WZ scattering~\cite{Denner:2019tmn} at the LHC.
In both cases, jet production via low-virtuality photon splitting
happens only on a very small fraction of phase space, so that the overall
contribution of the photon-to-jet conversion part to the cross section
is very small as well.


\subsection*{Acknowledgements}
The work of LC was financially supported by the European Union's
Horizon 2020 research and innovation programme under the Marie
Sk\l{}odowska-Curie grant agreement No. 754496 - FELLINI.  CS is
supported by the European Research Council under the European Union's
Horizon 2020 research and innovation Programme (grant agreement
ERC-AdG-740006).

\let\Herwig\undefined
\let\Pythia\undefined
\let\Sherpa\undefined
\let\Rivet\undefined
\let\Professor\undefined
\let\NNLOJET\undefined
\let\DIPHOX\undefined
\let\JETPHOX\undefined
\let\openloops\undefined
\let\tgNNLO\undefined
\let\eps\undefined
\let\mc\undefined
\let\mr\undefined
\let\mb\undefined
\let\tm\undefined
\let\rd\undefined

\newcommand{\Herwig}{H\protect\scalebox{0.8}{ERWIG}\xspace}
\newcommand{\Pythia}{P\protect\scalebox{0.8}{YTHIA}\xspace}
\newcommand{\Sherpa}{S\protect\scalebox{0.8}{HERPA}\xspace}
\newcommand{\Rivet}{R\protect\scalebox{0.8}{IVET}\xspace}
\newcommand{\Recola}{R\protect\scalebox{0.8}{ECOLA}\xspace}
\newcommand{\Amegic}{A\protect\scalebox{0.8}{MEGIC}\xspace}
\newcommand{\Professor}{P\protect\scalebox{0.8}{ROFESSOR}\xspace}
\newcommand{\OpenLoops}{O\protect\scalebox{0.8}{PENLOOPS 2}\xspace}
\newcommand{\Collier}{C\protect\scalebox{0.8}{OLLIER}\xspace}
\newcommand{\Madgraph}{M\protect\scalebox{0.8}{G5\_aMC@NLO}\xspace}
\newcommand{\eps}{\varepsilon}
\newcommand{\mc}[1]{\mathcal{#1}}
\newcommand{\mr}[1]{\mathrm{#1}}
\newcommand{\mb}[1]{\mathbb{#1}}
\newcommand{\tm}[1]{\scalebox{0.95}{$#1$}}
\newcommand{\vp}{\ensuremath{\vphantom{\int_a^b}}}
\newcommand{\vP}{\ensuremath{\vphantom{\int\limits_a^b}}}

\section{NLO QCD and electroweak corrections to off-shell WWW production~\protect\footnote{
  S.~Dittmaier,
  G.~Knippen,
  M.~Sch{\"o}nherr,
  C.~Schwan}{}}

\label{sec:SM_WWW}


\subsection{Introduction}
\label{sec:SM_WWW:intro}

Owing to its high scattering energy and luminosity, the LHC is able 
to explore particle processes up to energy scales of several TeV 
even if the corresponding cross sections are in the range of
femtobarns only. 
For the experimental investigation of electroweak (EW) interaction,
this means that the LHC can observe the phenomenologically highly
interesting processes of EW vector-boson scattering (VBS) and
of triple EW vector-boson production (TVP) for the first time.
The analysis of those process classes is particularly interesting
because of their direct sensitivity to quartic gauge self-interactions
and to off-shell Higgs-boson exchange. The latter property renders
those processes an alternative window to EW symmetry
breaking, complementary to processes with direct (on-shell)
Higgs-boson production.
In this contribution we focus on triple W-boson production, which
was analyzed by ATLAS~\cite{Aaboud:2016ftt,Aad:2019dxu}
and CMS~\cite{CMS:2019mpq} with 4.1$\sigma$ evidence by ATLAS in Run~2.

Taking into account the decays of the EW massive vector bosons,
the VBS and TVP processes are of the types 
$\Pp\Pp\to 4\text{leptons}+2\text{jets}+X$
and $\Pp\Pp\to 6\text{leptons}+X$, respectively, and thus involve
already six particles at leading order (LO).
The calculation of radiative corrections to processes of such complexity is 
rather demanding. 
For TVP processes with leptonically decaying vector bosons, the calculation
of NLO QCD corrections (up to the phase-space integration) has only the complexity
of a $2\to3$ particle process, so that in particular 
the NLO QCD corrections to WWW production with \cite{Campanario:2008yg} 
and without \cite{Binoth:2008kt} leptonic decays have been known for more than ten years.
The NLO EW corrections to WWW production were first calculated for stable W~bosons
and for on-shell W~bosons with decays treated in the narrow-width approximation
in Refs.~\cite{Yong-Bai:2016sal,Dittmaier:2017bnh,Frederix:2018nkq}, before a 
treatment of the full off-shell $2\to6$ process became possible.
This last step poses the challenge of one-loop diagrams with up to eight
particles in the loop (8-point functions), the evaluation of which was
only possible with great advances in the automated calculation of one-loop
amplitudes (see, e.g., Ref.~\cite{Denner:2019vbn} for details and references).

Recently, two independent evaluations of WWW production processes at NLO EW with
leptonically decaying W~bosons, including all off-shell effects, have been
presented in the literature: 
the calculation of Ref.~\cite{Schonherr:2018jva} based on \Sherpa~\cite{Bothmann:2019yzt}
with \Recola~1.2 \cite{Actis:2012qn,Actis:2016mpe} as one-loop matrix element provider 
on the one hand and the two more process-specific calculations of 
Ref.~\cite{Dittmaier:2019twg} based on
\OpenLoops~\cite{Cascioli:2011va,Kallweit:2014xda,Buccioni:2019sur} and 
\Recola~1.4 \cite{Actis:2012qn,Actis:2016mpe} on the other.
Unfortunately, the results of Refs.~\cite{Schonherr:2018jva,Dittmaier:2019twg} initially did not
agree in all parts, rendering a detailed comparison of individual components of the two
calculations necessary.
In this contribution, we briefly report on the methods and tools used in the two calculations,
on the salient features of the NLO corrections, and on the comparison of results that shows good
agreement between the results of Ref.~\cite{Dittmaier:2019twg} and revised results of
Ref.~\cite{Schonherr:2018jva}.

\subsection{Calculational details, methods, and tools}
\label{sec:SM_WWW:methods}

In detail, the NLO calculation of Ref.~\cite{Schonherr:2018jva} for WWW production
employs a combination of \Sherpa 
\cite{Bothmann:2019yzt,Gleisberg:2008ta,Bothmann:2016nao,Hoeche:2014rya} 
and \Recola \cite{Actis:2012qn,Actis:2016mpe}, where
\Sherpa provides the tree-level matrix elements, 
infrared subtraction, process management, and phase-space 
integration through its matrix element generator \Amegic 
\cite{Krauss:2001iv,Gleisberg:2007md,Schonherr:2017qcj}. 
\Recola is interfaced \cite{Biedermann:2017yoi} to provide 
all renormalized virtual corrections, where the loop integrals are evaluated with the
\Collier library~\cite{Denner:2016kdg},
which in turn is based on the results of 
Refs.~\cite{Denner:2002ii,Denner:2005nn,Denner:2010tr}.

On the other hand, the results of Ref.~\cite{Dittmaier:2019twg}, called DKS in the following,
were produced and double-checked using two private codes: 
The first one was developed specifically for this process, 
and the second one is a generic code that was already used to calculate other EW 
processes~\cite{Ballestrero:2018anz,Denner:2019tmn,Denner:2019zfp}.
The first code uses dipole subtraction as presented in Ref.~\cite{Catani:1996vz} for QCD corrections and 
in Refs.~\cite{Dittmaier:1999mb,Dittmaier:2008md} for EW corrections.
Matrix elements were generated using \Madgraph \cite{Alwall:2014hca} and \Recola,
and the loop integrals were evaluated with \Collier.
The second code is able to use either dipole subtraction as presented in Ref.~\cite{Catani:1996vz} 
for both QCD and EW corrections, the latter using the trivial substitutions for the Casimir operator 
(see, e.g., in Sect~3.2 of Ref.~\cite{Kallweit:2014xda}) or, alternatively, 
the EW dipole subtraction of Refs.~\cite{Dittmaier:1999mb,Dittmaier:2008md} 
as in the first code.
The matrix elements are provided by \OpenLoops \cite{Cascioli:2011va,Kallweit:2014xda,Buccioni:2019sur}, 
which uses \Collier for the evaluation of rank~0 and 1 tensor one-loop integrals.
Both codes use multi-channel Monte Carlo techniques \cite{Hilgart:1992xu,Kleiss:1994qy} for the phase-space integration with phase-space mappings similar to the ones presented in Ref.~\cite{Dittmaier:2002ap}.

\begin{figure}[t]
\centering
\begin{tabular}{ccccc}
\includegraphics[width=0.28\textwidth]{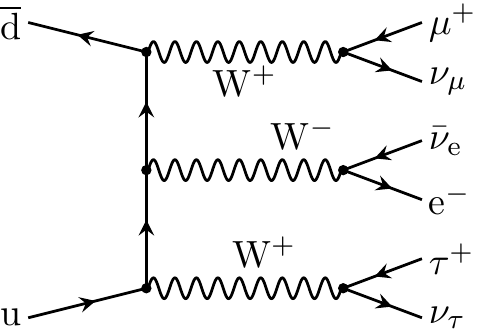} &&
\includegraphics[width=0.28\textwidth]{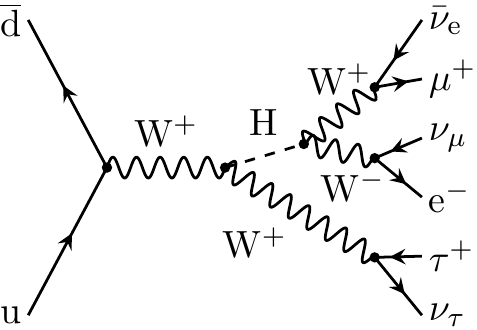} &&
\includegraphics[width=0.28\textwidth]{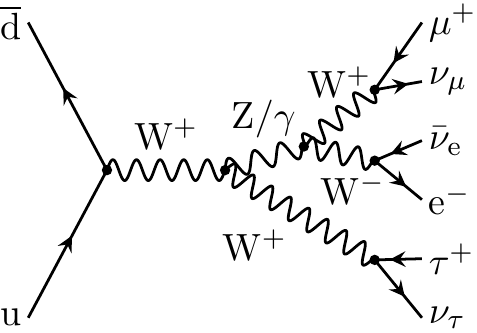}
\\[1.5em]
\centering
\includegraphics[width=0.28\textwidth]{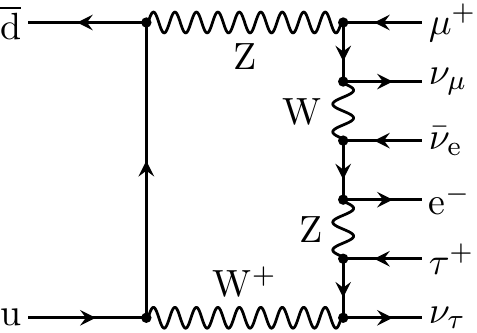} &&
\includegraphics[width=0.28\textwidth]{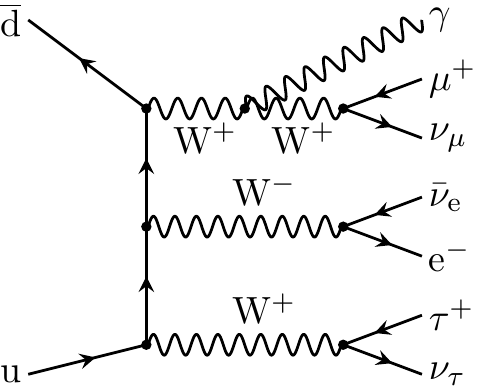} &&
\includegraphics[width=0.28\textwidth]{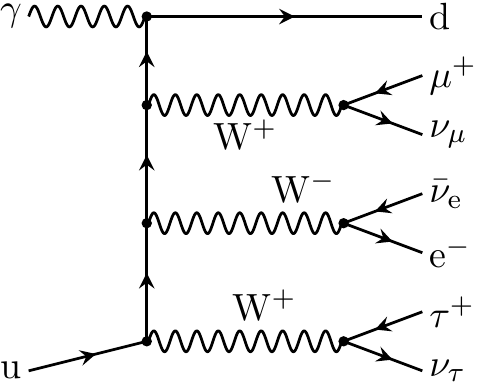}
\end{tabular}
\caption{Sample Feynman diagrams contributing to $\Pp\Pp\rightarrow\Pe^-\bar\nu_{\Pe}\mu^+\nu_\mu\tau^+\nu_\tau+X$
at LO (top row) and at NLO EW (bottom row).}
\label{fig:SM_WWW:www-diagrams}
\end{figure}

Figure~\ref{fig:SM_WWW:www-diagrams} illustrates the various types of diagrams
that occur in the calculation of LO and NLO EW contributions to the cross sections of the
WWW production process $\Pp\Pp\rightarrow\Pe^-\bar\nu_{\Pe}\mu^+\nu_\mu\tau^+\nu_\tau+X$.
At LO, we can distinguish three basic classes of diagrams according to their different resonance structure:
\begin{enumerate}
\item diagrams with three simultaneously resonant \PW bosons (left diagram in the top row of 
Fig.~\ref{fig:SM_WWW:www-diagrams}), 
\item Higgs production in association with a \PW boson (middle diagram in the top row of            
Fig.~\ref{fig:SM_WWW:www-diagrams}), where the produced Higgs boson further decays into an on- and an off-shell \PW boson, and
\item $\mathrm{WZ}$ production, where the \PZ boson either decays into an on- and an off-shell \PW boson 
(right diagram in the top row of Fig.~\ref{fig:SM_WWW:www-diagrams})
or into a four-fermion state via a resonant \PW~boson (not shown in the figure).
\end{enumerate}
All other diagrams show less resonance enhancement.
The production of $\mathrm{WZ}$ is strongly suppressed because of the four-body decay of the \PZ boson, 
while associated Higgs production and triply-resonant $\PW\PW\PW$ contributions dominate the 
cross sections of the given processes.
Due to the extremely narrow width of the Higgs boson and the fact that the Higgs-boson mass is smaller 
than twice the \PW-boson mass, associated Higgs production is well separated from the triply-resonant 
WWW contributions in phase space and therefore can, in principle, be isolated by phase-space cuts.

The bottom row of Fig.~\ref{fig:SM_WWW:www-diagrams} shows representative diagrams for the three contributions to
the NLO EW corrections: one-loop contributions (left), photonic bremsstrahlung contributions (middle),
and contributions from photon-induced channels (right). 
Note that both the evaluation of virtual and real corrections is technically challenging,
the former in view of a fast and numerically stable evaluation of loop diagrams up to
8-point complexity, the latter owing to the complicated resonance structures in the 7-particle
phase space.
We finally mention that in all the calculations presented in
Refs.~\cite{Schonherr:2018jva,Dittmaier:2019twg} the resonances are treated in the 
complex-mass scheme \cite{Denner:2005fg} (see also \Ref.~\cite{Denner:2019vbn}),
i.e.\ complex gauge-boson masses defined by 
\begin{equation}
  \mu_V^2=M_V^2-\mathrm{i}M_V\Gamma_V, \qquad V=\PW,\PZ,
\end{equation}
are used in all propagators and couplings consistently, in order to ensure
gauge independence and NLO precision in resonant and non-resonant phase-space regions.

\subsection{Tuned comparison of results from the different NLO calculations}
\label{sec:SM_WWW:comparison}

To compare the independent
calculations laid out in the previous section we 
choose the following setup.
The fiducial cross section for the process 
$\Pp\Pp\rightarrow\Pe^-\bar\nu_{\Pe}\mu^+\nu_\mu\tau^+\nu_\tau+X$
and its charge conjugate counterpart is defined by the phase-space cuts 
on the charged leptons summarized in Tab.~\ref{tab:SM_WWW:cuts}.
\begin{table}[t]
  \centering
  \begin{tabular}{c|c}
    Kinematical variable & fiducial range\\\hline
    $p_\mathrm{T}(\ell)$ & $[20,\infty]\,\text{GeV}$ \\
    $\eta(\ell)$ & $[-2.5,2.5]$ \\
    $p_\mathrm{T}(\ell_1)$ & $[27,\infty]\,\text{GeV}$ \\
    $\Delta R(\ell_i,\ell_j)$ & $[0.1,\infty]$
  \end{tabular}
  \caption{
    Definition of the fiducial region. Lepton requirements relate 
    to dressed leptons using a cone algorithm with $\Delta R_\text{dress}=0.1$.
    \label{tab:SM_WWW:cuts}
  }
\end{table}
The gauge-boson masses and widths are defined by their on-shell 
values provided by the Particle Data Group \cite{Tanabashi:2018oca}, 
\begin{center}
  \begin{tabular}{ll}
    $M_{\PW}^\text{OS}=80.379\,\text{GeV}$\;,\qquad & $\Gamma_{\PW}^\text{OS}=2.085\,\text{GeV}$\;, \\
    $M_{\PZ}^\text{OS}=91.1876\,\text{GeV}$\;,\qquad & $\Gamma_{\PZ}^\text{OS}=2.4952\,\text{GeV}$\;.
  \end{tabular}
\end{center}
They are then converted to pole masses using 
\begin{equation}
  M_V=\frac{M_V^\text{OS}}{\sqrt{1+\left(\frac{\Gamma_V^\text{OS}}{M_V^\text{OS}}\right)^2}}\;,
  \qquad
  \Gamma_V=\frac{\Gamma_V^\text{OS}}{\sqrt{1+\left(\frac{\Gamma_V^\text{OS}}{M_V^\text{OS}}\right)^2}}.
\end{equation}
In addition, we set the Higgs-boson and top-quark masses
and widths to 
\begin{center}
  \begin{tabular}{ll}
    $M_{\PH}=125\,\text{GeV}$\;, & $\Gamma_{\PH}=0.004088\,\text{GeV}$\;, \\
    $M_{\mathrm{t}}=173\,\text{GeV}$\;, & $\Gamma_{\mathrm{t}}=0$\;. \\
  \end{tabular}
\end{center}
All remaining quarks and leptons, in particular the bottom quark and the $\tau$-lepton,
are considered massless.
The CKM matrix is parametrized using the Cabibbo angle 
\begin{equation}
  \theta_\text{C}=0.22731\;,\nonumber
\end{equation}
neglecting mixing with the third generation. 
All parameters of the EW part of the Standard Model 
are fixed using the $G_\mu$ scheme \cite{Dittmaier:2001ay} with
\begin{equation}
  G_\mu=1.1663787\cdot 10^{-5}\,\text{GeV}^{-2}\;,\nonumber
\end{equation}
with the electromagnetic coupling fixed through the 
real parts of the complex masses, i.e.\
\begin{equation}
  \alpha=\frac{\sqrt{2}}{\pi}\,G_\mu\,M_{\PW}^2\left(1-\frac{M_{\PW}^2}{M_{\PZ}^2}\right)\;.
\end{equation}
The EW parameters are accordingly renormalized using the 
complex version of the EW on-shell renormalization scheme~\cite{Denner:2005fg,Denner:2019vbn}.

The parton densities of the proton are parametrized using the 
NNPDF 3.1 QCD LO PDF set \cite{Ball:2017nwa} for the LO cross 
section $\sigma^\text{LO}$, NNPDF 3.1 QCD+QED NLO PDF set 
\cite{Bertone:2017bme} for the Born contribution $\sigma_1^\text{LO}$ to the 
NLO calculation and all genuine NLO corrections. 
We choose the PDF sets with the strong coupling set to
\begin{equation}
  \alpha_{\mathrm{s}}(M_{\PZ})=0.118\nonumber
\end{equation}
and \textsc{Lhapdf}\footnote{
  In particular, we use \textsc{Lhapdf} 6.2.1 with PDF sets 
  \texttt{NNPDF31\_lo\_as\_0118} and 
  \texttt{NNPDF31\_nlo\_as\_0118\_luxqed}.
} to evaluate them, and
set the renormalization and factorization scale according to 
\begin{equation}
  \mu_\text{R/F}^2=\left(\vp3\,M_{\PW}\right)^2+\left(\sum\limits_{i\in S}\vec{p}_{\mathrm{T},i}\right)^2
\end{equation}
where $S$ denotes all colour-neutral final-state particles.

The relative NLO corrections are defined according to
\begin{equation}
  \delta_{q\bar{q}}^\text{EW}
  =\frac{\Delta\sigma_{q\bar{q}}^\text{NLO EW}}{\sigma_1^\text{LO}}
  \;,\qquad
  \delta_{q\gamma}^\text{EW}
  =\frac{\Delta\sigma_{q\gamma}^\text{NLO EW}}{\sigma^\text{LO}}
  \;,\qquad
  \delta^\text{QCD}
  =\frac{\sigma_1^\text{LO}-\sigma^\text{LO}+\Delta\sigma^\text{NLO QCD}}{\sigma^\text{LO}}\,,
\end{equation}
where the subscripts of the corrections $\Delta \sigma^\text{NLO}$
indicate the class of parton luminosities that contribute.
Furthermore, $\sigma^\text{LO}$ denotes the LO integrated cross
section evaluated with LO PDFs, whereas $\sigma_1^\text{LO}$ denotes
the LO integrated cross section evaluated with NLO PDFs (as part of
the NLO prediction).
With these definitions, both the QCD correction and the photon-induced 
EW correction are defined w.r.t.\ the pure LO calculation,
while $\delta_{q\bar{q}}^\text{EW}$ is almost
entirely insensitive to the actual PDF chosen, and thus 
universal.

The results obtained with both calculations for this setup 
for both channels, $\mathrm{W^+W^+W^-}$ and $\mathrm{W^+W^-W^-}$, for the LHC 
at both 13 and 14\,TeV centre-of-mass (CM) energy are detailed 
in Tabs.\ \ref{tab:SM_WWW:xsecs13} and \ref{tab:SM_WWW:xsecs14}. 
\begin{table}[t]
  \centering
  (i) $\mathrm{pp}\to e^-\mu^+\tau^+\bar{\nu}_e\nu_\mu\nu_\tau+X$\\
  \begin{tabular}{l|c|c|c|c|c}
    \hline
    13\,TeV \vP
    & LO [fb] & NLO [fb] 
    & $\delta_{q\bar{q}}^\text{EW}$ [\%]
    & $\delta_{q\gamma}^\text{EW}$ [\%]
    & $\delta^\text{QCD} [\%]$\\\hline
    \hfill DKS \vp
    & 0.194990(19) & 0.2626(10) & $-7.70(40)$ & 7.220(5) & 38.02(04) \\
    \hfill\Sherpa{}+\Recola \vp
    & 0.195118(83) & 0.2649(21) & $-7.38(57)$ & 7.217(3) & 38.11(10) \\\hline
  \end{tabular}\\[2mm]
  (ii) $\mathrm{pp}\to e^+\mu^-\tau^-\nu_e\bar{\nu}_\mu\bar{\nu}_\tau+X$\\
  \begin{tabular}{l|c|c|c|c|c}
    \hline
    13\,TeV \vP
    & LO [fb] & NLO [fb] 
    & $\delta_{q\bar{q}}^\text{EW}$ [\%]
    & $\delta_{q\gamma}^\text{EW}$ [\%]
    & $\delta^\text{QCD} [\%]$\\\hline
    \hfill DKS \vp
    & 0.118411(12) & 0.1597(06) & $-7.00(30)$ & 7.260(5) & 37.17(4) \\
    \hfill\Sherpa{}+\Recola \vp
    & 0.118420(73) & 0.1584(14) & $-6.73(51)$ & 7.267(3) & 37.07(9) \\\hline
  \end{tabular}
  \caption{
    Comparison of cross sections and relative NLO corrections at the LHC CM energy of 13\,TeV.
    \label{tab:SM_WWW:xsecs13}
  }
\vspace{1em}
  \centering
  (i) $\mathrm{pp}\to e^-\mu^+\tau^+\bar{\nu}_e\nu_\mu\nu_\tau+X$\\
  \begin{tabular}{l|c|c|c|c|c}
    \hline
    14\,TeV \vP
    & LO [fb] & NLO [fb] 
    & $\delta_{q\bar{q}}^\text{EW}$ [\%]
    & $\delta_{q\gamma}^\text{EW}$ [\%]
    & $\delta^\text{QCD} [\%]$\\\hline
    \hfill DKS \vp
    & 0.209820(20) & 0.2872(12) & $-7.80(40)$ & 7.780(5) & 40.04(04) \\
    \hfill\Sherpa{}+\Recola \vp
    & 0.209962(85) & 0.2898(23) & $-7.47(59)$ & 7.793(4) & 40.10(11) \\\hline
  \end{tabular}\\[2mm]
  (ii) $\mathrm{pp}\to e^+\mu^-\tau^-\nu_e\bar{\nu}_\mu\bar{\nu}_\tau+X$\\
  \begin{tabular}{l|c|c|c|c|c}
    \hline
    14\,TeV \vP
    & LO [fb] & NLO [fb] 
    & $\delta_{q\bar{q}}^\text{EW}$ [\%]
    & $\delta_{q\gamma}^\text{EW}$ [\%]
    & $\delta^\text{QCD} [\%]$\\\hline
    \hfill DKS \vp
    & 0.129986(13) & 0.1779(07) & $-7.20(40)$ & 7.730(5) & 39.15(04) \\
    \hfill\Sherpa{}+\Recola \vp
    & 0.130016(76) & 0.1766(15) & $-6.81(55)$ & 7.738(4) & 39.18(10) \\\hline
  \end{tabular}
  \caption{
    Comparison of cross sections and relative NLO corrections at the LHC CM energy of 14\,TeV.
    \label{tab:SM_WWW:xsecs14}
  }
\end{table}

\begin{figure}[t]
  \centering
  \includegraphics[width=0.5\textwidth]{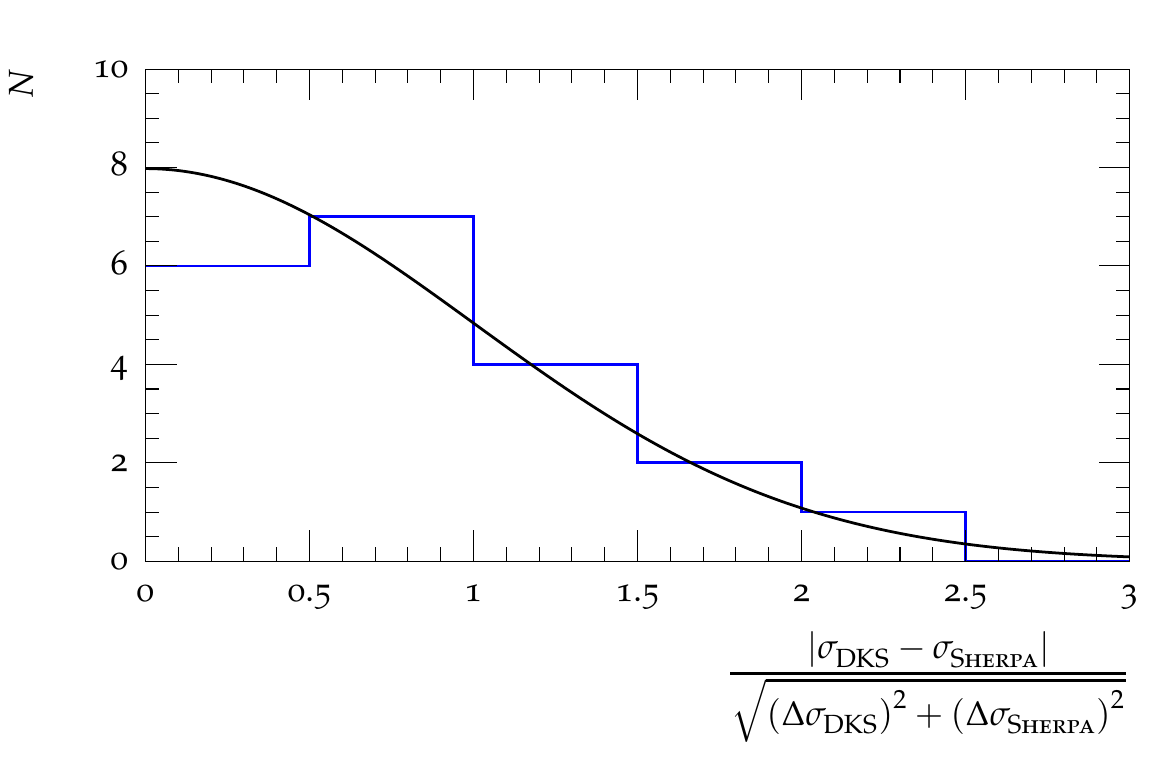}
  \caption{
    Comparison of computed cross sections and relative 
    corrections of DKS and \Sherpa{}+\Recola. 
    For reference the black line 
    shows a properly normalized normal distribution.
    \label{fig:SM_WWW:xscomp}
  }
\end{figure}

We generally find good agreement, see Fig.~\ref{fig:SM_WWW:xscomp},
despite convergence issues owing to the presence
of multiple narrow resonances.

\subsection{Summary of salient features of WWW production
cross sections at NLO}

Having validated the NLO predictions for the integrated
cross sections of WWW production at the LHC, we briefly
summarize the salient features of the integrated and differential
cross sections based on the results presented in
Ref.~\cite{Dittmaier:2019twg}:

\begin{itemize}
\item Similarly to the case of $\PW\PW\PW$ production with stable \PW bosons, 
a strong but accidental cancellation among the 
quark--antiquark and the remarkably large
quark--photon-induced EW corrections is observed.
For the chosen event setup at LHC energies of 13--14~TeV, 
they are of similar size ($\sim$\,7--8\,\%) but different in sign, 
so that the total EW corrections are below the percent level.
\item QCD corrections at the LHC CM energies of 13--14\,TeV amount to approximately $40\,\%$.
As the analyzed process is independent of $\alpha_{\mathrm{s}}$ at LO, there is no decrease of the residual scale 
dependence from LO to NLO.
To obtain a reduction of the scale uncertainty, next-to-next-to-leading order (NNLO) QCD calculations 
or multi-jet merging would be necessary.
\item Differential distributions that are sensitive to the momentum transfer in the process
show a strong impact of the EW high-energy logarithms, 
which reach 20--30\,\% in the TeV range, but
angular distributions are only slightly modified in shape by NLO EW corrections.
\begin{figure}[pt]
\centering
  \includegraphics{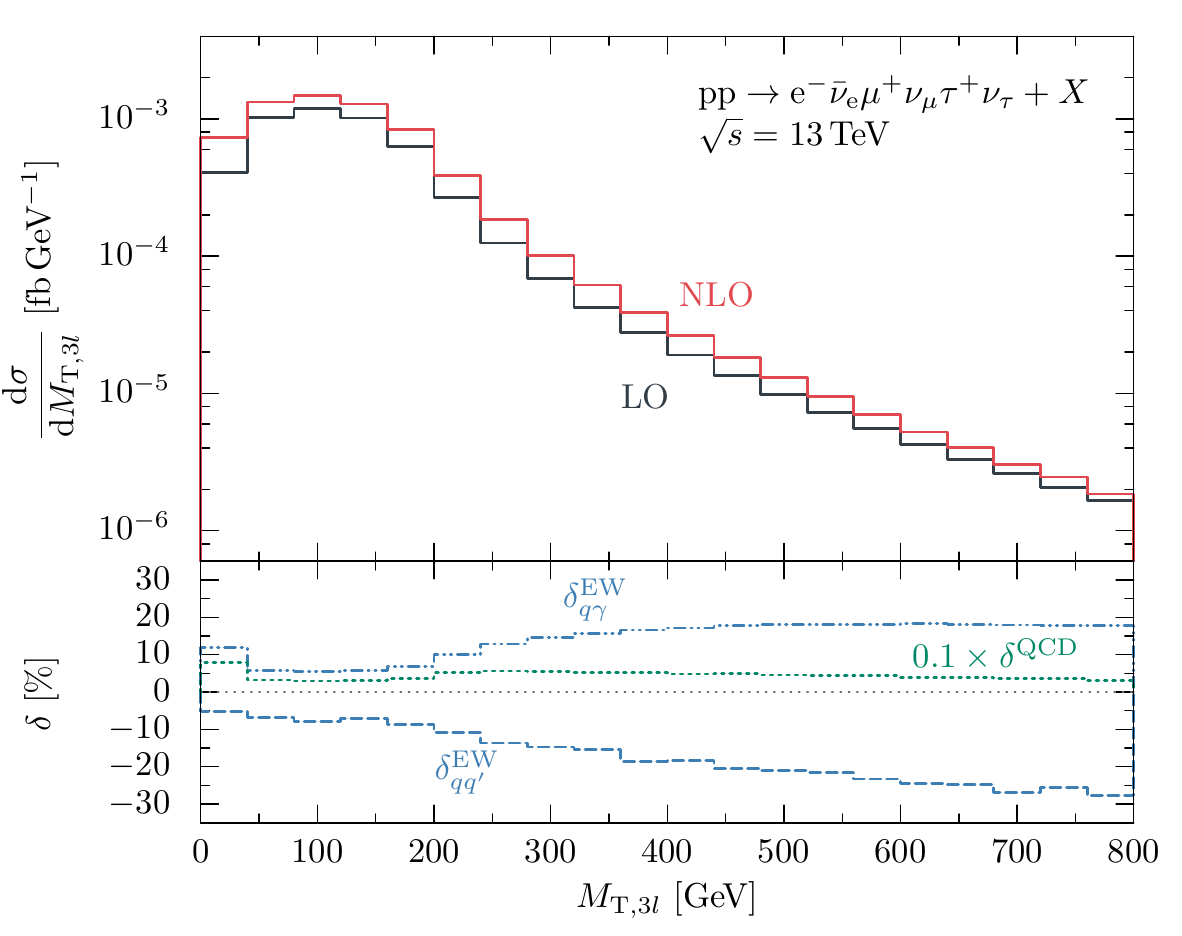}
  \caption{Differential distribution in the transverse mass $M_{\mathrm{T},3\ell}$ of the three-lepton system. 
  The NLO QCD correction $\delta^{\mathrm{QCD}}$ is scaled down by a factor of 10 for better readability.}
  \label{fig:SM_WWW:results_mt3l}
  \vspace*{\floatsep}
  \centering
  \includegraphics{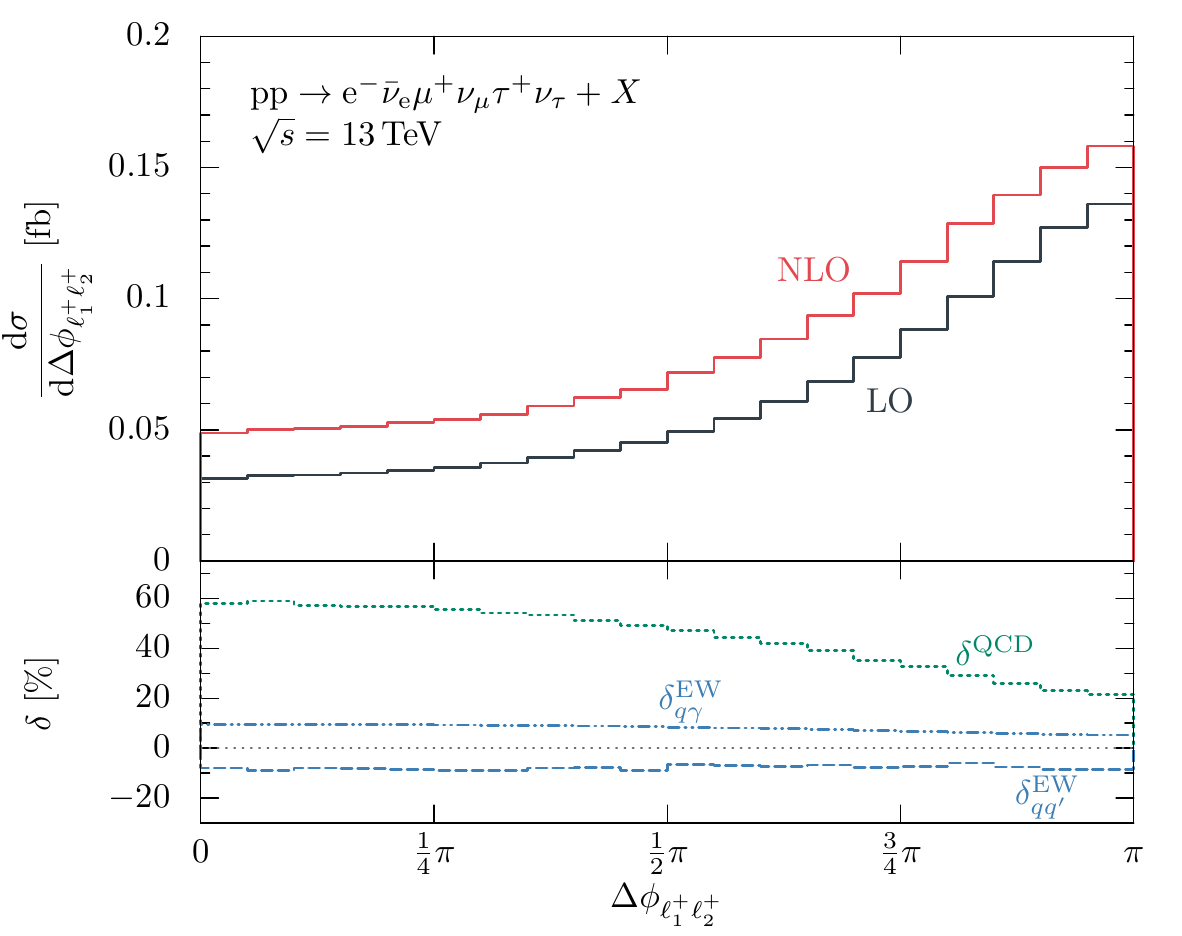}
  \caption{Differential LO and NLO cross section and relative NLO corrections in the difference in the 
  azimuthal angle of the two positively charged leptons, $\Delta\phi_{\ell^+_1\ell^+_2}$.}
  \label{fig:SM_WWW:results_deltaphi}
\end{figure}
In Figs.~\ref{fig:SM_WWW:results_mt3l} and \ref{fig:SM_WWW:results_deltaphi}
these features are illustrated for the distributions in the 
transverse mass $M_{\mathrm{T},3\ell}$ of the three-lepton system and in the
difference in the azimuthal angle of the two positively charged leptons, $\Delta\phi_{\ell^+_1\ell^+_2}$, respectively.
Distortions induced by QCD corrections strongly depend on the type of observable,
especially on their sensitivity to jet recoil effects.
Figures~\ref{fig:SM_WWW:results_mt3l} and \ref{fig:SM_WWW:results_deltaphi}, for instance, show that 
$\Delta\phi_{\ell^+_1\ell^+_2}$ is quite sensitive to those recoil effects, but
$M_{\mathrm{T},3\ell}$ is not.
In summary, we conclude that the inclusion of NLO corrections is important in any analysis
that constrains anomalous gauge couplings.
\item
Apart from the full off-shell calculation, Ref.~\cite{Dittmaier:2019twg} presents results
on the NLO corrections to WWW production within a triple-pole approximation (TPA),
which is based on the leading term in the expansion of the one-loop matrix elements 
around the resonances of the three \PW bosons.
For a consistent comparison of TPA and fully off-shell results, the Higgs-strahlung subprocess
has to be excluded by phase space cuts,
which is possible due to the good separation originating from 
the small Higgs width and the mass hierarchy $M_{\PH}<2M_{\PW}$.
The TPA performs very well in integrated cross sections and in angular and rapidity distributions, 
which are insensitive to off-shell effects.
For some observables, however, that become sensitive to non-resonant contributions, 
like the missing transverse momentum at high scales, the TPA is not a sufficient approximation.
Sizeable deviations can be observed in these regions.
Nevertheless, the size of the TPA uncertainty can be estimated reasonably well to identify 
those regions by analyzing TPA results only.
\end{itemize}
In summary, NLO results for EW corrections based on the full off-shell matrix elements 
are certainly sufficient for the analyses of WWW production at the LHC.
For integrated cross sections, even NLO EW corrections in the TPA will be sufficiently precise.

\subsection*{Acknowledgements}
CS is supported by the European Research Council under the European
Union's Horizon 2020 research and innovation Programme (grant
agreement ERC-AdG-740006).

\let\Herwig\undefined
\let\Pythia\undefined
\let\Sherpa\undefined
\let\Rivet\undefined
\let\Recola\undefined
\let\Professor\undefined
\let\Amegic\undefined
\let\OpenLoops\undefined
\let\Collier\undefined
\let\eps\undefined
\let\mc\undefined
\let\mr\undefined
\let\mb\undefined
\let\tm\undefined
\let\vp\undefined
\let\vP\undefined

\let\Madgraph\undefined

\chapter{Parton distribution functions}
\label{cha:pdf}
\def\lsim{\mathrel{\rlap{\lower4pt\hbox{\hskip1pt$\sim$}}
    \raise1pt\hbox{$<$}}}  

\section{Consistency of LHC top pair production data and their impact on parton distributions~\protect\footnote{
S.~Forte,
E.~R.~Nocera,
J.~Rojo}{}}


  We revisit the impact of the ATLAS and CMS top pair production measurements
  at $\sqrt{s}=8$ TeV on a
  global determination of parton distribution functions (PDFs).
  Our analysis includes all the differential distributions
  from the  ATLAS 8~TeV $t\bar{t}$ lepton+jet
  data set, together with their cross-correlations, in a
PDF determination
akin to the published NNPDF3.1  set.
We study the mutual consistency of these distributions
and their consistency with the rest of the data sets of
the global fit.
We specifically address the relative impact of the
normalized and unnormalized data, the consequences of fitting the charm
PDF, the role of the top quark transverse momentum distributions, and the
effects of partially decorrelating experimental systematic
uncertainties.

\subsection{Top pair production data and parton distributions}
\label{sec:SM_ttbar_for_PDFs:toppdfs}

The set of processes used for the accurate determination of the parton
distribution functions (PDFs)~\cite{Gao:2017yyd} has been steadily widening over time,
beyond the traditional 
combination of deep-inelastic scattering (DIS), Drell-Yan (DY) and jet 
production data which has now been used for more than thirty
years~\cite{Martin:1987vw}.
Top pair production data were suggested as
an effective way to constrain the gluon distribution at large $x$
since the early days of the LHC (see e.g. Ref.~\cite{Forte:2013wc});
and their impact on PDF fits was studied both at the level of total
cross-sections~\cite{Czakon:2013tha} and, subsequently, of differential
distributions~\cite{Czakon:2016olj}.
Several measurements  
have been published by both ATLAS and CMS at center-of-mass energies
of $\sqrt{s}=5.02$, 7, 8 and 13 TeV: total cross-sections and
differential distributions with respect to a variety
of kinematic variables,  including the top transverse momentum 
$p_T^t$, the top rapidity $y_t$, the rapidity of the top pair $y_{t\bar t}$ and 
the invariant mass of the top pair $m_{t\bar t}$, both normalized and 
unnormalized with respect to the total cross-section (see Sect.~67.3.1 in 
Ref.~\cite{Tanabashi:2018oca} for an updated review of all of the available 
measurements).
\begin{table}[!t]
  \centering
  \tiny
  \begin{tabular}{llrccccccccc}
\toprule
Dataset &
& $N_{\rm dat}$ & Fit 1 & Fit 2 & Fit 3 & Fit 4 & Fit 5 & Fit 6
              & Fit 7 & Fit 8 & Fit 9 \\
\midrule
ATLAS $t\bar{t}$ norm. diff. (cor.)~\cite{Aad:2015mbv} &
&   21 & [2.74] & [2.90] & [2.64] &  2.28  & [4.60] &  2.29  & [3.49] &  2.23  & [2.31] \\
ATLAS $t\bar{t}$ norm. diff. (unc.)~\cite{Aad:2015mbv} &
&   21 & [2.08] & [2.07] & [2.05] & [1.94] & [4.16] & [1.89] & [3.07] & [1.92] & [1.97] \\
\quad ATLAS $1/\sigma d\sigma/dp_T^t$ &
&    7 & [3.50] & [3.57] & [3.25] &  2.94  & [2.45] &  2.95  & [2.54] &  2.92  & [3.10] \\
\quad ATLAS $1/\sigma d\sigma/dy_t$ & \ \ \ \ \ \ $\vdash$
&    4 &  1.45  &  1.33  &  1.08  &  1.20  & [4.81] &  1.10  & [2.98] &  1.10  & [1.05] \\
\quad ATLAS $1/\sigma d\sigma/dy_{t\bar{t}}$ &
&    4 & [1.26] & [1.13] & [1.66] &  1.55  & [10.2] &  1.40  & [6.30] &  1.31  & [1.59] \\
\quad ATLAS $1/\sigma d\sigma/dm_{t\bar{t}}$ &
&    6 & [1.78] & [1.83] & [1.67]  & 1.59  & [1.36] &  1.61  & [1.42] &  1.61  & [1.61] \\
ATLAS $t\bar{t}$ unnor. diff. (cor.)~\cite{Aad:2015mbv} &
&   25 & [7.96] & [8.33] & [7.52] & [6.88] &  5.76  & [7.18] &  5.23  & [2.25] &  2.16  \\ 
ATLAS $t\bar{t}$ unnor. diff. (unc.)~\cite{Aad:2015mbv} &
&   25 & [2.09] & [2.08] & [2.09] & [2.06] & [2.28] & [2.25] & [2.17] & [2.07] & [2.02] \\ 
\quad ATLAS $d\sigma/dp_T^t$ & $\dag$ $\ddag$
&    8 & [2.41] & [2.46] & [2.50] & [2.43] &  2.50  & [2.46] &  2.54  & [2.50] &  2.42  \\
\quad ATLAS $d\sigma/dy_t$ & $\dag$
&    5 & [0.87] & [0.78] & [0.73] & [0.76] &  1.14  & [0.73] &  0.87  & [0.73] &  0.66  \\
\quad ATLAS $d\sigma/dy_{t\bar{t}}$ & $\dag$
&    5 & [1.21] & [1.11] & [1.32] & [1.19] &  2.36  & [1.15] &  1.86  & [1.14] &  1.16  \\
\quad ATLAS $d\sigma/dm_{t\bar{t}}$ & $\dag$ $\ddag$
&    7 & [3.27] & [3.30] & [3.18] & [3.24] &  2.85  & [3.25] &  2.94  & [3.25] &  3.16  \\
\midrule
Fixed target DIS (NC) & 
& 1039 &  1.21  &  1.20  &  1.20  &  1.20  &  1.20  &  1.27  &  1.27  &  1.26  &  1.26  \\
HERA DIS (NC) & 
& 1064 &  1.15  &  1.14  &  1.15  &  1.14  &  1.15  &  1.20  &  1.20  &  1.20  &  1.20  \\
Fixed target DIS (CC) &
&  908 &  1.08  &  1.09  &  1.08  &  1.09  &  1.08  &  1.11  &  1.11  &  1.11  &  1.10  \\
HERA DIS (CC) &
&   81 &  1.19  & 1.18   &  1.19  &  1.18  &  1.19  &  1.15  &  1.14  &  1.14  &  1.15  \\
Fermilab DY &
&  189 &  1.24  &  1.22  &  1.23  &  1.24  &  1.23  &  1.14  &  1.14  &  1.15  &  1.14  \\
Tevatron DY &
&   74 &  1.29  &  1.26  &  1.28  &  1.27  &  1.30  &  1.25  &  1.25  &  1.25  &  1.24  \\
ATLAS DY &
&   75 &  1.55  &  1.50  &  1.53  &  1.48  &  1.49  &  1.80  &  1.81  &  1.81  &  1.77  \\
\quad ATLAS W/Z rap. 2011 & 
&  34  &  2.14  &  2.19  &  2.15  &  2.07  &  2.10  &  2.69  &  2.72  &  2.71  &  2.63  \\
CMS DY &
&  154 &  1.23  &  1.23  &  1.22  &  1.23  &  1.22  &  1.24  &  1.23  &  1.24  &  1.25  \\
LHCb DY &
&   85 &  1.47  &  1.52  &  1.57  &  1.49  &  1.55  &  1.43  &  1.47  &  1.41  &  1.42  \\
ATLAS jets &
&   31 &  0.90  &  1.13  &  1.09  &  1.08  &  1.01  &  1.11  &  1.05  &  1.11  &  1.12  \\
CMS jets &
&  133 &  0.88  &  0.96  &  0.93  &  0.95  &  1.05  &  0.94  &  0.99  &  0.93  &  0.93  \\
ATLAS $Z\,p_T$ &
&   92 &  0.90  &  0.92  &  0.93  &  0.91  &  0.95  &  0.95  &  0.97  &  0.94  &  0.96  \\
CMS $Z\,p_T$ &
&   28 &  1.33  &  1.30  &  1.31  &  1.35  &  1.33  &  1.30  &  1.31  &  1.31  &  1.28  \\
ATLAS $\sigma_{t\bar{t}}$~\cite{Aad:2014kva,Aaboud:2016pbd} &
&  3/2 &  0.86  &  0.84  &  0.77  &  0.74  &  0.68  &  0.72  &  0.83  &  0.72  &  0.71  \\
CMS $\sigma_{t\bar{t}}$~\cite{Khachatryan:2016mqs,Khachatryan:2015uqb} &
&    3 &  0.20  &  0.21  &  0.23  &  0.29  &  0.56  &  0.35  &  0.34  &  0.28  &  0.27  \\
CMS $1/\sigma d\sigma/dy_{t\bar{t}}$~\cite{Khachatryan:2015oqa} &
&   9 &  0.94  &  1.08  &  1.05  &  0.98  &  1.01  &  0.97  &  1.46  &  0.97  &  1.01  \\
\midrule
Total & & & 1.148 &  1.163 & 1.163  & 1.171  & 1.195  & 1.204  & 1.227 & 1.204 &  1.203 \\
\bottomrule
\end{tabular}

  \caption{\small Value of $\chi^2$ per data point for all of the ATLAS
    top pair-production distributions, with respect to the top transverse 
    momentum $p_T^t$, the top rapidity $y_t$, the rapidity of the top pair 
    $y_{tt}$ and the invariant mass of the top pair $m_{tt}$, both normalized and
    unnormalized, and for all of the above combined; $\chi^2$ values are
    also shown for all other data from the NNPDF3.1
    data set~\cite{Ball:2018iqk}, which are included in all the PDF fits 
    considered here: charged- and neutral-current (CC and NC) DIS structure 
    functions from both fixed-target and HERA combined experiments;
    fixed-target and collider DY rapidity and invariant mass distributions;
    single-jet inclusive cross-sections; $Z$ transverse momentum distributions;
    total top-pair cross-sections; and the CMS top rapidity distribution. 
    For each data set the total number of data points is shown; note
    that indented data sets are subsets of the preceding non-indented data set. 
    For the ATLAS top-pair differential distributions we indicate whether they 
    were included in NNPDF3.1~\cite{Ball:2017nwa} ($\vdash$), 
    in CT18~\cite{Hou:2019efy} ($\ddag$), or in the MMHT-based study of 
    Ref.~\cite{Bailey:2019yze} ($\dag$). Each column corresponds to a different 
    PDF fit (see text); $\chi^2$ values for the data sets not included in the 
    fit are quoted in brackets.
    \label{tab:SM_ttbar_for_PDFs:chi2}
}\end{table}

Differential cross-sections for top-quark pair production measured by 
ATLAS~\cite{Aad:2015mbv} and CMS~\cite{Khachatryan:2015oqa} in the lepton+jets
channel at a center-of-mass energy of 8 TeV were used
in the NNPDF3.1 global PDF determination~\cite{Ball:2017nwa}, complementing
selected total cross-sections (already used in the
ABM12~\cite{Alekhin:2013nda} and NNPDF3.0~\cite{Ball:2014uwa}  PDF fits).
In order to include these observables in the PDF fit, the choice 
of a specific kinematic distribution had to be made because the information
on correlations across distributions was not available at that time:
their simultaneous inclusion would have otherwise amounted to double counting, 
as they come from the same underlying data.
The particular choice of 
observables adopted for NNPDF3.1 (see Table~\ref{tab:SM_ttbar_for_PDFs:chi2}),
namely the normalized rapidity distribution of the top quark (for ATLAS) and
of the top pair (for CMS), was based on the results of a previous
study~\cite{Czakon:2016olj}, which analyzed the impact of different observables 
on PDFs and their  consistency.

In the NNPDF3.1 global analysis the impact of the inclusion
of the top rapidity distributions at 8~TeV on the resulting 
PDFs was assessed, and
found to be significant on the gluon PDF (in the region 
$0.1\lsim x\lsim 0.5$) and negligible for other PDFs.
The consistency of the constraint imposed on the large-$x$
gluon by this data with those coming from other data
included in NNPDF3.1 was further studied in
Ref.~\cite{Nocera:2017zge}, where top, $Z$ transverse momentum
and single-inclusive
jet distributions were added in turn to a baseline global data set.
Excellent consistency was found, with all data sets pulling the
gluon in the same direction, and the top and jet data having the
biggest impact.

There are a number of reasons why the impact of top data on PDF determination
is worth revisiting.

\begin{itemize}
    \item Since the publication of the original ATLAS
    paper~\cite{Aad:2015mbv}, the 
    covariance matrix of the individual measurements was
    updated~\cite{ATL-PHYS-PUB-2018-017,Giuli:2019wtu}: it is advisable to check
    whether the previous results of
    Ref.~\cite{Ball:2017nwa} are affected by this update. 
    \item In the same Refs.~\cite{ATL-PHYS-PUB-2018-017,Giuli:2019wtu} full information on
    the correlation between pairs of different kinematic distributions was made
    available: it is now possible to include all distributions
    at once and check the comparative impact on PDFs, and how it
    affects the conclusions of Refs.~\cite{Ball:2017nwa,Nocera:2017zge}.
    \item Some recent studies, specifically an
    ATLAS study~\cite{ATL-PHYS-PUB-2018-017,Giuli:2019wtu} within the 
    {\sc xFitter}~\cite{Alekhin:2014irh} framework, and a
    study~\cite{Bailey:2019yze} based  on the
    MMHT~\cite{Harland-Lang:2014zoa} framework, found that there
    are serious difficulties in simultaneously including  all of
    the differential distributions
    from the ATLAS 8 TeV lepton+jets top data set in a PDF
    determination: it is worth
    investigating whether similar conclusions also apply when
    analyzing these data within the current NNPDF
    framework~\cite{Ball:2017nwa}.
    \item The recently published CT18 PDF set~\cite{Hou:2019efy}
    also includes top-quark pair differential distributions, but with a
    different choice of observables in comparison to NNPDF3.1
    (see Table~\ref{tab:SM_ttbar_for_PDFs:chi2}). It is interesting to compare and
    assess the impact of different choices at the PDF level.
\end{itemize}
          
We will address all these issues by performing a number of
PDF determinations based on the NNPDF3.1 methodology and
data set, adding the full set of ATLAS top data either in
normalized or unnormalized form to the
baseline data set in various ways, and studying the fit quality
and the impact on PDFs.

\subsection{The ATLAS top production data and their impact on PDFs}
\label{sec:SM_ttbar_for_PDFs:fits}

All the PDF fits presented here are based on the NNPDF3.1
methodology of Ref.~\cite{Ball:2017nwa}, with the slightly modified
data set used in Ref.~\cite{Ball:2018iqk}.
The latter differs from the
original NNPDF3.1 data set in that only processes for which full NNLO
computations are available are
included (in particular, in the NNPDF3.1 fit some jet data were included
using NLO theory).
This data set will be supplemented with a number of top pair
differential distributions measured in the lepton+jet channel at a 
center-of-mass energy of 8 TeV by ATLAS (see Table~\ref{tab:SM_ttbar_for_PDFs:chi2}).
We refer to Ref.~\cite{Ball:2017nwa} for a detailed discussion of the 
NNPDF3.1 data set and the associated fitting methodology.
In Sect.~\ref{sec:SM_ttbar_for_PDFs:pdffits} we will present the various PDF sets and 
discuss how their features vary as the underlying data set is changed,
in a series of pairwise comparisons between PDFs.
In Sect.~\ref{sec:SM_ttbar_for_PDFs:disc} we will then discuss these
results and our best
understanding of them, also using information from PDF determinations
in which methodological changes are made, either in the treatment of
theory or of correlated experimental uncertainties, as well as
a few further
auxiliary PDF determinations based on special subsets of data.

\subsection{The impact of the data set choice on PDFs}
\label{sec:SM_ttbar_for_PDFs:pdffits}

In Table~\ref{tab:SM_ttbar_for_PDFs:chi2} we list all
the ATLAS top-quark pair observables corresponding to the 8 TeV lepton+jets data set
whose inclusion we consider and
compare.
In each case we provide the total number of data points, and we indicate
whether the observable was included in NNPDF3.1~\cite{Ball:2017nwa}
($\vdash$),
CT18~\cite{Hou:2019efy} ($\ddag$) or the recent MMHT-based
study~\cite{Bailey:2019yze} ($\dag$).
In the case of the  CMS measurements,
the same top-quark data as in the NNPDF3.1 PDF determination
are included: on the one hand, full
correlations are not available for CMS, hence the inclusion of all
differential distributions at once is not possible; also, there appear
to be no
specific issues with the CMS 8 TeV top-quark data, with 
consensus~\cite{Czakon:2016olj,Hou:2019gfw,Bailey:2019yze} that all
observables show a similar pull on the 
gluon distribution, with only the invariant mass distribution
providing a poor fit quality.
Hence, as in the related study of Ref.~\cite{Bailey:2019yze},
here we will focus our attention on the ATLAS data.

Our results in terms of fit quality are summarized in Table~\ref{tab:SM_ttbar_for_PDFs:chi2},
where we show the $\chi^2$ per datapoint
for the ATLAS top normalized and unnormalized (or, equivalently, absolute) distributions, as well as
(indented) the breakdown of each of them into the four individual
observables that make up the normalized or absolute top data sets: transverse momentum distribution,
rapidity distribution,  pair rapidity distribution, and pair invariant
mass distribution.
Note that the normalized distributions always have
one fewer datapoint, because the last data bin is fixed by the
normalization condition.
Also, to avoid double counting, the corresponding measurement
of the total cross-section is removed from the fit whenever unnormalized
distributions are included (while other total cross-sections at 7 and 13 TeV
are retained). Note that in principle, when fitting the normalized
distribution, the correlation between the differential distribution
and total cross-section should be included. This is not  done because
this information is not available to us. In principle, for full consistency
we should therefore exclude the total cross-section from the data set:
note however that because this is a single data point, with
$\chi^2\lsim 1$, so this exclusion
would in practice make no difference.
We have also computed  
$\chi^2$ values for the full ATLAS top normalized or
unnormalized data set decorrelating different 
distributions, i.e., using  a block-diagonal covariance matrix that
only correlates data points that belong to the same distribution: these
values are provided
for illustration as a separate row in the Table (and not used for
fitting).

In Table~\ref{tab:SM_ttbar_for_PDFs:chi2} we also provide  $\chi^2$ values  for all
other data sets in the global fit: DIS structure functions from both 
fixed-target and HERA combined experiments; fixed-target and collider DY 
rapidity and invariant mass distributions; single-jet inclusive cross-sections; 
$Z$ transverse momentum distributions; total top-pair cross-sections; 
the CMS top rapidity distribution; and finally the global fit quality for the 
complete fitted data set.
In view of the discussion in Sect.~\ref{sec:SM_ttbar_for_PDFs:disc} 
below for the ATLAS DY data we also show (indented) the $\chi^2$ value corresponding
to the specific subset of this data corresponding to the 2011 $W/Z$ rapidity
distribution~\cite{Aaboud:2016btc}.
Fit quality is always shown both for data which are and for those
which are not included in the fit. All $\chi^2$ values not used for
fitting are shown in square brackets in Table~\ref{tab:SM_ttbar_for_PDFs:chi2}.

Each column in Table~\ref{tab:SM_ttbar_for_PDFs:chi2} corresponds to a separate PDF
fit.
All these fits have been carried out using NNLO theory, and they only differ in the treatment of
the ATLAS top data, with all the rest of the data set being identical
to Ref.~\cite{Ball:2018iqk}.
We also have performed the corresponding
NLO fits, but these are not shown here because they do not appear to
add anything to
the discussion: they merely exhibit somewhat worse fit quality but with
identical qualitative features.

The fits included in Table~\ref{tab:SM_ttbar_for_PDFs:chi2} and which we will discuss below are
the following:
\begin{description}
\item[Fit 1] NNPDF3.1: this simply reproduces for reference the
  published~\cite{Ball:2017nwa}  NNPDF3.1 results.

  \item[Fit 2] Baseline: this is our baseline fit, which only differs from
  the published NNPDF3.1 because it is based on the slightly different data set 
  already adopted in Ref.~\cite{Ball:2018iqk}.\footnote{Note that the number of
  data points for the ATLAS top rapidity distribution is $N_{\rm  dat}=10$ 
  in Table~3 of Ref.~\cite{Ball:2017nwa}, while it is $N_{\rm dat}=5$ 
  in Table~\ref{tab:SM_ttbar_for_PDFs:chi2}. This is due to the fact that both the
  distribution with respect to rapidity ($N_{\rm  dat}=10$) and with respect to 
  the absolute value of the rapidity   ($N_{\rm  dat}=5$) were
  published by ATLAS in Ref.~\cite{Aad:2015mbv}. The former was
  used in Ref.~\cite{Ball:2017nwa}, but information on
  correlations was only made available in Ref.~\cite{ATL-PHYS-PUB-2018-017,Giuli:2019wtu} 
  for the latter, which is therefore used here.}

  \item[Fit 3] Baseline, corrected: this is the same as the baseline, but
  now using the new, updated covariance matrix for the top data
  from Ref.~\cite{ATL-PHYS-PUB-2018-017,Giuli:2019wtu}.

  \item[Fit 4] All, normalized: this includes all of the different ATLAS
  top observables, and correlations across different 
  distributions~\cite{ATL-PHYS-PUB-2018-017,Giuli:2019wtu,Mandy,Lucian}.\footnote{We
  have checked with the authors~\cite{Lucian} of Ref.~\cite{Bailey:2019yze}
  that our implementation of correlations is in agreement with their own.}

  \item[Fit 5] All, unnormalized: this is the same as \#4,
  but now including all of the observables in the
  absolute, rather than normalized version.

 \item[Fit 6] Perturbative charm, normalized: this is the same as
 \#4, but now with the charm PDF being generated through
 perturbative matching (as e.g. CT18 and MMHT do), rather
 than independently parametrized and fitted~\cite{Ball:2016neh} as in the
 default NNPDF3.1 set.

  \item[Fit 7] Perturbative charm, unnormalized: this is the same
  as \#6, but now using absolute observables.

  \item[Fit 8] Perturbative charm, normalized and decorrelated:
  this is the same as  \#6, but now decorrelating
  parton-shower systematic uncertainties across bins belonging to different
  distributions, as suggested in Ref.~\cite{Bailey:2019yze}.

  \item[Fit 9] Perturbative charm, unnormalized and decorrelated:
  this is the same as  \#8, but now using absolute observables.

\end{description}

\begin{figure}[!t]
\begin{center}
\includegraphics[scale=0.5]{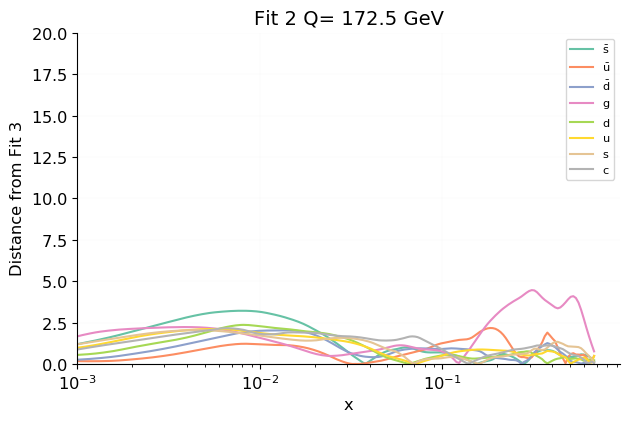}
\includegraphics[scale=0.5]{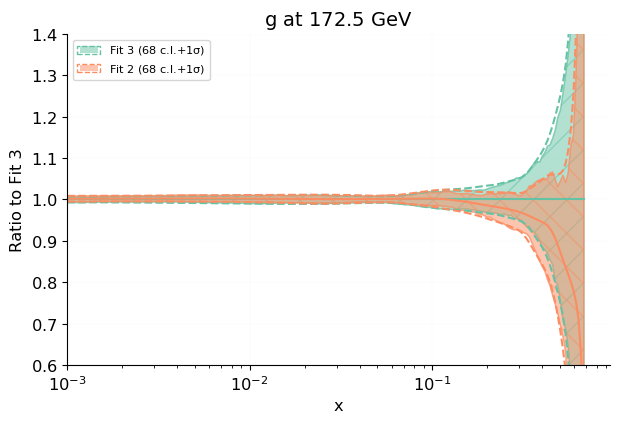}\\
\caption{\small Comparison between the baseline (set \#2) and PDFs determined
  using the same data set but the updated covariance matrix (set
  \#3). The distances between PDFs (left) and the gluon distributions
  (right) are shown.}
\label{fig:SM_ttbar_for_PDFs:comp23}
\end{center}
\end{figure}

We now discuss and compare the PDF determinations \#1-\#5, which
correspond to different choices of underlying data set,  in order to
address the various issues listed at the end of
Sect.~\ref{sec:SM_ttbar_for_PDFs:toppdfs}; these comparisons have been generated using the {\sc ReportEngine}
software~\cite{zahari_kassabov_2019_2571601}.
For each comparison, we show (as a function of
$x$ and  at the scale corresponding to the top quark mass,
$Q=172.5$~GeV) the distances between all PDFs and we compare the gluon
PDF, which is mostly affected by the top data.
Recall that the distance $d$ is defined
as the difference in units of the standard deviation of the mean, so
for a sample of 100 replicas $d\sim1$ corresponds to statistically
identical PDFs (replicas extracted from the same underlying
distribution) and $d\sim 10$ corresponds to PDFs that differ by
one-$\sigma$.
In Sect.~\ref{sec:SM_ttbar_for_PDFs:disc} we will discuss the PDF
determinations \#6-\#9, which correspond to changes in methodology which we
have performed in order to correctly interpret these results.

\begin{figure}[!t]
\begin{center}
\includegraphics[scale=0.5]{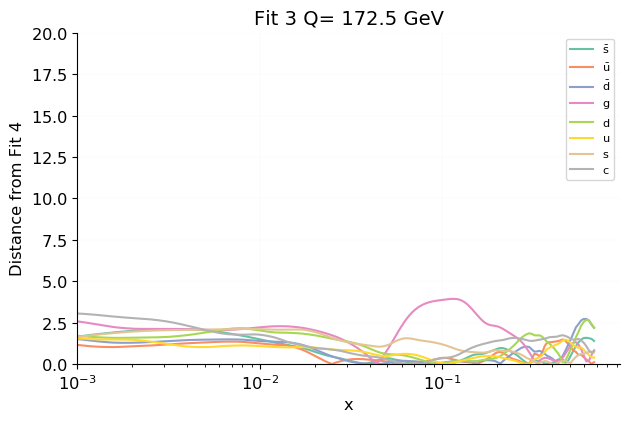}
\includegraphics[scale=0.5]{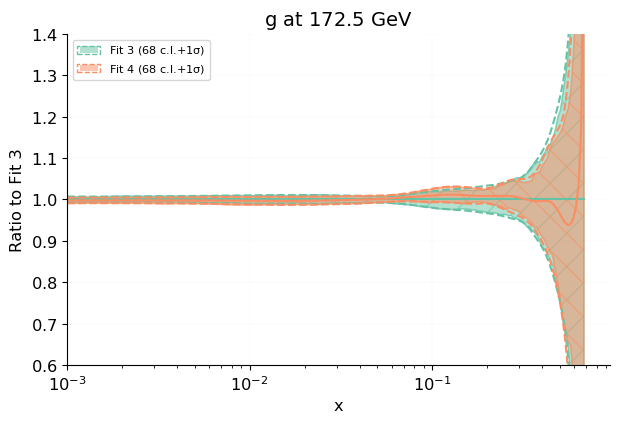}\\
\caption{\small As Fig.~\ref{fig:SM_ttbar_for_PDFs:comp23} but now comparing  PDFs
  determined from  the baseline data set
  (but improved covariance matrix, set \#3) and PDFs determined
  including  the full set
  of ATLAS normalized distributions  (set \#4).}
\label{fig:SM_ttbar_for_PDFs:comp34}
\end{center}
\end{figure}

First, we assess the impact of the update in the ATLAS 8 TeV lepton+jets covariance matrix
presented in  Ref.~\cite{ATL-PHYS-PUB-2018-017,Giuli:2019wtu} on the results of
Ref.~\cite{Ball:2017nwa}.
We start from the baseline PDF set \#2,
which is essentially the same as NNPDF3.1, as can be seen from the
$\chi^2$ values in Table~\ref{tab:SM_ttbar_for_PDFs:chi2}. In Fig.~\ref{fig:SM_ttbar_for_PDFs:comp23} we
compare the baseline to PDFs determined using the same  data set (i.e. essentially the 
NNPDF3.1 data set) but with the new updated covariance matrix. 
 It is clear from the figure that the two sets of PDFs are very close
 to being  statistically indistinguishable: the updated covariance
 matrix has essentially
 no effect on the PDF
 determination.
 Interestingly, it does however lead to an improved
 value of the $\chi^2$ for the top rapidity distribution, which now corresponds to a near-perfect fit,
 $\chi^2=1.08$.

\begin{figure}[!t]
\begin{center}
\includegraphics[scale=0.5]{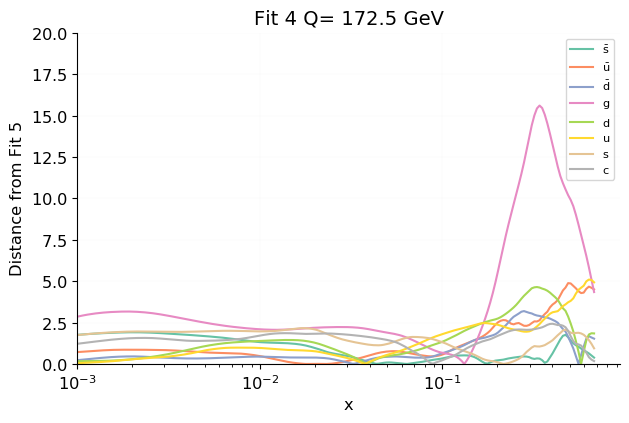}
\includegraphics[scale=0.5]{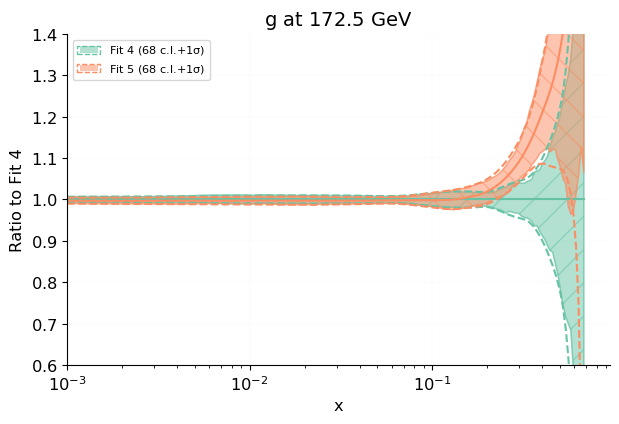}\\
\caption{\small As Fig.~\ref{fig:SM_ttbar_for_PDFs:comp23} but now comparing  PDFs
  determined
  including  the full set
  of ATLAS normalized distributions respectively in normalized (set
  \#4) or unnormalized (set \#5) form.}
\label{fig:SM_ttbar_for_PDFs:comp45}
\end{center}
\end{figure}

Next, we enlarge the top data set to include all of the ATLAS
 distributions, in normalized form; PDFs before and after this
 enlargement of the baseline data set are compared in
 Fig.~\ref{fig:SM_ttbar_for_PDFs:comp34}.
 It is clear that also in this case no
 significant effect is seen: the simultaneous inclusion of the four
 differential distributions 
 carries effectively  no new information
 as compared to fitting only the $y_t$ distribution.
 The fit quality for individual observables is poor for
 the transverse momentum distribution ($\chi^2=2.94$) but fair to good
 for all other distributions, with the rapidity distribution being
 fitted best; the fit quality
 to the whole set of top data is fair, and it does not significantly
 improve by decorrelating systematic uncertainties.
 The fit quality to the global
 data set is essentially the same as that of the baseline, as it must
 be, given that the PDFs are unchanged. The reasons for the poor fit of the transverse momentum
 distributions will be discussed in Sect.~\ref{sec:SM_ttbar_for_PDFs:disc}. Note that the
 fit quality to each of the unnormalized distributions is similar (and
 sometimes better) to that of the corresponding normalized
 ones, despite the fact that these distributions are
 not being fitted; only the fit
 quality to the invariant mass distribution exhibits a significant deterioration.
 However, the fit quality to the full set of unnormalized
 distributions is very poor ($\chi^2>6$).  Nonetheless,
 if we recompute this $\chi^2$ value decorrelating uncertainties
 as discussed above
 (by taking a block-diagonal covariance matrix and thus neglecting
 cross-correlations between distributions), it becomes fair
 ($\chi^2=2.06$).
 This  suggests an issue
 with correlated uncertainties for unnormalized observables. We will
 revisit this point when discussing PDF sets \#8 and \#9.

 We now repeat the PDF determination with all ATLAS top distributions
 included, but using the unnormalized distributions.
 The resulting PDFs
 are compared to those obtained using the normalized distributions in
 Fig.~\ref{fig:SM_ttbar_for_PDFs:comp45}.
 It is clear that now  a shift by more
 than one $\sigma$ is observed between the two gluon PDFs
 in the large-$x$ region $0.1\lsim x\lsim 1$, with
 some smaller shift also seen for some quark PDFs:
 the absolute top-pair distributions
 appears to pull the large-$x$ gluon upwards
 in comparison to the normalized distributions.
 The fit quality to the individual absolute distributions however turns out to be
 similar (and sometimes even worse) in comparison to the case in which they were
 not fitted, and the pattern is unchanged: it is only the fit quality
 to the correlated set of top observables that improves somewhat (from
 $\chi^2=6.88$ to $\chi^2=5.76$), though it  remains very poor.
 Just like in the case in which normalized distributions were fitted,
 we find that this value improves considerably
 if it is recomputed decorrelating experimental
 systematics: from  $\chi^2=5.76$ to
 $\chi^2=2.28$, a value 
 similar (in fact slightly worse) to
 the value found when the normalized distributions were fitted.
 It is important to observe that in fit \#5, in which top observables
 are included in unnormalized form, 
 the fit quality to the global data set deteriorates somewhat in
 comparison to fit \#4, in which normalized observables were used.

 This concludes our presentation of  results from PDF fits
 corresponding to the variations of underlying data set that we
 consider here.
 We now turn to their interpretation.

\subsubsection{Interpretation and dependence of PDFs on the methodology}
\label{sec:SM_ttbar_for_PDFs:disc}

The PDF determinations presented in the previous section lead to the
following immediate conclusions:
\begin{itemize}
\item The parton distributions
  determined using the NNPDF3.1 data set and methodology are
    unaffected if the ATLAS normalized $y_t$ distribution is
    supplemented by the full set of ATLAS normalized differential distributions;
    the fit quality is generally good except for the transverse momentum
    distribution which is poorly fitted.
    \item  If the normalized distributions are replaced by the absolute
      ones, the large-$x$ gluon is pushed upwards in the large $x$ region.
      The fit quality to the
      individual ATLAS top distributions is similar to the one found in the
      normalized case, but the fit quality to  the full set of
      correlated observables is very poor, and the fit quality to the
      rest of the global data set deteriorates somewhat. 
\end{itemize}

\begin{figure}[!t]
\begin{center}
\includegraphics[scale=0.5]{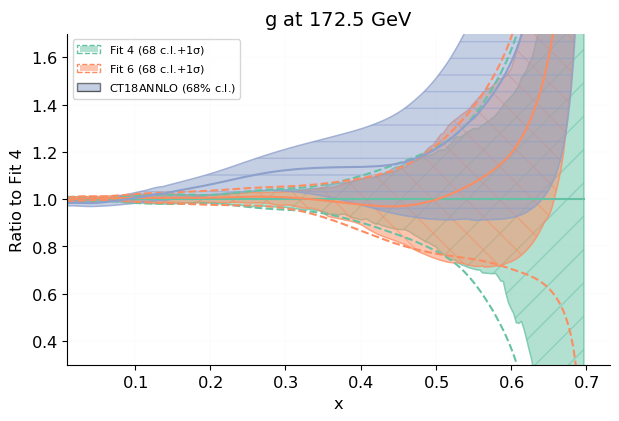}
\includegraphics[scale=0.5]{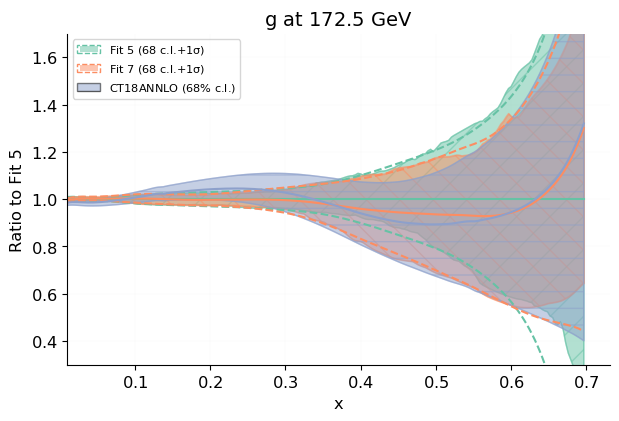}
\caption{\small Comparison between the gluon PDFs in the sets in
  which all distributions are fitted, with fitted charm
   or perturbative charm. The comparison is shown both for fits to
   normalized distributions (left: PDF set \#4 vs. set \#6) and to
   unnormalized distributions (right: PDF set \#5 vs. set \#7). 
   The gluon from the
  CT18A PDF set is also shown for comparison.}
\label{fig:SM_ttbar_for_PDFs:compCT}
\end{center}
\end{figure}
\noindent
In order to compare with the results obtained by other groups,
it is important to recall that a notable aspect of the NNPDF3.1 methodology is that the charm PDF is
fitted, instead of being obtained from perturbative matching
conditions.
In Ref.~\cite{Ball:2017nwa}
this choice was found to be crucial in
order to achieve a reasonable fit to the high-precision ATLAS 2011 $W/Z$ rapidity
distributions~\cite{Aaboud:2016btc}.
In view of the fact that the
ATLAS study of Refs.~\cite{ATL-PHYS-PUB-2018-017,Giuli:2019wtu} appears to find tension
between their 2011 $W,Z$ data set and their 8 TeV top observables, it is interesting to
investigate whether these conclusions are affected if the NNPDF3.1
methodology is modified by deriving the charm PDF from the corresponding perturbative
matching conditions, rather than being fitted.
This choice corresponds to
fits \#6 and \#7.

It is apparent from the $\chi^2$ values of Table~\ref{tab:SM_ttbar_for_PDFs:chi2} that, if the
charm PDF is no longer fitted, the quality to the fit to the ATLAS 2011 $W/Z$ rapidity
data significantly deteriorates, consistently with the results of
Ref.~\cite{Ball:2017nwa}.
However, the fit quality to the top observables
remains essentially the same as that found in the corresponding
fitted-charm PDF sets.
Interestingly, however, the quality of the
global $\chi^2$ in all these fits somewhat deteriorates, and it is
similar (though somewhat worse) to that found using PDF set \#5, namely when
using top absolute rather than normalized distributions. The origin of
this state of affairs can be understood by comparing the gluon
distribution which is found in each of these cases. The comparison is
displayed in Fig.~\ref{fig:SM_ttbar_for_PDFs:compCT}, where the CT18 gluon PDF is also
shown for reference.
The gluon from the CT18A set is shown, because it is based on a
data set which  also includes the ATLAS 2011 $W/Z$ data, which are excluded
in the baseline CT18 determination.

Recall from Fig.~\ref{fig:SM_ttbar_for_PDFs:comp45} that we found that, when fitting the absolute
top distributions, the gluon PDF was pushed upwards somewhat, and that this led to
some deterioration of the global fit quality, suggesting that this
enhanced gluon is disfavored by the global fit.
It appears from
Fig.~\ref{fig:SM_ttbar_for_PDFs:compCT} that when fitting the normalized distribution,
and
replacing fitted charm with 
perturbative charm, the gluon is similarly pushed upwards. If
the unnormalized distribution is fitted instead,  it makes
essentially no difference whether charm is fitted or not: a similar
fit quality is found with either choice. Also, the gluon determined
from a fit to unnormalized distributions is found to be in very good
agreement with the CT18 gluon.

From this comparison it therefore appears that the good simultaneous fit of the ATLAS
top data set and the global fit achieved in fit \#4 (which is in turn
essentially identical to NNPDF3.1) relies on two ingredients: using
the normalized distributions, and fitting charm. If the unnormalized
distributions are used, an enhanced gluon is found, with a worse global
fit quality, and very little dependence on whether charm is fitted or
not. This gluon is in excellent agreement with the CT18 gluon.
If charm PDF is not fitted, but rather generated from perturbative matching, a poor fit to the ATLAS 2011 $W/Z$
rapidity distribution is obtained. 

In the MMHT-based study of Ref.~\cite{Bailey:2019yze}, where only top-pair absolute distributions were
considered, it was suggested that the poor fit quality to these
distributions could be improved by decorrelating the parton shower (PS)
uncertainties, and it was found that such a decorrelation affects the
gluon. We have therefore checked whether our results would also be
affected by decorrelating uncertainties as suggested in
Ref.~\cite{Bailey:2019yze}, by producing  PDF sets \#8 and
\#9. These PDF sets
differ only in the treatment
of correlated uncertainties
from  PDF sets \#6 and \#7 respectively. Note that this decorrelation
is milder than that used in the computation of the decorrelated rows of
Table~\ref{tab:SM_ttbar_for_PDFs:chi2} (not used for fitting), in which the covariance
matrix for the ATLAS top distributions was taken to be
block-diagonal. The PDF sets   \#8 and
\#9, with perturbative charm and decorrelated uncertainties, are directly
comparable to Ref.~\cite{Bailey:2019yze}.

\begin{figure}[!t]
\begin{center}
\includegraphics[scale=0.75]{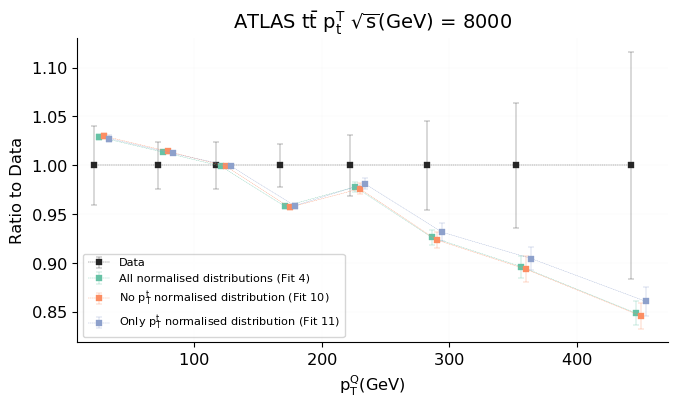}
\caption{\small Comparison between data and theory predictions for the
  ATLAS top transverse momentum distribution. The predictions shown
  correspond to PDF sets \#4, \#10 and \#11 (see text).
 } \label{fig:SM_ttbar_for_PDFs:ptcomp}
\end{center}
\end{figure}

As should be clear from Table~\ref{tab:SM_ttbar_for_PDFs:chi2},
we find that indeed decorrelating uncertainties as suggested in
Ref.~\cite{Bailey:2019yze} does lead to an acceptable fit quality for
the full set of unnormalized top data, both when they are fitted, or
when the absolute data are fitted instead. Interestingly, the value of
the $\chi^2$ found for the full set of  unnormalized top data in
these fits is almost identical to the value found in the corresponding
fits in which the correlations were kept when fitting, but the
$\chi^2$ was fully decorrelated: for the unnormalized
data in fit \#9 we find
$\chi^2=2.16$ , while in fit \#7 the uncorrelated $\chi^2$ value 
is $\chi^2=2.17$. This suggests that  the bulk of the
correlation is indeed coming from the PS uncertainties singled out in
Ref.~\cite{Bailey:2019yze}: removing them leads to same answer as
removing correlations between different observables altogether.
Also, it suggests that whether one fits
the correlated or uncorrelated quantities makes very little difference
at the level of  PDFs, since the uncorrelated $\chi^2$ value remains  the
same. This conclusion is supported by the fact that indeed all the
$\chi^2$ values for other data sets are essentially unchanged by having
performed the 
decorrelation of Ref.~\cite{Bailey:2019yze}:  $\chi^2$ values for the other data sets are the same in
fit \#7 and fit \#9, and also the same in fit  \#6 and fit \#8. We 
have also checked explicitly that this is the case at the PDF
level: when decorrelating uncertainties PDFs change very little.
We conclude  that fit results obtained in our framework are stable upon
 decorrelation.

We finally turn to our preferred PDF set \#4, which, as mentioned, achieves
good fit quality to both the ATLAS top data set and the global fit. We
have seen that this PDF set is extremely stable, in that fitting just
the top rapidity distribution, or the whole data set, leads to essentially
the same
PDFs. One may however note that 
even though the fit quality to the full data set is fair, the fit
quality to the transverse momentum distribution remains poor. One may
then  ask  first, whether this is again due to issues with the
correlation matrix, and furthermore,  if these data might favor a
different PDF shape. In order to answer this question, we have
performed two more PDF fits:
\begin{description}
\item[Fit 10] Same as Fit 4, but excluding the transverse
  momentum distribution from the ATLAS top data set.
\item[Fit 11] Same as Fit 4, but now only including  the transverse
  momentum distribution in the ATLAS top data set.
\end{description}

The ATLAS data for the transverse momentum distribution are compared
to predictions obtained using our preferred Fit \#4 as well as these
two PDF sets in Fig.~\ref{fig:SM_ttbar_for_PDFs:ptcomp}. It is clear that the poor fit
quality to these data is due to the fact that they have a genuinely
different shape in comparison to the theory prediction, and thus
it cannot be due to a treatment of correlations. However, when
excluding these data, or only including them, nothing changes: the
global fit remains perfectly stable upon their inclusion or exclusion,
as we have also verified at the PDF level. Indeed, interestingly, even
if only the top transverse momentum distribution is fitted, the
best-fit PDFs are indistinguishable from those obtained fitting all of
the (normalized) distributions.
In sum, while the reason of this
data-theory discrepancy is unclear, it seems to be immaterial for the purposes of
PDF determinations.

\subsection{Conclusions}
\label{sec:SM_ttbar_for_PDFs:conc}

We have studied the effect of including the full set of
differential top-quark pair distributions from the ATLAS 8~TeV lepton+jets data set
 in the NNPDF3.1 global PDF
 determination.
 Our main conclusions are the following:
\begin{itemize}
  \item Inclusion of the normalized observables yields results which
    are essentially identical to those of the NNPDF3.1 PDF
    determination,
    in which only the top rapidity
    distribution was included.
    \item Good fit quality to all top observables except the
      transverse momentum distribution is found.
      \item The top transverse momentum distribution appears to have
        a somewhat different shape in comparison to the theory
        prediction; however, fit results are stable upon its inclusion
        or exclusion, and in fact even a fit in which only the
        transverse momentum distribution is included leads to PDFs
        which are the same as those when all distributions are
        fitted.
        \item Fitting the charm PDF, rather than obtaining it from
          perturbative matching, is crucial in order to achieve
          compatibility of the top production data and the ATLAS 2011 $W/Z$
rapidity distribution data; if charm is not fitted the gluon PDF is
affected and the global fit quality deteriorates.
\item If unnormalized observables are used instead, the gluon PDF is
  somewhat enhanced in the large $x$ region; this leads to a
  deterioration of the global fit quality. The fit quality to the full
  set of top observables is extremely poor, but it can be brought to
  be similar to what is found when fitting normalized observables by
  decorrelating different distributions; PDFs are stable upon this decorrelation.
\item If unnormalized observables are fitted, it makes
  little difference to the gluon PDF if charm is fitted or not, though
  if charm is not fitted,
  all the light quark PDFs change by an amount which is small but
  sufficient to lead to considerable deterioration of the fit to the ATLAS 2011 $W/Z$
  rapidity distribution data.
 \end{itemize} 

We conclude that normalized top observables, together with fitted
charm, are necessary ingredients in order to achieve good fit quality to
the ATLAS 8 TeV lepton+jets
top production data within the framework of the NNPDF3.1 global determination. Best-fit results
obtained with these choices are extremely stable upon variations of the
data set and treatment of uncertainties.
A detailed benchmarking against results found in a CT,
MMHT and ATLAS-{\sc xFitter} framework would be extremely beneficial
for a complete understanding and validation of our findings.
Also, it will be interesting to see to what extent these conclusions remain
true when
additional top-quark pair data sets are included in the fit, in particular the
ATLAS and CMS $\sqrt{s}=13$ TeV measurements, as well as other
gluon-sensitive observables such as the jet and dijet
cross-sections. 

\subsection*{Acknowledgements}

We are grateful to Shaun Bailey and Lucian Harland-Lang  for
correspondence and detailed clarifications
concerning Refs.~\cite{Bailey:2019yze,ATL-PHYS-PUB-2018-017}
and for comments. We thank Joey Huston, Robert Thorne and Mandy
Cooper-Sarkar
for discussions and comments on the manuscript.\\
Stefano Forte is supported by the
European Research Council under the European Union's 
Horizon 2020 research and innovation Programme (grant agreement ERC-AdG-740006).
Emanuele R. Nocera is supported by the European Commission through the Marie 
Sk\l odowska-Curie Action ParDHonS FFs.TMDs (grant number 752748).
Juan Rojo is partially supported by the Dutch National Science
Foundation (NWO).\\



\newcommand{\CTHERAII}{CT14$_{\textrm{HERAII}}$}

\section{Assessing the compatibility of experimental pulls on LHC parton luminosities with the $L_2$ sensitivity~\protect\footnote{
  T.~J.~Hobbs,
  J.~Huston,
  P.~Nadolsky
  }{}}

\label{sec:SM_PDFs_L2}


\subsection{Introduction}
\label{sec:SM_PDFs_L2:intro}
The present lack of complete knowledge of the proton's parton distribution functions (PDFs) forms one of the most significant uncertainties for crucial physics processes at the LHC, such as the $gg$ fusion cross section for Higgs boson production. Information on the PDFs comes from global fits to a wide variety of high-energy data, including those taken by various LHC experiments. Global PDF analyses involve a subtle interplay among all of the fitted data sets in order to determine an optimal set of central PDFs and their associated uncertainties. Complicating the realization of these optimal PDFs are systematic tensions, which can exist among data sets, and which tend to resist the reduction in PDF uncertainties suggested by the precision of the fitted data. 

Any program to comprehend and resolve these tensions necessarily requires a set of tools to determine the PDF sensitivities and pulls of the data in a given
global analysis. The CTEQ-TEA (CT) group has pioneered a number of techniques to establish the sensitivity of a particular data set for constraining a particular PDF or observable, such
as the Lagrange Multiplier (LM) scans \cite{Stump:2001gu}, which, though robustly informative, are computationally costly \cite{Hou:2019efy}, and evaluated for specific, fixed values of
the parton momentum fraction, $x$ and factorization scale, $Q$. On the other hand, the outcomes of popular fast techniques based on Monte-Carlo PDF reweighting \cite{Giele:1998gw,Ball:2010gb,Ball:2011gg,Sato:2013ika}, or Hessian profiling \cite{Paukkunen:2014zia} and updating \cite{Schmidt:2018hvu}, sensitively depend
on the choice of either statistical weights or tolerance. 

A technique that does not involve the computational overhead of the LM method or the ambiguities of the reweighting approach is the $L_2$ sensitivity technique, as defined in Ref.~\cite{Hobbs:2019gob} and deployed in the recently-released CT18 global fit~\cite{Hou:2019efy}. 
%
%
%
The $L_2$ sensitivity is inexpensive to compute
and provides an informative approximation to the $\Delta \chi^2$
trends in a given global analysis. Moreover, the $L_2$ sensitivity can also be
readily calculated across a wide range of $x$, allowing the $\Delta \chi^2$
variations shown in the LM scans to be visualized and interpreted
for multiple, simultaneous $x$ values. We stress that the qualitative conclusions
revealed by consideration of the $L_2$ sensitivities, discussed and
presented below, are consistent with the picture based on the LM
scans themselves.
Although the $L_2$ sensitivities do not always provide the same
numerical ordering as the LM scans for the subdominant experiments,
they offer complementary information over broader reaches of $x$
that are not completely captured by the LM scans.

While the $L_2$ sensitivity was used to analyze the pulls of data on the PDFs themselves
in recent CT fits, the method is sufficiently flexible that
it may be applied to other phenomenologically relevant quantities, including
the parton-parton luminosities used in predictions for processes
at hadron colliders.  In this note, we demonstrate this application, highlighting
a number of phenomenological consequences.

\subsection{Definition of the $L_2$ sensitivity}
\label{sec:SM_PDFs_L2:definition}
We work in the Hessian formalism \cite{Pumplin:2002vw,Nadolsky:2008zw,Pumplin:2001ct} and compute the $L_2$ sensitivity, $S_{f, L2}(E)$, for each experiment, $E$, as
\begin{equation}
S_{f, L2}(E) = \vec{\nabla} \chi^2_E \cdot \frac{ \vec{\nabla} f } { |\vec{\nabla} f| }
             = \Delta \chi^2_E\, \cos \varphi (f, \chi^2_E)\ ,
\label{eq:SM_PDFs_L2:L2}
\end{equation}
which yields the variation of the log-likelihood function $\chi^2_E$ due to a unit-length
displacement of the fitted PDF parameters away from the global minimum $\vec{a}_0$ of
$\chi^2(\vec{a})$  in the direction of $\vec{\nabla}f$. 
The PDF parameters $\vec a$ are normalized so that a unit displacement
from the best fit in any direction corresponds to the default
confidence level of the Hessian error set (90\% for CT18,
on average corresponding to slightly less than
$\Delta\chi^2_{\mathit{tot}}=100$ in a given direction.)

This displacement increases the
PDF $f(x,Q)$ by its Hessian PDF error, $\Delta f$, and, to the extent its PDF variation is
correlated with that of $f(x,Q)$ through the correlation angle
\begin{equation}
	\varphi(f, \chi^2_E) = \cos^{-1} \left( \frac{\vec{\nabla} f}{|\vec{\nabla} f |} \cdot \frac{\vec{\nabla} \chi^2_E}{|\vec{\nabla} \chi^2_E |} \right)\ ,
\end{equation}
it changes $\chi^2_E$ by $\Delta \chi^2_E (\hat{a}_f) = \Delta \chi^2_E\, \cos \varphi (f, \chi^2_E) = S_{f, L2}(E)$.
The $L_2$ sensitivity, $S_{f, L2}(E)$, therefore quantifies the impact that uncertainty-driven
variations of PDFs at fixed $x$ and $Q$ have upon the description of fitted data sets.  Plotting $S_{f, L2}(E)$
against $x$ furnishes useful information regarding the pulls of the CT18(Z) data sets 
upon the PDFs fitted in the global analysis, as well as various PDF combinations of interest.
This also permits the rapid visualization of possible tensions within the global fit, since
the PDF variations of some parton densities of given flavor are correlated with the
variation of $\chi^2_E$ ({\it i.e.}, $S_{f, L2}(E) > 0$), while others are
anti-correlated ($S_{f, L2}(E) < 0$), at similar values of $(x, Q)$.

\begin{figure}[t]
\centering
\includegraphics[width=1\textwidth]{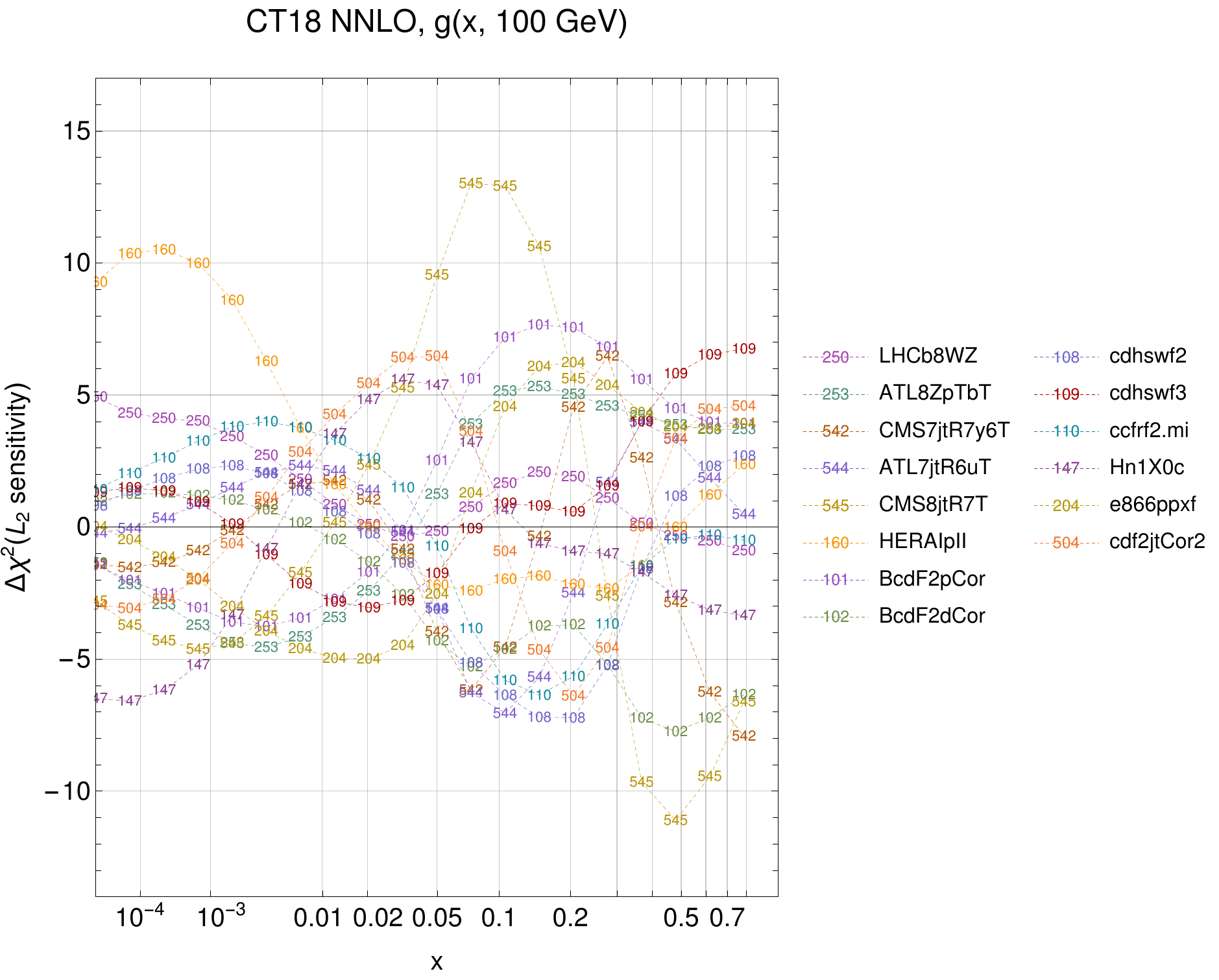}
\caption{The $L_2$ sensitivity of the most important experiments in the CT18 global PDF fit for the gluon distribution, $g(x,Q\!=\!100\, \mathrm{GeV})$,
as a function of parton momentum fraction, $x$.}
\label{fig:SM_PDFs_L2:CT18_gluon_L2_sensitivity}
\end{figure}

The terms on the right-hand side of Eq.~(\ref{eq:SM_PDFs_L2:L2}) for $S_{f,L2}$
are computed as
\begin{equation}
\Delta X=\left\vert \vec{\nabla}X\right\vert
=\frac{1}{2}\sqrt{\sum^{N_\mathit{eig}}_{i=1}\left(X_{i}^{(+)}-X_{i}^{(-)}\right)^{2}},\label{eq:SM_PDFs_L2:masterDX}
\end{equation}
and
\begin{equation}
\cos\varphi=\frac{\vec{\nabla}X\cdot\vec{\nabla}Y}{\Delta X\Delta
  Y}=\frac{1}{4\Delta X\,\Delta
  Y}\sum^{N_\mathit{eig}}_{i=1}\left(X_{i}^{(+)}-X_{i}^{(-)}\right)\left(Y_{i}^{(+)}-Y_{i}^{(-)}\right),\label{eq:SM_PDFs_L2:cosphi}
\end{equation}
from the values $X_{i}^{(+)}$ and $X_{i}^{(-)}$ that a quantity
$X$ takes for the parameter displacements
along the ($\pm$) direction of the $i$-th
eigenvector. With these symmetric master formulas, the sum of
$S_{f,L2}(E)$ over all experiments $E$ should be within a
few tens from zero, since the tolerance boundary for the total $\chi^2$ is close to being spherically
symmetric. The $S_{f,L2}(E)$ variables for individual experiments tend to
cancel among themselves to this accuracy; the order of magnitude 
of $S_{f,L2}(E)$ can be also interpreted as a measure  of tension of
$E$ against the rest of the experiments. 

\subsection{Application to parton luminosities}
\label{sec:SM_PDFs_L2:luminosities}
The $L_2$ sensitivity was explored in the CT18 paper~\cite{Hou:2019efy}, from which the description above is largely borrowed. It was applied to the determination of experimental sensitivities to
specific PDFs, $f_a(x,Q)$, at a chosen factorization scale, $Q$, and as a function of the parton momentum fraction, $x$. For example, the $L_2$ sensitivity for the gluon distribution at a $Q$ value of
100 GeV is shown in Fig.~\ref{fig:SM_PDFs_L2:CT18_gluon_L2_sensitivity}. The pulls of a particular experiment on the gluon distribution can vary as a function of $x$. As stated previously, the larger the absolute
value of $S_{f,L2}(E)$, the greater the sensitivity of that experiment to the determination of the PDF at that $x$ (and $Q$) value. The $L_2$ sensitivity can be positive or negative. If positive,
the upward variation of $f_a(x,Q)$ leads to an increase in the $\chi^2$ for the specified experiment. A negative $L_2$ sensitivity indicates that the variation will lead to a decrease in $\chi^2$ for
this experiment. 
A large collection of figures illustrating $L_2$ sensitivities for various flavors of PDFs and parton luminosities in the CT18 and CT18Z NNLO analyses can be viewed online at \cite{CT18L2Sensitivity}.

So, for example, in Fig.~\ref{fig:SM_PDFs_L2:CT18_gluon_L2_sensitivity}, at an $x$ value near 0.01, a region sensitive to Higgs boson production through gluon-gluon fusion at 14 TeV, the strongest preference for
a smaller gluon density [signaled by $\Delta \chi^2\! >\! 0$ when $g(x,Q)$ is increased] comes from CDF jet data, $F_2$ measurements from CCFR, H1 heavy-flavor production, and the combined HERA1+II 
inclusive DIS data, followed by the ATLAS 7 TeV jet data. Descriptions of these and other quoted experimental data sets can be found in Ref.~\cite{Hou:2019efy}. The two most
important experiments that pull in the opposite direction are E866/NuSea, a fixed-target Drell-Yan experiment from Fermilab, and the ATLAS 8 TeV Z-boson $p_T$ measurement. Note that only the most sensitive of the experiments for the
determination of the gluon distribution have been plotted. There are 39 experimental data sets in the CT18 fit. 

\begin{figure}[t]
  \begin{center}
    \begin{tabular}{ccc}
      \includegraphics[width=0.45\textwidth]{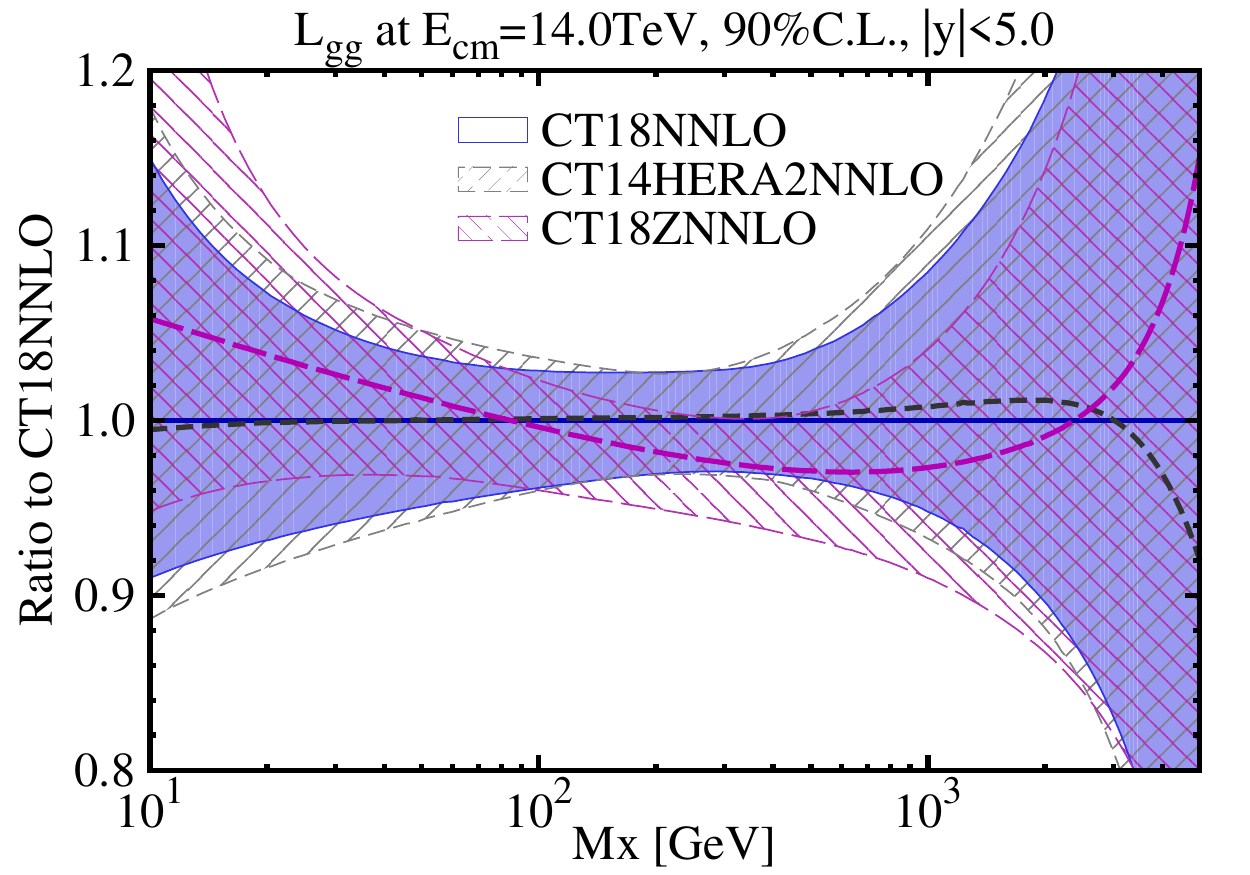} &&
      \includegraphics[width=0.45\textwidth]{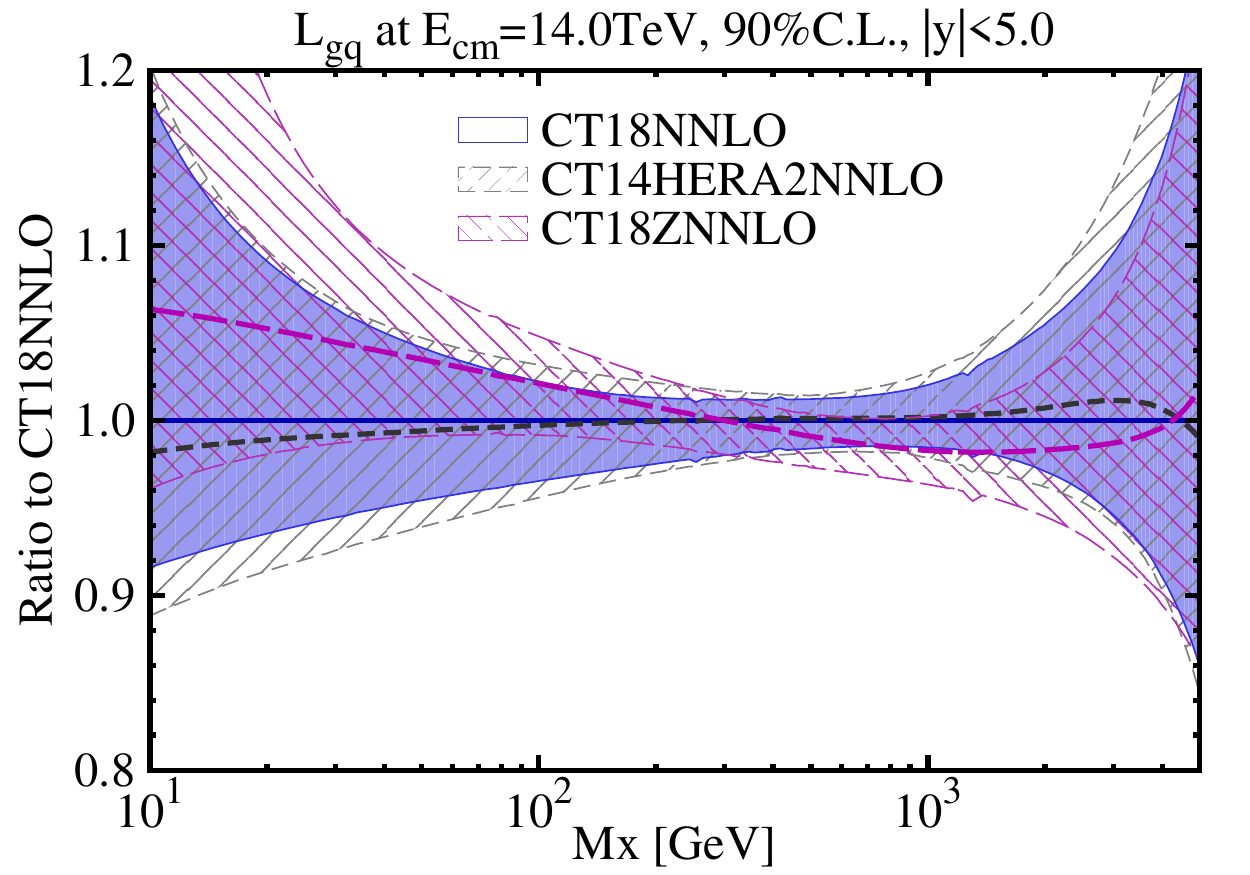}\\
      \includegraphics[width=0.45\textwidth]{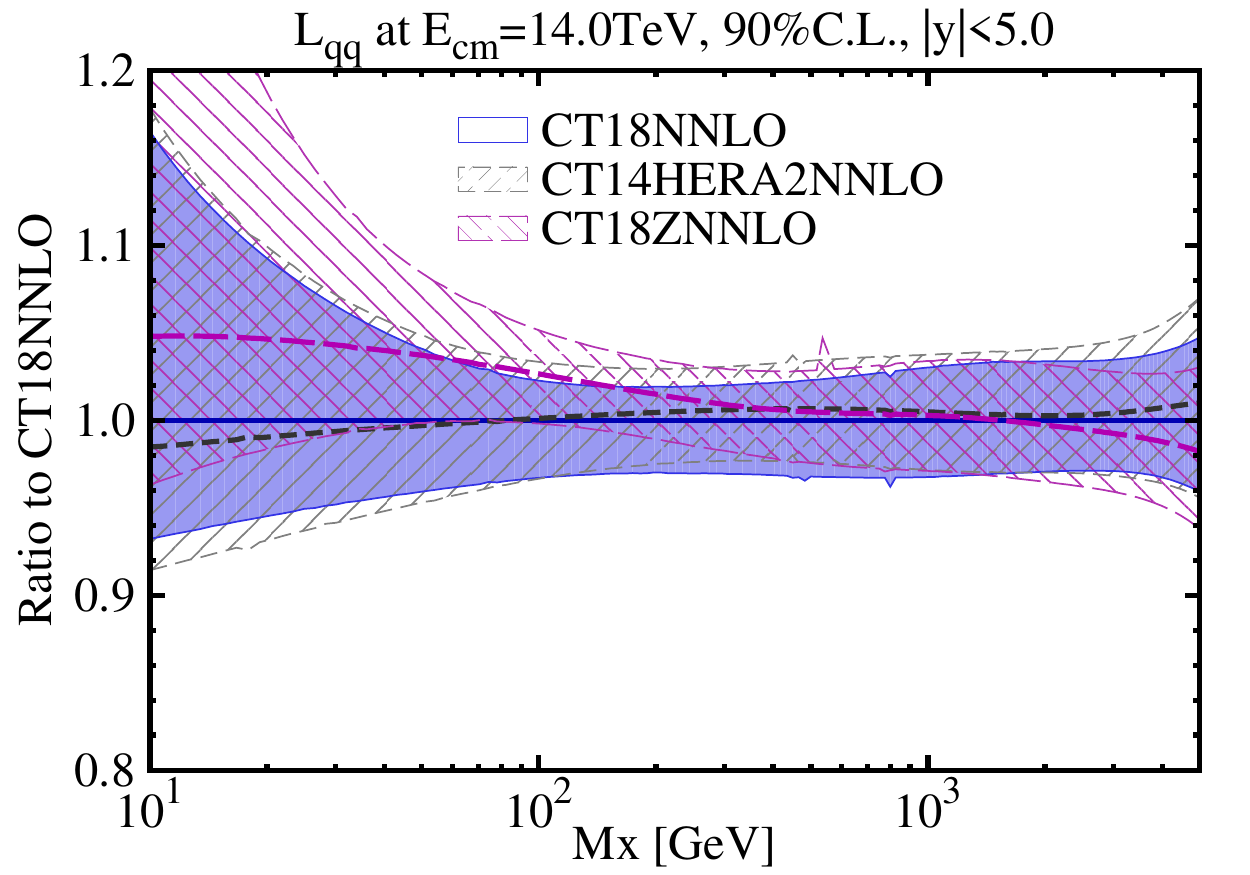} &&
      \includegraphics[width=0.45\textwidth]{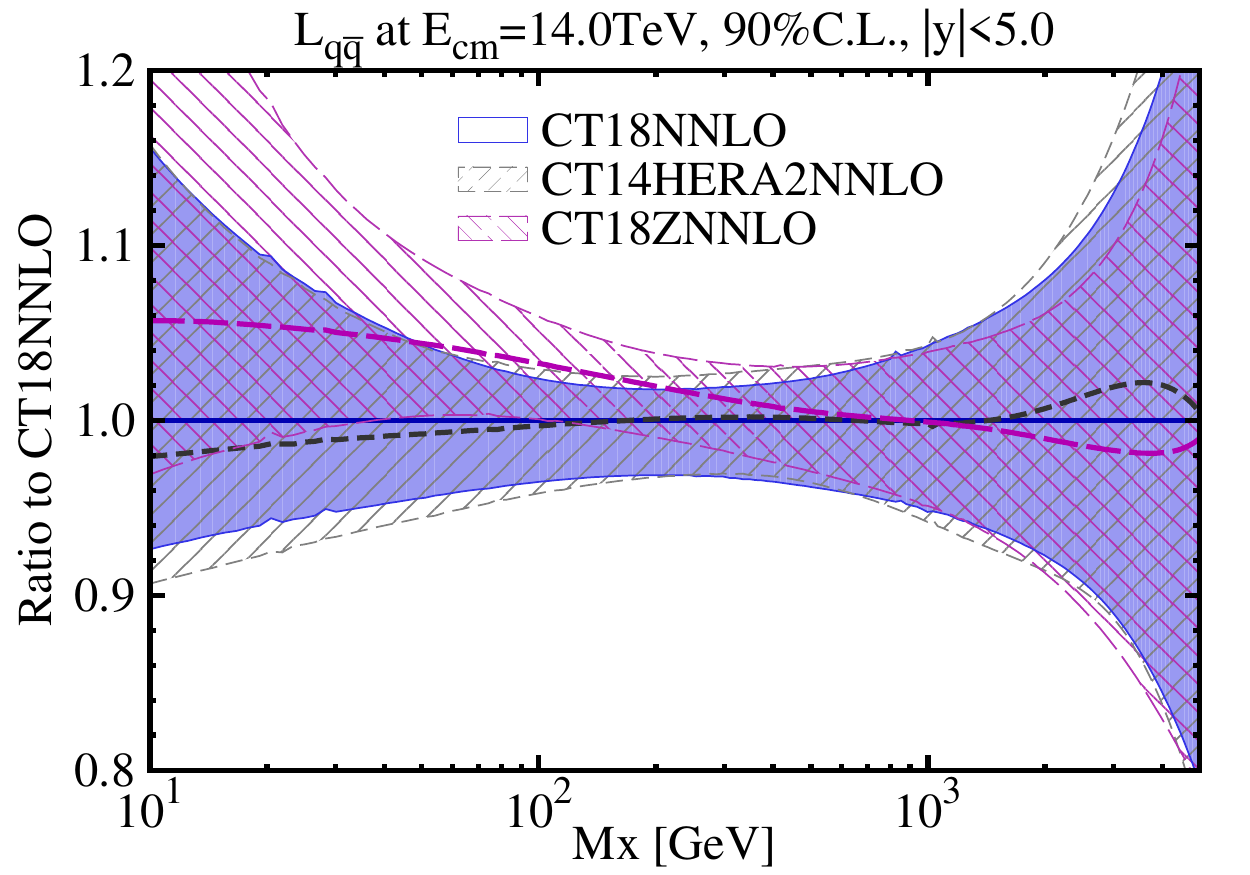}
    \end{tabular}
  \end{center}
	\vspace{-2ex}
	\caption{
		Parton luminosities for processes
		at the LHC at $\sqrt{s} = 14$ TeV, in the central
                rapidity region $|y|<5$: 
                $L_{gg}$ (upper-left), $L_{gq}$ (upper-right),
                $L_{qq}$ (lower-left), and $L_{q\bar{q}}$ (lower-right);
		evaluated using the CT18 (solid violet), CT18Z (short-dashed
		gray), and \CTHERAII~(long-dashed magenta) NNLO PDFs. In each instance, we
		display the luminosity ratios normalized to CT18.
	}
\label{fig:SM_PDFs_L2:lumia}
\end{figure}

The CT18 analysis demonstrated using the $L_2$ sensitivity and other methods that the combination of the most extensive DIS experimental data sets -- HERA, BCDMS, NMC, CCFR,... -- at the moment imposes the dominant constraints on the CT18 gluon PDF $g(x,Q)$ through $Q$ dependence of DIS cross sections over a wide region of $x$ and $Q$. At hadron-hadron colliders, the most sensitive measurements of $g(x,Q)$ are provided by inclusive jet production, especially by CMS and ATLAS. 
There can be no further improvement of the HERA data, or of E866/NuSea, but the importance shown by the LHC measurements provides an indication of where future, more precise, measurements at the
LHC may improve the PDF uncertainties for the Higgs boson cross section, or for any other LHC measurement. 
Fig.~\ref{fig:SM_PDFs_L2:CT18_gluon_L2_sensitivity} indicates the $L_2$ sensitivities only at particular $x$ values. This would correspond to one particular rapidity value for the Higgs boson, near zero. As Higgs bosons are produced over a reasonably wide rapidity range, production will be sensitive to a wide partonic $x$ range, approximately, over $0.001\! \lesssim\! x\! \lesssim\! 0.1$. 

\begin{figure}[t]
\centering
\includegraphics[height=0.42\textheight]{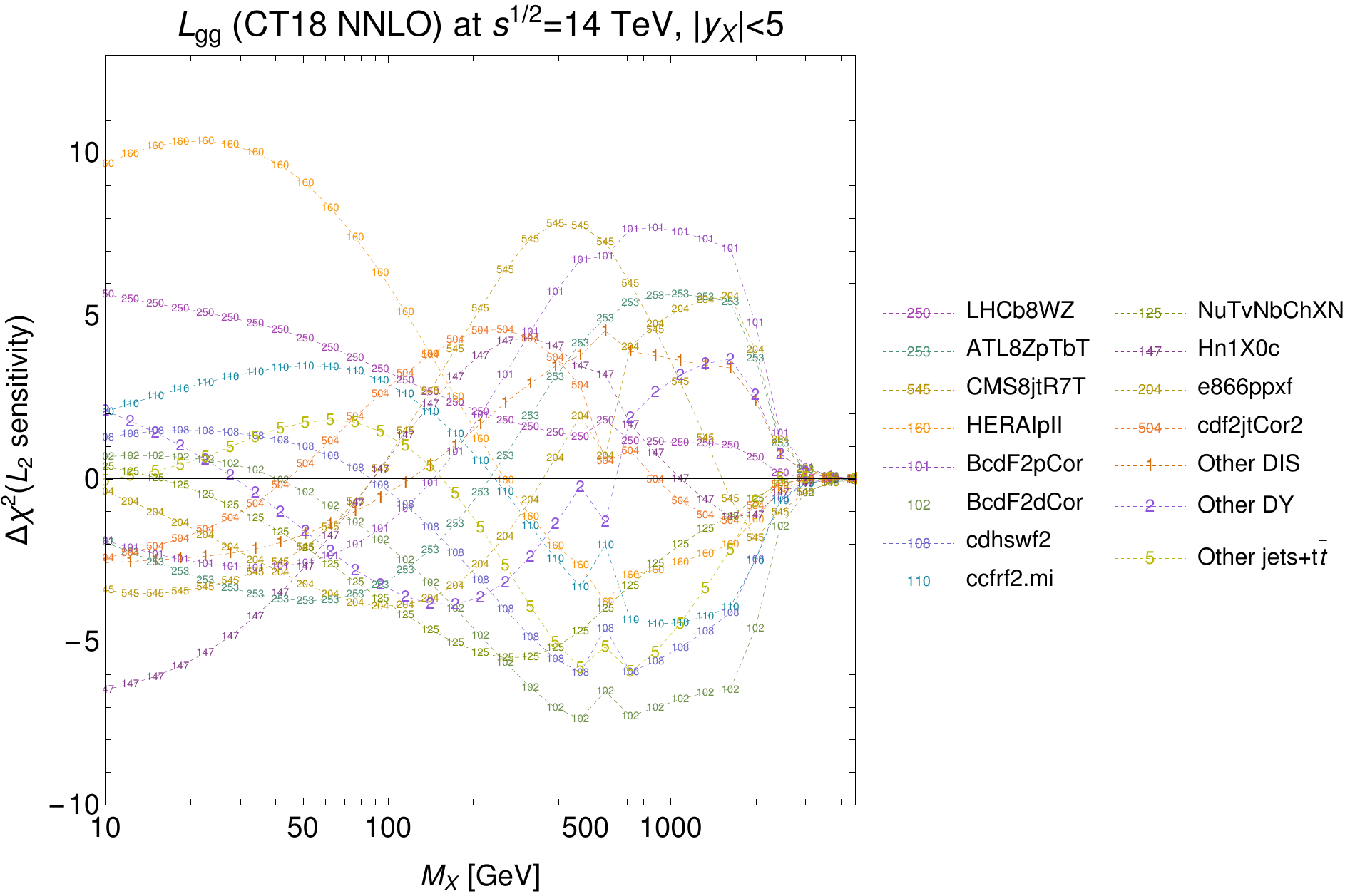}
\caption{The $L_2$ sensitivity of the most important experiments in the CT18 global PDF fit for the $gg$ parton luminosity as a function of the mass of the final state.
Here, we show the experimental pulls on the parton luminosity computed with a less restrictive rapidity cut, $|y_\mathit{cut}|\! <\! 5$, as compared with the
$|y_\mathit{cut}|\! <\! 2.5$ selection more appropriate for LHC measurements shown in subsequent plots.
}
\label{fig:SM_PDFs_L2:CT18_gg_y5}
\end{figure}

A more succinct understanding of the importance of each experiment to the production of a particle of a particular mass can be gained by showing the $L_2$ sensitivity to the parton-parton luminosity for a pair of initial partons $a$ and $b$,
defined as in \cite{Campbell:2006wx} for production of a final state with invariant mass $M_X$ at collider energy $\sqrt{s}$;  we apply an additional constraint that the rapidity of the final state, $y = \frac{1}{2} \ln (x_{2}/x_{1})$,
does not exceed $y_{cut}$ in its absolute value, resulting in the parton luminosity definition:
\begin{equation}
L_{ab}(s,M^{2}_X,y_{cut})=\frac{1}{1+\delta_{ab}}\left[\int_{\frac{M_X}{\sqrt{s}}e^{-y_{cut}}}^{\frac{M_X}{\sqrt{s}}e^{y_{cut}}}\frac{d\xi}{\xi}f_{a}(\xi,M_X)f_{b}\left(\frac{M_X}{\xi\sqrt{s}},M_X\right)+\left(a\leftrightarrow b\right)\right]\ .
\end{equation}
The uncertainty bands for the gluon-gluon, gluon-quark, quark-quark, and quark-antiquark luminosities at 14 TeV, as relevant for the LHC, are shown in Fig.~\ref{fig:SM_PDFs_L2:lumia} based upon the CT18, CT18Z, and \CTHERAII~NNLO PDFs.
The respective $L_2$ sensitivity for the $gg$ parton luminosity can be viewed in Fig.~\ref{fig:SM_PDFs_L2:CT18_gg_y5}. Again, a more complete collection of $L_2$ sensitivities for parton luminosities can be viewed at \cite{CT18L2Sensitivity}.

Integrating over a larger range of parton $x$ values, for the Higgs boson mass of 125 GeV, increases the importance for the HERAI+II data set (in the positive direction), with the $L_2$ sensitivity approaching a value of 6, and $\bar{\nu}$DIS dimuon production [NuTvNbChXN] and the E866pp data (in the negative direction), with a value on the order of -5. BCDMS data on $F^d_2$ and the ATLAS 8 TeV Z $p_T$ distribution are also important on the negative side in this case. 
An astute reader will notice that the plot above was made by applying a rapidity cut of $\pm5$ on the produced Higgs boson. However, the precision coverage for ATLAS and CMS does not run past a
rapidity of $|y|\! <\! 2.5$. A similar plot, but now imposing a rapidity cut of 2.5 is shown in Fig.~\ref{fig:SM_PDFs_L2:CT18z_gg_y2.5}. A comparison between the two plots shows little difference, because most
of the Higgs boson production in the $gg$ fusion channel occurs within a rapidity of 2.5 in any case. From now on, a maximal rapidity cut of 2.5 will be applied.

\begin{figure}[t]
\centering
\includegraphics[height=0.42\textheight]{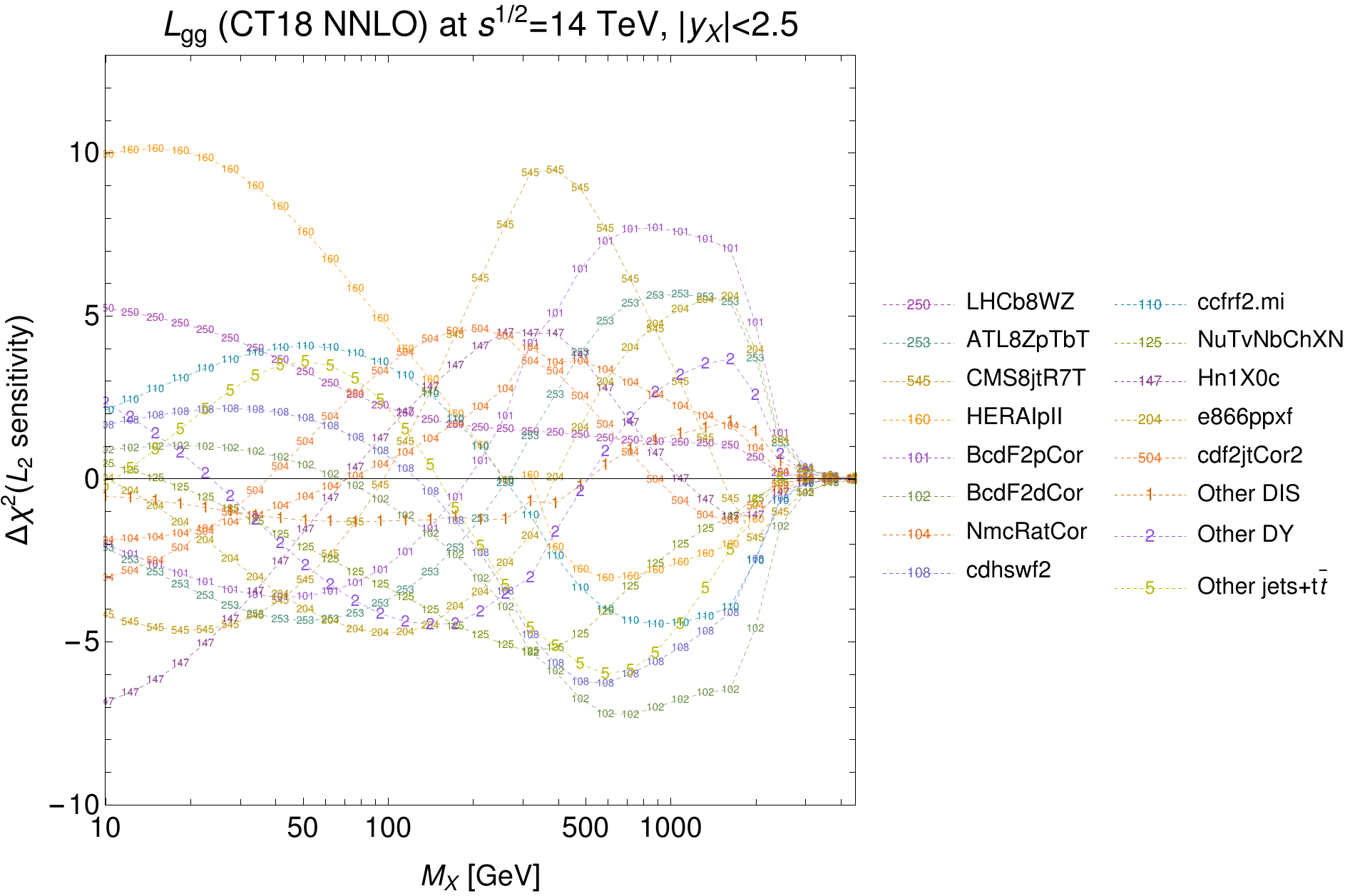}
\caption{The analog of the $L_2$ sensitivity plot for the $gg$ parton luminosity shown in Fig.~\ref{fig:SM_PDFs_L2:CT18_gg_y5}, but in this case,
calculated using a more restrictive rapidity cut of $|y_\mathit{cut}|\! <\! 2.5$.
}
\label{fig:SM_PDFs_L2:CT18_gg_y2.5}
\end{figure}

All of the above $L_2$ sensitivity plots have been computed using the CT18 PDFs. It is instructive to also examine similar plots with CT18Z, which adds the precision ATLAS 7 TeV $W/Z$ boson data to the fit, and, most importantly for the purposes of the $gg$ parton luminosity, changes the scale used for low-$x$ DIS production \cite{Hou:2019efy}. This has the impact of significantly increasing the low-$x$ gluon distribution. The $gg$ parton luminosity for CT18Z is shown in Figure~\ref{fig:SM_PDFs_L2:CT18z_gg_y2.5}, and the changes leading to CT18Z have a marked effect on the pulls of the CT18Z experiments upon $L_{gg}$.
Most notably, this is true of the HERAI+II data, which under CT18Z exhibit pulls on the glue-glue luminosity with significantly different dependence on $M_X$ as compared to CT18.
\begin{figure}[t]
\centering
\includegraphics[height=0.42\textheight]{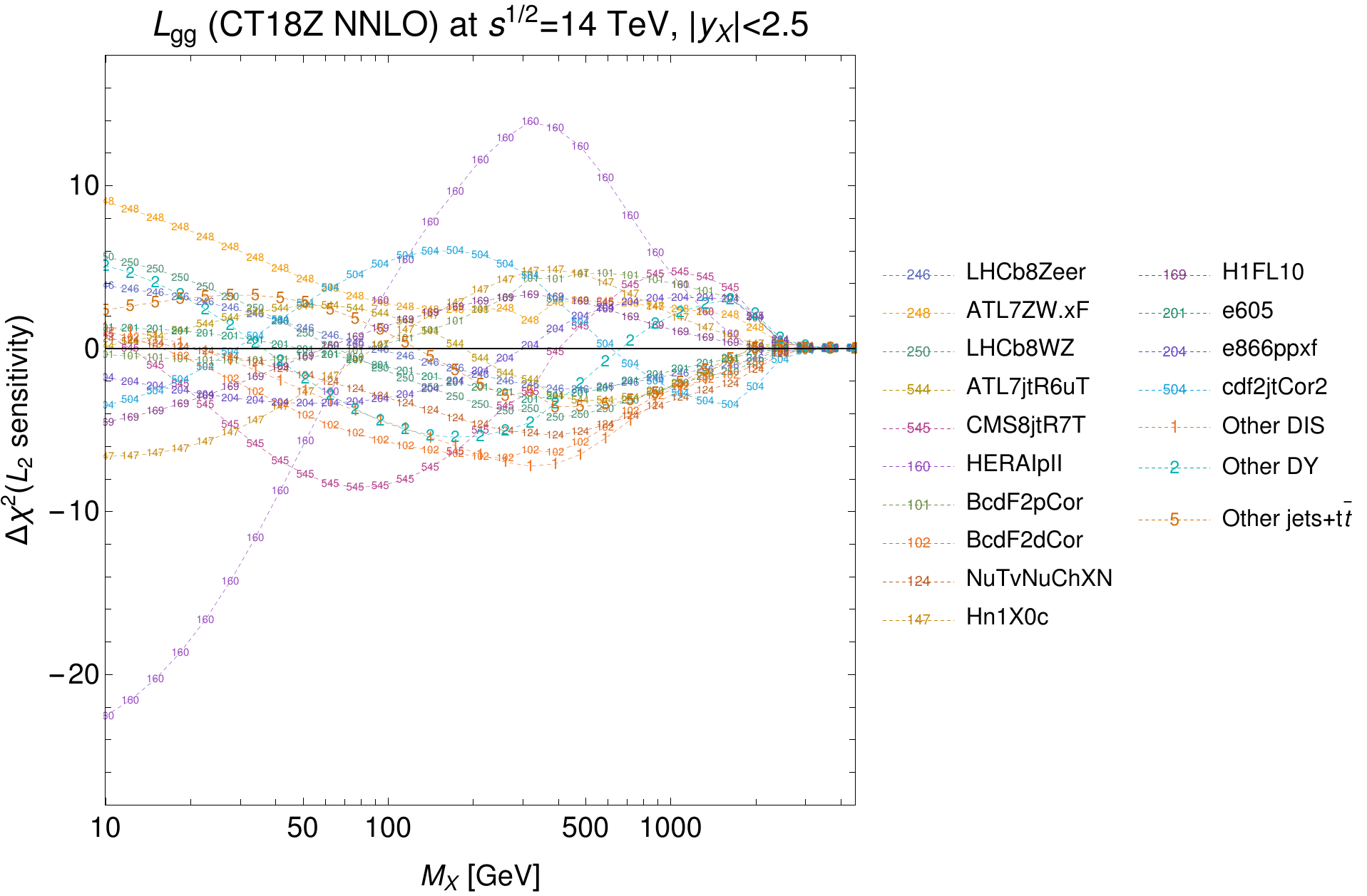}
\caption{The $L_2$ sensitivity of the most important experiments in the CT18Z global PDF fit for the $gg$ parton luminosity.
}
\label{fig:SM_PDFs_L2:CT18z_gg_y2.5}
\end{figure}
For instance, whereas the HERAI+II data resisted increases to the gluon distribution relevant for the lighter-mass, $M_X \lesssim 100$ GeV, region under CT18, for CT18Z, 
these pulls are essentially reversed, with the HERA data preferring the larger gluon at low $x$, leading to reductions in $\chi^2_E$ in this light mass region.
This feature is consistent with the large rise observed for the gluon PDF at low $x$ with CT18Z relative to CT18 shown in Ref.~\cite{Hou:2019efy}. 
In the immediate, $M_X\! \sim\! 125$ GeV, neighborhood of the Higgs production region, the HERA information has an $L_2$ sensitivity of approximately $+5$ under both fits.

\begin{figure}[t]
\centering
\includegraphics[height=0.42\textheight]{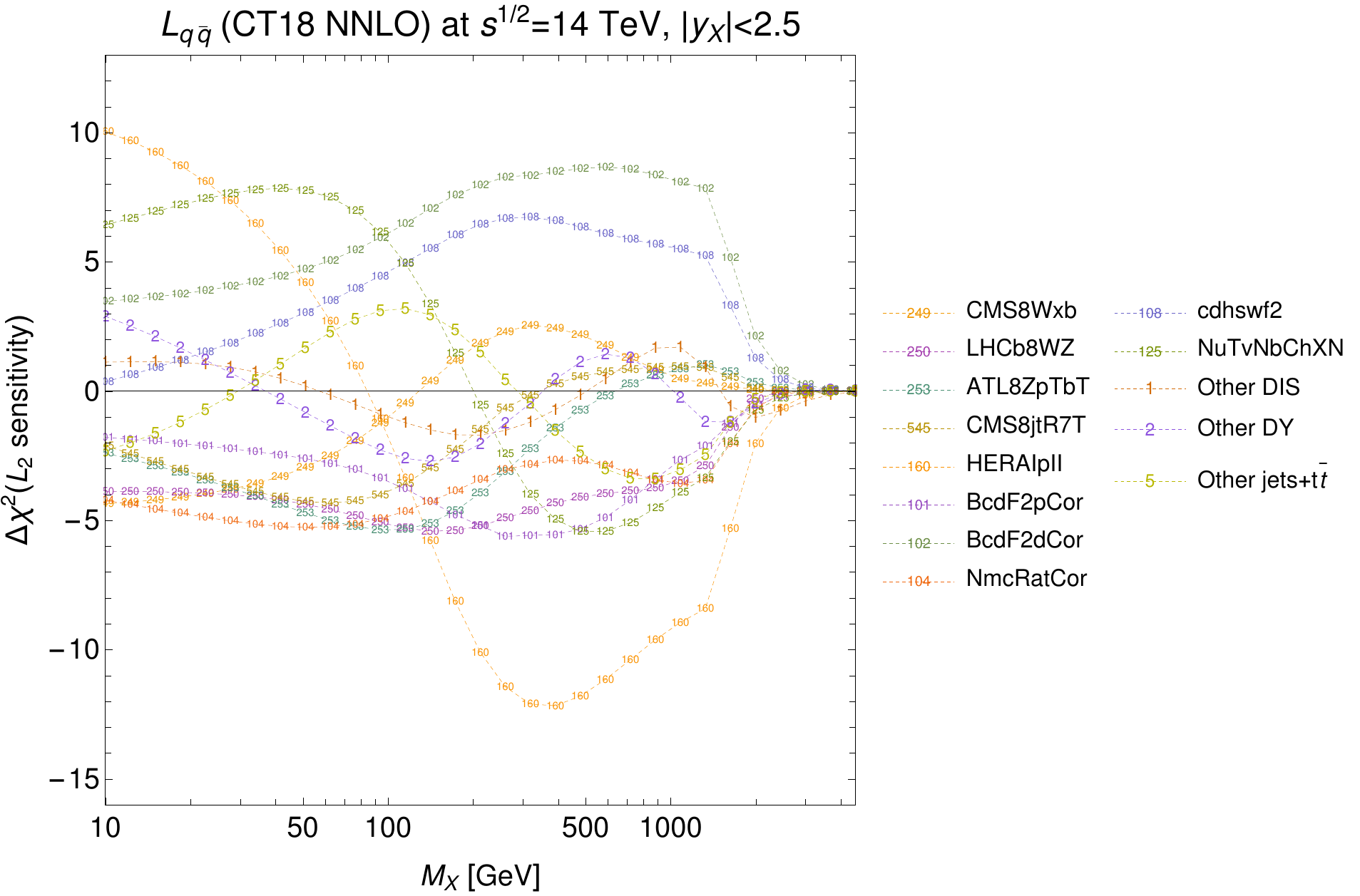}
\caption{The $L_2$ sensitivity of the most important experiments in the CT18 global PDF fit, as in Fig.~\ref{fig:SM_PDFs_L2:CT18_gg_y2.5}, but for the $q\bar{q}$ parton luminosity.
}
\label{fig:SM_PDFs_L2:CT18_qqb_y2.5}
\end{figure}

Of course, we have mainly concentrated on the $gg$ parton luminosity and its impact on the Higgs boson production. The technique can provide useful information for other mass values for the $gg$ parton luminosity, and for other processes using other PDF luminosities. For example, the $q\bar{q}$ parton luminosity is plotted in Fig.~\ref{fig:SM_PDFs_L2:CT18_qqb_y2.5}. At the mass of the $W/Z$ boson, the primary influences in the positive direction are the NuTeV $\bar{\nu}$ data (NuTvNbChXN), BCDMSF2d, and CDHSWF2, and in the negative direction, the ATLAS 8 TeV $Z$ $p_T$ distribution, LHCb 8 TeV $W/Z$ data, NMC structure function ratios, and CMS 8 TeV
jet data. For higher masses, on the order of 1 TeV, BCDMSF2d and CDHSWF2 are again most important on the positive side, while the HERA1+II experiment dominates in the negative $L_2$ direction. 

\begin{figure}[t]
\centering
\includegraphics[height=0.42\textheight]{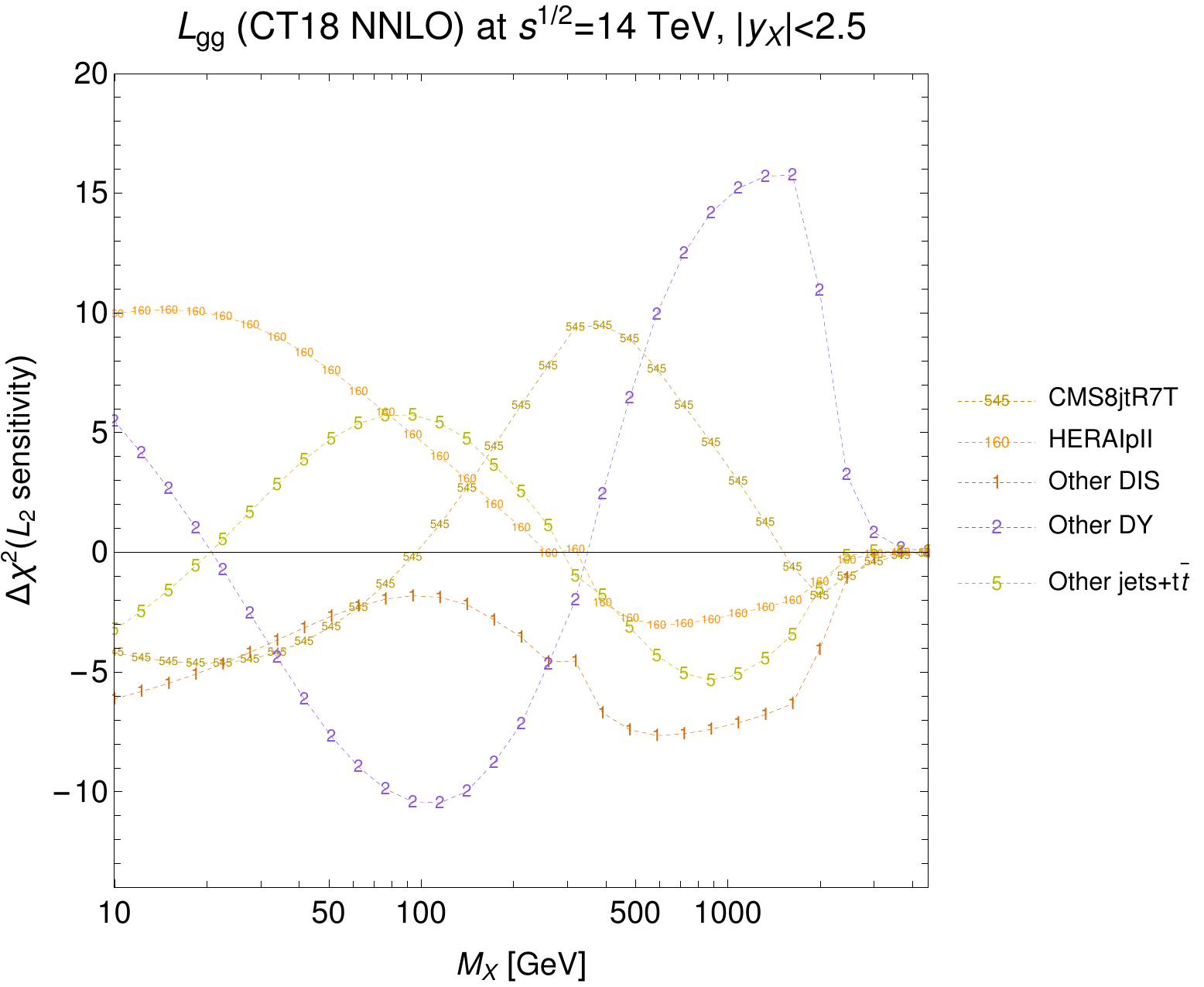}
\caption{The $L_2$ sensitivity to the $gg$ parton luminosity of all experiments with an $L_2$ sensitivity greater than 8, plus the combination of all other DIS data, all Drell-Yan data, and all
$t\bar{t}$ + jets data.
}
\label{fig:SM_PDFs_L2:CT18_gg_y2.5_Sfcut8}
\end{figure}
\begin{figure}[t]
\centering
\includegraphics[height=0.42\textheight]{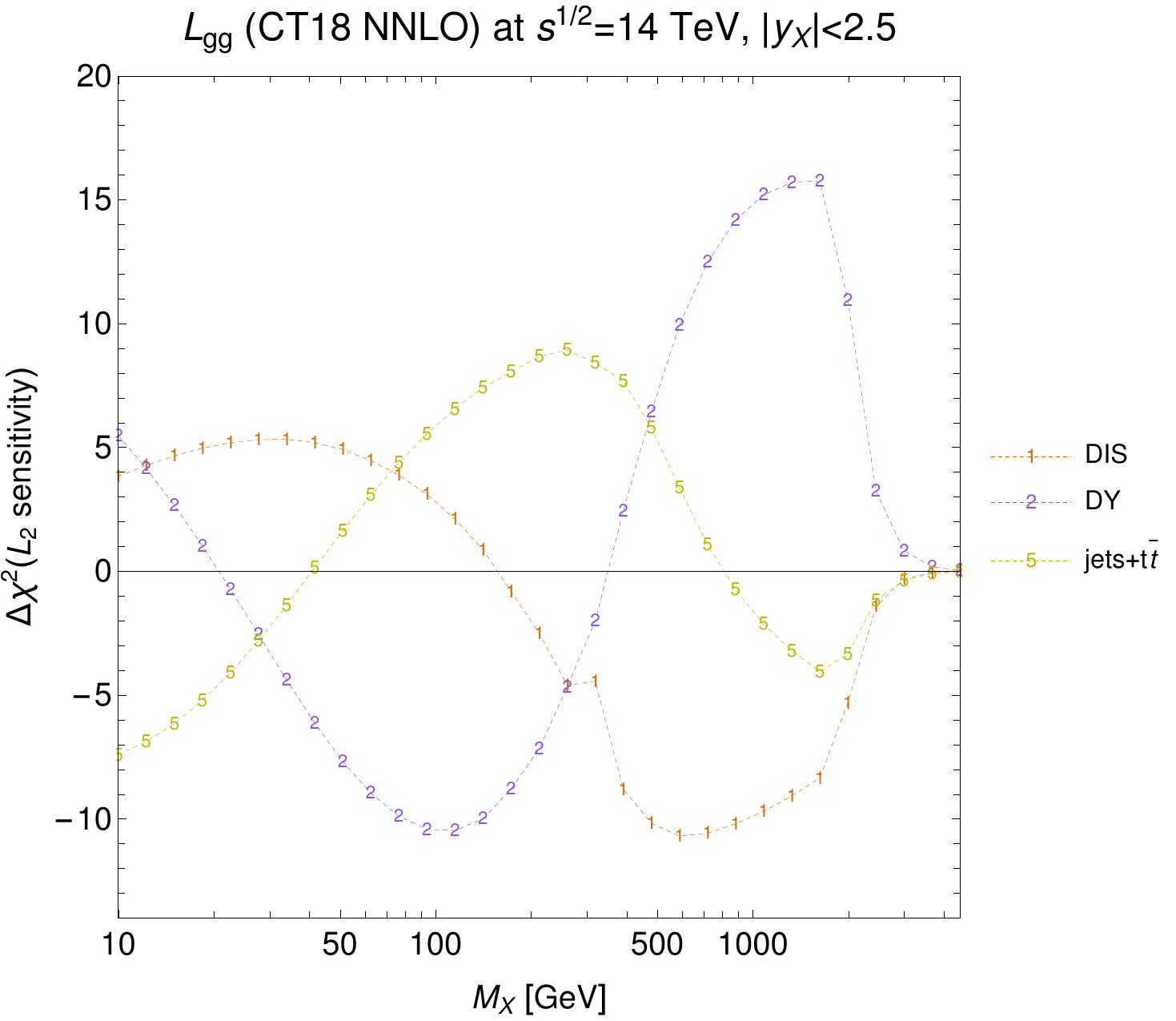}
\caption{The $L_2$ sensitivity to the $gg$ parton luminosity of all the data fitted in CT18, now collected into categories for the DIS, Drell-Yan, and jets+$t\bar{t}$ data.
}
\label{fig:SM_PDFs_L2:CT18_gg_y2.5_SfCut20}
\end{figure}

It is also useful to examine the $L_2$ sensitivities for different categories of data. For example, in Fig.~\ref{fig:SM_PDFs_L2:CT18_gg_y2.5_Sfcut8}, the $L_2$ sensitivities are shown for all experiments with an $L_2$ sensitivity exceeding 8 in some interval of $M_X$. It turns out that there are only two such experiments, the HERAI+II data and the CMS 8 TeV inclusive jet data, CMS8jtR7T. The other DIS data are added together (labeled as 1), as are all of the Drell-Yan data (2) and all of the $t\bar{t}$ and jets data (5). At the Higgs boson mass, the sum of all Drell-Yan data (2) has a pronounced pull in the negative direction, in contrast to the sum of all
$t\bar{t}$ + jets data (5) and HERAI+II, which pull more moderately in the opposing, positive direction. The other combined DIS data (excluding the inclusive HERA data) and CMS jet information have more
modest pulls in this region. If we add all DIS data together, all Drell-Yan data together and all $t\bar{t}$+jets data together, we get the result in Fig.~\ref{fig:SM_PDFs_L2:CT18_gg_y2.5_SfCut20}.

\subsection{Conclusion}

The $L_2$ sensitivity, plotted as a function of the invariant mass of the final state, is a useful indicator to understand the pulls on parton luminosity combinations from different experimental inputs, and the size of any tensions that may exist between experimental data sets, especially those from the LHC. The studies shown here have been created for the CT18 and CT18Z PDF sets. Comparable constructions for the other global PDF sets will help with the combination of such PDFs for the ongoing PDF4LHC20 benchmarking exercise. Ameliorating the tensions examined in this discussion will be critical to achieving the PDF precision required
for the discovery program at the High-Luminosity LHC and beyond.

In addition to community benchmarking and other explorations in PDF fitting, future high-precision experiments will also be helpful. As
an example, the Electron-Ion Collider (EIC) \cite{Accardi:2012qut} will perform extremely precise measurements that are likely to substantially supersede the current fixed-target experimental data fitted in
CT18. Independent EIC measurements will be valuable, for instance, given the competing pulls especially evident in Figs.~\ref{fig:SM_PDFs_L2:CT18_gg_y5} and~\ref{fig:SM_PDFs_L2:CT18_gg_y2.5} of the $F^p_2$ and $F^d_2$
data from BCDMS on $L_{gg}$ in the
$300\,\mathrm{GeV} \lesssim M_X \lesssim 2\,\mathrm{TeV}$ region. Precision measurements from the EIC will also have the potential to extend sensitivity to the higher-mass $M_X \gtrsim 2\,\mathrm{TeV}$
region, where the $L_2$ sensitivities of current experiments are rapidly vanishing. By measuring inclusive cross sections with high precision over a wide sweep of $x$ and $Q$, DIS experiments have
the capacity to constrain scaling violations and provide access to the gluonic structure of the nucleon or of nuclei. For the EIC, the expected luminosities ($10^2\!-\!10^3$ times that of HERA)
are sufficiently great that the resulting improvements in the gluon PDF can in turn significantly reduce the PDF uncertainty on the LHC $gg\!\to\!$ Higgs production. This conclusion has been demonstrated by computing the $\mathit{L_1}$~{\it sensitivity} of the EIC pseudodata to the PDF uncertainty of the 14 TeV Higgs-production cross section, as presented in the right panel of Fig.~2 in Ref.~\cite{Hobbs:2019sut}.

This work is partially supported by the U.S.~Department of Energy under Grant No.~DE-SC0010129 and by the U.S.~National Science Foundation under Grant No.~PHY-1719914. T.~J.~Hobbs acknowledges support from a JLab EIC Center Fellowship.

\let\CTHERAII\undefined

\chapter{Jet substructure studies}
\label{cha:jets}
\newcommand{\order}[1]{{\cal O}\left(#1\right)}
\newcommand{\avg}[1]{\left\langle\smash{#1}\right\rangle}
\newcommand{\as}{\alpha_s}
\newcommand{\ycut}{y_{\text{cut}}}
\newcommand{\zcut}{z_{\text{cut}}}
\newcommand{\fcut}{f_{\text{cut}}}
\newcommand{\ftrim}{f_{\text{trim}}}
\newcommand{\Rtrim}{R_{\text{trim}}}
\newcommand{\rtrim}{r_{\text{trim}}}
\newcommand{\zprune}{z_{\text{prune}}}
\newcommand{\Rprune}{R_{\text{prune}}}
\newcommand{\rprune}{r_{\text{prune}}}
\newcommand{\e}{\varepsilon}
\newcommand{\nf}{n_{F}}
\newcommand{\MSb}{\overline{\rm MS}}
\newcommand{\W}{{\rm W}}
\newcommand{\TiTj}{{\bf T}_i \cdot {\bf T}_j}
\newcommand{\ord}{\mathcal{O}}
\newcommand{\gstrong}{g_s}
\newcommand{\muNP}{\mu_\text{NP}}
\newcommand{\amax}{a_2}
\newcommand{\amin}{a_1}
\newcommand{\tlambda}{\tilde{\lambda}}

\renewcommand{\d}{\mathrm{d}}

\newcommand{\SD}{SoftDrop\xspace}
\newcommand{\ttt}[1]{{\small\texttt{#1}}}
\newcommand{\fastjet}{\texttt{FastJet}\xspace}
\newcommand{\fjcontrib}{\texttt{fjcontribtJet}\xspace}

\definecolor{darkgreen}{rgb}{0,0.5,0}
\definecolor{darkblue}{rgb}{0,0,0.7}
\definecolor{darkred}{rgb}{0.5,0,0.0}

\newcommand{\sm}[1]{\textbf{\color{darkgreen}  [#1 -- sm]}}

\section{Jet Studies: Four decades of gluons~\protect\footnote{S.~Marzani and B.~Nachman (section coordinators); S.~Amoroso, P.~Azzurri, H.~Brooks, S.~Forte, P.~Gras, Y.~Haddad, J.~Huston, A.~Larkoski, M.~LeBlanc, P.~Loch, K.~Long, E.~Metodiev, D.~Napoletano, S.~Prestel, P.~Richardson, F.~Ringer, J.~Roloff, D.~Soper, G.~Soyez, V.~Theeuwes.}}

Studies related to gluon jets have played a key role in particle and nuclear physics since their discovery at PETRA exactly (to the day!) \textbf{four decades} prior to the 2019 Les Houches workshop.  This section investigates gluon fragmentation at the LHC, covering nearly \textbf{four decades} in energy scales.  Low energy scales involving gluon (sub)jets are studied from the point of view of hadronization and Monte Carlo tuning.  Higher-order effects in parton shower programs are investigated using deep learning.  Gluon jet rejection is considered in the context of vector boson fusion/scattering processes.  One of the main studies at this Les Houches was an investigation into the usefulness of a gluon jet differential cross section measurement in the context of parton distribution functions.  Gluon jet identification was also briefly discussed for searches at the highest energies accessible at the LHC.

\subsection{Introduction}
\label{sec:SM_jetsub_gluons:intro}

Jets are collimated sprays of hadrons that emerge from high energy quarks and gluons and are an important asset or significant nuisance in the majority of collider particle physics analyses.  Understanding jets and their internal structure (jet substructure~\cite{Abdesselam:2010pt,Altheimer:2012mn,Altheimer:2013yza,Adams:2015hiv,Asquith:2018igt,Larkoski:2017jix,Marzani:2019hun}) will directly or indirectly address a variety of fundamental questions in particle and nuclear physics.  One of the first studies related to jet substructure occurred nearly four decades ago, with the direct discovery of the gluon at PETRA~\cite{Brandelik:1979bd,Barber:1979yr,Berger:1979cj,Bartel:1979ut,Ellis:2014rma}.  It was of paramount importance at the time to study differences between jets initiated by quarks (quark jets) and jets initiated by gluons (gluon jets) in order to categorize the properties of the new boson.  This complex topic is still an active area of research in the present day and was the subject of the 2015 Les Houches report on jets~\cite{Badger:2016bpw,Gras:2017jty}.   The goal of this report is to study gluon jets at all relevant energies at the LHC, from non-perturbative scales all the way to the highest accessible energies.   Traversing nearly four decades in energy scales will reveal a plethora of interesting phenomena.  

At the lowest energies, jets are dominated by non-perturbative effects.  While there has been significant progress in understanding jet formation when fixed-order or resummed perturbation theory is accurate, there has been much less progress outside these regions of phase space.  While such contributions are small for many observables of interest, they are relevant for any precision program involving hadronic final states.  One example is the determination of the strong coupling constant, $\alpha_s$, from hadronic event shapes~\cite{Abbate:2010xh,Hoang:2015hka,Heister:2003aj,Abreu:1996mk,Abdallah:2004xe,Biebel:1999zt,Abbiendi:2004qz,Buskulic:1992hq}.   After lattice determinations, the most precise extractions of $\alpha_s$ use thrust and the $C$-parameter from $e^+e^-$ data.  One of the biggest challenges of this extraction is that the non-perturbative corrections are nearly degenerate with changes to $\alpha_s$~\cite{Abbate:2010xh}.  The 2017 Les Houches report on jets studied the possibility of using jet substructure at the LHC to determine $\alpha_s$~\cite{Bendavid:2018nar}.  A key ingredient to this study is jet grooming, which is a set of tools to systematically remove soft and wide angle radiation within a jet.  Well-designed grooming algorithms allow for precise theory predictions of certain observables in part because the magnitude and the onset of non-perturbative effects can be parametrically suppressed with respect to the un-groomed case.  While jet grooming may not be enough to eliminate the need to estimate non-perturbative effects, grooming may provide a unique opportunity to isolate these effects for further study.   While the perturbative regions of phase space have received significant attention from the community~\cite{Frye:2016aiz,Frye:2016okc,Marzani:2017mva,Marzani:2017kqd,Kang:2018vgn,Kang:2018jwa,Baron:2018nfz,Kardos:2018kth,Kardos:2020ppl,Kardos:2020gty}, the non-perturbative regions have only recently been investigated~\cite{Hoang:2019ceu}.   One of the goals of this report is to explore the non-perturbative region of groomed jets using phenomenological tools for guidance. 

Both perturbative and non-perturbative regions of phase space at low energy can be important inputs to Parton Shower Monte Carlo (PSMC) parameter tuning.  In particular, there is a need for data enriched in gluon jets as many of the existing tunes are either based solely on or are anchored based on $e^+e^-$ data.  While those data are free from many nuisances like the underlying event, they are dominated by quark jets.  Various tuning campaigns at the LHC have found potential sources of tension between tunes that use jet substructure from the LHC and those that use jet and event shapes from LEP~\cite{ATL-PHYS-PUB-2014-021,Aad:2016oit,CMS:2013kfa,CMS:2017wyc}.  It is therefore critical to collect new measurements with unique and overlapping sensitivity to a variety of phase space regions.  The community repository for storing measurements is \textsc{HepData}~\cite{Buckley:2010jn,Maguire:2017ypu} and the standard for encoding an analysis for reinterpretation is \textsc{Rivet}~\cite{Buckley:2010ar}.  In the preparation of this report, new routines have been added to the existing databases and a list of jet substructure measurements from the LHC experiments has been tabulated.

While many aspects of PSMC programs are built on phenomenological models that must be tuned to data, there are also a variety of components that are based on fundamental aspects of strong interactions and can be systematically improved.  Various MC programs such as \textsc{Dire}~\cite{Hoche:2015sya}, \textsc{Vincia}~\cite{Giele:2007di}, and \textsc{Deductor}~\cite{Nagy:2014mqa} include various subleading resummation, helicity, and color corrections.  In particular, the \textsc{Dire} program, which is a plugin to both \textsc{Pythia}~\cite{Sjostrand:2014zea,Sjostrand:2007gs} or \textsc{Sherpa}~\cite{Gleisberg:2008ta,Bothmann:2019yzt} now includes all of the next-to-leading order components of the QCD splitting functions including the triple-collinear and double-soft splittings.~\footnote{While these proceedings were finalized, a new proposal for a next-to-leading logarithmic parton shower appeared~\cite{Dasgupta:2020fwr}.}
The 2017 Les Houches report briefly discussed an investigation of standard jet substructure observables (such as the two-prong tagger $N_2$~\cite{Moult:2016cvt}) to the triple collinear splitting function~\cite{Bendavid:2018nar}.  A non-exhaustive list of such observables showed no sensitivity to this splitting function.  In order to know if any jet observable is sensitive to the extended physics modelling, deep neural network classifiers were constructed using the full observable jet phase space (kinematics and particle types).   This study confirmed that the triple-collinear splitting function is essentially non-observable, but the neural networks were able to significantly detect the double-soft splitting function.  Future work is required to construct simple observables that may become near-future measurements for probing this in data.

Jet classification techniques have been used for a variety of other tasks, including quark versus gluon (q/g) jet tagging.  In the context of Les Houches 2019, the focus of q/g tagging was on the isolation of vector boson fusion (VBF) and vector boson scattering (VBS) processes.  These processes are distinguished in part by two moderate $p_T$ forward quark jets, and q/g jet tagging has been employed to suppress gluon-initiated backgrounds 
for example in the context of electroweak VBF measurements ~\cite{Chatrchyan:2013jya,Khachatryan:2014dea,Sirunyan:2017jej,Sirunyan:2019dyi}. In the context of Higgs production, quark/gluon tagging is useful both for separating the Higgs from other Standard Model backgrounds as well as separating different Higgs production modes.   The usefulness of q/g tagging to distinguish the gluon fusion (ggH) and VBF Higgs production modes was first investigated by CMS~\cite{Khachatryan:2015bnx}.  This report will show additional studies to understand the interplay between q/g tagging and other analysis selections such as requiring a large dijet invariant mass ($m_{jj}$).

While q/g tagging has traditionally been used to \textit{reject quarks}, there may also be a physics case for \textit{tagging gluons}.  One possibility in particular is the prospect of using gluon jets to constrain the gluon parton distribution function (PDF).  The gluon PDF has a large uncertainty at high $x$ ($m_{jj}\sim 1$~TeV) because the existing inclusive jet data are dominated by $qq$ initial states and $gg$ constraints from $t\bar{t}$ production become statistically limited.  What if one could directly measure the $gg$ reaction cross section?  The first step in answering this question is to establish a strong correlation between the initial and final state flavors.  This means that gluon tagging the final state will bias the initial state to be more gluonic.   The second step is to identify the tradeoff between tagging performance and uncertainties.  If the current PDF uncertainty can be made larger than the statistical and systematic uncertainties, new data would be useful in constraining the gluon PDF.  Lastly, it is important that the gluon tagging strategy is theoretically well-understood so that it can be simultaneously calculated with the $p_T$ spectrum as input to the PDF fits.  For this purpose, a new observable is considered - the \textit{Les Houches multiplicity} $n_\text{LH}$, first proposed in Ref.~\cite{Marzani:2019hun} as a variation on the iterative SoftDrop multiplicity~\cite{Frye:2017yrw}.

At the kinematic limit of LHC jets, gluon tagging may also have an important role for searches for new particles.  While not studied extensively at Les Houches, there was a general brainstorming session for gluon tagging applications and one promising example is the search for $X\rightarrow gg$.  At high $m_{jj}$, the SM background is dominated by valence quark scattering.  Furthermore, gluon tagging is more useful at high $p_T$ where counting observables like $n_{LH}$ have a better q/g tagging performance.  For these reasons, gluon tagging has an interesting potential to increase the sensitivity of the high mass dijet search.

This remainder of this chapter is organized from low to high energy.  Section~\ref{sec:SM_jetsub_gluons:np} begins with studies related to non-perturbative aspects of jets after grooming.  Then, Sec.~\ref{sec:SM_jetsub_gluons:mc} investigates the potential for jet substructure observables for PSMC tuning.  Next, section~\ref{sec:SM_jetsub_gluons:psmc} presents methods for probing higher-order effects in PSMCs.  A brief study of q/g tagging in the context of VBF/VBS is highlighted in Sec.~\ref{sec:SM_jetsub_gluons:vbsbvf}.  At higher energies, the feasibility of Suppressing QUarks in the Region of RElatively Large-$x$ (\textsc{Squirrel}) is studied for the gluon PDF in Sec.~\ref{sec:SM_jetsub_gluons:pdf}.  The kinematic limit is briefly described in Sec.~\ref{sec:SM_jetsub_gluons:highest} and the chapter ends with conclusions and outlook in Sec.~\ref{sec:SM_jetsub_gluons:conclusion}.

\subsection{Non-Perturbative effects at low jet mass}
\label{sec:SM_jetsub_gluons:np}

Jet grooming is the systematic removal of soft and wide-angle emission from inside a jet.  Advances in jet grooming have resulted in algorithms like SoftDrop and the modified mass drop tagger (mMDT)~\cite{Larkoski:2014wba,Dasgupta:2013ihk} that are amenable to high order resummation.   Both ATLAS~\cite{Aaboud:2017qwh,Aad:2019vyi} and CMS~\cite{Sirunyan:2018xdh} have measured the SoftDrop jet mass and compared the differential cross section with theoretical calculations~\cite{Frye:2016aiz,Frye:2016okc,Marzani:2017mva,Marzani:2017kqd,Kang:2018vgn,Kang:2018jwa,Baron:2018nfz,Kardos:2018kth}.  As the level of both experimental and theoretical precision improves, it is natural to consider what are the next challenges.  The reconstruction and prediction of jet masses near the non-perturbative regime $k_T\lesssim \Lambda_\text{QCD}$ is becoming comparably important to improving the experimental and theoretical precision at medium and high masses.

\begin{figure}[t]
\centering
\includegraphics[width=0.62\textwidth]{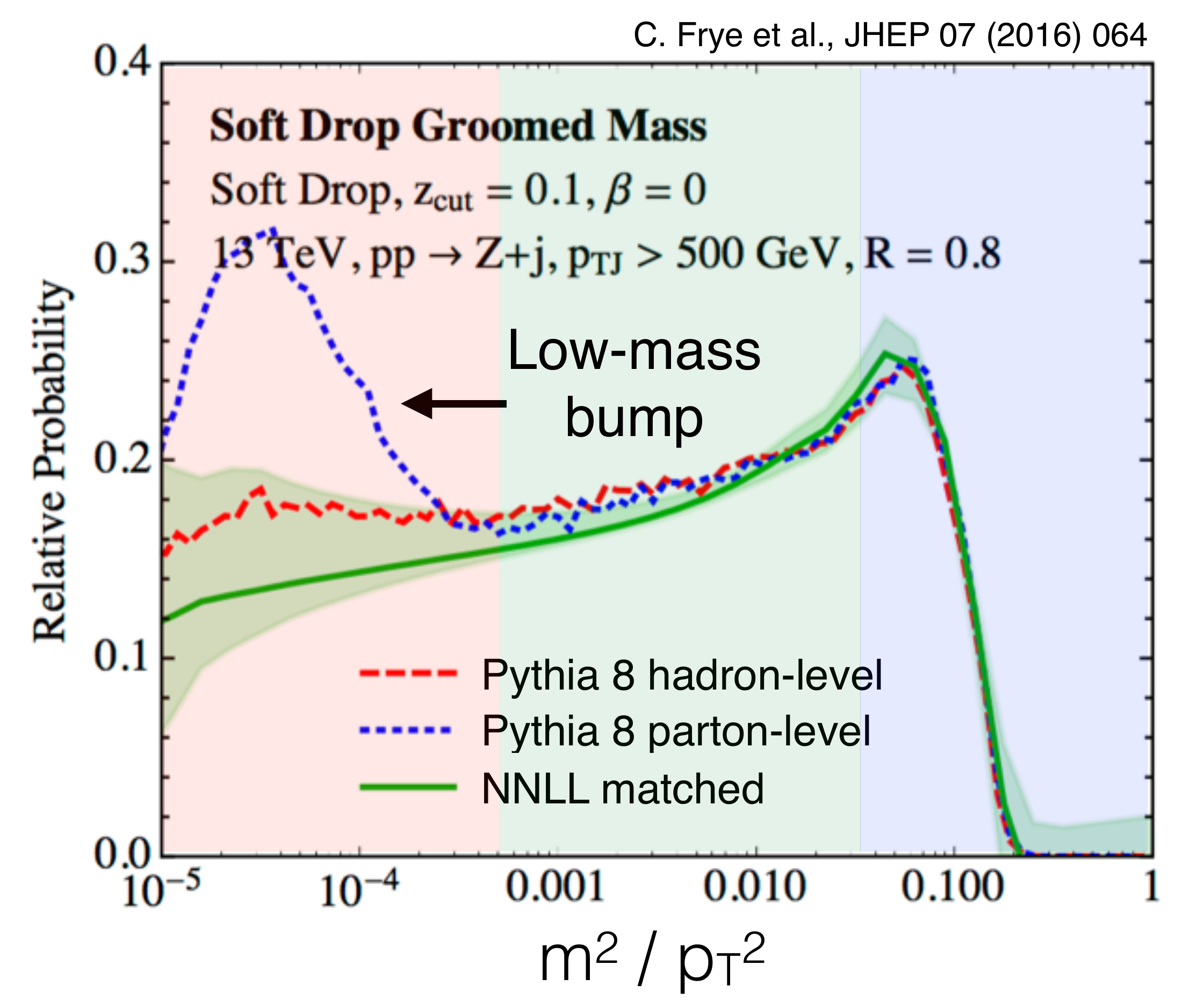}
\caption{Figure adapted from Ref.~\cite{Frye:2016aiz}.  The differential cross-section has three regimes: the left part (red) where non-perturbative effects dominate, the middle (green) regime where resummation is most accurate, and the right (blue) regime where fixed-order effects are the most relevant.}
\label{fig:SM_jetsub_gluons:np:illustration}
\end{figure}

At the same time, jet grooming also offers an exciting opportunity to study non-perturbative effects in isolation.  In particular, non-perturbative corrections to the jet mass are localized after jet grooming.  This is illustrated in Fig.~\ref{fig:SM_jetsub_gluons:np:illustration}.  Turning on and off non-perturbative modeling in \textsc{Pythia} lead to an $\mathcal{O}(1)$ effect in the differential cross section, but only at low jet masses.  The goal of this section is to use phenomenological tools to investigate this structure in more detail.  This region of the phase space is particularly hard to measure experimentally and has thus far largely been avoided.  However, innovations in mass reconstruction will improve the precision at low mass and this region may be well-suited for non-perturbative studies including phenomenological model tuning.  Analytic approaches to also probe this region are also an important complement to these studies and have begun with the recent work in Ref.~\cite{Hoang:2019ceu}.

\begin{figure}[t]
\centering
\includegraphics[width=0.62\textwidth]{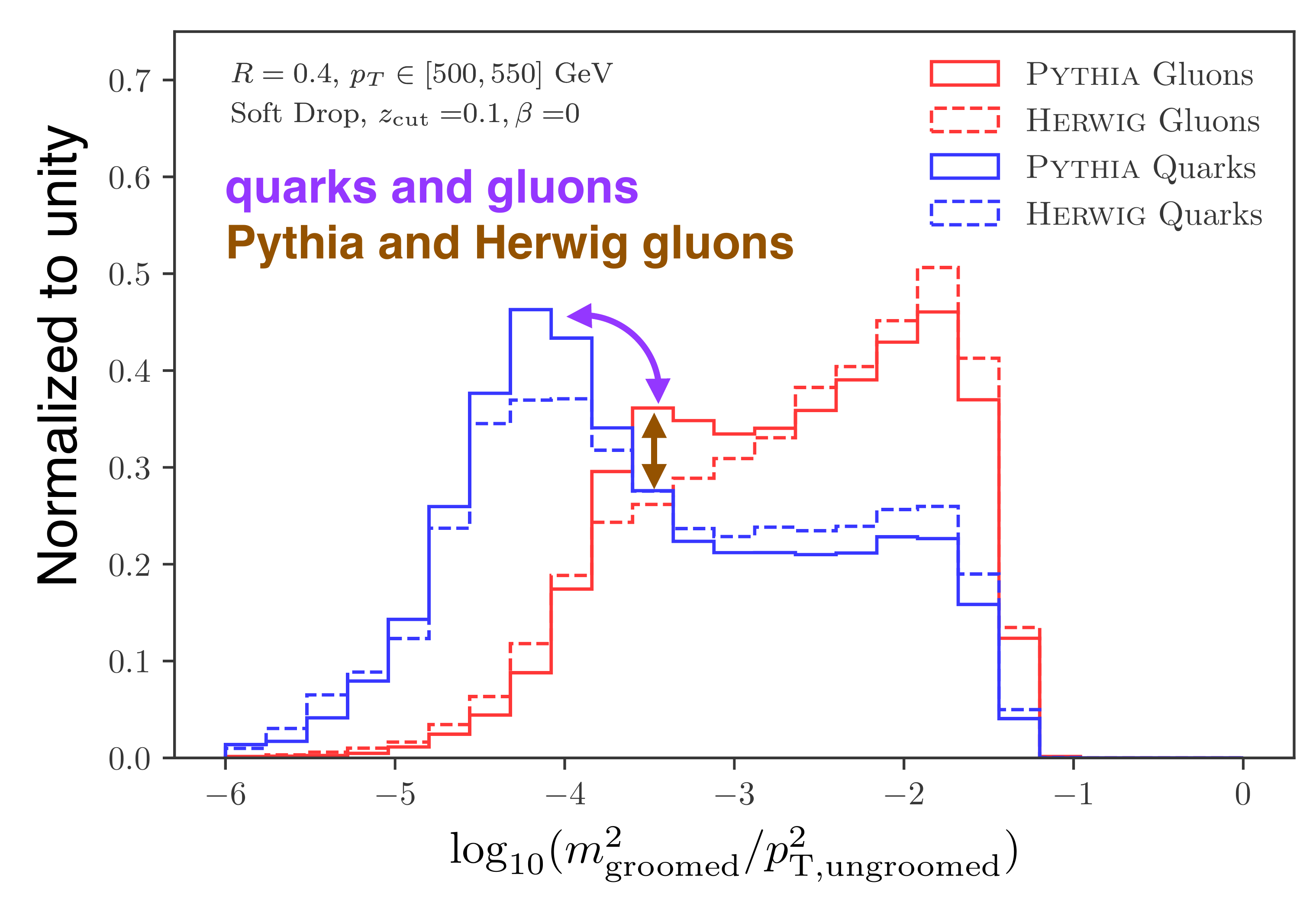}
\caption{The binned differential cross-section for the groomed jet mass using \textsc{Pythia} and \textsc{Herwig} and separately for quark and gluon jets.}
\label{fig:SM_jetsub_gluons:np:qg}
\end{figure}

Figure~\ref{fig:SM_jetsub_gluons:np:qg} shows the groomed jet mass predicted for dijet events using \textsc{Pythia} and \textsc{Herwig}~\cite{Bahr:2008pv,Bellm:2015jjp,Bellm:2019zci} and separated into quark jets and gluon jets.  The SoftDrop grooming algorithm is used with the most aggressive grooming parameter $\beta=0$ (this also corresponds to mMDT).  The leading logarithm prediction for the differential cross section in the intermediate mass region ($-3\lesssim\log_{10}(m^2/p_T^2)\lesssim -2$) is a roughly linear distribution, the slope of which is different for quarks and gluons and it is nearly flat for quarks because of the value $z_\text{cut}=0.1$ chosen in this study~\cite{Dasgupta:2013ihk}. In the absence of non-perturbative effects, these trends would continue down to arbitrarily small masses. Interestingly, the non-perturbative corrections are very different for quarks and gluons, with a much larger bump for quarks.  Furthermore, the corrections for \textsc{Pythia}and \textsc{Herwig} gluons are qualitatively different, with a clear peak for \textsc{Pythia} gluons and the absence of a peak for \textsc{Herwig}.

\begin{figure}[t]
\centering
\includegraphics[width=0.95\textwidth]{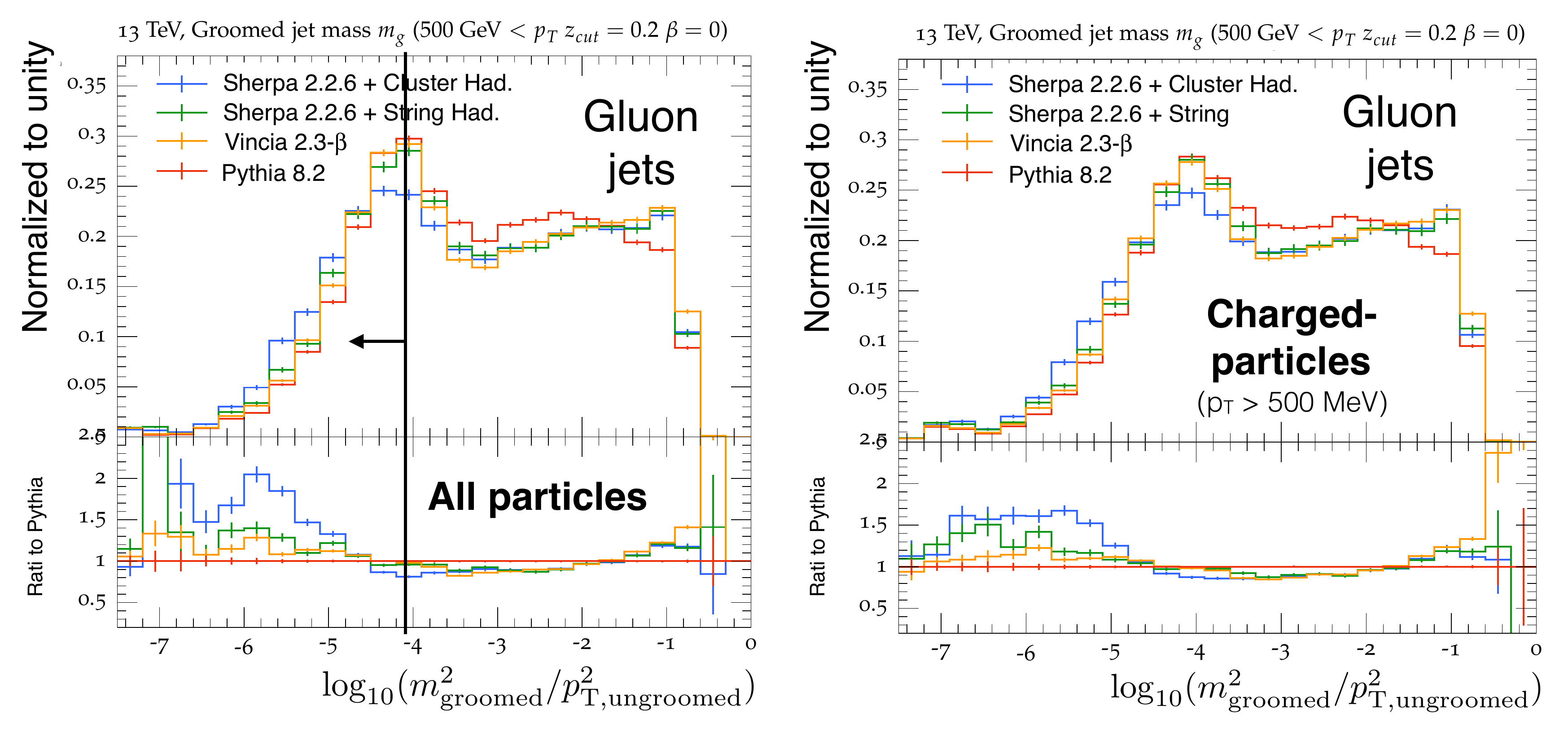}
\caption{The binned differential cross-section for the groomed jet mass using various Monte Carlo models and for all particle (left) and charged-particles only (right).  The vertical line with the arrow in the left plot shows the bin boundary for the lowest mass bin in the recent ATLAS measurements~\cite{Aaboud:2017qwh,Aad:2019vyi}.  Gluon jets are simulated by generating $Z+g$.}
\label{fig:SM_jetsub_gluons:np:tracks}
\end{figure}

The large model-dependence for gluon jets in Fig.~\ref{fig:SM_jetsub_gluons:np:qg} motivated further studies.  It was found that the differences between quarks and gluons and between models were unrelated to gluon splitting to heavy flavor quarks and to specific hadronic resonances.  The effects are nearly 100\% correlated with jet constituent multiplicity.  Furthermore, the results depend strongly on the grooming parameters ($\beta$ and $z_\text{cut}$).  Additional studies of the dependence on the hadronization model are illustrated in Fig.~\ref{fig:SM_jetsub_gluons:np:tracks}. In this figure a comparison is made for gluon jets originating from $pp\to Zg$ for 3 different showers and making use of 2 different hadronization models.  In particular, the blue and green lines are produced using \textsc{Sherpa}~2.2.6~\cite{Bothmann:2019yzt} and fix everything about the simulation except for the hadronization model, which is switched between the default Cluster hadronization model~\cite{Winter:2003tt} and the Lund String model~\cite{Sjostrand:1982fn}. In addition lines for \textsc{Pythia}~8.2~\cite{Sjostrand:2014zea} and Vincia~2.3-$\beta$~\cite{Fischer:2016vfv} are included, both of which make use of the Lund String model. For $\lesssim\log_{10}(m^2/p_T^2)\lesssim -5$, there are $\mathcal{O}(1)$ differences between the different hadronization models while the rest of the spectrum is nearly unchanged, as expected.  However, this region is not probed differentially in the most recent ATLAS and CMS jet mass measurements.  The left plot of Fig.~\ref{fig:SM_jetsub_gluons:np:tracks} shows that the entire region of interest for the large non-perturbative effects is covered by a single measurement bin.  The experimental jet mass resolution is poor in this region due to finite calorimeter granularity.  The right plot of Fig.~\ref{fig:SM_jetsub_gluons:np:tracks} shows that the trends are largely preserved when only using tracking information, which has the potential to provide the necessary precision to probe the non-perturbative region in detail.

The non-perturbative region is often avoided, but these studies indicate that there may be interesting and useful insight to learn from future studies that probe the differential cross section for the mass and potentially related observables.


\subsection{Monte Carlo tuning with jet substructure observables}
\label{sec:SM_jetsub_gluons:mc}

While the jet mass differential cross section holds great potential for Monte Carlo tuning, it is only one of many observables.  The last several years have produced numerous measurements of jet and jet substructure observables at the LHC. 
Some of these measurements have been motivated by improving parton distribution functions and testing perturbative QCD predictions, 
while others have been designed to improve jet modeling by providing new and better inputs to Monte Carlo tuning.
Some of these, like the fragmentation functions~\cite{Aad:2019onw}, have been used for years to tune Monte Carlo predictions, 
while others, like the Lund Jet Plane~\cite{ATLAS:2019sol}, are measurements of observables which have only recently been proposed~\cite{lundPlane}.

With all of these observables, it is useful to consider which measurements are the most constraining for tuning jet modeling.
This is important for motivating future measurements, and can also be used to understand characteristics of the most effective observables.
This study compares several classes of variables in order to provide a more in-depth understanding of the interplay between observables and tuning.
Several simplifications were made in order to ease the comparisons. 
To remove any dependence on topologies, only measurements in dijet events are considered, and only 13 TeV measurements are used.
While both ATLAS and CMS have produced several measurements of jet substructure observables, only ATLAS measurements are considered, since there are more available measurements.
Currently, two measurements are considered: the SoftDrop mass measurement~\cite{Aaboud:2017qwh} and a measurement of a variety of jet substructure observables~\cite{Aaboud:2019aii}.

These studies scan a similar set of parameters as the ATLAS A14 tune~\cite{ATL-PHYS-PUB-2014-021}, focusing on parameters which are sensitive to parton showers and hadronization.
In addition to parameters which were considered for A14, one additional parameter, \texttt{StringPT:sigma}, is included due to its sensitivity to hadronization.
The list of parameters and their ranges of allowed values are shown in Table~\ref{tab:SM_jetsub_gluons:parameterSpace}.
The parameter space is scanned using a sampling of 300 different configurations determined by \textsc{Professor2}~\cite{Buckley:2009bj}, and the results are fit with a 3rd order polynomial.
Each configuration is run with \texttt{PhaseSpace:pTHatMin} of 400 GeV, using 200,000 events per configuration. 
This allows sufficient sampling of the parameter space for the specific $p_{T}$ cuts of each analysis. 

\begin{table}[t]
\centering\begin{tabular}{ | c | | c | c | } \hline
                                     & Min. Value   & Max. Value    \\ \hline
\texttt{SigmaProcess:alphaSvalue}             &  0.12        & 0.15    \\ \hline
\texttt{BeamRemnants:primordialKThard}        &  1.5         & 2.0     \\ \hline
\texttt{SpaceShower:pT0Ref}                   &  0.75        & 2.0     \\ \hline
\texttt{SpaceShower:pTmaxFudge}               &  0.5         & 1.5     \\ \hline
\texttt{SpaceShower:pTdampFudge}              &  1.0         & 1.5     \\ \hline
\texttt{SpaceShower:alphaSvalue}              &  0.10        & 0.15    \\ \hline
\texttt{TimeShower:alphaSvalue}               &  0.10        & 0.15    \\ \hline
\texttt{StringPT:sigma}                       &  0.3         & 0.37    \\ \hline
\texttt{MultipartonInteractions:pT0Ref}       &  1.5         & 3.0     \\ \hline
\texttt{MultipartonInteractions:alphaSvalue}  &  0.1         & 0.15    \\ \hline
\end{tabular}
\caption{Choices of parameters to tune, and their maximum and minimum values.}
\label{tab:SM_jetsub_gluons:parameterSpace}
\end{table}

Several sets of tunes are compared in order to disentangle different effects from the measurements. 
Additionally, two sets of tunes are compared in order to disentangle the impact of different effects on the tuning.
In all cases, the results are compared for four different observables:
the SoftDrop jet mass distribution for $\beta=0$, the SoftDrop jet mass distribution for $\beta=2$, 
the number of subjets in a soft-dropped jet $N_{\mathrm{subjets}}$, and $\mathrm{ECF}_2^{\mathrm{norm}}$.
The two mass measurements provide insight into different aspects of the tune; $\beta=0$ is more sensitive to the perturbative parton shower information, 
while $\beta=2$ is more affected by hadronization. The number of subjets is sensitive to the hard splittings within a jet, and $\mathrm{ECF}_2^{\mathrm{norm}}$ is more sensitive to the 
distribution of energy within the jet. The event and jet selection is defined in the respective papers, and is taken from their \textsc{Rivet} routines~\cite{Buckley:2010ar}.

\begin{figure}[t]
\begin{center}
\includegraphics[width=0.49\textwidth]{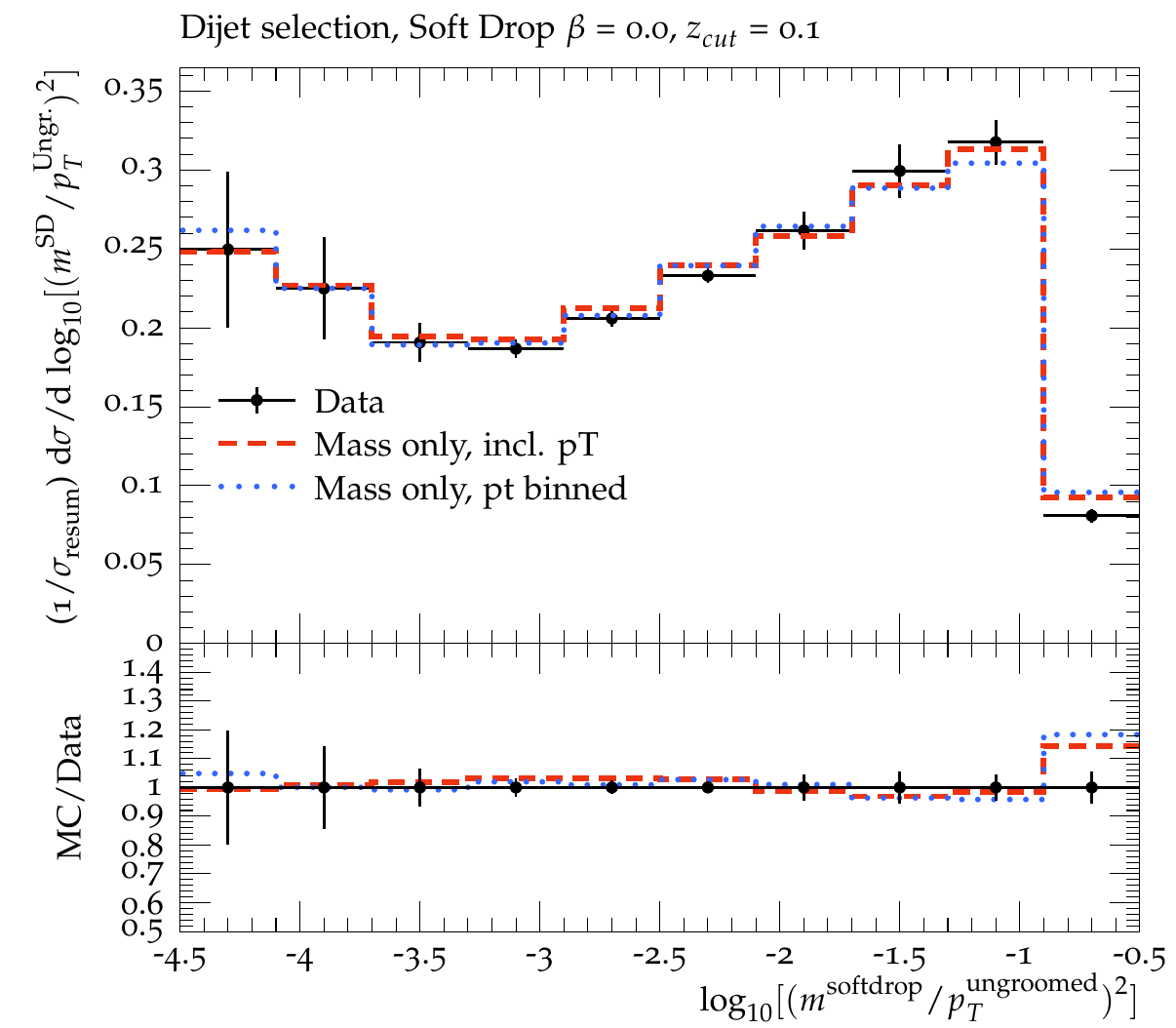} \hfill
\includegraphics[width=0.49\textwidth]{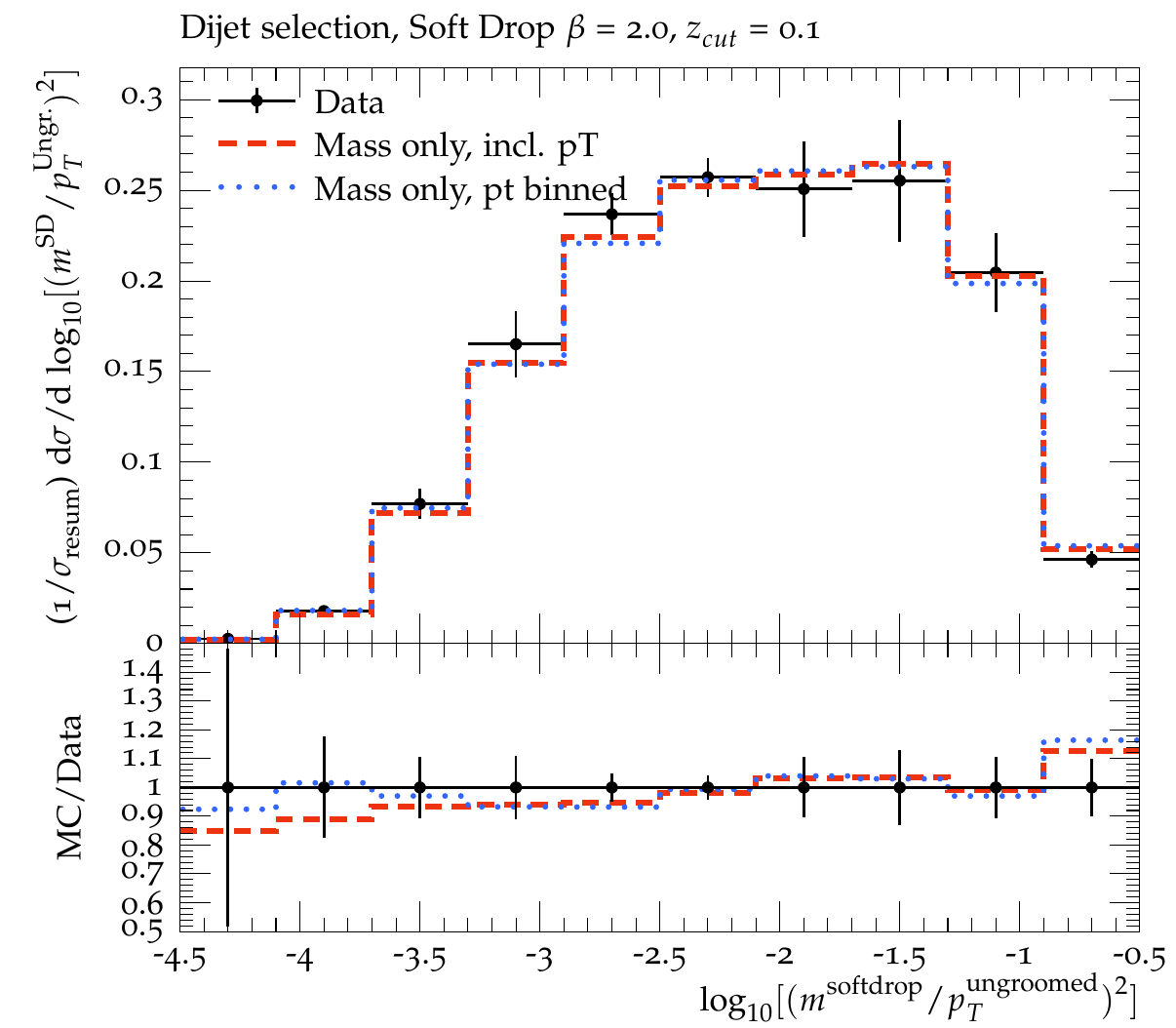} \hfill
\includegraphics[width=0.49\textwidth]{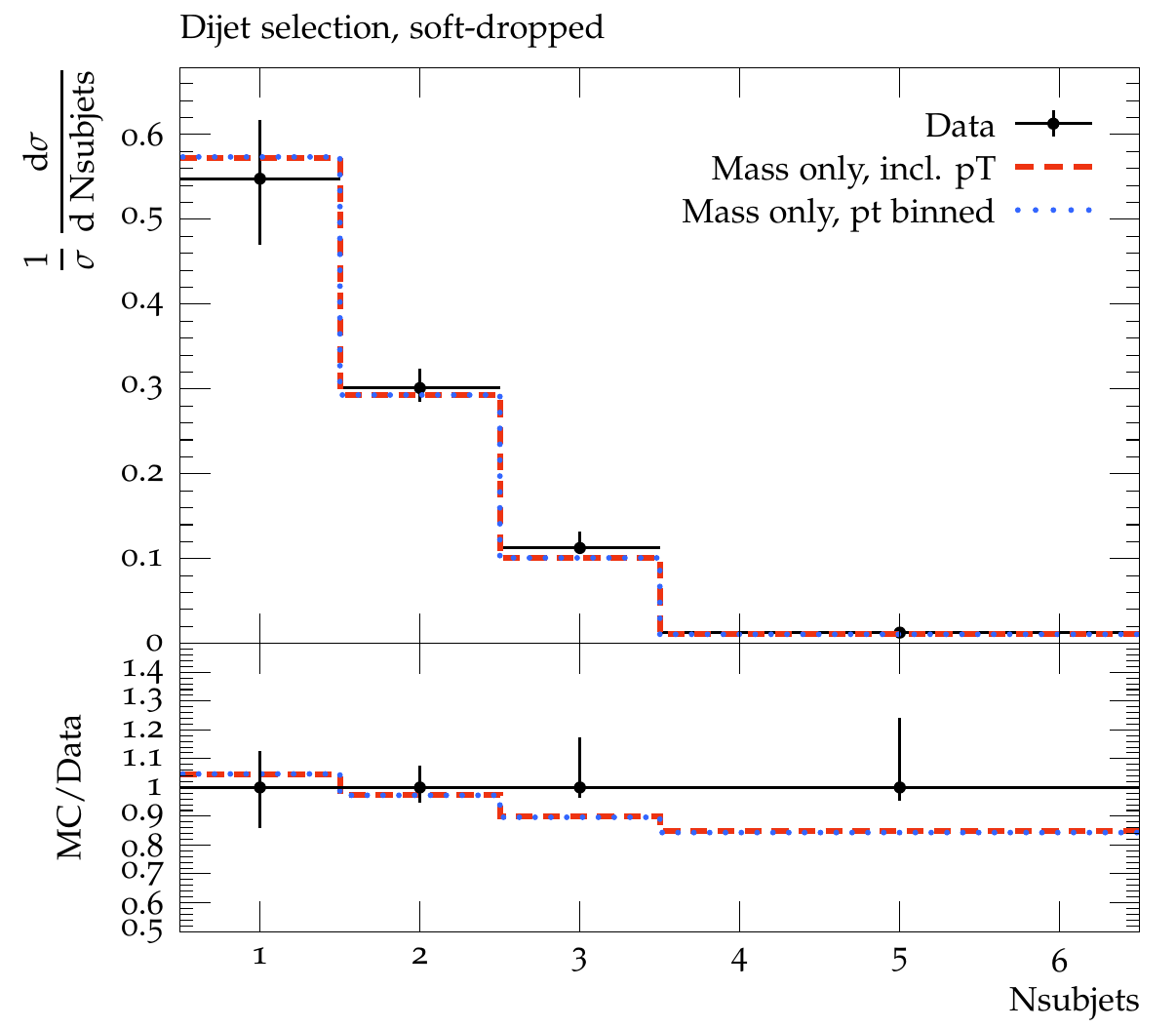} \hfill
\includegraphics[width=0.49\textwidth]{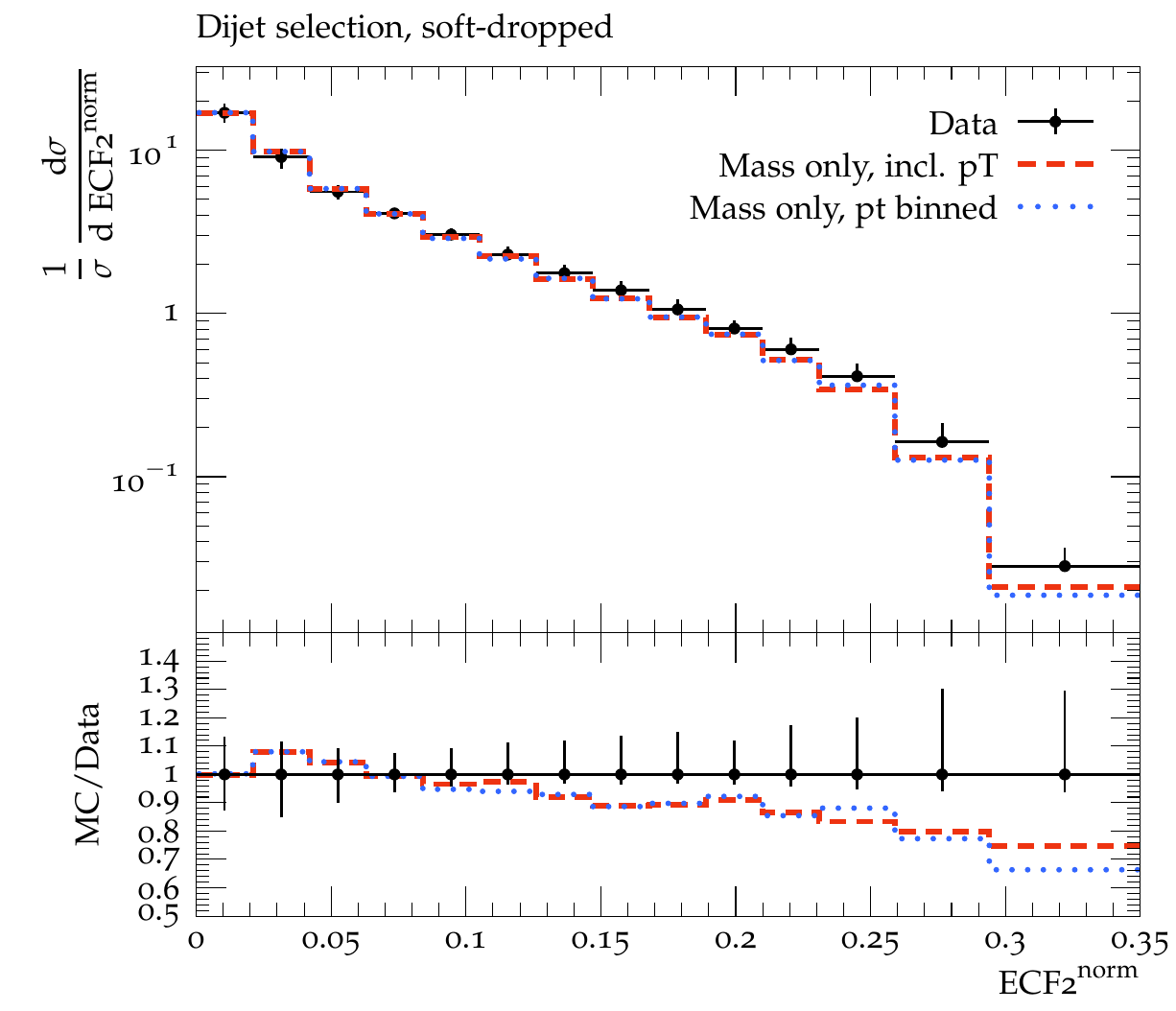} \hfill
\end{center}
\caption{Examples of results obtained with the mass-only tune.}
\label{fig:SM_jetsub_gluons:massOnlyTune}
\end{figure}

The first tune comparison studies the sensitivity to the $p_T$ selection and binning used for the measurements. Two different tunes are performed, each using only the jet mass as input.
The first tune uses the three jet mass measurements with different SoftDrop parameters from Ref.~\cite{Aaboud:2017qwh}, using the inclusive $p_T$ binning. 
The second uses the $p_T$-binned measurements from the same measurement. 
The results of these three tunes are shown in Figure~\ref{fig:SM_jetsub_gluons:massOnlyTune}. 
The two tunes which use the high-$p_T$ measurement produce similar tunes, with similar uncertainties on their parameters. 
While the $p_T$-binned result in principle provides more access to information such as quark-gluon differences or scaling with $p_T$, the impact on the results is small.

\begin{figure}[t]
\begin{center}
\includegraphics[width=0.49\textwidth]{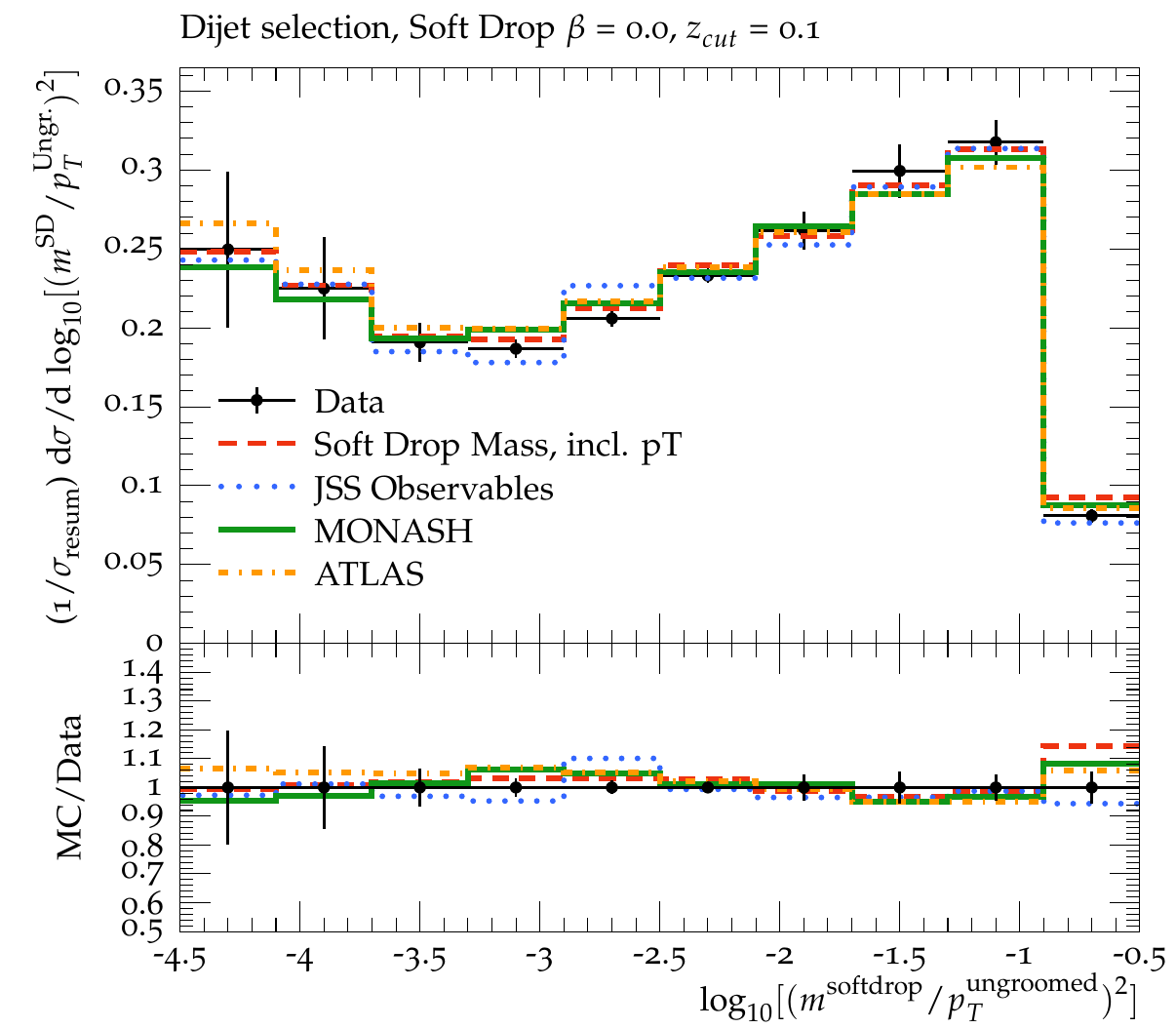} \hfill
\includegraphics[width=0.49\textwidth]{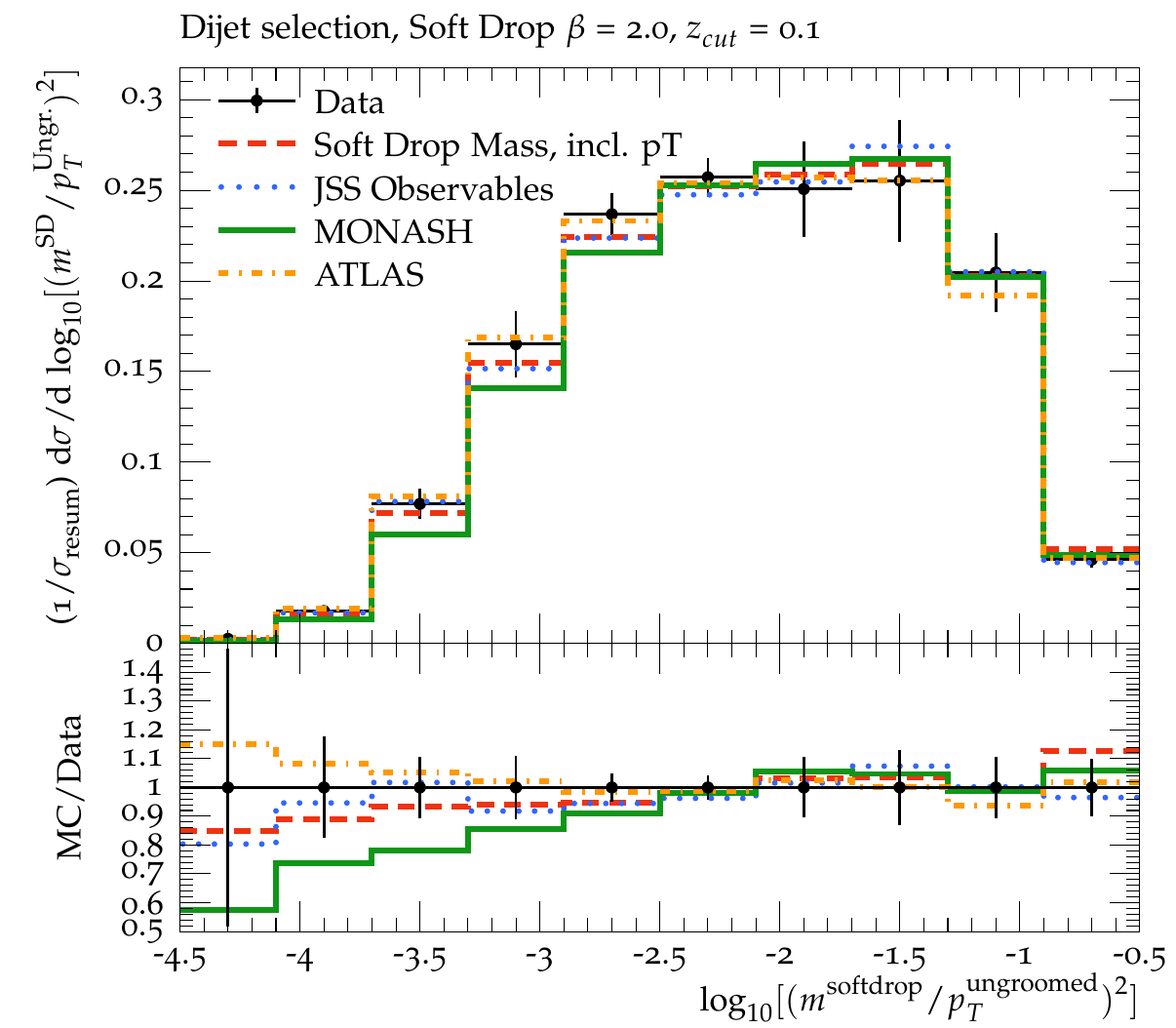} \hfill
\includegraphics[width=0.49\textwidth]{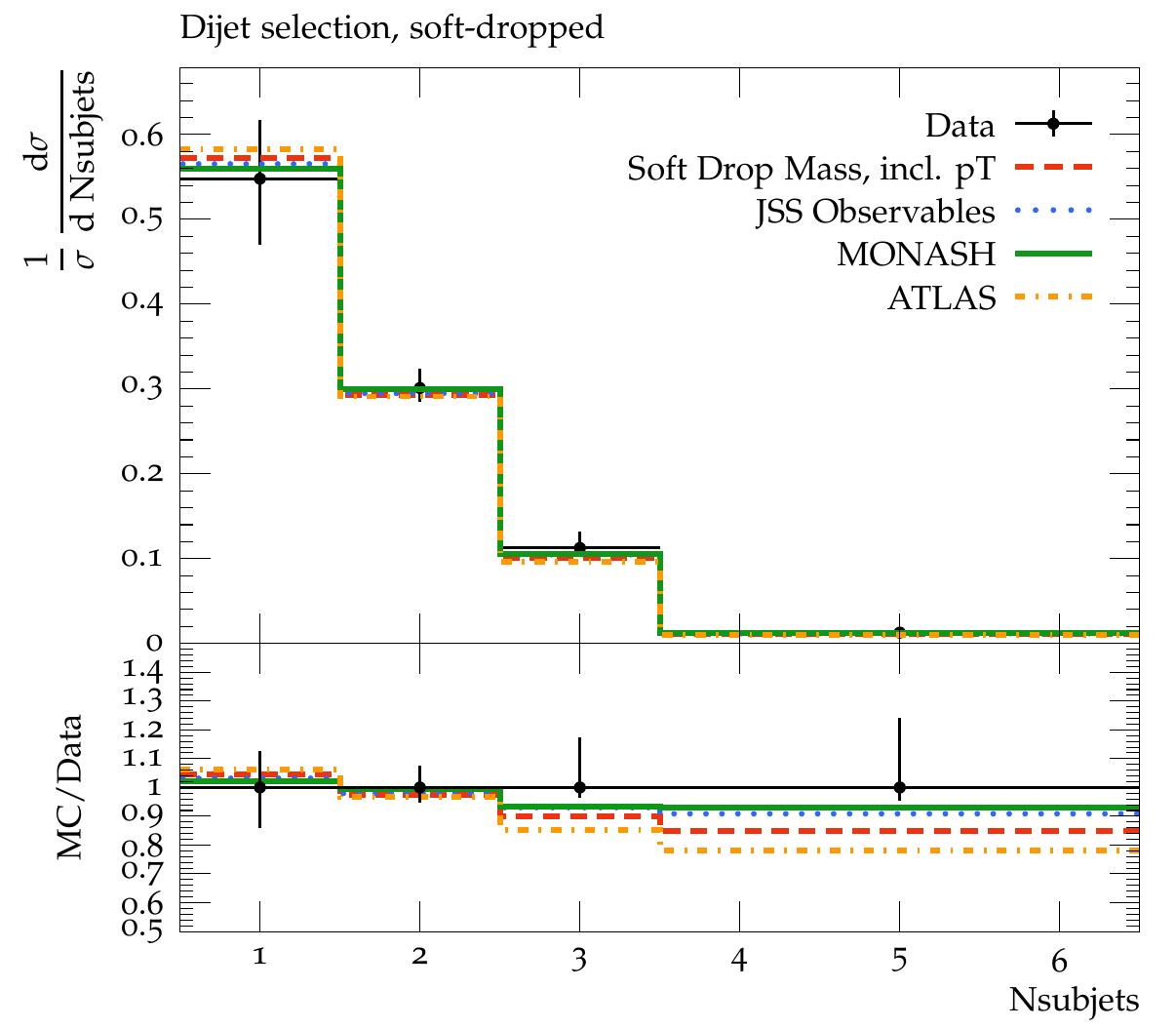} \hfill
\includegraphics[width=0.49\textwidth]{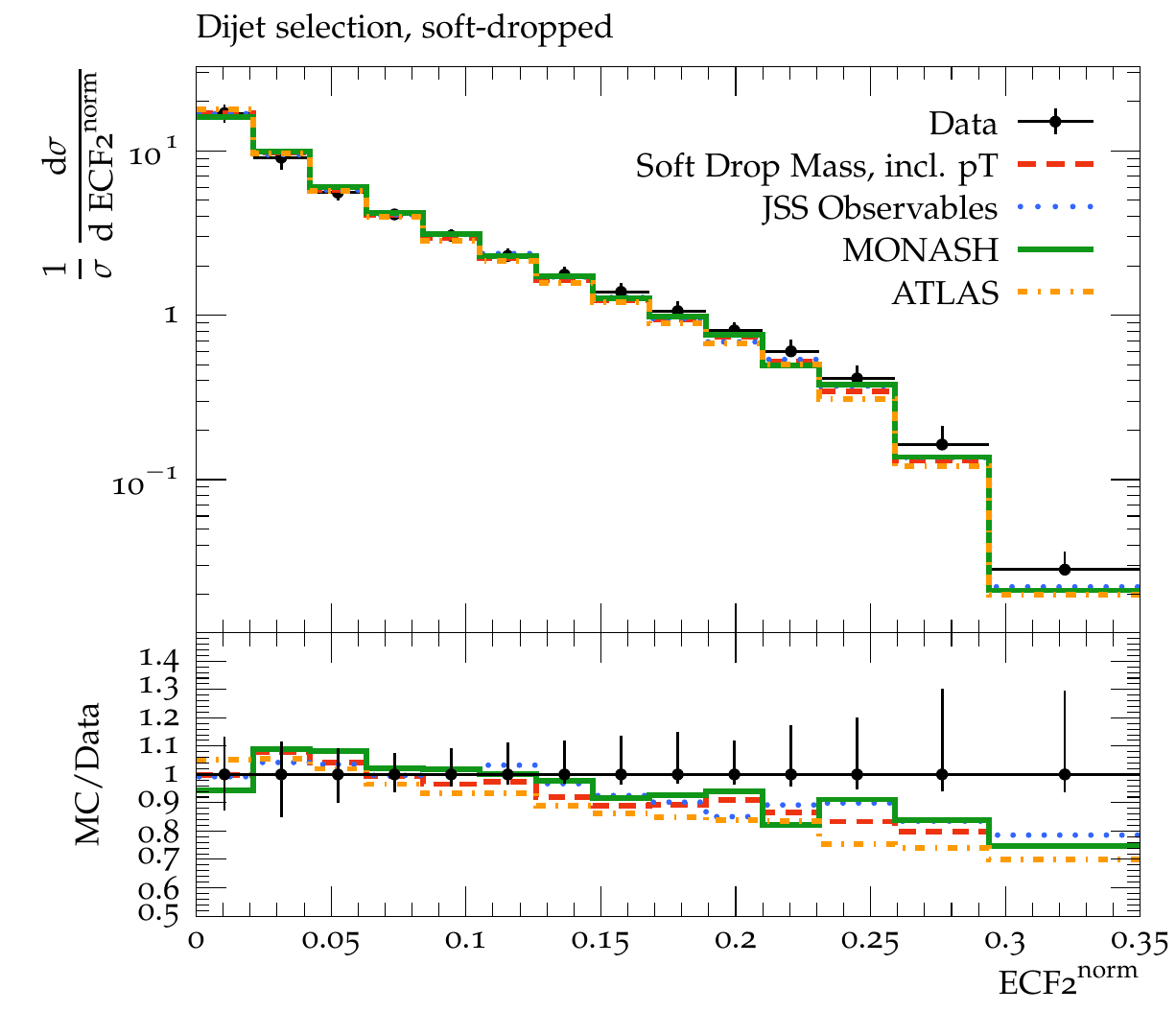} \hfill
\end{center}
\caption{Examples of results obtained with the tune using six jet substructure measurements.}
\label{fig:SM_jetsub_gluons:allTune}
\end{figure}

The second set of tunes compares the results from each individual measurement. The first of these is the same tune as before, using the $p_T$ inclusive SoftDrop jet mass measurements.
The second tune uses the measurements of six jet substructure measurements in jets groomed with the SoftDrop algorithms, as measured in Ref.~\cite{Aaboud:2019aii}.
These measurements are compared to two standard tunes: the ATLAS A14 tune and the MONASH tune.\footnote{The ATLAS tune is slightly different than the standard ATLAS tune, since it uses the ATLAS tune results, but does not use the recommended PDF set.}
The results of these are shown in Figure~\ref{fig:SM_jetsub_gluons:allTune}. 
In general, the agreement of substructure observables is improved by the use of substructure measurements compared to the ATLAS A14 tune or the MONASH tune.
As demonstrated in the mass distribution, the tunes from this study seem to lack some information about the fixed-order tune, 
and so they likely need to be combined with other measurements for full accuracy.

The full set of tuned parameters and their uncertainties is shown in Table~\ref{tab:SM_jetsub_gluons:tuneResults}.

\renewcommand{\baselinestretch}{1.2}
\begin{table}[t]
\resizebox{\textwidth}{!}{

\centering\begin{tabular}{ | c | | c | c | c | c | c |} \hline
                                     & MONASH   & ATLAS  & SDM Incl Pt   & SDM Pt Binned & JSS Observables  \\ \hline
\texttt{SigmaProcess:alphaSvalue}             &  0.130   & 0.144  & 0.126 $\pm$ 0.003 & 0.126 $\pm$ 0.001 & 0.133 $\pm$ 0.002 \\ \hline
\texttt{BeamRemnants:primordialKThard}        &  1.8     & 1.72   & 1.825 $\pm$ 0.055 & 1.794 $\pm$ 0.011 & 1.785 $\pm$ 0.048 \\ \hline
\texttt{SpaceShower:pT0Ref}                   &  2.0     & 1.30   & 1.668 $\pm$ 0.100 & 1.744 $\pm$ 0.078 & 1.721 $\pm$ 0.121 \\ \hline
\texttt{SpaceShower:pTmaxFudge}               &  1.0     & 0.95   & 1.150 $\pm$ 0.054 & 1.071 $\pm$ 0.014 & 1.036 $\pm$ 0.034 \\ \hline
\texttt{SpaceShower:pTdampFudge}              &  1.0     & 1.21   & 1.214 $\pm$ 0.058 & 1.157 $\pm$ 0.011 & 1.284 $\pm$ 0.040 \\ \hline
\texttt{SpaceShower:alphaSvalue}              &  0.1365  & 0.125  & 0.123 $\pm$ 0.003 & 0.126 $\pm$ 0.001 & 0.130 $\pm$ 0.003 \\ \hline
\texttt{TimeShower:alphaSvalue}               &  0.1365  & 0.126  & 0.132 $\pm$ 0.001 & 0.131 $\pm$ 0.001 & 0.133 $\pm$ 0.000 \\ \hline
\texttt{StringPT:sigma}                       &  0.335   & 0.335  & 0.348 $\pm$ 0.003 & 0.350 $\pm$ 0.003 & 0.333 $\pm$ 0.006 \\ \hline
\texttt{MultipartonInteractions:pT0Ref}       &  2.28    & 1.98   & 2.000 $\pm$ 0.100 & 2.181 $\pm$ 0.049 & 2.441 $\pm$ 0.148 \\ \hline
\texttt{MultipartonInteractions:alphaSvalue}  &  0.130   & 0118   & 0.116 $\pm$ 0.003 & 0.126 $\pm$ 0.002 & 0.128 $\pm$ 0.003 \\ \hline
\end{tabular}}
\renewcommand{\baselinestretch}{1.0}
\caption{Values of tuned parameters.}
\label{tab:SM_jetsub_gluons:tuneResults}
\end{table}
\renewcommand{\baselinestretch}{1.0}

More study is needed in order to fully understand the implications of these results, but there are a few interesting comments. 
The SoftDrop jet mass distribution was designed to study perturbative QCD, but also had several bins sensitive to non-perturbative effects. 
With these preliminary studies, it appears to be as effective in tuning as Ref.~\cite{Aaboud:2019aii}, even though it uses fewer observables. 
This is likely due to the factorization of different effects in the SoftDrop mass distribution, allowing it to be simultaneously sensitive to fixed order effects, the parton shower, and hadronization. 
This shows that factorization is important, and that it is important to be sensitive to a variety of effects when creating these tunes.

\subsection{Probing higher-order effects in PSMC}
\label{sec:SM_jetsub_gluons:psmc}


This study aims at exposing, with the help of energy flow 
networks~\cite{Komiske:2018cqr}, new features of jets induced by higher-order
corrections to parton showers, using the \textsc{Dire} parton shower~\cite{Hoche:2015sya}
as a test case.

Parton shower programs are an an important aspect
of LHC phenomenology, since they imprint perturbative all-order effects onto the
jet structure produced by event generators. As such, they serve two primary
purposes: $1)$ the distribution of low-multiplicity
(hard-scattering) states over states of arbitrarily high multiplicity, as well
as $2)$ generating the effect of resummation for observable depending only on 
low-multiplicity configurations. These two goals often lead to conflicting
requirements, as e.g.\ choices to improve
$2)$ often limit the potential to improve $1)$ -- and vice versa. Luckily,
these conflicts are not directly apparent at lowest (i.e.\ leading) order. This
has resulted in parton showers being ``stuck" at leading-order 
(leading-logarithmic) accuracy in their description of emission and 
no-emission rates. With increased demand for more precise event generation, 
improved parton showers will become necessary at the LHC and beyond. For 
example, the use of NLO PDFs (as e.g.\ mandated in NLO+PS
matching) in the parton shower in principle requires parton showers beyond
lowest order.

\begin{figure}[t]
\centering
\subfigure[]{
\includegraphics[width=0.2\textwidth]{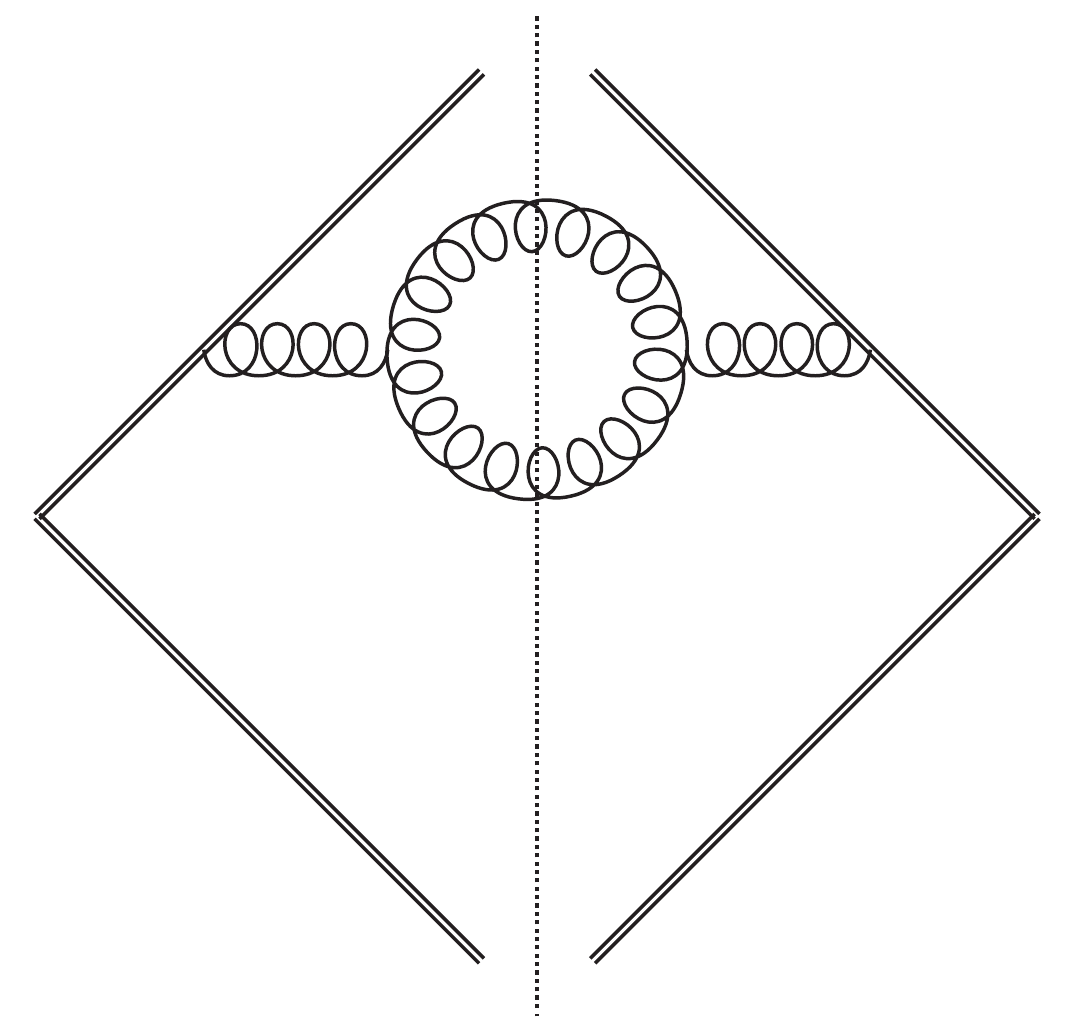}
\label{fig:SM_jetsub_gluons:np:triplecollineardiagrams1}
}
\subfigure[]{
\includegraphics[width=0.2\textwidth]{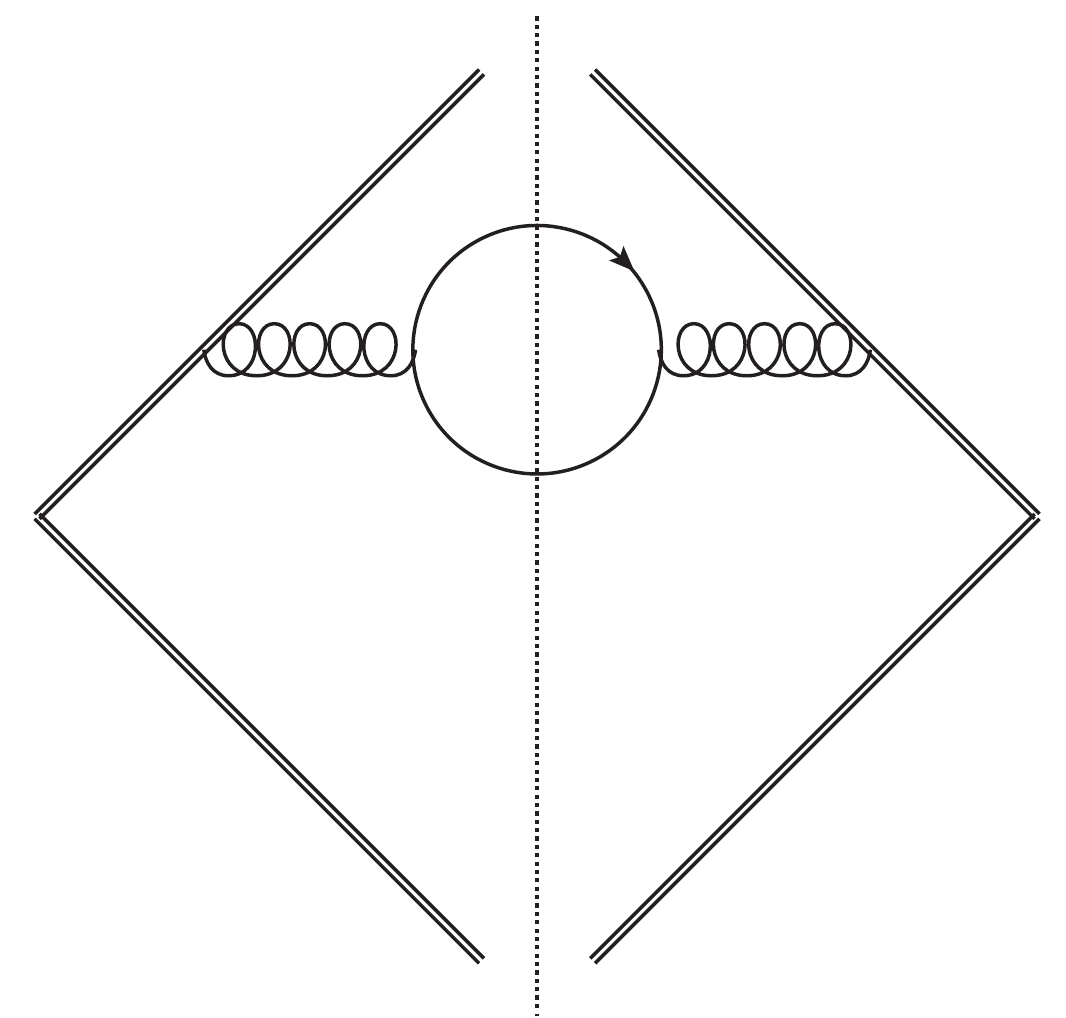}
\label{fig:SM_jetsub_gluons:np:triplecollineardiagrams2}
}
\subfigure[]{
\includegraphics[width=0.2\textwidth]{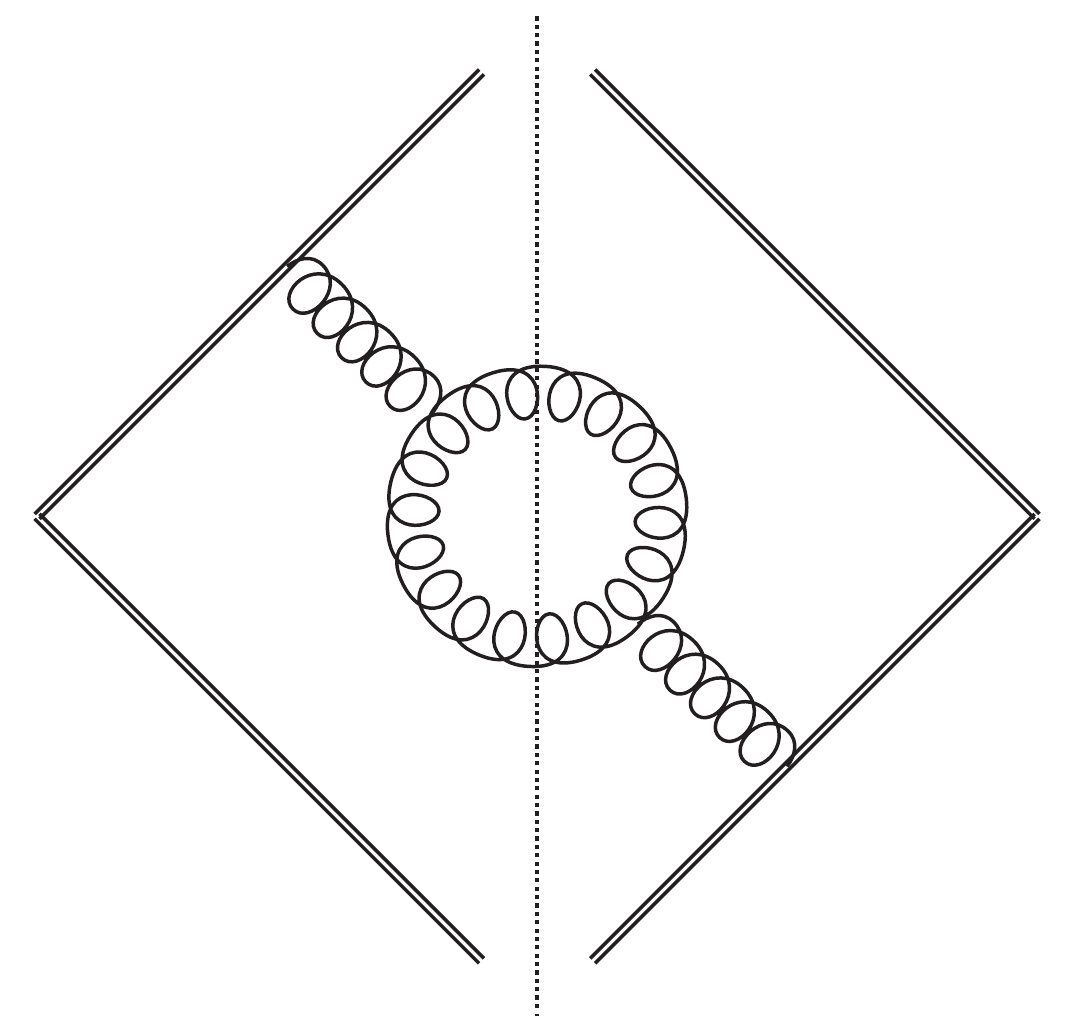}
\label{fig:SM_jetsub_gluons:np:triplecollineardiagrams3}
}
\subfigure[]{
\includegraphics[width=0.2\textwidth]{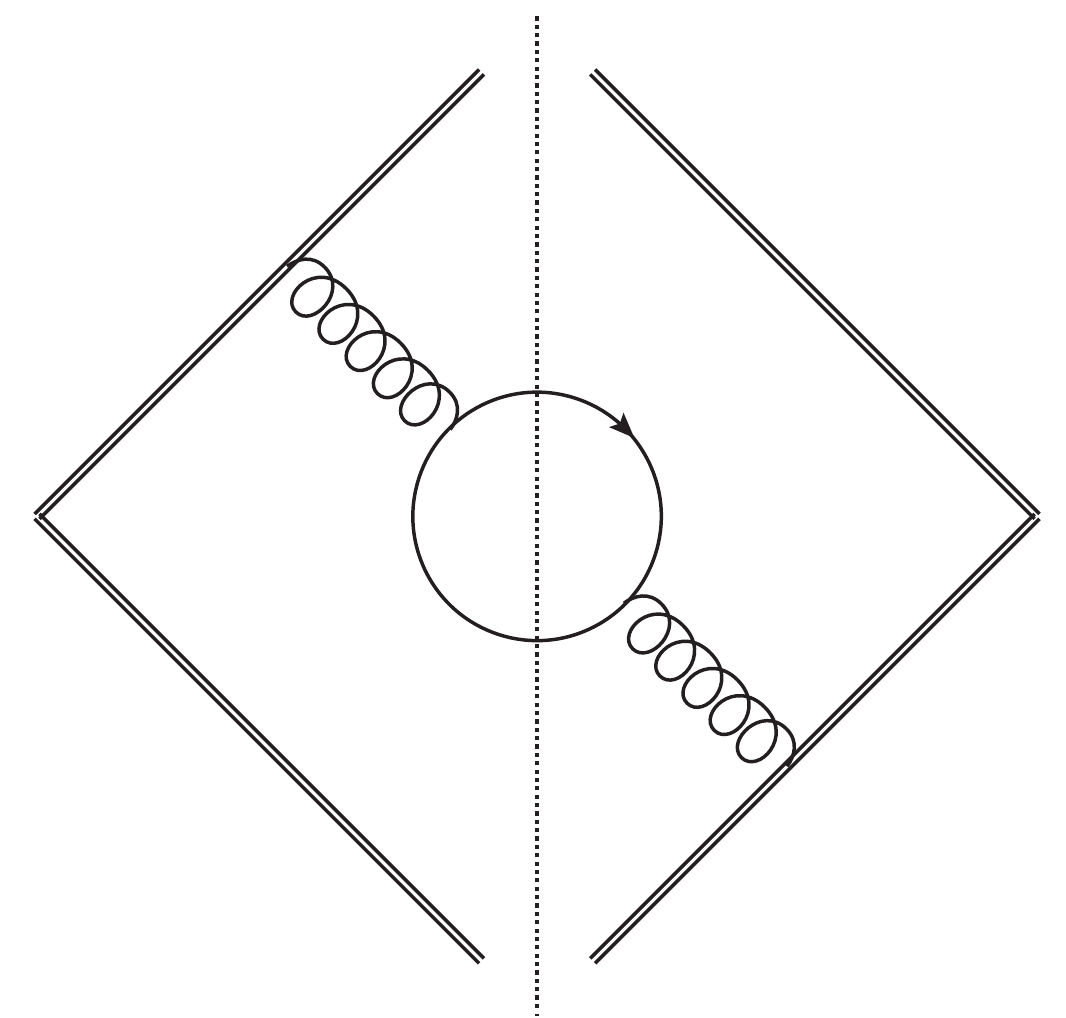}
\label{fig:SM_jetsub_gluons:np:triplecollineardiagrams4}
}
\\
\subfigure[]{
\includegraphics[width=0.2\textwidth]{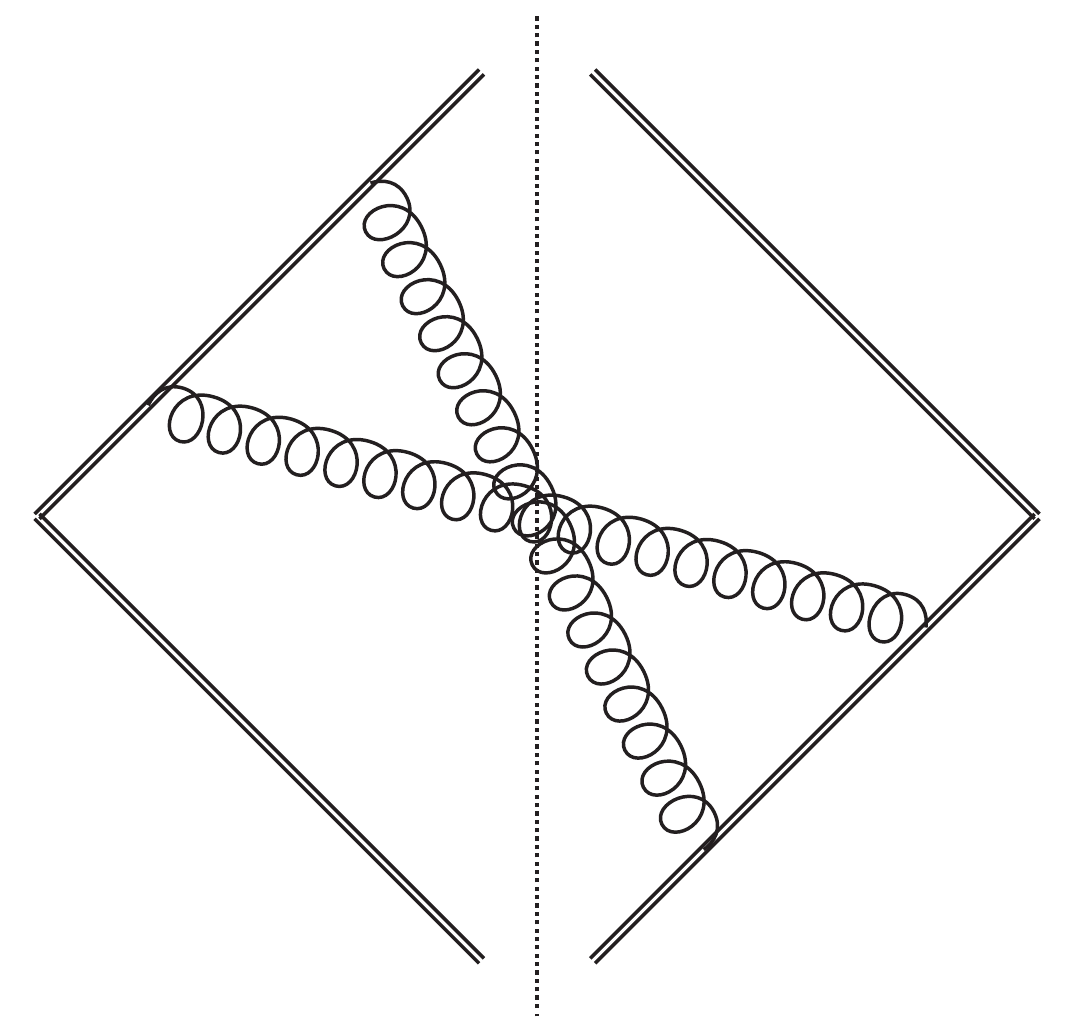}
\label{fig:SM_jetsub_gluons:np:triplecollineardiagrams5}
}
\subfigure[]{
\includegraphics[width=0.2\textwidth]{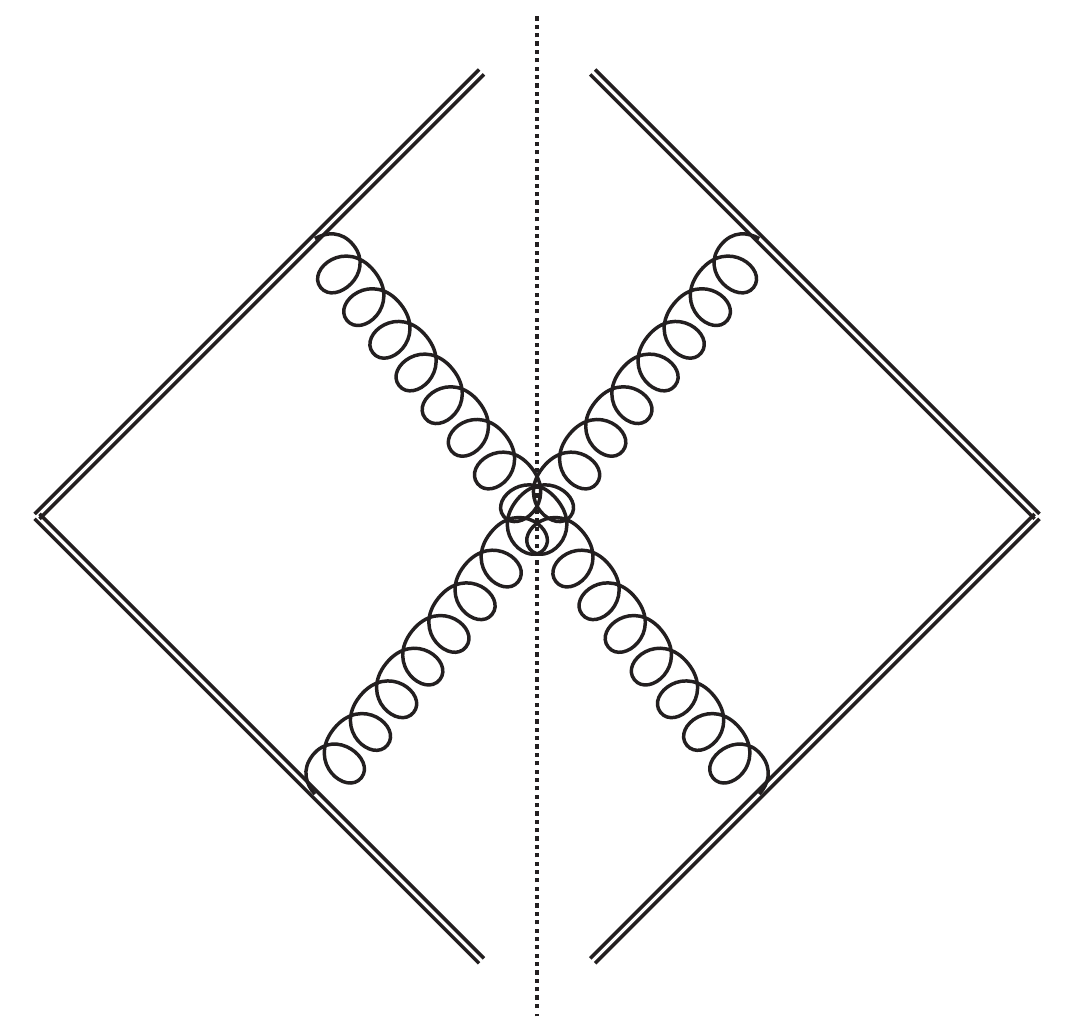}
\label{fig:SM_jetsub_gluons:np:triplecollineardiagrams6}
}
\subfigure[]{
\includegraphics[width=0.2\textwidth]{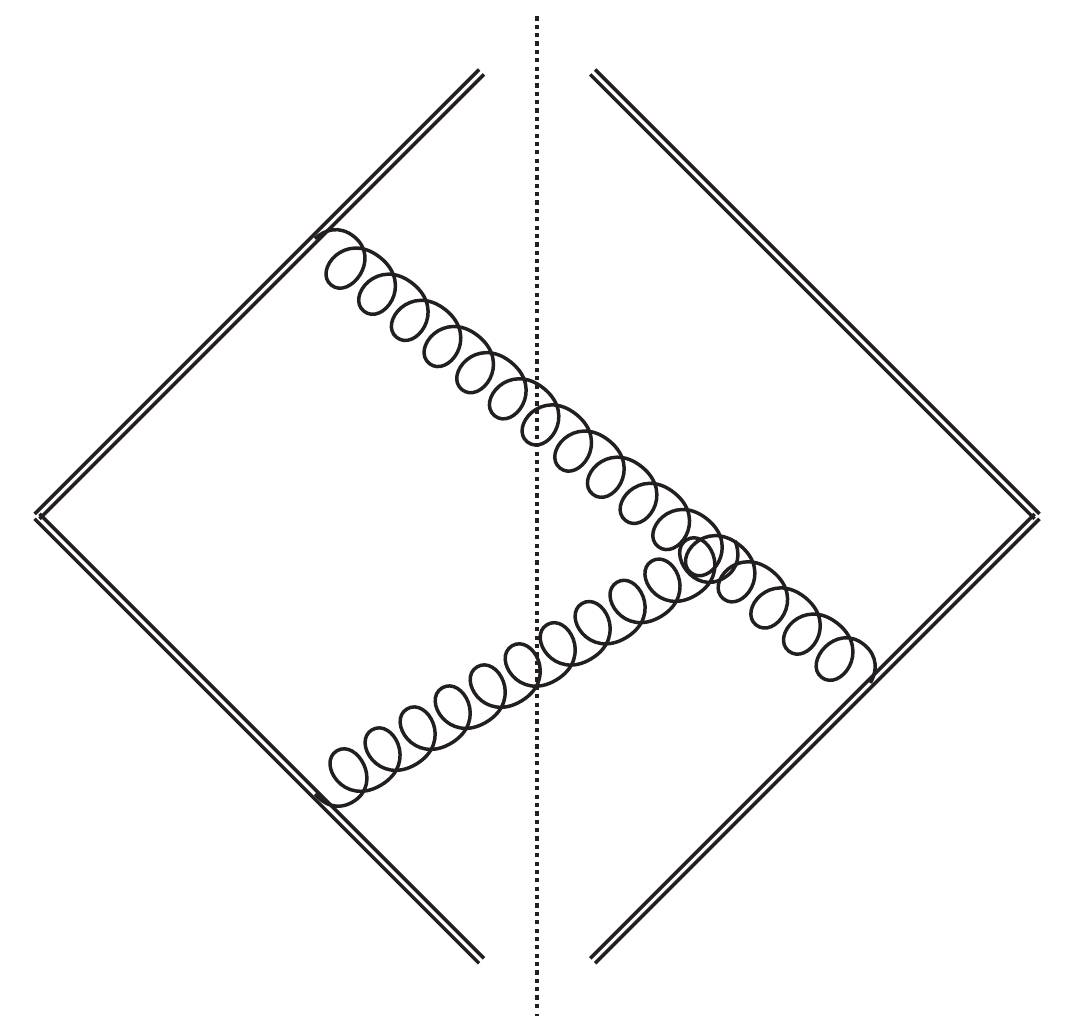}
\label{fig:SM_jetsub_gluons:np:triplecollineardiagrams7}
}
\subfigure[]{
\includegraphics[width=0.2\textwidth]{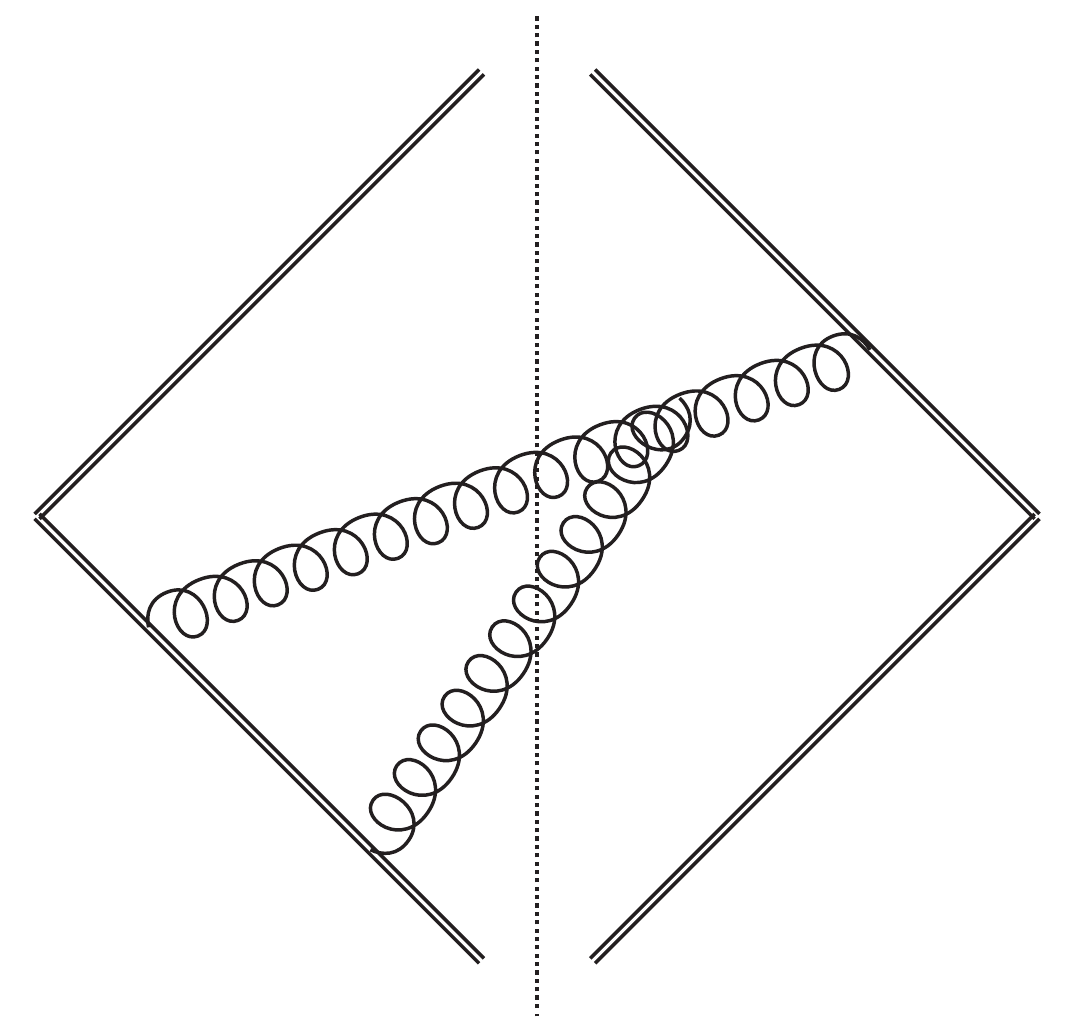}
\label{fig:SM_jetsub_gluons:np:triplecollineardiagrams8}
}
\caption{Examples of parton shower configurations required to go beyond leading order.}
\label{fig:SM_jetsub_gluons:np:triplecollineardiagrams}
\end{figure}

One way to improve the all-order behavior of the parton shower (point $2)$ 
above while preserving a systematically improvable state distribution (point $1)$ 
above is to consistently include higher-order and higher-multiplicity 
splitting functions in the parton 
shower~\cite{Li:2016yez, Hoche:2017iem,Dulat:2018vuy}. Some of the 
necessary ingredients at NLO are shown in 
Fig.~\ref{fig:SM_jetsub_gluons:np:triplecollineardiagrams}. 
At leading order, these configurations are approximated through the iterated 
application of leading-order splittings. This approximation however may
yield an incorrect distribution\footnote{\dots e.g.\ if the polarisation of 
the intermediate gluon in 
Figs.~\ref{fig:SM_jetsub_gluons:np:triplecollineardiagrams1}-\ref{fig:SM_jetsub_gluons:np:triplecollineardiagrams4},
\ref{fig:SM_jetsub_gluons:np:triplecollineardiagrams7}-\ref{fig:SM_jetsub_gluons:np:triplecollineardiagrams8}
is omitted, leading to an incorrect modulation of azimuthal angles,
or by disregarding the interference between
$C_F$-type and $C_A$-type color structures from 
Figs.~\ref{fig:SM_jetsub_gluons:np:triplecollineardiagrams7}-\ref{fig:SM_jetsub_gluons:np:triplecollineardiagrams8}.}
or may be limited to phase-space regions constrained by successive 
ordering requirements\footnote{\dots easily leading to the 
incorrect overall phase space volume, and thus failing to recover known anomalous
dimensions upon integration.}.
The correct final result is obtained by including the 
complete configurations in Fig.~\ref{fig:SM_jetsub_gluons:np:triplecollineardiagrams}
as new rates in the parton shower, and subtracting from these new rates
the leading-order result. This subtraction will, given a suitable definition
of the leading-order shower, act to ensure local finiteness of the new, 
subtracted, rates. We will call these subtracted rates ``NLO corrections".

In~\cite{Hoche:2017iem}, triple-collinear corrections from diagrams
similar to~\ref{fig:SM_jetsub_gluons:np:triplecollineardiagrams2} were considered, with
the difference that instead of the primary parton (indicated with a double 
line), one of the quarks in the loop was considered as ``hard". Such
configurations give rise, upon integration, to the flavor-changing 
DGLAP kernels $P_{qq'}$ and $P_{q\bar q}$. We will call this 
``triple-collinear" correction. The calculation of these corrections has
helped define a method to use the overlap with lowest order to construct 
locally finite splitting rates at NLO. Numerically however, this correction is 
expected to be small to modest.
The soft limits of all the diagrams in Fig.~\ref{fig:SM_jetsub_gluons:np:triplecollineardiagrams}
was considered in~\cite{Dulat:2018vuy}, which also included all necessary 
virtual corrections obtained by moving the cuts in the individual diagrams in 
all possible ways. We will call this ``double-soft" correction\footnote{
It should be noted that there is overlap between the
triple-collinear and double-soft limits. A complete differential calculation 
that consistently (i.e.\ without overlap) includes all components has yet
to be produced. Thus, we assess the potential to find observables
that discriminate between leading-order and next-to-leading order
results separately, for triple-collinear, and for double-soft corrections.}. 
The numerical effect of these double-soft corrections is expected to be appreciable.

In this study, we use the implementation of the triple-collinear and 
double-soft corrections in \textsc{Dire} to produce NLO pseudo-data, with the aim
of highlighting the characteristic new features of either correction.

Events are treated as sets of particles, with each particle $p_i$ specified by its momentum $\vec p_i^\mu$, mass, and particle-type.
The events are rotated to a consistent orientation by vertically aligning the second moment of the energy flow~\cite{Komiske:2019asc}.
This is accomplished by diagonalizing the spatial component of $\mathcal I^{\mu\nu} = \sum_{i=1}^M E_i v_i^\mu v_i^\nu$, where $v_i^\mu = p_i^\mu/E_i$ is the particle velocity.
As a machine learning architecture to process the entire events in their natural representation as sets of particles, we use Particle Flow Networks (PFNs)~\cite{Komiske:2018cqr} (see also Ref.~\cite{DBLP:conf/nips/ZaheerKRPSS17}).
Intuitively, PFNs learn a collection of additive observables which are processed by a fully-connected network.
A PFN acts on an event with $M$ particles $p_i$ as $\text{PFN}(\{p_i\}_{i=1}^M) = F\left(\sum_{i=1}^M \Phi(p_i)\right)$, where $F$ and $\Phi$ are parameterized by dense networks.
The network sizes of $F$ and $\Phi$ are identical to those in Ref.~\cite{Komiske:2018cqr}, with a latent space dimension of 256.
The train, validation, and test set sizes were 175k, 10k, and 15k, respectively.
The PFN classifiers were trained for 25 epochs with a batch size of 500.

\begin{figure}[t]
\centering
\includegraphics[width=0.85\textwidth]{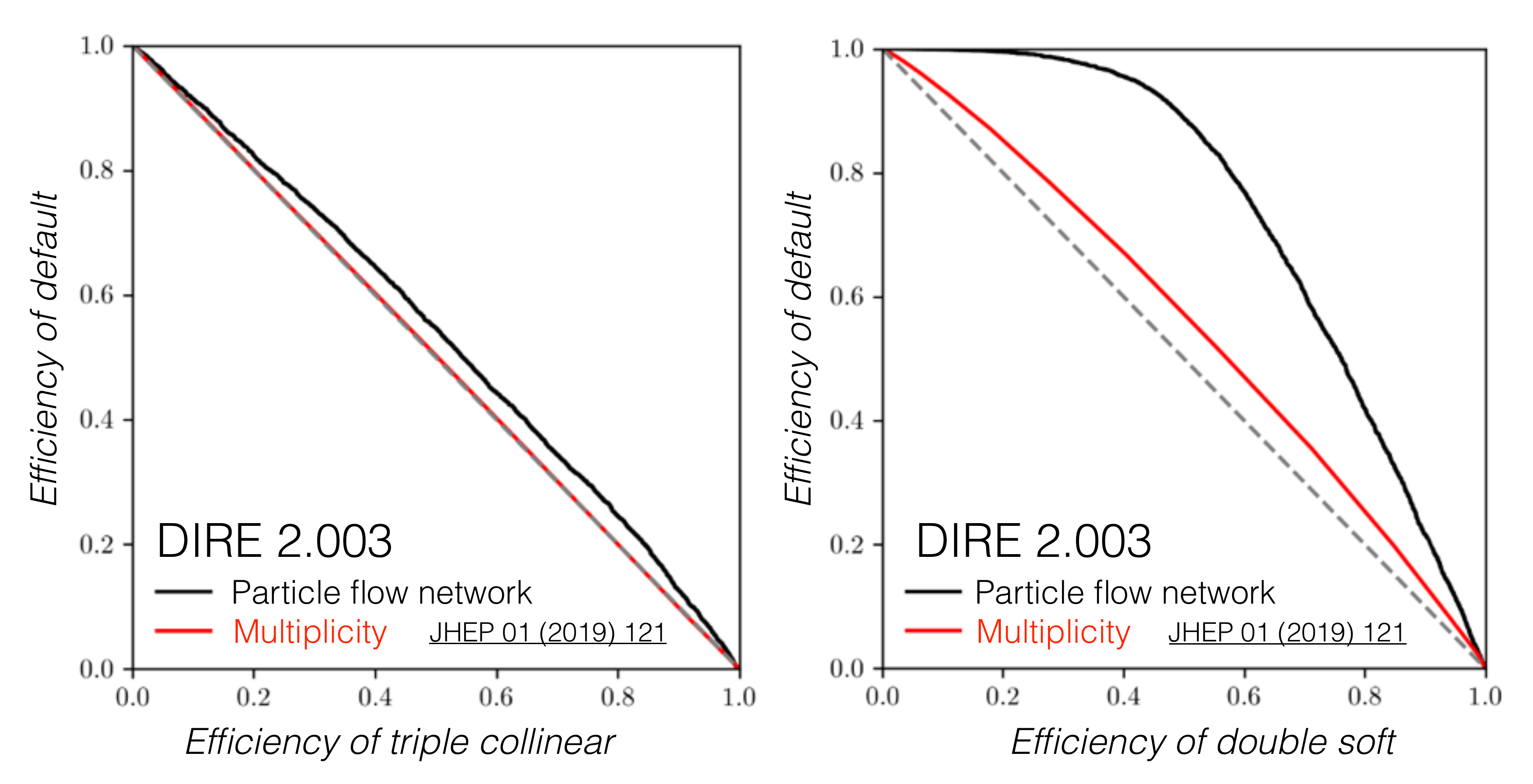}
\caption{Receiver operating characteristic (ROC) curves for pseudo-data with and without the triple collinear splitting functions (left) and with and without the double soft splitting functions (right).  The performance of a classifier using just the jet constituent multiplicity is compared with a deep neural network acting on the full observable jet phase space.  For reference, a classifier that cannot distinguish between the two models is depicted with a dashed line.  Better classifiers are up and to the right.}
\label{fig:SM_jetsub_gluons:np:triplecollinearNN}
\end{figure}

Receiver operating characteristic (ROC) curves from the machine learning classifiers are presented in Fig.~\ref{fig:SM_jetsub_gluons:np:triplecollinearNN}.  These curves show the performance of a classifier designed to distinguish the default simulation from one that includes either the triple collinear splitting function or the double soft splitting function.  A neural network is compared with a simple classifier that only uses the jet constituent multiplicity.  We find that the triple-collinear corrections (which integrate to
the DGLAP kernels $P_{qq'}$ and $P_{q\bar q}$) is difficult to pinpoint. 
It is somewhat surprising that the impact is almost vanishing.  Furthermore, we find that double-soft corrections have sizable impact, and can
easily be filtered out of the data. This is expected, since the theoretical
description of soft gluons changes significantly.  It is currently unclear what features the neural network is using to distinguish the default simulation from the one that includes the double soft splitting function.  Figure~\ref{fig:SM_jetsub_gluons:np:triplecollinearNN} indicates that the network is using more than just the jet constituent multiplicity.  Future studies will be required to identify a suitable observable to measure (perhaps the neural network itself).  

\subsection{q/g tagging in VBF and VBS}
\label{sec:SM_jetsub_gluons:vbsbvf}

Quark/gluon tagging is a key benchmark for jet substructure studies and has been extensively studied elsewhere (see e.g. Ref.~\cite{Badger:2016bpw,Gras:2017jty}).  Tagging quark jets in the context of VBF/VBS analyses has also been explored recently by CMS~\cite{Khachatryan:2015bnx}.  With recent advances in q/g tagging and with upcoming detector upgrades to extend q/g tagging capabilities in the forward regions of ATLAS and CMS, it is prudent to revisit this important topic for VBF/VBS analyses.  In general, this task has two aspects: (1) using q/g tagging to distinguish electroweak signals from continuum QCD backgrounds~\cite{Chatrchyan:2013jya,Khachatryan:2014dea,Sirunyan:2017jej,Sirunyan:2019dyi} and (2) using these techniques to differentiate signal production mechanisms.
In particular, VBF Higgs production can have a similar phenomenology to gluon-gluon fusion Higgs production (ggH) produced in association with two jets.  For various global fits, it is important to be able to statistically differentiate the various production modes.

A further complication to q/g studies in general is that usually other analyses selections are optimized first and then q/g tagging is applied near the end of a selection chain.  This can make the use of q/g tagging suboptimal and one may gain from relaxing other traditional requirements (e.g. $m_{jj}$ or $\Delta\eta_{jj}$) while tightening the q/g tagging selection.  At Les Houches 2019, this interplay was investigated in the context of separating ggH from VBF Higgs production.

The simulated samples of Higgs boson events are generated using the \textsc{Powheg} program for both ggH \cite{Campbell:2012am} and VBF \cite{Nason:2009ai} at NLO accuracy. To properly simulate the recoil of the final state particles caused by additional QCD radiation, \textsc{Powheg} generator is interfaced with the \textsc{Pythia} program \cite{Sjostrand:2014zea}. The Higgs boson is left undecayed and its mass is set at $M_H=125~\rm GeV$. The Parton Distribution Functions (PDF) used in this note are NNPDF3.0 \cite{Ball:2017nwa} for both the matrix element and parton shower simulations. The final states particles are then clustered into anti-$k_T$\cite{Cacciari:2008gp} jets of radius $R = 0.4$ using FastJet3 \cite{Cacciari:2011ma}. All jet constituents are required to have $|\eta| < 4.7$ and $p_T > 1\rm ~GeV$. Events with at least two jets with $p_T> 30\rm ~GeV$ and $|\eta|<4.7$ are kept.

The selected events are then used to train boosted decision trees using the \textsc{XGBoost} library~\cite{Chen:2016btl}. Two sets of jet-substructure variables, defined below, are computed to increase effectiveness in distinguishing VBF from ggF:

\begin{itemize}
    \item Jet angularities \cite{Larkoski:2014pca} can be written as
    \begin{equation}
        \lambda_{\alpha} = \sum_{i \in \rm jet} z_i \theta_{i}^\alpha
    \end{equation}
    where $z_i$ is the momentum fraction of jet constituent i, and $\theta_i$ is the normalised rapidity-azimuth angle to the jet axis\footnote{$z_i = p_{T,i}/\sum_{i\in \rm jet} p_{T,i}$, $\theta_i = \Delta R_i/R$ where R is the jet radius and $\Delta R_i$ is the rapidity-azimuth distance from constituent $i$ to the jet axis.}.
    The Energy Correlation Functions (EFC) \cite{Moult:2016cvt} variables are defined by
    \begin{equation}
        C_{\alpha} = \sum_{i < j} z_i z_j \theta_{ij}^\alpha
    \end{equation}
    $\alpha$ is set to $0.5$, $1$ and $2$ for both $C_{\alpha}$ and $\lambda_{\alpha}$. It easy to note that $\lambda_1$ represent the width and $\lambda_2$ the mass of the jet.
    \item Jet multiplicity variables, which is the count of number of charged  tracks (within $|\eta|<2.5$) and the total number of particles within a jet.
\end{itemize}

In addition to jet-substructure variables, event-level kinematic variables are defined using the VBF tagging jets in order to isolate VBF event from other SM processes. These are listed below
\begin{itemize}
    \item Invariant mass $m_{jj}$, $\Delta\eta_{jj}$ and $\Delta\phi_{jj}$ separation of the two jets
    \item Transverse momentum of the two jets
\end{itemize}

The variables defined above are combined in three separate \textsc{XGBoost} models. The first model uses kinematic variables only. The second uses jet constituent multiplicities alongside the kinematic variables. The third and final model uses all the previously mentioned variables while adding jet substructure angularities. Every model is trained\footnote{\textsc{XGBoost} is used with the default training parameters.} in 3 different kinematic bins, $0\leq m_{jj}<350~\rm GeV$, $350\leq m_{jj} <700~\rm GeV$ and $m_{jj}\geq 700~\rm GeV$.

\begin{figure}[h!]
  \centering
\includegraphics[width=0.3\textwidth]{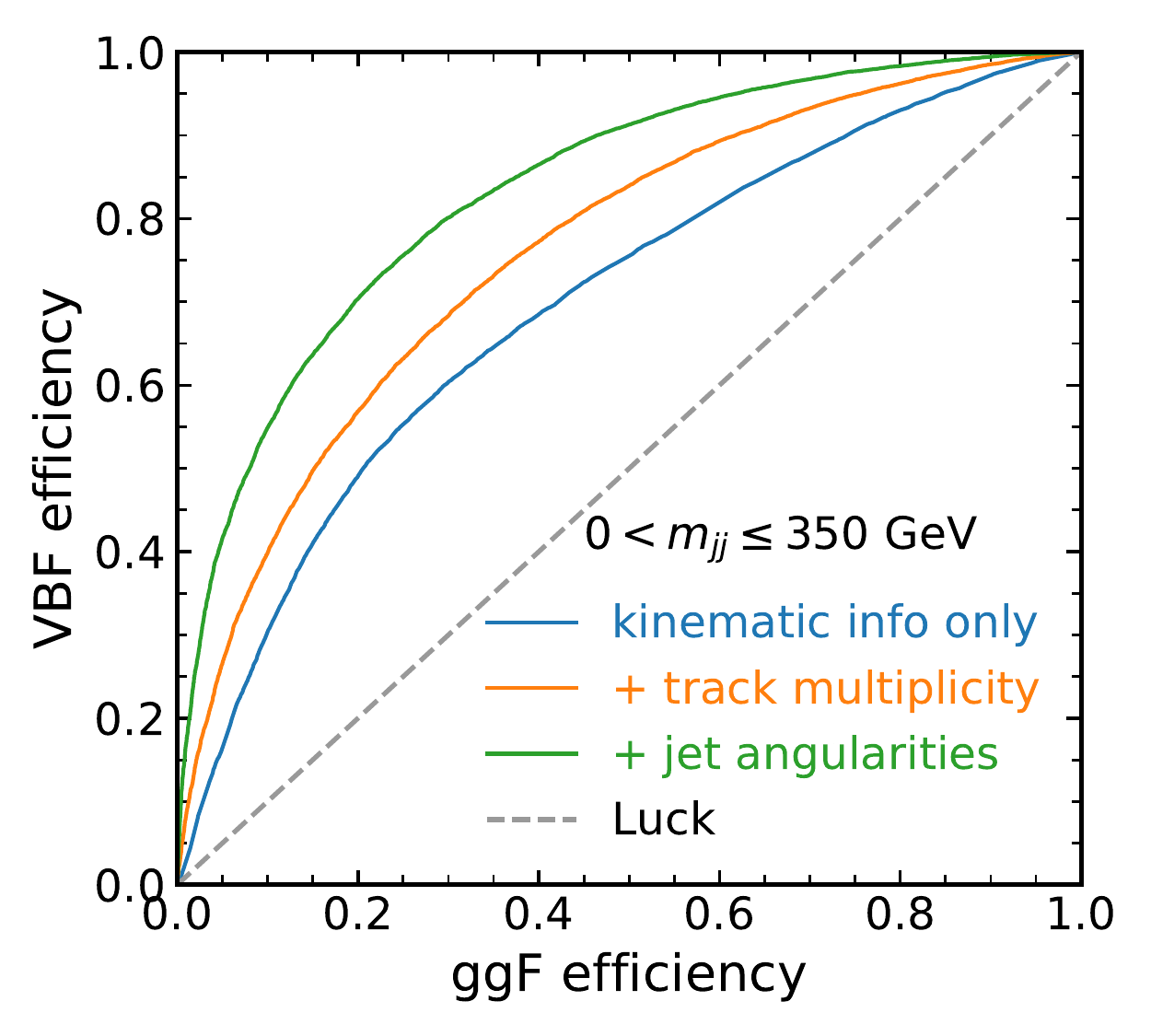}
\includegraphics[width=0.3\textwidth]{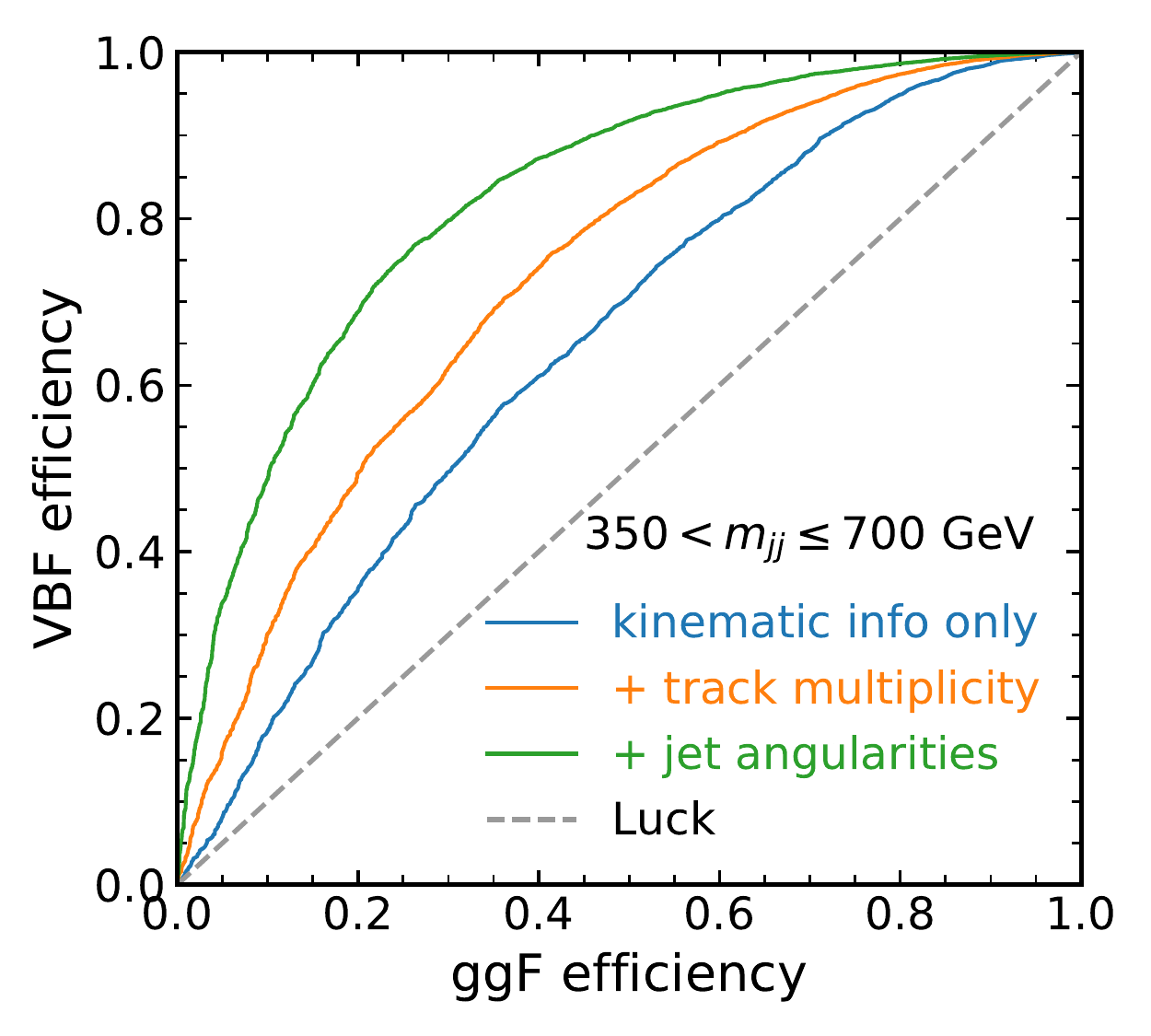}
\includegraphics[width=0.3\textwidth]{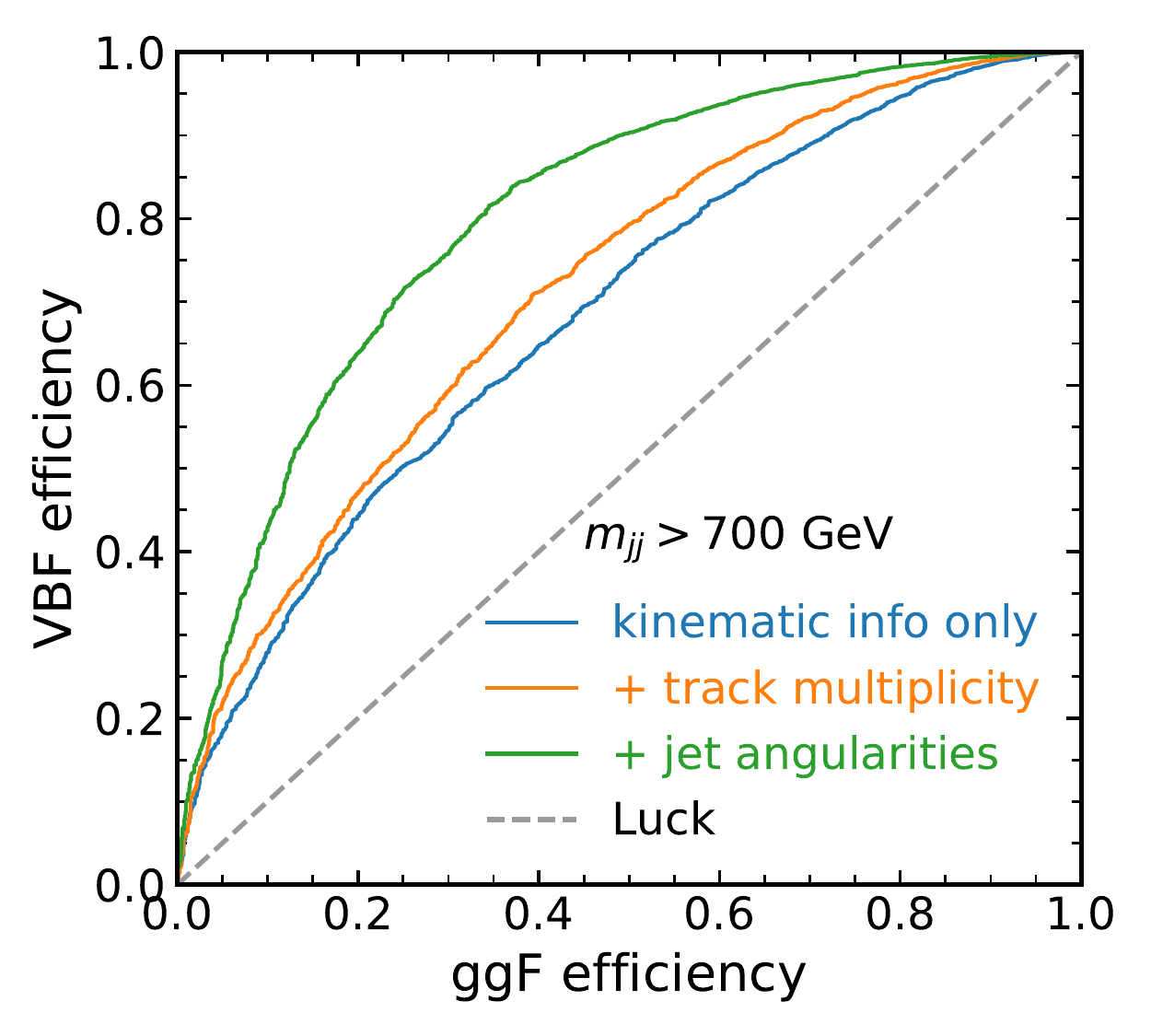}
  \caption{ROC curves for separating VBF Higgs production from ggH Higgs production.  The three plots are distinguished by their $m_{jj}$ values, indicated above.  The three colored lines in each plot correspond to ROC curves for various levels of information used in the classifiers: jet kinematic information only (blue), additionally including jet constituent track multiplicity (orange), and also including a suite of angularities for the two jets (green).}
  \label{fig:jets:qg}
\end{figure}
The results are summarised in Fig.~\ref{fig:jets:qg}.  Since tracking information is only available in the central region, various levels of information were used to train classifiers.  Higher $m_{jj}$ requirements bias the jet $\eta$ distribution to be more forward and thus there was less performance gain from adding track observables.  The interplay with $m_{jj}$ is non-trivial and so gains may be possible by considering a simultaneous optimisation.

The interplay between q/g tagging and other event selections is an important area of research for further study in the future.  This includes both the correlation between q/g tagging and event kinematic features as well as with other jet substructure observables such as subjet q/g tagging and $\tau_{21}$ and jet mass.

\subsection{\textsc{Squirrel} for the Gluon PDF}
\label{sec:SM_jetsub_gluons:pdf}

Parton Distribution Functions (PDFs) describe the non-perturbative dynamics of quarks and gluons in the protons that take part in high-energy collisions. Therefore, they are a key ingredient for every theoretical prediction that aims to describe particle interactions at high-energy colliders such as the LHC. As a consequence, their precise determination is of utmost importance for LHC phenomenology. 
The non-perturbative nature of PDFs hampers their determination from first principles.
However, for inclusive enough processes, they are universal, i.e.\, up to power corrections, they do not depend on the particular process, and they can be determined by fitting data from previous experiments. Moreover, although they are themselves non-perturbative objects, their dependence on the energy is governed by the DGLAP equation and the evolution kernels can be computed as a power expansion in the strong coupling. This implies that data collected at past experiments, at different energies, can be used to constrain PDFs. 

Traditionally, the main source of uncertainties assigned to the determination of PDFs arises from the experimental error of the data that enter the fit.\footnote{Very recently, the inclusion of theory uncertainties in PDF determination has also been achieved~\cite{Harland-Lang:2018bxd,AbdulKhalek:2019ihb,AbdulKhalek:2019bux}.} In extreme regions of phase-space, for instance at small- or large-$x$, the experimental uncertainties typically deteriorate and one has to face a reduced number of data points. This is reflected in PDFs which are largely unconstrained in these regions. 
For instance, the large PDF uncertainty in the $x\to 1$ region, also known as the threshold region, has a negative impact on searches for new and heavy states.
 Although this will probably not wash out a potential discovery, it will definitely obscure the nature and the properties of the new state, such as its mass and its couplings. 
The way to reduce this PDF uncertainty is to include in the fit data at larger $x$. However, this raises interesting theoretical issues, because fixed-order perturbation theory becomes less reliable as $x$ becomes close to unity and one should supplement theoretical predictions with threshold resummation, as studied for instance in~\cite{Corcella:2005us,Sato:2013wea,Westmark:2013vea,Bonvini:2015ira,Accardi:2014qda}.

In this study, we focus on the gluon PDF in the region of relatively large longitudinal momentum fraction, $x\sim 10^{-1}$. The datasets that mostly constrain the gluon in this region are the inclusive jet spectra, in the region of the jet transverse momentum above 1~TeV and the production of top quark pairs. From a theoretical point of view, both processes are known to very high accuracy, i.e. next-to-next-to-leading order (NNLO)~\cite{Czakon:2015owf,Currie:2016bfm}. Phenomenologically, the two processes have pros and cons. Inclusive jet production features high statistics across a wide kinematical range and, consequently, even in the high $p_T$ region we are interested the experimental uncertainties do not exceed 10\%. However, because one measures inclusive jets, one cannot distinguish the flavour content and the cross section is dominated by quark-quark scattering, which bears little information about the gluon PDF. 
On the other hand, at LHC energies, top pair production is dominated by gluon fusion and therefore offers a direct probe of the gluon luminosity. In this case, however, we pay a much higher price in terms of experimental uncertainties, essentially because we run out of statistics for values of the top transverse momentum much smaller than what is reached in the case of inclusive jets. 
Ideally, we would like to exploit the vast jet samples collected by the LHC experiments to tease out more information about the gluon PDF. We immediately realise that one way of achieving this scope would be to supplement the inclusive jet $p_T$ spectrum with some information about the jet flavour. 
Therefore, in this section, we are going to explore the possibility of using the \emph{inclusive gluon-jet $p_T$} spectrum to extract parton densities, rather than its flavour-blind version. Properly defining quark jets versus gluon jets is a very active area of jet substructure (for a review, see for instance~\cite{Marzani:2019hun}) and indeed it was one of the focus of a past edition of the Les Houches proceedings~\cite{Badger:2016bpw} (see also the follow-up study~\cite{Gras:2017jty}).

\begin{figure}[t]
\begin{center}
\includegraphics[width=0.49\textwidth, page=9]{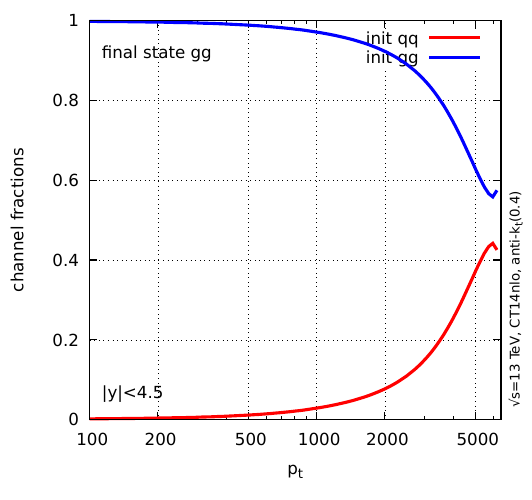} \hfill
\includegraphics[width=0.49\textwidth, page=10]{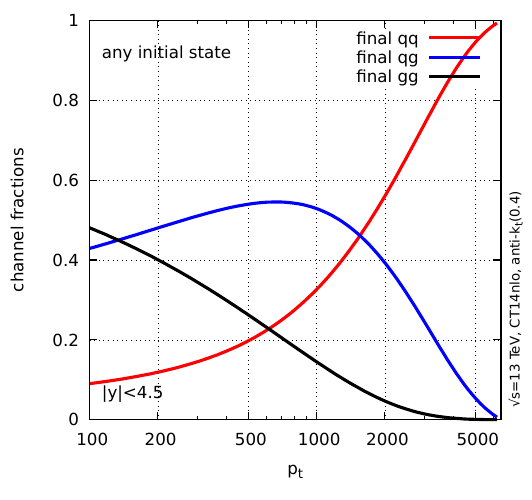}
\caption{Born-level studies of the flavour composition of dijet events at $\sqrt{s}=13$~TeV, as a function of the jet transverse momentum. The plot on the left shows the fractions of quark-initiated and gluon-initiated processes that contribute to a $gg$ final state. The plot of the right instead shows the fractional composition of the final state for any initial state.}
\label{fig:born_studies} 
\end{center}
\end{figure}

Before discussing how we can sensibly attach a flavour tag to a jet, let us perform a zeroth order test of this idea. Let us assume that we can indeed tag a gluon jet in the final state. Then, the obvious question we should ask ourselves is how strongly the flavour of the final state, which we measure, is correlated with the flavour of the initial state, which intimately related to the parton densities we want to study. 
We can easily assess this correlation at Born level by explicitly considering $2 \to 2$ parton scattering and focussing on the two-gluon ($gg$) final state. The left-hand plot of Fig.~\ref{fig:born_studies} shows the fraction of the $gg$ final state that originates from quark-anti-quark initial state ($q \bar q \to gg$) in red and the one from gluon-gluon initial state ($gg \to gg$)  in blue, as a function of the final-state transverse momentum for proton-proton collisions at $\sqrt{s}=13$~TeV (the plot uses the NLO PDF set CT14~\cite{Dulat:2015mca}). 
The result of this very first study is rather encouraging: in the region $p_T=1-2$~TeV we are interested in, there is indeed very strong correlation between the initial- and final-state flavours. This is, of course, only a Born-level study and we can reasonably expect this correlation to deteriorate at higher-orders mostly due to wide-angle radiation. 
Although a quantitative estimate of these effects goes beyond the scope of these proceedings, we do not expect them to be dramatic. In any case, one could in principle reduce such contributions with jet grooming. 
With the same Born-level setup, we can study how the different partonic final-states contribute to the inclusive cross section. This is shown on the right-hand plot of Fig.~\ref{fig:born_studies}. As $p_T$ increases, the fraction of final-state quarks rapidly increases. Indeed, in the region of interest, the $gg$ final state represents less than 10\% of the inclusive sample. This makes the enterprise of enhancing the $gg$ contributions (or, equivalently, suppressing the quarks) particularly challenging.

\begin{figure}[t]
\begin{center}
\includegraphics[width=0.42\textwidth, page=1]{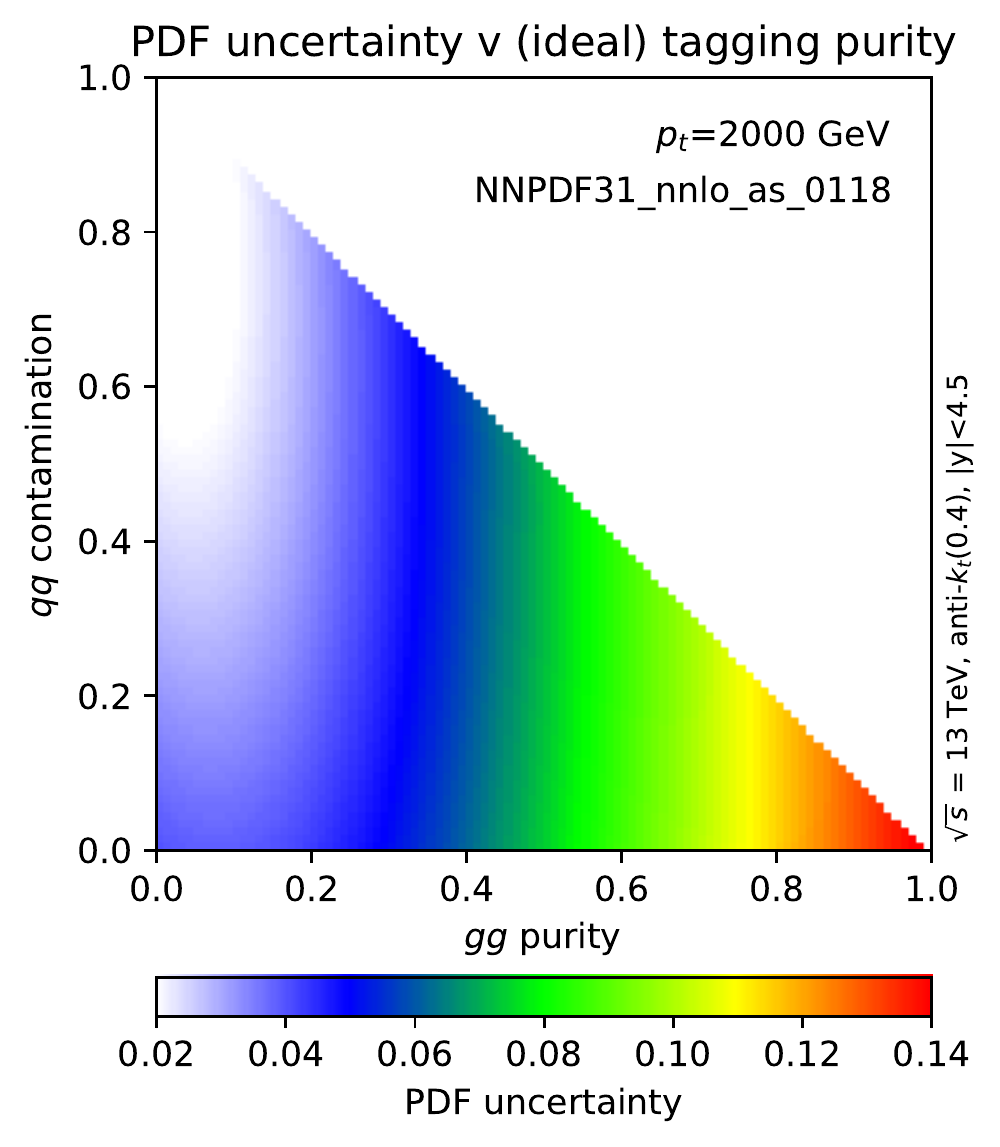} \hfill
\includegraphics[width=0.49\textwidth, page=2]{figures/performance-plots.pdf}
\caption{The plot on the left shows the PDF uncertainty, evaluated using the NNLO set from NNPDF3.1 as a function of $gg$ purity and $qq$ contamination, as defined in the text.
The plot on the right shows how the PDF uncertainty compared to experimental systematic and statistical uncertainties, as a function of the $gg$ purity, assuming that  gluon jets have been identified using a tagger with efficiency $\varepsilon_g=0.42$.}
\label{fig:pdf_unc_studies} 
\end{center}
\end{figure}

The next step in our study is to evaluate the current PDF uncertainties, as a function of the final-state flavour composition. 
In order to do so, we imagine as a fist step to have at our disposal an idealised tagging procedure that allows us to freely enhance or depress the different partonic components of the final state. We will come back to actual realisations of this tagger later. 
In this context, we find useful to define the gluon-gluon ($gg$) purity as
\begin{equation}\label{gg-purity}
gg \, \text{purity}= \frac{\sigma_{gg}}{\sigma_{qq}+\sigma_{qg}+\sigma_{gg}},
\end{equation}
where $\sigma_{ij}$ is the cross section for producing parton $i$ and $j$, evaluated at Born level.  In an analogous way, we can also define the $q q$ contamination, while $q g$ is then fixed by unitarity. 
For instance, we already know from Fig.~\ref{fig:born_studies}, that the inclusive, i.e.\ untagged, case correspond to $gg$ purity of approximately 5\% at $p_T=2$~TeV, while $qq$ and $qg$ makes up roughly 55\% and 40\% of the inclusive sample, respectively. 

The PDF uncertainty on the jet cross section, for given values of $gg$ purity and $qq$ is shown in Fig.~\ref{fig:pdf_unc_studies}, on the left. The plot is obtained using the NNLO PDF set from NNPDF3.1~\cite{Ball:2017nwa} for jets at $2$~TeV.
 %
%
%
From the plot, we see that PDF uncertainty for 5\% $gg$ purity and 55\% $qq$ contamination, which roughly corresponds to the inclusive, i.e.\ untagged, jet cross section at 2~TeV, then is about a few percent.
This reflects the fact that the quark parton densities are fairly-well constrained in the region of interest. 
As we move to higher values of the $gg$ purity, the less constrained gluon PDF starts to play a more significant role and, as a consequence, the overall uncertainty goes up. For instance, if we were able to devise a tagger that purifies the $gg$ final state to 80\%, we would increase the PDF uncertainty from 2\% to 12\%. 

We can now attempt to assess how good a tagger we should devise in order for the gluon-jet $p_t$ spectrum to be able to constrain the gluon PDF at relatively large $x$. For this purpose, we would like to achieve a situation where the PDF uncertainty is the largest uncertainty, i.e.\ it dominates over the other theoretical and experimental uncertainties. For this feasibility study, we have decided to neglect uncertainties related to the tagging procedure, which can be evaluated, for a given algorithm, using standard scale-variation based, methods. Instead, we concentrate on experimental systematic and statistical uncertainties. These uncertainties are shown on the plot in Fig.~\ref{fig:pdf_unc_studies}, on the right, as function fo the $gg$ purity, for a (yet-to-be-defined) tagging procedure that works at $\varepsilon_g=0.42$ gluon efficiency\footnote{Instead of fixing the gluon efficiency, we could have specified the $qq$ contamination, which directly corresponds to an horizontal slice of the left-hand plot of Fig.~\ref{fig:pdf_unc_studies}. However, in view of the discussion about taggers that will follow, we find the efficiency more informative.}. 
The experimental systematic uncertainty (in dotted red)  is assumed to be a half of the one reported by the LHC collaboration in 2015~\cite{Aaboud:2017jcu}, i.e.\ approximately 5\%. The statistical uncertainty (in dashed green) correspond to an integrated luminosity of 300 fb$^{-1}$, i.e.\ roughly the amount of date collected by the end of Run~III of the LHC.
We conclude that the PDF uncertainty becomes the dominant one if the $gg$ purity is above 0.3, for $\varepsilon_g=0.42$. Thus, this sets the goal for our tagger.

\begin{figure}[t]
\begin{center}
\includegraphics[width=0.49\textwidth, page=4]{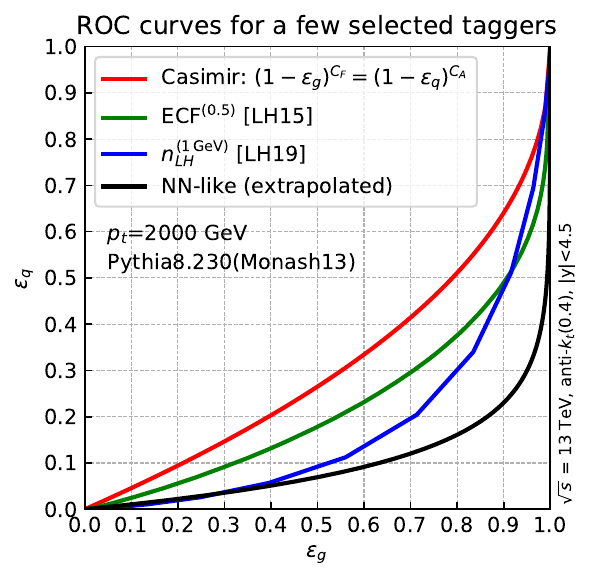} \hfill
\includegraphics[width=0.49\textwidth, page=5]{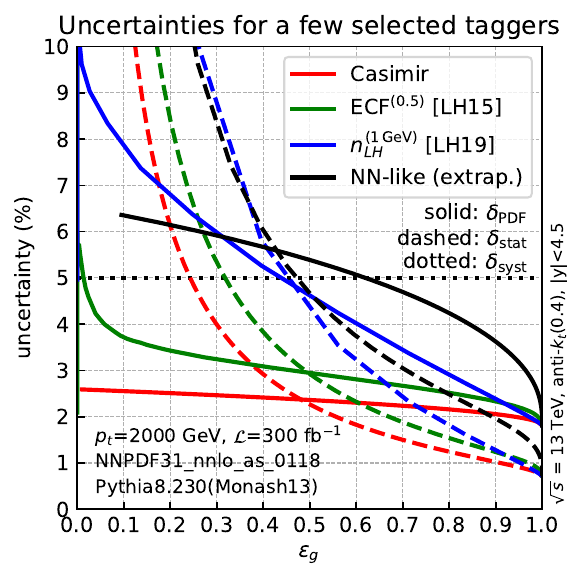}
\caption{The plot on the left show the tagging performance of the Les Houches energy correlation function and Les Houches multiplicity discussed in this study. 
The ROC curve are obtained using a numerical simulation with the Monte Carlo parton shower \textsc{Pythia}~8.230~\cite{Sjostrand:2014zea}, with the Monash13 tune~\cite{Skands:2014pea} and they are compared to Casimir scaling and the extrapolated performance of a neural-network based tagger that makes use of jet topics~\cite{Metodiev:2018ftz,Komiske:2018vkc}.
The plot on the right show, for each of the tagger (or idealised tagger) we have have considered here, the different sources of uncertainties (namely, PDF, statistical and systematic uncertainties, as a function of the gluon efficiency. 
}
\label{fig:performance_studies} 
\end{center}
\end{figure}
A variety of techniques to define and discriminate quark-initiated versus gluon-initiated jets have been proposed and studied in the literature. It is customary to express a tagger performance in terms of ROC curves, i.e. plots that exhibits the algorithm ability of identify the signal, i.e.\ its efficiency, versus its mis-tag rate. ROC curves for a handful of quark/gluon taggers are shown in Fig.~\ref{fig:performance_studies}, on the left.
The plot shows the signal (gluon) efficiency $\varepsilon_g$ on the horizontal axis and the background (quark) efficiency ($\varepsilon_q$) on the vertical axis. The red-line can be take as the reference and it corresponds to so-called Casimir scaling, and it is related to the universal scaling between the colour factor of the fundamental ($C_F$) and adjoint ($C_A$) representations~\cite{Larkoski:2014pca}. 

Because of the different colour factors characterising quark and gluon radiation, gluons tend to radiate more than quarks. Jet shapes such as generalised angularities~\cite{Larkoski:2014pca} and energy-correlation functions (EFCs)~\cite{Larkoski:2013eya} are a probe of such radiation and therefore by selecting jets which exhibit values of the jet shape above a certain threshold, we can enrich our gluon-jet sample. 
Furthermore, jet shapes are fairly-well understood observables and precision-calculations exploiting both fixed-order and resummed perturbation theory are possible, thus systematic reduction of the tagger theoretical uncertainties is, in principle possible. Following the Les Houches studies performed in 2015, the best quark/gluon separation is achieved for the so-called Les Houches ECF, which is characterised by an angular exponent $\alpha=0.5$.
The ROC curve for this tagger is shown in green on the left-hand plot of  Fig.~\ref{fig:performance_studies}. We note that despite the fact that jet shapes exhibit Casimir scaling at their lowest order (leading logarithmic accuracy) their discriminating power is increased if higher-order effects are included.
The curve has been obtained using a numerical simulation with the Monte Carlo parton shower \textsc{Pythia}~8.230~\cite{Sjostrand:2014zea}, with the Monash13 tune~\cite{Skands:2014pea}.

Given the above consideration, it is natural to wonder if it is possible to find substructure tools
which have a different behaviour already at leading-logarithmic
accuracy.
It is well known that counting observables, such as the particle multiplicity in a jet, or the charged-track multiplicity, typically outperform jet shapes as gluon taggers. However, these observables are not infra-red and collinear (IRC) safe. Instead we would like to employ  a counting observable that, unlike the aforementioned multiplicities exhibits IRC safety and therefore can be calculated using perturbation theory. This requirement is particularly important the context we are discussing as one would have to provide a theoretical calculation for a fit of parton densities. 
An observable that satisfies all these properties is the Iterated SoftDrop (ISD) multiplicity, which was introduced in Ref.~\cite{Frye:2017yrw}.
This algorithm applies the SoftDrop procedure~\cite{Larkoski:2014wba} multiple
times, following the hardest branch in
the recursion procedure. This gives a list of
branchings which pass the SoftDrop condition
$(z_1,\theta_1), \dots, (z_n,\theta_n)$. The multiplicity is simply the number of such branchings.
It was immediately noticed that for the Iterated SoftDrop multiplicity to IRC safe, one needs either to take a negative value of the SoftDrop angular exponent $\beta$ or impose an explicit cut on the angular separation  $\theta_\text{cut}$.
For this study we employ a variant of the SoftDrop iterated multiplicity which is built imposing a minimum relative transverse momentum cut ($k_t=1$~GeV), rather than an angular one. We name this variant of the Iterated SoftDrop multiplicity, the Les Houches multiplicity ($n_\text{LH}$).
The ROC curve for this tagger is shown in blue on the left-hand plot of  Fig.~\ref{fig:performance_studies}. We notice the gain in the performance, while maintaining full calculability. 
Finally, the ROC curve shown in black corresponds to an extrapolation of the behaviour obtained with a neural-network (NN) architecture exploiting jet topics~\cite{Metodiev:2018ftz,Komiske:2018vkc}. This idea originates from techniques employed in text-classification and, as the plot shows, outperforms the Les Houches multiplicity at high gluon efficiencies $\varepsilon_g > 0.5$. Measurements of jet topics have already been performed~\cite{Aad:2019onw}, however their theoretical understanding is still in their infancy and whether one can perturbatively predict their behaviour is still work in progress.

For each tagger we want to study, we can now pick an efficiency working point $\varepsilon_g$. Then, the corresponding ROC curve will give us the corresponding miss-tag rate $\varepsilon_q$ and with these two inputs we can estimate a realistic $gg$ purity using Eq.~(\ref{gg-purity});
\begin{equation}\label{gg-purity-after-tagging}
gg \, \text{purity}\Big|_\text{after tagging}= \frac{\sigma_{gg} \varepsilon_g^2}{\sigma_{qq}\varepsilon_q^2+\sigma_{qg}\varepsilon_q \varepsilon_g +\sigma_{gg} \varepsilon_g^2},
\end{equation}
and analogously for $qq$ and $qg$. With this information, we are now ready to compile the final plot of this study, which is shown in Fig.~\ref{fig:performance_studies}, on the right. 
This plot is similar in spirit to the right-hand plot of Fig.~\ref{fig:pdf_unc_studies} but now for actual quark/gluon taggers, rather than an idealised one.
As before, we show the different uncertainties: from PDFs ($\delta_\text{PDF}$, solid), statistical uncertainty ($\delta_\text{stat}$, dashed) and systematic one ($\delta_\text{syst}$, dotted). As discussed before, the systematic uncertainty is taken to be constant and equal to 5\%. The statistical uncertainty instead is the square root of the inverse number of the events, and so it  depends on $\varepsilon_g$ (and $\varepsilon_q$). For this, an integrated luminosity of 300~fb$^{-1}$ is assumed. Finally,  the PDF uncertainty of jet cross-section for $p_T>2$~TeV, after tagging is evaluated with NNPDF3.1, as a function of the tagger efficiency. 
The plot shows the different uncertainties $\delta_i$ for the taggers mentioned before: a tagger exhibiting simple Casimir scaling (red curves), the Les Houches ECFs (green curves), the Les Houches multiplicity (blue curves) and the neural-network tagger (black curves). The systematic uncertainty is assumed to be the same for each tagger. 
The plot shows that for jets with transverse momentum above 2~TeV a pure Casimir-scaling tagger or a ECF-based tagger are never good enough to enrich the final-state gluon content such that $\delta_\text{PDF}> \delta_\text{syst}, \delta_{stat}$.
Instead, if we pick $\epsilon_g\simeq 0.5$ the $n_\text{LH}$ tagger and the NN one do provide $\delta_\text{PDF}$ which are comparable, if not definitely bigger, then statistical and systematic uncertainties. This is definitely more marked for the NN tagger but, as mentioned before, it is currently unclear how to perform perturbative calculations for this. On the other hand, the jet transverse momentum distribution with a cut on $n_\text{LH}$ is well-defined and calculable in perturbation theory. 
However, in order to reach firmer conclusions of about the ability of these taggers to effectively discriminate between quark-like and gluon-like jets, we would have to add to this study an assessment of the tagging uncertainties. In a Monte Carlo study, this can be estimated by looking at ROC curves obtained with different Monte Carlo event generators, while in the case of an analytic study one can vary the perturbative scales. We leave such studies for future work.

\subsection{The highest energy gluons at the LHC}
\label{sec:SM_jetsub_gluons:highest}

In addition to their use for precision QCD, high $p_T$ gluons may also be a powerful tool for BSM searches.  In particular, Fig.~\ref{fig:born_studies} showed that the fraction of gluon-gluon final states decreases rapidly near the kinematic limit.  Suppose that there is a new particle at high invariant mass which decays to gluons.  Such particles are present in many BSM models, including a significant fraction of those that were proposed to explain the early Run 2 diphoton excess~\cite{Khachatryan:2016hje,Aaboud:2016tru}.  While the jets from this signal would be gluons, the dominant background at high $m_{jj}$ is from quark jets.  Therefore, a powerful gluon tagger may be able to significantly improve the sensitivity of a search to these di-gluon resonances\footnote{During the preparation of these proceedings, this was also pointed out by Ref.~\cite{Nayak:2019quy}.}.  In fact, the theoretical efficacy of gluon tagging improves the higher the mass of the new resonance: the background becomes more quark-like and quark and gluon jets become more different due to their increasing constituent multiplicity.  These theoretical gains are limited by experimental challenges related to the reconstruction of high energy constituents inside the dense core of high $p_T$ jets.  For a typical working point of 50\% gluon jet efficiency and a quark jet rejection of 10, the in-principle gain in significance to high-mass gluon resonances is $0.5^2/\sqrt{0.1^2}=2.5$.

\subsection{Conclusion and Outlook}
\label{sec:SM_jetsub_gluons:conclusion}
The jet subgroup at Les Houches 2019 concentrated their effort on the study of gluonic jets across four decades of energy, which are explored by the LHC. 
In the low-energy regime jets are sensitive to the non-perturbative dynamics of QCD and their description is beyond the jurisdiction of the standard first-principle approach based on perturbative field theory. Therefore, phenomenological models are usually employed to describe the non-perturbative parton-to-hadron transition. 
In our study, we have first addressed some of qualitative features of the groomed jet mass distribution in the non-perturbative regime and then turned our attention to the possibility of employing jet substructure variables to test and, eventually, improve on, the modelling of non-perturbative corrections in Monte Carlo event generators. 

As we go up in energy, we enter the regime where we expect parton-shower algorithms to correctly capture the relevant physics. In this context, we have investigated the impact of higher-order corrections to the splitting kernels that are at the core of any parton-branching algorithm. In particular, we have exploited modern machine-learning techniques in order to design observables that are sensitive to triple-collinear and double-soft corrections. 
At high energy, the issue of determining whether a given jet can be labelled as quark-initiated or gluon-initiated becomes central in Higgs physics and in the context of searches for particles and interactions beyond the Standard Model. Therefore, we have studied the performance of quark/gluon tagging in vector-boson-fusion and vector-boson-scattering analyses. Inspired by quark/gluon tagging in searches, we have explored the possibility of measuring gluon-jet transverse momentum distributions as a probe of parton distribution functions at high, i.e.\ 1~TeV, scale.
Finally, we have discussed how to probe the most energetic gluons at the LHC.

This report testifies that the jet subgroup have enjoyed a fruitful workshop, characterised by lively discussions and cross-pollination of ideas not only between theory and experimental communities but also between different methodologies, such as first-principle calculations in field theory, Monte Carlo simulations and machine-learning techniques. 
We are confident that these results, in some cases still preliminary, are already seeding new ideas and research projects which we look forward to further developing at the next edition of this workshop.

\subsection*{Acknowledgments}

We thank the participants of Les Houches 2019 for a lively environment and useful discussions.
BN is supported in part by the Office of High Energy Physics of the U.S. Department of Energy under Contract No. DE-AC02-05CH11231.
SM is also supported by the curiosity-driven grant ``Using jets to challenge the Standard Model of particle physics" from Universit\`a di Genova.


\chapter{Standard Model Higgs}
\label{cha:higgs}
\newcommand{\Herwig}{H\protect\scalebox{0.8}{ERWIG}\xspace}
\newcommand{\Pythia}{P\protect\scalebox{0.8}{YTHIA}\xspace}
\newcommand{\Sherpa}{S\protect\scalebox{0.8}{HERPA}\xspace}
\newcommand{\Rivet}{R\protect\scalebox{0.8}{IVET}\xspace}
\newcommand{\Professor}{P\protect\scalebox{0.8}{ROFESSOR}\xspace}
\newcommand{\eps}{\varepsilon}
\newcommand{\mc}[1]{\mathcal{#1}}
\newcommand{\mr}[1]{\mathrm{#1}}
\newcommand{\mb}[1]{\mathbb{#1}}
\newcommand{\tm}[1]{\scalebox{0.95}{$#1$}}
\newcommand{\ggH}{\ensuremath{gg H}\xspace}
\newcommand{\ttH}{\ensuremath{t\bar{t} H}\xspace}
\newcommand{\pTH}{\ensuremath{p_T(H)}\xspace}


\section{Progress on Simplified Template Cross Section (STXS) framework~\protect\footnote{
  M.~D\"{u}hrssen-Debling,
  J.\,A.~Mcfayden,
  J.\,K.\,L.~Michel,
  M.~Moreno Ll\'{a}cer,
  F.\,J.~Tackmann,
  H.\,T.~Yang
}{}}

\label{sec:projname}

This note discusses recent developments on the Simplified Template Cross Section (STXS) framework.
Bins in the transverse momentum of Higgs boson have been introduced for the associated production of a Higgs boson with a top-anti-top-quark pair.
In addition, possible improvements to the high transverse momentum bins for the gluon fusion production mode are discussed.

\subsection{Exploit kinematics of \ttH\ production mode under STXS framework}
\label{sec:ttH_STXS}
While other Higgs boson production modes have been divided into kinematic regions, the Higgs boson produced in association with a top-anti-top-quark pair (\ttH) remains inclusive in the most recent Stage~1.1 Simplified Template Cross Section (STXS) framework~\cite{Berger:2019wnu}. 
With the observation of the \ttH\ production mode by CMS and ATLAS experiments using LHC Run~2 data~\cite{Sirunyan:2018hoz,Aaboud:2018urx},
there is now sufficient precision to warrant splitting the \ttH\ mode into different kinematic regions in the STXS framework.

Several observable choices, such as the invariant mass of the \ttH\ system or the scalar sum of the transverse-momenta of the jets in the final state ($H_T$), have been studied. In the end, the Higgs boson transverse momentum (\pTH) is selected to divide up the phase space. Besides containing rich physics information (e.g. potential CP-mixing in the top Yukawa coupling or an anomalous Higgs self-coupling will modify the \pTH\ spectrum), compared with other options, \pTH\ can be reconstructed with high resolution in the diphoton decay channel (which is currently the most sensitive channel for studying \ttH) and does not require a fiducial definition for a top quark and its decay products.

\begin{figure}[t]
    \centering
    \includegraphics[width=0.5\textwidth]{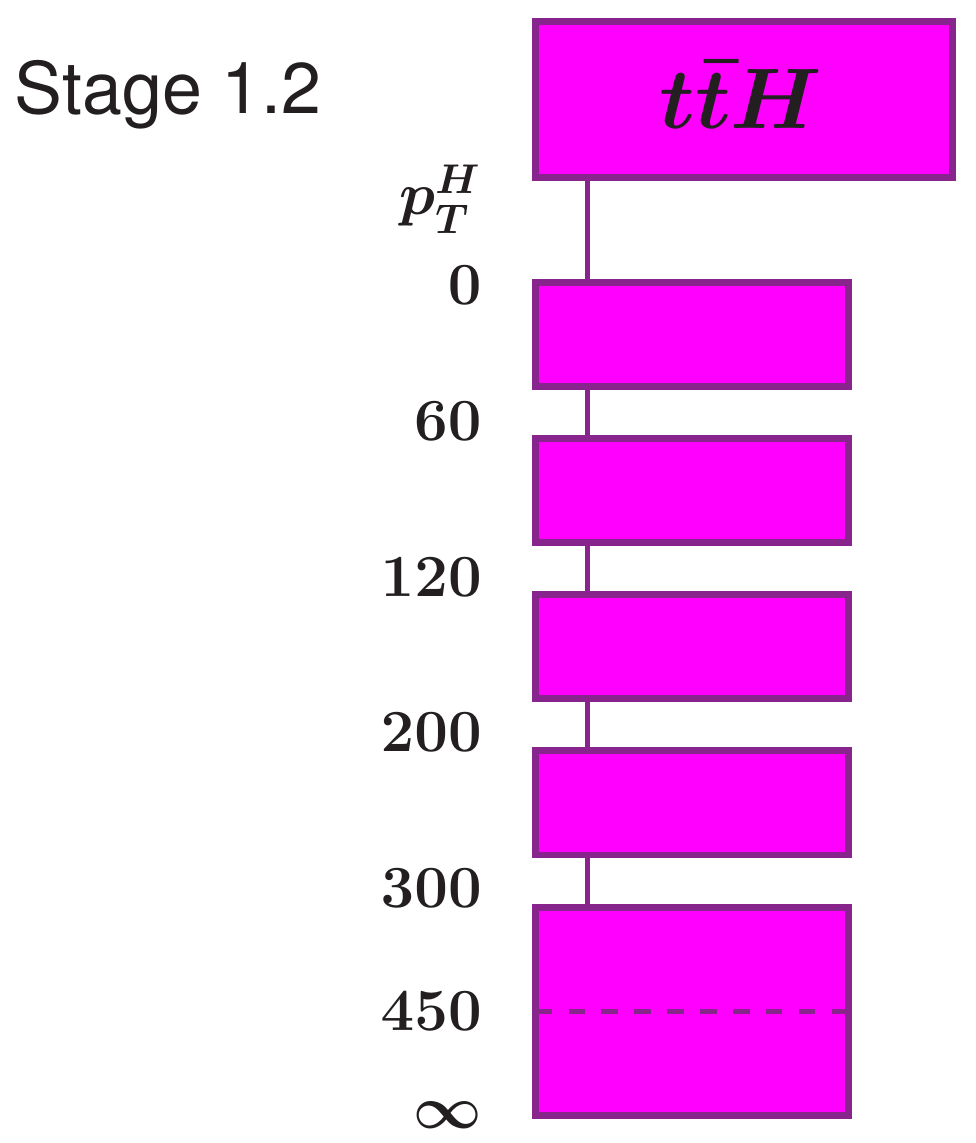}
    \caption{Definition of STXS binning in the \ttH\ production mode.}
    \label{fig:STXS_ttH}
\end{figure}

The proposed STXS bins for \ttH\ production mode in Stage~1.2 framework are summarized in Figure~\ref{fig:STXS_ttH}. It will supersede the inclusive bin in Stage~1.1. The proposed binning has been studied in Higgs decaying to diphoton and $b\bar{b}$ channels. The sensitivity to the lowest \pTH\ bin of 0~--~60~\GeV\ is driven by the diphoton channel. As \pTH\ goes higher, the bb and multi-lepton channels start to play a more important role. 
The bin boundaries are defined to also match those proposed for the gluon fusion production mode, including the high \pTH\ bins to be discussed in the next sub-section, to facilitate the merging of bins between the two production modes.

\let\GeV\undefined

\subsection{Exploring jet binning for the \ggH\ production mode at large Higgs transverse momentum}
\label{sec:H1jet_veto}

\newcommand{\df}{\mathrm{d}}
\newcommand{\pTcut}{p_T^\mathrm{cut}}
\newcommand{\rcut}{r_\mathrm{cut}}
\newcommand{\trcut}{\tilde{r}_\mathrm{cut}}
\newcommand{\GeV}{\,\mathrm{GeV}}
\newcommand{\Ymax}{Y_\mathrm{max}}
\newcommand{\FO}{\mathrm{FO}}

We consider the production of a Higgs boson via gluon fusion at $p_{T,H} \gtrsim 200 \GeV$
and investigate the utility of jet binning under the aspect of initial-state discrimination,
which on its own could motivate splitting the large-$p_{T,H}$ STXS bins further by jet multiplicity $N_j$.
For color-singlet final states (where $N_j = 0$ at Born level),
it is well known that jet binning greatly enhances the sensitivity
to the underlying partonic production channel~\cite{Ebert:2016idf}
because initial-state gluons tend to emit harder radiation than quarks
due to their larger color charge $C_A > C_F$.
For this reason, bins with $N_j \geq 1$ jet passing a given transverse momentum threshold $\pTcut$
are enriched with gluon-induced events,
while the gluon-induced contribution to the bin with exactly $N_j = 0$ jets is depleted.
Interpretation of the observed rates then requires precise theory predictions
for the 0-jet efficiency in a given production mode~\cite{Banfi:2012jm, Becher:2013xia, Stewart:2013faa, Banfi:2015pju}.
Theory predictions for the efficiency can also be used during the design stage of the analysis
to pick values of $\pTcut$ that maximize the discrimination power between channels
and maintain roughly equal sample sizes in all bins.

In the following, we provide NLL theory predictions for the exclusive 1-jet efficiency at large $p_{T,H}$,
where already at Born level we have $N_j = 1$ jet initiated by the recoiling hard parton.
We then use our results to explore the possible gain in discrimination power and recommend a choice of $\pTcut$.
We consider the breakdown of the inclusive $p_{T,H}$ spectrum
into an exclusive 1-jet and an inclusive $\geq 2$-jet contribution,
\begin{equation}
\frac{\df \sigma}{\df p_{T,H}} = \frac{\df \sigma_{1j}(\pTcut)}{\df p_{T,H}} + \frac{\df \sigma_{\geq 2j}(\pTcut)}{\df p_{T,H}}
\,.\end{equation}
with a corresponding 1-jet efficiency defined point by point in $p_{T,H}$ by
\begin{equation} \label{eq:H1jet_veto_def_efficiency}
\eps_{1j}(p_{T,H}, \pTcut) \equiv \frac{\df \sigma_{1j}(\pTcut)/\df p_{T,H}}{\df \sigma/\df p_{T,H}}
\,.\end{equation}
At tree level, the following three partonic configurations $ab \to Hj$ contribute to the inclusive spectrum:
\begin{equation} \label{eq:H1jet_veto_channels}
gg \to Hg \,, \qquad gq \to Hq \,, \qquad q\bar{q} \to Hg
\,,\end{equation}
where a sum over quark flavors, $q \leftrightarrow \bar{q}$, and $ab \leftrightarrow ba$ is understood.
It is interesting to ask whether jet binning can improve the discrimination between these configurations,
in particular because different gluon fusion mechanisms can be expected to lead to different admixtures
of partonic channels. E.g., in the Standard Model it is expected
that at large $p_{T,H}$, the relative $gg \to Hg$ contribution decreases because it proceeds through both a top triangle (like the other channels)
and a top box contribution (which becomes suppressed at energies much larger than the top mass).
On the other hand, the channel breakdown for a generic gluon-Higgs contact operator
induced by heavy BSM at a scale $\Lambda \gg p_{T,H}$ does not have this feature.
Therefore one may hope to gain additional information about the production mechanism beyond the total rate
by discrimating the dominant $gg$ and $gq$ channels and inspecting their $p_{T,H}$ dependence.

At leading order in the strong interaction, the inclusive spectrum is given by
\begin{equation} \label{eq:H1jet_veto_lo_cross_section}
\frac{\df \sigma_\text{LO}}{\df p_{T,H}}
= \sum_{ab} \int \! \df Y_H \, \int \! \df \eta_j \,
\frac{\df \hat{\sigma}_\text{LO}^{ab}(\mu)}{\df p_{T,H} \, \df Y_H \, \df \eta_j} \, f_a(x_a, \mu) \, f_b(x_b, \mu) \,
\Bigl[ 1 + \mathcal{O}(\alpha_s) \Bigr]
\,,\end{equation}
where $\df \hat{\sigma}_\text{LO}^{ab}$ is the Born partonic cross section for $ab \to Hj$
and $f_{a,b}(x_{a,b}, \mu)$ are the PDFs at momentum fractions $x_{a,b}$ set by the kinematics of the process.
Here we have integrated over the rapidity $Y_H$ of the Higgs
and over the pseudorapidity $\eta_j$ of the final-state parton (the jet).
In practice, we evaluate this formula at $\mu = \mu_\FO = m_{T,H}$
to obtain the denominator in Eq.~\eqref{eq:H1jet_veto_def_efficiency},
with $m_{T,H}$ the transverse mass of the Higgs.
We have checked these LO results separately for the three channels in Eq.~\eqref{eq:H1jet_veto_channels}
against \texttt{NNLOjet}.

On the other hand, at leading power in $\pTcut/p_{T,H} \ll 1$,
all emissions beyond the leading jet must be either soft or collinear.
In this limit the exclusive 1-jet cross section factorizes as~\cite{Liu:2012sz, Liu:2013hba}
\begin{align} \label{eq:H1jet_veto_1jet_cross_section}
\frac{\df \sigma_{1j}(\pTcut)}{\df p_{T,H}}
= \sum_{ab} \int \! \df Y_H \, \int \! \df \eta_j \, &
H_{abj}(s, t, u, \mu) \, B_a(x_a, \pTcut, \mu, \nu) \, B_b(x_b, \pTcut, \mu, \nu)
\nonumber \\ &\times
J_j(p_T^j, R, \mu) \, S_{abj}(\pTcut, R, \mu, \nu) \,
\Bigl[ 1 + \mathcal{O}\Bigl( \frac{\pTcut}{p_{T,H}}, R \Bigr) \Bigr]
\,.\end{align}
The hard function $H_{abj}$ encodes the underlying hard process $ab \to Hj$
and depends on the Mandelstam invariants $s,t,u$ of the Born kinematics and the color charges of $abj$.
The beam functions $B_{a,b}(x_{a,b}, \pTcut, \mu, \nu)$ describe collinear radiation off the initial-state partons below $\pTcut$.
The jet function $J_j$ describes collinear radiation initiated by the final-state parton $j$ with $p_T^j = p_{T,H}$
and clustered into a jet with jet radius parameter $R$.
The soft function $S_{abj}$ describes isotropic soft radiation below $\pTcut$,
excluding an unconstrained region of size $R$ around the jet,%
\footnote{We note that contributions from nonglobal logarithms are neglected in Eq.~\eqref{eq:H1jet_veto_1jet_cross_section},
which in principle become relevant at NLL for this observable.}
and implicitly depends on the color charges of $abj$ and the angles between them.
Eq.~\eqref{eq:H1jet_veto_1jet_cross_section} encodes the factorization of physics at the high scale $\mu_H \sim p_{T,H} \sim \sqrt{s}, \sqrt{t}, \sqrt{u}$
from soft and collinear dynamics at the much lower scales $\mu_B \sim \mu_S \sim \pTcut$ and $\mu_J \sim R\, p_{T,H}$.
By performing the renormalization group evolution between the two scales,
the dominant next-to-leading logarithmic (NLL) perturbative corrections are resummed to all orders,
leading to a Sudakov suppression of the 1-jet cross section in the numerator
of Eq.~\eqref{eq:H1jet_veto_def_efficiency} as $\pTcut \to 0$.
When the resummation is switched off, i.e., when all scales are set equal,
no additional emissions are generated and the 1-jet cross section simply becomes equal to the LO inclusive cross section.
Specifically, the ingredients in Eq.~\eqref{eq:H1jet_veto_1jet_cross_section} satisfy
\begin{equation} \label{eq:H1jet_veto_tree_level_relations}
H_{abj}(s, t, u, \mu) = \frac{\df \hat{\sigma}_\text{LO}^{ab}(\mu)}{\df p_{T,H} \, \df Y_H \, \df \eta_j} \,
\Bigl[ 1 + \mathcal{O}(\alpha_s) \Bigr]
\,, \qquad
B_i(x, \pTcut, \mu, \nu) = f_i(x, \mu) \,
\Bigl[ 1 + \mathcal{O}(\alpha_s) \Bigr]
\,,\end{equation}
and $J_j = S_{abc} = 1 + \mathcal{O}(\alpha_s)$.
By choosing the resummation scales as a function of $\pTcut$ and $p_{T,H}$ such
that they asymptote to $\mu_\FO$ as $\pTcut \to p_{T,H}$~\cite{Stewart:2013faa},
we ensure that $\eps_{1j} \to 1$ as $\pTcut \to \infty$ as it must.

\begin{figure*}
\centering
\includegraphics[width=0.48\textwidth]{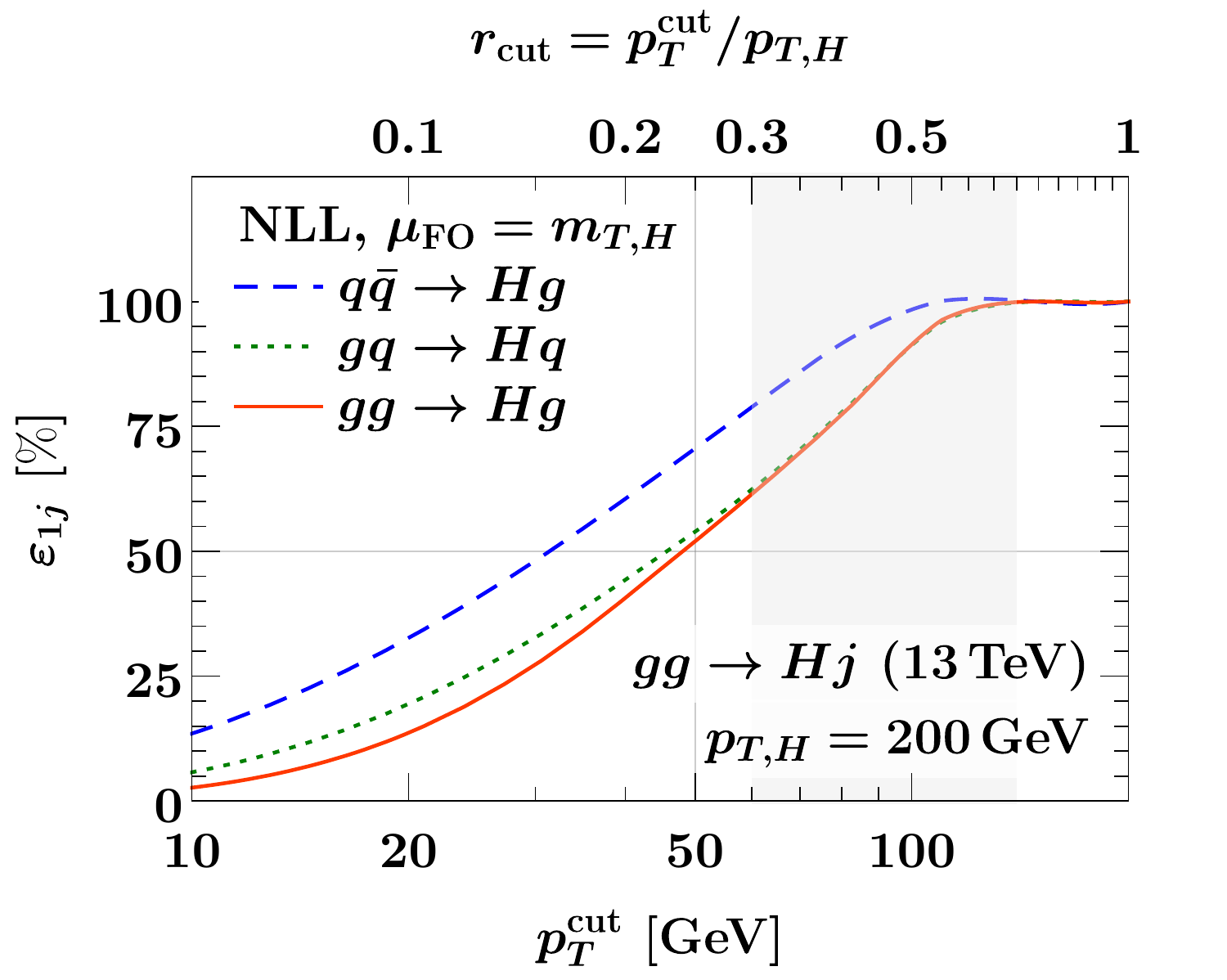}%
\hfill%
\includegraphics[width=0.48\textwidth]{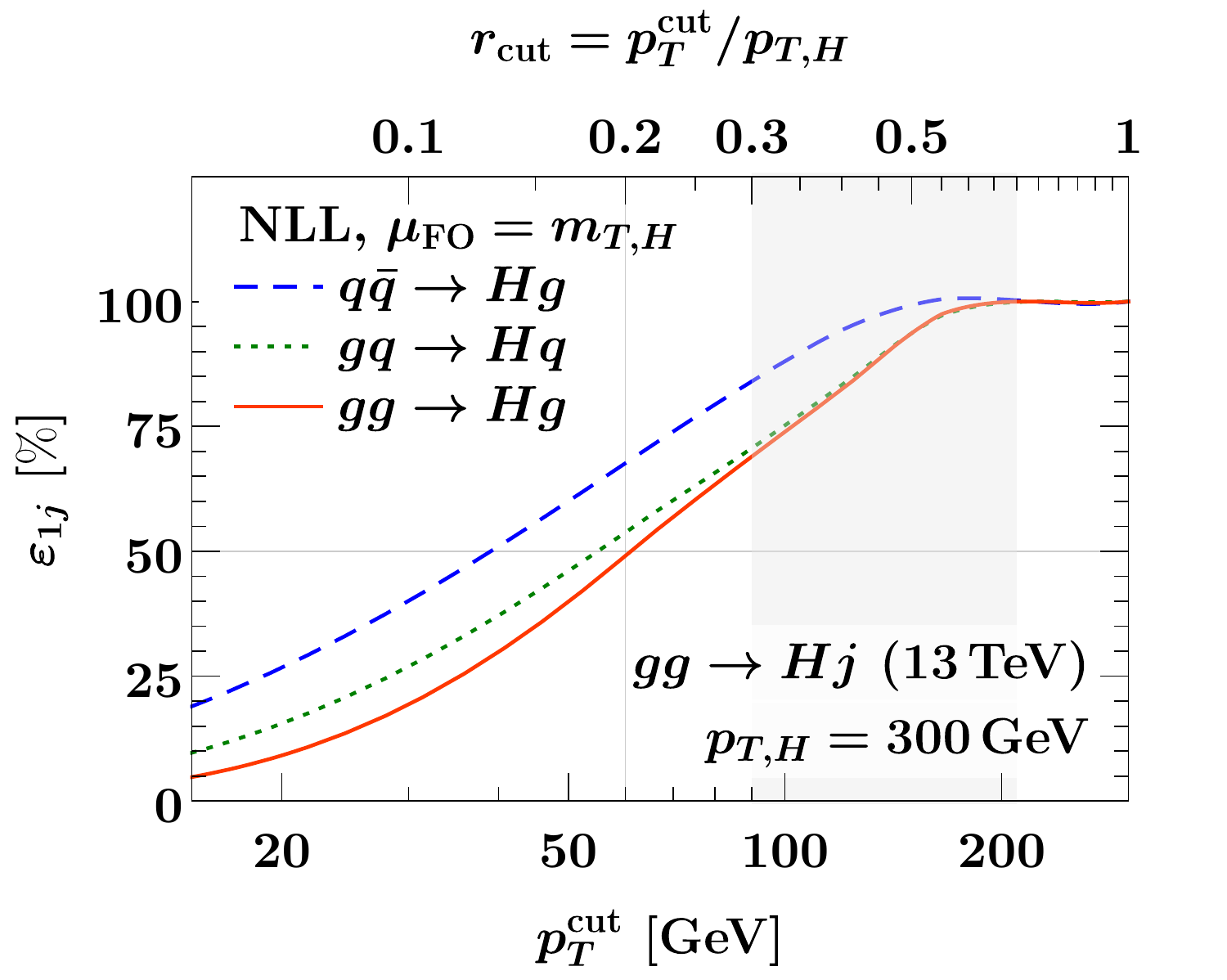}%
\\
\includegraphics[width=0.48\textwidth]{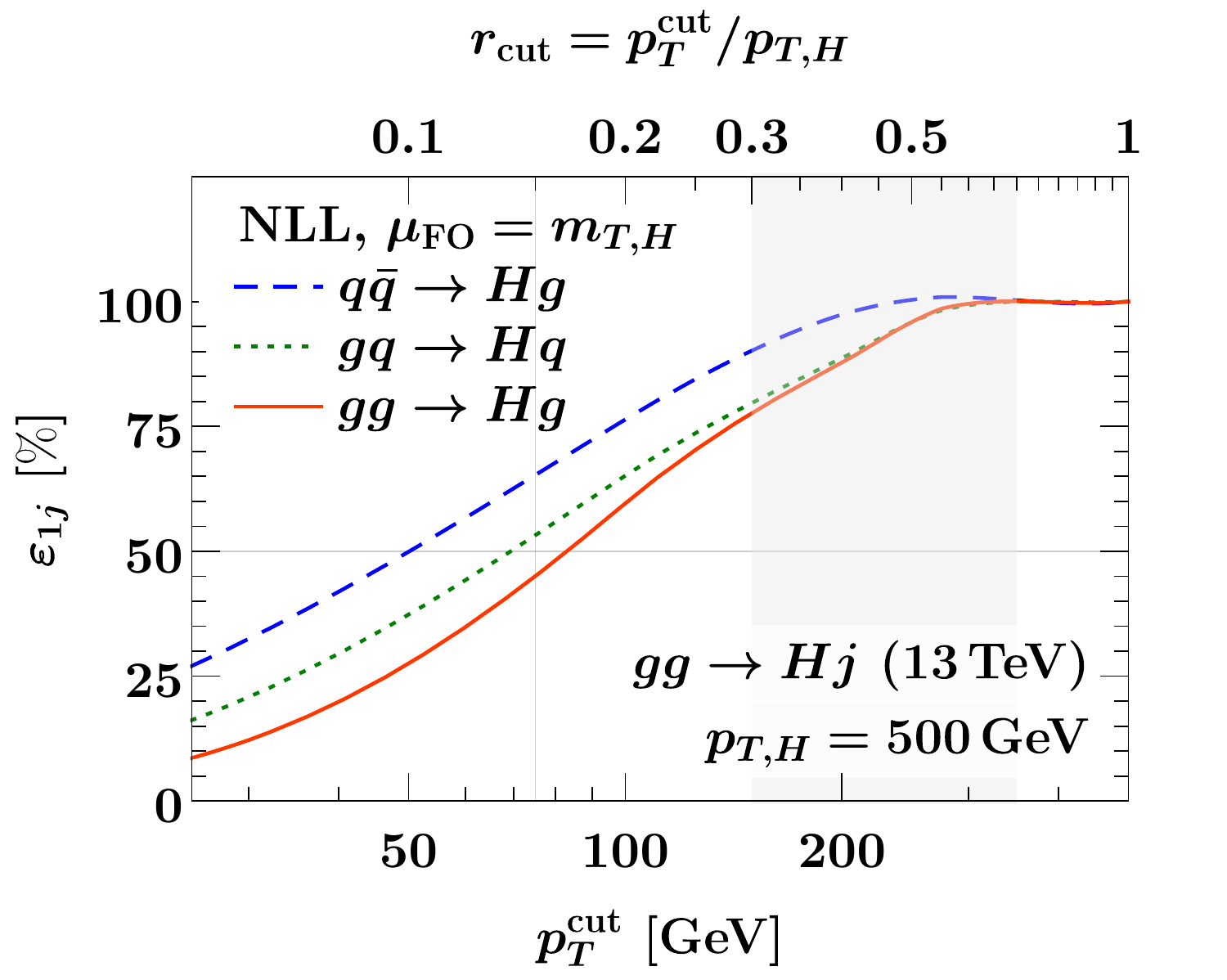}%
\hfill%
\includegraphics[width=0.48\textwidth]{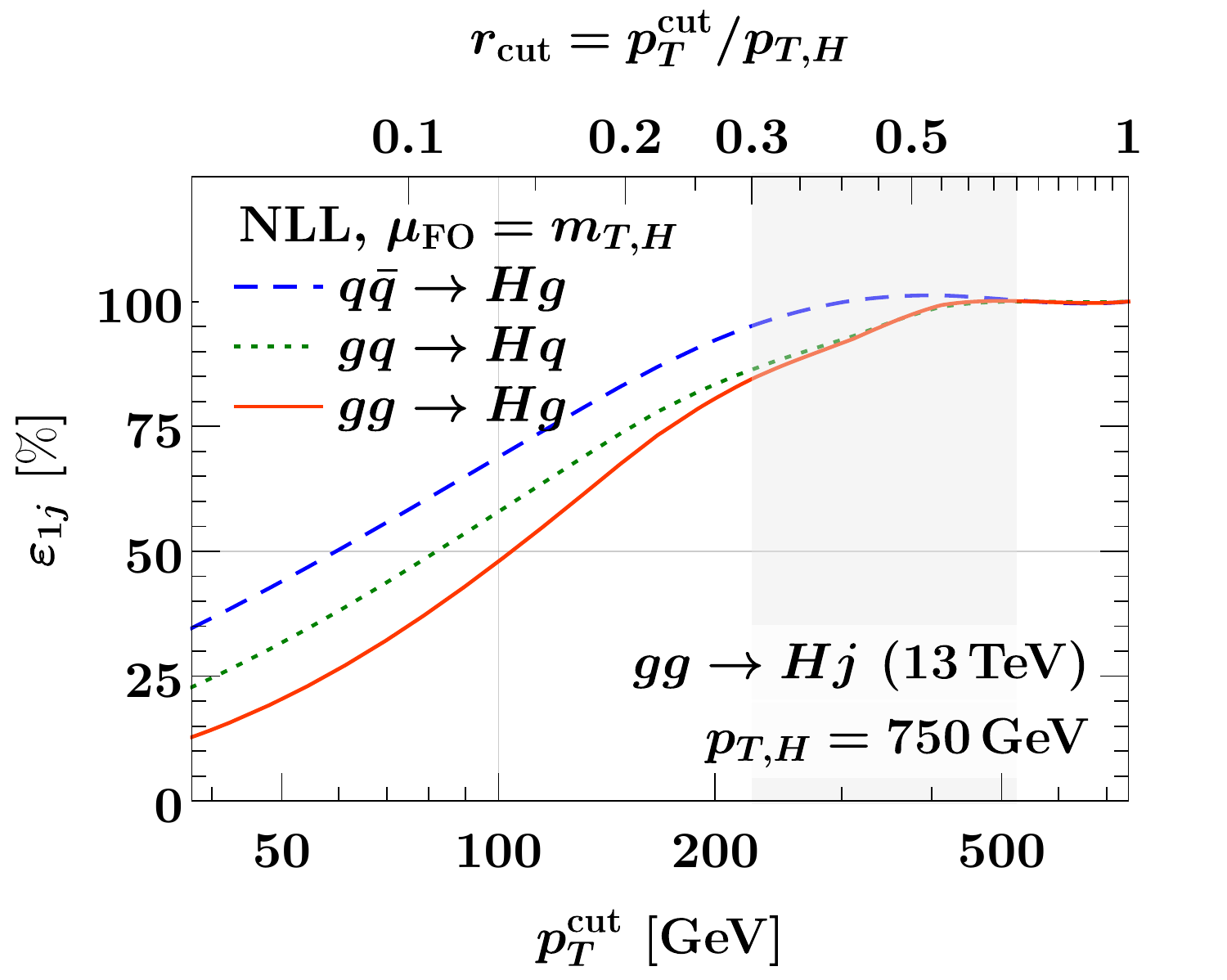}%
\caption{Exclusive 1-jet efficiencies for Higgs production in gluon fusion at different values of $p_{T,H}$
as a function of the jet bin boundary $\pTcut$ (bottom axis) and the ratio $\rcut = \pTcut/p_{T,H}$ (top axis).
We quote separate efficiencies for each color channel in Eq.~\eqref{eq:H1jet_veto_channels}.
A reliable prediction in the gray-shaded region would require fixed-order matching.
Grid lines indicate an efficiency of $\approx 50 \%$ in the $gg$ and $gq$ channels and the corresponding $\pTcut$ value.}
\label{fig:H1jet_veto_efficiency_pTcut}
\end{figure*}

Results for the 1-jet efficiency as a function of $\pTcut$
at fixed representative values of $p_{T,H}$ are given in Figure~\ref{fig:H1jet_veto_efficiency_pTcut}.
We use $\alpha_s(m_Z) = 0.118$, the \texttt{PDF4LHC15\_nnlo\_mc} PDF set~\cite{Butterworth:2015oua, Dulat:2015mca,
Harland-Lang:2014zoa, Ball:2014uwa, Carrazza:2015hva}, and set $R = 0.4$.
We restrict the $Y_H$ integral in Eqs.~\eqref{eq:H1jet_veto_lo_cross_section} and~\eqref{eq:H1jet_veto_1jet_cross_section}
to the range $|Y_H| < 2.4$.
(Note that this has virtually no impact on the efficiency.)
Here we have exploited that to our working order,
the sums over parton channels in Eqs.~\eqref{eq:H1jet_veto_lo_cross_section} and~\eqref{eq:H1jet_veto_1jet_cross_section}
are in one-to-one correspondence, i.e., higher-order corrections do not yet mix partonic channels.
We can therefore take the three color channels in Eq.~\eqref{eq:H1jet_veto_channels}
to be three independent subprocesses, each with a different all-order Sudakov structure
encoded in the color charges going into Eq.~\eqref{eq:H1jet_veto_1jet_cross_section},
and quote separate 1-jet efficiencies for each of them.
Similarly, while we have assumed a contact operator in our numerical results
(specifically, the SM in the heavy-top limit),
we stress that the model dependence of the efficiency cancels point by point in $(p_{T,H}, Y_H, \eta_j) \leftrightarrow (s, t, u)$
due to the correspondence between the hard function and the partonic cross section in Eq.~\eqref{eq:H1jet_veto_tree_level_relations},
and a mild model dependence is only reintroduced by different weights under the $Y_H$ and $\eta_j$ integrals.
The physical reason is that the resummed soft and collinear emissions below $\pTcut \ll p_{T,H} \sim m_t$ cannot resolve the top loop.

We observe that the jet binning has a very similar effect
on the $gg \to Hg$ and $gq \to Hq$ channels, with only the $q\bar{q} \to Hg$ channel
being discriminated slightly by the jet binning.
This is reasonable (if unfortunate)
because we expect $\pTcut$ to be most sensitive to the initial state color configuration,
which still contains a gluon in the case of $gq \to Hq$,
i.e., several of the color charge operators driving the Sudakov in $\pTcut$ are still $\propto C_A$.
A fairly pure $q\bar{q}$ sample could be obtained by making $\pTcut$ as tight as $\lesssim 20 \GeV$
for $p_{T,H} = 200 \GeV$, which however comes at the cost
of a strongly reduced rate in the exclusive 1-jet bin.
For the $gg$ and $gq$ channels, an even split between $N_j = 1$ and $N_j \geq 2$ is reached
at $\pTcut \approx 50 \GeV$ for $p_{T,H} = 200 \GeV$.
For higher values of $p_{T,H}$, this optimal value shifts up to values of $\pTcut \approx 100 \GeV$ for $p_{T,H} = 750 \GeV$,
indicating that holding $\pTcut$ fixed over all values of $p_{T,H}$ is impractical.
Indeed, the Sudakov exponentiation of $\ln(\mu_H/\mu_{B,S})$
with $\mu_H \sim p_{T,H}$ and $\mu_{B,S} \sim \pTcut$ suggests that
\begin{equation} \label{eq:H1jet_veto_def_rcut}
r \equiv \frac{p_T^\text{subl.\ jet}}{p_{T,H}}
\leq \frac{\pTcut}{p_{T,H}}
\equiv \rcut
\,,\end{equation}
might be a useful variable to cut on to obtain an even split for any $p_{T,H}$.
In the remainder of this note we will explore bin definitions like Eq.~\eqref{eq:H1jet_veto_def_rcut} further;
corresponding values of $\rcut$ are also indicated in Figure~\ref{fig:H1jet_veto_efficiency_pTcut}.
In the gray-shaded region in Figure~\ref{fig:H1jet_veto_efficiency_pTcut},
the assumption $\pTcut \ll p_{T,H}$ breaks down, so a reliable prediction in this region
would require matching to a fixed-order calculation for $H+2j$ production.
(In this region the quoted efficiencies can also exceed one for this reason.)
From a comparison to the result at leading-logarithmic order (LL),
we in general estimate that missing higher-order uncertainties on the results in Figure~\ref{fig:H1jet_veto_efficiency_pTcut}
can shift the results for $0.1 \leq \rcut \leq 0.25$ by $\mc{O}(10)$ percent points,
but do not affect the ordering between channels or the overall trend.

\begin{figure*}
\centering
\includegraphics[width=0.48\textwidth]{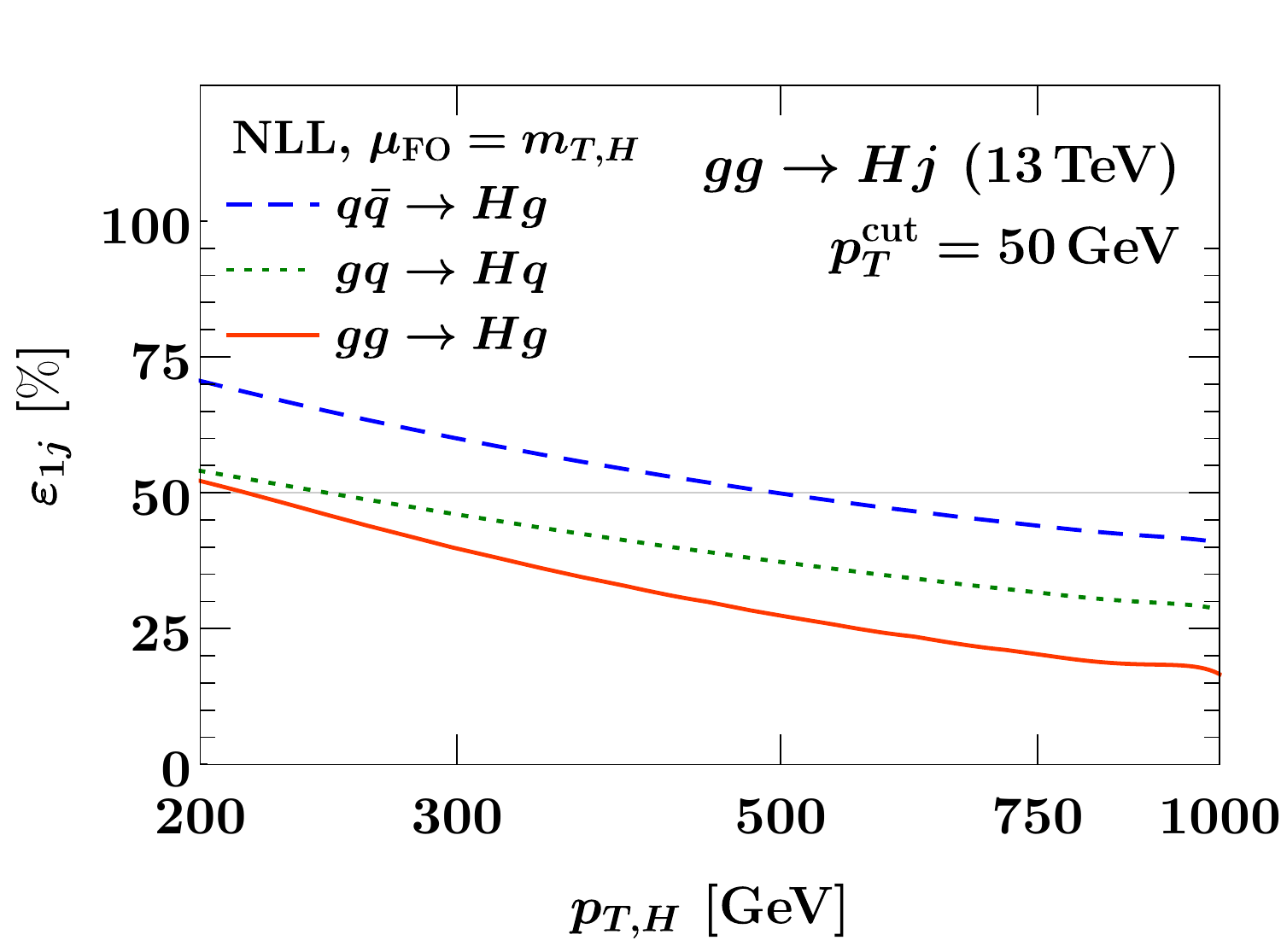}%
\hfill%
\includegraphics[width=0.48\textwidth]{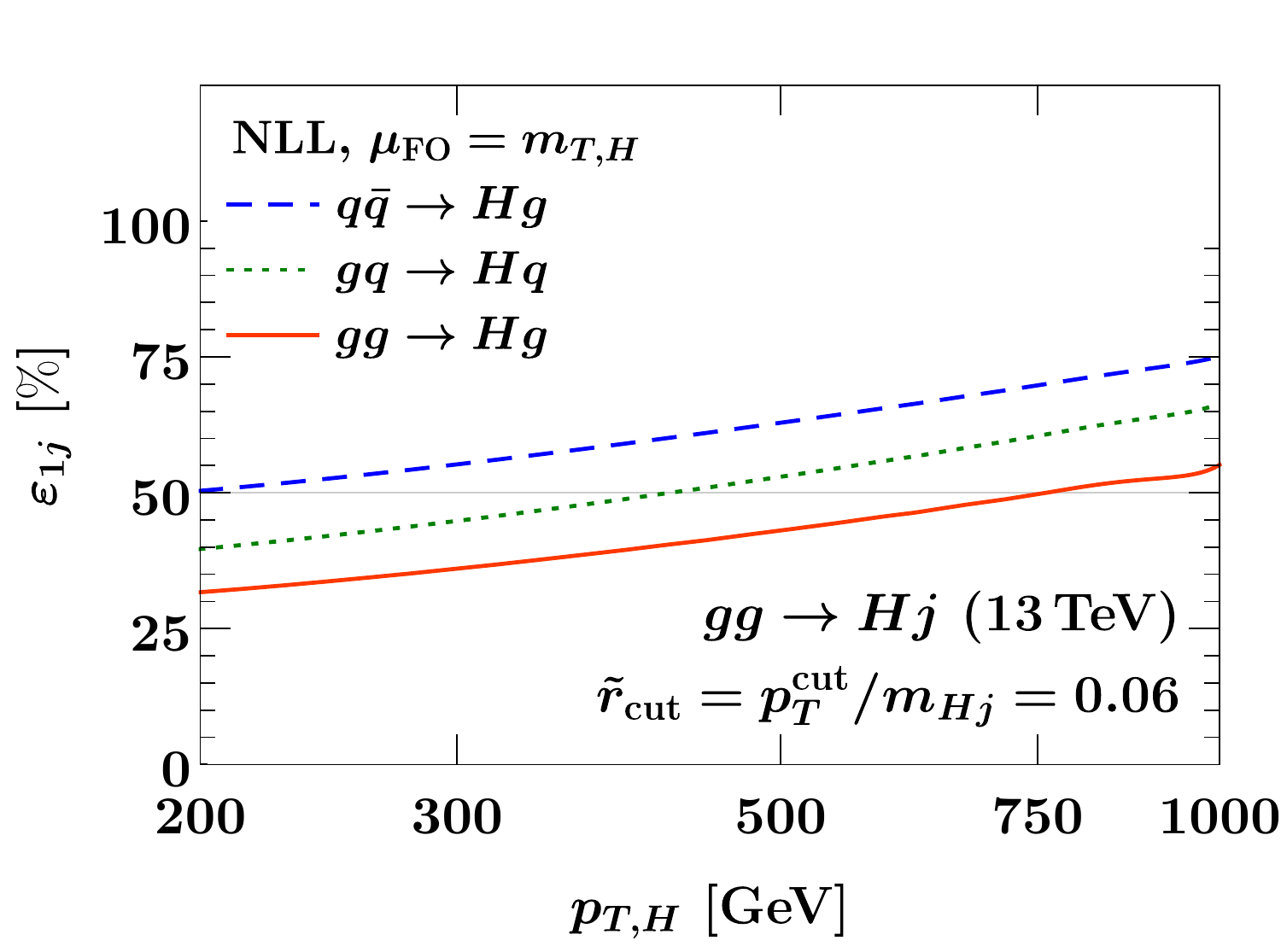}%
\\
\includegraphics[width=0.48\textwidth]{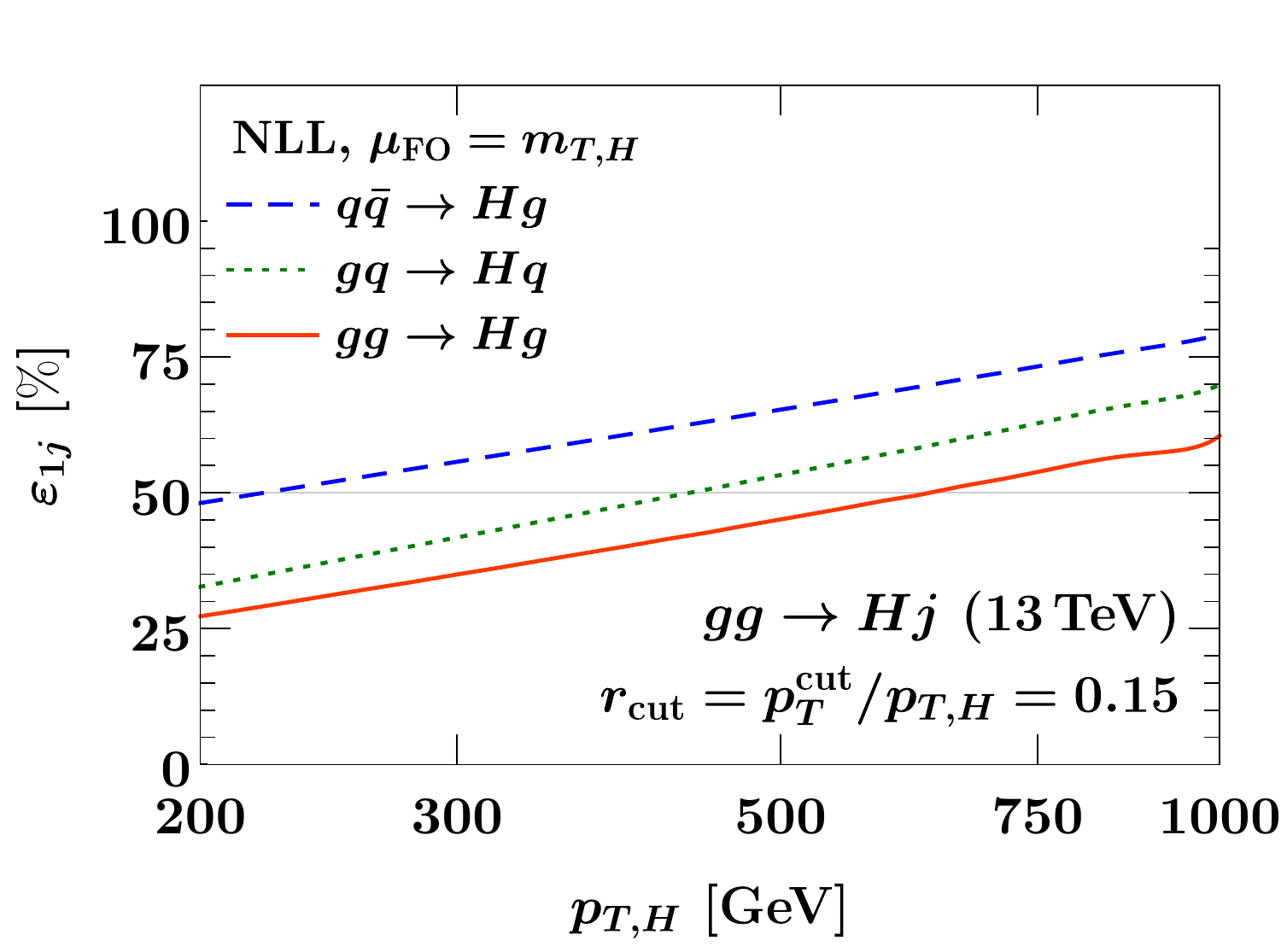}%
\hfill%
\includegraphics[width=0.48\textwidth]{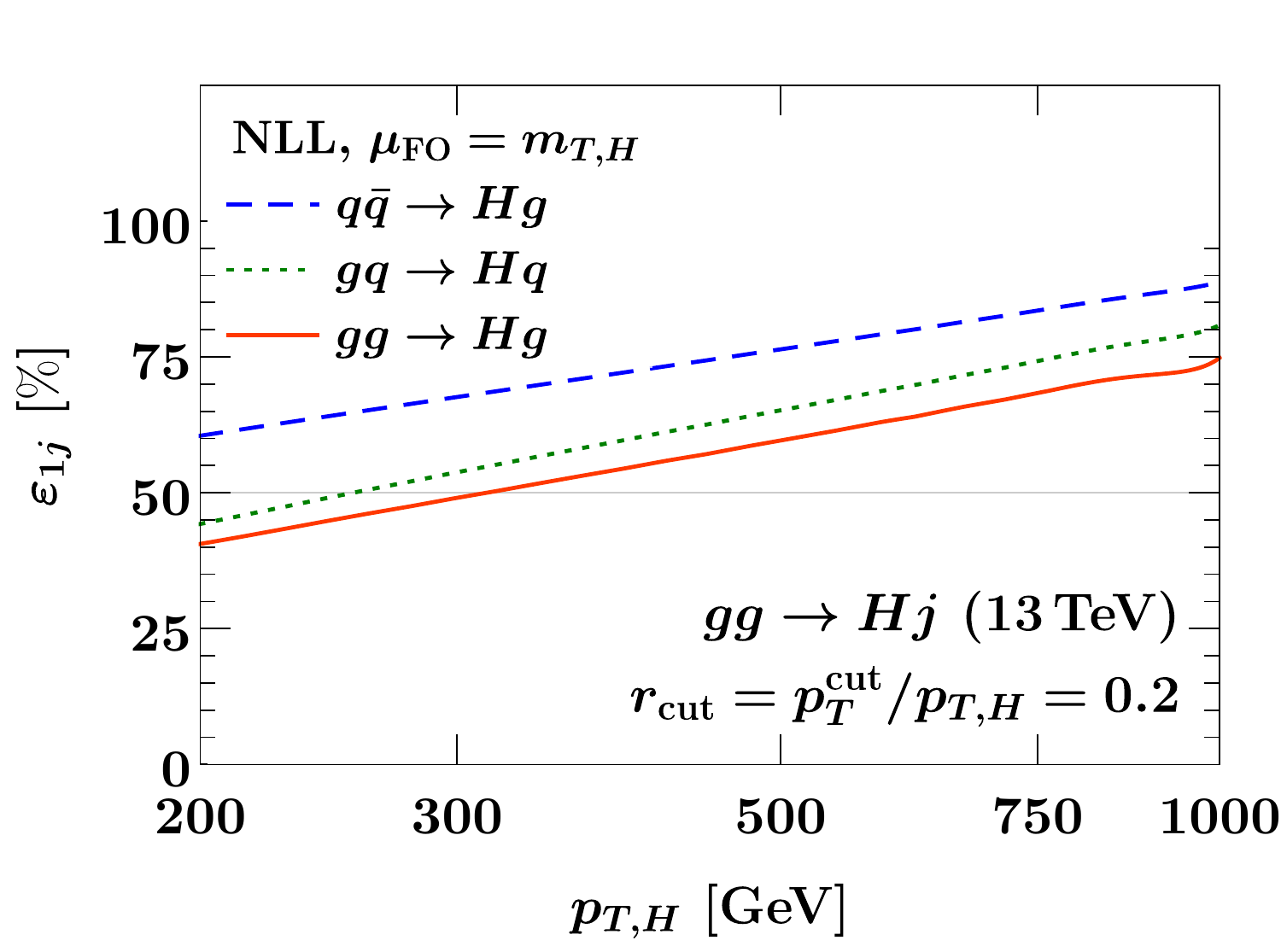}%
\caption{Exclusive 1-jet efficiencies per color channel for gluon-fusion Higgs production
using different jet bin definitions as a function of the Higgs transverse momentum $p_{T,H}$.
We compare a fixed value of $\pTcut = 50 \GeV$ (top left),
a cut at $\rcut = \pTcut/p_{T,H} = 0.2$ (bottom left) and $\rcut = 0.15$ (bottom right),
and a cut at $\trcut = \pTcut/m_{Hj} = 0.06$ (top right),
with $m_{Hj}$ the invariant mass of the $H+j$ system.
}
\label{fig:H1jet_veto_efficiency_other_variables}
\end{figure*}

In Figure~\ref{fig:H1jet_veto_efficiency_other_variables} we compare the efficiency
for different bin definitions as a function of $p_{T,H}$.
In the top left panel, we hold $\pTcut = 50 \GeV$ fixed.
As discussed earlier, while this choice is optimal for $p_{T,H} = 200 \GeV$,
the efficiency drops to just $\approx 25 \%$ for $p_{T,H} = 500 \GeV$.
On the other hand, we expect on theoretical grounds that by holding $\rcut$ fixed as defined in Eq.~\eqref{eq:H1jet_veto_def_rcut},
we should find a more stable efficiency as a function of $p_{T,H}$.
We see that this is not quite the case for $\rcut = 0.2$ (bottom left) and $\rcut = 0.15$ (bottom right),
with the 1-jet efficiency now drifting to larger values as $p_{T,H}$ increases.
This can be understood from the running of the QCD coupling,
which decreases as both the hard scale ($p_{T,H}$) and the emission scale ($\pTcut = \rcut \,p_{T,H}$) increase,
leading to softer emissions that preferentially populate the 1-jet bin.
Finally, we consider another variant of Eq.~\eqref{eq:H1jet_veto_def_rcut},
where we cut on
\begin{equation} \label{eq:H1jet_veto_def_rcut_mHj}
\tilde{r} \equiv \frac{p_T^\text{subl.\ jet}}{m_{Hj}}
\leq \frac{\pTcut}{m_{Hj}}
\equiv \trcut
\,,\end{equation}
with $m_{Hj} = \sqrt{s}$ the total invariant mass of the $H+j$ system.
Note that $m_{Hj}$ also changes as a function of $Y_H$ and $\eta_j$ underneath the integral in Eq.~\eqref{eq:H1jet_veto_1jet_cross_section}.
This choice is equally justified as a scaling variable since parametrically $m_{Hj} \sim p_{T,H}$ for large $p_{T,H}$.
We find that $\trcut = 0.06$ (top right panel of Figure~\ref{fig:H1jet_veto_efficiency_other_variables})
is roughly optimal for $p_{T,H} \approx 500 \GeV$,
but otherwise exhibits similar drift of the efficiency with $p_{T,H}$ as for $\rcut$.
In all four cases in Figure~\ref{fig:H1jet_veto_efficiency_other_variables},
we find that the $gq \to Hq$ efficiency closely traces the $gg \to Hg$ one,
making it difficult to discriminate these two important color configurations using any of the variants of jet binning discussed.

In conclusion, we find that unlike the color-singlet case,
jet binning for $H + j$ production in gluon fusion
seems to provide little discrimination power on the partonic initial state (the color channel).
It follows that jet binning alone is unlikely
to provide additional information about the underlying production mechanism,
even under the assumption that different gluon-fusion mechanisms strongly differ in their breakdown into partonic channels.
Here, it would be very interesting to combine the jet binning discussed above (effectively an ISR tagger)
with jet substructure methods like quark/gluon jet discrimination (FSR taggers)
in order to exploit the full color flow information between all legs of the hard scattering.
This could break the degeneracy between the $gg \to Hg$ and $gq \to Hq$ color channels,
but clearly is beyond the scope of the STXS framework.

In the STXS context, it is worthwhile to point out that jet binning
is relevant in many places other than initial-state discrimination for the signal process, which we focused on here.
In particular, (tentatively) splitting a given bin further into jet bins allows
for an improved handle on systematic uncertainties of the combined bin,
and in some cases is mandatory to suppress the background, e.g.\ in $H \to WW$.
In these cases, the above results are useful to make an informed choice for the veto parameter $\pTcut$
that ensures an even split of the signal sample, i.e., a 1-jet efficiency of $\eps_{1j} \approx 50 \%$.
We stress the importance of increasing $\pTcut$ along with $p_{T,H}$
in order to avoid a Sudakov suppression (depletion) of the exclusive 1-jet bin;
however, strictly keeping the ratio $\pTcut/p_{T,H}$ fixed as in the bottom row of Figure~\ref{fig:H1jet_veto_efficiency_other_variables}
leads to no clear improvement.
Likewise, cutting on the subleading jet $p_T$ normalized to the total invariant mass of the $H+j$ system
leads to no clear improvement, and in addition loses the clear association with the $p_{T,H}$ spectrum.
Instead, $\pTcut$ can for simplicity be picked to be a constant over each bin in $p_{T,H}$,
with the appropriate value of $\pTcut$
to be read off from Figures~\ref{fig:H1jet_veto_efficiency_pTcut} and~\ref{fig:H1jet_veto_efficiency_other_variables}.
Because the $p_{T,H}$ spectrum is steeply falling,
picking $\pTcut$ as a function of the lower bin boundary $p_{T,H}^\mathrm{min}$ is recommended to capture the main characteristics.
Values of $\pTcut$ that lead to $\eps_{1j} \approx 50 \%$ at NLL for the $gg$ and $gq$ color channels
are compiled in Table~\ref{tab:H1jet_veto_suggested_pTcut} for representative values of $p_{T,H}^\mathrm{min}$.

\begin{table}
\centering
\begin{tabular}{lll}
\hline\hline
\multicolumn{1}{c}{$p_{T,H}^\mathrm{min}~[\!\GeV]$}
& \multicolumn{1}{c}{$\pTcut~[\!\GeV]$}
& \multicolumn{1}{c}{$\pTcut/p_{T,H}^\mathrm{min}$}
\\
\hline
$200$ & $50$ & $0.25$ \\
$300$ & $60$ & $0.2$ \\
$500$ & $75$ & $0.15$ \\
$750$ & $100$ & $0.133$ \\
\hline\hline
\end{tabular}%
\caption{Suggested values of $\pTcut$ for different representative lower bin boundaries $p_{T,H}^\mathrm{min}$ in the $p_{T,H}$ spectrum.
Based on the NLL results presented here, these are expected to lead to 1-jet efficiencies of $\approx 50\%$ for the dominant $gg$ and $gq$ color channels
in gluon-fusion Higgs production. Note that $\pTcut$ should increase with $p_{T,H}^\mathrm{min}$
to avoid a Sudakov depletion of the exclusive 1-jet bin.
Due to running-coupling effects, the ratio is not constant.}
\label{tab:H1jet_veto_suggested_pTcut}
\end{table}

\let\df\undefined
\let\pTcut\undefined
\let\rcut\undefined
\let\trcut\undefined
\let\GeV\undefined
\let\Ymax\undefined
\let\FO\undefined

\let\Herwig\undefined
\let\Pythia\undefined
\let\Sherpa\undefined
\let\Rivet\undefined
\let\Professor\undefined
\let\eps\undefined
\let\mc\undefined
\let\mr\undefined
\let\mb\undefined
\let\tm\undefined
\let\ttH\undefined
\let\ggH\undefined
\let\pTH\undefined
\let\GeV\undefined

\newcommand{\Herwig}{H\protect\scalebox{0.8}{ERWIG}\xspace}
\newcommand{\JHUGEN}{JHUG\protect\scalebox{0.8}{EN}\xspace}
\newcommand{\Powheg}{P\protect\scalebox{0.8}{OWHEG}\xspace}
\newcommand{\Pythia}{P\protect\scalebox{0.8}{YTHIA}\xspace}
\newcommand{\Sherpa}{S\protect\scalebox{0.8}{HERPA}\xspace}
\newcommand{\Rivet}{R\protect\scalebox{0.8}{IVET}\xspace}

\newcommand{\Professor}{P\protect\scalebox{0.8}{ROFESSOR}\xspace}
\newcommand{\eps}{\varepsilon}
\newcommand{\mc}[1]{\mathcal{#1}}
\newcommand{\mr}[1]{\mathrm{#1}}
\newcommand{\mb}[1]{\mathbb{#1}}
\newcommand{\tm}[1]{\scalebox{0.95}{$#1$}}
\newcommand{\dphi}{\Delta\phi_{jj}}

\section{STXS CP-sensitive binning options for VBF production modes\protect\footnote{Y. Haddad, P. Francavilla}{}}

\label{sec:Higgs_STXS_VBFCP}
The azimuthal angle correlation of the jets in Higgs boson production via vector-boson fusion, provide a general experimental probe of the CP structure of Higgs boson interactions to gauges bosons. As such, we propose to extend the definition of the Simplified Template Cross Section framework to include the CP sensitive bins.

\subsection{Introduction}
\label{sec:Higgs_STXS_VBFCP:intro}
The experimental observation of the Higgs boson at the Large Hadron Collider (LHC) has led the way for detailed studies of its properties. Although some of its parameters have been already determined, such as the mass or the spin-parity, one question concerning the charge (C) and the parity (P) symmetries remains. In the Standard Model (SM) of particles physics, the Higgs boson interactions preserve the CP symmetries, any deviation from this prediction would, therefore, be considered as a manifestation of physics beyond the SM (BSM).

Vector Boson Fusion (VBF) production process is the second most copious production channel at the LHC. It has the advantage of a distinct kinematic structure with two forward tagging jets resulting from the scattered quarks. This feature allows for a better background rejection and, hence, for a fairly clean signal sample. Due to the fusion of the two gauge bosons, this channel allows the test of the tensor structure of the HVV vertex ($V=W, Z$), which is sensitive to the CP properties of the Higgs boson. The most general tensor structure describing the interaction of the Higgs boson with two spin-one gauge boson which contributes to the VBF production mode can be written as

\begin{equation}
    \begin{aligned} 
    T^{\mu\nu} (q_1,q_2)  &=a_1(q_1,q_2) ~g^{\mu\nu} \\ 
                          &+ a_2(q_1,q_2)~[q_1\cdot q_2 g^{\mu\nu} - q_1^{\mu}q_2^{\nu}]\\
                          &+ a_3(q_1,q_2)~\epsilon^{\mu\nu\alpha\beta} q_{1,\alpha}q_{2,\beta}
    \end{aligned}
    \label{eq:Higgs_STXS_VBFCP:hvvtensor}
\end{equation}

where $q_1$ and $q_2$ are the four-momenta of the two fusing gauge bosons $V$. The scalar $a_1$ represents the SM contribution to the coupling, while the form factors $a_{2}$ and $a_3$ represent CP-even and CP-odd amplitudes respectively. An anomaly in couplings can manifest itself as a deviation of these parameters from their SM values (which are $a_1=1$ and $a_2, a_3=0$).

To study the tensor structure of the HVV couplings, the azimuthal angle between the two tagging jets $\Delta\phi_{jj}$, that characterise the VBF process, is proven to be an important tool, independently on the Higgs decay mode. It was shown in Ref. \cite{Hankele:2006ma, Klamke:2007cu} that the absolute value of this observable is sensitive to the form factor effects and it provides an excellent distinction between the 3 form factors in Eq. \ref{eq:Higgs_STXS_VBFCP:hvvtensor}. However, when both CP-even and CP-odd couplings of similar strength are present their effects cancel out and the resulted distribution is very similar to the SM prediction. This issue can be tackled by redefining the azimuthal angle to include its sign such as it exhibits the interference effect between $a_2$ and $a_3$. If $b_{+}$ and $b_{-}$ are the four-momenta of the two proton-beams and $p_{+}$ and $p_{-}$ the four momenta of the two tagging jets from VBF, then 

\begin{equation}
    \begin{aligned} 
        \epsilon_{\mu\nu\rho\sigma}b_{+}^\mu p_{+}^\nu b_{-}^\rho b_{-}^\sigma &= 2p_{T,+}p_{T,-}\sin(\phi_{+} - \phi_{-})\\
        &= 2p_{T,+}p_{T,-}\sin(\Delta\phi_{jj})
    \end{aligned}
    \label{eq:Higgs_STXS_VBFCP:dphijj}
\end{equation}

\begin{figure}[t]
  \begin{center}
    \includegraphics[height=6cm]{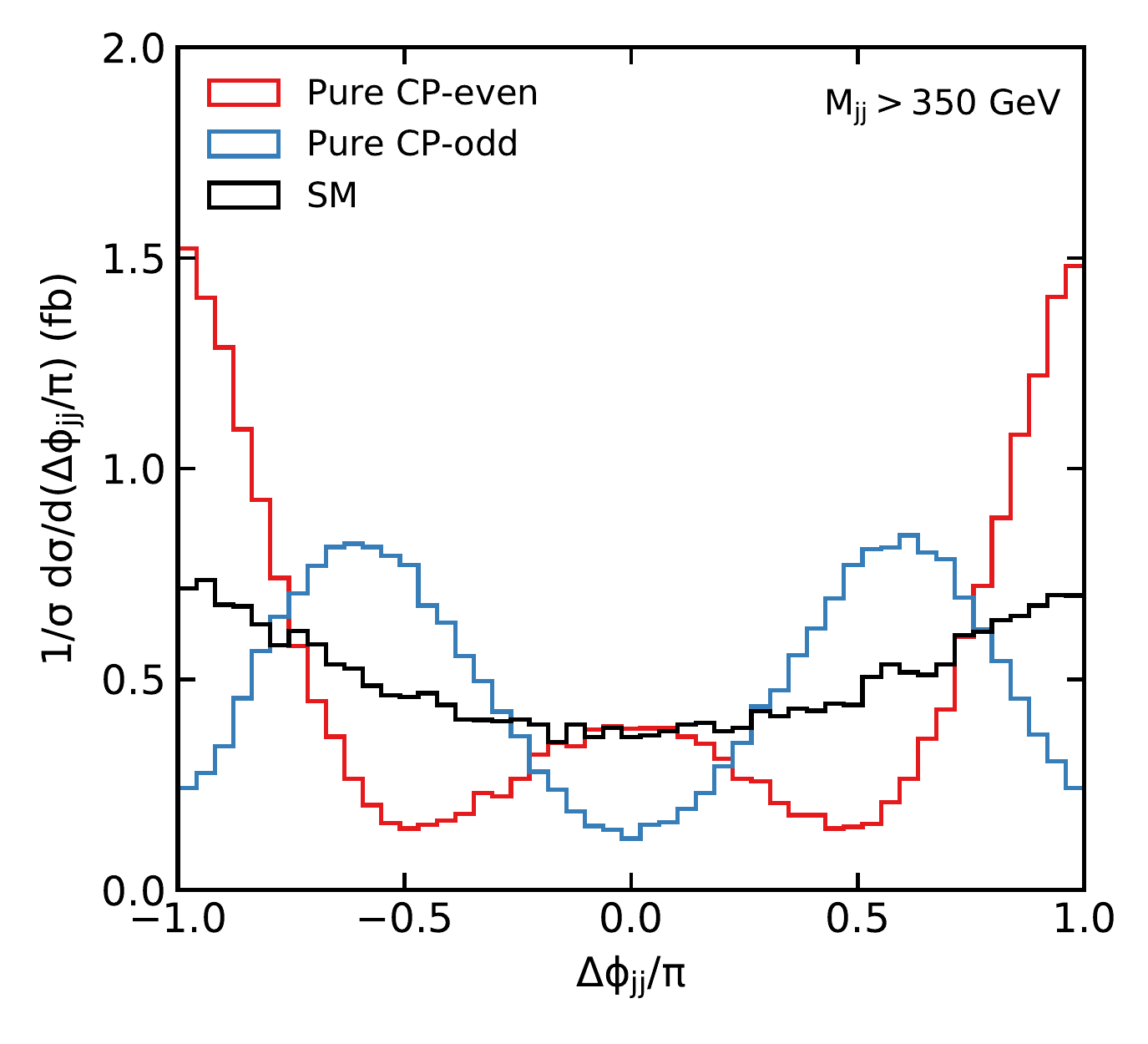}
    \caption{$\Delta\phi_{jj}$ observable for pure CP-odd, CP-even and SM-like couplings.}
    \label{fig:Higgs_STXS_VBFCP:pure_CP}
  \end{center}
\end{figure}

$p_{+}~(\phi_{+})$ and $p_{-}~(\phi_{+})$ denotes the four-momenta (azimuthal angles) of the two tagging jets, where $p_{+}$ ($p_{-}$) points to the same detector hemisphere as $b_{+}$ ($b_{-}$). Such ordering removes ambiguity in the standard definition of $\Delta\phi_{jj}$. The ratio of the form factors can directly be measured by determining the minimum of the $\Delta\phi_{jj}$ distribution. The typical distribution of purely CP-odd, CP-even and SM couplings are shown in Fig. \ref{fig:Higgs_STXS_VBFCP:pure_CP}. 

The Simplified Template Cross Sections (STXS) framework \cite{Badger:2016bpw} is considered as the natural evolution of the signal strength measurement performed during Run1. Various results from ATLAS and CMS have been already released with a limited number of bins. However, as more data are been taken at the LHC, it becomes important to further explore the phase-space by probing narrow kinematic regions that are sensitive BSM deviations. This study proposes to refine the Stage 1.1 bins \cite{Berger:2019wnu} using the signed $\Delta\phi_{jj}$ observable to test the CP properties of the Higgs boson at the next LHC measurements.

\subsection{Calculation tools and results}
\label{sec:Higgs_STXS_VBFCP:tools}
Simulated samples of Higgs boson events produced via anomalous HVV couplings are generated for both scalar and pseudo-scalar hypotheses using the generator \JHUGEN 7.0.2 \cite{Anderson:2013afp,Bolognesi:2012mm,Gao:2010qx}. To properly simulate the recoil of the final state particles caused by additional QCD radiation, the \JHUGEN generator is interfaced with the \Pythia\cite{Sjostrand:2014zea} program. The Higgs boson is left undecayed and its mass is set at $M_H=125\UGeV$. The Parton Distribution Functions (PDF) used in this note are NNPDF30 \cite{citeNNPDF} for both the matrix element and parton shower simulations. The \JHUGEN samples produced with the SM couplings are compared with the equivalent samples generated by the \Powheg \cite{Nason:2009ai} event generator at NLO QCD, with parton showering with \Pythia applied in both cases, and the kinematic distributions are found to agree.

The Simplified Template Cross-Section binning for stage 1.1 described in \cite{Berger:2019wnu}, identifies the VBF topology with an invariant mass threshold starting at $M_{jj}>350~\rm GeV$. Then, events are divided into regions in the Higgs boson transverse momentum. Further split is then done on $M_{jj}$ and on $p_{T}^{Hjj}$. The later split targets the identification of exclusive 2-jets and inclusive 3-jets categories.

To estimate the non-SM contributions, the signal distribution $p_{\rm sig}$, for a given observable $x$, can be parameterised as a linear combination of the terms originating from the SM-like and anomalous amplitudes and their interference \cite{Anderson:2013afp, Khachatryan:2014kca}

\begin{equation}
    p_{\rm sig}(x) = (1-f_{\rm mix}) ~p_{\rm a_1}(x) + f_{\rm mix}~p_{\rm a_n}(x) + \sqrt{f_{\rm mix}(1-f_{\rm mix})} ~p^{\rm int}_{\rm a_1, a_n}
\end{equation}

where $p_{\rm a_n}$ is template histogram or probability of pure $\rm a_n$ term, and $p^{\rm int}_{\rm a_1, a_n}$ describes the interference between the two terms. Three values of the anomalous coupling are introduced 0.1\%, 1\% and 10\%.

\begin{figure}[t]
\centering
\begin{subfigure}[]{
\includegraphics[width=.45\linewidth] {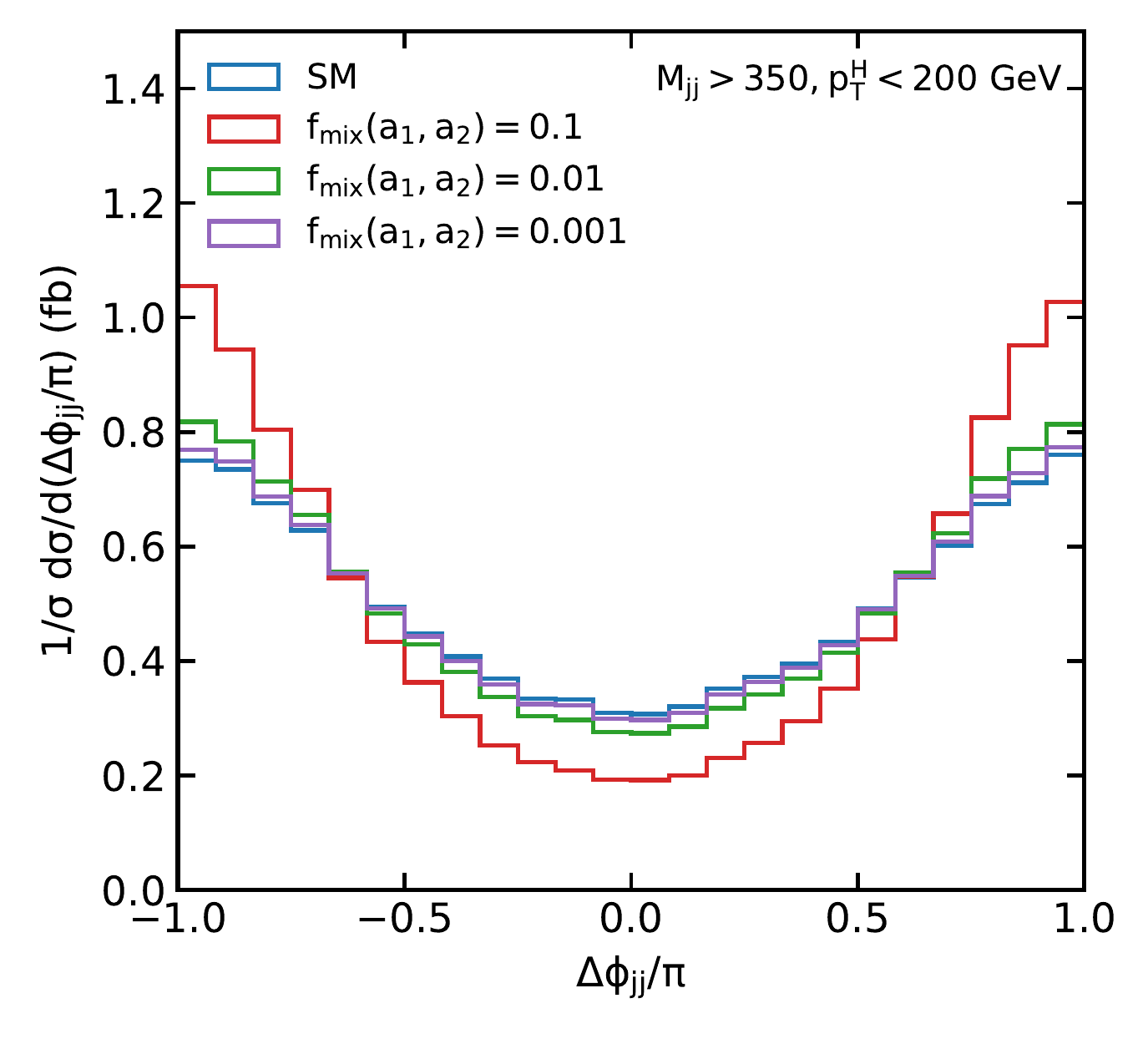}   
\label{fig:Higgs_STXS_VBFCP:CP-bins:a} }
\end{subfigure}
\hfill
\begin{subfigure}[]{
\includegraphics[width=.45\linewidth] {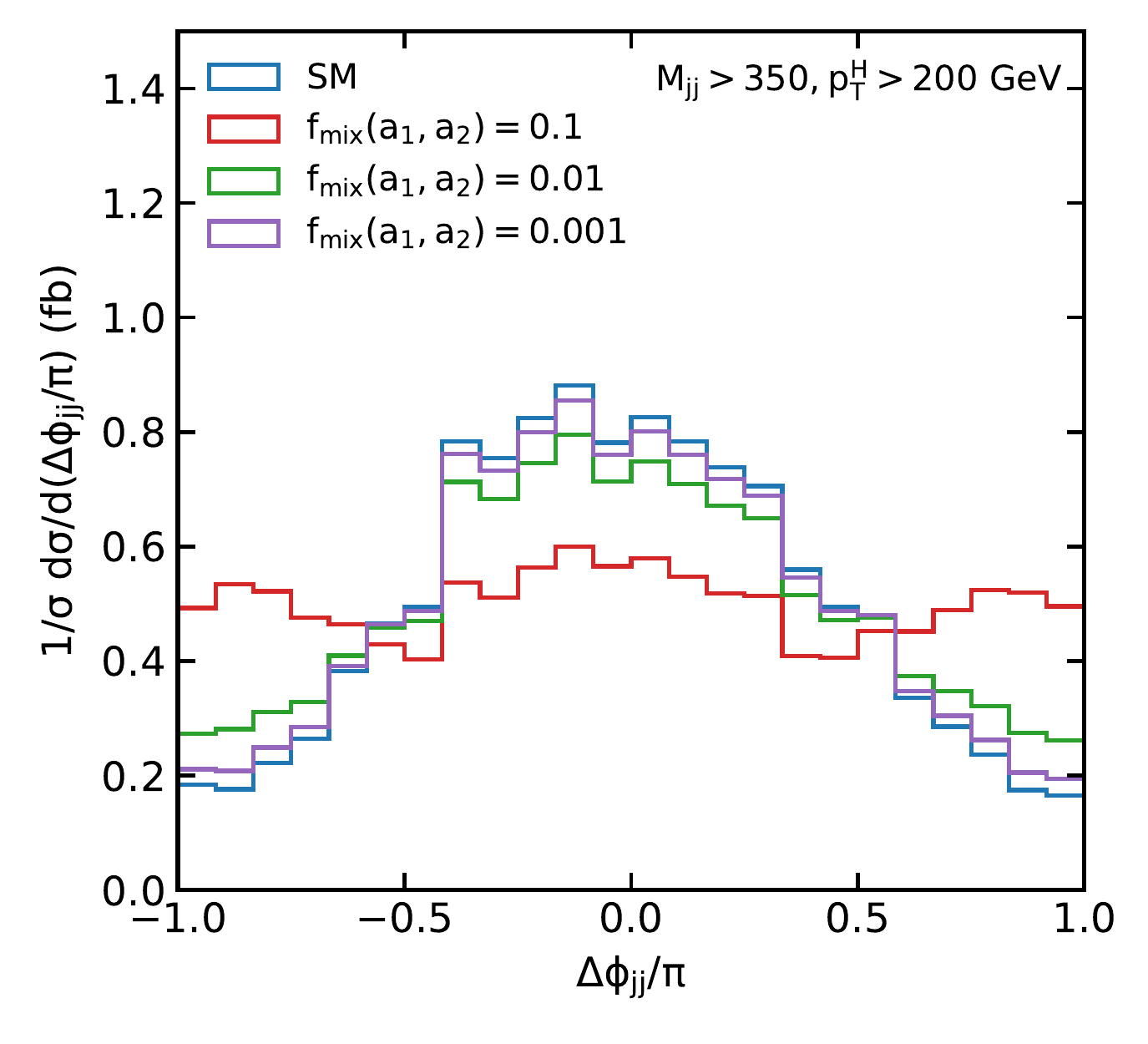}   
\label{fig:Higgs_STXS_VBFCP:CP-bins:b} }%
\end{subfigure}
\hfill
\begin{subfigure}[]{
\includegraphics[width=.45\linewidth] {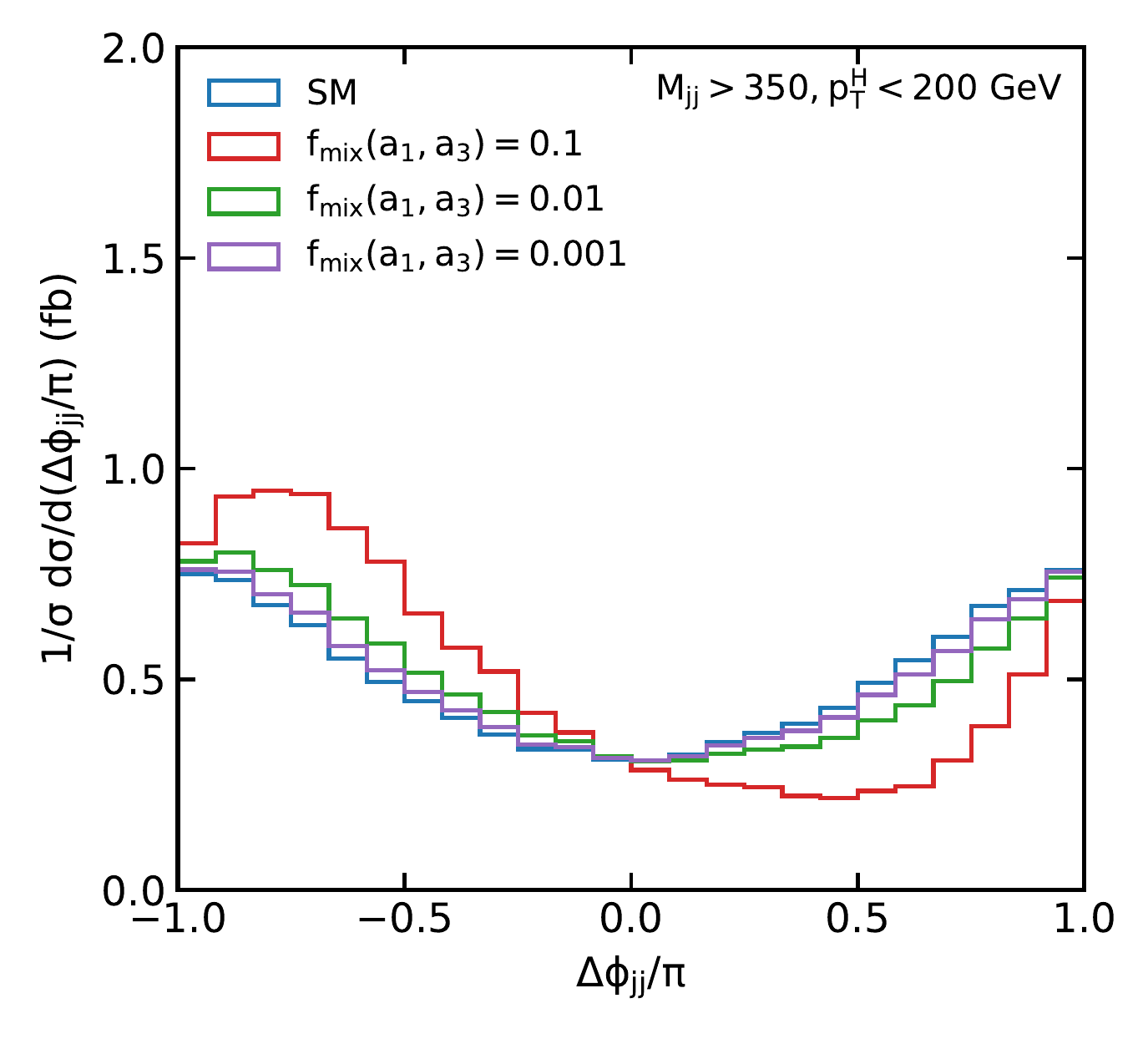}   
\label{fig:Higgs_STXS_VBFCP:CP-bins:c} }%
\end{subfigure}%
\hfill
\begin{subfigure}[]{
\includegraphics[width=.45\linewidth] {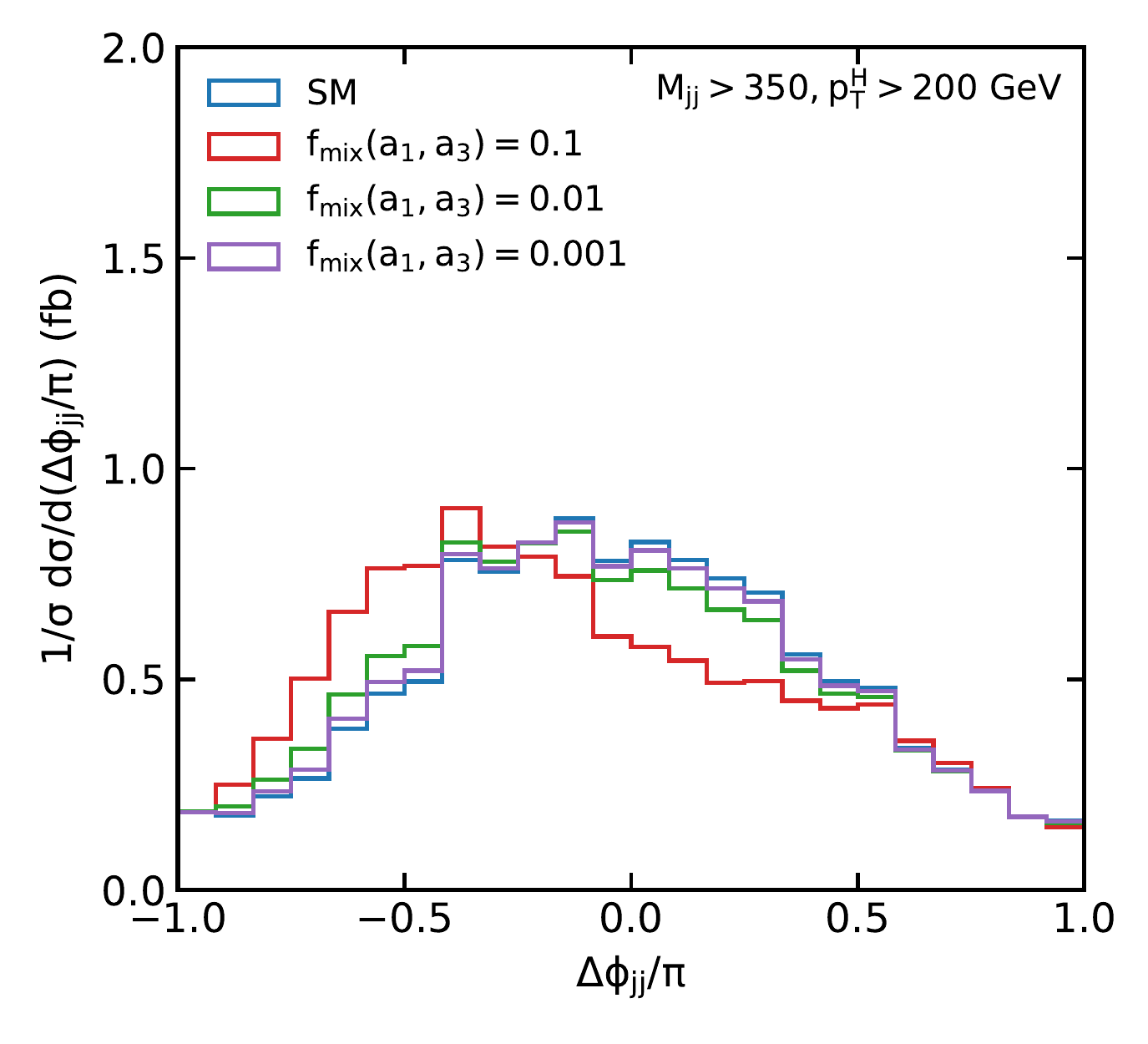}   
\label{fig:Higgs_STXS_VBFCP:CP-bins:d} }%
\end{subfigure}%
 
\caption{ Normalised distributions of the jet-jet azimuthal angle difference as defined in Eq. \ref{eq:Higgs_STXS_VBFCP:dphijj} for various mixed CP scenarios between SM and CP-even (a, b) and CP-odd (c, d) couplings. Colours represents different mixing strength values $f_{\rm mix}=0.1\%, 1\%, 10\%$. }
\label{fig:Higgs_STXS_VBFCP:CP-bins}
\end{figure}

We would like to look at signed $\Delta\phi_{jj}$ in different bins of the STXS stage 1.1 bins definition. For this study, we decided to merge the $M_{jj}$ bins above $350\UGeV$, as the CP-phase has low dependency on the $M_{jj}$ as well as for $p_{T}^{Hjj}$. The $p_{T}^{H}$ split at $200\UGeV$ is however kept as it was proven to be sensitive to BSM \cite{Berger:2019wnu}. The corresponding azimuthal angle distributions are shown in Fig. \ref{fig:Higgs_STXS_VBFCP:CP-bins}.

The interference term in the case of the CP-even coupling is even in term of $\dphi$, so it has the same phase as the SM distribution, see Fig. \ref{fig:Higgs_STXS_VBFCP:CP-bins:a} and Fig. \ref{fig:Higgs_STXS_VBFCP:CP-bins:b}. The maximum deviations occur at the edges and in the centre of the distribution. This deviation is further enhanced by looking at the high $p_{T}^{H}>200$ GeV bin (Fig. \ref{fig:Higgs_STXS_VBFCP:CP-bins:b}). As the amplitude flips around $\pm\pi/2$ a binning of $[-\pi,-\pi/2, \pi/2, \pi]$ would capture any CP-even deviation from the SM expectation in both high and low  $p_{T}^{H}$ bins. 

On the other hand, the CP-odd interference introduces a shift of the $\dphi$ phase toward the negative value. Since the SM distribution is symmetric in $\dphi$, the parity violation must originate from a parity-odd coupling, namely $a_3$ term in Eq. \ref{eq:Higgs_STXS_VBFCP:hvvtensor}. If such coupling occurs at the same time as the CP-even from SM amplitudes or from $a_2$ term it would imply CP-violation in the Higgs sector. This is demonstrated in Fig. \ref{fig:Higgs_STXS_VBFCP:CP-bins:c} and Fig. \ref{fig:Higgs_STXS_VBFCP:CP-bins:d} where we see that the amplitude changes phase at $\dphi=0$ in the low $p_{T}^{H}$ bin. This suggests slight modification of the previous binning by introducing a split at $\dphi=0$, such that the presence of any asymmetries in the cross-section measurement between the positive and the negative bins would indicate the presence of a CP-odd anomaly.

CP-sensitive bins could be also defined for other production modes in stage 1.1 in particular for the gluon-fusion and the Higgsstrahlung processes. For the latter, the same behaviours rise for the HVV vertex, where $\dphi$ could be defined for Higgs + 2 jets events with $M_{jj} < 120\UGeV$. 

\subsection{Conclusion}
\label{sec:Higgs_STXS_VBFCP:conclusion}

On summary, we have presented an extension to the VBF STXS stage 1.1 to include CP-sensitive bins based on the signed $\dphi$ observable. We have demonstrated that the usage for such observable is largely independent of the form factor and allows the probe of the HVV coupling. A binning of $[-\pi,-\pi/2, 0, \pi/2, \pi]$ is proposed under $M_{jj}$ bins above $350\UGeV$ in both high and low $p_{T}^{H}$ branches.

\let\Herwig\undefined
\let\JHUGEN\undefined
\let\Powheg\undefined
\let\Pythia\undefined
\let\Sherpa\undefined
\let\Rivet\undefined
\let\Professor\undefined
\let\eps\undefined
\let\mc\undefined
\let\mr\undefined
\let\mb\undefined
\let\tm\undefined
\let\dphi\undefined

\section{Uncertainties originating from the top mass in Higgs Processes at
the LHC~\protect\footnote{
S.~P.~Jones,
M.~Spira}{}}

\label{sec:higgs_msu}

Loop-induced Higgs-boson production and decay processes are discussed
within the Standard Model with particular emphasis on the uncertainties
induced by the scheme and scale dependence of the top-quark mass. This
uncertainty has quite often been neglected or underestimated in the
past. It turns out to be relevant for off-shell Higgs production and
decay and Higgs boson production at large transverse momenta.

\subsection{Introduction}
The discovery of the scalar resonance with a mass of 125 GeV at the LHC
\cite{Aad:2012tfa,Chatrchyan:2012xdj}, which is compatible with the
Standard Model (SM) Higgs boson \cite{Higgs:1964ia, Higgs:1964pj,
Englert:1964et, Guralnik:1964eu, Higgs:1966ev, Kibble:1967sv}, completed
the SM of strong and electroweak interactions. However, the consistency
with the predictions for a SM Higgs boson has to be tested in more
detail by determining the coupling strengths to the other SM particles
and the self-interactions of the Higgs bosons. This is
achieved by extracting the corresponding couplings from Higgs boson
production and decay processes at the LHC \cite{Khachatryan:2016vau} and
will be pursued in the future runs. The extraction of the basic Lagrange
parameters from the physical observables is plagued by experimental and
theoretical uncertainties that have to be analysed in detail. Focusing
on the theoretical uncertainties, the usual procedure is to study the
factorization and renormalization scale dependences originating from the
parton densities and the strong coupling involved in most production
processes for the QCD uncertainties. In some processes also the
renormalization scale dependence of the Yukawa coupling is included as
e.g.~in $H\to q\bar q$ ($q=b,c$). In addition, the uncertainties due to
the unknown higher-order electroweak corrections have to be added to
obtain a complete estimate of the theoretical uncertainties.

However, in loop-induced Higgs-boson production processes, such as
gluon-fusion $gg\to H, HH$ or the transverse-momentum distribution of
the Higgs particle, another uncertainty plays a relevant role as soon as
a large kinematical energy scale enters the top-quark loops, namely the
uncertainties due to the scheme and scale dependence of the virtual
top-quark mass. This uncertainty has been analysed in on-shell Higgs
production via gluon fusion and turned out to be small. The reason for the 
small uncertainty is that the Higgs mass of 125 GeV is small compared to
the top-quark mass and thus the production of an on-shell SM Higgs boson
is well described by the heavy-top limit (HTL) in which finite top-quark
mass effects and their related uncertainties are suppressed in a natural way.
The same feature is true also for Higgs bosons at
smaller transverse momenta up to $\sim 200-300$ GeV. At large transverse
momenta, however, large momentum scales enter the top-quark loops so
that top mass effects play a relevant role \cite{ptLO1,ptLO2}.

Another important kinematical range of Higgs-boson production is
provided by off-shell Higgs production which -- in combination with
on-shell Higgs production -- allows the total decay width
of the SM Higgs boson to be constrained \cite{Caola:2013yja, Campbell:2013una}. Off-shell
Higgs production is characterized by particular final states that are
mediated both by diagrams with and without an intermediate Higgs-boson exchange.  In this
context it is important to include interference effects rigorously. Far
off-shell Higgs-boson exchange contributions in the $s$-channel are
related to the gluon-fusion process and, at large momenta $Q$, the top-quark loops
are probed above threshold. For these processes, top-quark mass
effects are relevant and thus so are the related mass scheme uncertainties.

\subsection{Off-shell Higgs-boson production via gluon fusion}
Off-shell Higgs-boson production is defined in terms of the full process
$gg\to X$, where $X$ is a particular final state that the Higgs boson
couples to. The differential cross section is built up of three parts,
\begin{equation}
\frac{d\sigma}{d Q^2} = \frac{d\sigma_H}{d Q^2} + \frac{d\sigma_{int}}{d
Q^2} + \frac{d\sigma_{cont}}{d Q^2}
\end{equation}
where $\sigma_H$ denotes the part with an $s$-channel Higgs exchange,
$\sigma_{cont}$ the continuum contribution without $s$-channel Higgs
exchange and $\sigma_{int}$ the interference part between both
contributions. Closed top-quark loops yield the dominant contribution to
$\sigma_H$ and $\sigma_{int}$ due to the diagrams that coincide with the
usual diagrams of on-shell Higgs-boson production via gluon fusion.
Since the Higgs boson is a scalar ${\cal CP}$-even state, no spin
information is transferred through the Higgs propagator and the Higgs
part $\sigma_H$ splits into a production and a decay part (if
higher-order corrections are neglected),
\begin{equation}
\frac{d\sigma_H}{d Q^2} = \frac{Q}{\pi} \frac{\sigma(gg\to H^*)
\times \Gamma(H^*\to X)}{(Q^2-M_H^2)^2 + M_H^2 \Gamma_H^2}
\end{equation}
where the off-shell cross section and decay width are given by
\begin{equation}
\sigma(gg\to H^*) = \left.\sigma(gg\to H)\right|_{M_H^2 \to Q^2}, \qquad
\Gamma(H^*\to X) = \left.\Gamma(H\to X)\right|_{M_H^2 \to Q^2}
\end{equation}
The Standard-Model (SM) Higgs boson production cross section via
gluon-fusion $gg\to H$ is known up to N$^3$LO in QCD
\cite{Djouadi:1991tka, Dawson:1990zj, Graudenz:1992pv, Spira:1995rr,
Harlander:2005rq, Anastasiou:2009kn, Catani:2001ic, Harlander:2001is,
Harlander:2002wh, Anastasiou:2002yz, Ravindran:2003um, Gehrmann:2011aa,
Anastasiou:2013srw, Anastasiou:2013mca, Kilgore:2013gba, Li:2014bfa,
Anastasiou:2014lda, Anastasiou:2015ema, Anastasiou:2015yha,
Anastasiou:2016cez, Mistlberger:2018etf} in the limit of heavy top
quarks and up to NLO QCD \cite{Graudenz:1992pv, Spira:1995rr,
Harlander:2005rq, Anastasiou:2009kn} and NLO electroweak
\cite{Djouadi:1994ge, Chetyrkin:1996wr, Chetyrkin:1996ke,
Aglietti:2004nj, Degrassi:2004mx, Aglietti:2006yd, Actis:2008ug,
Actis:2008ts, Anastasiou:2008tj} including finite quark mass effects
supplemented by soft and collinear gluon resummation up to the N$^3$LL
level \cite{Kramer:1996iq, Catani:2003zt, Moch:2005ky, Ravindran:2005vv,
Ravindran:2006cg, Idilbi:2005ni, Ahrens:2008nc, deFlorian:2009hc,
deFlorian:2012yg, deFlorian:2014vta, Bonvini:2014joa, Bonvini:2014tea,
Catani:2014uta, Schmidt:2015cea}.

\begin{figure}[t]
\begin{center}
\includegraphics[width=0.6\textwidth]{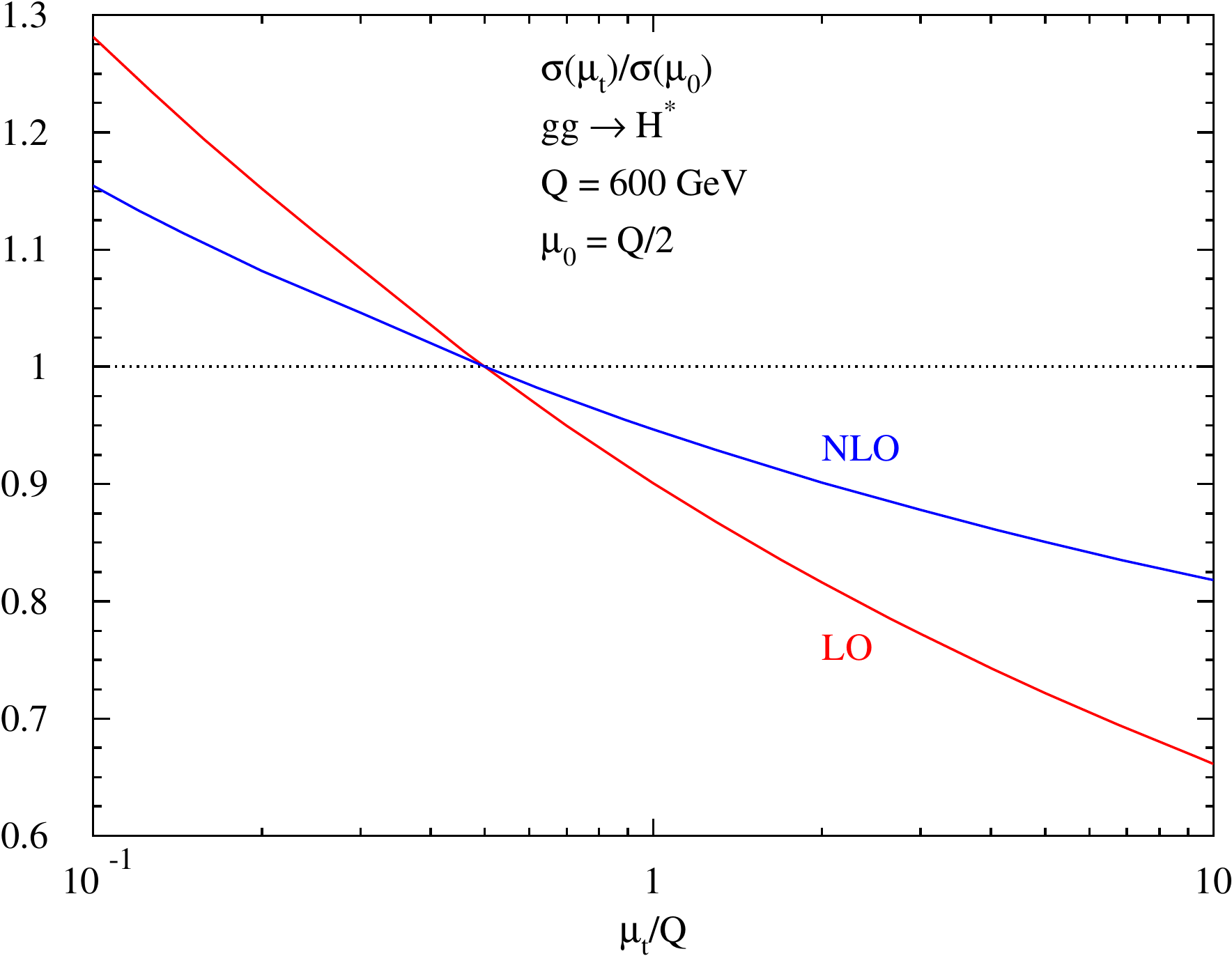}
 \caption{$\mu_t$-scale dependence of the off-shell Higgs production cross
section via gluon fusion for an invariant mass of $Q=600$ GeV normalized
to the cross section at the scale $\mu_t=Q/2$.}
\label{fg:mtscale}
\end{center}
\end{figure}

Since the full quark-mass dependence is only known at NLO QCD, the
uncertainties due to the scheme and scale dependence of the top mass can
only be studied at LO and NLO in depth. We will compare the production
cross section $\sigma(gg\to H)$ in terms of the top pole mass $m_t$ and
the top $\overline{\rm MS}$ mass $\overline{m}_t(\mu_t)$ for different
scale choices of $\mu_t$. 
In this work, we take for the top pole mass $m_t = 172.5\ \mathrm{GeV}$. 
We use the N$^3$LO relation between the top pole
and $\overline{\rm MS}$ mass \cite{Gray:1990yh, Chetyrkin:1999ys,
Chetyrkin:1999qi, Melnikov:2000qh},
\begin{eqnarray}
{\overline{m}}_{t}(m_{t}) & = & \frac{m_{t}}{\kappa(m_t)} \nonumber \\
\kappa(m_t) & = & 1+\frac{4}{3} \frac{\alpha_{s}(m_t)}{\pi} + K_2
\left(\frac{\alpha_s(m_t)}{\pi}\right)^2 + K_3
\left(\frac{\alpha_s(m_t)}{\pi}\right)^3
\label{eq:mspole}
\end{eqnarray}
with $K_2\approx 10.9$ and $K_3\approx 107.1$. The scale dependence of
the $\overline{\rm MS}$ mass is treated at N$^3$LL,
\begin{eqnarray}
{\overline{m}}_{t}\,(\mu_t)&=&{\overline{m}}_{t}\,(m_{t})
\,\frac{c\,[\alpha_{s}\,(\mu_t)/\pi ]}{c\, [\alpha_{s}\,(m_{t})/\pi ]}
\label{eq:msbarev}
\end{eqnarray}
with the coefficient function \cite{Tarasov:1982gk, Chetyrkin:1997dh}
\begin{eqnarray*}
c(x)=\left(\frac{7}{2}\,x\right)^{\frac{4}{7}} \, [1+1.398x+1.793\,x^{2}
- 0.6834\, x^3]
\end{eqnarray*}
Transforming the top pole to the $\overline{\rm MS}$ mass, the
dependence of the gluon-fusion cross section on the scale $\mu_t$ is
shown in Fig.~\ref{fg:mtscale}. A sizable scale dependence can be
inferred from this Figure indicating that this contributes significantly
to the total theoretical uncertainty in addition to the usual
renormalization and factorization scale dependence, which are at the
level of 20\% at NLO (and 5\% at N$^3$LO in the HTL). \\

For the determination of the total contribution to the theoretical
uncertainty, the scale $\mu_t$ will be chosen as either the
$\overline{\rm MS}$ mass $\overline{m}_t$ itself or varied in the range
between $Q/4$ and $Q$ to determine the maximal and minimal cross section
for each value of $Q$, for the central prediction the pole mass is used.
At LO the uncertainties obtained in this way
are just the parametric dependences on the value of the top mass
involved in the loop function,
\begin{eqnarray}
\sigma(gg\to H^*)\Big|_{Q=125~{\rm GeV}} & = 18.43^{+0.8\%}_{-1.1\%}\,
\mathrm{pb}, \qquad
\sigma(gg\to H^*)\Big|_{Q=300~{\rm GeV}} & = 4.88^{+23.1\%}_{-1.1\%}\,
\mathrm{pb} \nonumber \\[0.5cm]
\sigma(gg\to H^*)\Big|_{Q=400~{\rm GeV}} & = 4.94^{+1.2\%}_{-1.8\%}\,
\mathrm{pb}, \qquad
\sigma(gg\to H^*)\Big|_{Q=600~{\rm GeV}} & = 1.13^{+0.0\%}_{-26.2\%}\,
\mathrm{pb} \nonumber \\[0.5cm]
\sigma(gg\to H^*)\Big|_{Q=900~{\rm GeV}} & = 0.139^{+0.0\%}_{-36.0\%}\,
\mathrm{pb}, \quad
\sigma(gg\to H^*)\Big|_{Q=1200~{\rm GeV}} & = 0.0249^{+0.0\%}_{-41.1\%}\,
\mathrm{pb} \nonumber \\[0.5cm]
~&~&
\end{eqnarray}
with the numbers obtained for a c.m.~energy of 14 TeV and using
PDF4LHC15 NLO parton densities with a NLO strong coupling normalized to
$\alpha_s(M_Z)=0.118$\footnote{Note that these choices are incompatible
with a consistent LO prediction, but the relative uncertainties
related to the scheme and scale choice of the top mass will be hardly
affected by this inconsistency.}. At NLO the results for the production
cross section at different values of $Q$ are given by
\begin{eqnarray}
\sigma(gg\to H^*)\Big|_{Q=125~{\rm GeV}} & = 42.17^{+0.4\%}_{-0.5\%}\,
\mathrm{pb}, \qquad
\sigma(gg\to H^*)\Big|_{Q=300~{\rm GeV}} & = 9.85^{+7.5\%}_{-0.3\%}\,
\mathrm{pb} \nonumber \\[0.5cm]
\sigma(gg\to H^*)\Big|_{Q=400~{\rm GeV}} & = 9.43^{+0.1\%}_{-0.9\%}\,
\mathrm{pb}, \qquad
\sigma(gg\to H^*)\Big|_{Q=600~{\rm GeV}} & = 1.97^{+0.0\%}_{-15.9\%}\,
\mathrm{pb} \nonumber \\[0.5cm]
\sigma(gg\to H^*)\Big|_{Q=900~{\rm GeV}} & = 0.230^{+0.0\%}_{-22.3\%}\,
\mathrm{pb}, \quad
\sigma(gg\to H^*)\Big|_{Q=1200~{\rm GeV}} & = 0.0402^{+0.0\%}_{-26.0\%}\,
\mathrm{pb} \nonumber \\[0.5cm]
~&~&
\end{eqnarray}
These results indicate that the uncertainties related to the scheme and
scale dependence of the top-quark mass drop by roughly a factor of two
from LO to NLO, but not more, i.e.~about one half of the parametric
dependence is compensated by the logarithms involving the scale $\mu_t$
and the scheme-transformation part of the NLO corrections. The reason
for the sizable residual scale and scheme dependence is the large size
of the NLO corrections that will be compensated further by the mass
effects of the NNLO corrections. The latter are only approximately known
in terms of a large top-mass expansion at present which cannot be used
for large values of $Q$ \cite{Harlander:2009bw, Pak:2009bx,
Harlander:2009mq, Pak:2009dg, Davies:2019nhm, Czakon:2020vql}.

\subsection{Off-shell $H^*\to\gamma\gamma$}
\begin{figure}[t]
\begin{center}
\includegraphics[width=0.9\textwidth]{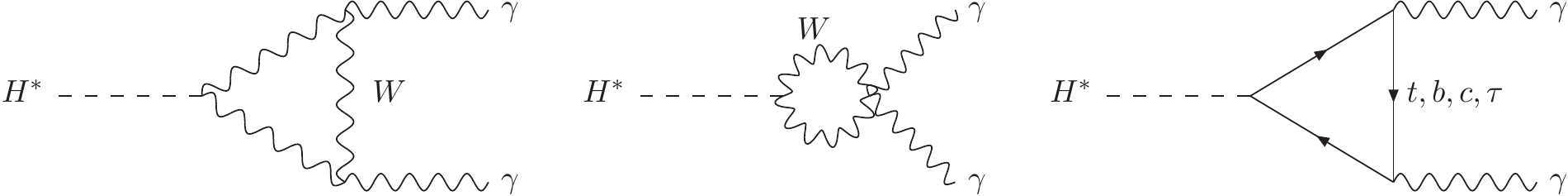}
\caption{\label{fg:hgagadia} \it Diagrams contributing to (off-shell)
Higgs-boson decays into photon pairs.}
\end{center}
\end{figure}
\begin{figure}[t]
\begin{center}
\includegraphics[width=0.6\textwidth]{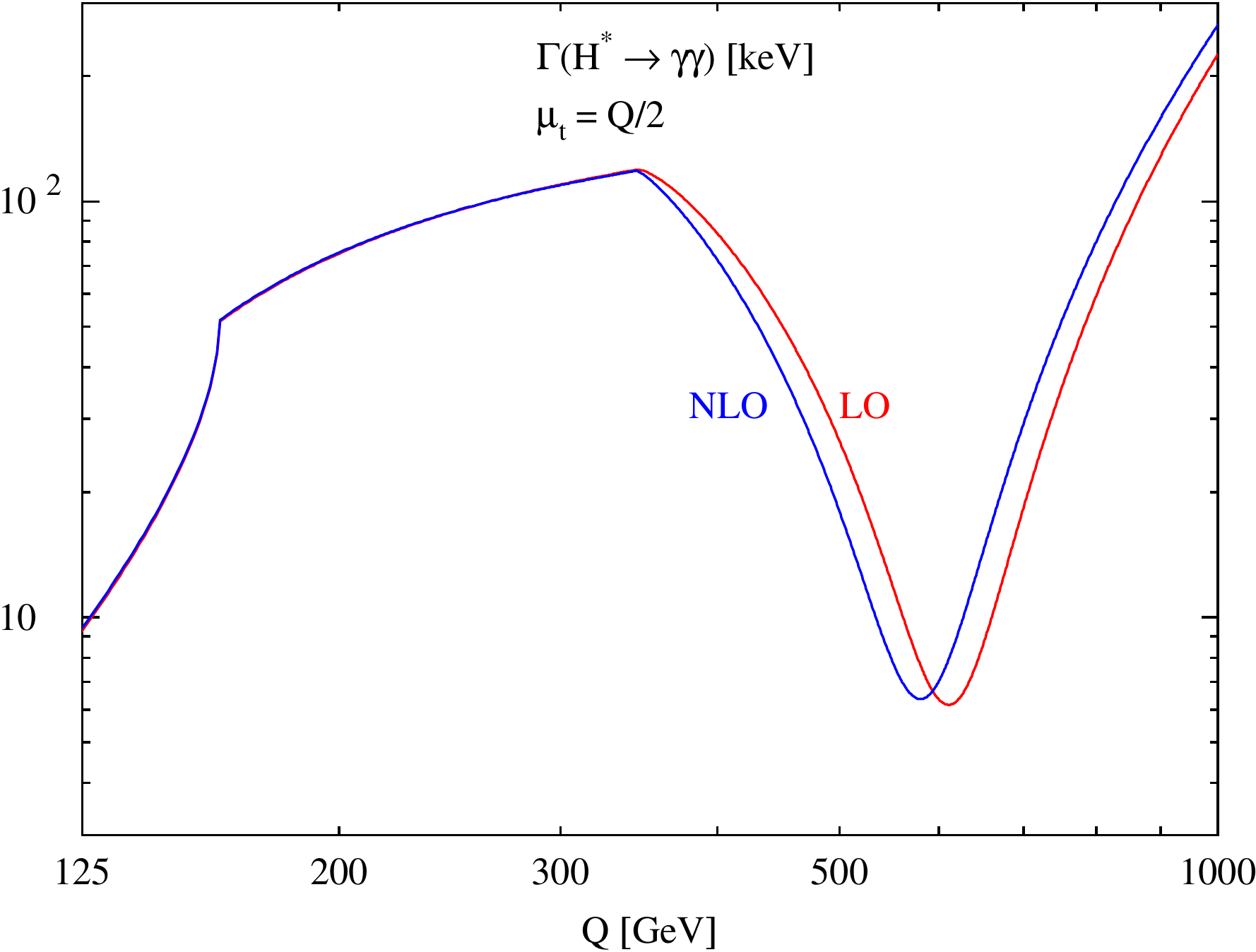}
 \caption{Partial decay width of off-shell Higgs bosons into photon
pairs as a function of the virtuality $Q$.}
\label{fg:hgaga}
\end{center}
\end{figure}

\noindent
Uncertainties due to the scheme and scale choice of the top mass will
also play a role for off-shell Higgs decays $H^*\to\gamma\gamma$, since
above the $t\bar t$ threshold destructive interference effects between
the $W$ and the top loops (see Fig.~\ref{fg:hgagadia}) become sizeable.
In fact, the off-shell decay width nearly vanishes for an off-shellness
of $Q\sim 600$ GeV as can be inferred from Fig.~\ref{fg:hgaga}, which
shows the off-shell decay width into photon pairs at LO and NLO QCD
\cite{Spira:1995rr, Djouadi:1990aj, Melnikov:1993tj, Djouadi:1993ji,
Inoue:1994jq, Fleischer:2004vb}.  Electroweak corrections are neglected
for consistency. In order to keep the NLO QCD corrections at a moderate
level throughout the full range in $Q$ the top mass has been defined as
the running quark mass
\begin{equation}
m_t(\mu_t) = \kappa(m_t) \overline{m}_t(\mu_t)
\end{equation}
with $\kappa(m_t)$ defined in Eq.~(\ref{eq:mspole}). The default choice
of the scale $\mu_t$ is adopted as $\mu_t = Q/2$ for the central
prediction along the lines of defining the partial width
$\Gamma(H\to\gamma\gamma)$ in {\tt Hdecay} \cite{Djouadi:1997yw,
Djouadi:2018xqq} so that for the case $Q=2m_t$ the running top mass
coincides with the pole mass and the virtual $t\bar t$ threshold is not
shifted. In our numerical analysis we also include the bottom- and
charm- as well as $\tau$-lepton loops at their default choices.

For the related uncertainties induced by the scheme and scale choice of
the top mass we derive the minimal and maximal partial width for the
pole mass and the running top mass at scales varied between $Q/4$ and
$Q$. The residual uncertainties at NLO read
\begin{eqnarray}
\Gamma(H^*\to \gamma\gamma)\Big|_{Q=125~{\rm GeV}} & = 9.43^{+0.1\%}_{-0.4\%}\,
\mathrm{keV}, \qquad
\Gamma(H^*\to \gamma\gamma)\Big|_{Q=300~{\rm GeV}} & = 109.4^{+0.5\%}_{-2.2\%}\,
\mathrm{keV} \nonumber \\[0.5cm]
\Gamma(H^*\to \gamma\gamma)\Big|_{Q=400~{\rm GeV}} & = 72.3^{+9.9\%}_{-35\%}\,
\mathrm{keV}, \qquad
\Gamma(H^*\to \gamma\gamma)\Big|_{Q=600~{\rm GeV}} & = 7.03^{+156\%}_{-35\%}\,
\mathrm{keV} \nonumber \\[0.5cm]
\Gamma(H^*\to \gamma\gamma)\Big|_{Q=900~{\rm GeV}} & = 158.7^{+16\%}_{-1.5\%}\,
\mathrm{keV}, \quad
\Gamma(H^*\to \gamma\gamma)\Big|_{Q=1200~{\rm GeV}} & = 572.3^{+3.4\%}_{-0\%}\,
\mathrm{keV} \nonumber \\[0.5cm]
~&~&
\end{eqnarray}
where the very large uncertainty at $Q=600$ GeV is generated by the
strong cancellation between the $W$ and top loops as indicated by the
strong dip in Fig.~\ref{fg:hgaga}. In analogy to the production cross
section these results indicate that the top-mass scheme and scale
uncertainties are sizeable for off-shell Higgs bosons and have to be
included in analyses to derive the total Higgs width from the interplay
between on-shell and off-shell Higgs production and decay.

\subsection{Higgs-pair production}
Higgs-pair production via gluon fusion is mediated by triangle and box
diagrams involving closed top-quark loops at LO \cite{Glover:1987nx,
Plehn:1996wb}, the decomposition is therefore very similar to the usual
off-shell Higgs-boson production corresponding to the triangle diagrams
and the continuum contribution in terms of the box diagrams. Both
contributions develop a relevant top-mass dependence so that the related
uncertainties have to be included in the total theoretical uncertainty.

\begin{figure}[t]
\begin{center}
\includegraphics[width=0.9\textwidth]{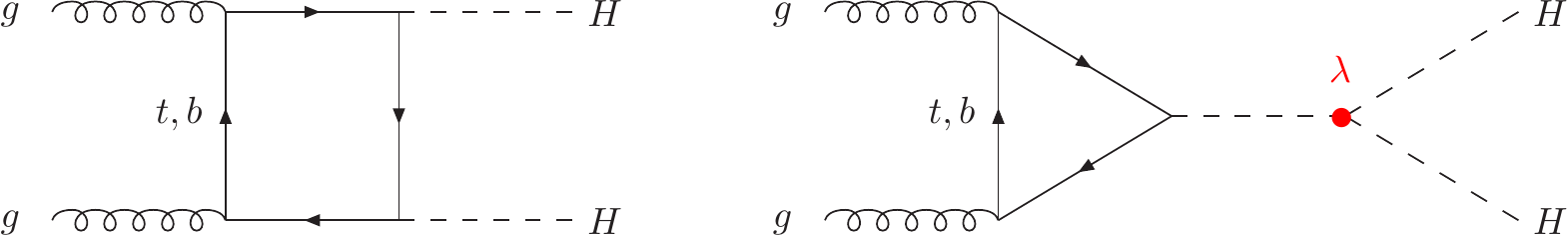}
\caption{\label{fg:hhdia} \it Diagrams contributing to Higgs-boson pair
production via gluon fusion. The contribution of the trilinear Higgs
coupling is marked in red.}
\end{center}
\end{figure}
Higgs-boson pairs are mainly produced via the gluon-fusion mechanism
$gg\to HH$ which is primarily mediated by top-quark loops and receives only a small contribution
from bottom-quark loops, see Fig.~\ref{fg:hhdia}. There are box (left
diagram) and triangle (right diagram) diagrams, with the latter involving
the trilinear Higgs coupling $\lambda$ \cite{Glover:1987nx,Plehn:1996wb}, which
interfere destructively. The dependence of the cross section on the size
of the trilinear coupling can roughly be estimated  as
$\Delta\sigma/\sigma \sim -\Delta\lambda/\lambda$ in the vicinity of the
SM value of $\lambda$. Thus, the determination of the trilinear coupling
from Higgs pair production requires a reduction of the theoretical
uncertainties of the corresponding cross section, i.e.~the inclusion of
higher-order corrections becomes indispensable. The full QCD corrections
are known up next-to-leading order (NLO) \cite{Borowka:2016ehy,
Borowka:2016ypz, Baglio:2018lrj} and at next-to-next-to-leading order
(NNLO) in the limit of heavy top quarks \cite{deFlorian:2013uza,
deFlorian:2013jea, Grigo:2014jma}. Very recently, the N$^3$LO QCD
corrections have been computed in the limit of heavy top quarks
resulting in a small further increase of the cross section
\cite{Banerjee:2018lfq, Chen:2019lzz, Chen:2019fhs}. The QCD corrections
increase the total LO cross section by about a factor of two. Recently,
the full NLO results have been matched to parton showers
\cite{Heinrich:2017kxx, Jones:2017giv} and the full NNLO results in the
limit of heavy top quarks have been merged with the NLO mass effects and
supplemented by additional top-mass effects in the double-real
corrections \cite{Grazzini:2018bsd}. However, a reliable estimate of the
theoretical uncertainties is necessary, i.e.~considering the usual
renormalization and factorization scale dependences but in addition also
the uncertainties induced by the top-mass scheme and scale dependence.

This analysis has been performed in Ref.~\cite{Baglio:2018lrj} for the first
time including the full NLO QCD corrections. The final results look very
similar to the single off-shell Higgs case, i.e.~the top-mass scheme and
scale uncertainties drop by roughly a factor of two from LO to NLO. At
LO we obtain the uncertainties
\begin{eqnarray}
\frac{d\sigma(gg\to HH)}{dQ}\Big|_{Q=300~{\rm GeV}} & = &
0.01656^{+62\%}_{-2.4\%}\, \mathrm{fb/GeV},\nonumber\\
\frac{d\sigma(gg\to HH)}{dQ}\Big|_{Q=400~{\rm GeV}} & = &
0.09391^{+0\%}_{-20\%}\, \mathrm{fb/GeV},\nonumber\\
\frac{d\sigma(gg\to HH)}{dQ}\Big|_{Q=600~{\rm GeV}} & = &
0.02132^{+0\%}_{-48\%}\, \mathrm{fb/GeV},\nonumber\\
\frac{d\sigma(gg\to HH)}{dQ}\Big|_{Q=1200~{\rm GeV}} & = &
0.0003223^{+0\%}_{-56\%}\, \mathrm{fb/GeV}
\end{eqnarray}
where the full spread of the cross sections including the top pole mass
(central values) and the scale $\mu_t$ of the $\overline{\rm MS}$ mass
$\overline{m}_t(\mu_t)$ chosen as either the $\overline{\rm MS}$ mass
$\overline{m}_t$ itself or varied in the range between $Q/4$ and $Q$ as
in the single (off-shell) Higgs case considered earlier.  The final NLO
results read \cite{Baglio:2018lrj}
\begin{eqnarray}
\frac{d\sigma(gg\to HH)}{dQ}\Big|_{Q=300~{\rm GeV}} & = &
0.02978(7)^{+6\%}_{-34\%}\, \mathrm{fb/GeV},\nonumber\\
\frac{d\sigma(gg\to HH)}{dQ}\Big|_{Q=400~{\rm GeV}} & = &
0.1609(7)^{+0\%}_{-13\%}\, \mathrm{fb/GeV},\nonumber\\
\frac{d\sigma(gg\to HH)}{dQ}\Big|_{Q=600~{\rm GeV}} & = &
0.03204(9)^{+0\%}_{-30\%}\, \mathrm{fb/GeV},\nonumber\\
\frac{d\sigma(gg\to HH)}{dQ}\Big|_{Q=1200~{\rm GeV}} & = &
0.000435(6)^{+0\%}_{-35\%}\, \mathrm{fb/GeV}
\end{eqnarray}
Since these uncertainties are similar in size to the renormalization
and factorization scale dependences they constitute an important
contribution to the total theoretical uncertainties.

\begin{figure}[t]
\begin{center}
\includegraphics[width=0.6\textwidth]{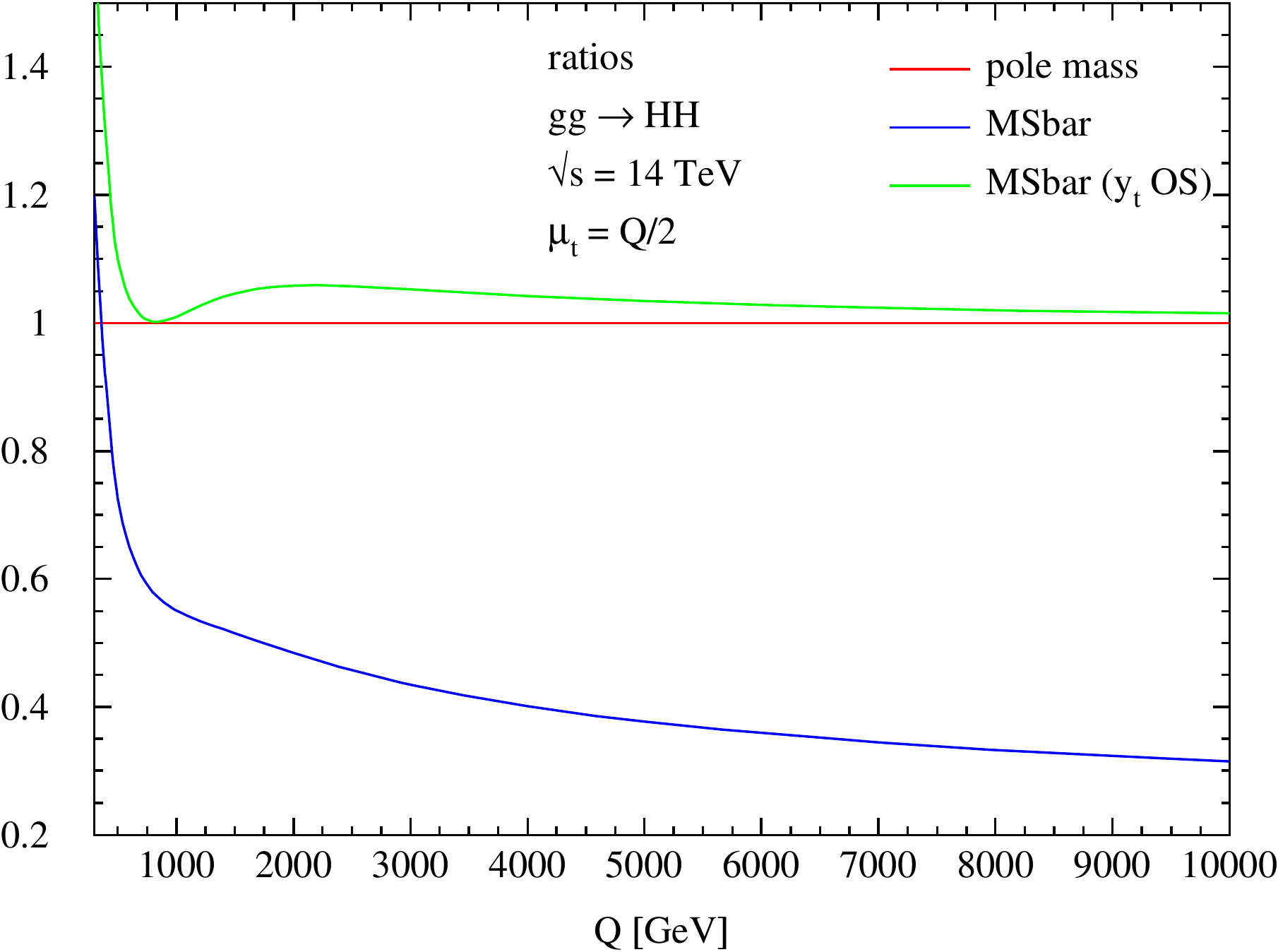}
\caption{\label{fg:hhrat} \it 
Ratio of the Higgs boson pair invariant mass distribution computed in various mass schemes to the on-shell result, at LO. The $\overline{\mathrm{MS}}$ ($y_t^\mathrm{OS}$) curve is produced by fixing the top Yukawa coupling to its on-shell value but using the $\overline{\mathrm{MS}}$ scheme for the top-quark mass elsewhere in the calculation. 
}
\end{center}
\end{figure}

In Fig.~\ref{fg:hhrat} we display the LO ratio of the
$\overline{\mathrm{MS}}$ calculation to the result obtained using the
pole mass scheme.  The difference between the result in the two schemes
is visible at all values of the invariant mass of the Higgs pair, $Q$,
and grows at large invariant mass.
The Higgs boson pair production amplitude depends on the top-quark mass both via the Yukawa coupling, $y_t \propto m_t$, and the mass appearing in the quark propagators.
After integration over the loop momentum, the mass appearing in the quark propagators can give rise to logarithms involving a ratio of the quark mass and the other relevant scales in the problem.
As a technical exercise, we may investigate at LO how the amplitude
behaves when the Yukawa coupling is fixed to its value in the on-shell
scheme but the mass appearing elsewhere in the calculation is left
scheme dependent. We find that with the Yukawa coupling fixed, the
difference between the schemes at high energy is significantly reduced,
see Fig.~\ref{fg:hhrat} (green curve).  

The asymptotic convergence at LO of the differential cross section with
the Yukawa coupling in terms of the top pole mass, but the propagator
mass in the $\overline{\rm MS}$ scheme, towards the differential cross
section defined entirely in terms of the top pole mass can be understood
immediately from the asymptotic expansions of
Ref.~\cite{Davies:2018qvx}. The amplitude may be written as the sum of
two form factors, $F_1$ and $F_2$, describing the scattering of incoming gluons with the
same helicity and opposite helicities, respectively. The contribution of the box
diagrams to the two form factors dominates at high energy.
Expanding the LO and NLO results of Ref.~\cite{Davies:2018qvx} around
large invariant Higgs-pair mass, $s$, we have in the on-shell scheme and
in the notation of Ref.~\cite{Davies:2018qvx},
\begin{eqnarray}
F_\mathrm{box,i} & = & F_\mathrm{box,i}^{(0)} +
\frac{\alpha_s(\mu_R)}{\pi} F_\mathrm{box,i}^{(1)} \qquad (i=1,2) \nonumber \\
F_\mathrm{box,i}^{(0)} & = & \frac{m_t^2}{s} c_{0,i} +
\mathcal{O}\left(\frac{1}{s^2}\right)  \nonumber \\
F_\mathrm{box,i}^{(1)} & = & 2 F_\mathrm{box,i}^{(0)} \log\frac{m_t^2}{s}
+ \frac{m_t^2}{s}  c_{1,i} +
\mathcal{O}\left(\frac{1}{s^2}\right)
\label{eq:hhexp}
\end{eqnarray}
where the coefficients $c_{0,i}$ and $c_{1,i}$ do not depend on the
top-quark mass. The overall factor of $m_t^2$ for the box contribution
originates entirely from the Yukawa couplings and examining $F_\mathrm{box,i}^{(0)}$,  
we find that the form factors are independent of the propagator top mass in the high-energy limit.
Therefore, for a fixed Yukawa coupling, we expect the results in different schemes to asymptote.
Transforming the top pole-mass $m_t$ into the $\overline{\rm MS}$ mass
$\overline{m}_t(\mu_t)$ the explicit expressions above are modified to
\cite{Baglio:future}
\begin{eqnarray}
F_\mathrm{box,i}^{(0)} & = & \frac{\overline{m}_t^2(\mu_t)}{s} c_{0,i}
+ \mathcal{O}\left(\frac{1}{s^2}\right) \qquad (i=1,2) \nonumber \\
F_\mathrm{box,i}^{(1)} & = & 2 F_\mathrm{box,i}^{(0)}
\left[\log\frac{\mu_t^2}{s} + \frac{4}{3}\right] +
\frac{\overline{m}_t^2(\mu_t)}{s} c_{1,i} +
\mathcal{O}\left(\frac{1}{s^2}\right)
\end{eqnarray}
This underlines that, in order to absorb the large logarithmic terms $\log\frac{m_t^2}{s}$, 
the scale choice $\mu_t=\sqrt{s}$ is the preferred {\it central} scale choice of the 
Yukawa couplings at large values of $s$. Since
for this scale choice the form factors are independent of the scale and
scheme of the propagator top mass, it is expected that the form factors
will approach each other only for a running top Yukawa coupling at a
large scale $\mu_t=\kappa \sqrt{s}$ with a coefficient $\kappa$ not too
far from unity.

Another interesting feature of Fig.~\ref{fg:hhrat} is the bump visible
at $Q \sim 2\ \mathrm{TeV}$, which exists due to an interplay of
different form factors.  The first form factor, $F_1$, dominates near to the Higgs 
pair production threshold, whilst at high energy $F_2$ dominates. 
At LO the contribution from the two form factors is equal at 
$Q \approx 1750\ \mathrm{GeV}$.  We find that the
$\overline{\mathrm{MS}}$ calculation with the Yukawa coupling fixed to
its on-shell value asymptotes more slowly to the on-shell scheme result
for form factor $F_2$. This leads to a bump in the ratio plot
around the $Q$ value at which latter form factor begins to dominate.

\subsection{Higgs boson transverse-momentum distribution}
\begin{figure}[t]
\begin{center}
\includegraphics[width=0.9\textwidth]{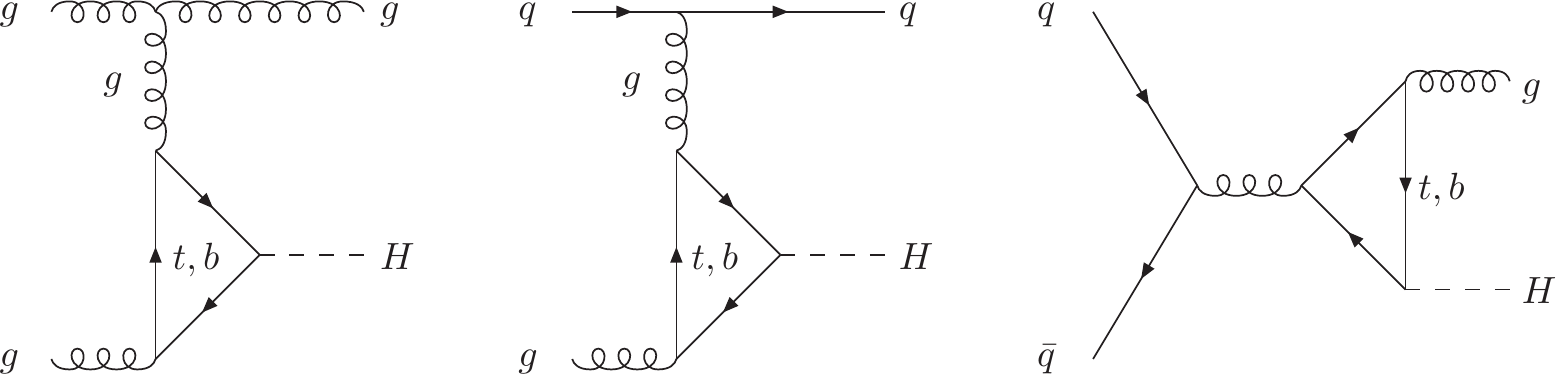}
\caption{\label{fg:hpt} \it Typical diagrams contributing to the
$p_T$ distribution of Higgs bosons via gluon fusion at LO.}
\end{center}
\end{figure}

Higgs-boson production via gluon fusion at large transverse momenta
$p_T$ forms part of the real corrections to the gluon fusion process, since
the Higgs boson has to be produced in association with a massless
parton, i.e.~through the processes $gg, q\bar q\to Hg, gq\to Hq, g\bar
q\to H\bar q$ (see Fig.~\ref{fg:hpt}). The quark-initiated
processes have to be summed over 5 flavours. A Higgs boson recoiling
against the parton at large transverse momenta introduces a large
kinematical momentum scale entering the top loops so that finite
top-mass effects start to be sizeable for larger transverse momenta.
Thus, the related uncertainties will be relevant for the total
theoretical uncertainties.

The leading-order (LO) Higgs transverse momentum distribution in gluon
fusion is known including the full top-mass dependence
\cite{ptLO1,ptLO2}. The NLO QCD corrections have first been determined
in the HTL and increase the distribution by roughly a factor of
two~\cite{ptNLO0,ptNLO1,ptNLO2,ptNLO3}. At NLO, top-mass effects have
been estimated by a large top-mass expansion of the NLO effects
\cite{Harlander:2012hf, Neumann:2014nha}. They have later been
supplemented by the inclusion of the full top-mass dependence in the
real corrections \cite{Neumann:2016dny}, similar to the approach used in
Ref.~\cite{Frederix:2016cnl}. 
The NLO corrections have also been computed in a small-mass expansion~\cite{Kudashkin:2017skd,Lindert:2018iug}, valid at large Higgs transverse momentum.
Subsequently, the full top-mass dependence was obtained by performing a numerical integration of the associated
two-loop diagrams \cite{Jones:2018hbb}. In the HTL the NNLO QCD
corrections have been derived with a moderate increase of the
transverse-momentum distribution and a considerable reduction of the
scale dependence~\cite{ptNNLO1,ptNNLO2,ptNNLO3}.

For small $p_T$ values the fixed-order results diverge thus requiring a
resummation of the (singular) logarithmic terms to all orders
\cite{Dokshitzer:1978hw, Parisi:1979se, Collins:1984kg, Bozzi:2005wk,
Catani:2010pd}. This necessitates the derivation of a resummed kernel
that is matched to the fixed-order result at large $p_T$.  This matching
introduces an unphysical matching scale.  Using the $b$-space formalism,
transverse-momentum resummation for Higgs production in gluon fusion in
the HTL has been derived at next-to-next-to-leading logarithmic accuracy
(NNLL) and matched to the NNLO cross section \cite{Bozzi:2005wk}, which
includes the NLO QCD-corrected expression for the transverse momentum
spectrum at large $p_T$.  Recently, a formulation of transverse momentum
resummation in direct space has been discussed in
Ref.~\cite{Monni:2016ktx}, which has been used to obtain the next-to-NNLL
(N$^3$LL) resummed prediction matched to the NNLO transverse-momentum
spectrum of the Higgs boson in the HTL
\cite{Bizon:2018foh,Bizon:2017rah}. Finite quark-mass effects have been
considered in the resummed spectrum up to NLL+NLO \cite{Mantler:2012bj,
Grazzini:2013mca, Hamilton:2015nsa}.

Considering Higgs boson plus jet production via a top loop at LO, we obtain the uncertainties:
\begin{eqnarray}
\frac{d\sigma(pp\to Hj)}{dm_{hj}} \Big|_{\substack{m_{hj}=700~{\rm GeV}\\ p_{T,j_1} > 30\ \mathrm{GeV} } } & = & 1.22^{+ 0.0\%}_{-2.3\%} \, \mathrm{fb/GeV},\nonumber\\
\frac{d\sigma(pp\to Hj)}{dm_{hj}} \Big|_{\substack{m_{hj}=700~{\rm GeV}\\ p_{T,j_1} > 300\ \mathrm{GeV} } } & =& 0.107^{+ 0.0\%}_{-12\%}\, \mathrm{fb/GeV}
\end{eqnarray}
where $m_{hj}$ is the invariant mass of the Higgs boson plus leading-jet system. For our central prediction we use the top-quark pole mass and to assess the uncertainty we compare to the $\overline{\rm MS}$ scheme with $\mu_t = m_{hj}/2$.

\begin{figure}[t]
\begin{center}
\includegraphics[width=0.49\textwidth]{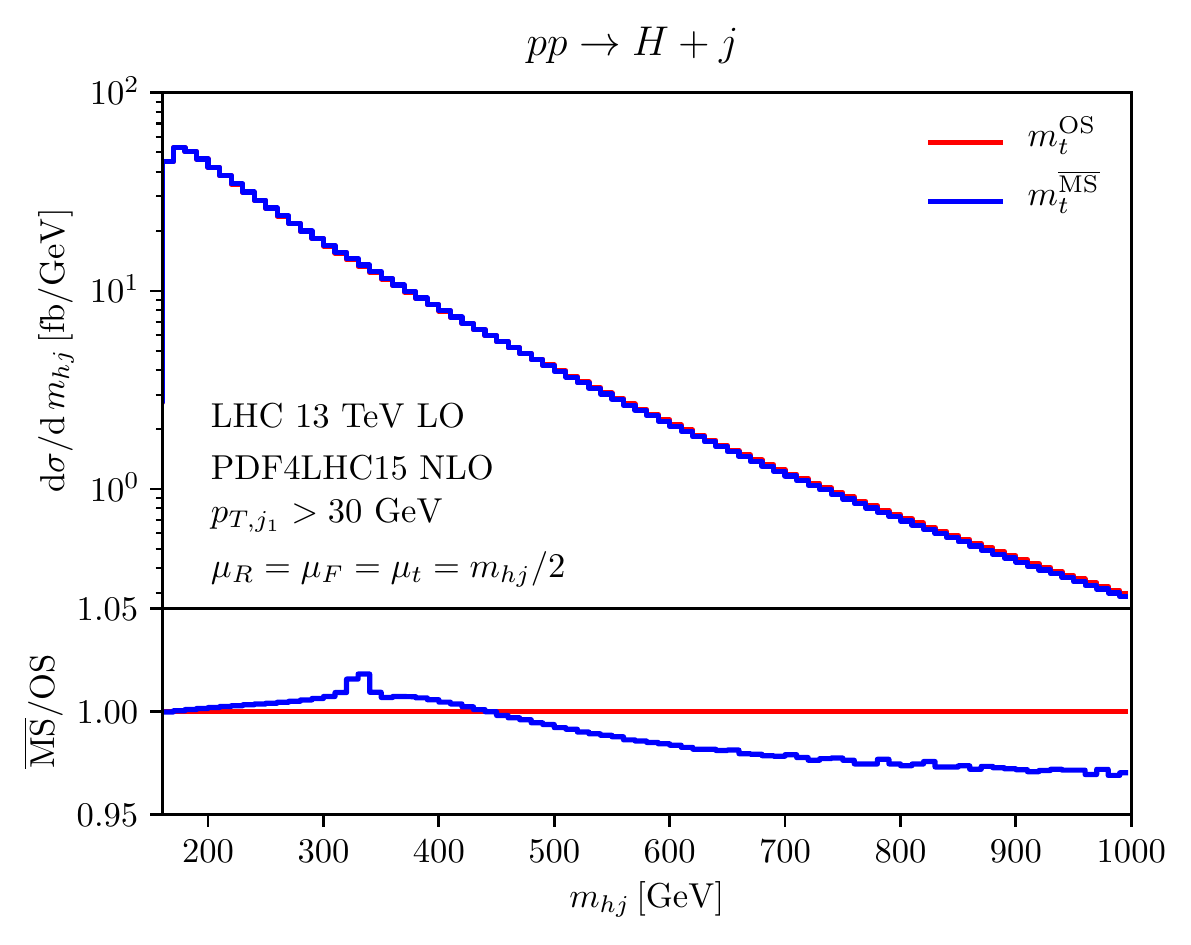}
\includegraphics[width=0.49\textwidth]{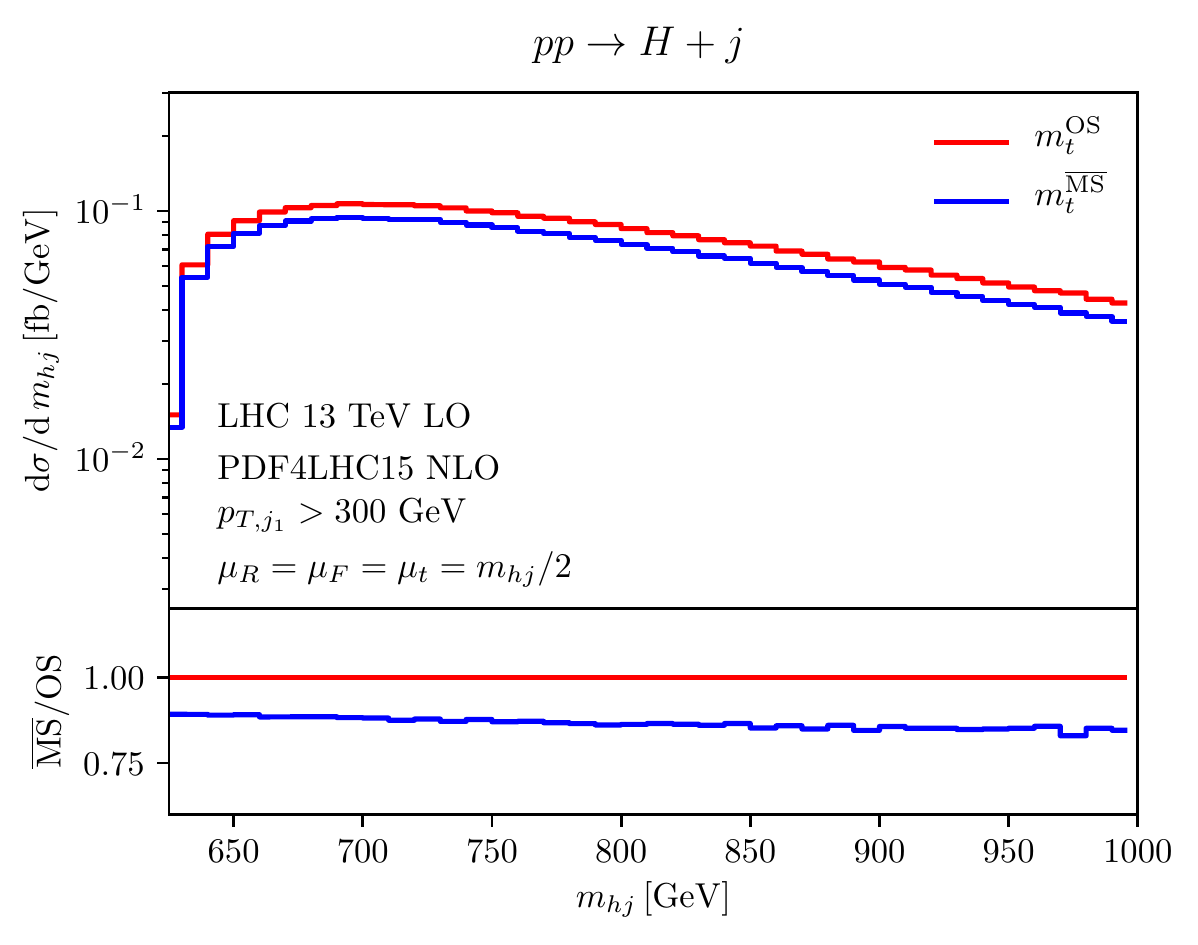}

\caption{\label{fg:hjmhj} \it Invariant mass distribution of the Higgs
boson plus jet system at LO using the on-shell top-quark mass scheme and
the $\overline{\rm MS}$ scheme with $\mu_t=m_{hj}$/2.  The left (right)
panel shows the distribution with a leading jet $p_T$ cut of $p_{T,j_1}
> 30\ \mathrm{GeV}$ ($p_{T,j_1} > 300\ \mathrm{GeV}$) applied.  Figure
produced using MCFM~\cite{Campbell:1999ah, Campbell:2011bn,
Campbell:2015qma}.  }

\end{center}
\end{figure}

In the left panel of Fig.~\ref{fg:hjmhj} we display the invariant mass
distribution of the Higgs boson plus leading-jet system including only
the dominant top-quark contribution.  Applying a regularising cut on the
leading jet of $p_{T,j_1} > 30\ \mathrm{GeV}$, we find that, contrary to
the previous processes considered, the top-quark mass scheme uncertainty
is small (below $\sim 3\%$ everywhere for $m_{hj} < 1\ \mathrm{TeV}$).
The reason for this small uncertainty is that, for all values of the
invariant mass, the distribution is dominated by contributions with a
small $p_T$ which do not probe the top-quark loop above threshold.  For
the $2 \to 2$ Born-like kinematics we have $p_{T,h}^2 = tu/s$, where
$s,t$ and $u$ are the Mandelstam invariants of the partonic process.
Therefore, even at large $m_{hj}$ we have a below threshold low-$p_T$
contribution from $t$- and $u$-channel gluon exchanges, see the left and
center diagrams of Fig.~\ref{fg:hpt}.  Note that, for the same reason,
the HTL approximates the full result reasonably well even for large
values of $m_{hj}$.

At large $p_{T,h}$ (= $p_{T,j_1}$ at LO) all Mandelstam invariants are
large and the top-quark loop is probed above threshold even by $t$- and
$u$-channel exchanges.  Applying a cut on the leading jet of $p_{T,j_1}
> 300\ \mathrm{GeV}$ we force the top-quark loop above threshold and
consequently we see a much larger mass scheme uncertainty of $\sim 10\%$
rising slightly as $m_{hj}$ increases, see the right panel of
Fig.~\ref{fg:hjmhj}.

\begin{figure}[t]
\begin{center}
\includegraphics[width=0.49\textwidth]{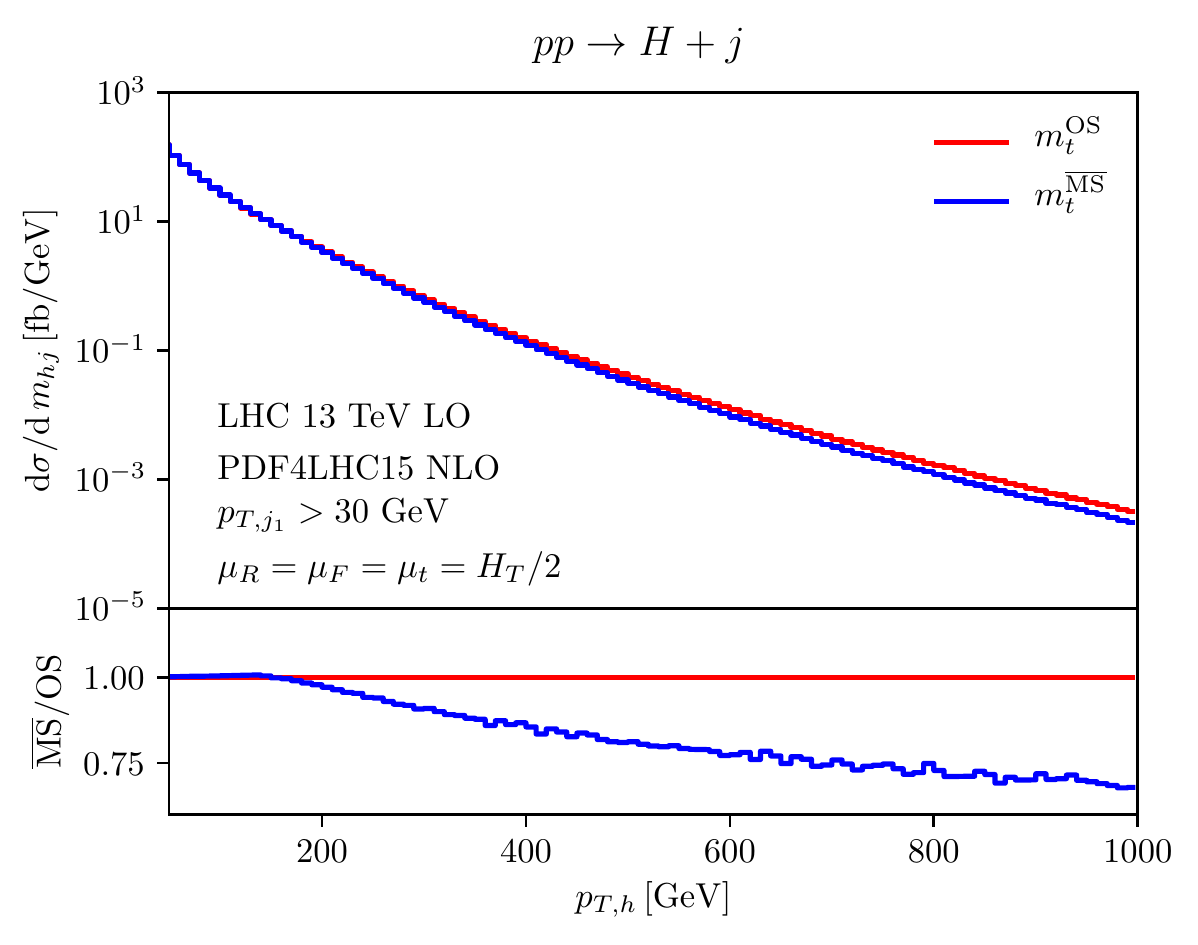}
\caption{\label{fg:hjpt} \it Transverse momentum distribution for Higgs
boson production in gluon fusion at LO using the on-shell top-quark mass
scheme and the $\overline{\rm MS}$ scheme with $\mu_t = H_T/2 = 1/2
\left( \sqrt{m_h^2 +p_{t,h}^2} + \sum_i | p_{t,i}| \right)$, where the
sum runs over the transverse momentum of all final state partons.
Figure produced using
MCFM~\cite{Campbell:1999ah,Campbell:2011bn,Campbell:2015qma}.}
\end{center}
\end{figure}

In Fig.~\ref{fg:hjpt} we display the Higgs boson transverse momentum
distribution.  For small values of $p_{T,h}$, where the top-quark loop
is probed below threshold, the mass scheme uncertainty is rather small.
As $p_{T,h}$ increases beyond $m_t$ the difference between the on-shell
and $\overline{\rm MS}$ scheme increases.  The size of the mass scheme
uncertainty for his process is therefore similar to that of the other
loop-induced processes which we have considered.

\subsection{Conclusions}
We have discussed and analysed the theoretical uncertainties of
Higgs-boson production and decay processes induced by the scheme and
scale dependence of the top-quark mass. This uncertainty turns out to be
sizeable in all gluon-induced processes of the Higgs particle and needs
to be taken into account for a full and rigorous estimate of the
residual theoretical uncertainties, in particular in kinematical regimes
in which a large momentum scale enters the corresponding top-quark loops.
We anticipate that this uncertainty could also play a role in
processes not involving the Higgs boson directly, for example, 
the gluon-induced continuum contribution to
$Z$ boson pair production. Similarly to Higgs boson pair and 
Higgs boson plus jet production, this contribution is also mediated by quark-box
diagrams involving the top quark.

\subsection*{Acknowledgments}

We would like to thank R.~R\"{o}ntsch for help and advice regarding the
use of MCFM.  We are also grateful to J.~Baglio, A.~Huss and M.~Kerner
for cross-checking part of our results.


\newcommand{\Herwig}{H\protect\scalebox{0.8}{ERWIG}\xspace}
\newcommand{\Pythia}{P\protect\scalebox{0.8}{YTHIA}\xspace}
\newcommand{\Sherpa}{S\protect\scalebox{0.8}{HERPA}\xspace}
\newcommand{\Rivet}{R\protect\scalebox{0.8}{IVET}\xspace}
\newcommand{\Professor}{P\protect\scalebox{0.8}{ROFESSOR}\xspace}
\newcommand{\eps}{\varepsilon}
\newcommand{\mc}[1]{\mathcal{#1}}
\newcommand{\mr}[1]{\mathrm{#1}}
\newcommand{\mb}[1]{\mathbb{#1}}
\newcommand{\tm}[1]{\scalebox{0.95}{$#1$}}
\newcommand{\SMEFTsim}{\texttt{SMEFTsim}}
\newcommand{\SMEFTatNLO}{\texttt{SMEFT@NLO}}
\newcommand{\Madgraph}{M\protect\scalebox{0.8}{ADGRAPH}\xspace}
\tikzset{
  particle/.style={thick, draw=black, postaction={decorate},
    decoration={markings, mark=at position 0.6 with {\arrow[black]{stealth}}}},
  virtualparticle/.style={thick, draw=black, postaction={decorate}},
  scalar/.style={densely dashed, thick, draw=black, postaction={decorate}},
  photon/.style={decorate, draw=black,
    decoration={coil, aspect=0, amplitude=3pt, segment length=6pt}},
  gluon/.style={decorate, draw=black,
    decoration={coil, segment length=5pt, amplitude=4pt}}
}

\newcommand{\cH}{\ensuremath{c_{H}}\xspace}
\newcommand{\cHbox}{\ensuremath{c_{H\square}}\xspace}
\newcommand{\cHDD}{\ensuremath{c_{HDD}}\xspace}
\newcommand{\cHG}{\ensuremath{c_{HG}}\xspace}
\newcommand{\cHW}{\ensuremath{c_{HW}}\xspace}
\newcommand{\cHB}{\ensuremath{c_{HB}}\xspace}
\newcommand{\cHWB}{\ensuremath{c_{HWB}}\xspace}
\newcommand{\ceHAbs}{\ensuremath{|c_{eH}|}\xspace}
\newcommand{\cuHAbs}{\ensuremath{|c_{uH}|}\xspace}
\newcommand{\cdHAbs}{\ensuremath{|c_{dH}|}\xspace}
\newcommand{\ceWAbs}{\ensuremath{|c_{eW}|}\xspace}
\newcommand{\ceBAbs}{\ensuremath{|c_{eB}|}\xspace}
\newcommand{\cuGAbs}{\ensuremath{|c_{uG}|}\xspace}
\newcommand{\cuWAbs}{\ensuremath{|c_{uW}|}\xspace}
\newcommand{\cuBAbs}{\ensuremath{|c_{uB}|}\xspace}
\newcommand{\cdGAbs}{\ensuremath{|c_{dG}|}\xspace}
\newcommand{\cdWAbs}{\ensuremath{|c_{dW}|}\xspace}
\newcommand{\cdBAbs}{\ensuremath{|c_{dB}|}\xspace}
\newcommand{\cHl}[1]{\ensuremath{c_{Hl#1}}\xspace}
\newcommand{\cHe}{\ensuremath{c_{He}}\xspace}
\newcommand{\cHq}[1]{\ensuremath{c_{Hq#1}}\xspace}
\newcommand{\cHu}{\ensuremath{c_{Hu}}\xspace}
\newcommand{\cHd}{\ensuremath{c_{Hd}}\xspace}

\newcommand{\cpDC}{\ensuremath{c_{pDC}}\xspace}
\newcommand{\cpG}{\ensuremath{c_{pG}}\xspace}
\newcommand{\cdp}{\ensuremath{c_{dp}}\xspace}
\newcommand{\cpe}{\ensuremath{c_{pe}}\xspace}
\newcommand{\cpl}[1]{\ensuremath{c_{pl#1}}\xspace}
\newcommand{\cpmu}{\ensuremath{c_{pmu}}\xspace}
\newcommand{\cpqi}{\ensuremath{c_{pq3i}}\xspace}
\newcommand{\ctpl}[1]{\ensuremath{c_{3pl#1}}\xspace}
\newcommand{\cpd}{\ensuremath{c_{pd}}\xspace}
\newcommand{\cpQ}{\ensuremath{c_{pQ3}}\xspace}
\newcommand{\cpQM}{\ensuremath{c_{pQM}}\xspace}
\newcommand{\cpqMi}{\ensuremath{c_{pqMi}}\xspace}
\newcommand{\cpt}{\ensuremath{c_{pt}}\xspace}
\newcommand{\cpu}{\ensuremath{c_{pu}}\xspace}
\newcommand{\ctG}{\ensuremath{c_{tG}}\xspace}
\newcommand{\ctp}{\ensuremath{c_{tp}}\xspace}
\newcommand{\cpW}{\ensuremath{c_{pW}}\xspace}
\newcommand{\cpBB}{\ensuremath{c_{pBB}}\xspace}
\newcommand{\cpWB}{\ensuremath{c_{pWB}}\xspace}
\newcommand{\ctB}{\ensuremath{c_{tB}}\xspace}
\newcommand{\ctW}{\ensuremath{c_{tW}}\xspace}
\newcommand{\cll}{\ensuremath{c_{ll}}\xspace}

\section{Study of EFT effects in loop induced Higgs processes~\protect\footnote{
  A.~Cueto,
  S.~Pigazzini}{}}

\label{sec:Higgs_LI_EFT}



\subsection{Introduction}
\label{sec:Higgs_LI_EFT:section1}
The Standard Model Effective Field Theory (SMEFT) approach is a powerful framework to look for hints of new physics. It allows to study large sets of experimental data without assuming that the theory used is valid to arbitrary high energies. In the SMEFT, the Standard Model (SM) as we know it is just an effective theory at energies around the electroweak scale. Beyond the Standard Model (BSM) physics manifests at higher scales, $\Lambda$, and is parameterised in terms of higher-dimensional operators that conserve the same fields and symmetries as the SM. At any mass dimension, a complete bases of non-redundant operators can be worked out and the full Lagrangian can be written as a power expansion
\begin{equation}
\mathcal{L}_{\textrm SMEFT} = \mathcal{L}_{\textrm SM} + \sum_{d>4}\sum_{i}\frac{c_i}{\Lambda^{d-4}}\mathcal{O}_{i}^{d},
\end{equation}  

where $\mathcal{L}_{\textrm SM}$ is the SM Lagrangian, $c_i$ are the Wilson coefficients and ${\mathcal{O}^{d}_i}$ the set of independent operators for dimension $d$. Operators with $d=5,7$  violate lepton and/or baryon number conservation~\cite{Degrande:2012wf,Kobach:2016ami}. Thus, dimension-6 operators represent the leading deviation from the SM and will be the focus of this work. The modification of a given cross section by the insertion of one dimension-6 operator in the amplitudes can be written as

\begin{equation}
\sigma = \sigma_{\textrm SM} + \sum_{i}\sigma_i^{\textrm int} \frac{c_i}{\Lambda^{2}} + \sum_{i,j}\sigma_{(i,j)}^{\textrm BSM} \frac{c_ic_j}{\Lambda^{4}},
\end{equation}  

where $\sigma_{\textrm SM}$ is the SM cross section of a given process, $\sigma_i^{\textrm int}$ is the interference between the SM and BSM amplitudes and $\sigma_{(i,j)}^{\textrm BSM}$ represents the pure BSM correction to the SM cross section. The leading term $\sigma_i^{\textrm int}$ is only suppressed by $\Lambda^{-2}$ and the one that will be investigated in this work. 

Several bases of independent operators can be found in the literature~\cite{Grzadkowski:2010es,Contino:2013kra,Gupta:2014rxa,Masso:2014xra}. In the context of the study of the Higgs boson, the SILH basis~\cite{Contino:2013kra} has been commonly used. However, it is not optimised, for example, for diboson processes. Even if the translation between bases is known and has been automated~\cite{Falkowski:2015wza,Aebischer:2017ugx}, experimental collaboration have started to publish their EFT interpretations in the Warsaw basis also in the Higgs sector~\cite{ATLAS:2019jst,ATL-PHYS-PUB-2019-042} to facilitate future global fits of electroweak, Higgs and top data.

The procedure to test the EFT effects for a given set of measurements can be tedious in practice and a big effort has been devoted to developing public code to perform this task in an automatic and generic way~\cite{Brivio:2019irc}. For the Warsaw basis, different Universal FeynRules Output (UFO)~\cite{Degrande:2011ua} models are available which can be interfaced with modern event generators.

The \SMEFTsim\ code~\cite{Brivio:2017btx} is a well documented UFO implementation of the full set of dimension-6 operators in the Warsaw basis. Its main scope is the estimation of the leading SMEFT corrections to the SM. The effective Lagrangian is strictly truncated at $\Lambda^{-2}$ and neither supports next-to-leading order (NLO) simulations nor loop induced processes apart from a few exceptions. For Higgs data interpretation this model has become of common use due to its completeness~\cite{Ellis:2018gqa,Falkowski:2019hvp,ATLAS:2019jst}.  To reproduce all the main Higgs production and decay channels in the SM, the loop-induced processes ($hgg$, $h\gamma\gamma$,$hZ\gamma$) are included as effective vertices. However, it is not meant for precise studies of the EFT effects in the Higgs plus jet production for the following reasons:

\begin{itemize}
\item Only operators with the same point-like structure as the effective vertices, included to reproduce loop-induced processes, can modify the cross sections of these processes. That means that, for example, a modification of the top Yukawa will not affect the gluon-gluon fusion Higgs production process.
\item Given the truncation of the Lagrangian, operators that enter though the shifts of input parameters or field redefinitions, and that will modify the cross section of any tree-level process, do not modify the cross section of loop-induced processes.
\item A reliable computation of the Higgs plus jet  production in gluon-gluon fusion requires top quark loop amplitudes at high $p_{\textrm T}$ and the implementation of $gggH$ vertices which are not included in \SMEFTsim\ .
\item The $gg\to ZH$ process cannot be simulated.
\end{itemize}

All these point go beyond \SMEFTsim\ scope. But, instead, the \SMEFTatNLO\ tool~\cite{SMEFTNLO} can be used for the loop induced Higgs processes. The tool includes a complete implementation of the SMEFT compatible with NLO QCD predictions. After comparing the predictions of \SMEFTsim\ and \SMEFTatNLO\ on $pp\to ZH$ and $pp\to t\bar{t}H$, we study the $gg\to ZH$ and $gg\to H$ processes using \SMEFTatNLO.

\subsection{Comparison between models}
\label{sec:Higgs_LI_EFT:section2}
The \SMEFTsim\ and \SMEFTatNLO\ tools have been validated against each other~\cite{Durieux:2019lnv} for the top sector. In this section, we compare both models at leading order (LO) by checking the cross sections of the $pp\to ZH$ and $pp\to t\bar{t}H$ processes. The comparison is made at the cross section level and, thus, not expected to be in perfect agreement since it will be affected by phase-space integration. The main goal of this comparison is to show the mapping between the different Wilson coefficients naming and to ensure that the setup used for both models is consistent.

For both models we use the $m_Z$, $m_W$, $G_F$ scheme of electroweak parameters\footnote{We use the \texttt{SMEFTsim\_A\_U35\_MwScheme\_UFO} model for \SMEFTsim\ and the \texttt{SMEFTatNLO\_U2\_2\_U3\_3\_cG\_4F\_LO\_UFO-LO} model for \SMEFTatNLO\ }. The latest versions of the models available in December 2019 are used.  The \Madgraph 2.6.6 generator is used to obtain the cross sections results. The definition of the $pp\to Z(l^{+}l^{-})H$  and $pp\to t\bar{t}H$  processes is as follows for the SM predictions in \SMEFTsim\ :

\noindent
  \texttt{ define p = p b b$\sim$ }\\
  \texttt{ generate p p $>$ h t t$\sim$ SMHLOOP=0 NP=0 }\\
and \\ 
  \texttt{ generate p p $>$ h l+ l- SMHLOOP=0  NP=0     }.\\

  The $\texttt{SMHLOOP}$ coupling setting is not needed for \SMEFTatNLO\ . The default values of several parameters like $m_W$, $mt$, $\alpha_S$ or $\Gamma_{H}$ are different between the models and they were set to the same values, namely:
  $m_W = 79.8244$~GeV (default value in \SMEFTatNLO), 
  $m_t = 172$~GeV (default value in \SMEFTatNLO),
  $\alpha_S = 0.1184$ (default value in \SMEFTatNLO) and 
  $\Gamma_{H} = 4.07$~MeV (default value in \SMEFTsim). 

  Throughout this note, the same definitions of operators and fields as provided in~\cite{SMEFTNLOdefs} are used. In this notation, $gs$ is the strong coupling constant and $v$ denotes the vacuum expectation value of the Higgs field $\psi$. $Q$ is the third generation left-handed quark $SU(2)$-doublet, $t$ is right-handed $SU(2)$-singlet top quark. $G_{\mu\nu}^{A}$, $B_{\mu\nu}$, $W^{I}_{\mu\nu}$ are the fields strength tensors. Finally, $T^{A}$ is the generator of the fundamental representation of $SU(3)$ and $\tau^{\mu\nu}=\frac{1}{2}[\gamma^{\mu},\gamma^{\nu}]$  with $\gamma^{\mu}$ being the Dirac gamma matrices.

\renewcommand{\baselinestretch}{1.5}
  \begin{table}[t]
    \centering
    \resizebox{\columnwidth}{!}{%
    \begin{tabular}{|l|l|c|c|}
      \hline
      \textbf{Operator} &\textbf{W. coefficient} & \textbf{SMEFTsim} & \textbf{SMEFTatNLO} \\
      \hline 
      &  SM-SM & 0.0251$\pm$ 0.0001& 0.0255$\pm$ 0.0003\\
      \hline 
      $\partial_{\mu}(\psi^{\dag}\psi)\partial^{\mu}(\psi^{\dag}\psi)$ & \cpd\ (\cHbox) & 0.00304$\pm$ 0.00001& 0.00308 $\pm$ 0.00003\\
      \hline
      $(\psi^{\dag}D_\mu\psi)^\dag(\psi^{\dag}D_\mu\psi)  $& \cpDC\ (\cHDD) & 0.00041$\pm$ 0.00001& 0.00043$\pm$ 0.00006\\
      \hline 
      $\left(\psi^\dag\psi-\frac{v^2}{2}\right)B^{\mu\nu}B_{\mu\nu}  $ & \cpBB\ (\cHB) & 0.00231$\pm$ 0.00001& 0.00229$\pm$ 0.00004\\
      \hline
      $\left(\psi^\dag\psi-\frac{v^2}{2}\right)W_{I}^{\mu\nu}W^{I}_{\mu\nu}$ & \cpW\ (\cHW) & 0.01818$\pm$ 0.00007& 0.0183$\pm$ 0.0002\\
      \hline 
      $\left(\psi^\dag\psi-\frac{v^2}{2}\right)B^{\mu\nu}W^{I}_{\mu\nu} $ & \cpWB\ (\cHWB) & 0.00838$\pm$ 0.00004& 0.0084$\pm$ 0.0001\\
      \hline  
      $ i(\psi^{\dag}\overleftrightarrow{D}_\mu \psi)(\bar d_i \gamma^\mu d_i)$ & \cpd\ (\cHd) &-0.0044$\pm$ 0.0002 & -0.00444$\pm$ 0.00004\\
      \hline
      $ i(\psi^{\dag}\overleftrightarrow{D}_\mu \psi)(\bar{e} \gamma^\mu e)$ & \cpe\ $+$\cpmu (\cHe) &-0.002853$\pm$ 0.000007& -0.00285$\pm$ 0.00001\\
      \hline
      $i(\psi^{\dag}\overleftrightarrow{D}_\mu \psi)(\bar{l_{1,2}}\gamma^\mu l_{1,2}) $ & \cpl1\ $+$ \cpl2\ (\cHl1) & 0.00324$\pm$ 0.00002 & 0.00327$\pm$ 0.00002\\
      \hline
      $i(\psi^{\dag}\overleftrightarrow{D}_\mu \tau_{I}\psi)(\bar{l_{1,2}}\gamma^\mu \tau^{I} l_{1,2}) $ & \ctpl1+\ctpl2 (\cHl3) & -0.00588$\pm$ 0.00002& -0.00590$\pm$ 0.00005\\
      
      \hline
    \end{tabular}
}    
\renewcommand{\baselinestretch}{1.0}
    \caption{ Comparison of the SM and interference predictions for the $pp\to Z(l^{+}l^{-})H$ process between the \SMEFTsim\ and \SMEFTatNLO. The operators definitions are consistent with those given in \SMEFTatNLO. The Wilson coefficients use an analogous definition to those provided  in the UFO model in \SMEFTatNLO\ and \SMEFTsim\ in parenthesis.}
    \label{tab:Higgs_LI_EFT:zhcompa}
  \end{table}
\renewcommand{\baselinestretch}{1.0}

The Tables~\ref{tab:Higgs_LI_EFT:zhcompa} and \ref{tab:Higgs_LI_EFT:tthcompa} show the comparison between the predictions obtained for SM in both models as well as the interference terms, obtained with the NP$^{\wedge}$2$==$1 (NP$^{\wedge}$2$==$2)  for the \SMEFTsim\ (\SMEFTatNLO) model, for the $pp\to Z(l^{+}l^{-})H$  and $pp\to t\bar{t}H$ processes respectively. The correspondence between the nomenclature of the Wilson coefficients, or combination of them, in the different models used for the comparison can be found in the ``W. coefficient'' columns of Tables~\ref{tab:Higgs_LI_EFT:zhcompa} and \ref{tab:Higgs_LI_EFT:tthcompa}. All the predictions agree within the statistical uncertainty for $pp\to Z(l^{+}l^{-})H$, but for  $pp\to t\bar{t}H$ a difference of a 20\% between the values is observed for the absolute value of \ctG. These differences are acknowledged by the authors of the models and reside in the absence of five-point interactions and higher in the \SMEFTsim\ model, which go beyond the LO truncation. They will be added in future versions of the model.

\renewcommand{\baselinestretch}{1.5}
\begin{table}[t]
    \centering
  \resizebox{\columnwidth}{!}{%
      \begin{tabular}{|l|l|c|c|}
      \hline
      \textbf{Operator}& \textbf{W. coefficient} & \textbf{SMEFTsim} & \textbf{SMEFTatNLO} \\
      \hline
      &  SM-SM & 0.402$\pm$ 0.001& 0.402$\pm$ 0.003\\
      \hline
      $\partial_{\mu}(\psi^{\dag}\psi)\partial^{\mu}(\psi^{\dag}\psi)$  & \cpd\ (\cHbox) & 0.049$\pm$ 0.001 & 0.04876$\pm$ 0.00002\\
      \hline
      $(\psi^{\dag}D_\mu\psi)^\dag(\psi^{\dag}D_\mu\psi)$ & \cpDC\ (\cHDD) & -0.01218$\pm$ 0.00002 & -0.01222$\pm$ 0.00008\\
      \hline
      $\left(\psi^\dag\psi-\frac{v^2}{2}\right)B^{\mu\nu}B_{\mu\nu}  $ & \cpBB\ (\cHB) & 0.0000893$\pm$ 0.0000002 & 0.0000897$\pm$ 0.0000008\\
      \hline
      $\left(\psi^\dag\psi-\frac{v^2}{2}\right)W_{I}^{\mu\nu}W^{I}_{\mu\nu}$ & \cpW\ (\cHW)& 0.00042$\pm$ 0.000001& 0.000423$\pm$ 0.000004\\
      \hline
      $\left(\psi^\dag\psi-\frac{v^2}{2}\right)B^{\mu\nu}W^{I}_{\mu\nu}$ &  \cpWB\ (\cHWB)& -0.0002499$\pm$ 0.0000005& -0.000253$\pm$ 0.000002\\
      \hline
      $ i(\psi^{\dag}\overleftrightarrow{D}_\mu \psi)(\bar d_i \gamma^\mu d_i)$ & \cpd\ (\cHd) & -0.0000761$\pm$ 0.0000003 & -0.000076$\pm$ 0.000002\\
      \hline
      $\left(\psi^{\dag}\psi-\frac{v^2}{2}\right)\bar{Q}t\tilde{\psi} + h.c.$ &  \ctp\ (\cuHAbs) & -0.0488$\pm$ 0.0001& -0.0494$\pm$ 0.0003\\
      \hline
      $i g_{s}\left(\bar{Q}\tau^{\mu\nu}T_{A}t\right)\tilde{\psi}G_{\mu\nu}^{A} + h.c.$ & \ctG\ (\cuGAbs) & -0.3393$\pm$ 0.0009 & 0.407 $\pm$ 0.002\\
      \hline
      $i(\psi^{\dag}\overleftrightarrow{D}_\mu \tau_{I}\psi)(\bar{l_{1,2}}\gamma^\mu \tau^{I} l_{1,2}) $ & \ctpl1+\ctpl2 (\cHl3) & -0.0489 $\pm$ 0.0001 & -0.0491 $\pm$ 0.0002\\
      
      
      
      \hline
    \end{tabular}
  }
\renewcommand{\baselinestretch}{1.0}
  \caption{ Comparison of the SM and interference predictions for the $pp\to t\bar{t} H$ process between the \SMEFTsim\ and \SMEFTatNLO. The operators definitions are consistent with those given in \SMEFTatNLO. The Wilson coefficients use an analogous definition to those provided  in the UFO model in \SMEFTatNLO\ and \SMEFTsim\ in parenthesis.}
  \label{tab:Higgs_LI_EFT:tthcompa}
\end{table}
\renewcommand{\baselinestretch}{1.0}


For the $\mathcal{O}_{uW}$ and $\mathcal{O}_{uB}$ operators defined as,

$$ \mathcal{O}_{tB} = i(\bar Q \sigma^{\mu\nu} t) \tilde \psi  B_{\mu\nu} + h.c.; \quad
   \mathcal{O}_{tW} =  i(\bar{Q} \tau^{\mu\nu}\tau_{I} t)  \tilde{\psi} \, W_{\mu\nu}^I + h. c.
$$

   there is no one-to-one correspondence between the models in their latest versions. The \SMEFTatNLO\ version released on 2019/04/03 was used instead to compare these two operators.
\renewcommand{\baselinestretch}{1.5}
  \begin{table}[t]
    \centering
    \resizebox{\columnwidth}{!}{%
      \begin{tabular}{|l|l|c|c|}
        \hline
        \textbf{Operator}& \textbf{W. coefficient} & \textbf{SMEFTsim} & \textbf{SMEFTatNLO} \\
        \hline
        \hline                                                                                                                                              
        $i(\bar Q \sigma^{\mu\nu} t) \tilde \psi  B_{\mu\nu} + h.c.$ & \ctB\ (\cuBAbs) & -0.000828$\pm$ 0.000002 & 0.00085$\pm$ 0.00001 \\
         \hline
        $i(\bar{Q} \tau^{\mu\nu}\tau_{I} t)  \tilde{\psi} \, W_{\mu\nu}^I +h.c$ & \ctW (\cuWAbs) & -0.002219$\pm$ 0.000006& 0.00223$\pm$ 0.00002\\
        \hline
 
\end{tabular}
    }
\renewcommand{\baselinestretch}{1.0}
    \caption{ Comparison of the SM and interference predictions for the $pp\to t\bar{t} H$ process between the \SMEFTsim\ and \SMEFTatNLO\ for \ctB\ (\cuBAbs) and \ctW\ (\cuWAbs). The operator definition are given in the way they are implemented in \SMEFTatNLO\ .}
    \label{tab:Higgs_LI_EFT:comptth2}
\end{table}
\renewcommand{\baselinestretch}{1.0}
The prediction for the operators shown in Table~\ref{tab:Higgs_LI_EFT:comptth2} agree in their absolute value within the statistical uncertainty but not in their sign. The way in which they are implemented in the model is also different. While in \SMEFTsim\ the absolute value and the phase of these complex operators can be changed by the user, only the real part can be tuned by the user in \SMEFTatNLO.

Other differences come from two-fermion operators involving quarks. In \SMEFTsim\ the couplings of all quarks are parametrized together in the same way, while in \SMEFTatNLO\ the top vertices are parameterised separately.

\subsection{$gg\to Z(l^{+}l^{-})H$}
\label{sec:Higgs_LI_EFT:ggzh}
The study of  $gg\to Z(l^{+}l^{-})H$ with $l=e,\mu$ is performed using the \SMEFTatNLO\ model. The renormalisation scale is set to $M_H=125$~GeV and the PDF set NNPDF2.3 for the parametrisation of the proton structure is used. The SM cross-section obtained for this process with the mentioned settings is 3.147$\pm$0.002 fb, for which the error only reflects the statistical uncertainty of the calculation. The generated events are passed through the \Pythia\ parton shower. A more in-depth study of the SMEFT effects for this process was performed in~\cite{Bylund:2016phk} using the main set of operators affecting the cross sections using a sample of NLO accuracy for $gg\to ZH$ and $gg\to ZHj$. Here we have considered all the operators available at NLO in \SMEFTatNLO\ which provide diagrams with a non-zero interference with the SM.

\begin{figure}[t]
\includegraphics[width=0.49\linewidth,page=7]{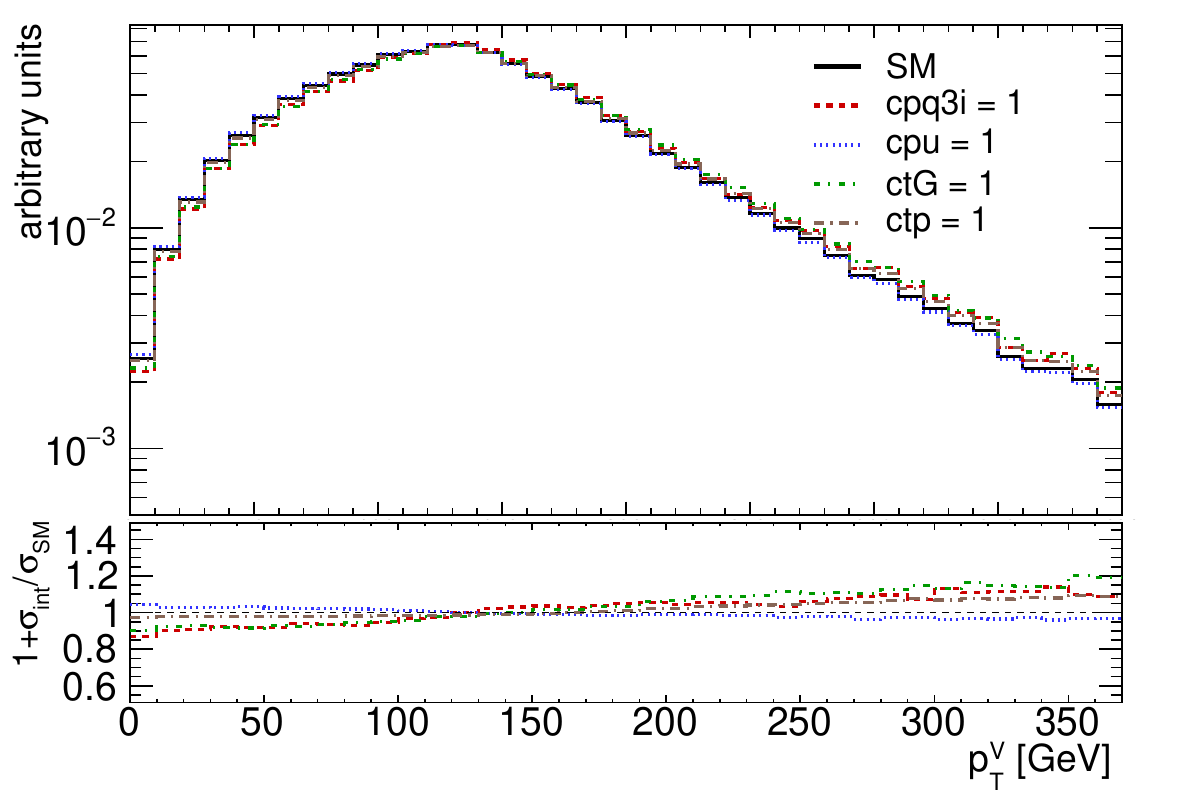}
\includegraphics[width=0.49\linewidth,page=10]{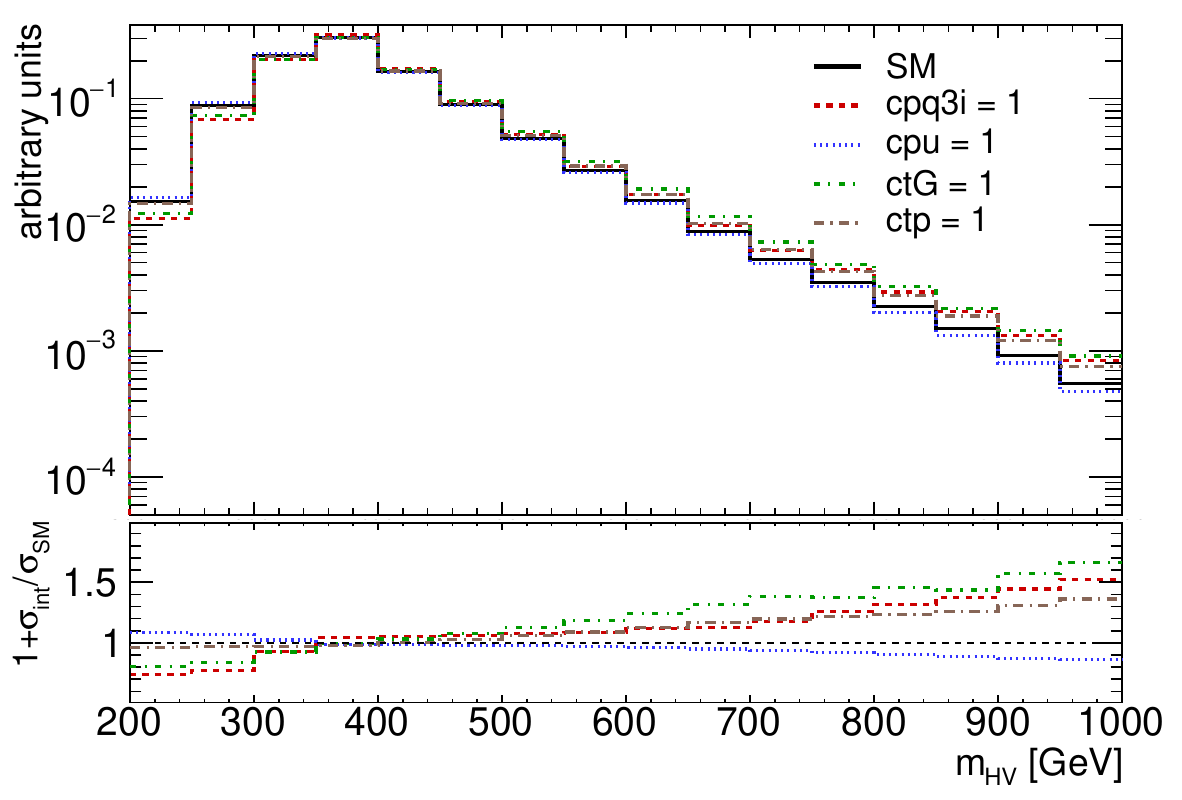}
\caption{Differential distributions as a function of $p_{T}^{V}$ and $m_{HV}$ for the SM predictions and its interference with operators with Wilson coefficients \ctG\ , \cpd\ , \cpu\ and \ctp\  at the lowest order in QCD. The value of $\Lambda$ was set to 1 TeV. The distribution was obtained using 150000 events.}
\label{fig:Higgs_LI_EFT:ggzh}
\end{figure}

In Figure~\ref{fig:Higgs_LI_EFT:ggzh}, differential distributions as functions of $p_{T}^{V}$ and $m_{HV}$  with BSM effects caused by \cpqi\ , \cpu\ , \ctG\ and \ctp\ are shown.  Many other operators modify the cross section of this process but only some examples of those that distort significantly the shape of the SM prediction for  $c_i=1$ are shown.

In addition to differential cross sections, measurements  of the Higgs couplings in terms of Simplified Template Cross Sections (STXS)~\cite{deFlorian:2016spz} also provide constraining power of the SMEFT parameters. A parametrisation in bins of the STXS in stage 1.2~\cite{Berger:2019wnu}  for $gg\to Z(l^{+}l^{-})H$  is provided in Table~\ref{tab:Higgs_LI_EFT:stxsggzh}. The results of	the SM cross section in each bin are shown for two reasons: It allows to recompute the parametrisation in merged scenarios and shows the statistical uncertainty that affects the computation of the parametrisation.

\renewcommand{\baselinestretch}{1.5}
\begin{table}[pt]
 \adjustbox{max width=\textwidth}{
  \begin{tabular}{p{0.42\textwidth} p{0.43\textwidth}p{0.15\textwidth}}
    \toprule
      \hline
      \textbf{Bin} & \textbf{Parametrisation} & \textbf{SM cross-section [nb]}\\
      \hline 
    $gg\to Hll (p_{\textrm T}^{V}<75$~GeV)&
     $-0.0012 \cpDC +0.121 \cdp -0.056 \cpe$\newline $+0.064 \cpl1 +0.064 \cpl2 -0.0566 \cpmu$\newline $-0.331 \cpqi -0.117 \ctpl1 -0.117 \ctpl2$\newline $+0.249 \cpd -0.166 \cpQ -0.129 \cpQM$\newline $-0.332 \cpqMi +0.047 \cpt +0.165 \cpu$\newline $+0.250 \ctG +0.0369 \ctp$ &  0.468 $\pm$ 0.003\\
     \hline 
     $gg\to Hll (75<p_{\textrm T}^{V}<150$~GeV)&
     $+0.0030 \cpDC +0.122 \cdp -0.057 \cpe$\newline $+0.065 \cpl1 +0.065 \cpl2 -0.0568 \cpmu$\newline $-0.285 \cpqi -0.118 \ctpl1 -0.118 \ctpl2$\newline $+0.213 \cpd -0.142 \cpQ -0.098 \cpQM$\newline $-0.283 \cpqMi +0.0262 \cpt +0.142 \cpu$\newline $+0.316 \ctG +0.0454 \ctp$ & 1.343 $\pm$ 0.005\\ 
     \hline
     $gg\to Hll $(0-jet,$150<p_{\textrm T}^{V}<250$~GeV)&
     $+0.025 \cpDC +0.120 \cdp -0.057 \cpe$\newline $+0.065 \cpl1 +0.065 \cpl2 -0.0561 \cpmu$\newline $-0.233 \cpqi -0.116 \ctpl1 -0.118 \ctpl2$\newline $+0.17 \cpd -0.115 \cpQ -0.029 \cpQM$\newline $-0.229 \cpqMi -0.027 \cpt +0.112 \cpu$\newline $+0.439 \ctG +0.084 \ctp$ & 0.250 $\pm$ 0.002\\
     \hline
     $gg\to Hll (\geq$ 1-jet,$150<p_{\textrm T}^{V}<250$~GeV)&
 $+0.016 \cpDC +0.122 \cdp -0.0569 \cpe$\newline $+0.065 \cpl1 +0.065 \cpl2 -0.0572 \cpmu$\newline $-0.244 \cpqi -0.118 \ctpl1 -0.117 \ctpl2$\newline $+0.183 \cpd -0.122 \cpQ -0.050 \cpQM$\newline $-0.245 \cpqMi -0.0111 \cpt +0.121 \cpu$\newline $+0.411 \ctG +0.072 \ctp$ & 0.699 $\pm$ 0.003\\
 \hline

      $gg\to Hll (p_{\textrm T}^{V}>250$~GeV)&
 $+0.049 \cpDC +0.120 \cdp -0.0585 \cpe$\newline $+0.066 \cpl1 +0.066 \cpl2 -0.0581 \cpmu$\newline $-0.197 \cpqi -0.116 \ctpl1 -0.116 \ctpl2$\newline $+0.153 \cpd -0.099 \cpQ +0.031 \cpQM$\newline $-0.199 \cpqMi -0.0820 \cpt +0.099 \cpu$\newline $+0.544 \ctG +0.134 \ctp$ & 0.285 $\pm$ 0.002\\
 \bottomrule
\end{tabular}
}
\renewcommand{\baselinestretch}{1.0}
 \caption{Parametrisation of the $gg\to Z(l^{+}l^{-})H$ bins of the STXS as defined in its stage 1.2 with the parameters definitions of the \SMEFTatNLO\ model. The numbers are rounded according to their statistical uncertainty.}
 \label{tab:Higgs_LI_EFT:stxsggzh}
\end{table}
\renewcommand{\baselinestretch}{1.0}

The potential uncertainties arising from the use of a different PDF, scales or any other different settings in the calculation are not carefully investigated. As a quick cross check, the parametrisation was re-derive using a different scale, namely $m_H/2$. The results are typically consistent within the statistical uncertainty. In a few cases, in which the statistical uncertainty does not cover the differences, they differ by at most 5\%.

\subsection{$gg\to H$}
\label{sec:Higgs_LI_EFT:section3}
The SMEFT effects in the Higgs production through gluon-gluon fusion is examined using the \SMEFTatNLO\ package. As in Section~\ref{sec:Higgs_LI_EFT:ggzh}, the 
study of this process is already available in the literature~\cite{Deutschmann:2017qum} for a limited set of operators. In this work we have considered all operators that have a non-zero interference with the SM. Those operators were found to be:
$\mathcal{O}_{\psi G}$, $\mathcal{O}_{tG}$, $\mathcal{O}_{t\psi}$, $\mathcal{O}_{d\psi}$, $\mathcal{O}_{\psi DC}$, $\mathcal{O}_{\psi l1}^{(3)}$, $\mathcal{O}_{ \psi l2}^{(3)}$ and $\mathcal{O}_{ll}$.
The last five operators enter in the process though shifts to the inputs parameters or the Higgs field redefinition and do not modify the shape of the SM prediction.

The predictions for the Higgs production in gluon-gluon fusion is obtained using $m_H/2$ as the renormalisation scale and the PDF4LHC15 PDF set. A cut of 20 GeV is applied by default to the transverse momentum of the parton at matrix-element level. Because \Madgraph\ cannot deal with the interference between loop-induced and tree-level processes, when the \cpG\ operator is considered, the reweighting module is used and the process is generated in three samples with different jet multiplicity, namely 0, 1 and 2 additional jets. The cross sections of the processes are
$14.082 \pm 0.003$~pb, $10.74\pm0.002$~pb and $5.598\pm0.008$~pb respectively for the 0, 1 and 2 additional jets cases. Additional multi-leg samples are produced for the SM and for all operators except for \cpG\  and used to cross check the results. These samples are merged with the CKKW-L~\cite{Lonnblad:2001iq} scheme using $30$~GeV as the merging scale.

\begin{figure}[t]
\includegraphics[width=0.49\linewidth,page=5]{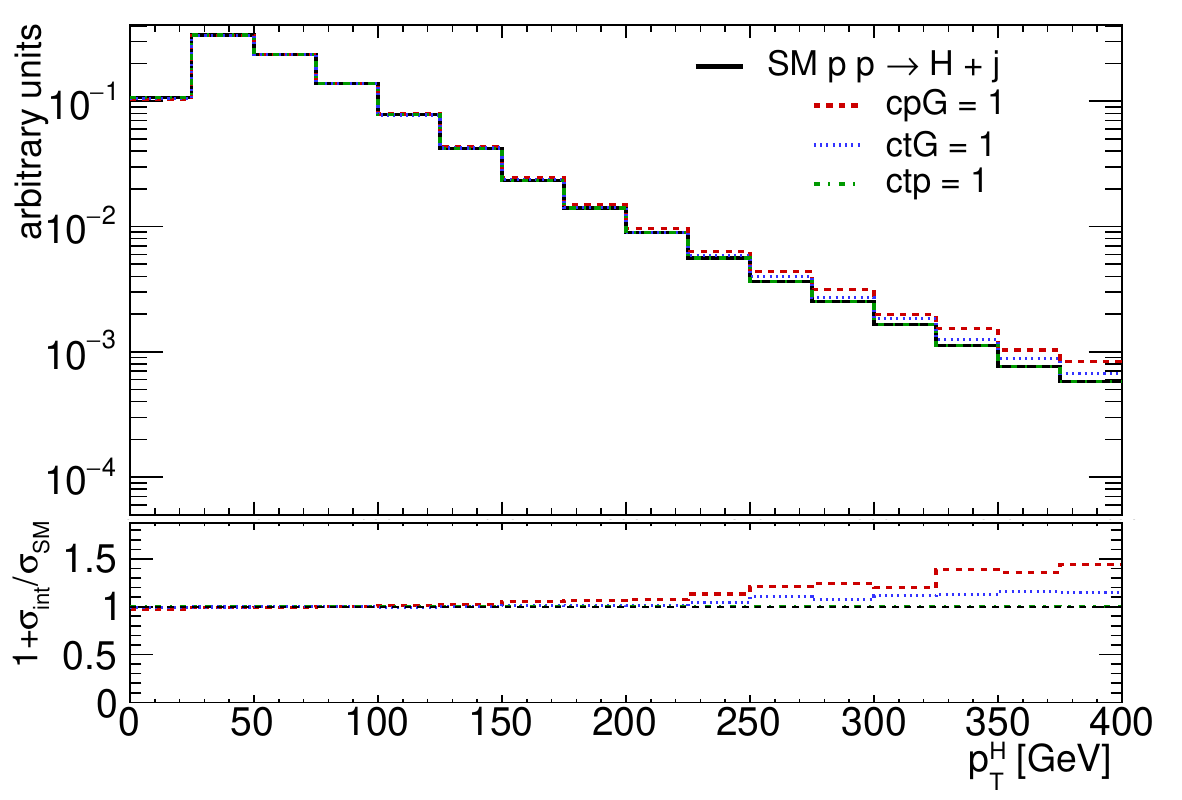}
\includegraphics[width=0.49\linewidth,page=5]{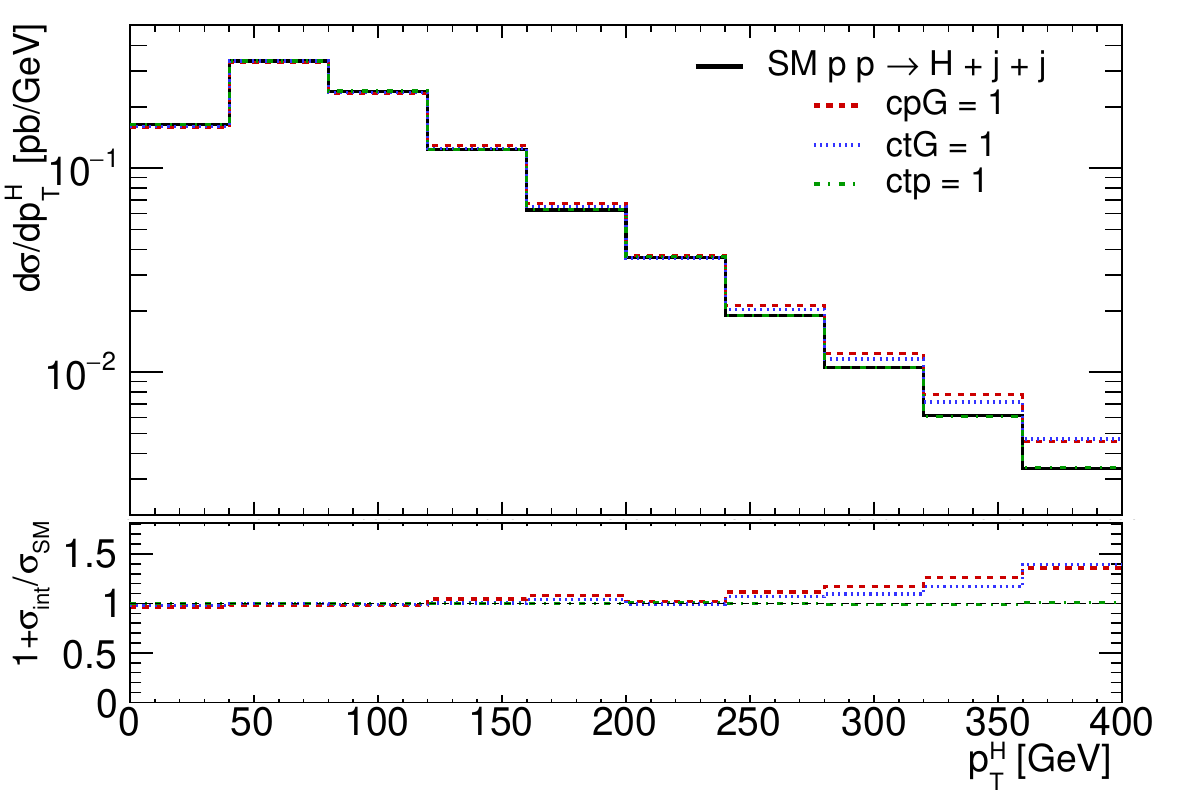}

\caption{Differential distributions normalised to unity as a function of $p_{T}^{H}$ for the SM prediction and its interference with operators with Wilson coefficients \cpG\ ,\ctG\ and \ctp\ for $p p \to H + j$ (left) and $p p \to H + j + j$ (right). The value of $\Lambda$ was set to 1 TeV. The left-hand-side (right-hand side) distribution is obtained using 400000 (50000) events.}
\label{fig:Higgs_LI_EFT:ggh}
\end{figure}

The differential distributions for the SM and the interference with the operators with Wilson coefficients \cpG ,\ctG\ and \ctp\ is shown in Figure~\ref{fig:Higgs_LI_EFT:ggh}. The value of the Wilson coefficients is set to unity and $\Lambda=1$~TeV is used. The distributions are normalised to unity so that only the shape differences induced by the different operators are displayed in the figure.

 In Tables~\ref{tab:Higgs_LI_EFT:stxsggh1} and \ref{tab:Higgs_LI_EFT:stxsggh2}, we provide the parametrisation of the $gg\to H$ STXS bins in stage 1.2. For reference and to give the needed inputs to obtain the parametrisation in other scenarios in which several STXS bins are merged, the SM cross section in each bin is provided. The results provided are cross-checked with the produced multi-leg samples.

\renewcommand{\baselinestretch}{1.5}
\begin{table}[t]
 \adjustbox{max width=\textwidth}{
  \begin{tabular}{p{0.38\textwidth} p{0.45\textwidth}p{0.17\textwidth}}
    \toprule
      \hline
      \textbf{Bin} & \textbf{Parametrisation} & \textbf{SM cross-section [pb]}\\      
      \hline
      $gg\to H$ ($200<p_{\textrm T}^{H}<300$~GeV)&
      $+1.8 \ctG -0.06 \ctpl1 -0.06 \ctpl2 +0.12 \cdp$\newline $-0.03 \cpDC -0.12 \ctp +45 \cpG + 0.061 \cll$  & 0.265 $\pm$ 0.009 \\
      \hline
      $gg\to H$ ($300<p_{\textrm T}^{H}<450$~GeV)&
      $+2.0 \ctG -0.06 \ctpl1 -0.06 \ctpl2 +0.12 \cdp$\newline $-0.03 \cpDC -0.12 \ctp +50 \cpG + 0.06 \cll$ & 0.068 $\pm$ 0.004  \\
      \hline
      $gg\to H$ ($450<p_{\textrm T}^{H}<650$~GeV)&
      $+2.5 \ctG -0.06 \ctpl1 -0.06 \ctpl2 +0.12 \cdp$\newline $-0.03 \cpDC -0.11 \ctp +65 \cpG + 0.06 \cll$ & 0.011 $\pm$ 0.002 \\
      \hline
      $gg\to H$ ($p_{\textrm T}^{H}>650$~GeV)&
      $+4.5 \ctG -0.07 \ctpl1 -0.06 \ctpl2 +0.12 \cdp$\newline $-0.03 \cpDC -0.12 \ctp +100 \cpG + 0.06 \cll$ & 0.0011 $\pm$ 0.0006\\
      \hline      
      $gg\to H$ (0-jet, $p_{\textrm T}^{H}<10$~GeV)&
      $+1.57 \ctG -0.060 \ctpl1 -0.060 \ctpl2 +0.121 \cdp$\newline $-0.030 \cpDC -0.122 \ctp +39.2 \cpG + 0.0605 \cll$ & 2.43 $\pm$  0.02\\
      \hline
      $gg\to H$ (0-jet, $p_{\textrm T}^{H}>10$~GeV)&
      $+1.58 \ctG -0.060 \ctpl1 -0.060 \ctpl2 +0.121 \cdp$\newline $-0.030 \cpDC -0.121 \ctp +39.2 \cpG + 0.0605 \cll$ & 6.37 $\pm$ 0.02\\
      \hline
      $gg\to H$ (1-jet,$p_{\textrm T}^{H}<60$~GeV)&
      $+1.59 \ctG -0.060 \ctpl1 -0.061 \ctpl2 +0.121 \cdp$\newline $-0.030 \cpDC -0.121 \ctp +40.0 \cpG + 0.061 \cll$ & 2.08 $\pm$ 0.01 \\ 
      \hline
      $gg\to H$ (1-jet,$60<p_{\textrm T}^{H}<120$~GeV)&
      $+1.60 \ctG -0.060 \ctpl1 -0.061 \ctpl2 +0.121 \cdp$\newline $-0.030 \cpDC -0.121 \ctp +40.3 \cpG + 0.061 \cll$ & 1.73 $\pm$ 0.01\\
      \hline
      $gg\to H$ (1-jet,$120<p_{\textrm T}^{H}<200$~GeV)&
      $1.64 \ctG -0.063 \ctpl1 -0.063 \ctpl2 +0.126 \cdp$\newline $-0.031 \cpDC -0.124 \ctp +42.3 \cpG +0.063 \cll$ & 0.310 $\pm$ 0.005\\
      \hline
      
       \bottomrule
\end{tabular}
}
\renewcommand{\baselinestretch}{1.0}
\caption{Parametrisation of the $gg\to H$ bins with no jet, 0-jet and 1-jet selection of the STXS as defined in its stage 1.2 with the parameters definitions of the \SMEFTatNLO\ model. The numbers are rounded according to their statistical uncertainty.}
\label{tab:Higgs_LI_EFT:stxsggh1}
\end{table}
\renewcommand{\baselinestretch}{1.0}

\renewcommand{\baselinestretch}{1.5}
\begin{table}[t]
 \adjustbox{max width=\textwidth}{
  \begin{tabular}{p{0.44\textwidth} p{0.41\textwidth}p{0.15\textwidth}}
    \toprule
      \hline
      \textbf{Bin} & \textbf{Parametrisation} & \textbf{SM cross-section [pb]} \\
      \hline
      $gg\to H$ ($\geq$ 2-jet, $m_{\textrm jj}<350$~GeV, $p_{\textrm T}^{H}<60$~GeV)&
      $+1.62 \ctG -0.061 \ctpl1 -0.061 \ctpl2$\newline $+0.126 \cdp -0.031 \cpDC -0.122 \ctp$\newline $+41 \cpG  + 0.061 \cll$ &0.66 $\pm$ 0.01\\
      \hline
      $gg\to H$ ($\geq$ 2-jet, $m_{\textrm jj}<350$~GeV,\hspace{5em} $60<p_{\textrm T}^{H}<120$~GeV)&
      $+1.63 \ctG -0.061 \ctpl1 -0.061 \ctpl2$\newline $+0.120 \cdp -0.031 \cpDC -0.121 \ctp$\newline $+40.8 \cpG  + 0.061\cll$ & 1.07 $\pm$ 0.02\\
      \hline
      $gg\to H$ ($\geq$ 2-jet, $m_{\textrm jj}<350$~GeV,\hspace{5em} $120<p_{\textrm T}^{H}<200$~GeV)&
      $+1.69 \ctG -0.062 \ctpl1 -0.062 \ctpl2$\newline $+0.120 \cdp -0.030 \cpDC -0.122 \ctp$\newline $+45 \cpG +0.062\cll$  & 0.62 $\pm$ 0.01 \\
      \hline
      $gg\to H$ ($\geq$ 2-jet, $350<m_{\textrm jj}<700$~GeV,\hspace{5em} $p_{\textrm T}^{H}<200$~GeV, $p_{\textrm T}^{Hjj}<25$~GeV)&
      $+1.5 \ctG -0.056 \ctpl1 -0.056 \ctpl2$\newline $+0.113 \cdp -0.027 \cpDC -0.113 \ctp$\newline $+42 \cpG + 0.058\cll$ & 0.095 $\pm$ 0.005\\
      \hline
      $gg\to H$ ($\geq$ 2-jet, $350<m_{\textrm jj}<700$~GeV,\hspace{5em} $p_{\textrm T}^{H}<200$~GeV, $p_{\textrm T}^{Hjj}>25$~GeV)&
      $+1.60 \ctG -0.060 \ctpl1 -0.060 \ctpl2$\newline $+0.117 \cdp -0.028 \cpDC -0.126 \ctp$\newline $+ 40 \cpG +0.06\cll$ & 0.334 $\pm$ 0.009 \\
      \hline
      $gg\to H$ ($\geq$ 2-jet, $m_{\textrm jj}>700$~GeV,\hspace{5em} $p_{\textrm T}^{H}<200$~GeV, $p_{\textrm T}^{Hjj}<25$~GeV)&  
      $+1.7 \ctG -0.058 \ctpl1 -0.058 \ctpl2$\newline $+0.12 \cdp -0.033 \cpDC -0.12 \ctp$\newline $+48 \cpG +0.058\cll$ & 0.035 $\pm$ 0.003\\
      \hline
      $gg\to H$ ($\geq$ 2-jet, $m_{\textrm jj}>700$~GeV,\hspace{5em} $p_{\textrm T}^{H}<200$~GeV, $p_{\textrm T}^{Hjj}>25$~GeV)&  
      $+1.7 \ctG -0.062 \ctpl1 -0.062 \ctpl2$\newline $+0.114 \cdp -0.031 \cpDC -0.118 \ctp$\newline $+44 \cpG +0.061\cll$ & 0.130 $\pm$ 0.005\\
      \hline

       \bottomrule
\end{tabular}
}
\renewcommand{\baselinestretch}{1.0}
\caption{Parametrisation of the $gg\to H$ bins with 2 or more jets selection  of the STXS as defined in its stage 1.2 with the parameters definitions of the \SMEFTatNLO\ model. The numbers are rounded according to their statistical uncertainty.}
\label{tab:Higgs_LI_EFT:stxsggh2}
\end{table}
\renewcommand{\baselinestretch}{1.0}

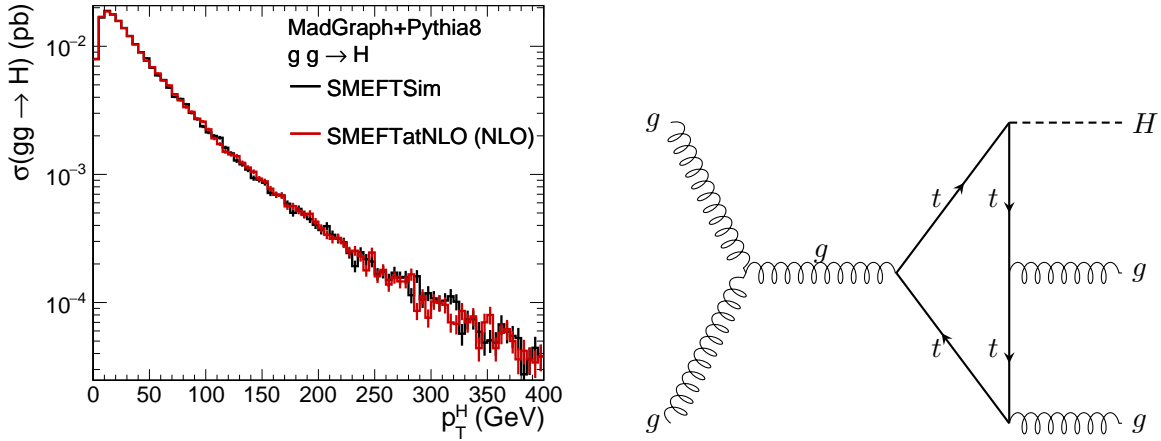
\begin{figure}[t]
\begin{tabular}{ccc}
  \includegraphics[width=0.48\linewidth]{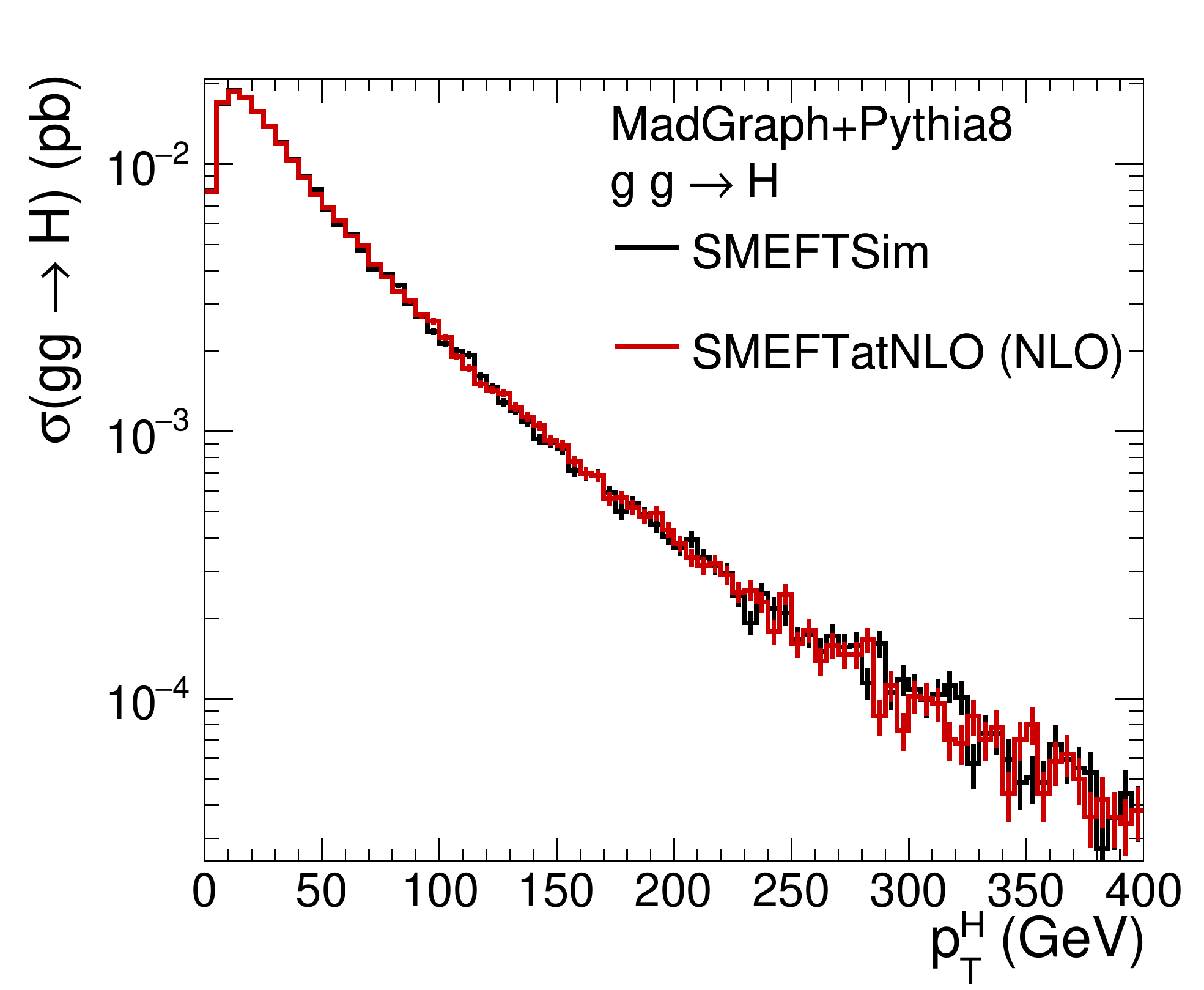} &&
  \begin{tikzpicture}[node distance=2cm and 0cm, scale=1.0]
      \coordinate[label=left:$g$] (q1) at (0,2);
      \coordinate[label=left:$g$] (q2) at (0,-2);
      \coordinate[] (V1) at (1,0);
      \coordinate[] (V2) at (3,0);
      \coordinate[] (V3) at (4.5,2);
      \coordinate[] (V4) at (4.5,0);
      \coordinate[] (V5) at (4.5,-2);
      \coordinate[label=right:$H$] (q3) at (6,2);
      \coordinate[label=right:$g$] (q4) at (6,0);
      \coordinate[label=right:$g$] (q5) at (6,-2);
      \draw[gluon] (q1) -- (V1);
      \draw[gluon] (q2) -- (V1);
      \draw[gluon] (V1) -- node[above]{$g$} (V2);
      \draw[particle] (V2) -- node[left]{$t$} (V3);
      \draw[particle] (V5) -- node[left]{$t$} (V2);
      \draw[particle] (V3) -- node[left]{$t$} (V4);
      \draw[particle] (V4) -- node[left]{$t$} (V5);
      \draw[scalar] (V3) -- (q3);
      \draw[gluon] (V4) -- (q4);
      \draw[gluon] (V5) -- (q5);
    \end{tikzpicture}    
\end{tabular}
\caption{ Left: Comparison of the differential cross section of  $gg\to H$ as a function of $p_{\textrm T}^{ \textrm H}$ in \SMEFTsim\ and \SMEFTatNLO. Right: Example diagram contributing to $gg\to H +j$ which is not considered in \SMEFTsim\ but it is implemented in \SMEFTatNLO.}
    \label{fig:Higgs_LI_EFT:ggHcomp}
\end{figure}

The parametrisation of \cpG\ for the $gg\to H$ production mode is different in the \SMEFTsim\ and \SMEFTatNLO\ for 1-jet and 2-jet . It has been checked that for the 0-jet case the values of the inclusive cross section in those models is the same and the differential distributions as a function of $p_{\textrm T}^{ \textrm H}$ are consistent within statistical uncertainty as shown in Figure~\ref{fig:Higgs_LI_EFT:ggHcomp}. In this case, also the same SMEFT effects for \cpG\ are observed.
However, when we add jets to the final state, the parametrisation changes significantly (it can be compared to the one shown in~\cite{ATL-PHYS-PUB-2019-042}). This is expected due to the different implementation of the process and different diagrams included. Additionally, this process also lacks 5- and 6-point interactions in \SMEFTsim\ which go beyond the LO truncation and are not included in the current public version. They will be added in the next versions of \SMEFTsim. In Figure~\ref{fig:Higgs_LI_EFT:ggHcomp} an example diagram which is included in \SMEFTatNLO\ and not considered in \SMEFTsim\ is depicted.

\subsection{Summary and conclusions}
In the absence of hints for new physics in the LHC, the SMEFT approach started to be widely adopted by the experimental collaborations for the interpretation of their measurements. In order to be able to have predictions for the SMEFT, implementation of the SM plus dimension-6 Lagrangian in the form of UFO files that can be interfaced with modern event generators is needed. Two different tools, \SMEFTsim\ and \SMEFTatNLO, have been used.

In this work, we have compared both tools for the $pp\to t\bar{t}H$ and $pp\to ZH$ production processes. The agreement between the predictions for the SM and interference terms is excellent except for the $\mathcal{O}_{\textrm tG}$ operator. Some other operators like $\mathcal{O}_{\textrm tW}$, $\mathcal{O}_{\textrm tZ}$, or two-fermion currents involving quarks cannot be directly compared. Even if the definition of each operator is available in both models, it would be helpful for the user to have a clear mapping between operators in the different tools.

The SMEFT effects have been studied by means of the distortion of the SM prediction shape and normalisation in differential cross sections as well as the parametrisation of STXS bins. Only the interference effects have been investigated. For $gg\to H$, the operators \cpG\ and \ctG\ have a different effect compared with the SM when different scales are proven, increasing at higher energies.
The parametrisation in terms of STXS bins for $\mathcal{O}_{\psi G}$ differs from others that can be found in the literature using \SMEFTsim\ due to the differences in the implementation of this process in both tools. The \SMEFTatNLO\ tool provides a reliable description of the Higgs plus jets production in gluon-gluon fusion.

For $gg\to ZH$, with $Z\to l^{+}l^{-}$,  many operators change the cross sections. However, most of them just introduce a deviation in the normalisation of the SM predictions at the interference level without distorting the SM shape. Among the ones that have an energy dependence we can find: $\mathcal{O}_{tG}$, $\mathcal{O}_{t\psi}$ or $\mathcal{O}_{\psi q_i}^{(3)}$ 


~\newline~

\let\Herwig\undefined
\let\Pythia\undefined
\let\Sherpa\undefined
\let\Rivet\undefined
\let\Professor\undefined
\let\Madgraph\undefined
\let\eps\undefined
\let\mc\undefined
\let\mr\undefined
\let\mb\undefined
\let\tm\undefined


\section{Improved NNLO Higgs pair production with EFT effects~\protect\footnote{
  D.~de~Florian,
  I.~Fabre,
  G.~Heinrich,
  J.~Mazzitelli}{}}

\label{sec:Higgs_HH_NNLO_EFT}


\subsection{Introduction}

The exploration of the Higgs potential represents one of the main goals of the future LHC runs and its high-luminosity upgrade.
To this end, an experimental determination of the Higgs cubic self-coupling $\lambda_{hhh}$ needs to be performed, and the major constraints will come from the measurement of the double-Higgs production cross section (see Ref.~\cite{DiMicco:2019ngk} for a review).

The dominant production mode of Higgs boson pairs at hadron colliders in the Standard Model (SM) is gluon fusion, mediated by a top-quark loop. 
In order to maximally profit from the experimental measurements, precise theoretical predictions are needed. The next-to-leading order (NLO) QCD corrections to this process have been computed in Refs.~\cite{Borowka:2016ehy,Borowka:2016ypz,Baglio:2018lrj} with full top-quark mass dependence, while higher-order corrections have been computed in the heavy top limit (HTL) up to third order in the strong coupling expansion~\cite{Dawson:1998py,deFlorian:2013jea,Grigo:2015dia,Chen:2019lzz,Chen:2019fhs}.
The large invariant mass of the final state, compared to the value of the top-quark mass, makes this approximation substantially less accurate than in the single-Higgs case, and improvements are needed in order to obtain sensible phenomenological results.
In this respect, the most advanced prediction available to date is the so-called NNLO$_\text{FTapprox}$~\cite{Grazzini:2018bsd}, which extends the FT$_\text{approx}$  introduced at NLO in Refs.~\cite{Frederix:2014hta,Maltoni:2014eza} to the next-to-next-to-leading order (NNLO).
In particular, this approximation consistently includes the full loop-induced double real corrections.

While the accurate prediction of the SM rates is of crucial importance, beyond the SM (BSM) scenarios will also present large QCD corrections, and in particular the differential K-factors can differ from the SM ones.
In addition it is desirable to match the theoretical prediction obtained for the SM, for instance in order to have a consistent treatment of the theoretical uncertainties.
Therefore, for the experimental determination of $\lambda_{hhh}$ precise predictions for $\lambda_{hhh} \neq  \lambda_{hhh}^\text{SM}$ are needed.
In addition to $\lambda_{hhh}$ scans, a more general parameterization of BSM effects in an effective field theory (EFT) approach is desirable.
Predictions including EFT operators have been obtained at NLO~\cite{Grober:2015cwa} and NNLO~\cite{deFlorian:2017qfk} in the HTL, and more recently with full top-quark mass dependence at NLO in Ref.~\cite{Buchalla:2018yce}.
Also, a NLO Monte Carlo generator allowing for $\lambda_{hhh}$ (and $y_t$) variations is publicly available~\cite{Heinrich:2019bkc}.
However, a proper combination of the full NLO and the approximate NNLO results beyond the SM has not been performed until now.
In these proceedings, we present a first combination of the full NLO results with the HTL NNLO predictions that includes $\lambda_{hhh}$ variations and other anomalous couplings, both for the total production cross section and the Higgs pair invariant mass distribution.
Our results allow us to obtain a precision very similar to the one available for the SM cross section, therefore permitting a more consistent treatment of potential deviations from the SM.

\subsection{Results}

We work within the so-called non-linear EFT framework~\cite{Dobado:1989ax,Buchalla:2015wfa} (also called HEFT), which respects all SM gauge symmetries, and assume CP-symmetry in the Higgs sector. The non-linear EFT Lagrangian
does not a priori assume a relation between the Higgs scalar $h$ and the Goldstone bosons $\phi_i$ of electroweak symmetry breaking, which means that the Higgs field $h$ is treated as an electroweak singlet.
The symmetry breaking pattern in the scalar sector is
$SU(2)_L \times SU(2)_R\to SU(2)_{L+R}$, such that the new physics preserves custodial symmetry which protects the $\rho$-parameter.
The Lagrangian relevant for Higgs boson pair production can be parameterized as~\cite{Buchalla:2018yce}
\begin{align}
{\cal L}\supset 
-m_t\left(c_t\frac{h}{v}+c_{tt}\frac{h^2}{v^2}\right)\,\bar{t}\,t -
c_{hhh} \frac{m_h^2}{2v} h^3+\frac{\alpha_s}{8\pi} \left( c_{ggh} \frac{h}{v}+
c_{gghh}\frac{h^2}{v^2}  \right)\, G^a_{\mu \nu} G^{a,\mu \nu}\;.
\label{eq:Higgs_HH_NNLO_EFT:ewchl}
\end{align}
This Lagrangian is very similar to the one in SMEFT~\cite{DiMicco:2019ngk}, the main difference being that in SMEFT there is a relation between $c_{ggh}$ and $c_{gghh}$, $c_{ggh}=2c_{gghh}$,
and $c_{tt}$ is suppressed compared to $c_t$, while in HEFT a priori no such relations hold. 

Within this framework, NLO predictions with full top-quark mass dependence have been obtained in Ref.~\cite{Buchalla:2018yce}, and at NNLO in the (Born-improved) HTL in Ref.~\cite{deFlorian:2017qfk}.
In order to profit from both calculations, we perform a combination based on a bin-by-bin reweighting of the Higgs pair invariant mass distribution, that is
\begin{equation}\label{eq:Higgs_HH_NNLO_EFT:NLO-i}
\Delta\sigma(\text{NNLO}_\text{NLO-i}) =
\Delta\sigma(\text{NLO}_\text{Full}) \times \frac{\Delta\sigma(\text{NNLO}_\text{B-i})}{\Delta\sigma(\text{NLO}_\text{B-i})}\,.
\end{equation}
The above reweighting is performed individually for each invariant mass bin of each point in the EFT parameter space.
The corresponding total cross sections are afterwards obtained by summing over the whole invariant mass range.
The exact definition of the Born-improved approximation can be found in Ref.~\cite{deFlorian:2017qfk}, and is based in replacing HTL form factors by their full LO counterparts, including in this way partial finite top-mass effects.

Of course, the results obtained by applying Eq.~(\ref{eq:Higgs_HH_NNLO_EFT:NLO-i}) do not fully agree with the NNLO$_\text{FTapprox}$ prediction, though this bin-by-bin reweighting was found to provide results very close to it at NNLO~\cite{Grazzini:2018bsd}.
Therefore, in order to provide a consistent prediction that behaves smoothly in the SM limit, we add a normalization factor (independent of both the invariant mass bin and the point in the EFT parameter space) to recover the correct NNLO$_\text{FTapprox}$ total cross section.
A similar procedure is performed for the scale variation (which in the present work is based on a 3-point variation, $\mu_R = \mu_F = \xi M_{hh}/2$ with $\xi = 1/2, 1, 2$).
Our results are for a centre-of-mass energy of $\sqrt{s}=13$\,TeV and are
computed using the
PDF4LHC15~\cite{Butterworth:2015oua}
parton distribution functions interfaced  via
LHAPDF~\cite{Buckley:2014ana}, along with the corresponding value for
$\alpha_s(\mu)$, with $\alpha_s(M_Z)=0.118$.  The masses of the Higgs boson and the top quark have been
set  to $m_h=125$\,GeV and $m_t=173$\,GeV, respectively. 

\begin{figure}[p]
\begin{center}
\includegraphics[width=.32\textwidth]{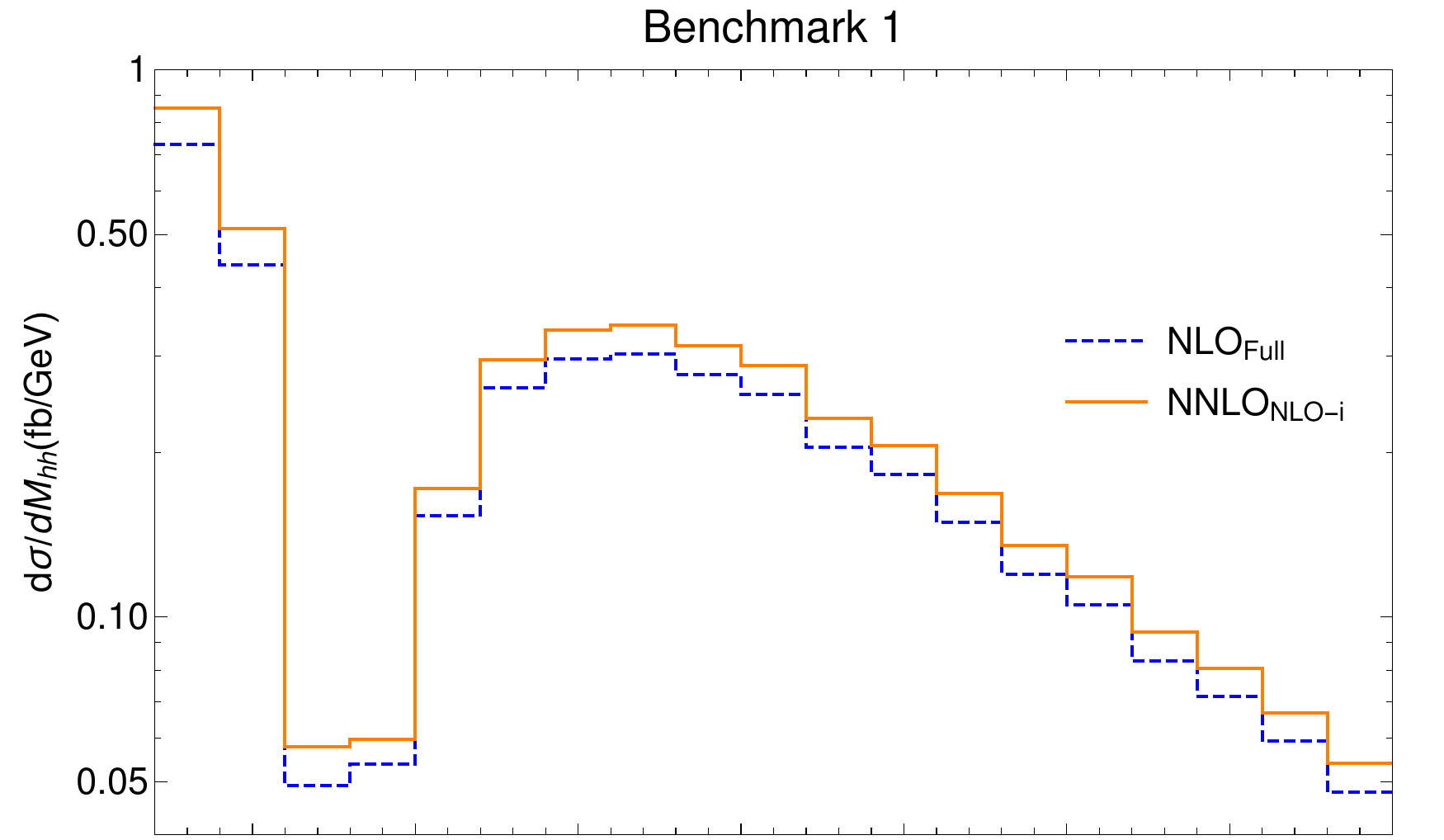}
\includegraphics[width=.32\textwidth]{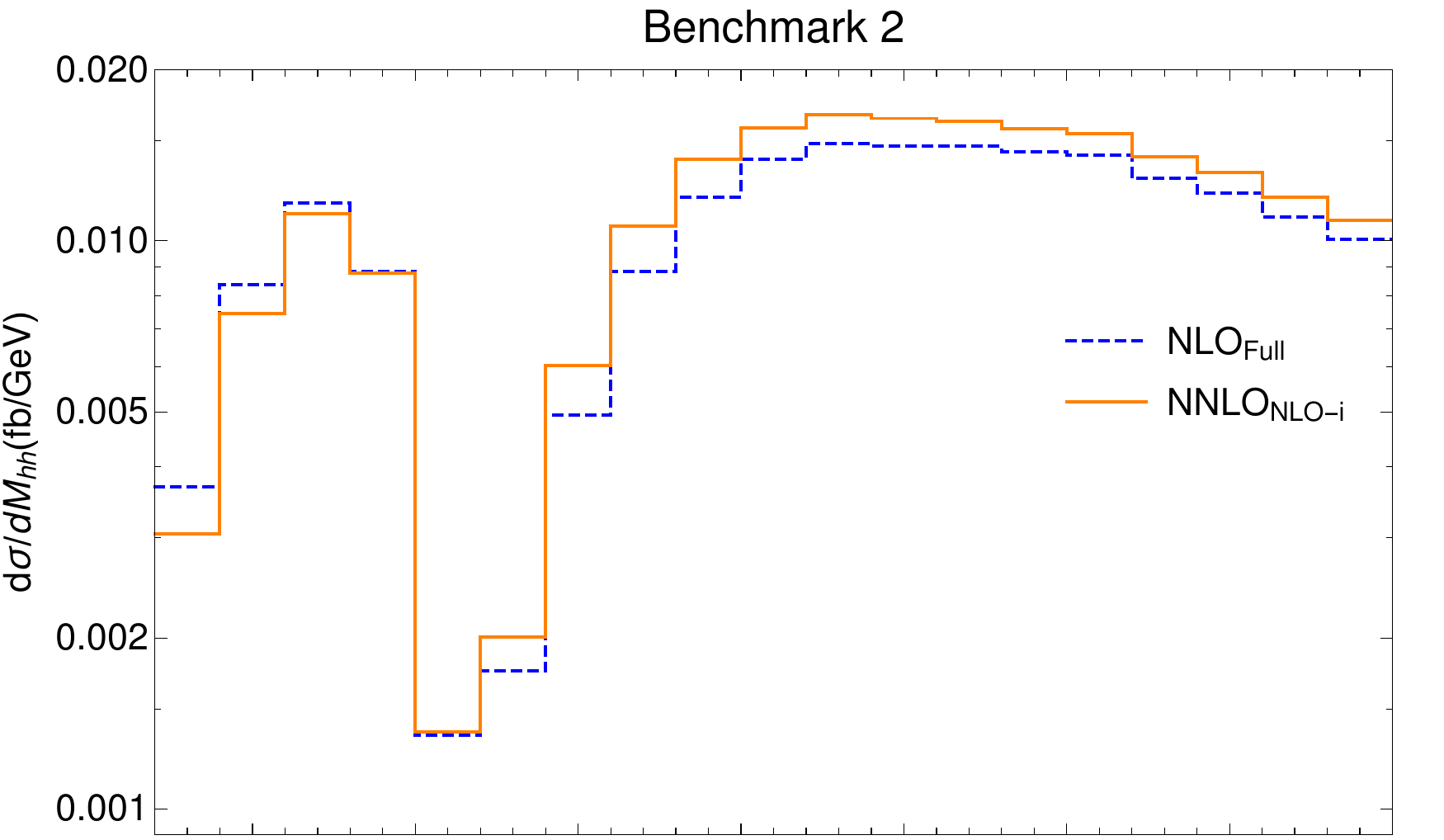}
\includegraphics[width=.32\textwidth]{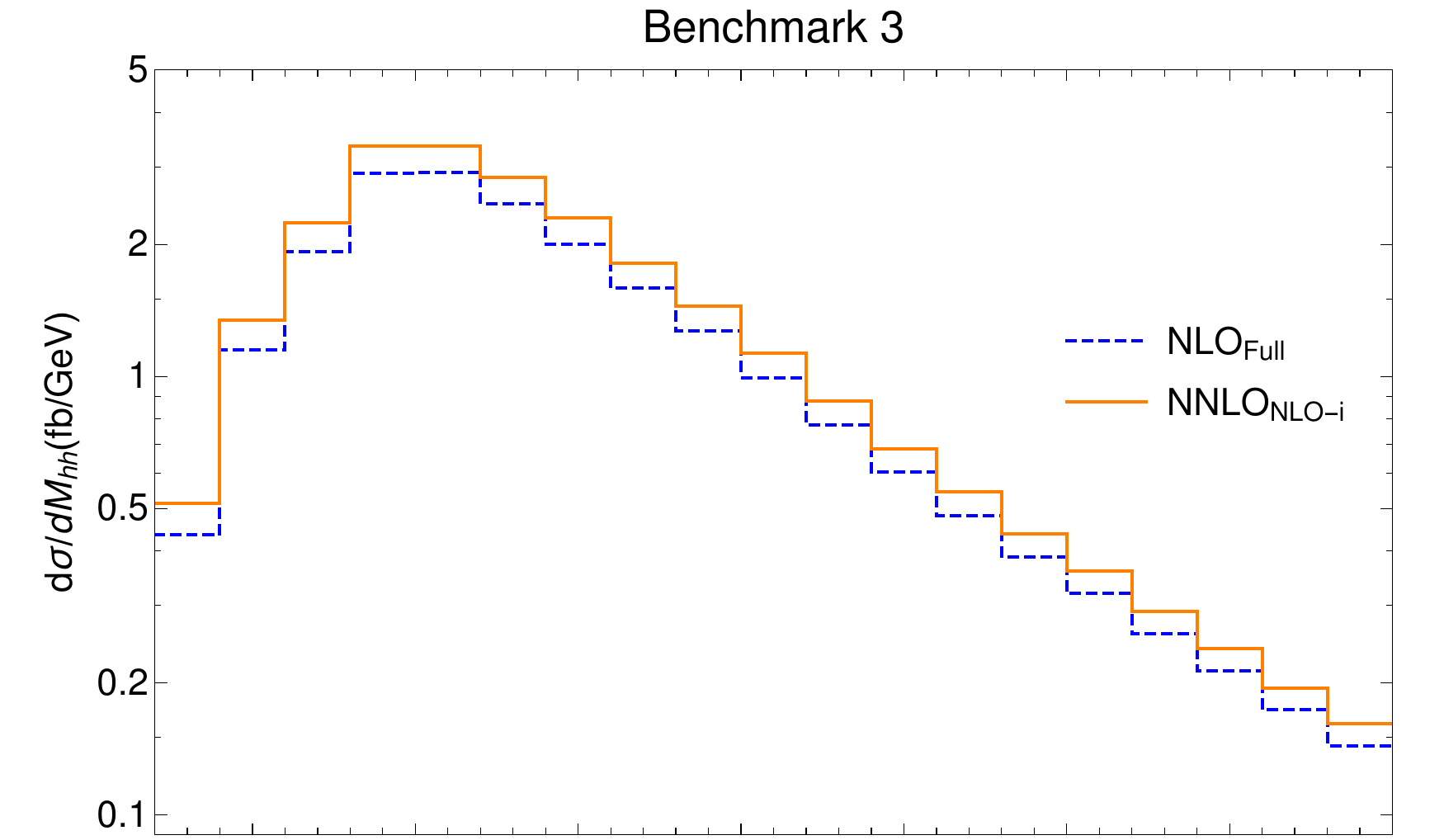}
\\
\includegraphics[width=.32\textwidth]{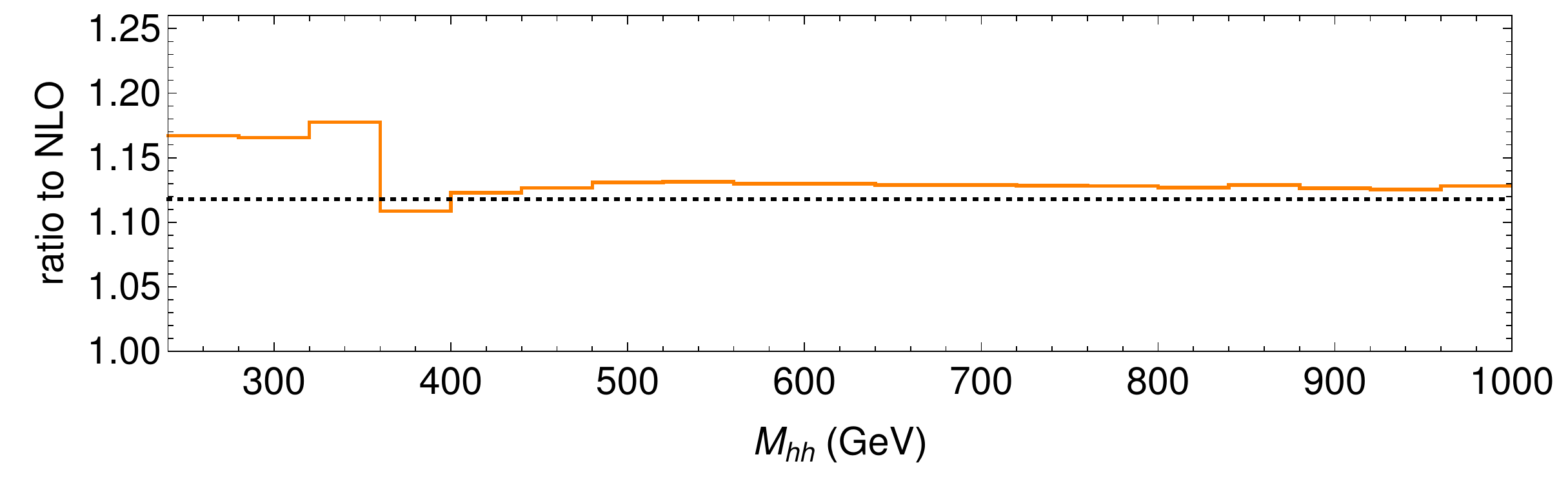}
\includegraphics[width=.32\textwidth]{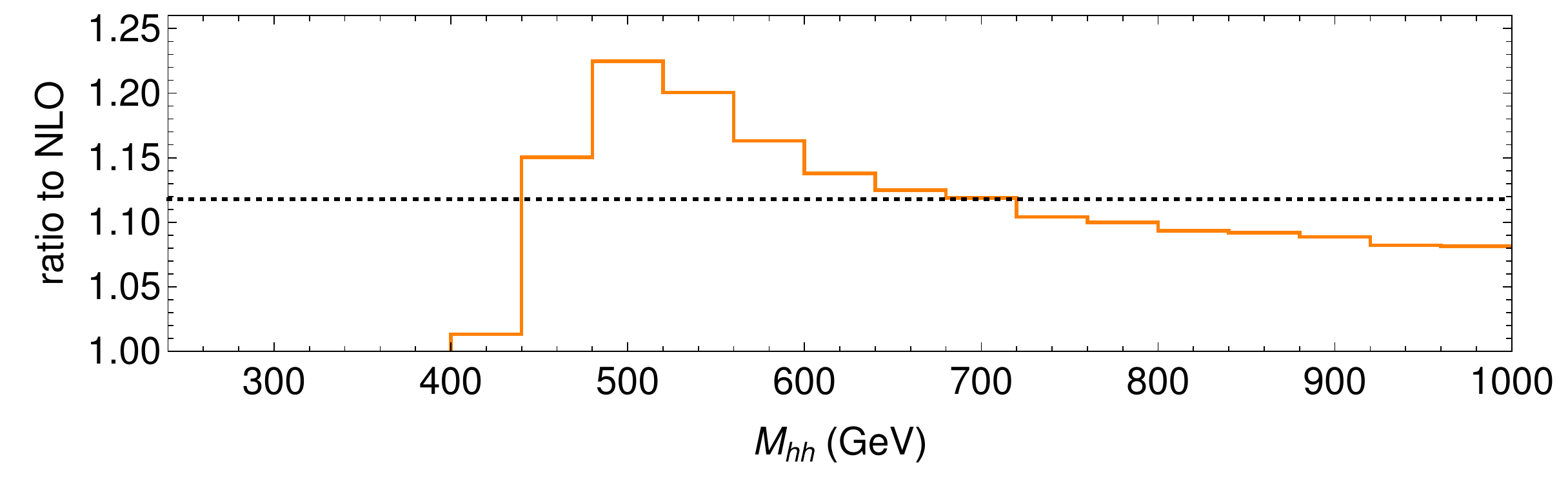}
\includegraphics[width=.32\textwidth]{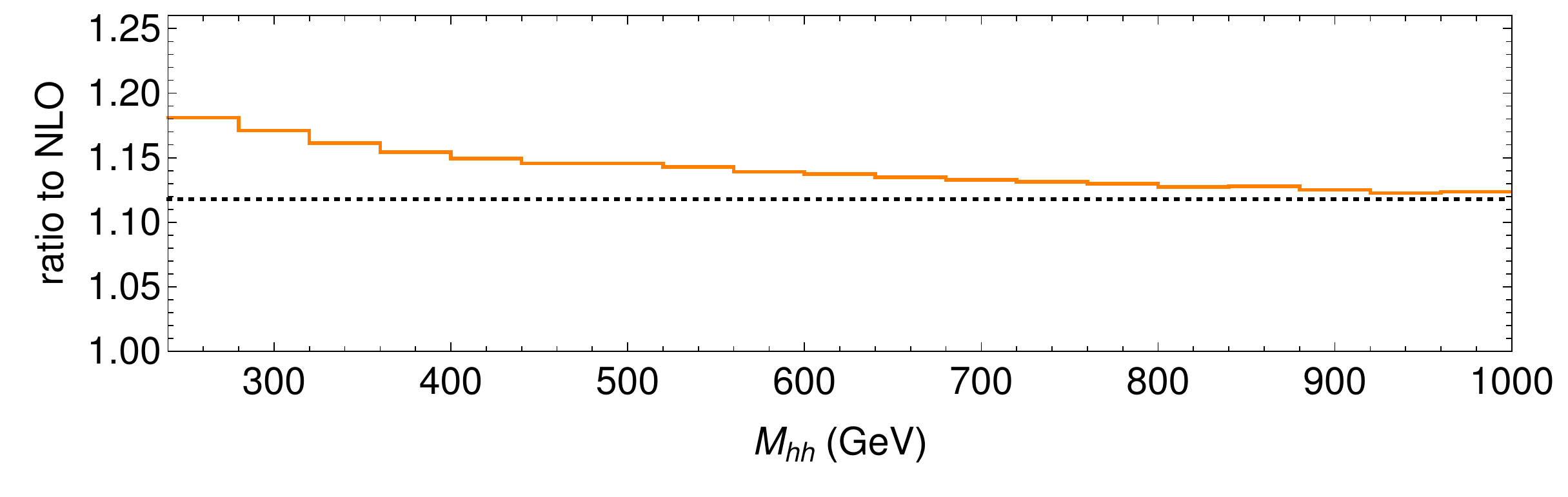}
\\
\vspace*{0.4cm}
\includegraphics[width=.32\textwidth]{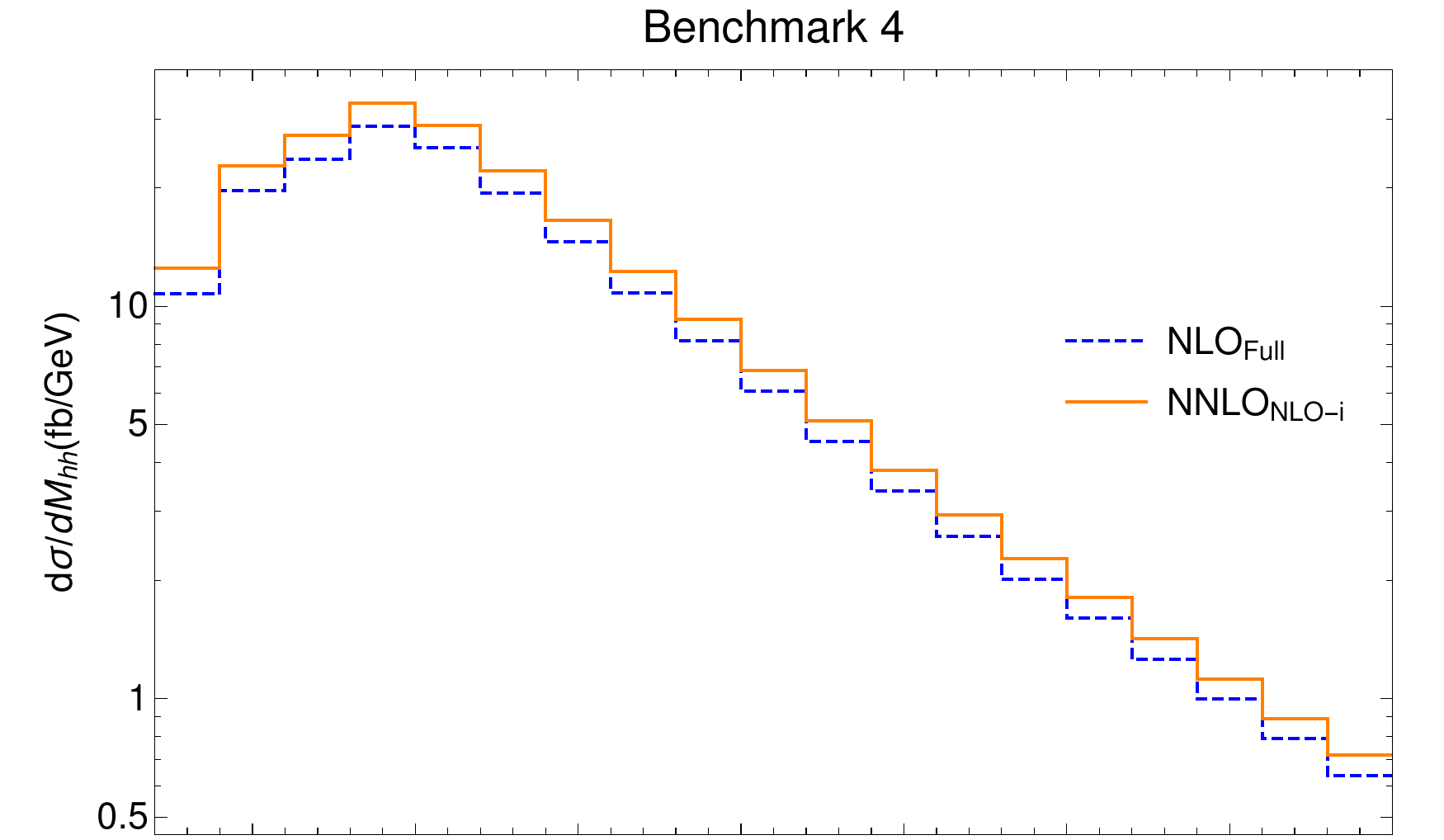}
\includegraphics[width=.32\textwidth]{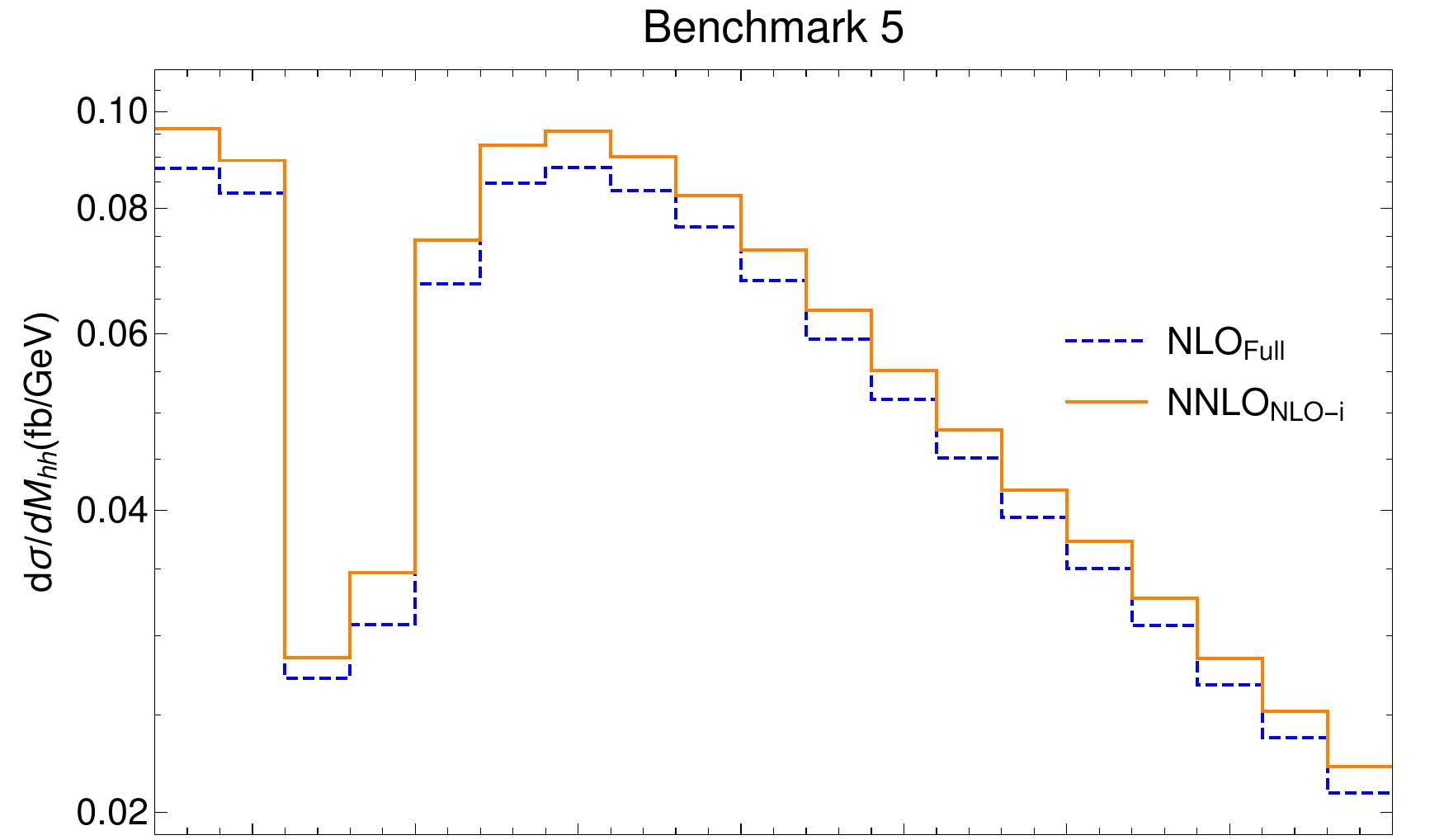}
\includegraphics[width=.32\textwidth]{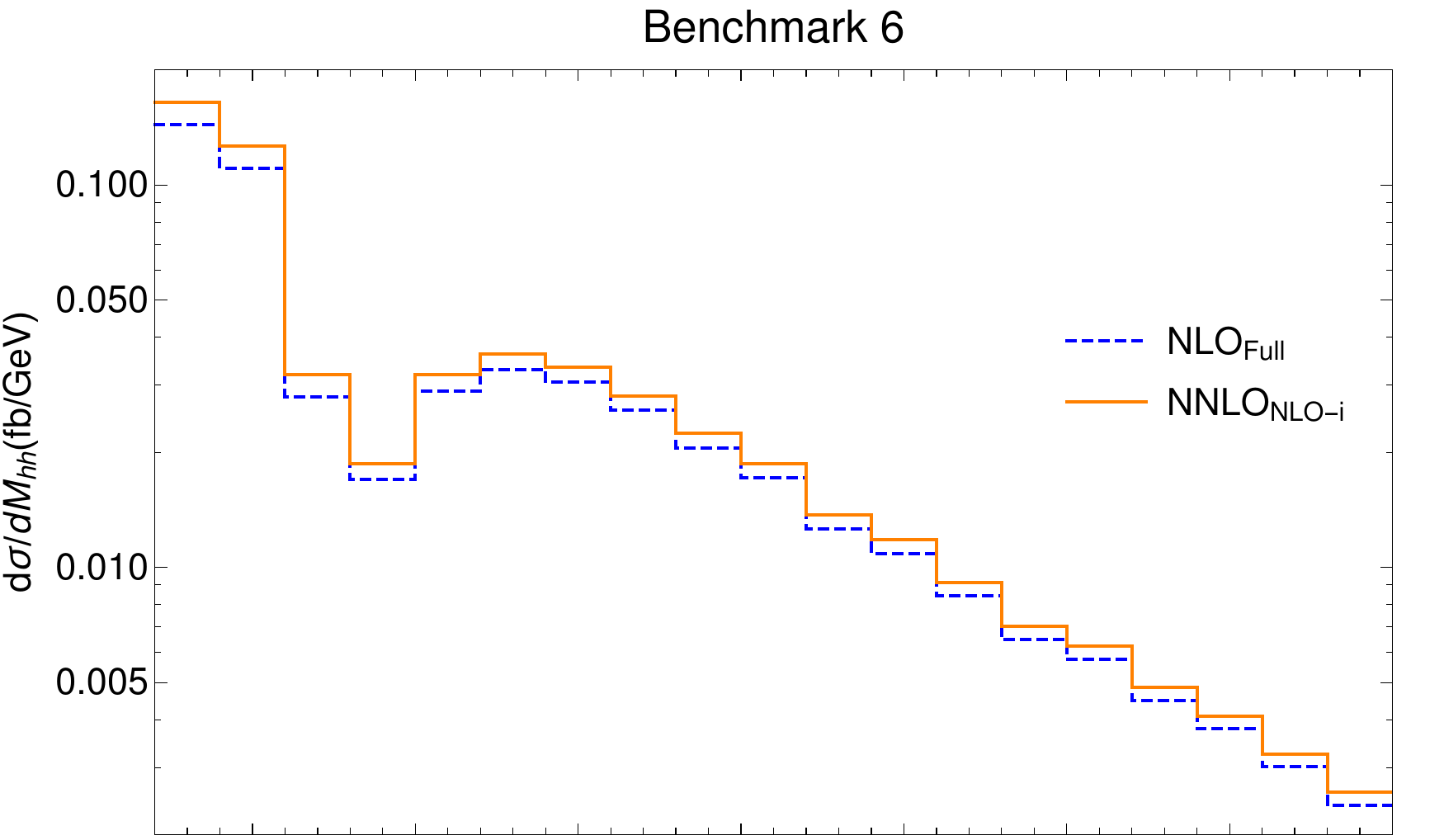}
\\
\includegraphics[width=.32\textwidth]{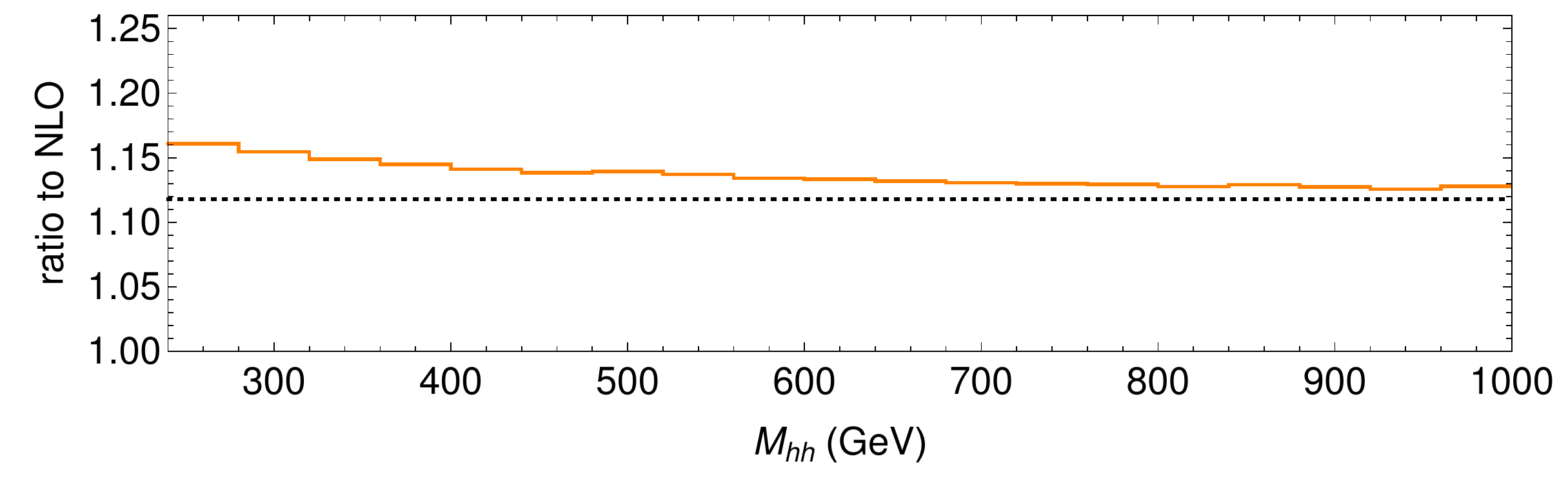}
\includegraphics[width=.32\textwidth]{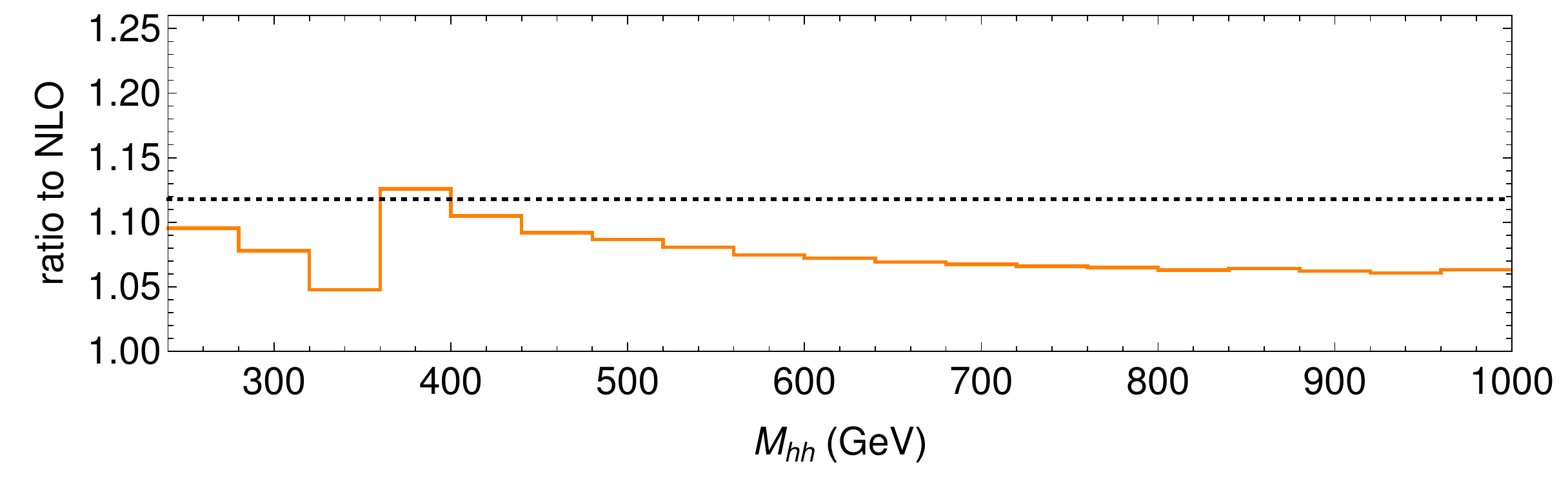}
\includegraphics[width=.32\textwidth]{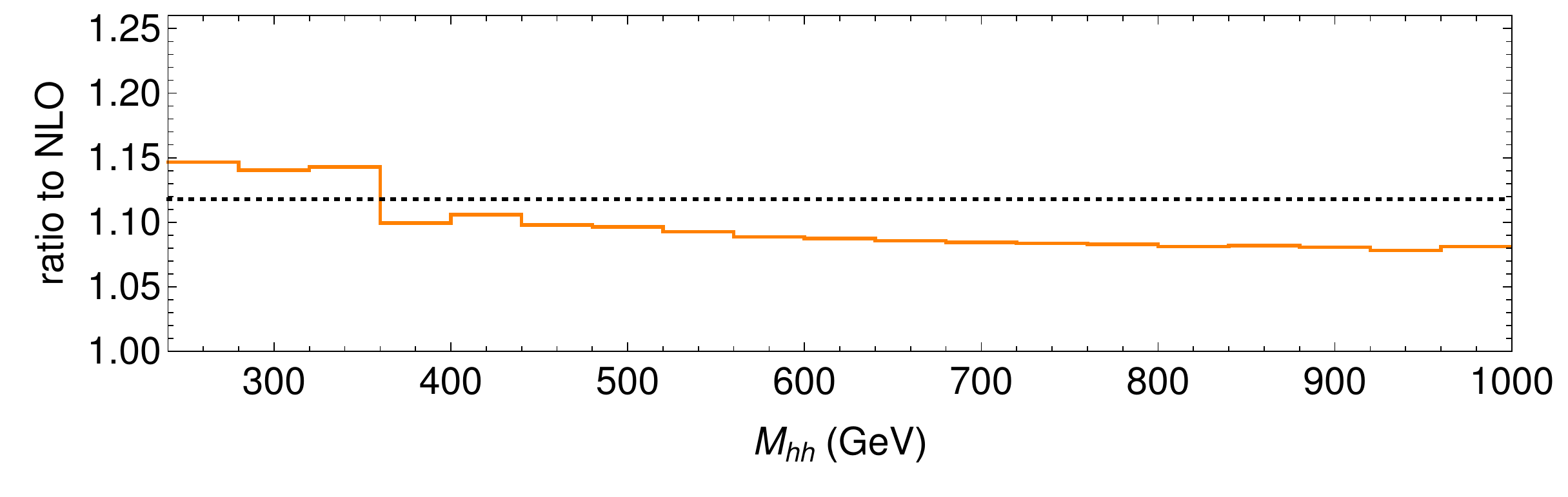}
\\
\vspace*{0.4cm}
\includegraphics[width=.32\textwidth]{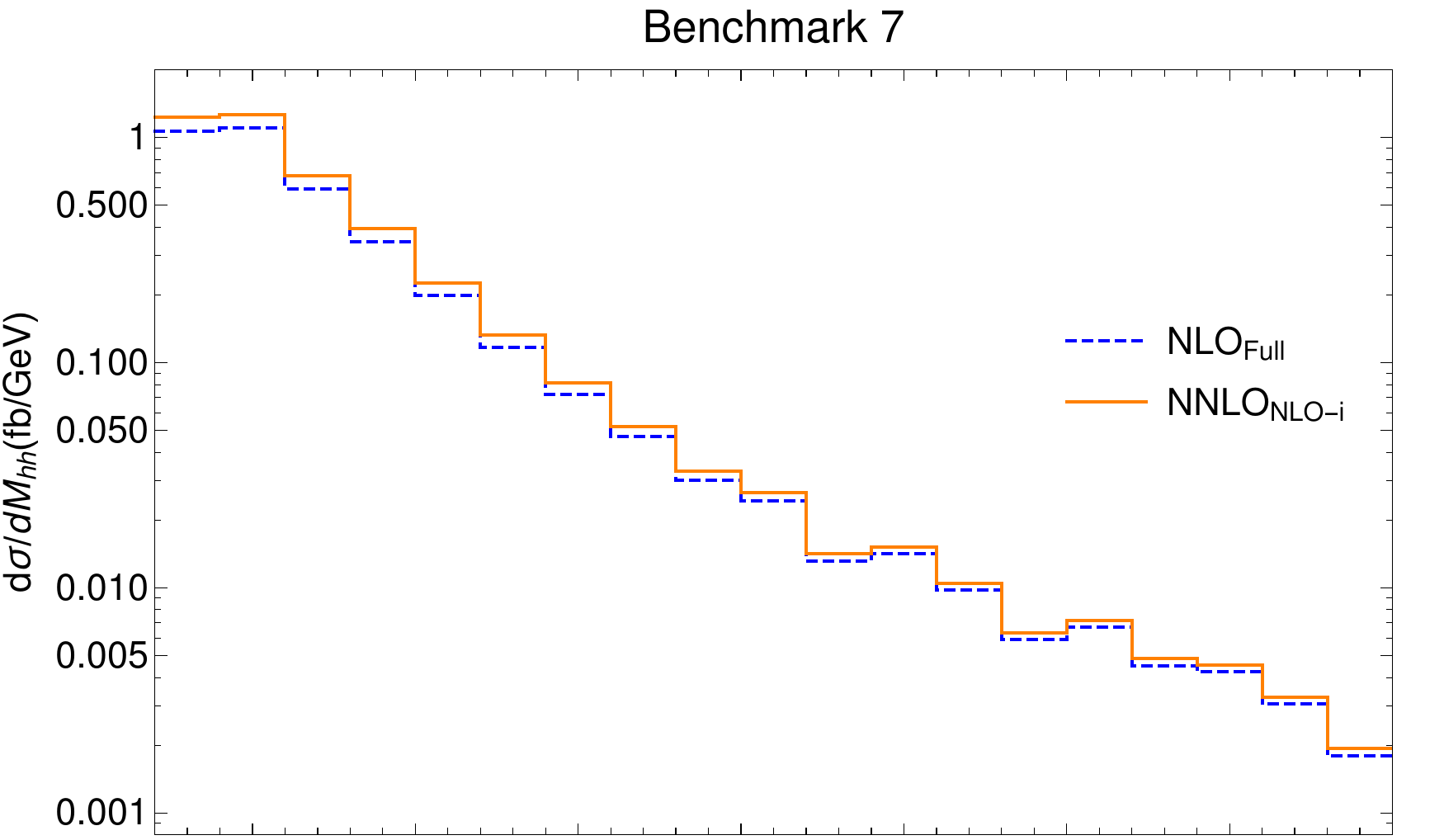}
\includegraphics[width=.32\textwidth]{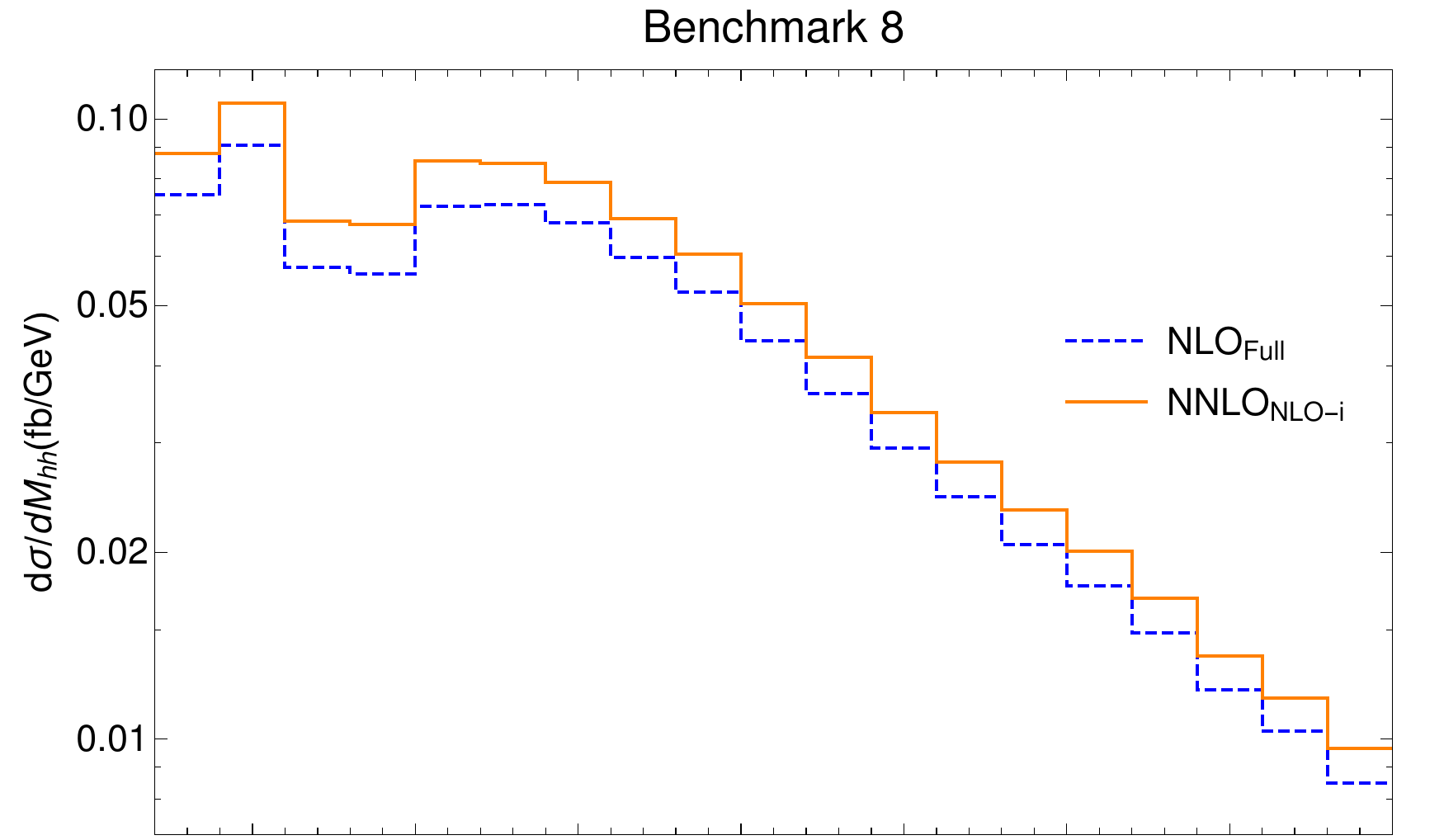}
\includegraphics[width=.32\textwidth]{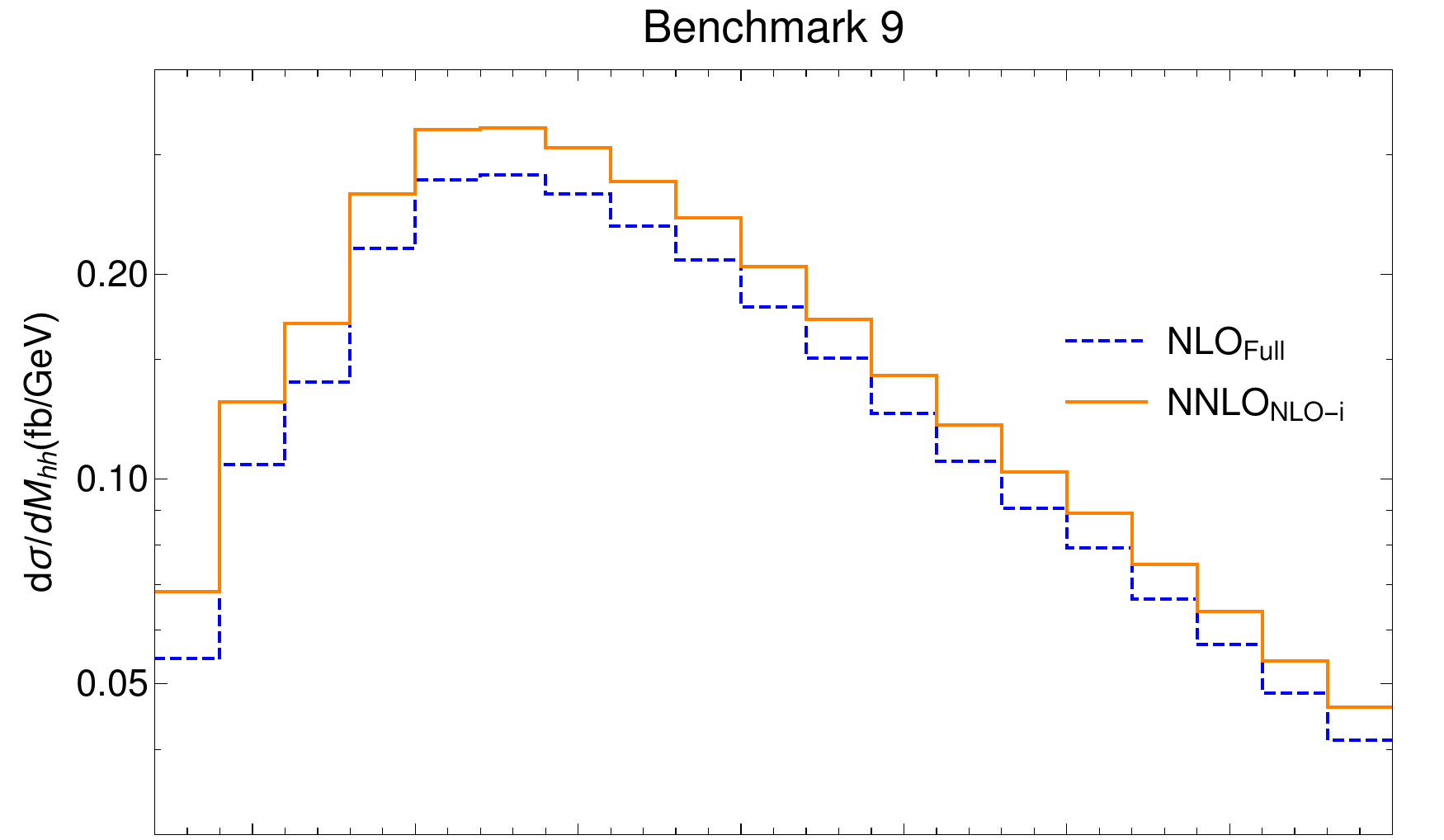}
\\
\includegraphics[width=.32\textwidth]{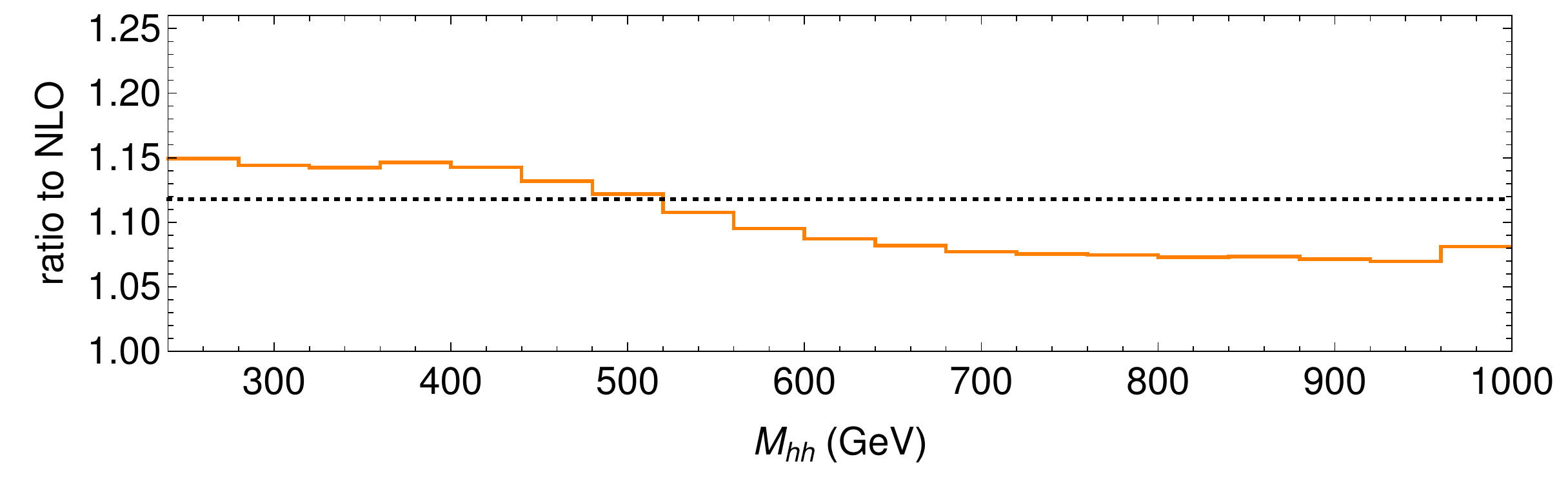}
\includegraphics[width=.32\textwidth]{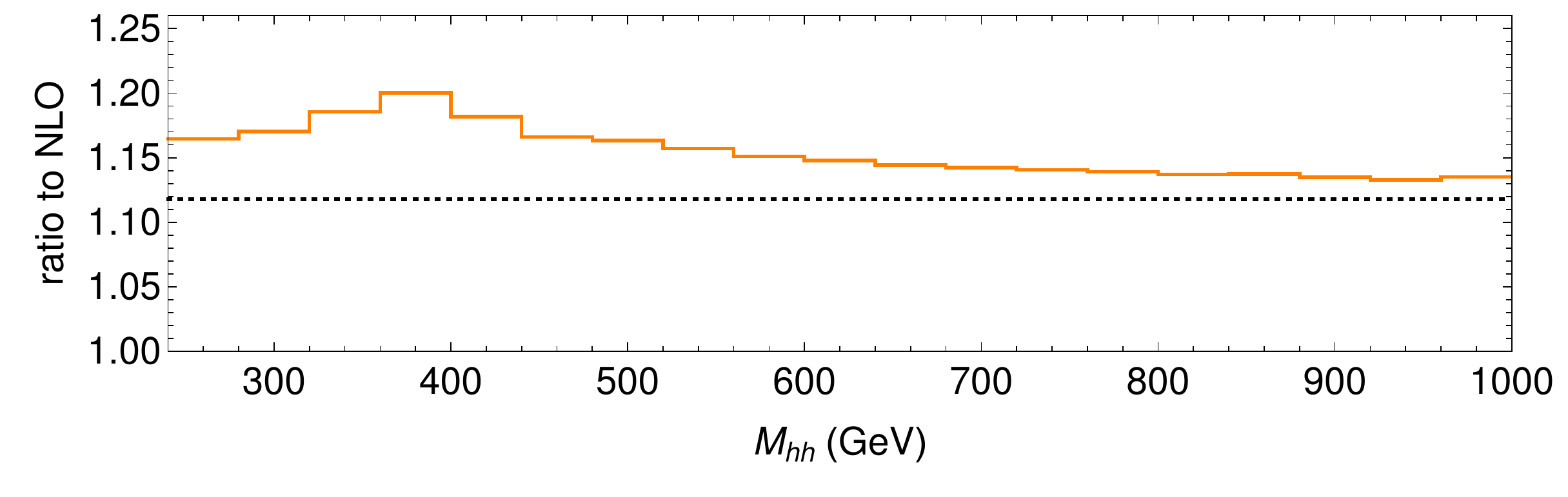}
\includegraphics[width=.32\textwidth]{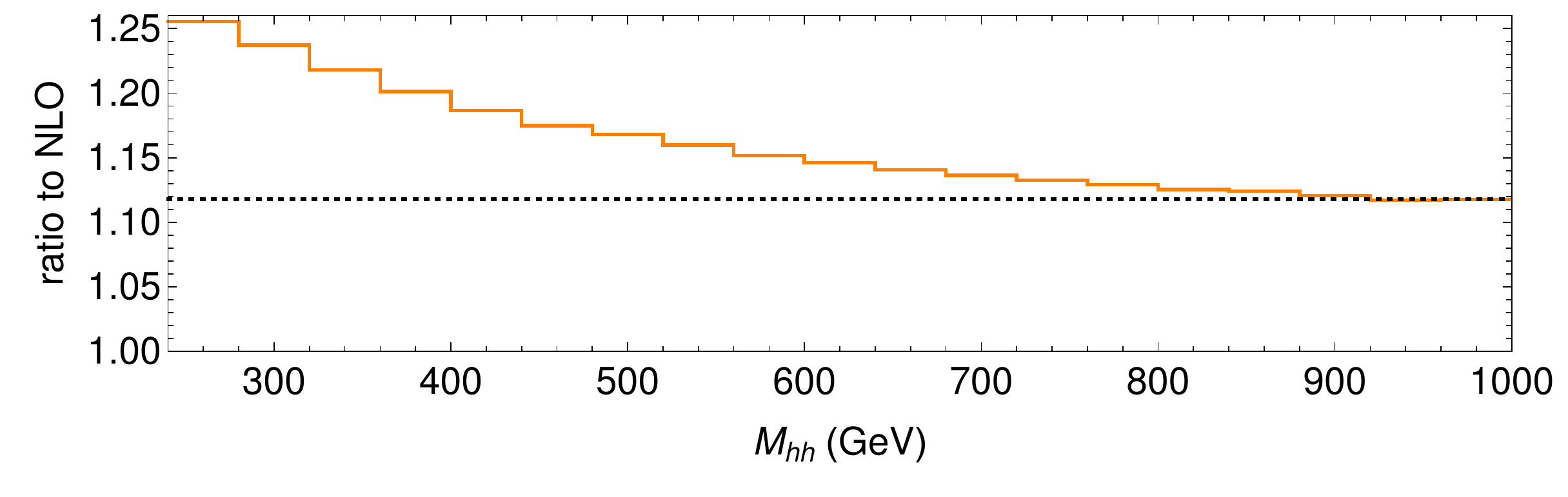}
\\
\vspace*{0.4cm}
\includegraphics[width=.32\textwidth]{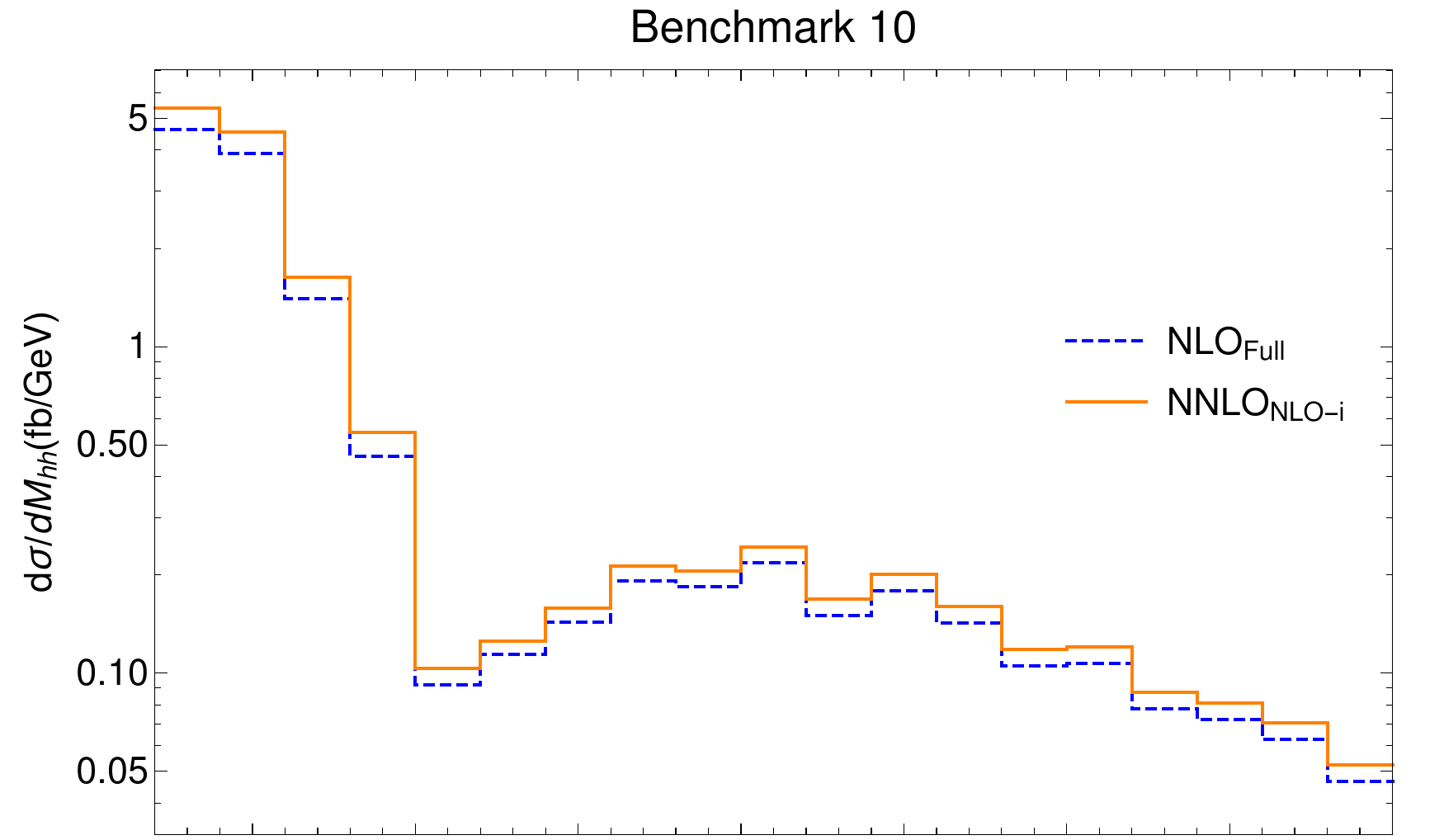}
\includegraphics[width=.32\textwidth]{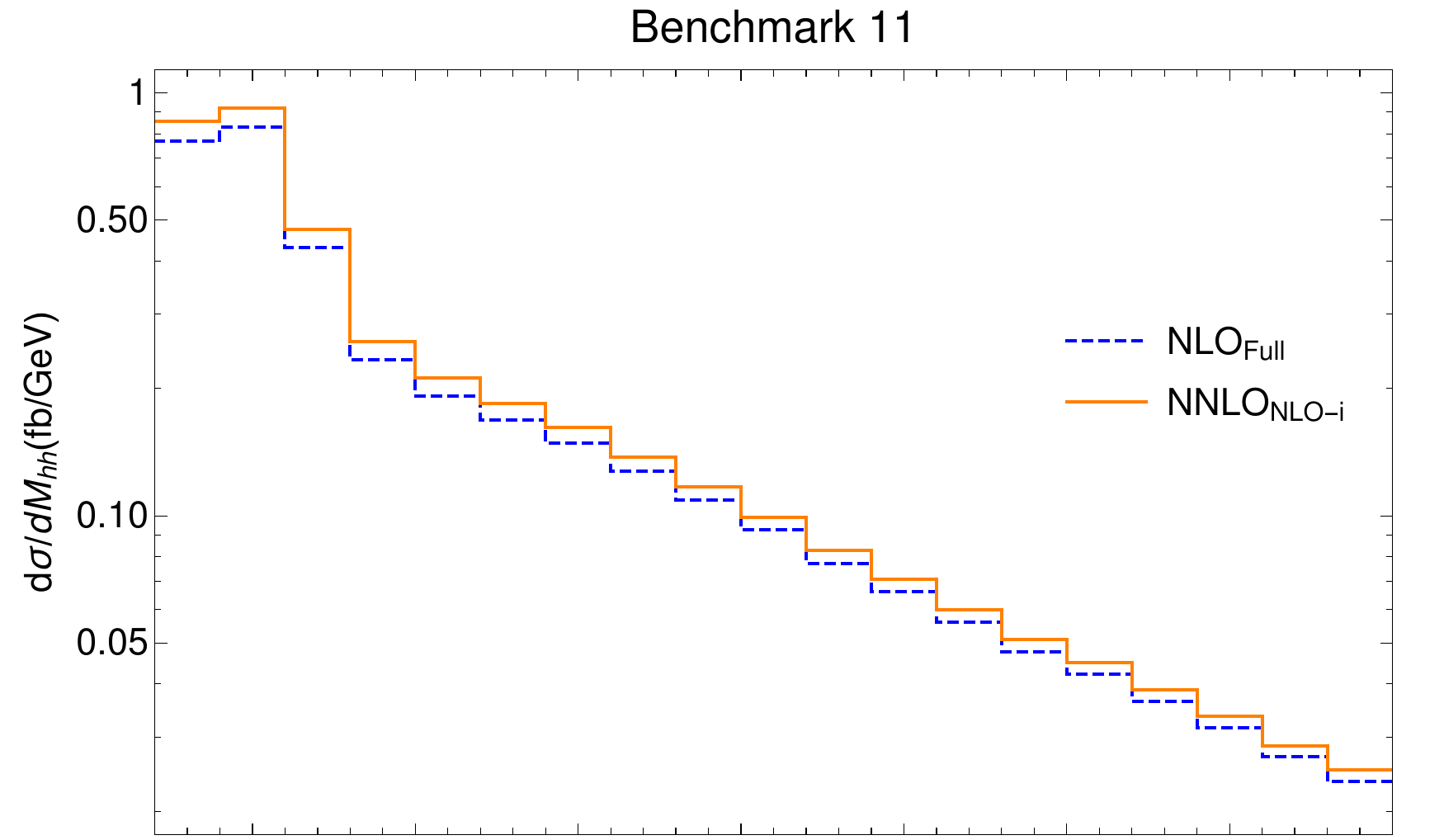}
\includegraphics[width=.32\textwidth]{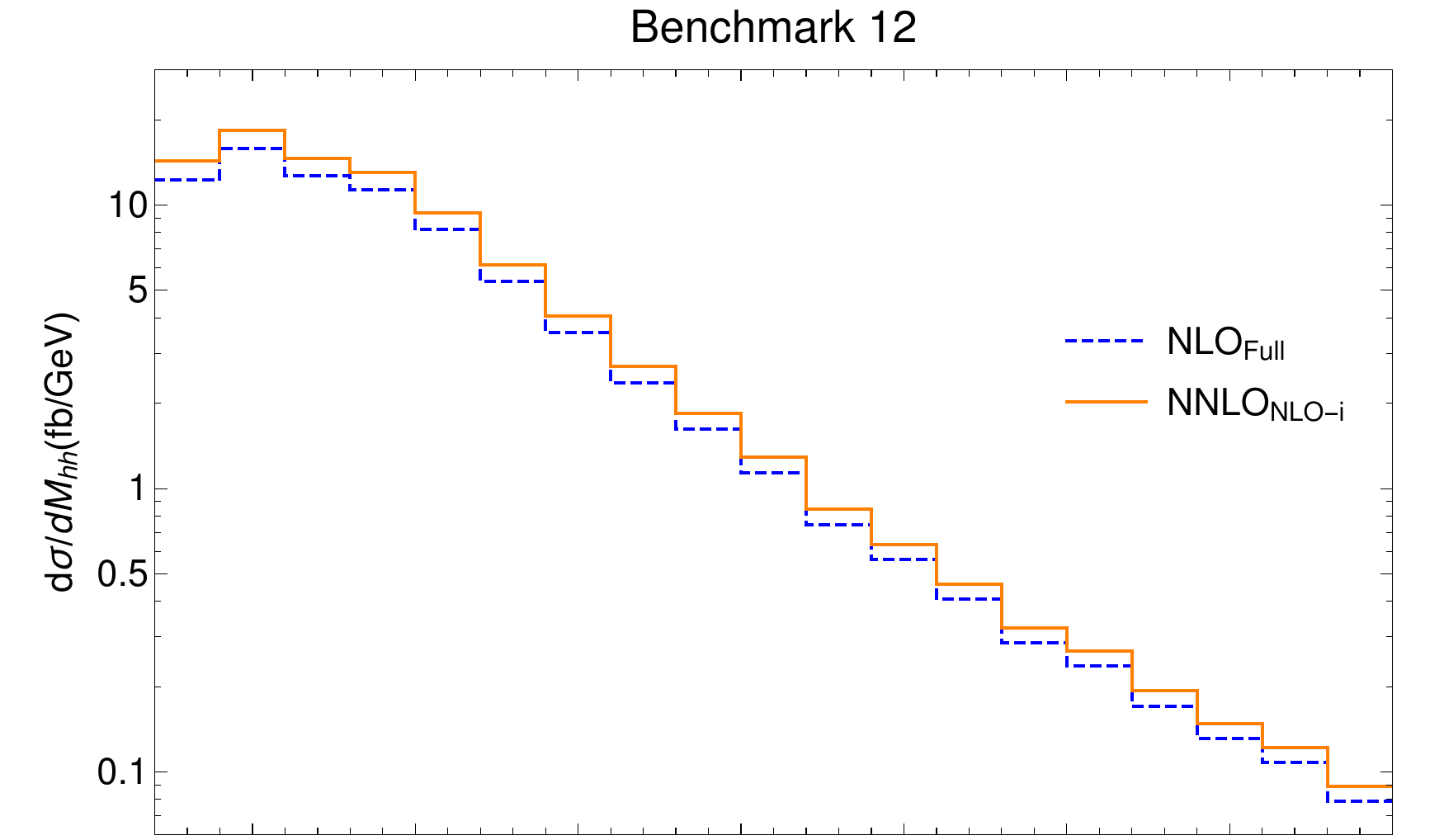}
\\
\includegraphics[width=.32\textwidth]{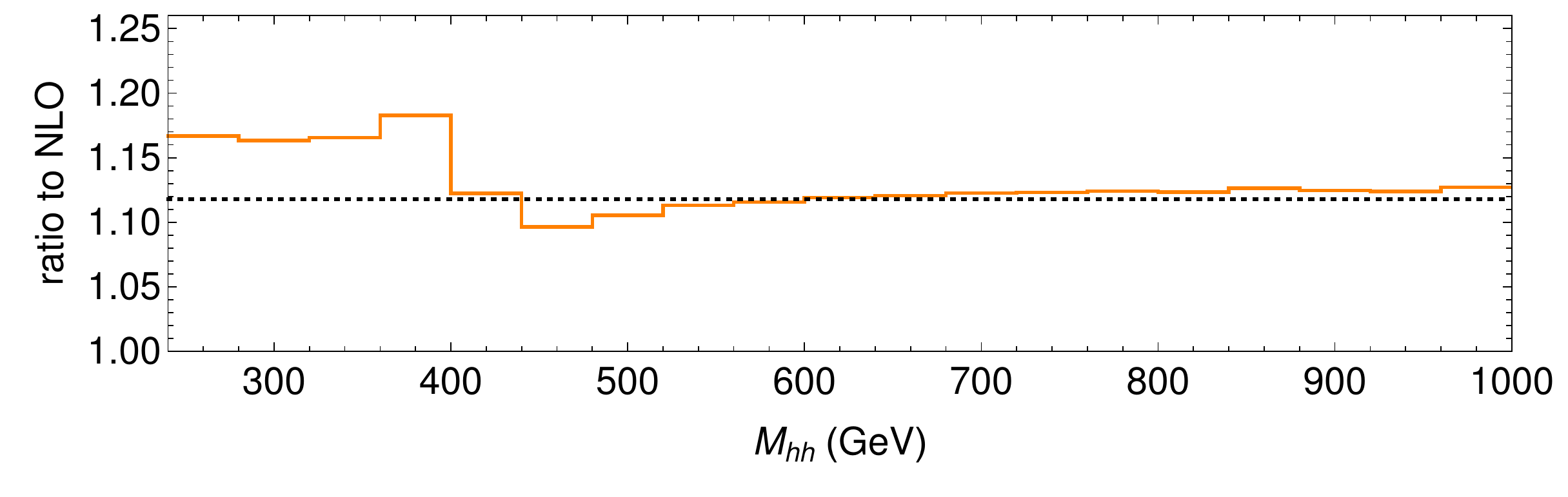}
\includegraphics[width=.32\textwidth]{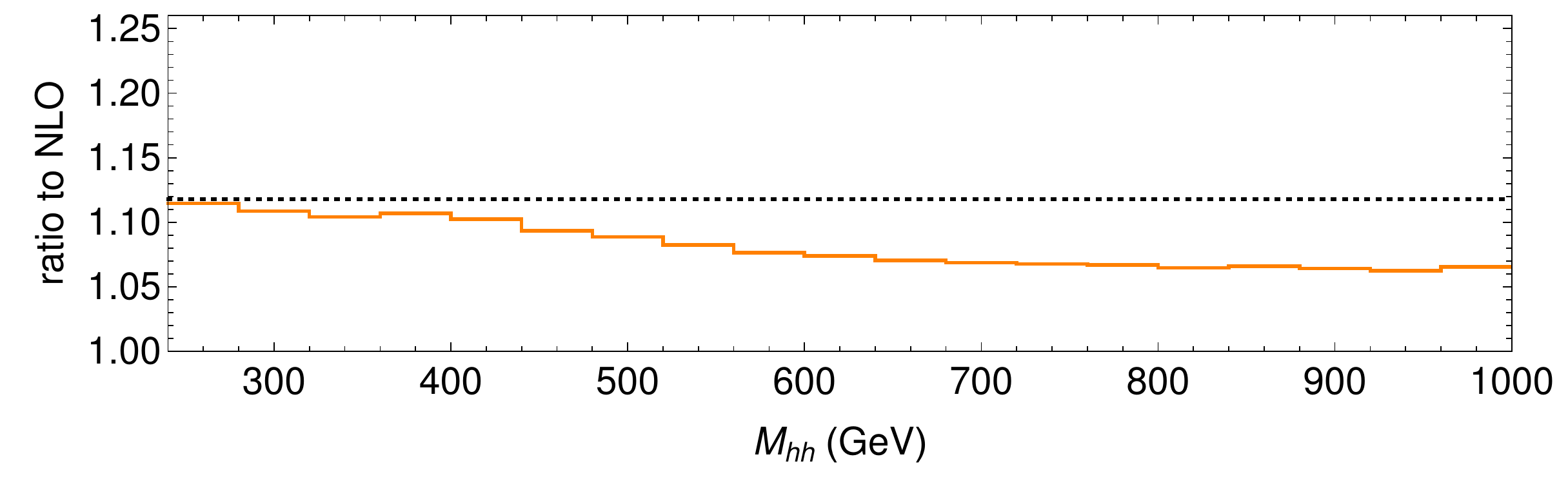}
\includegraphics[width=.32\textwidth]{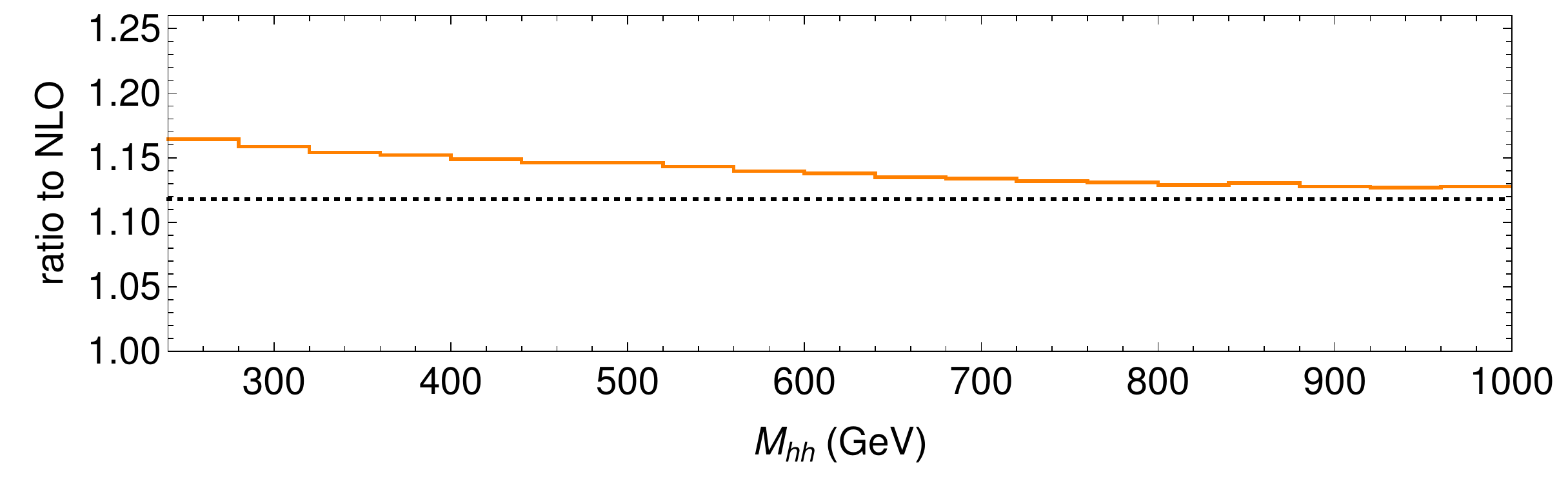}
\end{center}
\caption{\small Higgs-pair invariant mass distribution at 13~TeV for the different shape benchmarks, at NLO with full top mass dependence (blue-dashed) and NNLO HTL NLO-improved (orange-solid), the latter rescaled to the NNLO$_\text{FTapprox}$ total cross section in the SM limit. The lower panel shows the differential $K$-factor, defined as the ratio to the NLO prediction, together with the inclusive SM $K$-factor (black-dotted), $\sigma_\text{NNLO}^\text{FTapprox}/\sigma_\text{NLO}^\text{Full}$, as a reference.
\label{fig:Higgs_HH_NNLO_EFT:benchmarks_nnlo}}
\end{figure}

\begin{figure}[p]
\begin{center}
\includegraphics[width=.32\textwidth]{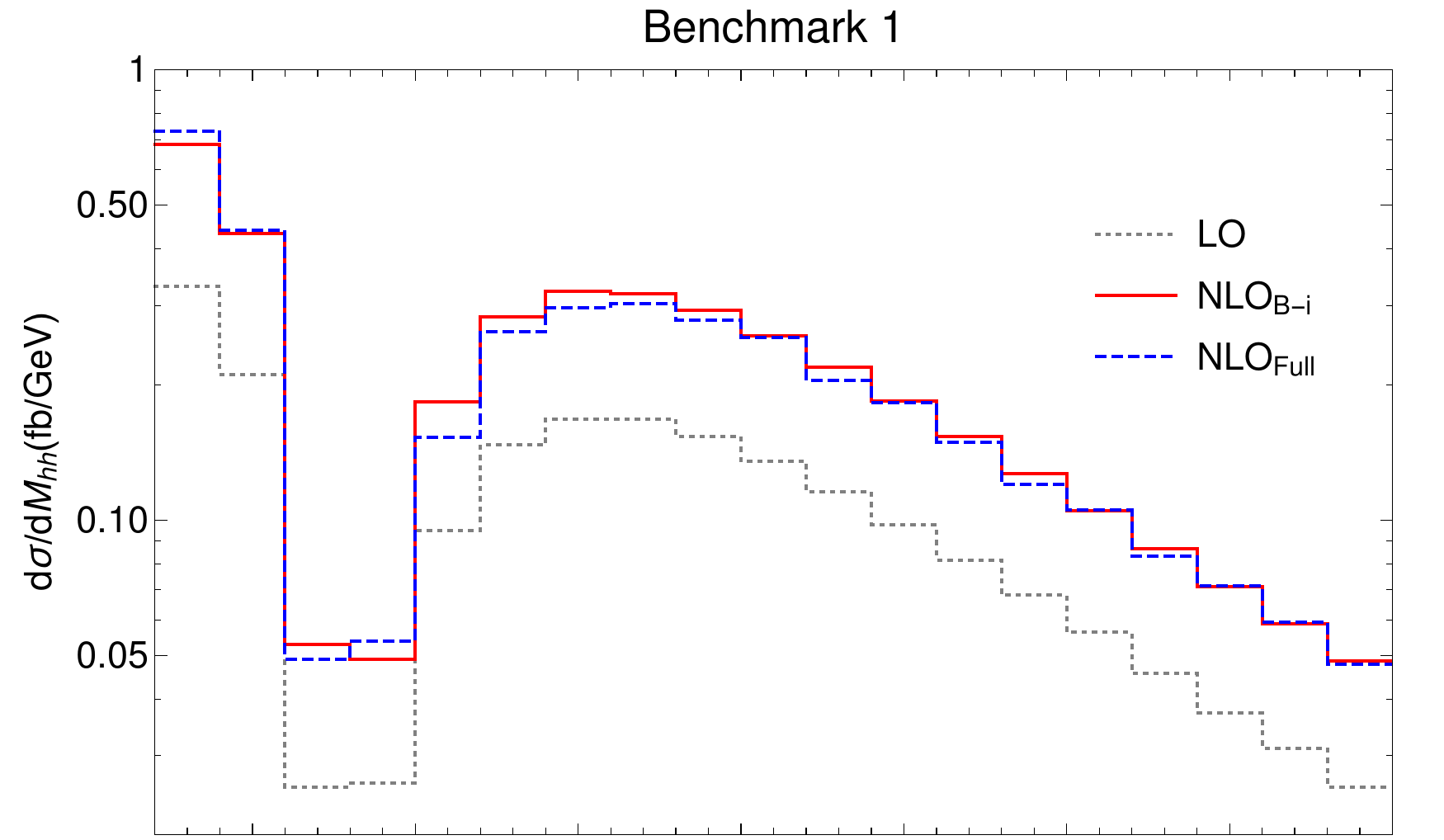}
\includegraphics[width=.32\textwidth]{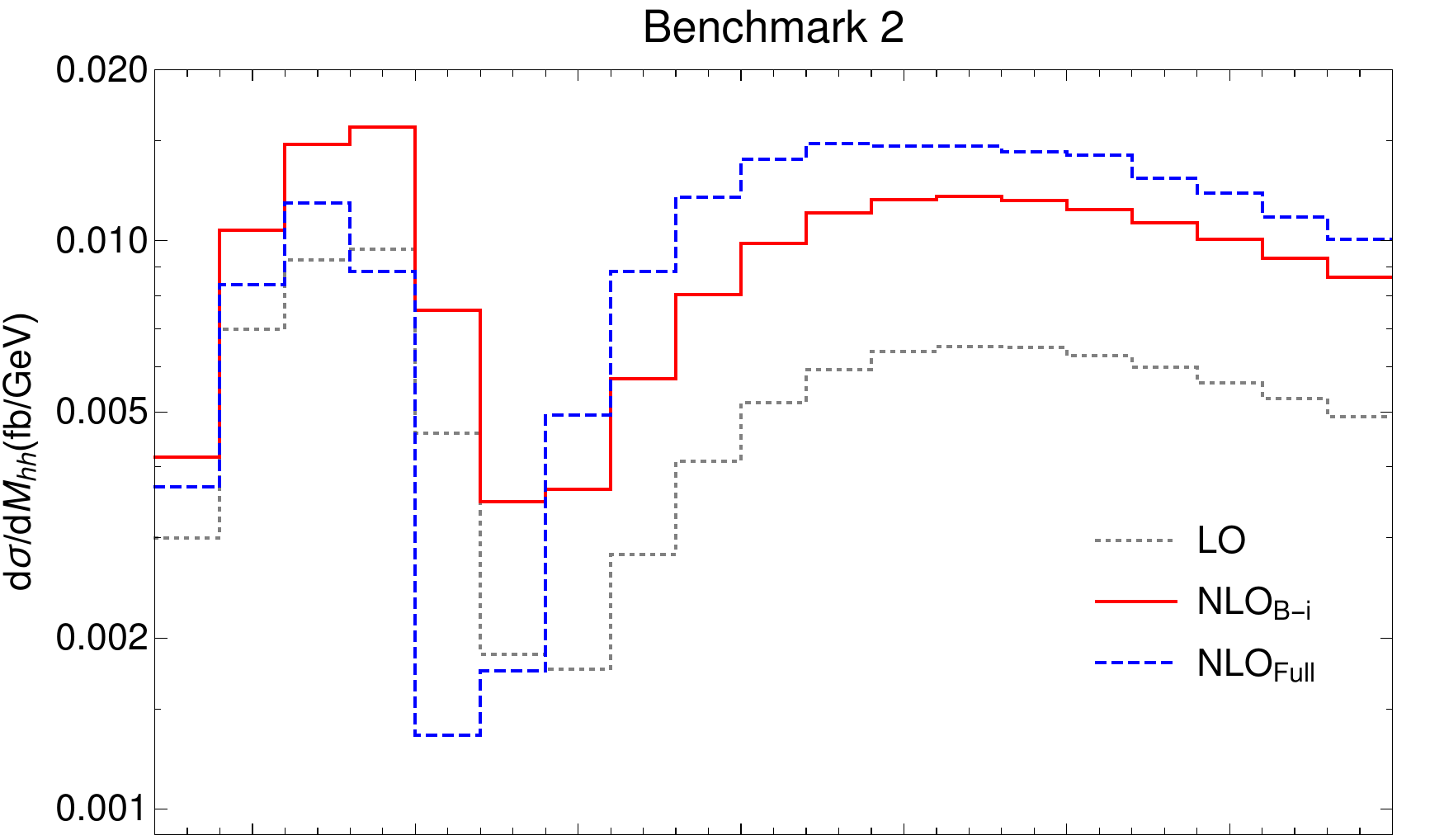}
\includegraphics[width=.32\textwidth]{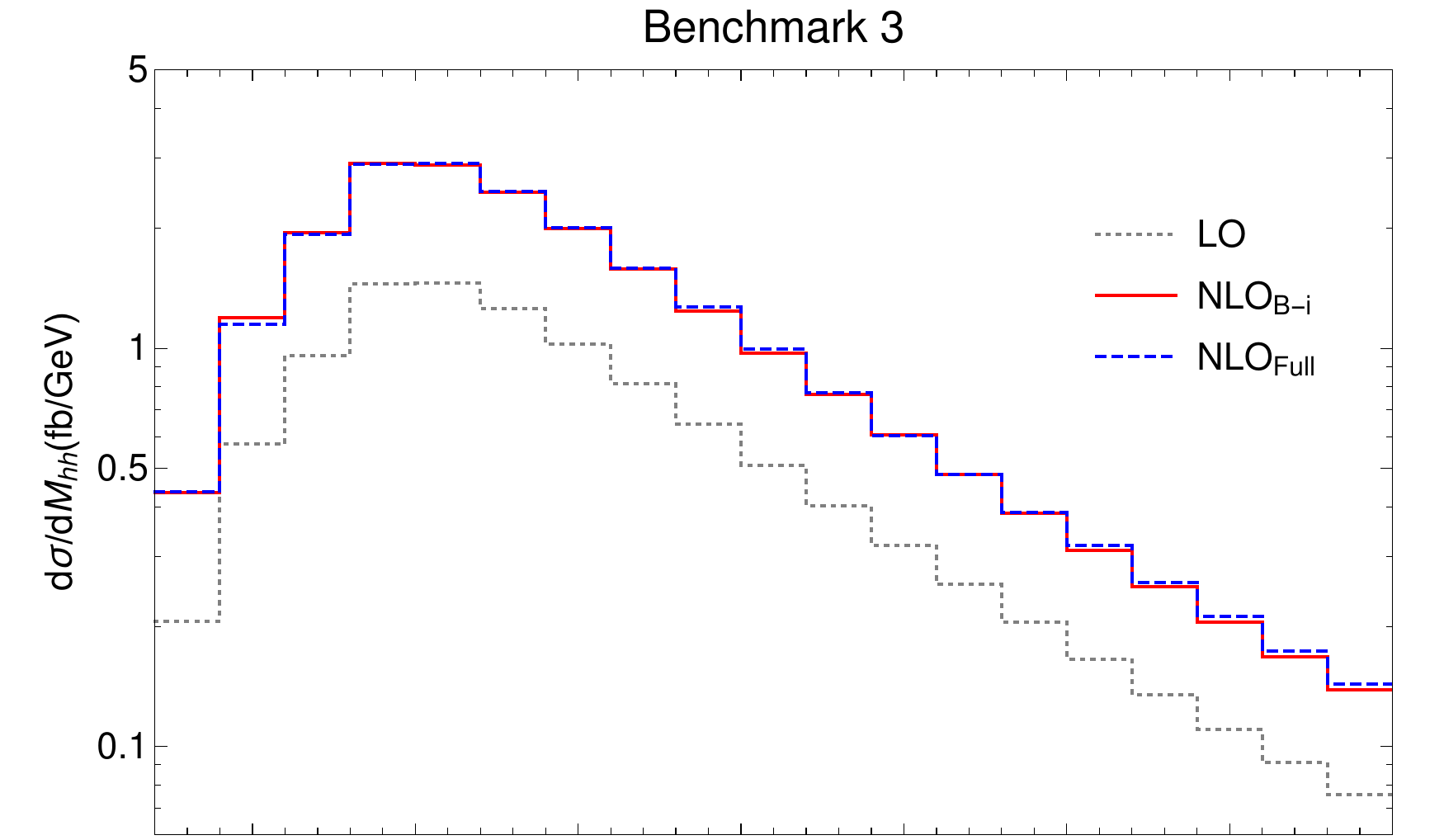}
\\
\includegraphics[width=.32\textwidth]{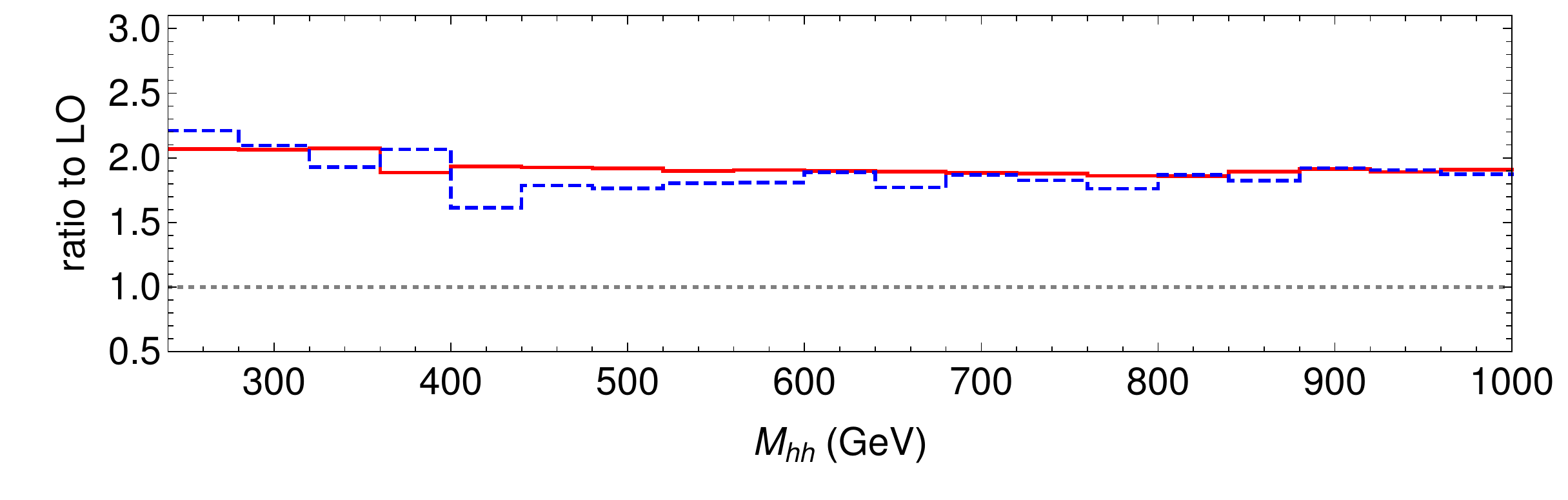}
\includegraphics[width=.32\textwidth]{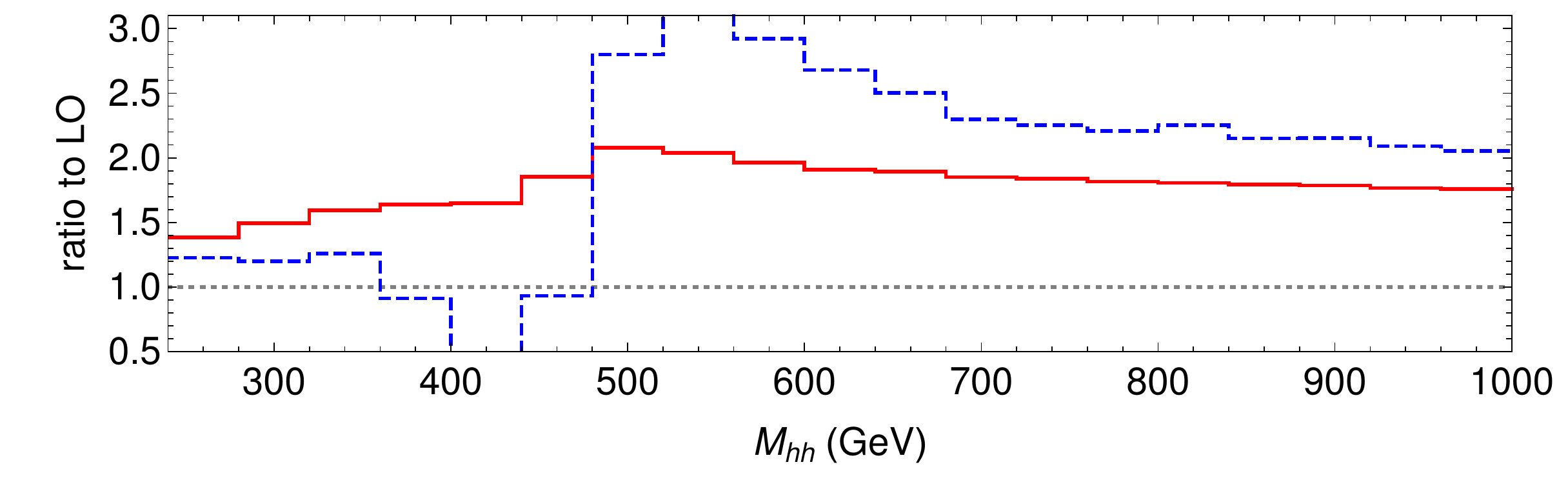}
\includegraphics[width=.32\textwidth]{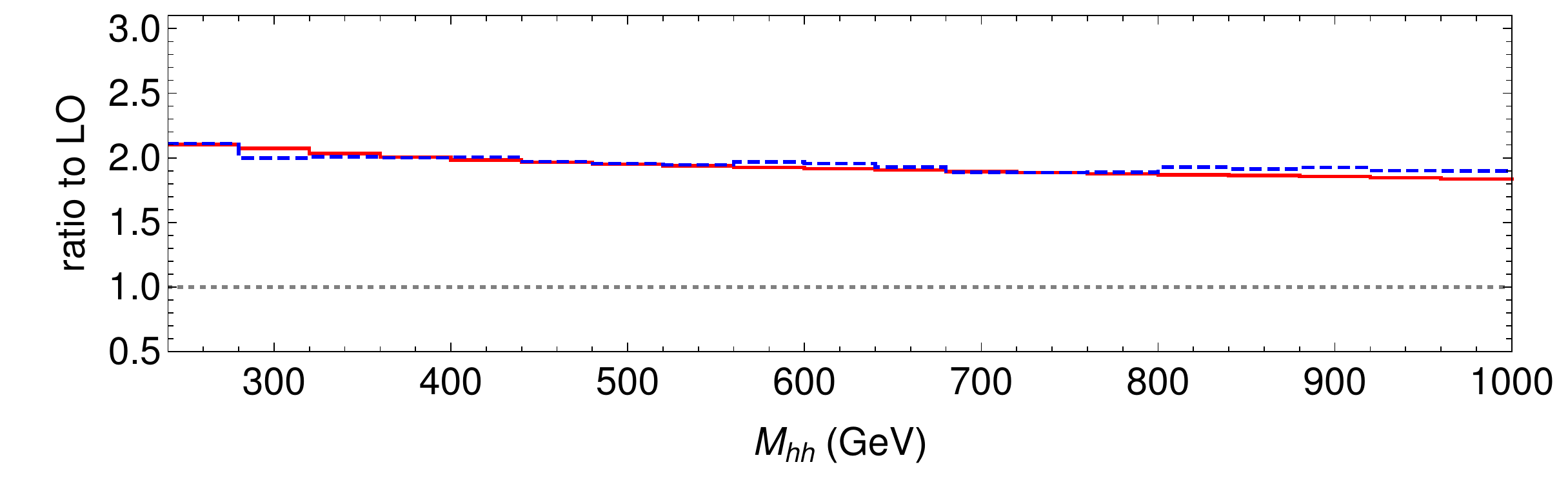}
\\
\vspace*{0.4cm}
\includegraphics[width=.32\textwidth]{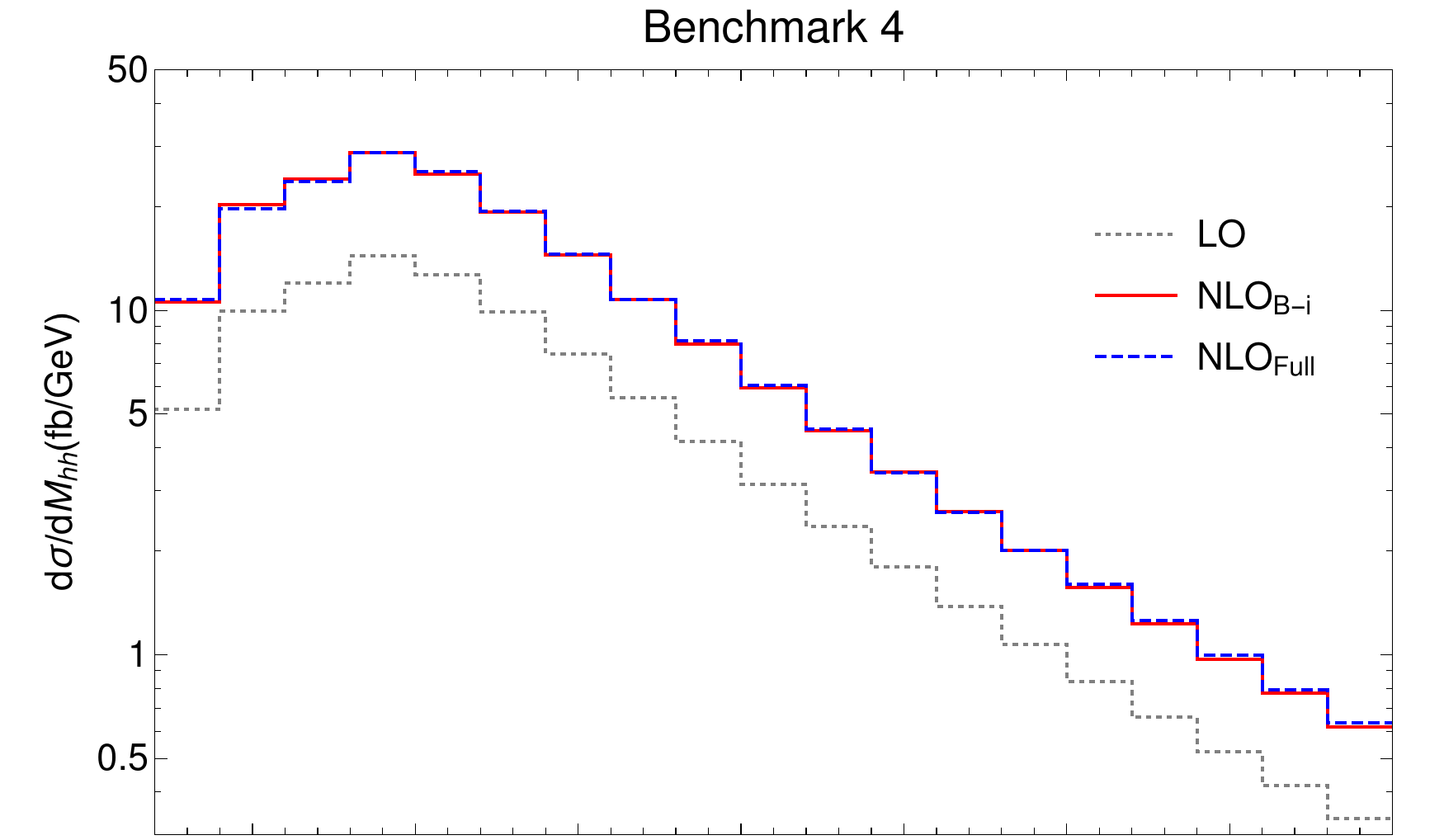}
\includegraphics[width=.32\textwidth]{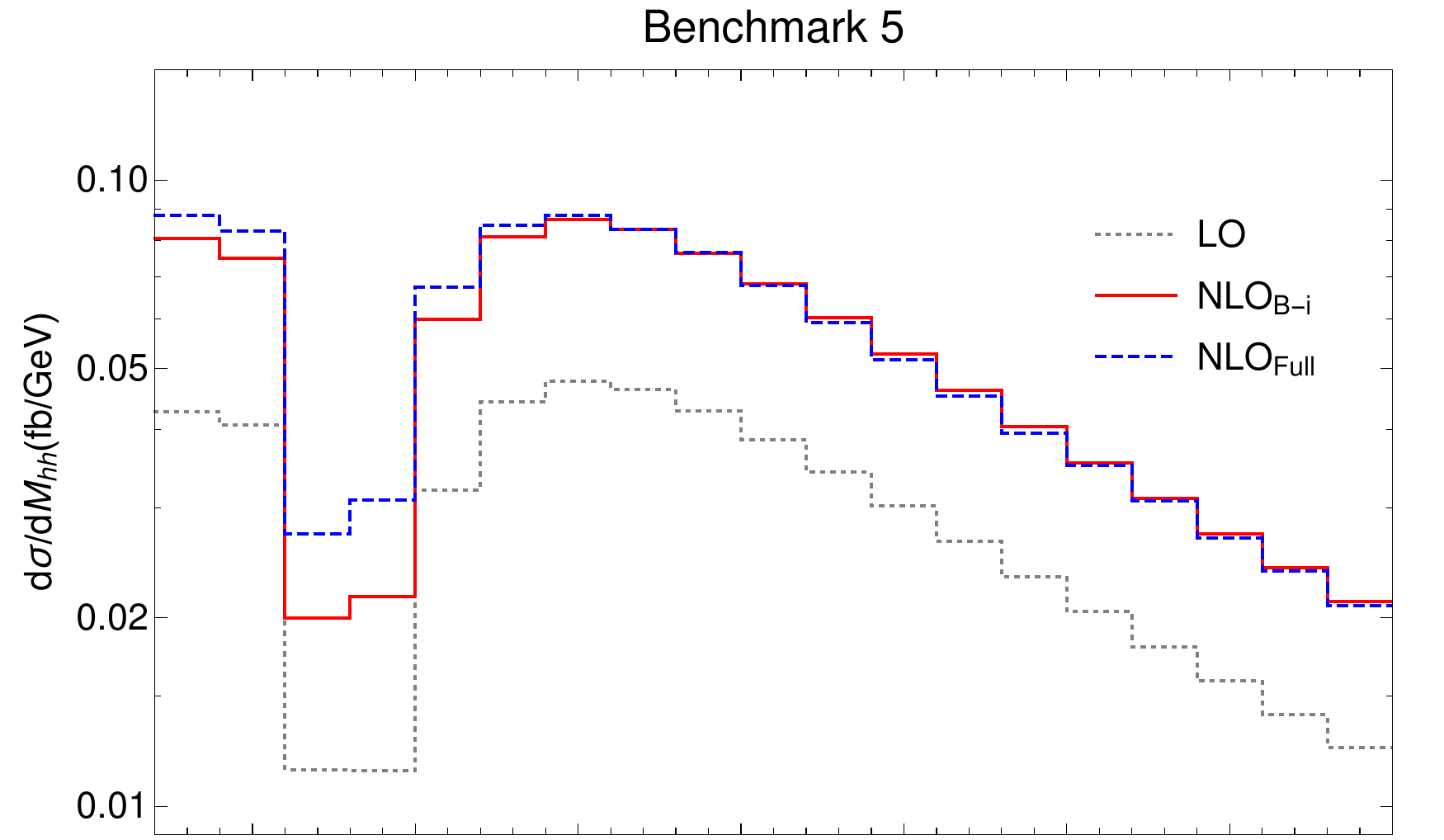}
\includegraphics[width=.32\textwidth]{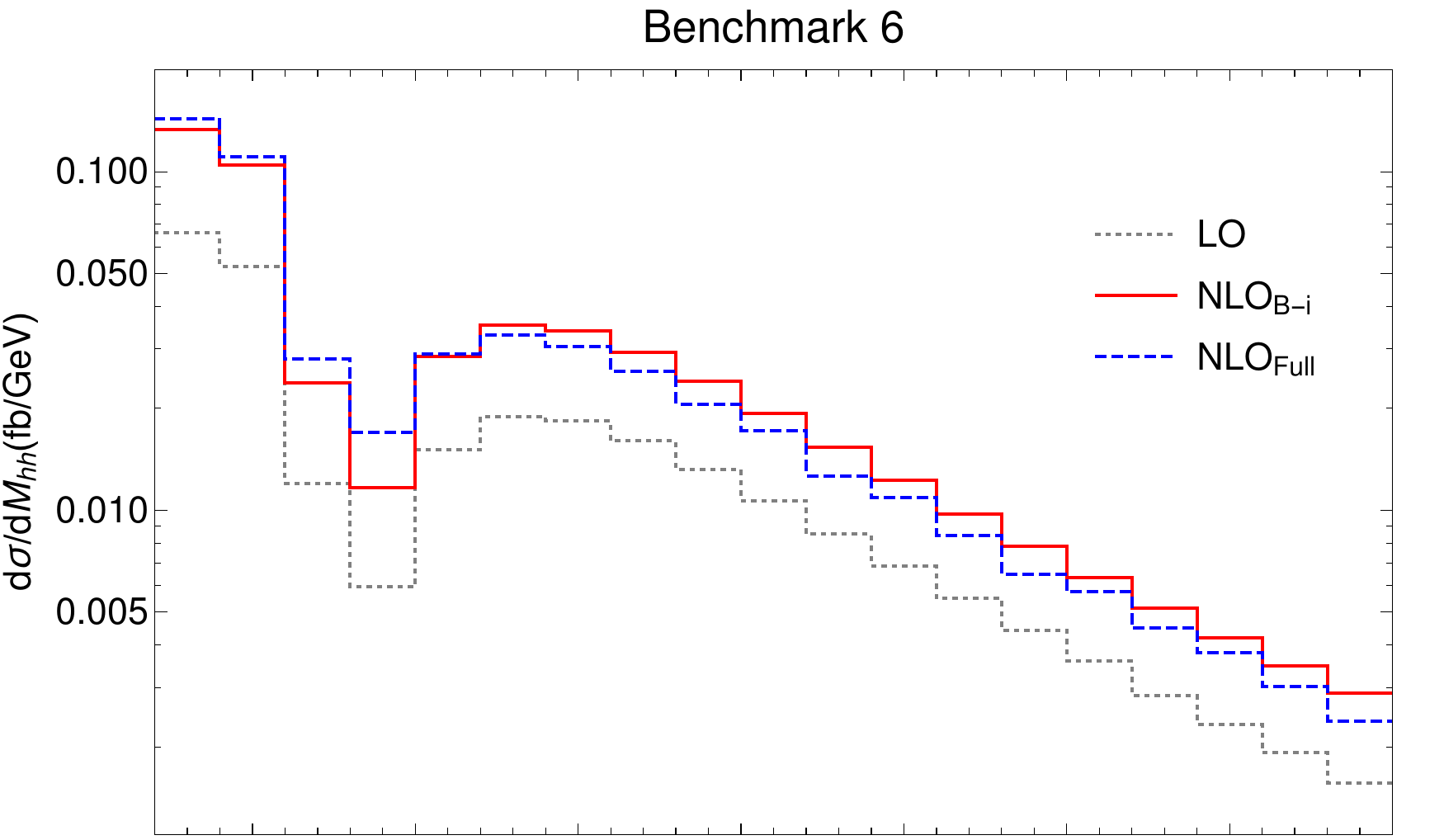}
\\
\includegraphics[width=.32\textwidth]{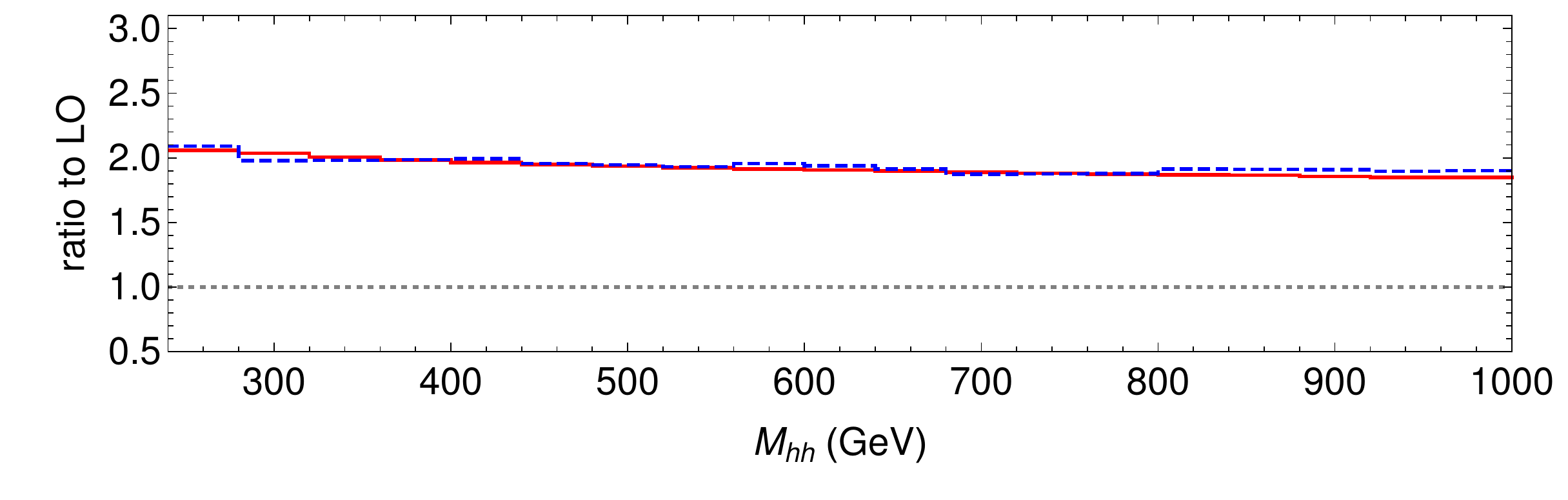}
\includegraphics[width=.32\textwidth]{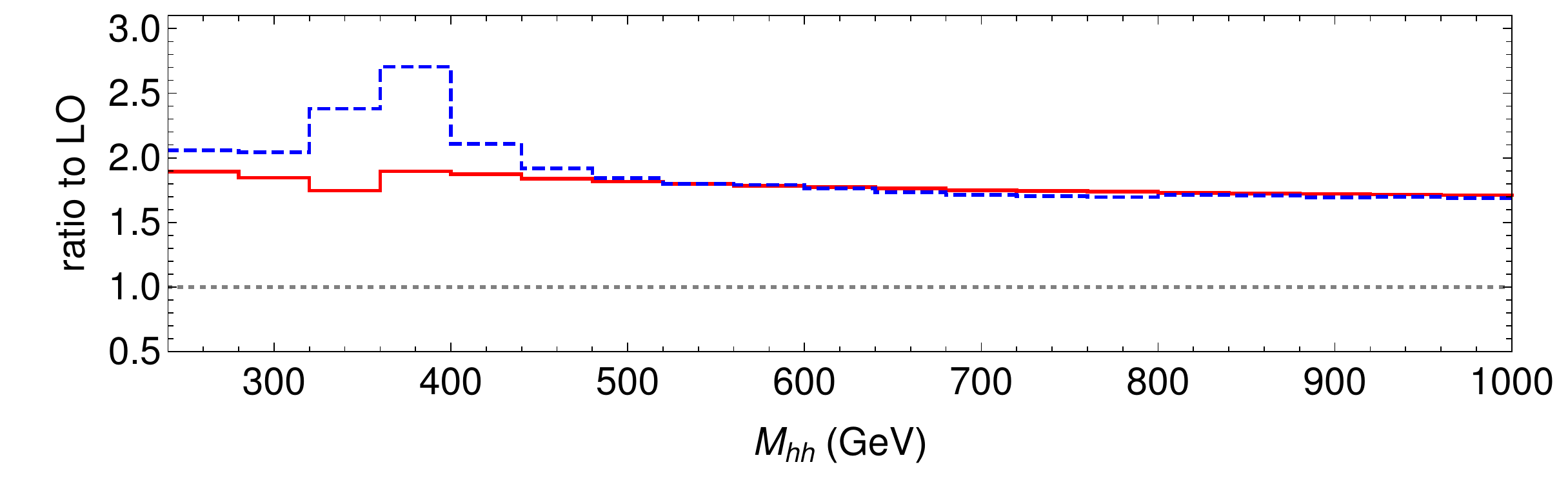}
\includegraphics[width=.32\textwidth]{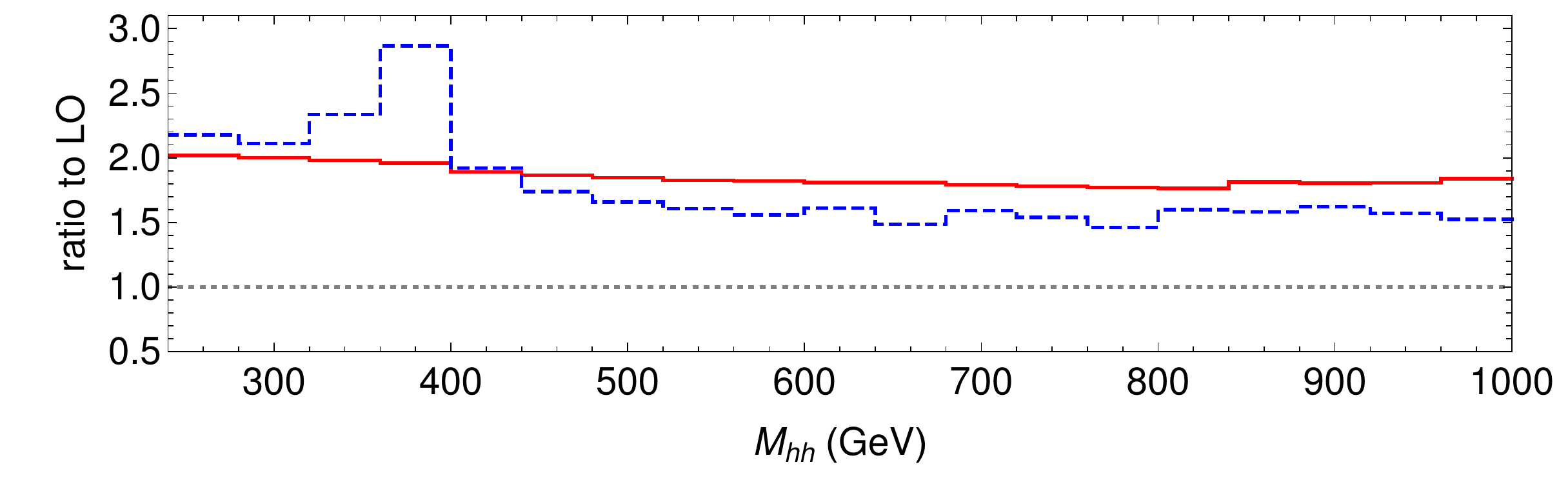}
\\
\vspace*{0.4cm}
\includegraphics[width=.32\textwidth]{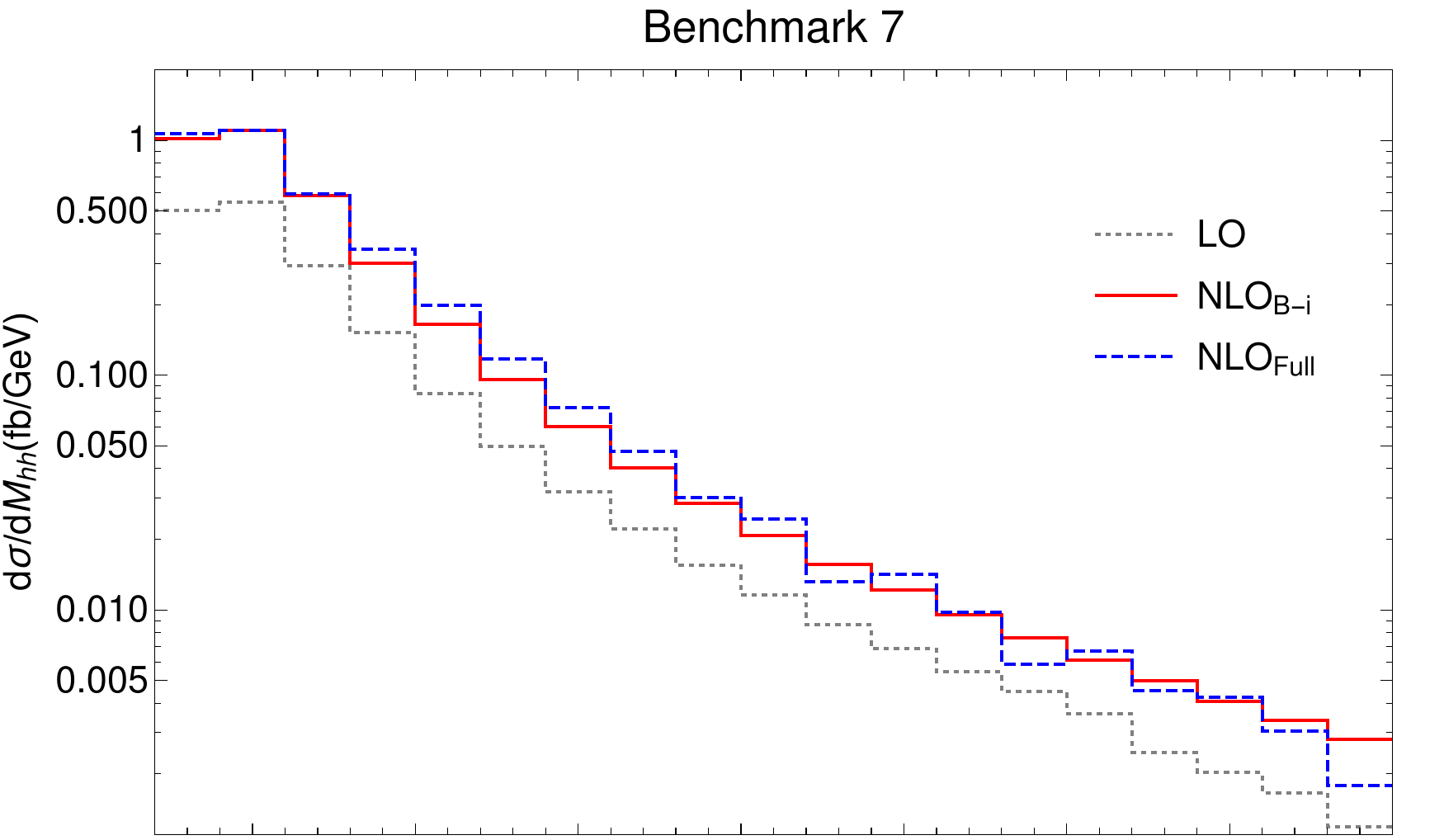}
\includegraphics[width=.32\textwidth]{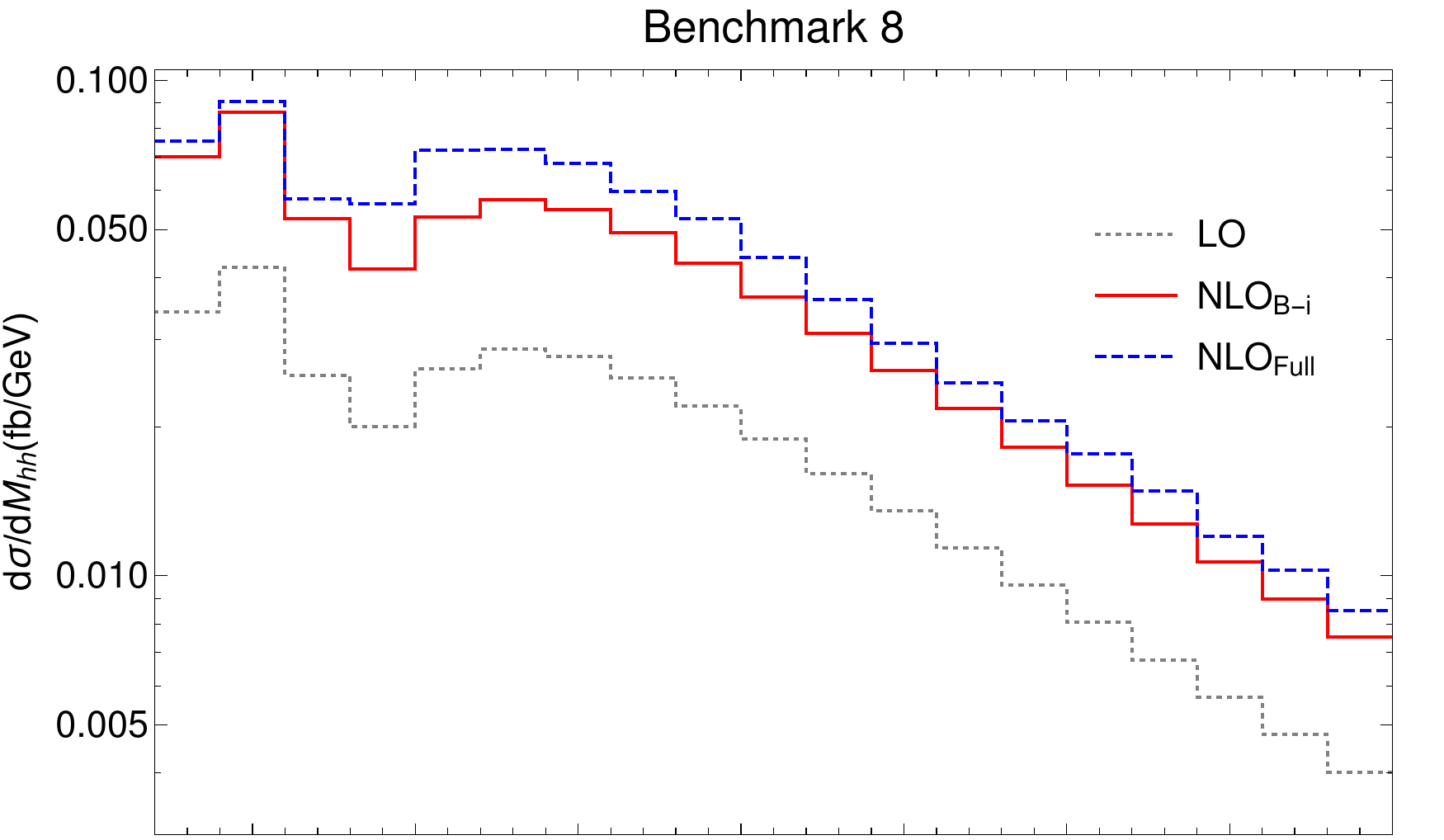}
\includegraphics[width=.32\textwidth]{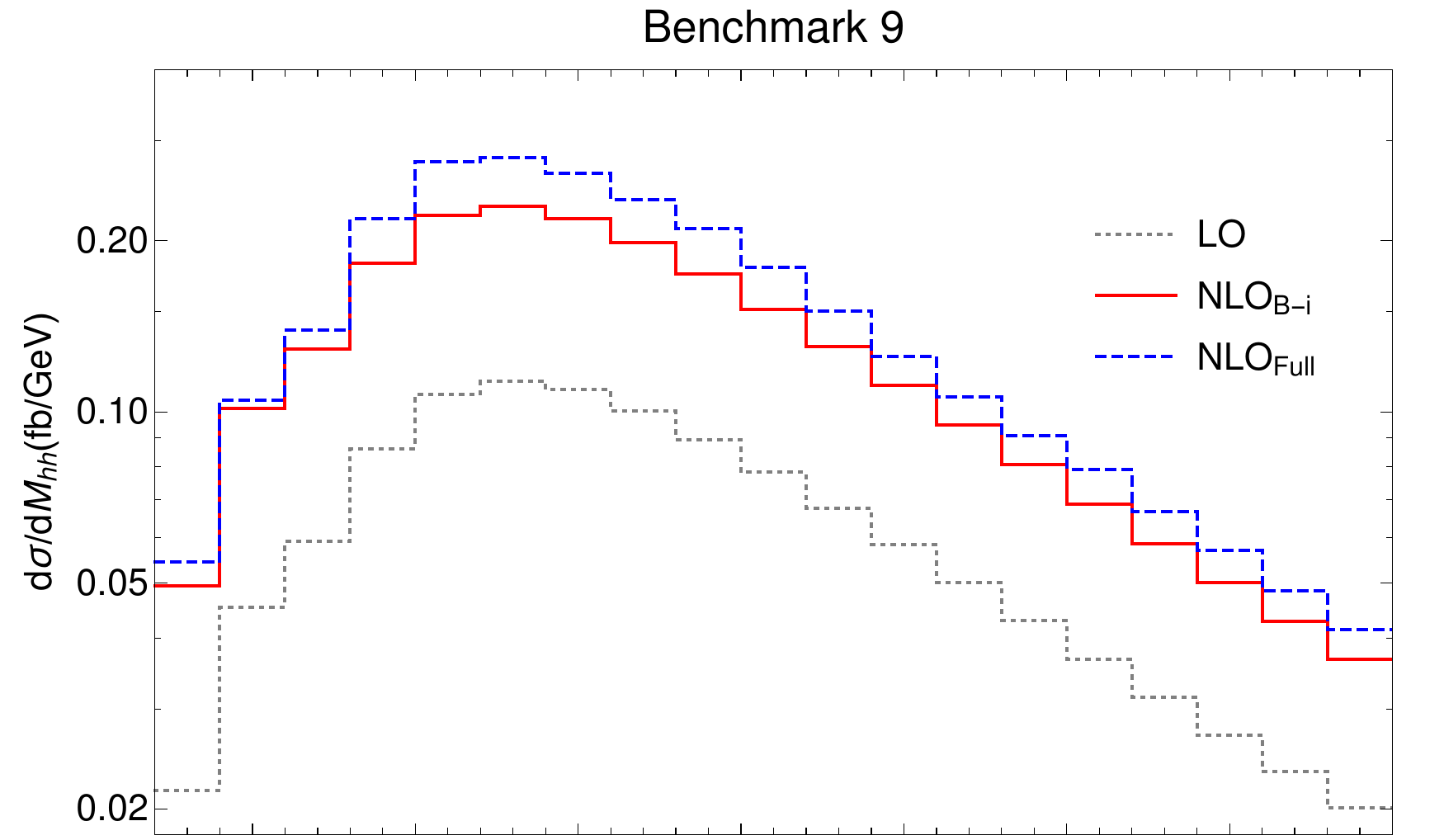}
\\
\includegraphics[width=.32\textwidth]{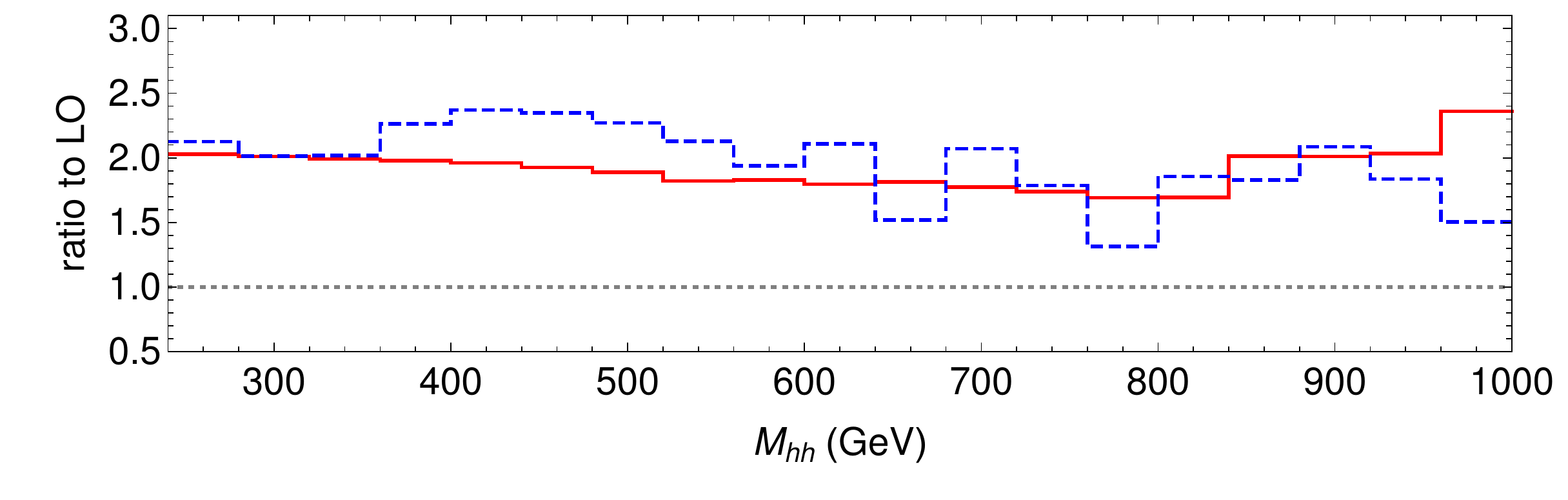}
\includegraphics[width=.32\textwidth]{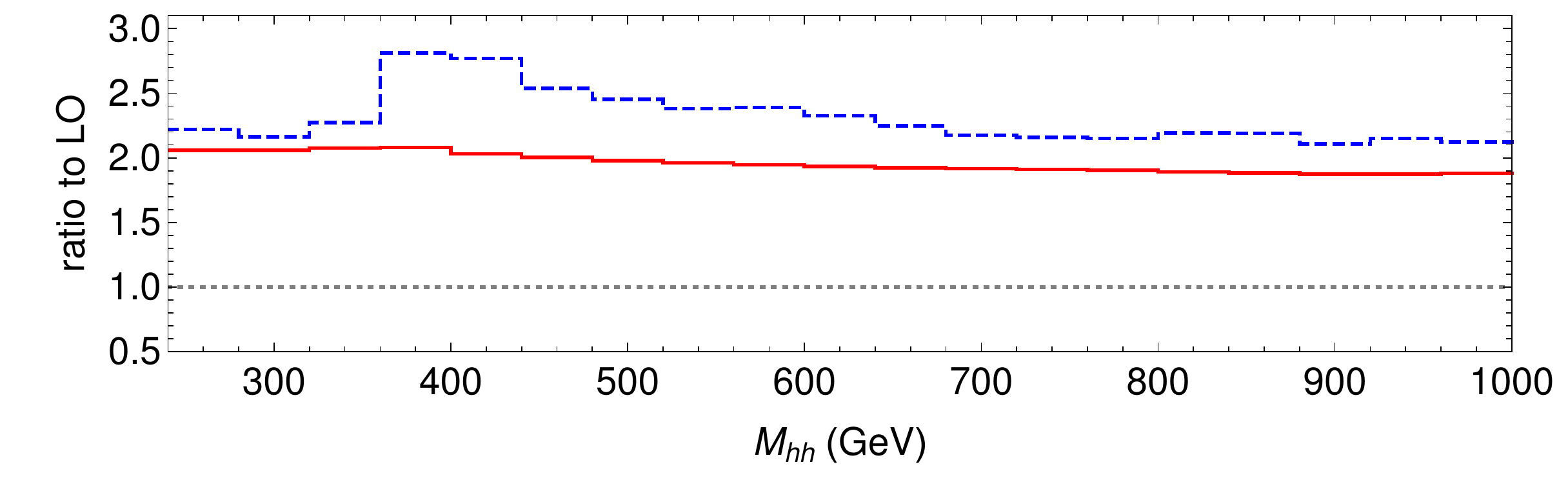}
\includegraphics[width=.32\textwidth]{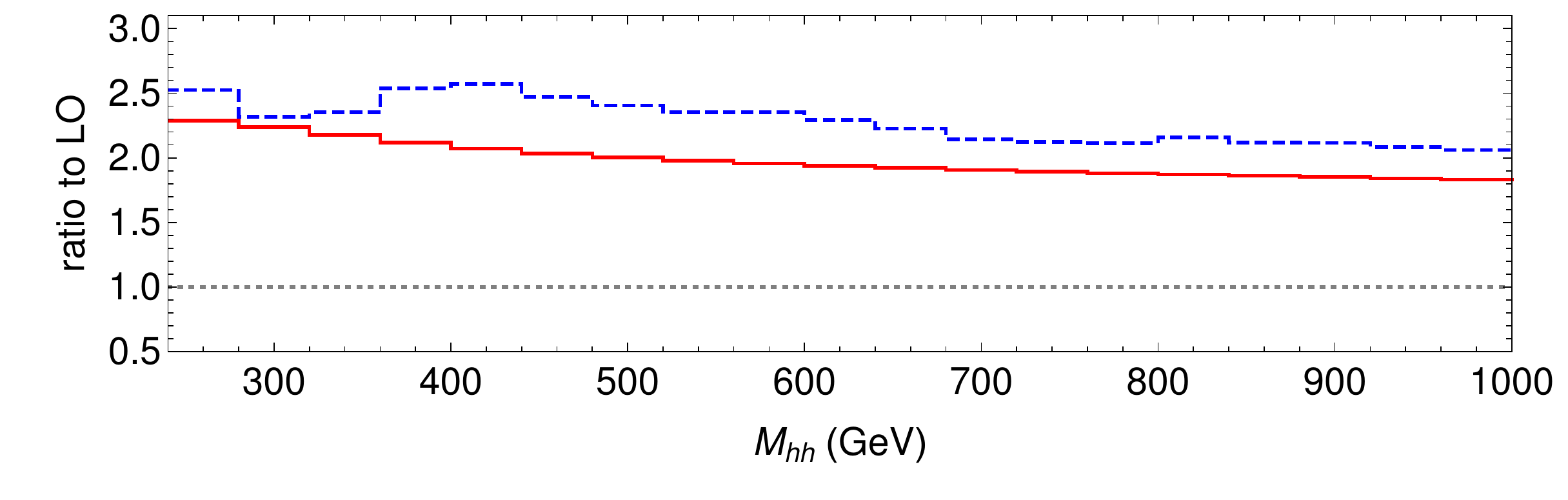}
\\
\vspace*{0.4cm}
\includegraphics[width=.32\textwidth]{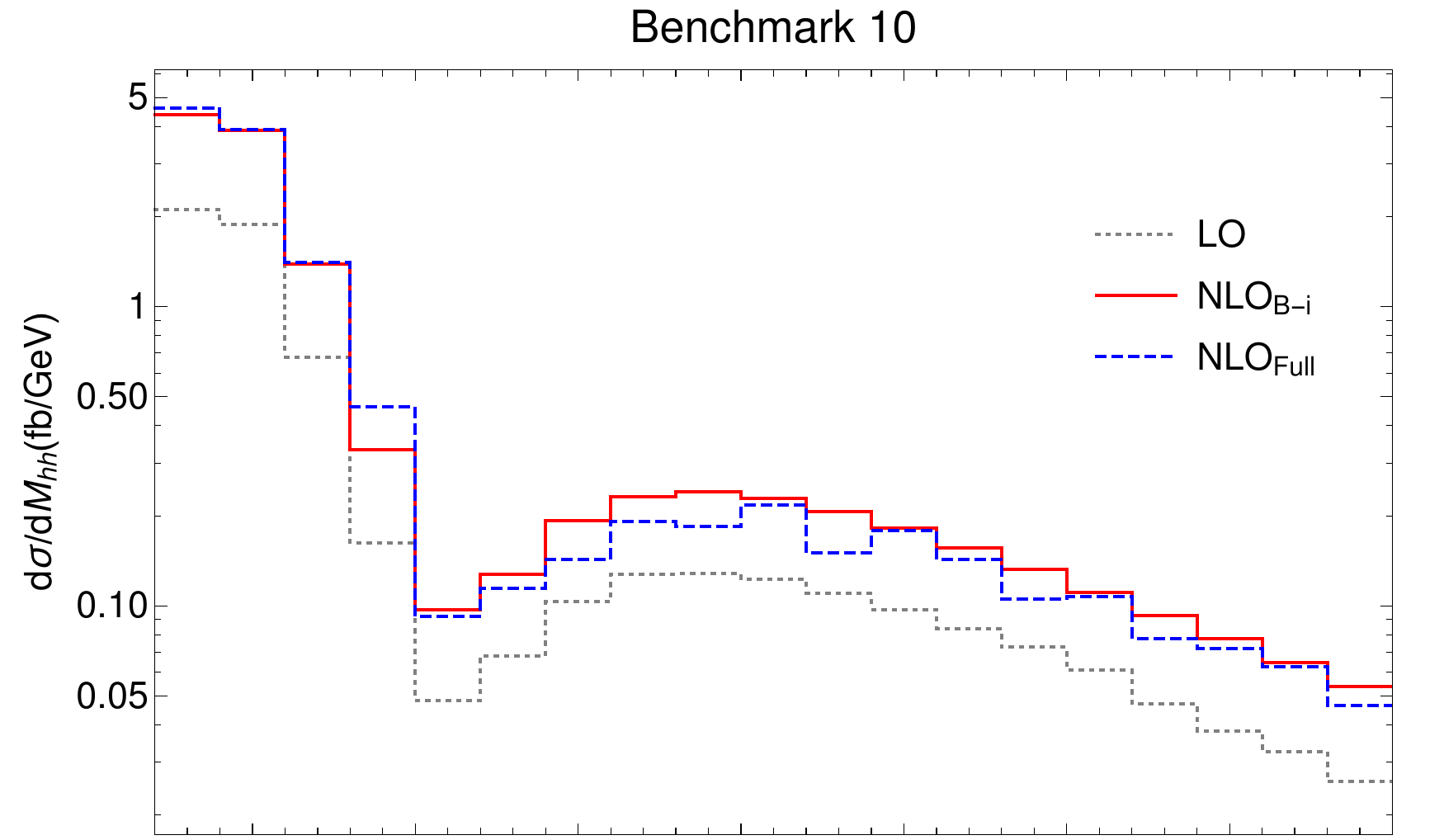}
\includegraphics[width=.32\textwidth]{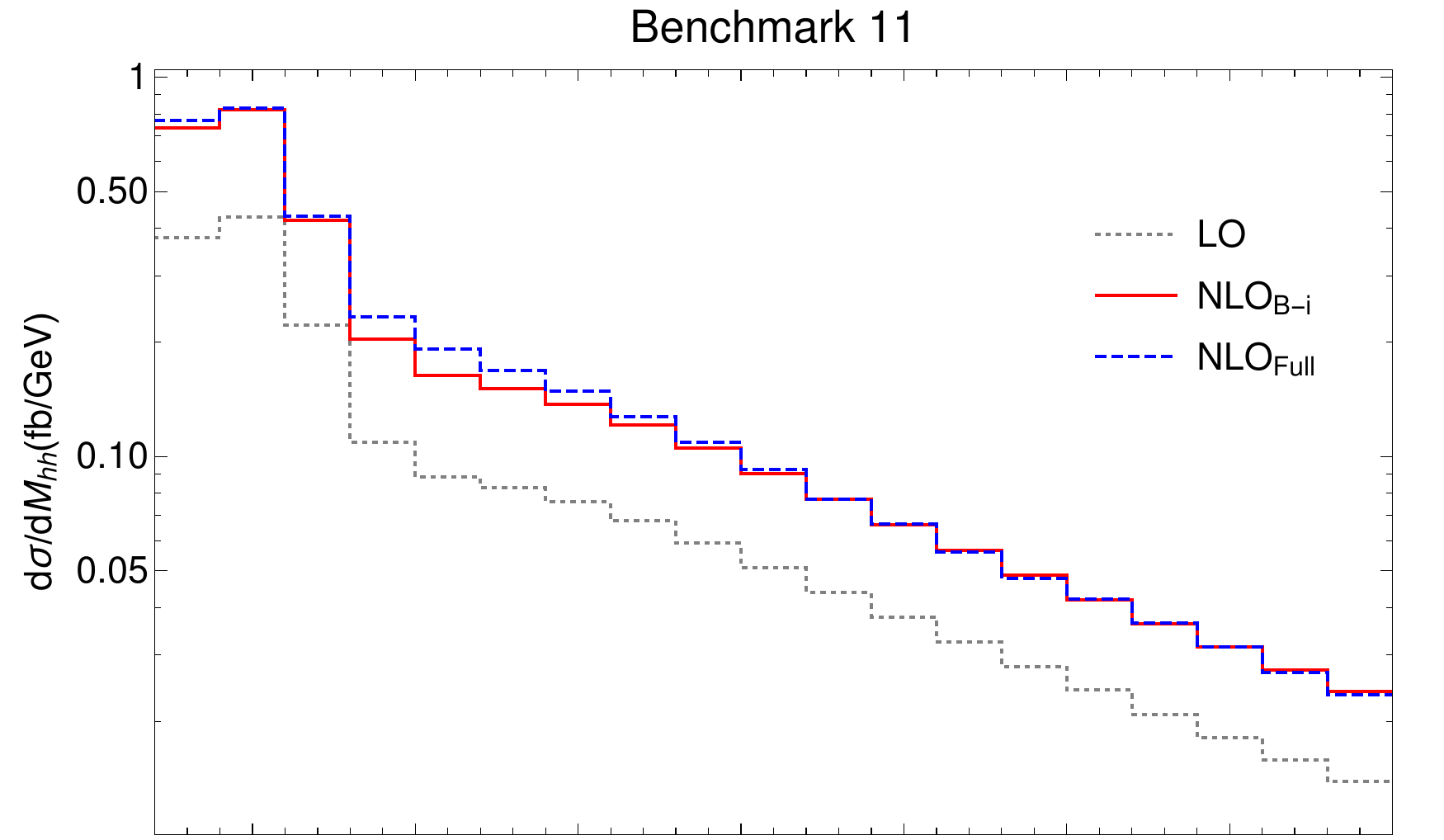}
\includegraphics[width=.32\textwidth]{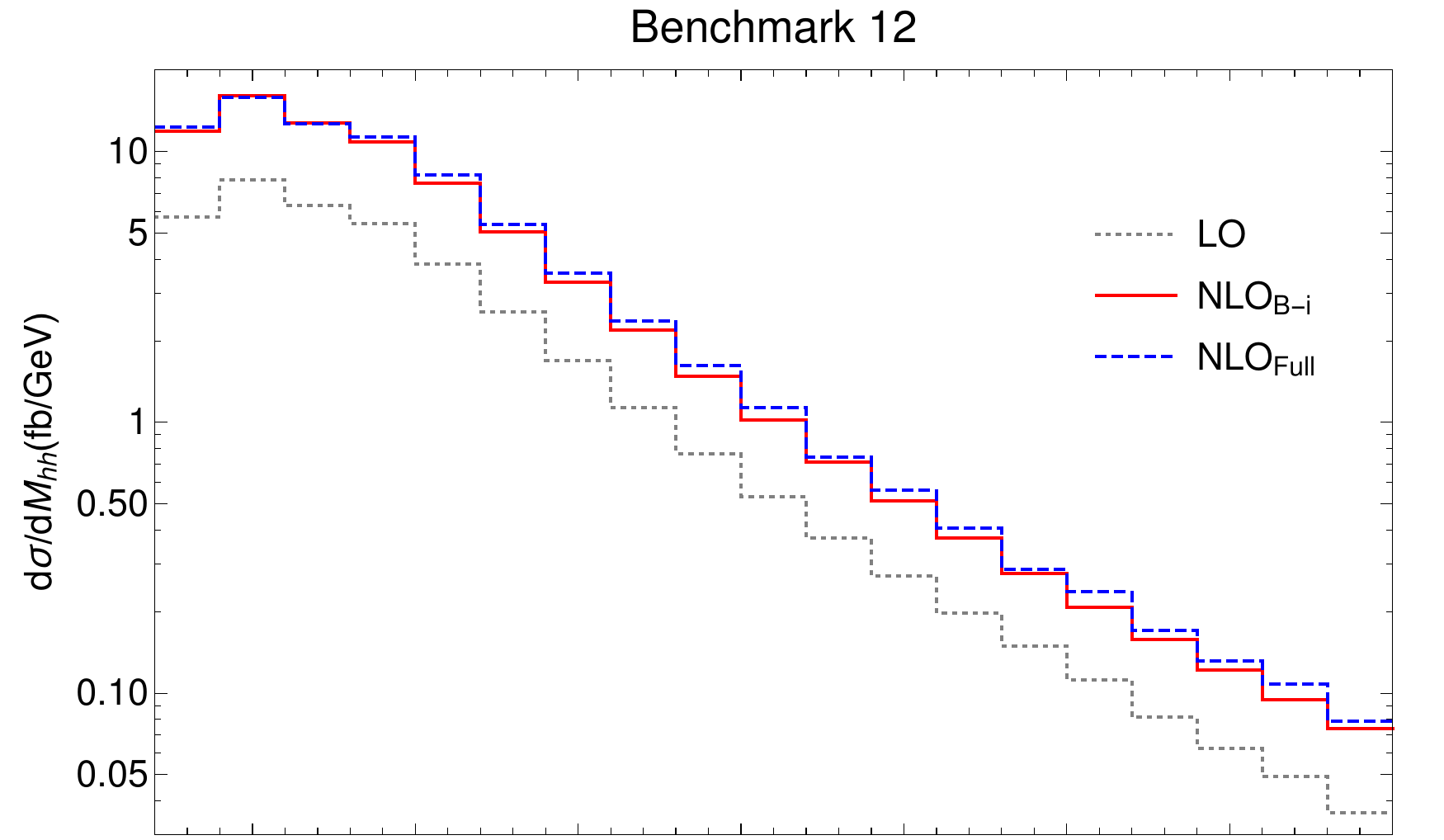}
\\
\includegraphics[width=.32\textwidth]{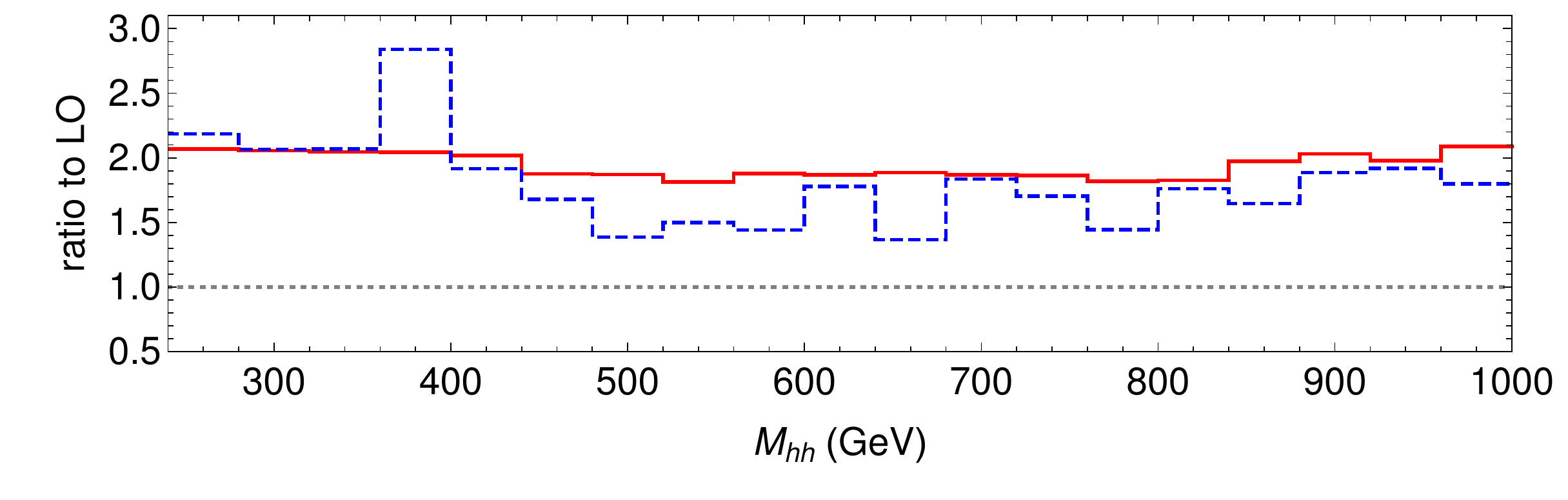}
\includegraphics[width=.32\textwidth]{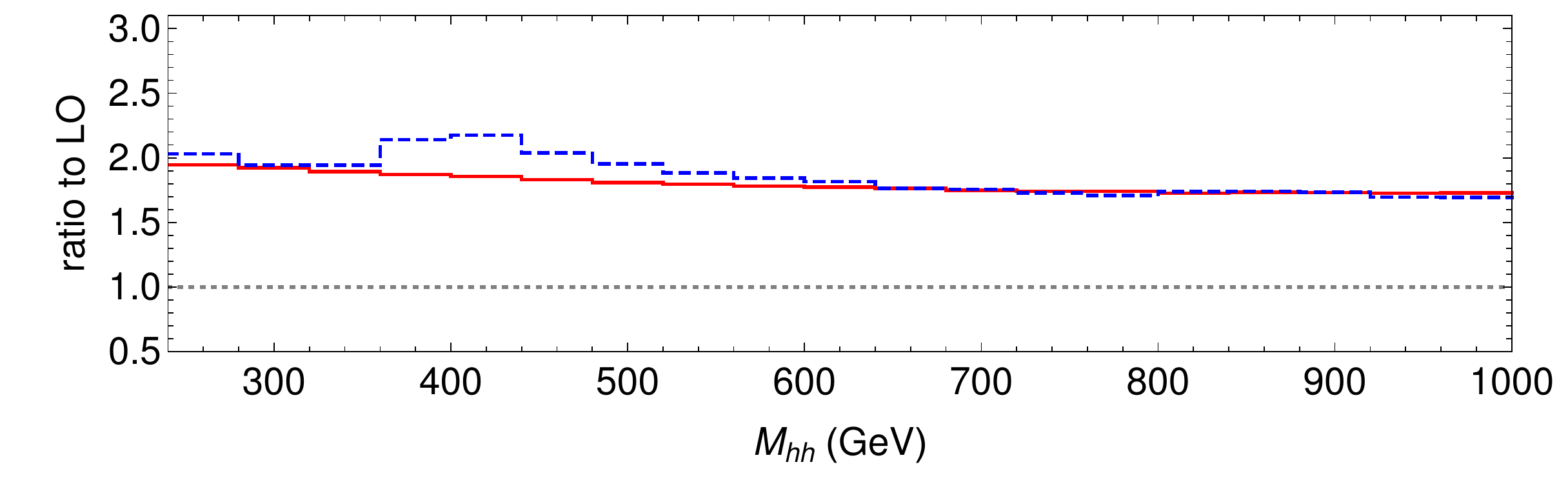}
\includegraphics[width=.32\textwidth]{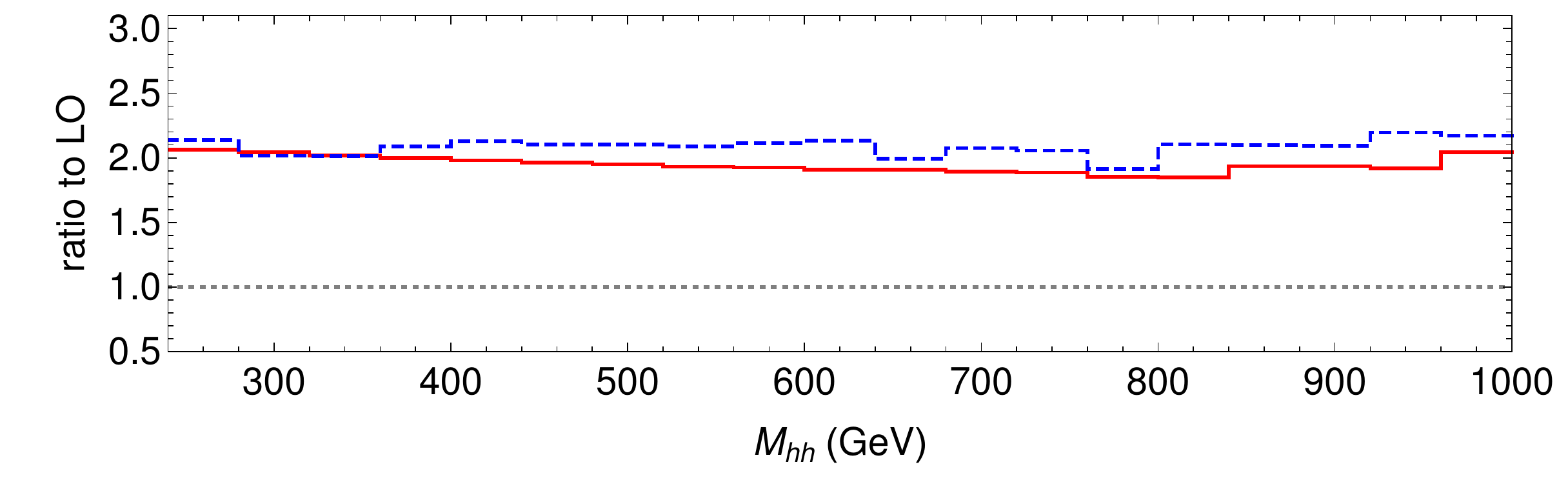}
\end{center}
\caption{\small Higgs-pair invariant mass distribution at 13~TeV for the different shape benchmarks, at NLO with full top mass dependence (blue-dashed), NLO Born-improved HTL (red-solid) and LO (gray-dotted). The lower panel shows the ratio to the LO prediciton.\label{fig:Higgs_HH_NNLO_EFT:benchmarks_nlo}}
\end{figure}

In order to show the impact of the QCD corrections, we present predictions for the benchmark points introduced in Ref.~\cite{Carvalho:2015ttv} (though we use the redefinition of the benchmark point number 8 presented in Ref.~\cite{Buchalla:2018yce}), see Table~\ref{tab:Higgs_HH_NNLO_EFT:benchmarks}.
The Higgs pair invariant mass distribution for these 12 benchmarks is shown in Fig.~\ref{fig:Higgs_HH_NNLO_EFT:benchmarks_nnlo}.
To assess the accuracy of the Born-improved HTL approximation defined in Ref.~\cite{deFlorian:2017qfk} and used here, we also present in Fig.~\ref{fig:Higgs_HH_NNLO_EFT:benchmarks_nlo} the corresponding results at NLO.

\begin{table}[t]
\begin{center}
{
\renewcommand{\arraystretch}{1.5}
\begin{tabular}{| c | c  c  c  c  c |}
\hline
Benchmark & $c_{hhh}$ & $c_t$ & $c_{tt}$ & $c_{ggh}$ & $c_{gghh}$ \\
\hline\hline
1 & 7.5 & 1.0 & $-1.0$ & 0.0 & 0.0 \\
\hline
2 & 1.0 & 1.0 & 0.5 & $-\frac{1.6}{3}$ & $-0.2$ \\
\hline
3 & 1.0 & 1.0 & $-1.5$ & 0.0 & $\frac{0.8}{3}$  \\
\hline
4 & $-3.5$ & 1.5 & $-3.0$ & 0.0 & 0.0 \\ 
\hline
5 & 1.0 & 1.0 & 0.0 &   $\frac{1.6}{3}$ & $\frac{1.0}{3}$\\
\hline
6 & 2.4 & 1.0 & 0.0 & $\frac{0.4}{3}$ & $\frac{0.2}{3}$  \\
\hline
7 & 5.0 & 1.0 & 0.0 & $\frac{0.4}{3}$ & $\frac{0.2}{3}$  \\
\hline
8a & 1.0 & 1.0 & 0.5 & $\frac{0.8}{3}$ & 0.0\\
\hline
9 & 1.0 & 1.0 & 1.0 &  $-0.4$ &  $-0.2$ \\
\hline
10 & 10.0 & 1.5 & $-1.0$ & 0.0 & 0.0 \\
\hline
11 & 2.4 & 1.0 & 0.0 & $\frac{2.0}{3}$ & $\frac{1.0}{3}$ \\
\hline
12 & 15.0 & 1.0 & 1.0 & 0.0 & 0.0 \\
\hline
SM & 1.0 & 1.0 & 0.0 & 0.0 & 0.0 \\
\hline
\end{tabular}
}
\end{center}
\caption{\small Benchmark points used for the distributions shown below.\label{tab:Higgs_HH_NNLO_EFT:benchmarks}}
\end{table}

From the distributions in Fig.~\ref{fig:Higgs_HH_NNLO_EFT:benchmarks_nnlo} we can observe that the NNLO corrections are sizeable and have a non-trivial dependence on the kinematics, and they can even become negative in some invariant mass and parameter space regions. For comparison, the inclusive SM $K$-factor, defined as $\sigma_\text{NNLO}^\text{FTapprox}/\sigma_\text{NLO}^\text{Full}$, is shown in the lower panels. Even if this constant $K$-factor does not reproduce all the features of our best prediction, it is worth noting that in most of the cases represents an improvement w.r.t. the NLO result.

\begin{figure}[p]
\begin{center}
\includegraphics[width=.35\textwidth]{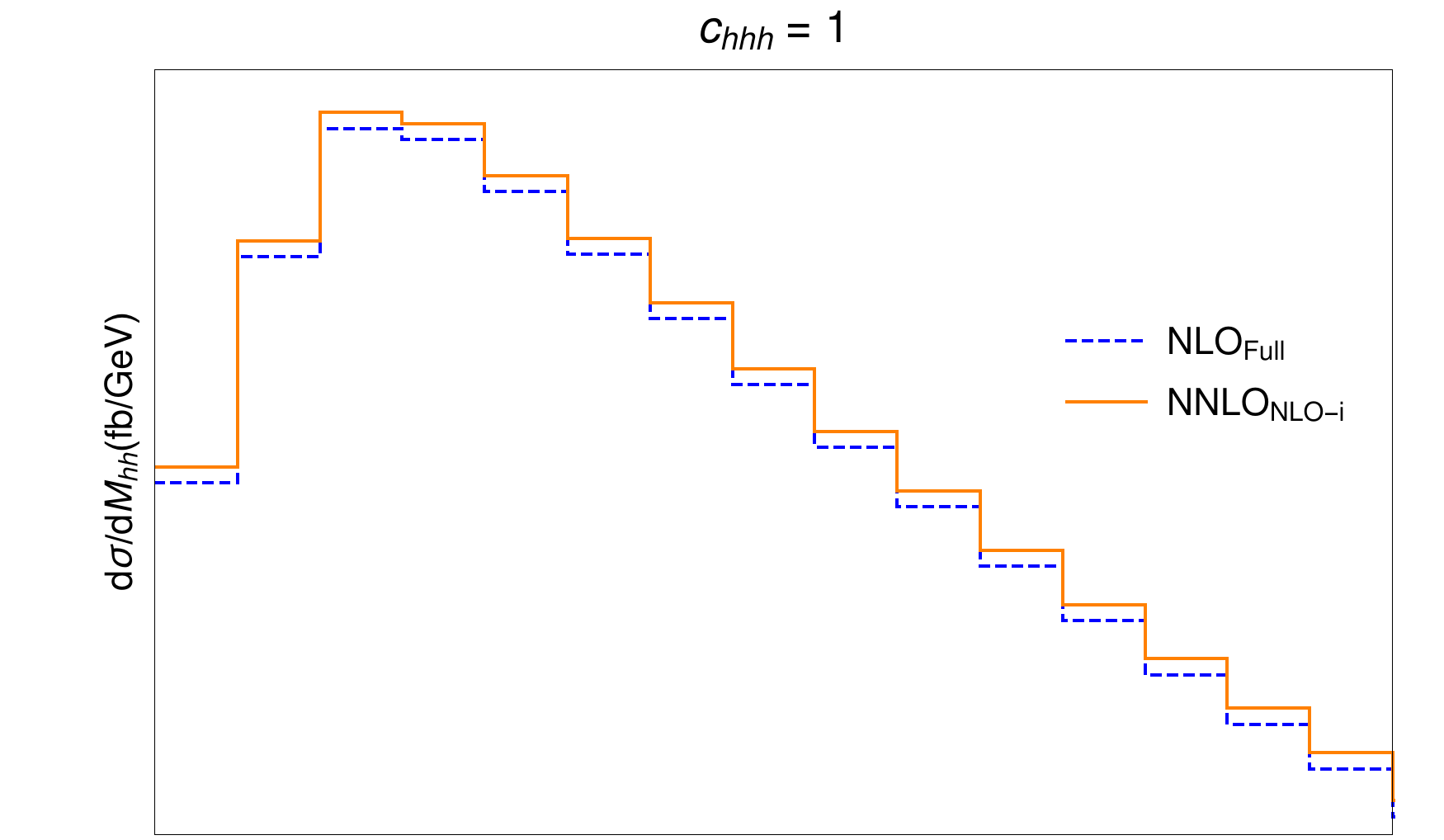}
\\
\includegraphics[width=.35\textwidth]{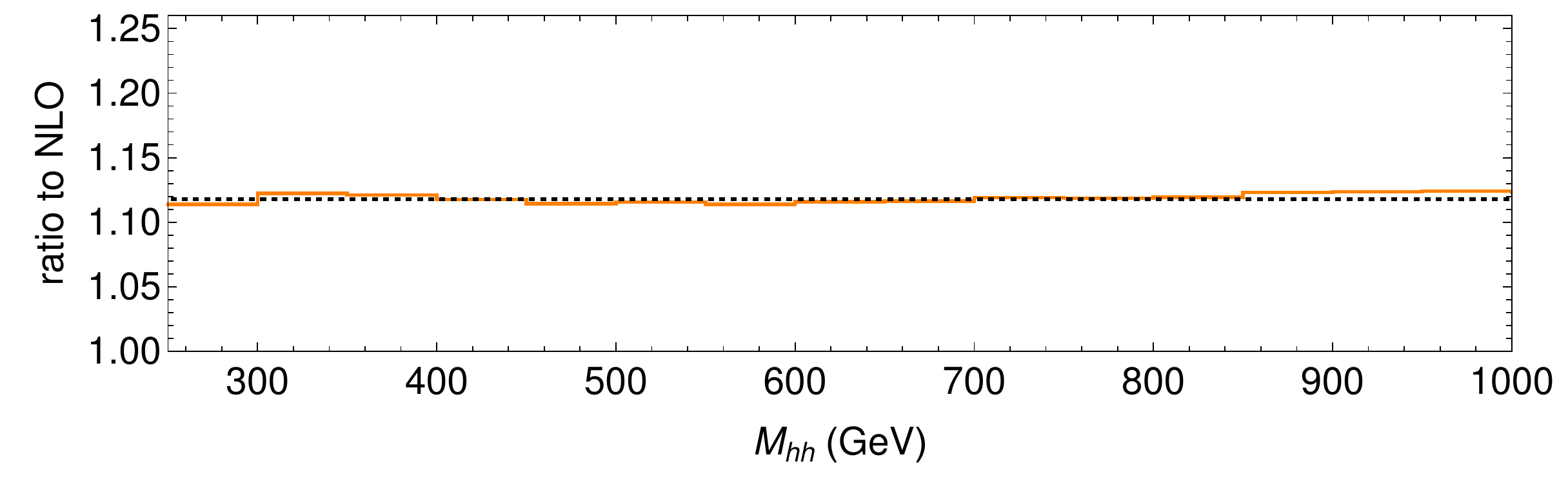}
\\
\includegraphics[width=.35\textwidth]{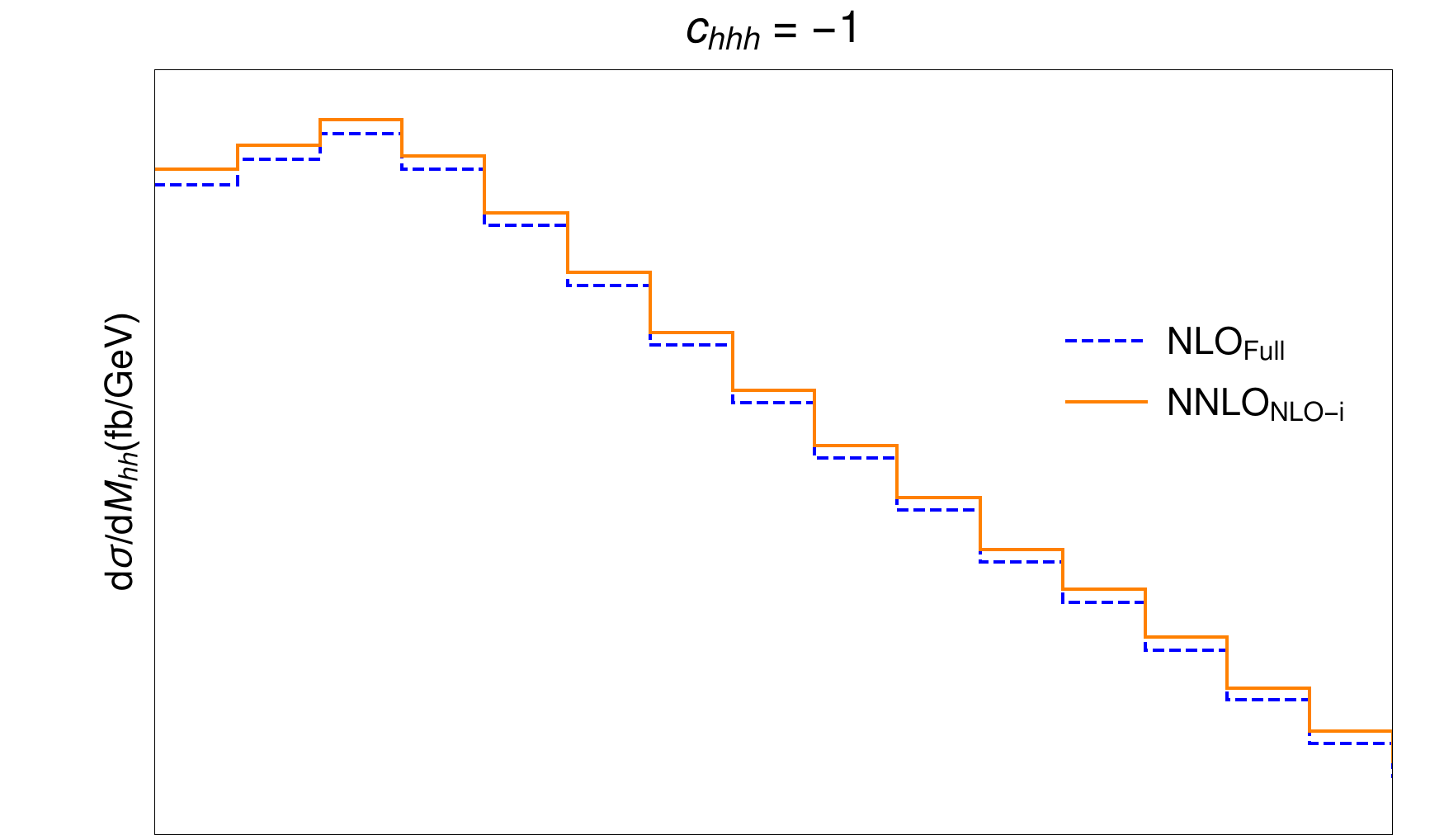}
\includegraphics[width=.35\textwidth]{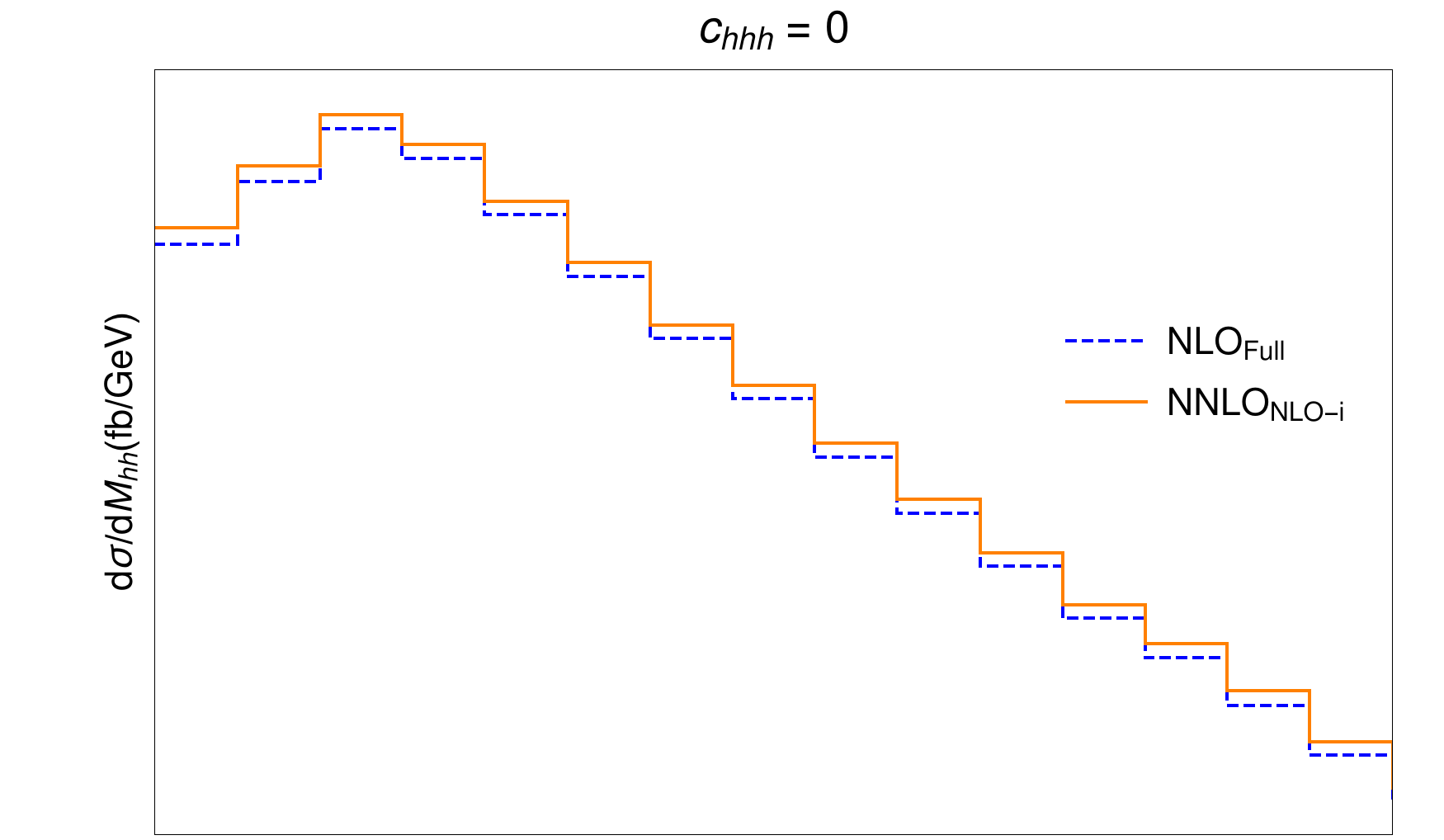}
\\
\includegraphics[width=.35\textwidth]{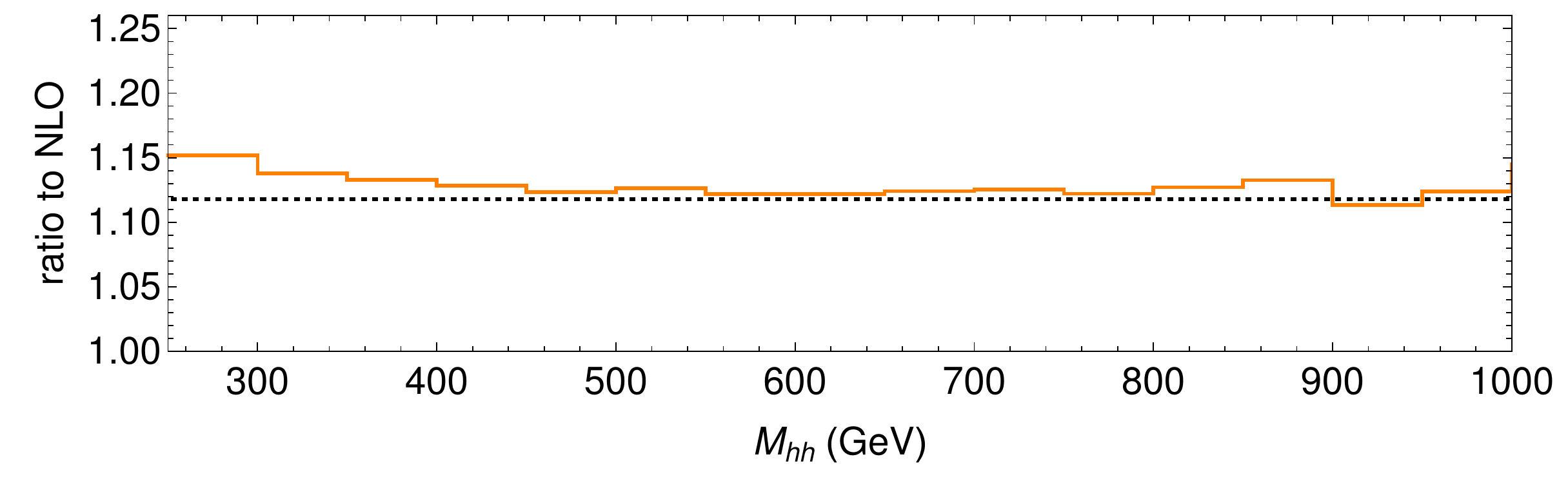}
\includegraphics[width=.35\textwidth]{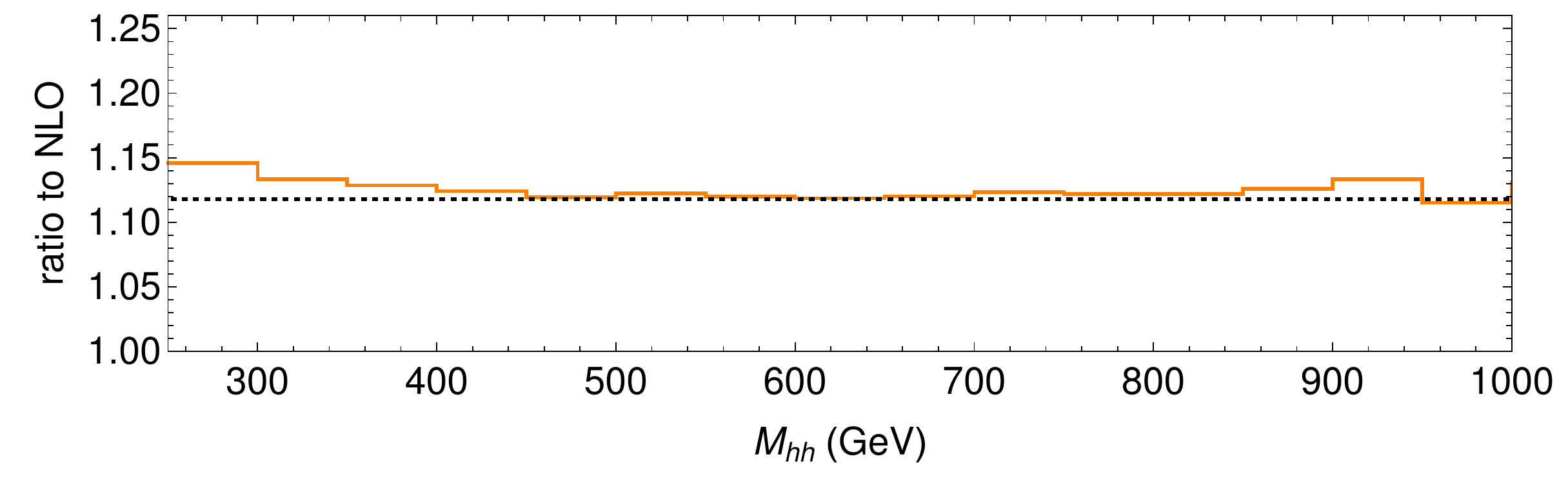}
\\
\includegraphics[width=.35\textwidth]{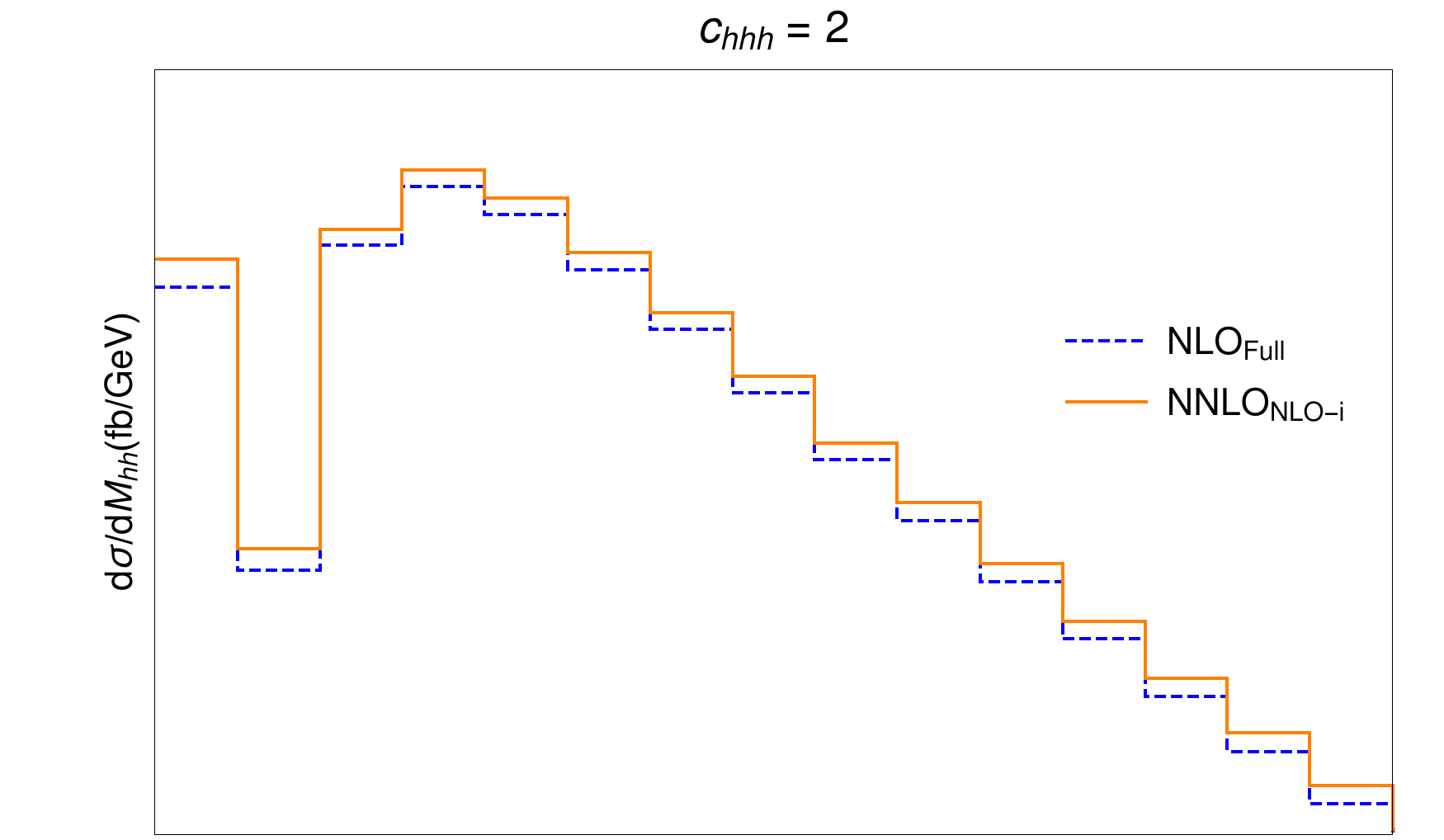}
\includegraphics[width=.35\textwidth]{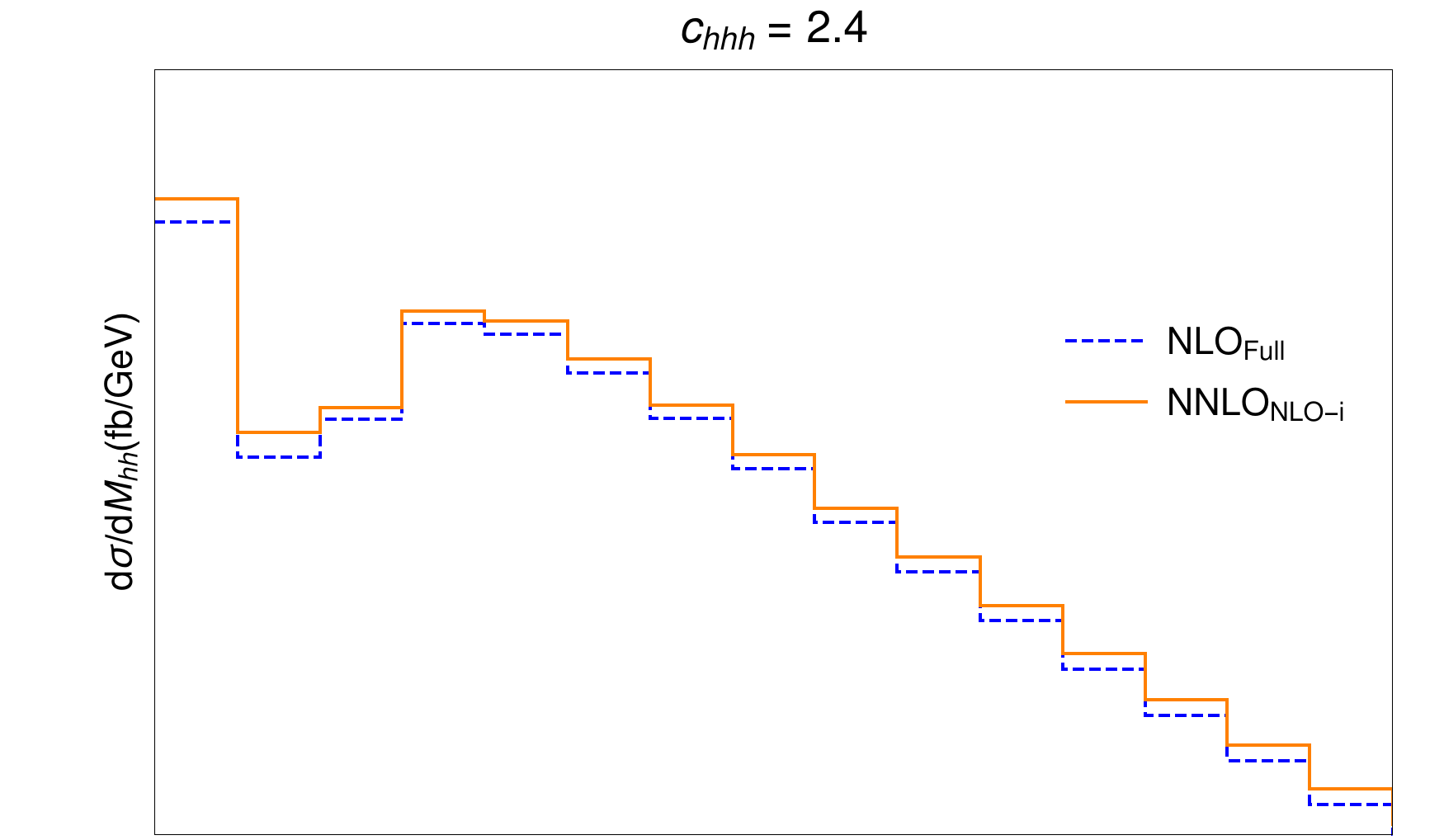}
\\
\includegraphics[width=.35\textwidth]{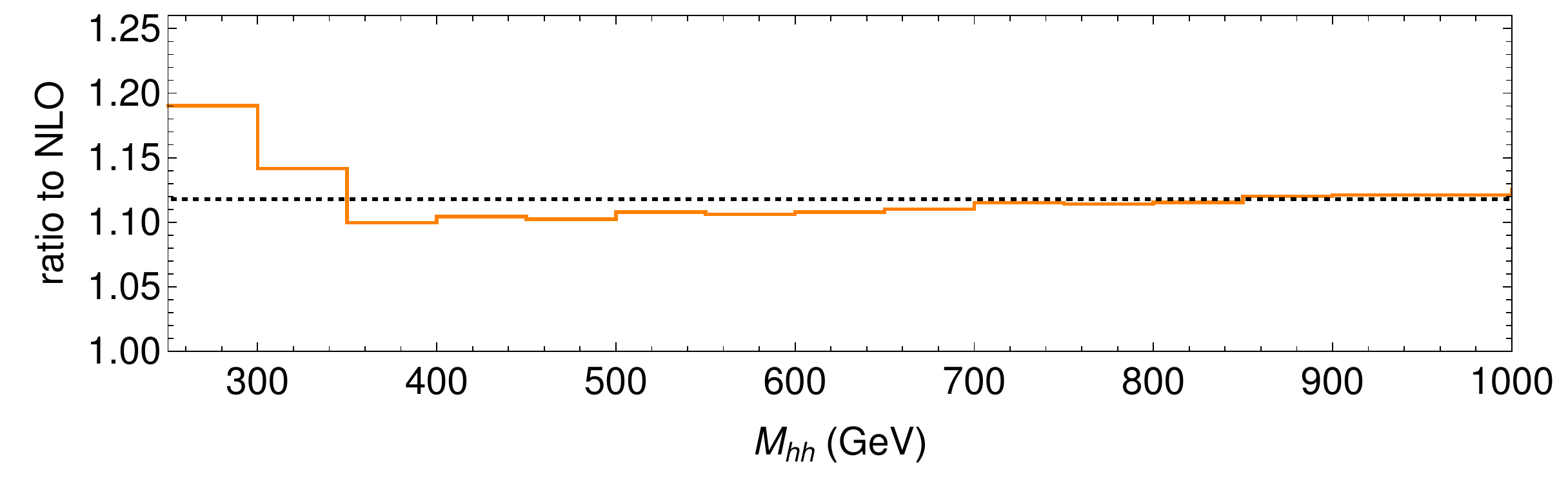}
\includegraphics[width=.35\textwidth]{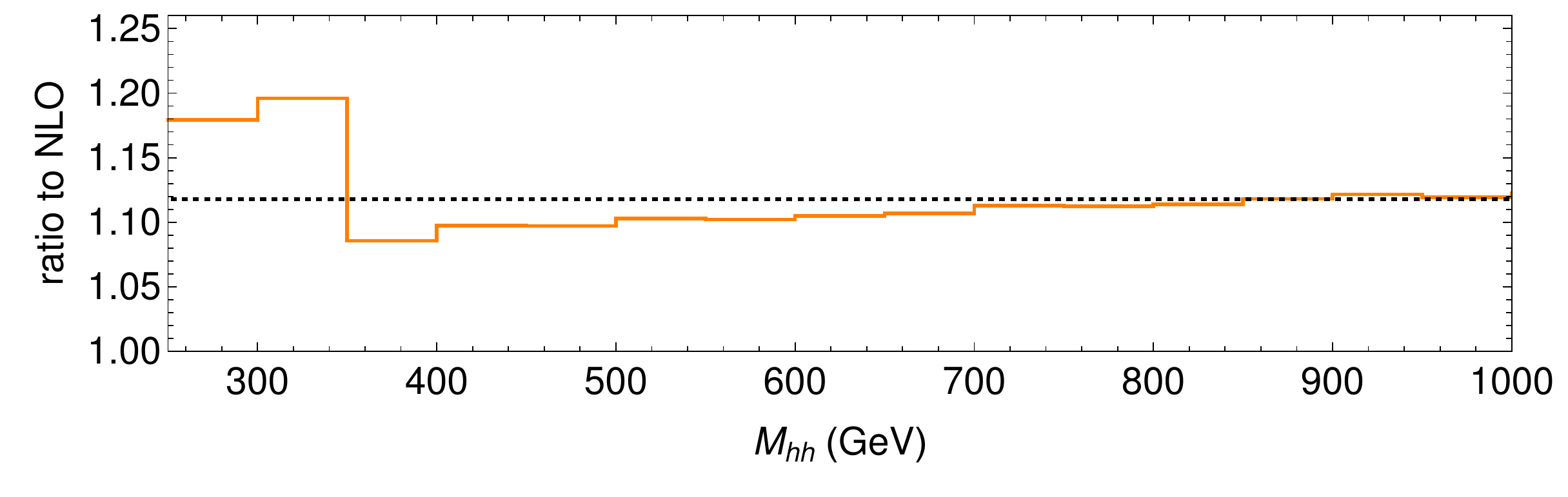}
\\
\includegraphics[width=.35\textwidth]{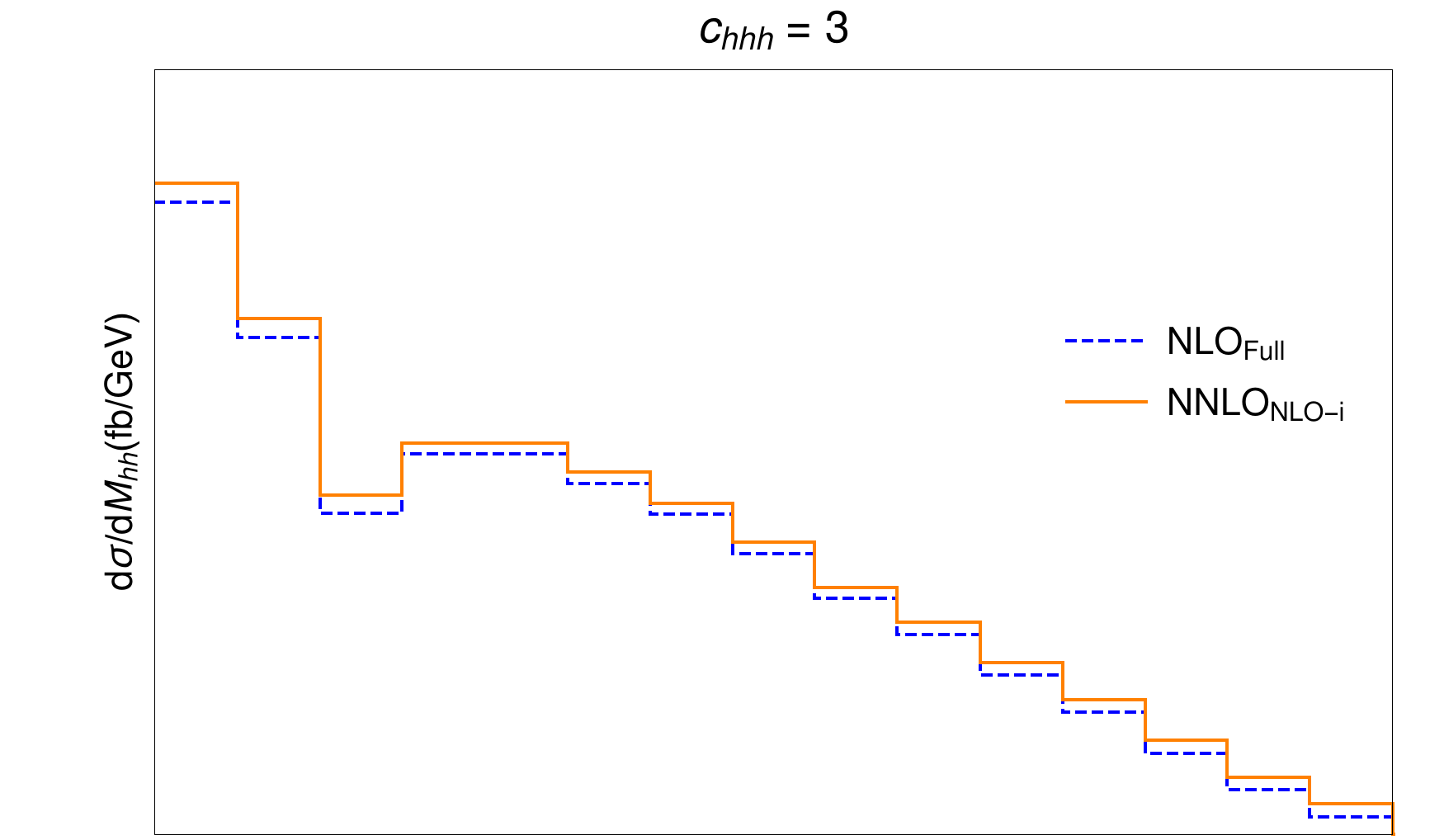}
\includegraphics[width=.35\textwidth]{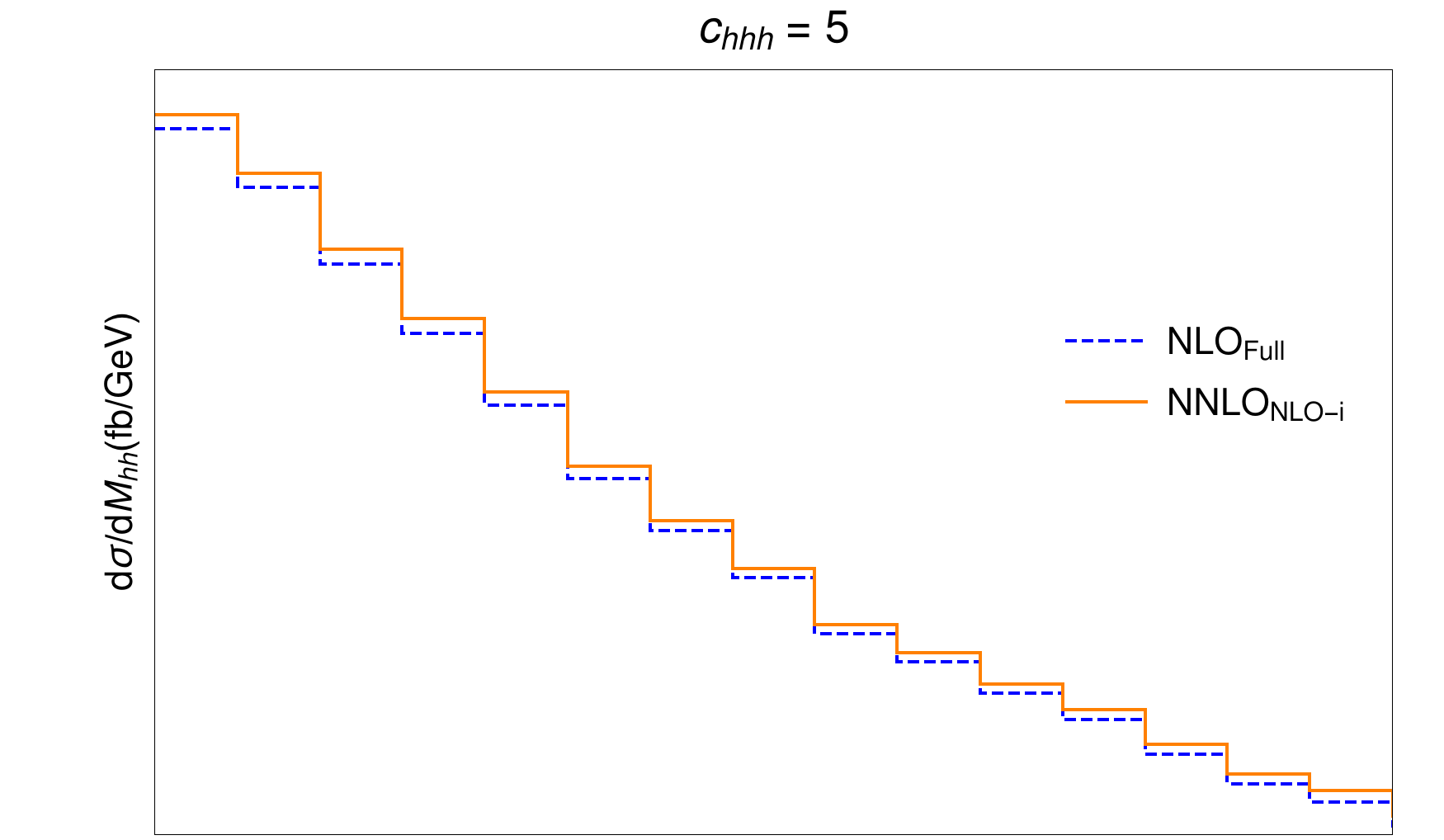}
\\
\includegraphics[width=.35\textwidth]{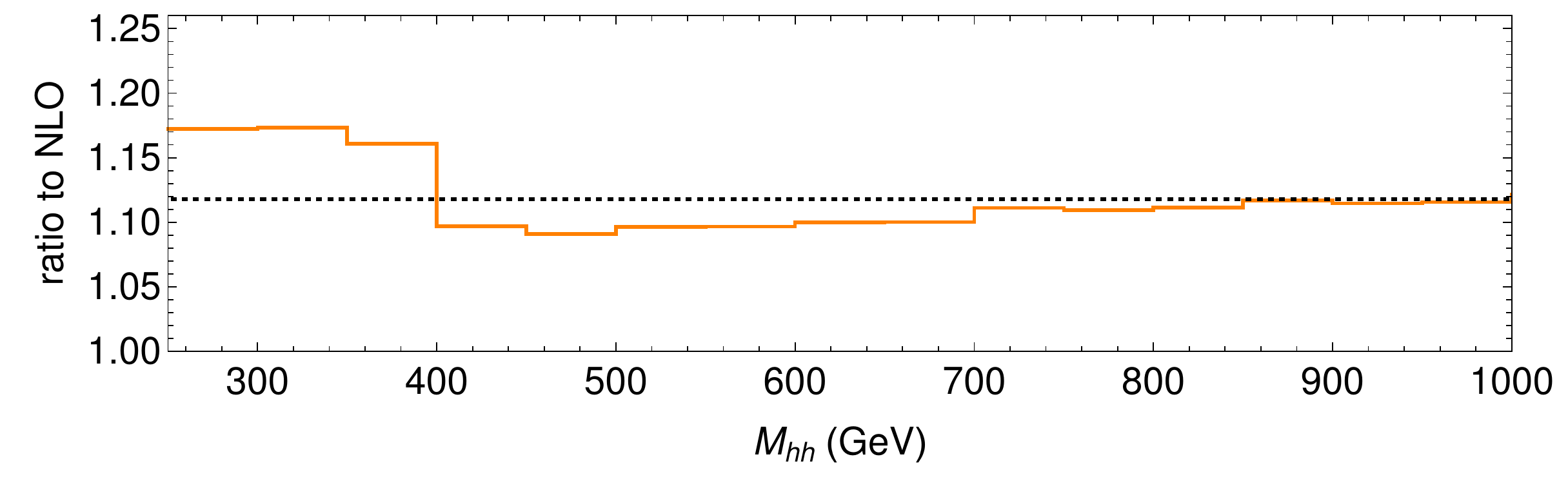}
\includegraphics[width=.35\textwidth]{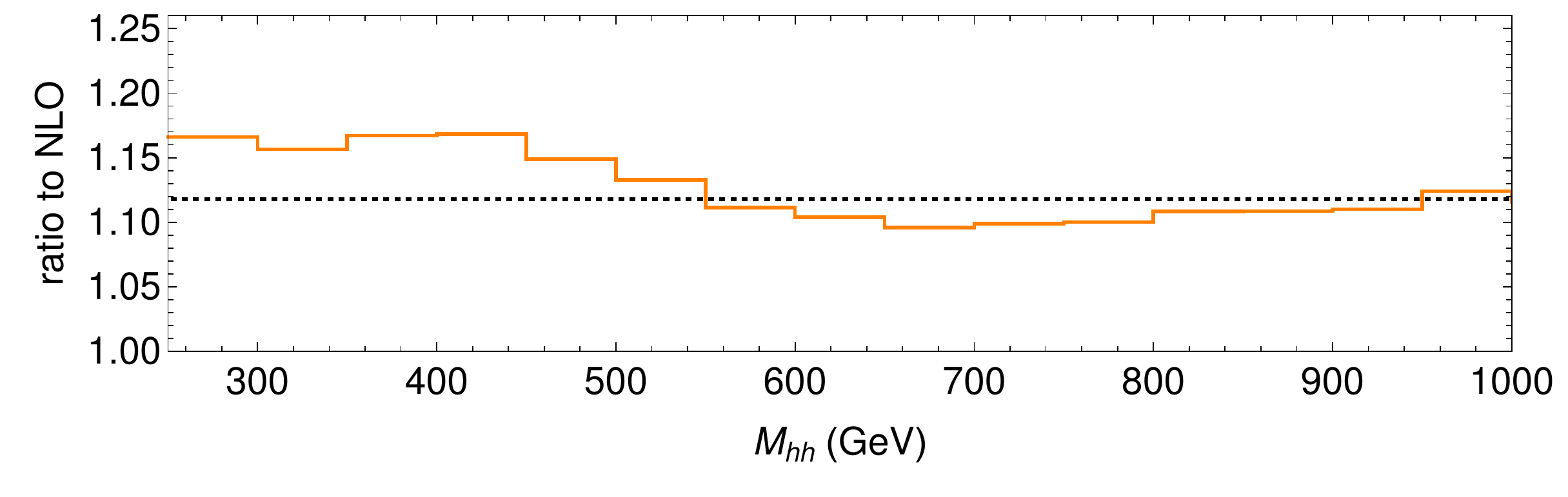}
\end{center}
\caption{\small Higgs-pair invariant mass distribution at 13~TeV for different values of the self-coupling, at NLO with full top mass dependence (blue-dashed) and NNLO HTL NLO-improved (orange-solid), the latter rescaled to the NNLO$_\text{FTapprox}$ total cross section in the SM limit.
The lower panel shows the ratio to the NLO prediction, together with the inclusive SM $K$-factor (black-dotted), $\sigma_\text{NNLO}^\text{FTapprox}/\sigma_\text{NLO}^\text{Full}$, as a reference.
\label{fig:Higgs_HH_NNLO_EFT:lambda}
}
\end{figure}

Finally, we focus on the results obtained for exclusive $\lambda_{hhh}$ variations.
The corresponding invariant mass distributions for different $\lambda_{hhh}$ values are presented in Fig.~\ref{fig:Higgs_HH_NNLO_EFT:lambda}.
In this case, compared to the more general EFT variations presented in the previous figures, we see a milder dependence of the corrections on the invariant mass values, and smaller deviations from the inclusive SM $K$-factor.

The total cross sections obtained for the different values of $\lambda_{hhh}$ are presented in Table~\ref{tab:Higgs_HH_NNLO_EFT:total_XS}.
As mentioned before, the scale uncertainties are adjusted by a normalization factor in order to match the ones of the NNLO$_\text{FTapprox}$ SM prediction.
In addition to the NNLO$_\text{NLO-i}$ uncertainties, the relative uncertainties of the NNLO$_\text{B-i}$ were also considered (again adjusted to match NNLO$_\text{FTapprox}$) and, in order to be conservative, the maximum between these two is the one reported in Table~\ref{tab:Higgs_HH_NNLO_EFT:total_XS}.

In line with what is observed at the differential level, we can see that the ratio of our NNLO$_\text{NLO-i}$ results to the corresponding NLO prediction is only mildly dependent on the value of $\lambda_{hhh}$, with corrections ranging between $11\%$ and $16\%$ in the range of $\lambda_{hhh}$ under study.

\begin{table}[t]
{
\renewcommand{\arraystretch}{1.6}
\begin{center}
\resizebox{\textwidth}{!}{ 
\begin{tabular}{ |c|c|c|c|c|c|c|c| } 
\hline
$c_{hhh}$ & -1 & 0 & 1 & 2 & 2.4 & 3 & 5 \\
 \hline\hline
 $\sigma\, [\text{fb}]$
 & $ 131.9^{+2.5\%}_{-6.7\%} $
 & $ 70.38^{+2.4\%}_{-6.1\%} $
 & $ 31.05^{+2.2\%}_{-5.0\%} $
 & $ 13.81^{+2.1\%}_{-4.9\%} $
 & $ 13.1^{+2.3\%}_{-5.1\%} $
 & $ 18.67^{+2.7\%}_{-7.3\%} $
 & $ 94.82^{+4.9\%}_{-8.8\%} $ \\
 \hline
 $\sigma / \sigma^\text{SM}$
 & 4.25
 & 2.27
 & 1
 & 0.445
 & 0.422
 & 0.601
 & 3.05
 \\
 \hline
  $\sigma / \sigma_\text{NLO}$
 & 1.13
 & 1.13
 & 1.12
 & 1.11
 & 1.12
 & 1.15
 & 1.16 \\ \hline
\end{tabular}
}
\end{center}
}
\caption{\small Higgs pair production total cross sections for a collider energy of 13~TeV at NNLO$_\text{NLO-i}$ (rescaled to the NNLO$_\text{FTapprox}$ total cross section in the $c_{hhh}\to 1$ limit) for different values of the self-coupling $\lambda_{hhh}$, together with the ratio w.r.t. the SM expectation and to the NLO prediction.\label{tab:Higgs_HH_NNLO_EFT:total_XS}}
\end{table}

\subsection{Summary}

We have performed a combination of the NLO results with full top-mass dependence with the NNLO predictions obtained in the Born-improved HTL for non-SM values of the trilinear self-coupling and, more generally, in the context of a non-linear EFT approach parameterizing BSM effects.
In particular, the results for the total cross sections and theoretical uncertainties at NNLO$_\text{NLO-i}$ in Table~\ref{tab:Higgs_HH_NNLO_EFT:total_XS} are a key ingredient for a more consistent treatment of $\lambda_{hhh}$ variations in experimental analyses.


\chapter{Monte Carlo studies}
\label{cha:mc}
\newcommand{\Herwig}{H\protect\scalebox{0.8}{ERWIG}\xspace}
\newcommand{\Pythia}{P\protect\scalebox{0.8}{YTHIA}\xspace}
\newcommand{\Sherpa}{S\protect\scalebox{0.8}{HERPA}\xspace}

\section{Self-consistency of backwards evolved initial-state parton showers~\protect\footnote{
  L.~Gellersen, D.~Napoletano, S.~Prestel}{}}

\label{sec:MC_PDFs_in_ISR}

In this contribution, we consider the dependence of initial-state parton showers
in the backward evolution formalism on non-perturbative parameters. The
backward evolution formalism leads to certain self-consistency relations
for initial-state parton showers, and in particular requires that some
products of parton-distribution functions and parton-shower no-emission
probabilities lead to results that are independent of the fraction $x$ of the
hadron momentum carried by the incoming parton. We test this assumption 
in a realistic parton shower environment.

\subsection{Background}
\label{sec:MC_PDFs_in_ISR:background}

Event generators are a crucial component of high-energy physics phenomenology~\cite{Buckley:2011ms}.
They help with detector and analysis design, while also providing precision
predictions that can be used as background baselines in direct and indirect
searches for new physics. Parton showers, in turn,  are a key part of event
generators, since they allow to bridge the large gap (in energy scales and
particle multiplicity) between partons produced in fixed-order
calculations of high-energy scatterings and the hadrons that are
registered in the detector. This is
necessary to e.g.\ describe the formation and evolution of jets, and to
ensure that the average parton energies before transitioning to a model
of confinement are moderate to small, irrespective of the hard scattering
-- such that the effects of hadron formation can be assumed ``universal", and
does not require refitting of parameters for individual measurements.
On top of this historical purpose, the quest for high precision at the LHC
has promoted the view of the parton shower as a tool to perform resummation of large logarithms, and
as an ingredient in combining multiple inclusive fixed-order calculations in an
overlap-free manner~\cite{Catani:2001cc,Mangano:2001xp,Mrenna:2003if,
Alwall:2007fs,Hamilton:2009ne,Hamilton:2010wh,Hoche:2010kg,Lavesson:2008ah,
Lonnblad:2012ng,Lavesson:2005xu,
Platzer:2012bs,Gehrmann:2012yg,Hoeche:2012yf,Lonnblad:2012ix,
Frederix:2012ps,Alioli:2012fc,Bellm:2017ktr,Giele:2011cb,Lonnblad:2001iq,Lonnblad:2011xx,Fischer:2017yja}. Thus, recent years have seen a renewed interest in defining,
assessing and improving the precision of parton showers, to prevent their
intrinsic choices from becoming a dominant source of uncertainties at the LHC.

In this contribution, we follow the same route, by asking how well
certain consistency relations for initial-state parton showers in the 
backward evolution formalism hold for the showers implemented in \Pythia{}~\cite{Sjostrand:2014zea}
and \Sherpa{}~\cite{Bothmann:2019yzt}. 

The backward evolution formalism forms the basis of all modern parton 
showers, and is constructed by rewriting the perturbative (DGLAP) evolution 
equations~\cite{Altarelli:1977zs} of non-perturbative structure functions: Instead of
performing the evolution from low to high energy 
scales (see~\cite{Fox:1979ag,Odorico:1982cu,Field:1982vz} for initial attempts
of ``forward" evolution), it is possible
to perform evolution from high to low scales~\cite{Sjostrand:1985xi,Gottschalk:1986bk}, 
allowing for numerically efficient calculations.
Thus, the starting point of this study is to test to which extent
fully realistic parton showers actually mimic the structure of the DGLAP
evolution equations.

We begin with the scaling violations of the parton distribution function
$f_a$ of an initial-state parton $a$ extracted from a colliding hadron,
as described by the DGLAP evolution equation,
\begin{equation}\label{eq:MC_PDFs_in_ISR:pdf_evolution}
  \frac{df_{a}(x,t)}{d \ln t}=
  \sum_{b=q,g}\int_x^1\frac{d  z}{z}\,\frac{\alpha_s}{2\pi}\left[P_{ba}(z)\right]_+\,f_{b}(x/z,t)\;,
\end{equation}
where $P_{ab}$ are the regularized evolution kernels. Assume that we define
$P_{ab}$ in terms of unregularized kernels, $\hat{P}_{ab}$, restricted to all
but an $\varepsilon$-environment around the soft-collinear pole, plus an endpoint contribution.
\begin{equation}
    P_{ba}(z,\varepsilon)=\;\hat{P}_{ba}(z)\,\Theta(1-z-\varepsilon)
    -\delta_{ab}\,\frac{\Theta(z-1+\varepsilon)}{\varepsilon}
    \sum_{c=q,g}\int_0^{1-\varepsilon}d \zeta\,\zeta\,\hat{P}_{ac}(\zeta)
\end{equation}
For finite $\varepsilon$, the endpoint subtraction can be interpreted as the approximate virtual plus 
unresolved real corrections, which are included in the parton shower by enforcing unitarity.
When ignoring momentum conservation, this cutoff can be taken to zero,
which allows us to identify $\left[P_{ba}(z)\right]_+$ as the $\varepsilon\to 0$ limit of $P_{ba}(z,\varepsilon)$.
For $0<\varepsilon\ll 1$, Eq.~\eqref{eq:MC_PDFs_in_ISR:pdf_evolution} changes to
\begin{equation}\label{eq:MC_PDFs_in_ISR:pdf_evolution-2}
  \frac{1}{f_{a}(x,t)}\,\frac{d  f_{a}(x,t)}{d \ln t}=
  -\sum_{c=q,g}\int_0^{1-\varepsilon}d \zeta\,\zeta\,\frac{\alpha_s}{2\pi}\hat{P}_{ac}(\zeta)\,
  +\sum_{b=q,g}\int_x^{1-\varepsilon}\frac{d  z}{z}\,
  \frac{\alpha_s}{2\pi}\,\hat{P}_{ba}(z)\,\frac{f_{b}(x/z,t)}{f_{a}(x,t)}\;.
\end{equation}
We can then define the Sudakov form factor $\Delta$ and the no-emission probability $\Pi$,
\begin{eqnarray}
  \Delta_a(t_1,t_0) &=&\exp\bigg\{-\int_{t_1}^{t_0}\frac{d  t}{t}
  \sum_{c=q,g} \int_0^{1-\varepsilon}d \zeta\,\zeta\,\frac{\alpha_s}{2\pi}\hat{P}_{ac}(\zeta)\bigg\}\\
  \Pi_a(t_1,t_0;x) &=& \exp\bigg\{- \int_{t_1}^{t_0}\frac{d  t}{t} \sum_{b=q,g}\int_x^{1-\varepsilon}\frac{d  z}{z}\,
                     \frac{\alpha_s}{2\pi}\,\hat{P}_{ba}(z)\,\frac{f_{b}(x/z,t)}{f_{a}(x,t)}\bigg\}
                     \label{eq:MC_PDFs_in_ISR:noemprob}
\end{eqnarray}
Thus, we can rewrite Eq.~\eqref{eq:MC_PDFs_in_ISR:pdf_evolution-2} (after a bit
of algebra and taking the exponential on both sides) as
\begin{equation}
  f_a(x,t)\Delta_a(t,\mu^2)=f_a(x,\mu^2)\,\Pi_a(t,\mu^2;x)\;.
\end{equation}
Given the validity of all the assumptions made so far, this indicates that the product
\begin{equation}
  D(t_0,t;x)\equiv\frac{f_a(x,\mu^2)}{f_a(x,t)}\,\Pi_a(t,\mu^2;x)
\end{equation}
has to be $x$-independent to correctly recover the DGLAP equation.
Indeed, in theory, if one trusts our derivation, it should be that
$D(t_0,t;x) = \Delta_a(t,\mu^2)$.
It is the aim of this study to test how well this equivalence, and
thus the $x$-independence of $D$, is 
reproduced by current parton showers\footnote{
It should be noted that the initial-state parton shower implemented
in Cascade~\cite{Jung:2010si} relies on dedicated PDF fits obtained through 
parton-branching methods~\cite{Jadach:2008nu,Lelek:2017dwb}. This most likely enforces 
the $x$-independence by construction, but also is designed to solve the CCFM
evolution equation. Thus, it would be interesting to confirm this thought.}.

Initial-state parton showers implement backward evolution by generating
emissions according to Eq.~\eqref{eq:MC_PDFs_in_ISR:noemprob}.
However, it is important to note here that solving the DGLAP equation for a single
initial-state parton is by far not the only task of parton showers. Parton
showers also simulate the evolution of further incoming
partons, as well as of final-state partons, and in general, should model
soft-gluon effects without double-counting.
Furthermore, at each evolution step, the parton showers used in this study enforce exact
four-momentum conservation,
such that
physical real-emission states can be generated (which can then be systematically
replaced by more accurate descriptions through matching and merging schemes). 
The main consequence of this is that the value of $\varepsilon$ depends on the 
evolution variable and the overall mass of the decaying system in a way that 
ensures that physical real-emission states can be reconstructed. 
Thus, $\varepsilon\ll 1$, as is the case for actual DGLAP evolution,
is not automatically fulfilled.

Real-emission phase space points are parametrized in terms of the evolution 
variable $t$, the energy sharing variable $z$ and an azimuthal angle $\phi$.
For the case at hand, an emission of an initial-state parton 
$\widetilde{ai}$ with an initial-state recoil partner $\widetilde{b}$
(i.e.\ the splitting $\widetilde{ai}+\widetilde{b}\rightarrow a+i+b$)
these are given by~\cite{Sjostrand:2004ef,Schumann:2007mg}
\begin{align*}
  z_{\mathrm{Pythia}} &=& \frac{(p_a-p_i+p_b)^2}{(p_a+p_b)^2}
                          \qquad\qquad t_{\mathrm{Pythia}} &=&  2(p_a\cdot p_i)\, (1-z)\\
  z_{\mathrm{Sherpa}} &=& \frac{(p_a-p_i+p_b)^2}{(p_a+p_b)^2}
                          \qquad\qquad t_{\mathrm{Sherpa}} &=&
                                                               (p_a\cdot p_i)\,\left[\delta_{ig}(1-z)
                                                               + \delta_{iq}\right]~.
\end{align*}
With these, and using $m_{\mathrm{D}}^2 = (p_a-p_i+p_b)^2$, momentum 
conservation implies that
\begin{eqnarray}
  \varepsilon_{\mathrm{Pythia}}
 &=& \frac{\sqrt{t_{\mathrm{Pythia}}}}{ m_{\mathrm{D}} }
   \left(\sqrt{1+ \frac{t_{\mathrm{Pythia}}}{4\,m_{\mathrm{D}}^2  } }- \frac{\sqrt{t_{\mathrm{Pythia}}}}{2\,m_{\mathrm{D}}}\right)\\  
  \varepsilon_{\mathrm{Sherpa}}
 &=& \frac{t_{\mathrm{Sherpa}}}{m_{\mathrm{D}}^2 + t_{\mathrm{Sherpa}}}\;.
\end{eqnarray}
In particular for small $m_{\mathrm{D}}$ and moderate $t$ values, these
$\varepsilon$ may be appreciably different from zero, which in turn
represents one the various departures of realistic parton showers from DGLAP
behavior.

To summarize this section, and the goal of the current study:
\begin{itemize}
\item Rewriting DGLAP evolution in a way that allows for backward evolution
from high to low energy scales leads to the self-consistency condition that
\begin{equation}
 D(t,\mu^2;x) = \frac{f_a(x,\mu^2)}{f_a(x,t)}\,\Pi_a(t,\mu^2;x)\quad\textnormal{is $x$-independent.}
\end{equation}
\item This relies on the use of DGLAP splitting kernels, and on assumptions
about phase-space boundaries that can be violated if physical intermediate
real-emission states are required. Both of these points are expected to
arise in most modern parton showers.
\item Thus we aim to test the accuracy of the $x$-independence,
  highlighting when this is expected to hold, and when not.
\end{itemize}
As a final note, in this study we are only interested in relative
$x$-shapes, rather than the actual value of $D(t,\mu^2;x)$ for fixed
$x,\mu,t$ values. 
Although this is interesting in its own right (and has indeed recently been studied
-- as a spin-off from the present study -- in~\cite{Nagy:2020gjv} in the context
of the Deductor parton shower), the value of $D(t,\mu^2;x)$ strongly depends on $t$,
as well as many other algorithmic choices such as ordering, $\alpha_s$ running
and recoil strategy, and the large change in value might be
distracting when verifying the $x$-independence property.
Thus, to compare the change in the shape of the $x$-distribution on plots
with a common scale, by investigating the normalized function
\begin{equation}
  \label{eq:MC_PDFs_in_ISR:norm-no-em-prob}
 d(t,\mu^2;x_i)  =  \frac{D(t,\mu^2;x_i)}{\sum_j D(t,\mu^2;x_j) }
 =
 \frac{\frac{f_a(x_i,\mu^2)}{f_a(x_i,t)}\,\Pi_a(t,\mu^2;x_i)}{\sum_j D(t,\mu^2;x_j) }\;,
\end{equation}
where the sum in the denominator runs over the contribution to all
$x_i$-bins in the histograms.
We use trial showering~\cite{Lonnblad:2001iq} to calculate the no-emission probabilities $\Pi_a(t,\mu^2;x)$.

\subsection{Results}
\label{sec:MC_PDFs_in_ISR:results}
In this section, we collect results on the $x$ distribution of 
$d(t,\mu^2;x_i)$. As discussed in the previous section, this is
reminiscent of the Sudakov form factor, and 
as such should be $x$-independent. Indeed, assuming that
the parton shower contains all the ingredients that enter PDF evolution, this 
distribution is expected to be flat. Again, it is worth stressing that we
do not want to test the overall normalization, but rather the shape
of the $x$-distribution. Thus, all of the curves in the figures below are 
normalized to the sum of their respective entries, as described in
Eq.~(\ref{eq:MC_PDFs_in_ISR:norm-no-em-prob}).

Results shown in this contribution are obtained using a parton shower with a 
single initial-initial dipole at fixed mass $m_{D}$ and flavour 
($d$, $s$ or $g$). The shower starting scale is determined by $t_0=k_0m_D^2$,
and the evolution is terminated at the scale $t_1=k_1t_1$. To not
over-complicate the presentation, we fix $k_0=0.75$ and vary 
$k_1\in[0.8,0.1,0.01]$. In addition, we compare two
different sets of PDFs, NNPDF23\_lo\_as\_0119\_qed and
NNPDF23\_nlo\_as\_0119\_qed~\cite{Ball:2012cx,Ball:2013hta},
with $\alpha_s^{\mathrm{PS}}(M_Z)=0.119$. These results might be interesting
because the formal equivalence between the backward evolution implemented
in the shower and exact DGLAP evolution is only expected to hold for
leading order (LO) evolution. In the current setup, we only consider
LO splitting kernels in the parton showers. It would be 
interesting to extend the test proposed in
this contribution to parton showers that implement higher order
corrections to the splitting
functions~\cite{Hoche:2015sya,Hoche:2017hno,Hoche:2017iem,Dulat:2018vuy}.

\begin{figure}[t]
\subfigure[$d$-quark evolution]{
  \includegraphics[width=0.31\textwidth]{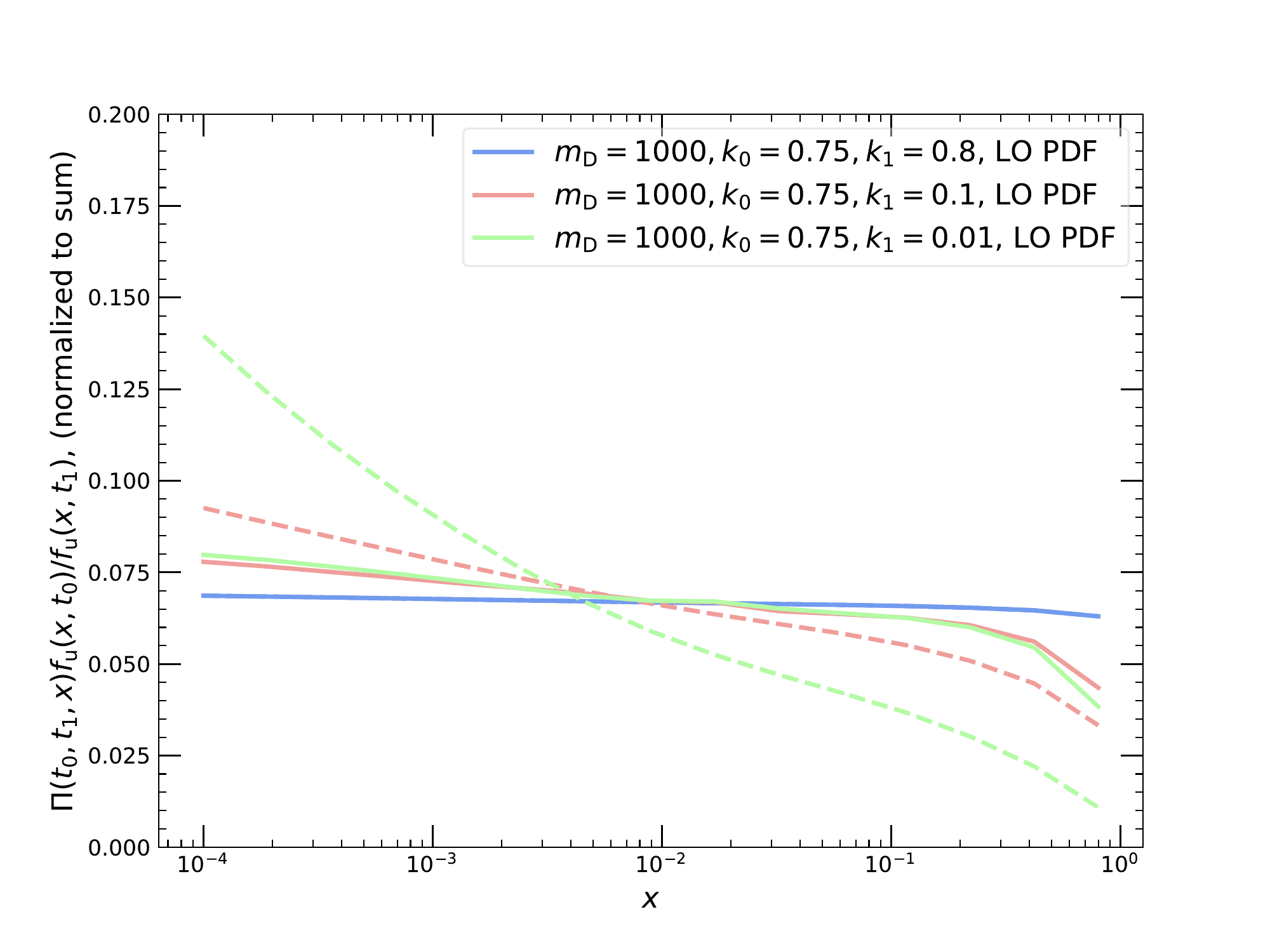}
  \label{fig:MC_PDFs_in_ISR:d-evol-1000}
}
\subfigure[$s$-quark evolution]{
  \includegraphics[width=0.31\textwidth]{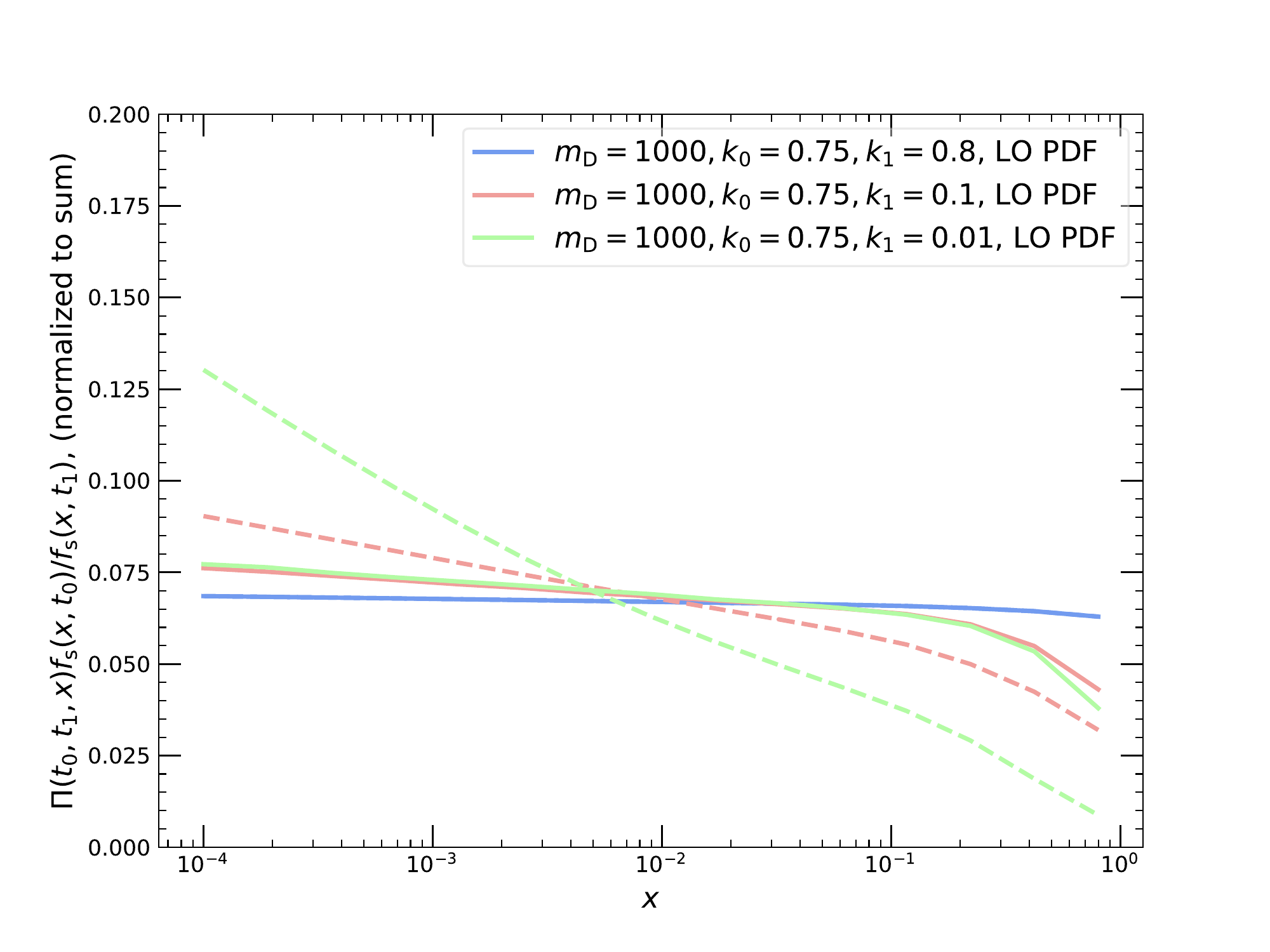}
  \label{fig:MC_PDFs_in_ISR:s-evol-1000}
}
\subfigure[gluon evolution]{
  \includegraphics[width=0.31\textwidth]{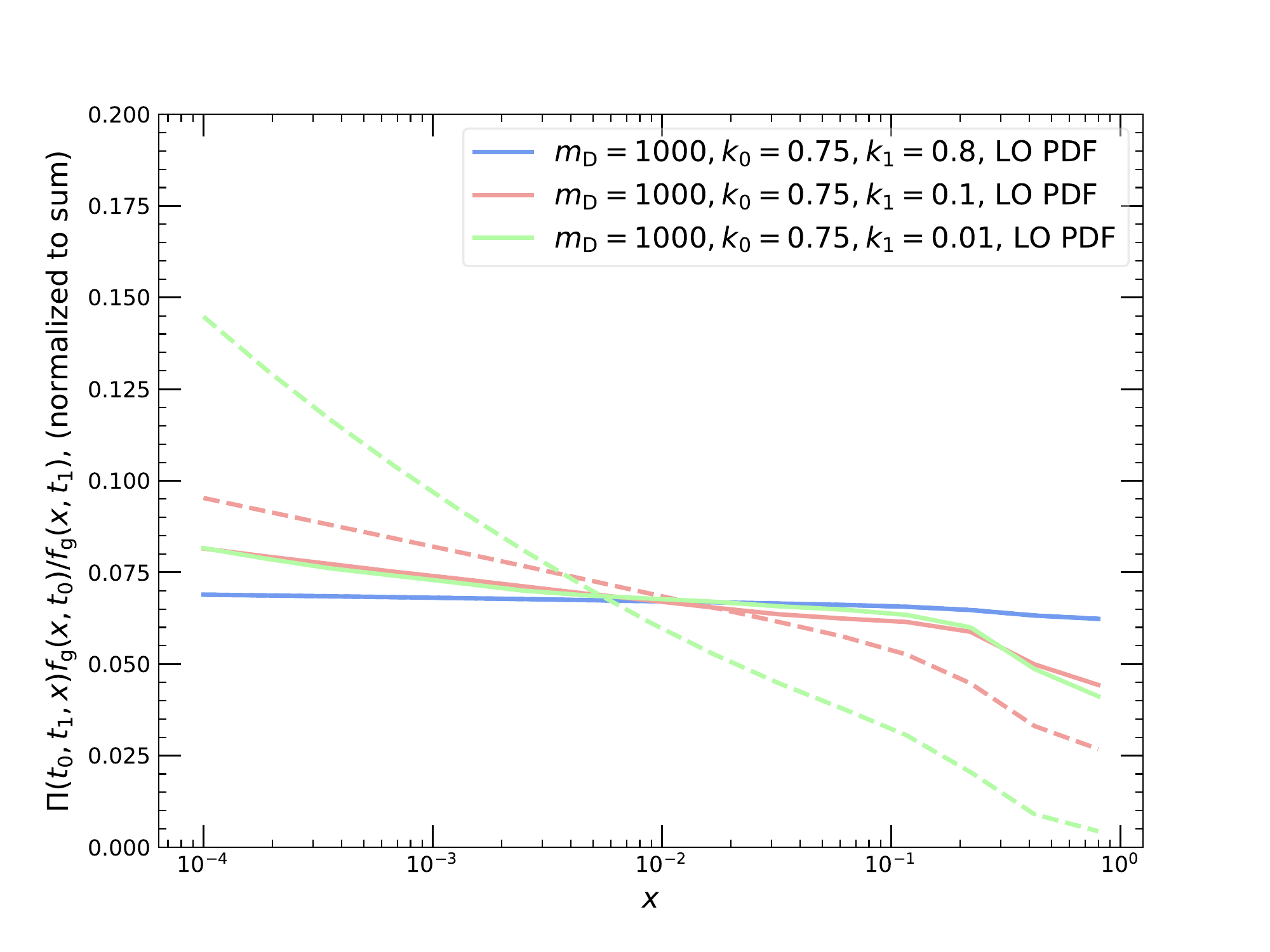}
  \label{fig:MC_PDFs_in_ISR:g-evol-1000}
}
\caption{
$x$-distribution for different length of parton-shower evolution, for
$m_{\mathrm{D}}=1000$ GeV, leading-order PDF set NNPDF23\_lo\_as\_0119\_qed, and for both 
\Pythia{} (solid curves) and \Sherpa{} (dashed curves)}
\label{fig:MC_PDFs_in_ISR:evol-1000}
%
\subfigure[$d$-quark evolution]{
  \includegraphics[width=0.31\textwidth]{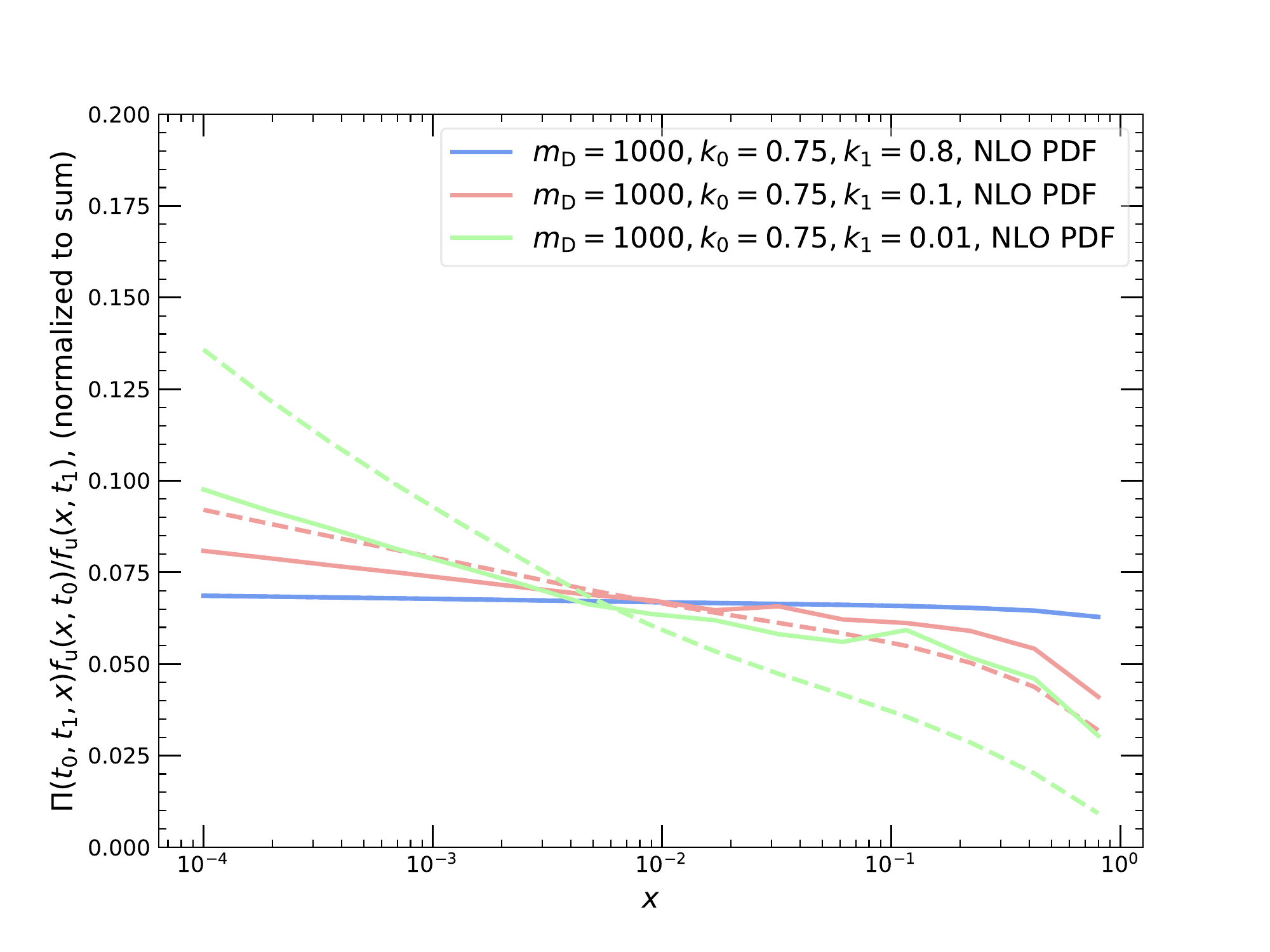}
  \label{fig:nlopdfs-in-isr:d-evol-1000}
}
\subfigure[$s$-quark evolution]{
  \includegraphics[width=0.31\textwidth]{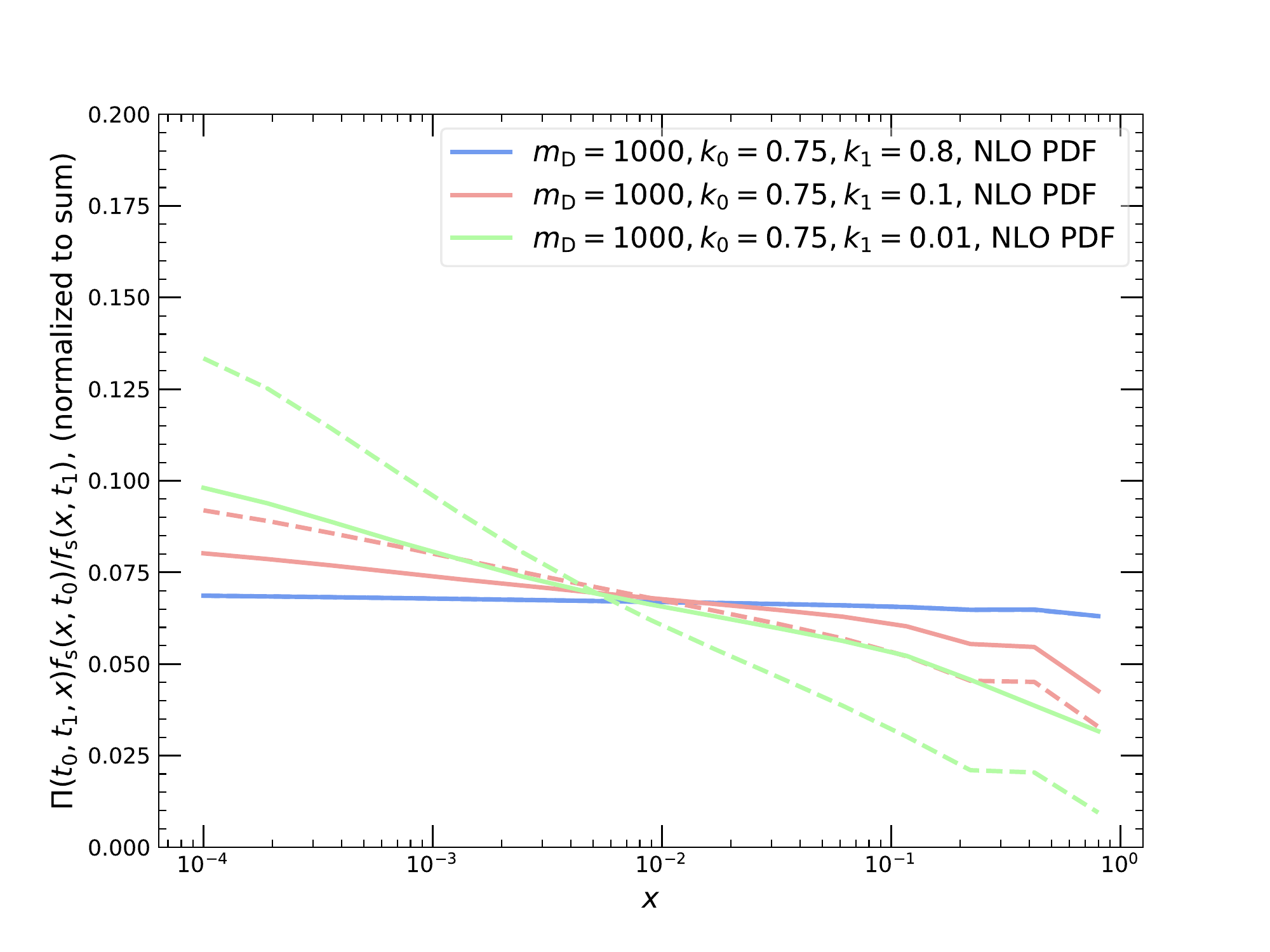}
  \label{fig:nlopdfs-in-isr:s-evol-1000}
}
\subfigure[gluon evolution]{
  \includegraphics[width=0.31\textwidth]{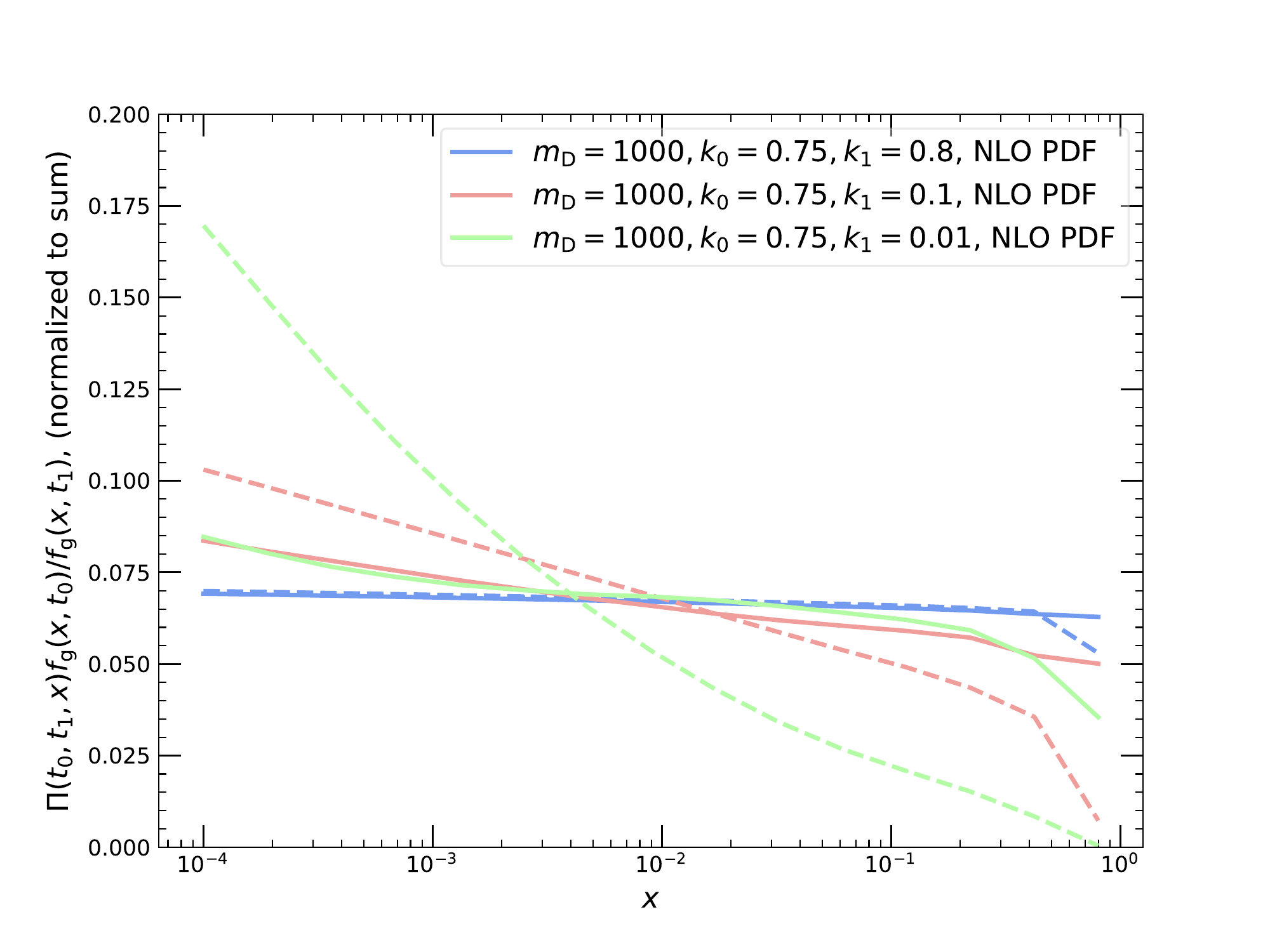}
  \label{fig:nlopdfs-in-isr:g-evol-1000}
}
\caption{
$x$-distribution for different length of parton-shower evolution, for
$m_{\mathrm{D}}=1000$ GeV, NLO PDF set NNPDF23\_nlo\_as\_0119\_qed, and for both 
\Pythia{} (solid curves) and \Sherpa{} (dashed curves)}
\label{fig:nlopdfs-in-isr:evol-1000}
\end{figure}

\begin{figure}[t]
\subfigure[$d$-quark evolution]{
  \includegraphics[width=0.31\textwidth]{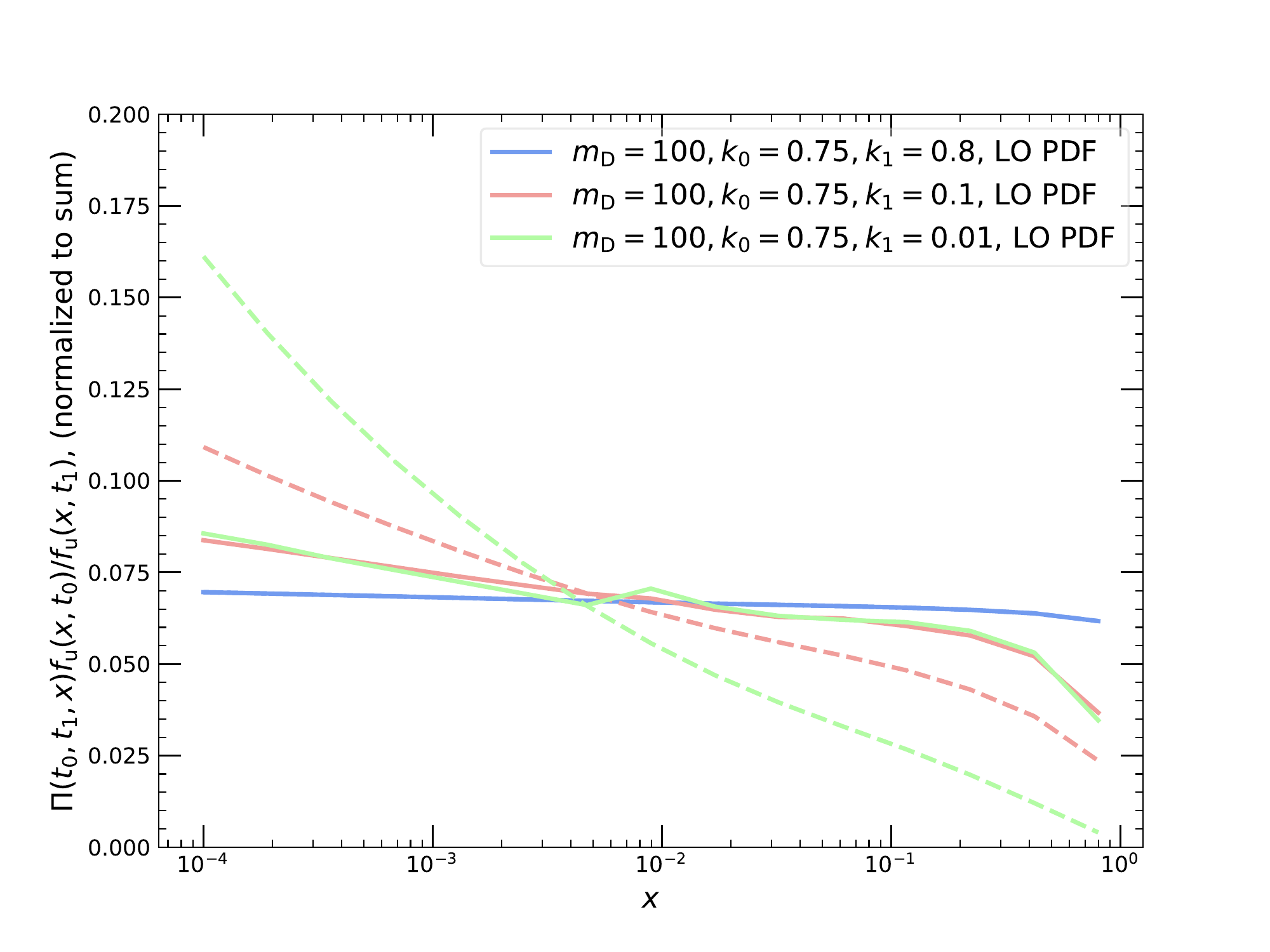}
  \label{fig:MC_PDFs_in_ISR:d-evol-100}
}
\subfigure[$s$-quark evolution]{
  \includegraphics[width=0.31\textwidth]{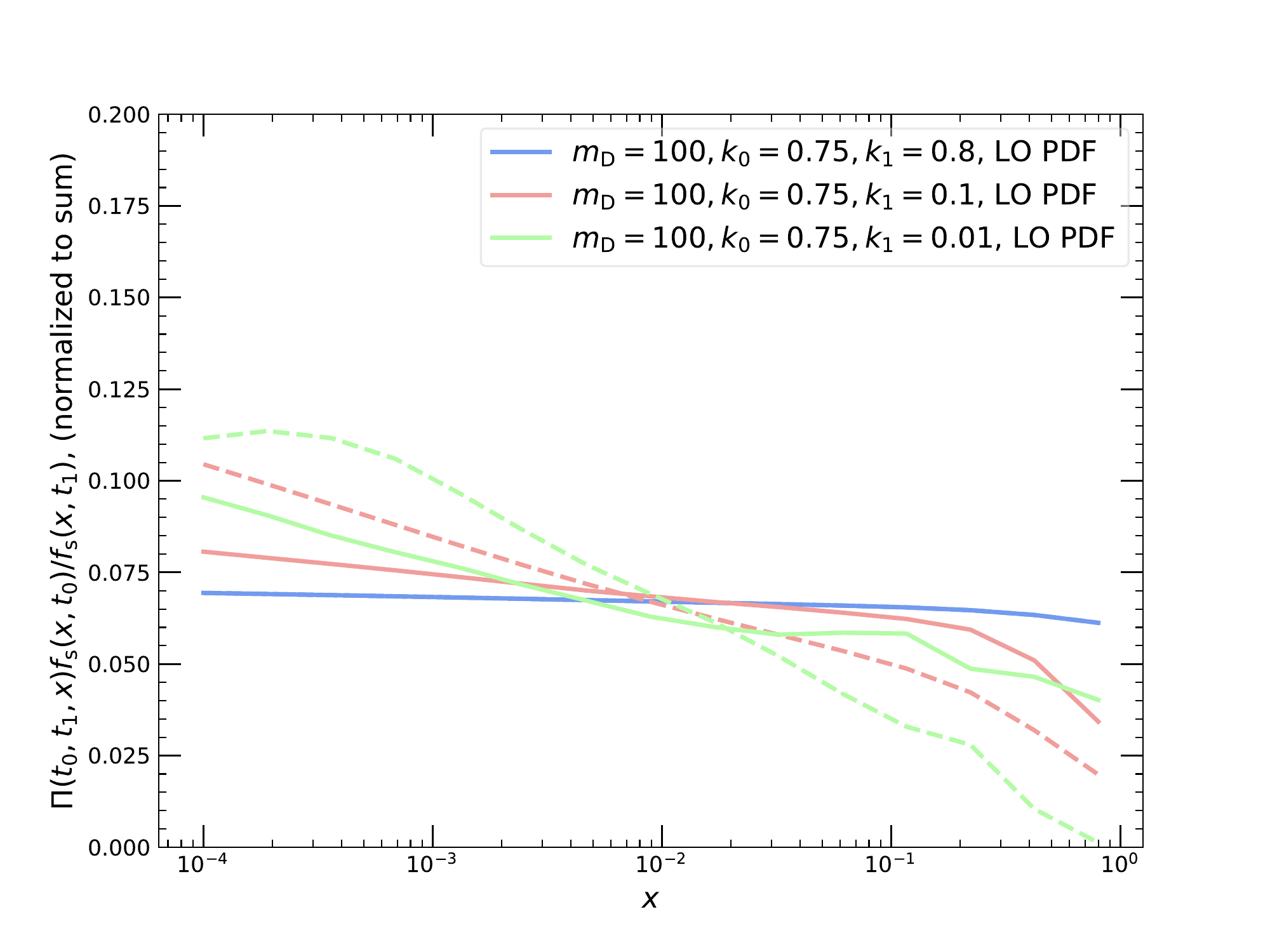}
  \label{fig:MC_PDFs_in_ISR:s-evol-100}
}
\subfigure[gluon evolution]{
  \includegraphics[width=0.31\textwidth]{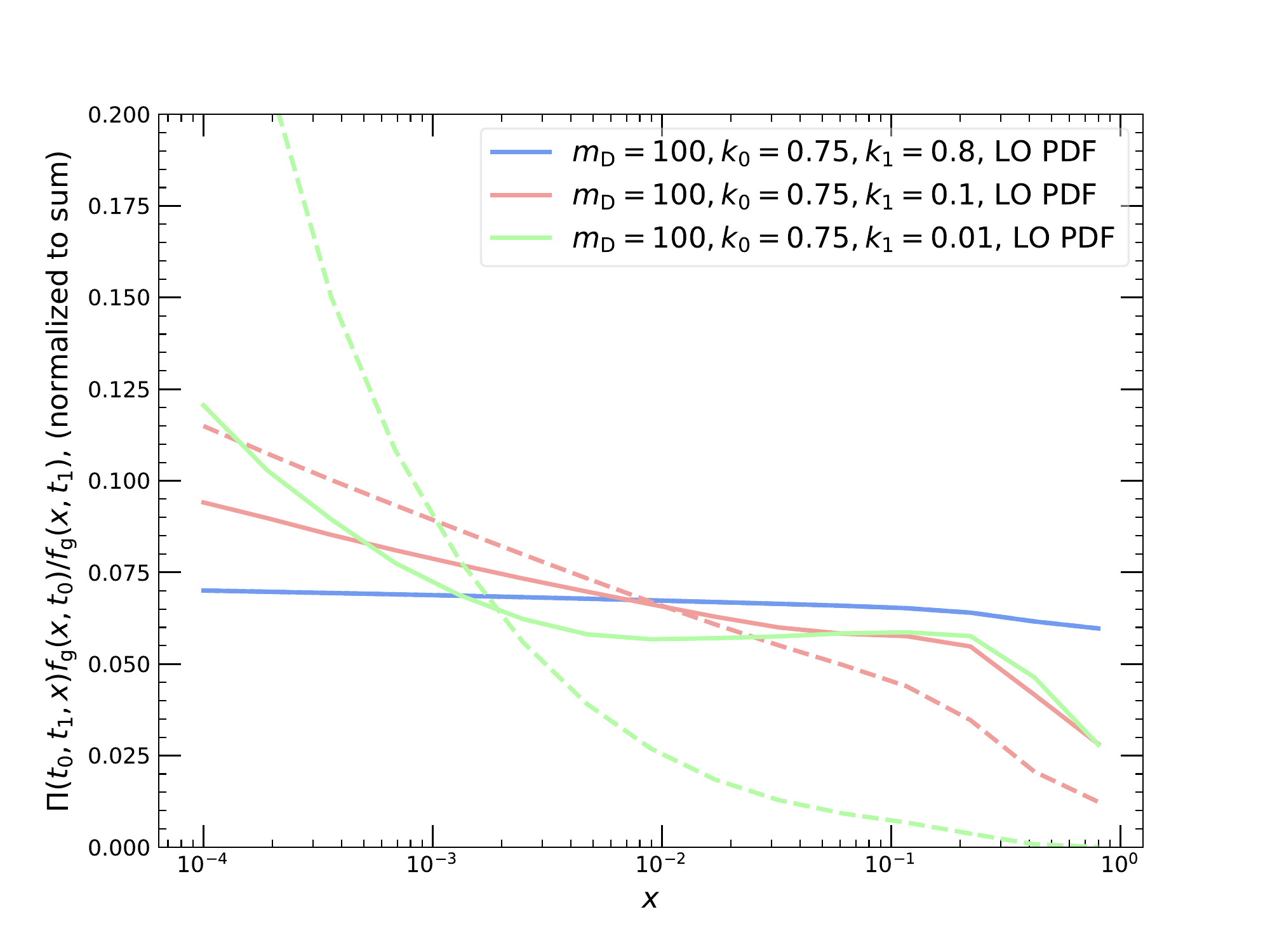}
  \label{fig:MC_PDFs_in_ISR:g-evol-100}
}
\caption{
$x$-distribution for different length of parton-shower evolution, for
$m_{\mathrm{D}}=100$ GeV, leading-order PDF set NNPDF23\_lo\_as\_0119\_qed, and for both 
\Pythia{} (solid curves) and \Sherpa{} (dashed curves)}
\label{fig:MC_PDFs_in_ISR:evol-100}
%
\subfigure[$d$-quark evolution]{
  \includegraphics[width=0.31\textwidth]{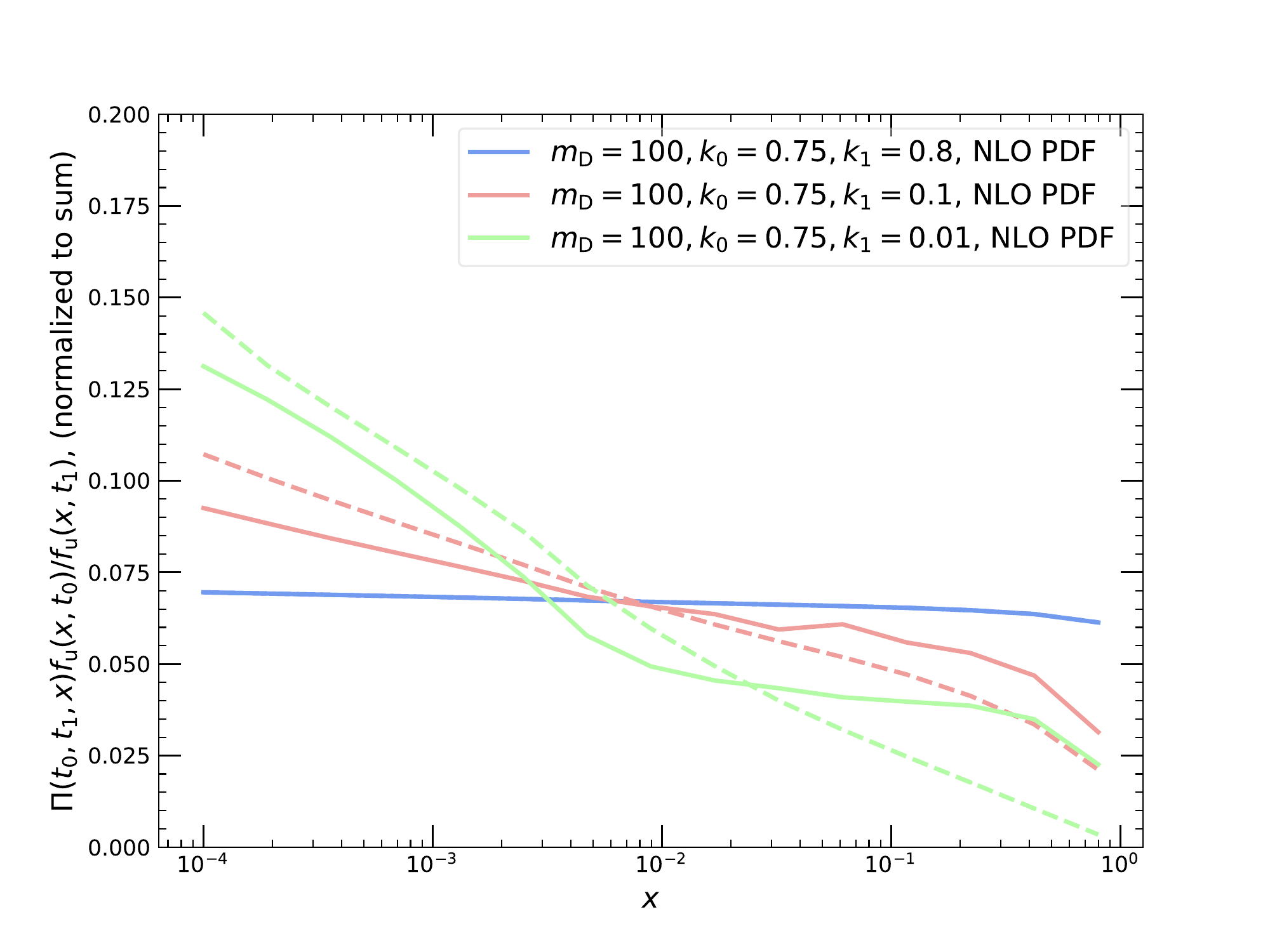}
  \label{fig:nlopdfs-in-isr:d-evol-100}
}
\subfigure[$s$-quark evolution]{
  \includegraphics[width=0.31\textwidth]{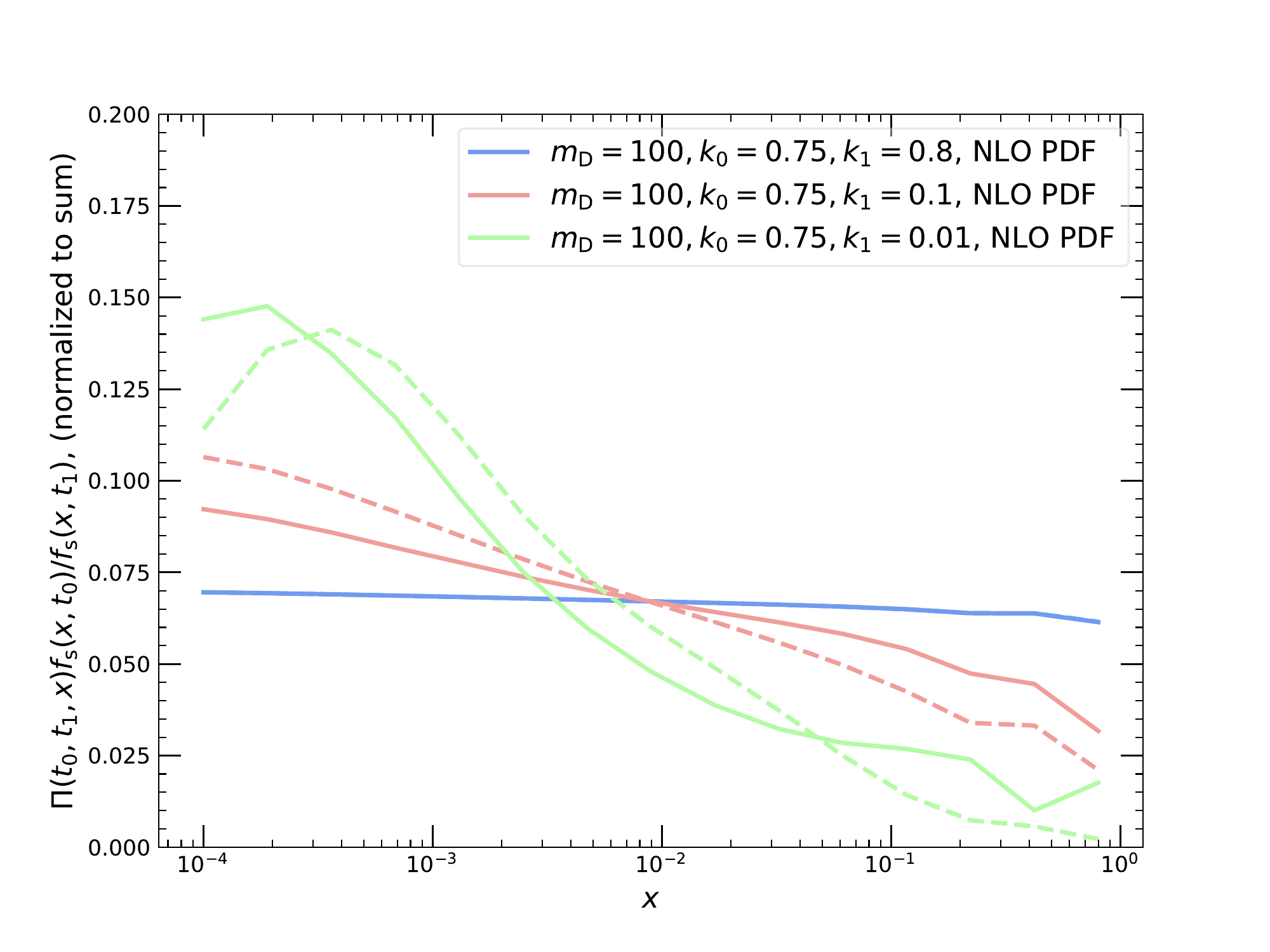}
  \label{fig:nlopdfs-in-isr:s-evol-100}
}
\subfigure[gluon evolution]{
  \includegraphics[width=0.31\textwidth]{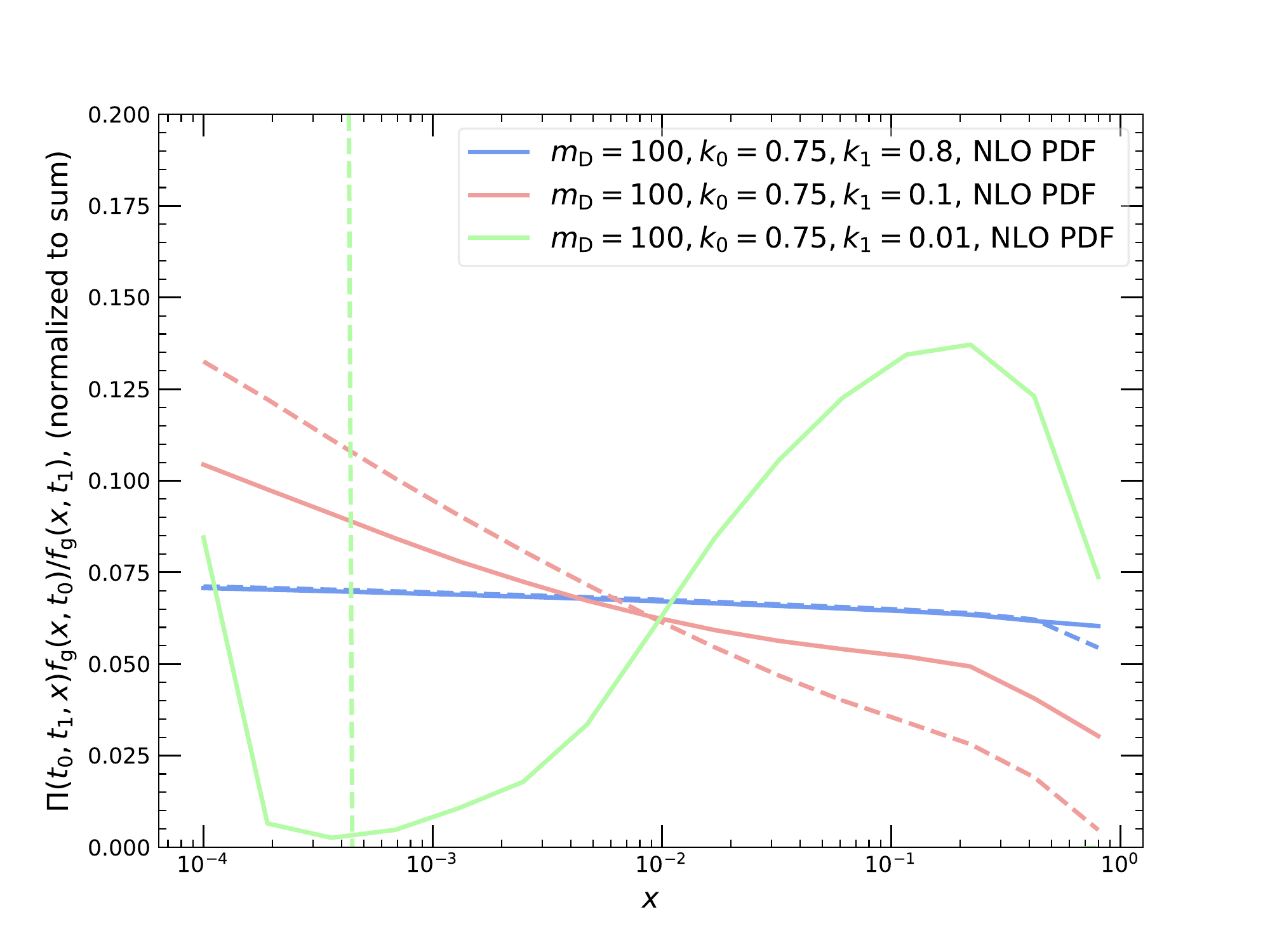}
  \label{fig:nlopdfs-in-isr:g-evol-100}
}
\caption{
$x$-distribution for different length of parton-shower evolution, for
$m_{\mathrm{D}}=100$ GeV, NLO PDF set NNPDF23\_lo\_as\_0119\_qed, and for both 
\Pythia{} (solid curves) and \Sherpa{} (dashed curves)}
\label{fig:nlopdfs-in-isr:evol-100}
\end{figure}

Results are reported in
Figs.~\ref{fig:MC_PDFs_in_ISR:evol-1000}-\ref{fig:nlopdfs-in-isr:evol-100}.
In particular, for a fixed flavour and $m_{\mathrm{D}}$, the variation of $k_0$
illustrates the effect of short, moderate or long parton-shower 
evolution (blue, red and green curves, respectively).
In addition, smaller (or larger) values of the dipole mass $m_{\mathrm{D}}$ 
further limit (increase) the available evolution phase space.
We start our discussion from a high dipole mass, $m_{\mathrm{D}} =
1000$~GeV. When using LO PDFs (Fig.~\ref{fig:MC_PDFs_in_ISR:evol-1000}),
we see that short evolution sequences (blue curves) do indeed show the
desired $x$-independence. For longer evolution, we observe a slight $x$-dependence
in \Pythia{}, and a larger dependence in \Sherpa{}, which becomes steeper
for longer evolution. For \Pythia{}, the
$x$-dependence is slightly more pronounced in gluon evolution than in
valence ($d$-quark) or sea ($s$-quark) evolution. It should be noted
that evolution sequences with no emission over two decades in $t$ are, in
typical LHC applications, extremely rare; the green curves should be
regarded as a worst-case scenario.

The LO response can be contrasted with
the use of next-to-leading-order (NLO) PDFs in Fig.~\ref{fig:nlopdfs-in-isr:evol-1000}.
In this case, it can be clearly seen that the $x$-independence in \Pythia{} is
violated more than for the corresponding LO results. As expected,
the violation becomes worse if PDFs that require NLO evolution are used. 
Interestingly, in \Sherpa{}, the shape of the $x$-dependence is very similar
at NLO and at LO. This feature needs to be studied with more
details. It has to be pointed out in this context, that the longer
evolution (for both $m_{\mathrm{D}} =100$ and $m_{\mathrm{D}} =100$~
GeV) window spills over the low scale $Q_0$ in the PDF, thus the way
PDFs are retrieved (or extrapolated) at these scales can play an important
role. In \Pythia{}, we have used a
dedicated interface to NNPDF written in the context of the Monash tune to
\Pythia{}~\cite{Skands:2014pea}, while \Sherpa{} uses an interface to
LHAPDF~\cite{Buckley:2014ana}. Difference in the
extrapolation region may well be responsible for the discrepancies
seen in this study.
In both \Pythia{} and \Sherpa{},
it is clearly visible that longer evolution sequences lead to a more
pronounced $x$-dependence. Comparing the LO and NLO PDF response of \Pythia{},
it seems prudent to favor LO PDFs in the evolution.

The effect of a more constrained phase space ($\varepsilon$ further away from 
zero) is assessed in Figs.~\ref{fig:MC_PDFs_in_ISR:evol-100} 
and~\ref{fig:nlopdfs-in-isr:evol-100}, by using $m_{\mathrm{D}}=100$ GeV. When using LO 
PDFs (Fig.~\ref{fig:MC_PDFs_in_ISR:evol-100}), we again observe a high
degree of $x$-independence for short evolution sequences. For longer evolution
sequences, a larger dependence on $x$ is observed, which becomes more
pronounced for the longest evolution (green curves). In \Pythia{}, this
trend is particularly visible for sea and gluon evolution. \Sherpa{} further
exhibits a large $x$-dependence of the valence content. Overall, an
$x$-dependence of above $\sim 30$\% is observed for moderate evolution
length of one decade in $t$ (orange curves). This, and the larger violations
at longer evolution, could indicate short-comings in the backward-evolution
formalism. However, it is again worth stressing that these scenarios are
most likely rare in LHC applications. Nevertheless, even effects 
of $\mathcal{O}(5\%)$ could be relevant for precision physics.

The use of NLO PDFs at $m_{\mathrm{D}}=100$ GeV (Fig.~\ref{fig:nlopdfs-in-isr:evol-100})
leads to surprisingly drastic results. It is interesting to note that,
as was the case at $m_{\mathrm{D}}=1000$ GeV, the size of the
$x$-dependence in \Sherpa{} is similar at LO and NLO. Both \Pythia{}
and \Sherpa{} show a large $x$-dependence for moderate evolution, and
a very large $x$-dependence for long evolution. The sea evolution shows
larger violations than that of the valence components. The gluon 
evolution shows a extreme behaviors for
long evolution, in particular for \Sherpa{}. It might be that small-$x$ effects (as
e.g.\ included in the PDF fit of~\cite{Bertone:2018dse}) could
play a role in improving the $x$-independence of the NLO gluon. From these
results, it again seems reasonable to favor LO PDFs for the evolution 
in \Pythia{}.

\subsection{Conclusions}
\label{sec:MC_PDFs_in_ISR:conclusions}

Initial-state parton showers are crucial components of event generators
for LHC physics. These parton showers aim to distribute the momenta of 
initial-state partons according to DGLAP evolution equations, by employing
backward evolution. This formalism leads to the requirement that
the function
\begin{equation*}
 D(t,\mu^2;x) = \frac{f_a(x,\mu^2)}{f_a(x,t)}\,\Pi_a(t,\mu^2;x)\;.
\end{equation*}
is $x$-independent. However, this $x$-independence is not automatically 
guaranteed when factoring in other requirements on the parton shower.
Thus, we have tested this assumed $x$-independence using
the parton showers implemented in \Pythia{} and \Sherpa{}. 
For these test cases, we find clear examples of the $x$-independence being
violated. We find small violations if the shower evolution is short (i.e.\
when splittings occur after short evolution intervals), for large
dipole masses (i.e.\ relatively wide phase space limits), and when using 
leading-order PDFs. These results are in line with our expectations.
The assumed $x$-independence is violated to a larger degree for tighter
phase-space limits due to a smaller dipole mass, and for longer evolution
sequences. When using leading-order PDFs and $m_{\mathrm{D}}=100$ GeV, we 
observe violations of above 30\% for very long evolution over two decades 
in $t$. The evolution of initial-state gluons in particular shows a large
$x$-dependence. It should however be noted that and evolution over two decades 
in $t$ is a rare, extreme occurrence; thus, this should be considered more as
a worst-case scenario rather than a disaster for LHC phenomenology.
Very pronounced effects are found when using NLO PDFs 
at $m_{\mathrm{D}}=100$ GeV, where violations of above 80\% are common.
This raises the question if NLO PDFs should be used with current parton
showers, as e.g.\ is the case in NLO matched calculations. A long-term
cure for this situation could be the construction of NLO parton showers.
In summary, we hope this study will be useful input and inspiration for 
future parton-shower developments and improvements. Discussions about the
current study during the Les Houches workshop have already inspired new
developments, as e.g.\ presented in~\cite{Nagy:2020gjv}.

\subsection*{Acknowledgements}
We thank the organizers for an inspiring workshop, and would like to thank 
Stefan H{\"o}che and Torbj{\"o}rn Sj{\"o}strand
for insightful discussions.
L.\,G.~and S.\,P.~have received funding from the European Union's Horizon 2020 
research and innovation program as part of the Marie Sk\l{}odowska-Curie 
Innovative Training Network MCnetITN3 (grant agreement no. 722104), and 
from the Swedish Research Council under contract number
2016-05996. D.\,N.~ is supported in part by the French Agence
Nationale de la Recherche, under grant 
ANR-15-CE31-0016 and by the ERC Starting Grant REINVENT-714788.

\let\Herwig\undefined
\let\Pythia\undefined
\let\Sherpa\undefined


\newcommand{\Herwig}{H\protect\scalebox{0.8}{ERWIG}\xspace}
\newcommand{\Pythia}{P\protect\scalebox{0.8}{YTHIA}\xspace}
\newcommand{\POWHEGplus}{P\protect\scalebox{0.8}{OWHEG}}
\newcommand{\POWHEG}{P\protect\scalebox{0.8}{OWHEG}\xspace}
\newcommand{\Sherpa}{S\protect\scalebox{0.8}{HERPA}\xspace}
\newcommand{\Openloops}{O\protect\scalebox{0.8}{PEN}L\protect\scalebox{0.8}{OOPS}\xspace}
\newcommand{\madgraphNLO}{M\protect\scalebox{0.8}{AD}G\protect\scalebox{0.8}{RAPH}5\protect\scalebox{0.8}{\_A}MC@NLO\xspace}
\newcommand{\Rivet}{R\protect\scalebox{0.8}{IVET}\xspace}
\newcommand{\Professor}{P\protect\scalebox{0.8}{ROFESSOR}\xspace}
\newcommand{\eps}{\varepsilon}
\newcommand{\mc}[1]{\mathcal{#1}}
\newcommand{\mr}[1]{\mathrm{#1}}
\newcommand{\mb}[1]{\mathbb{#1}}
\newcommand{\tm}[1]{\scalebox{0.95}{$#1$}}
\newcommand{\qT}{$p_{\perp}$\xspace}
\newcommand{\pTH}{$p_{\perp}^{H}$\xspace}
\newcommand{\pTZ}{$p_{\perp}^{Z}$\xspace}
\newcommand{\pTZH}{$p_{\perp}^{ZH}$\xspace}
\newcommand{\pTj}{$p_{\perp}^{j}$\xspace}
\newcommand{\njet}{$N_{\text{jets}}$\xspace}
\newcommand{\DPhiVH}{$\Delta \phi (Z, H)$\xspace}
\newcommand{\muR}{$\mu_\text{R}$\xspace}
\newcommand{\muF}{$\mu_\text{F}$\xspace}

\section{A study of loop-induced $ZH$ production with up to one additional jet~\protect\footnote{
  E.~Bothmann, 
  M.~Calvetti,
  P.~Francavilla,
  C.~Pandini,
  E.~Re,
  S.~L.~Villani
  }{}}

\label{sec:Higgs_ZH1j}


\begin{figure}[ht]
  \begin{center}
    \includegraphics[width=5cm]{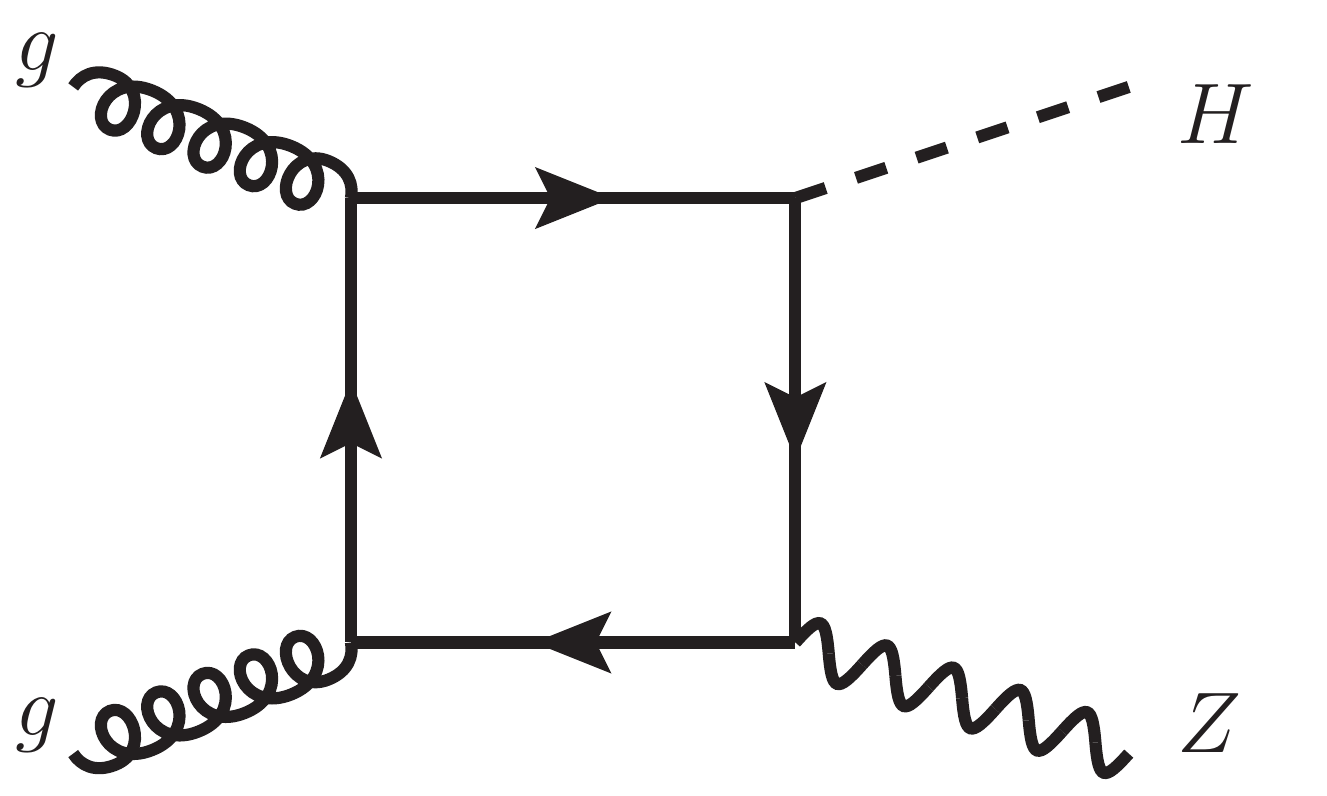}
    \hspace{1cm}
    \includegraphics[width=5cm]{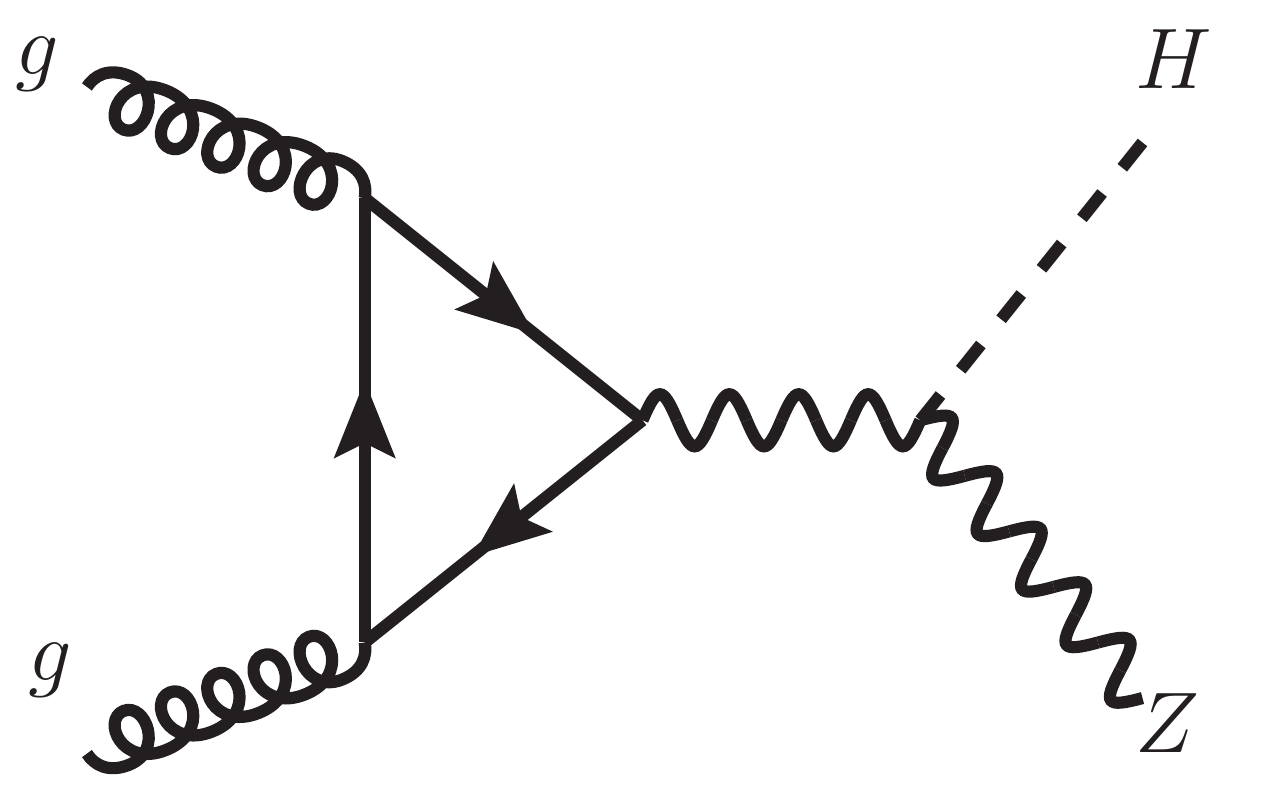}
    \caption{Representative Feynman diagrams that contribute to $gg \to ZH$ at LO.}
    \label{fig:Higgs_ZH1j:LOdiagrams}
  \end{center}
\end{figure}

\subsection{Introduction}
\label{sec:Higgs_ZH1j:introduction}
The $gg \rightarrow ZH$ production can be considered as a standalone process whose LO QCD contributions start at $\mathcal{O}(\alpha_S^2)$, corresponding to the Feynman diagrams in Fig.~\ref{fig:Higgs_ZH1j:LOdiagrams}. Calculations for the inclusive cross-section at LO QCD are available and are characterized by the destructive interference between the box and triangle diagrams~\cite{Englert:2013vua}. The two initial-state gluons lead to a rather strong renormalization and factorization scale dependence of about 30\,\%~\cite{deFlorian:2016spz}, thus increasing the theoretical uncertainty of $ZH$ relative to $WH$ production, where the gluonic channel does not contribute at LO. Experience from the gluon-fusion process $gg \rightarrow H$ (which has the same initial state and color structure as $gg \rightarrow ZH$) shows, however, that the LO scale uncertainty drastically underestimates the actual size of the higher-order corrections.

The NLO QCD $\mathcal{O}(\alpha_S^3)$ corrections to this process are beyond the technology currently available, due to the presence of massive multi-scale double-box integrals. However, a NLO perturbative correction factor $k_\text{NLO} = \sigma_\text{NLO} / \sigma_\text{LO}$ can be calculated in the limit of an infinite top quark mass and a vanishing bottom quark mass, known as the `effective field theory' NLO(EFT) approach~\cite{Altenkamp:2012sx}. The validity of this approximation holds well for $m_H=125$\,GeV at the center of mass energies considered at the LHC (from 7 to 13\,TeV), but of course worsens for larger $\sqrt{s}$ and in specific kinematic regimes (for instance in the boosted Higgs regime). The calculation yields $k_\text{NLO} \approx 2$ for $m_H=125$\,GeV, which is indeed not covered by the size of the LO scale uncertainty. The impact of a threshold resummed cross-section for $gg \rightarrow ZH$ at NLL has been considered~\cite{Harlander:2014wda}, matched to the NLO(EFT) result: the central value of the inclusive $\sigma^{ggZH}$ cross-section increases by 18\,\% at $\sqrt{s} = 13$\,TeV, while the uncertainty from scale variations decreases by a factor of three to four.

The NLO(EFT)+NLL $\sigma^{ggZH}$ contribution to the total $\sigma^{ZH}$ cross-section is of the order of 14\,\% at $\sqrt{s}=13$\,TeV, an already sizable contribution that becomes even more pronounced for large transverse momenta of the Higgs boson \pTH. This is a consequence of the threshold effect from the presence of top quark loops which makes the gluon-induced process especially important for \pTH$  \gtrsim m_{\text{top}}$, with a transverse momentum spectrum fundamentally different from the dominant quark-initiated contribution. Furthermore, the process has a peculiar sensitivity to new physics: through modified Higgs coupling to SM states, through new heavy colored states participating in the loops or through new $s$-channel pseudoscalars proposed in SM extensions.

A precise modeling of this process is thus key for the experimental analyses of LHC data performed by ATLAS and CMS targeting $VH$ final states. Both collaborations have so far relied on \POWHEG~\cite{Nason:2004rx, Frixione:2007vw, Alioli:2010xd} matched to the \Pythia~8~\cite{Sjostrand:2014zea} parton-shower (PS) to simulate these events at LO+PS, scaling the LO total cross-section to the state-of-the-art calculation at NLO(EFT)+NLL~\cite{deFlorian:2016spz}. While this approach allows to consider the normalization effect of the available higher-order corrections, we highlight that the modeling of differential distribution is included only at LO, and the perturbative QCD uncertainties induced by analysis selections and cuts (e.g. jet-vetoes, \pTH~cuts, etc.) fully rely on the LO+PS simulation.
While a full NLO calculation with finite top-mass effects remains to be performed, different Monte Carlo tools have been developed to allow the LO+PS simulation of $2\rightarrow3$ matrix elements for loop-induced $ZH$+jet processes, to obtain merged multi-leg samples of 0- and 1-jet multiplicities~\cite{Goncalves:2015mfa, Hespel:2015zea}. In this study we consider the impact of the simulation of higher jet multiplicities at matrix element level, by comparing the modeling of $gg \rightarrow ZH$ provided by the \Sherpa event generator with the \POWHEGplus+\Pythia setup normally used by the experimental analyses.

\subsection{Monte Carlo setup}
\label{sec:Higgs_ZH1j:MCsetup}
We consider different Monte Carlo (MC) tools providing LO predictions for the $gg\rightarrow ZH$ process with finite top-quark mass effects.
For the scope of this study we only consider the leptonic decay channel of the
Z boson to electrons for simplicity.
The decay is done on the matrix-element level, and hence finite-width effects and spin correlations are accounted for in the simulation. The Higgs is left undecayed. The simulation of multiple-parton interactions and hadronisation effects is disabled. The emission of additional QED radiation and other higher-order QED corrections are also turned off.
Only $gg \rightarrow Z(\rightarrow e^+e^-)H$ processes are thus considered in the following sections, evaluated
for proton-proton collisions at a centre-of-mass energy of $\sqrt s =
13\,\text{TeV}$.
The factorisation and the renormalisation scale are both set to
\[
   \mu_\text{F}=\mu_\text{R}=\hat{H}_{\perp}=\sqrt{M_{H}^{2}+{p_{\perp}^{H}}^{2}}+\sum_{i} p_{\perp}^{(i)} 
   ,\]
with the sum running over the two leptons. For some of the results, we show an envelope over 7-point variations of the renormalisation and factorisation scale, i.e.\ varying both scales independently by factors of two and one half, but omitting those where the two scale variation factors differ by a factor of 4. The shower starting scale is set to the invariant mass of the final-state system if not otherwise mentioned.
For the proton structure functions, the PDF4LHC15 set is used at NLO
accuracy~\cite{Butterworth:2015oua}.
The top-quark mass and its width are set to $m_t=172.5$\,GeV
and $\Gamma_t=1.32$\,GeV, respectively,
and the bottom-quark mass is given by $m_b=4.95$\,GeV.
Furthermore, we have $\alpha_\text{QED}=1/132.5070$.
The boson masses are $m_W=80.419$\,GeV, $m_Z=91.188$\,GeV and $m_H=125$\,GeV.

\subsubsection{\Sherpa}
\label{subsec:Higgs_ZH1j:SHERPA}
The first set of results is obtained with the \Sherpa event generator
v.2.2.8~\cite{Gleisberg:2008ta,Bothmann:2019yzt}, interfaced to \Openloops~2~\cite{Buccioni:2019sur} to provide the loop contributions.
The fixed-order sample is evolved further by the CS parton
shower~\cite{Schumann:2007mg}, which is the default shower implemented in \Sherpa.
For combining 0-jet and 1-jet samples into a single inclusive sample the
multi-jet merging at leading order implemented in \Sherpa is
used~\cite{Catani:2001cc,Hoeche:2009rj}, adapted for loop-induced processes
in~\cite{Cascioli:2013gfa}.
Factorisation and renormalisation scale variations are calculated on-the-fly~\cite{Bothmann:2016nao}.
In addition, we vary the shower starting scale up and down by factors of $\sqrt
2$ for the LO+PS sample, to give an estimate of the resummation uncertainty.
For the multijet-merged sample, we choose a merging cut of $Q_\text{cut}=20$\,GeV. This technical cut separates the phase-space regions populated by the matrix elements and the parton shower. Variations of this cut by a factor of 2, i.e.\ $Q_\text{cut}=10$\,GeV and $40$\,GeV, are also studied.
With the above specifications, we generate two \Sherpa samples:
\begin{itemize}
   \item loop-induced $gg \rightarrow Z(\rightarrow e^+e^-)H$ + 0-jet (LO+PS)
   \item loop-induced $gg \rightarrow Z(\rightarrow e^+e^-)H$ + 0,1-jets at LO with multijet-merging (MEPS 0,1j)
\end{itemize}
Note that the 1-jet matrix elements also include diagrams with initial-state quarks, cf.\ e.g.\ \cite{Hespel:2015zea,Goncalves:2015mfa}.
The corresponding squared loop amplitudes
form a finite and gauge-invariant subset of the NNLO corrections
for the $pp \to ZHj$ process.

\subsubsection{\POWHEGplus+\Pythia}
\label{subsec:Higgs_ZH1j:PP8}
The second set of results relies on the \POWHEG \texttt{ggHZ} event generator code to obtain an inclusive $gg \rightarrow Z(\rightarrow e^+e^-)H$ + 0-jet LO sample, interfaced to the \Pythia~8.2 parton-shower algorithm. The NNPDF3.0 PDF~\cite{Ball:2014uwa} set is used within the shower.
The PS matching relies on the \Pythia~8 `wimpy-shower' algorithm, in which the shower starting scale is set to the invariant mass of the $ZH$ system $m_{ZH}$, letting \Pythia~8 continue the showering process at the hardness scale at which \POWHEG leaves off. An alternative approach is considered, to assess the impact of the matching scheme on the results, by adopting the \POWHEG-specific \texttt{main31} \Pythia algorithm for a vetoed `power-shower', in which the shower starting scale is set to the kinematical limit ($\pT = \sqrt{\hat{s}}/2$), combined with an a-posteriori veto of emissions already covered by \POWHEG.

\subsection{Analysis and results}
\label{sec:Higgs_ZH1j:rivetresults}
The analysis selection applied for this study
has been implemented using the \Rivet 2 analysis
framework~\cite{Buckley:2010ar} and it is designed to obtain the simplest possible selection for the $gg \rightarrow Z(\rightarrow e^+e^-)H$ channel. Lepton and jet \qT and $\eta$ cuts are defined to be close to the ranges probed by experimental analyses~\cite{Aaboud:2018zhk}.
The event selection requires exactly 2 electrons with \qT$ > 7$ GeV and $|\eta|<2.7$, within a lepton-pair invariant mass window of $81$\,GeV $< m_{\ell\ell}< 101$\,GeV, in addition to exactly one undecayed Higgs candidate.
Jets are reconstructed and selected only to study their multiplicity and transverse momentum distribution, i.e.\ no event selection or veto is applied based on the jet activity. The reconstruction is done using the anti-$k_\text T$ clustering algorithm with a jet-radius parameter of $R=0.4$. The jet transverse momentum is required to be greater than $25$\,GeV, with $|\eta|<4.5$. Jet candidates are discarded if an electron is found within a cone of $\Delta R < 0.4$ around the jet axis. There is no requirement on the jet flavor.

\subsubsection{Total cross-sections}
\label{subsec:Higgs_ZH1j:xs}
We start by discussing the cross-section of the samples considered in this study. In Tab.~\ref{tab:Higgs_ZH1j:xs} we show the total cross-section obtained from the three setups, before and after the analysis selection detailed in Sec.~\ref{sec:Higgs_ZH1j:rivetresults}. The total cross-section is computed at LO in QCD in all cases, and it includes the branching ratio of the $Z$ boson decay to electrons. The uncertainty on the total cross-section comes from the 7-point ($\mu_R$, $\mu_F$) scale variations described in Sec.~\ref{sec:Higgs_ZH1j:MCsetup}. The total cross-section for the \Sherpa MEPS 0,1-jets setup includes a cut on the di-electron invariant mass of $m_{\ell\ell} > 66$\,GeV, to exclude the contribution of photon-mediated diagrams, which are not included in the other 0-jet LO calculations. The total cross-sections and their QCD uncertainties are well consistent with the $gg\rightarrow ZH$ calculations documented in the literature~\cite{deFlorian:2016spz}. The fiducial cross-sections show that the analysis selection has an acceptance of $85$--$90$\,\%, consistently among the three setups.

\begin{table}[t]
\begin{center}
\begin{tabular}{@{}llll@{}}
\toprule
    Cross-section [pb]        & \POWHEGplus+\Pythia & \Sherpa LO+PS &  \Sherpa MEPS 0,1j  \\ \midrule
total cross-section (before cuts) &0.001883(1)$^{+25\%}_{-19\%}$   &    0.001852(4)$^{+25\%}_{-19\%}$  & 0.001640(1)$^{+39\%}_{-26\%}$ \\
fiducial cross-section (after cuts)  &0.001575(3)  &    0.001661(3)  &   0.001384(1)  \\
\bottomrule
\end{tabular}
\end{center}
\caption[]{Total and fiducial $gg \rightarrow ZH$ cross-sections from the \POWHEGplus+\Pythia and \Sherpa 0-jet inclusive LO setups, and the \Sherpa MEPS 0,1-jets setup. The uncertainty quoted on the total cross-section is obtained from the 7-point ($\mu_R$, $\mu_F$) QCD scale variations.}
\label{tab:Higgs_ZH1j:xs} 
\end{table}

\subsubsection{Differential distributions}
\label{sec:Higgs_ZH1j:distributions}
\begin{figure}[t]
\begin{center}
\includegraphics[width=0.49\textwidth]{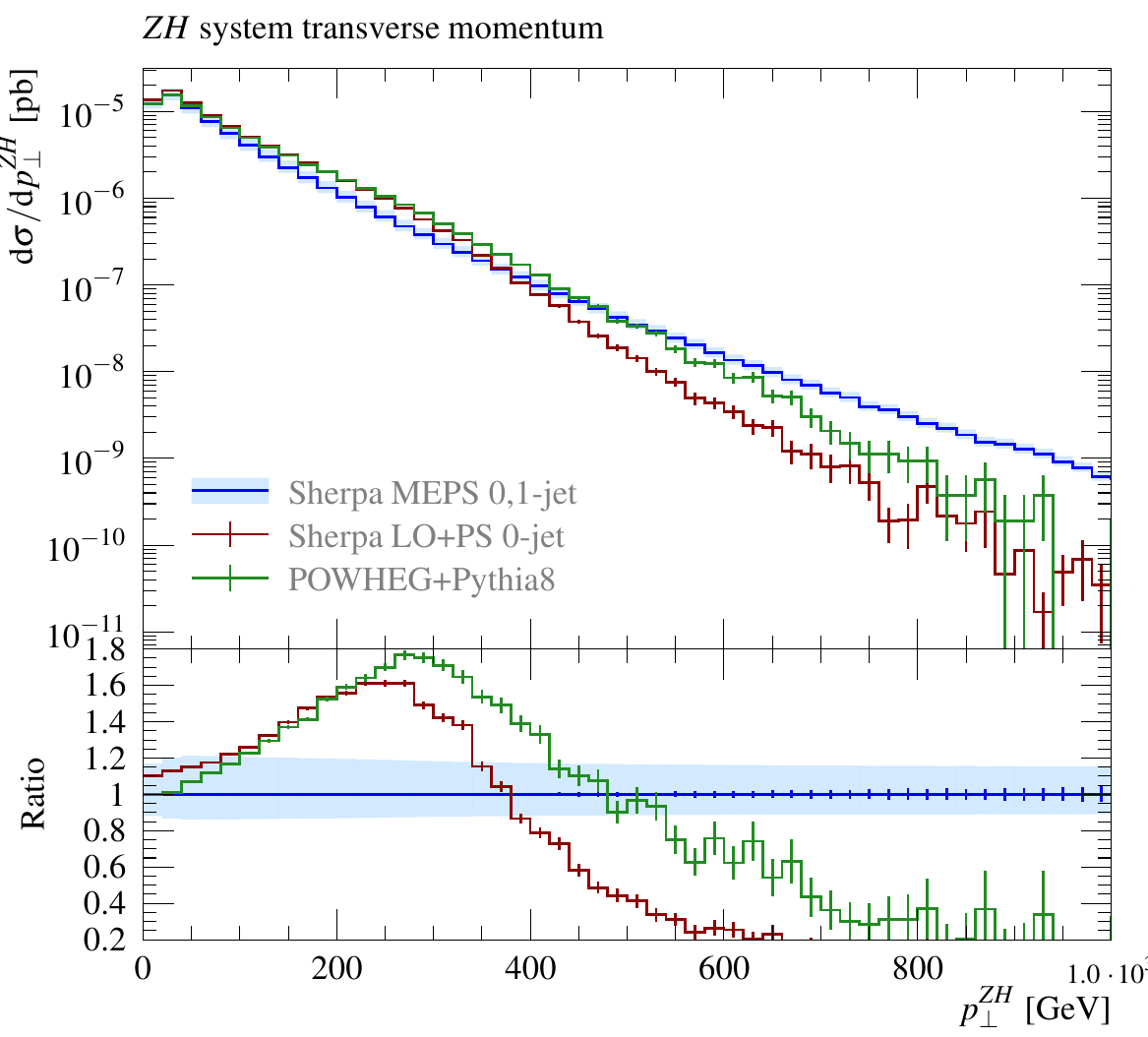}
\includegraphics[width=0.49\textwidth]{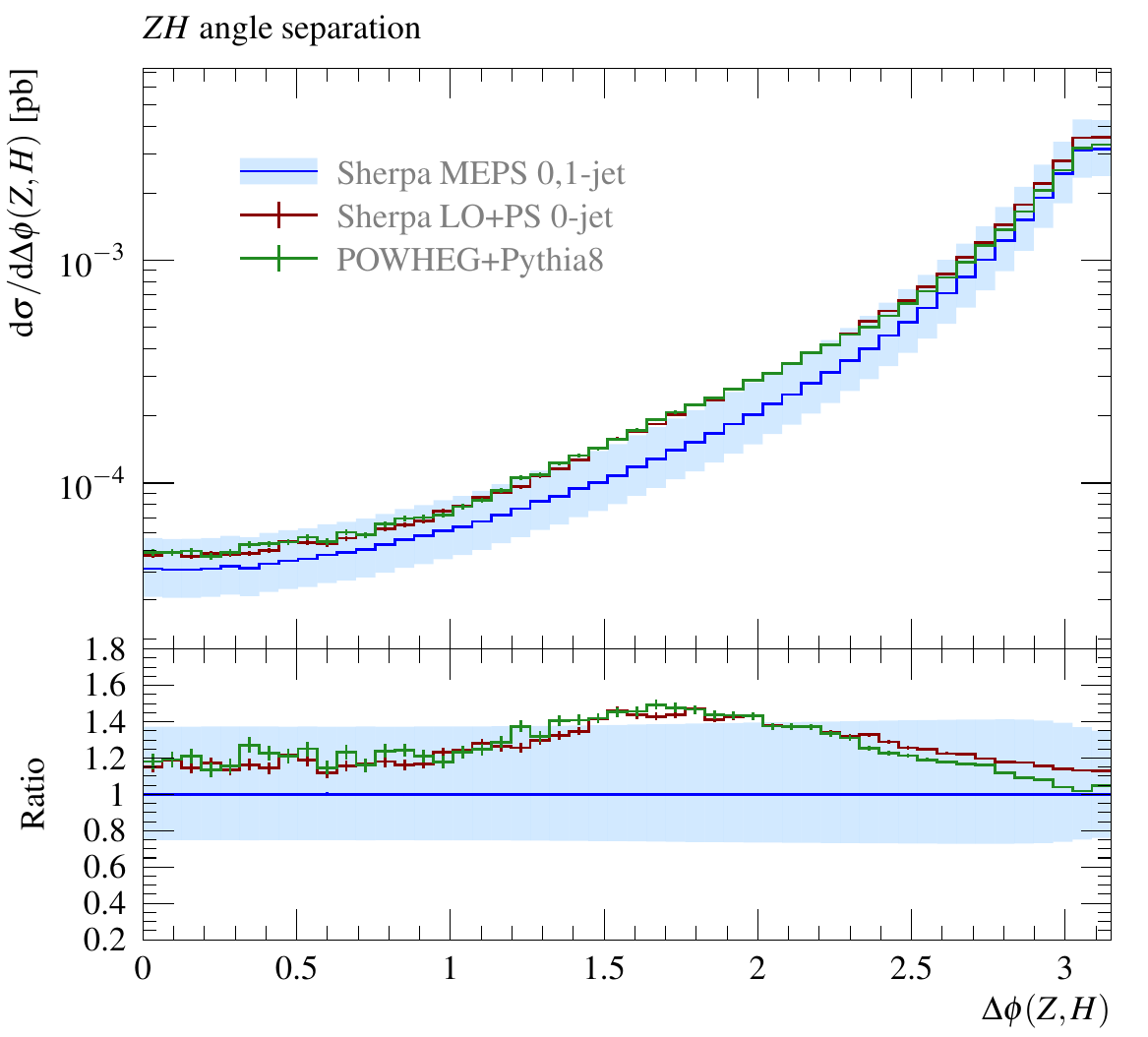}
\caption{Comparison of the 0-jet inclusive and 0,1-jets merged setups from \Sherpa and \POWHEGplus+\Pythia for the transverse momentum of the $ZH$ pair \pTZH, and the azimuthal separation between the bosons \DPhiVH. The blue error band represents the (\muR, \muF) QCD scale variations for the \Sherpa 0,1-jets sample.}
\label{fig:Higgs_ZH1j:plot1} 
\end{center}
\end{figure}

\begin{figure}[t]
\begin{center}
\includegraphics[width=0.49\textwidth]{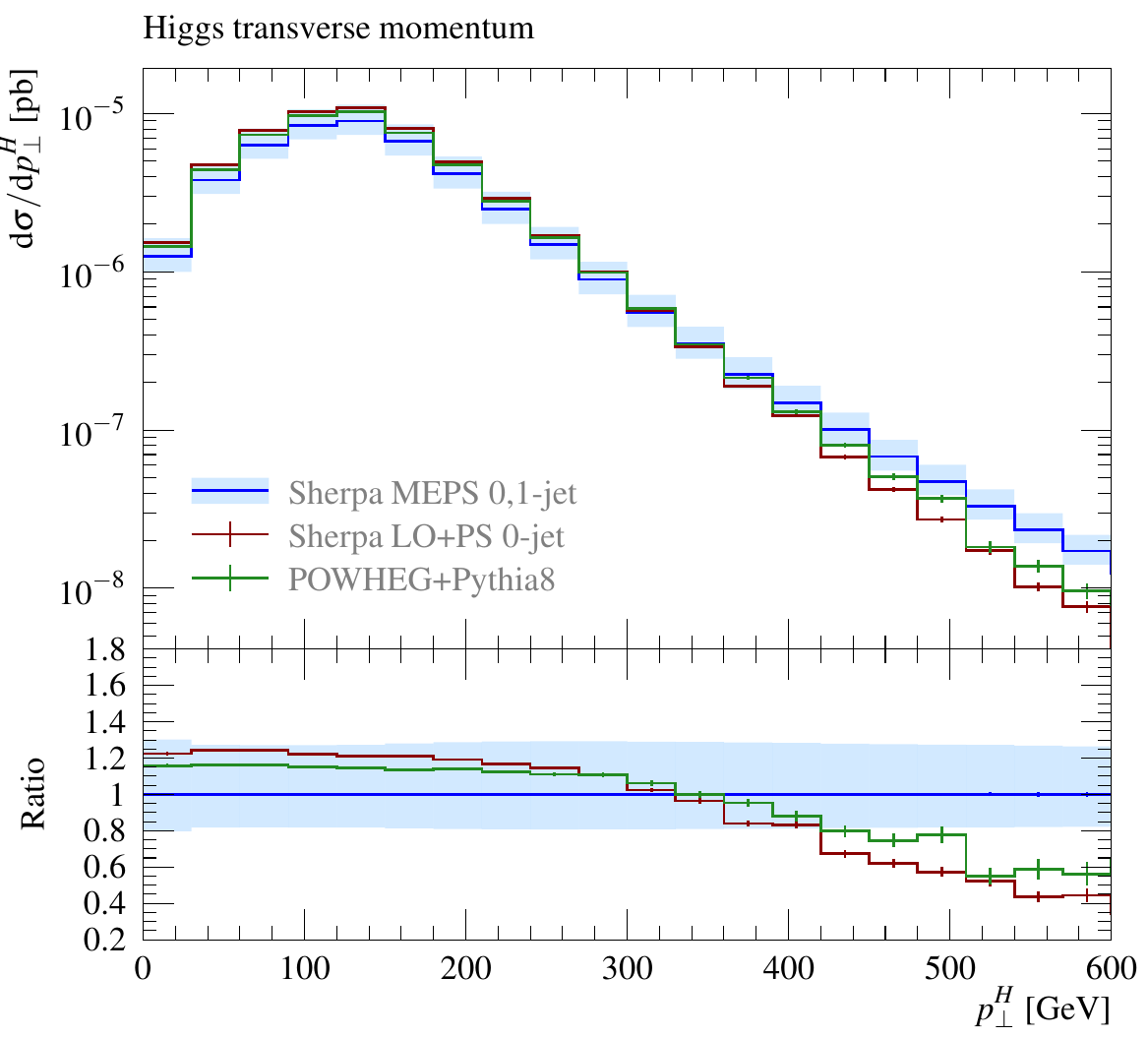}
\includegraphics[width=0.49\textwidth]{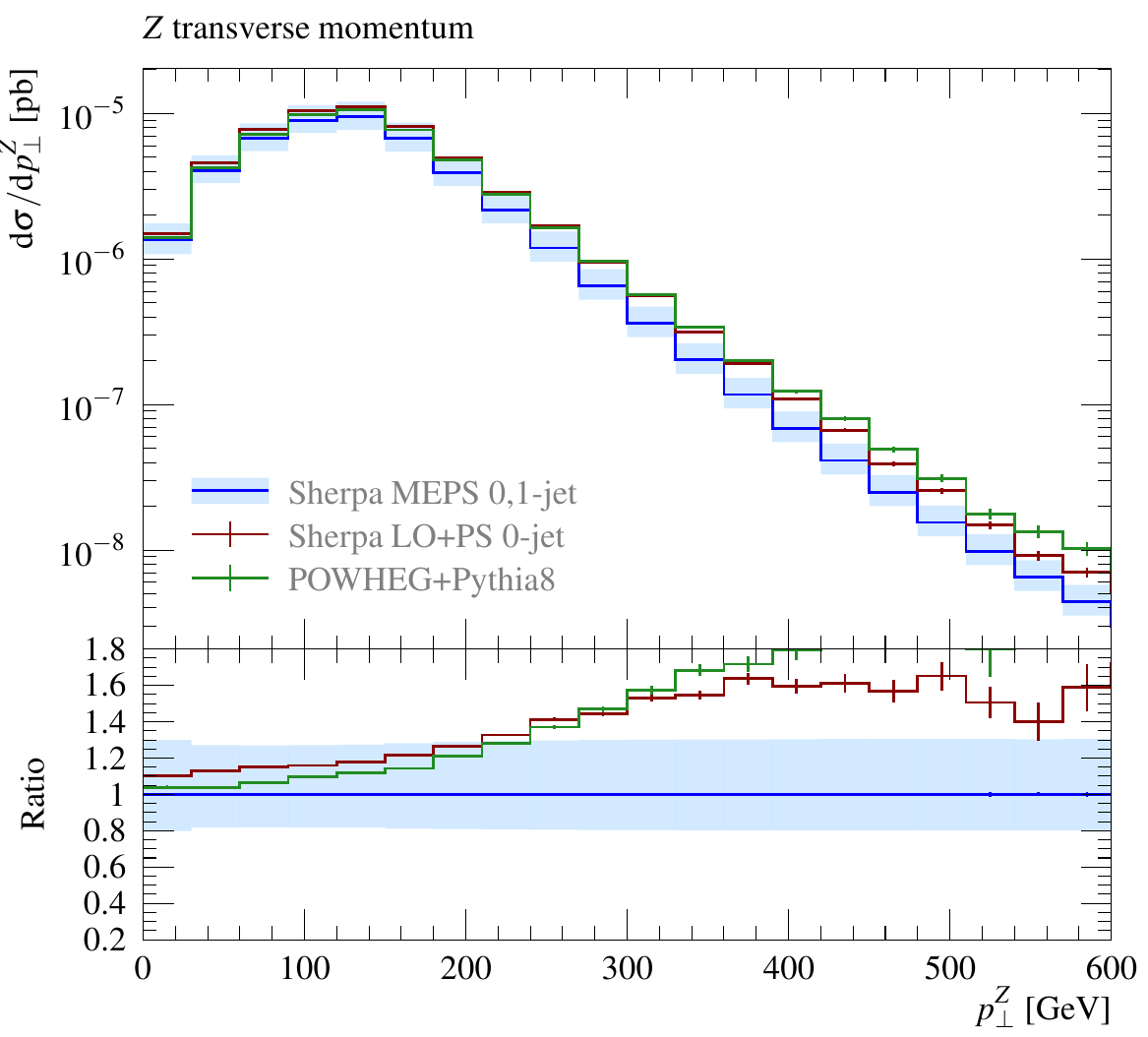}
\caption{Comparison of the 0-jet inclusive and 0,1-jets merged setups from \Sherpa and \POWHEGplus+\Pythia for the transverse momenta of the Higgs and the $Z$ boson, \pTH and \pTZ. The blue error band represents the (\muR, \muF) QCD scale variations for the \Sherpa samples.}
\label{fig:Higgs_ZH1j:plot2} 
\end{center}
\end{figure}

\begin{figure}[t]
\begin{center}
\includegraphics[width=0.49\textwidth]{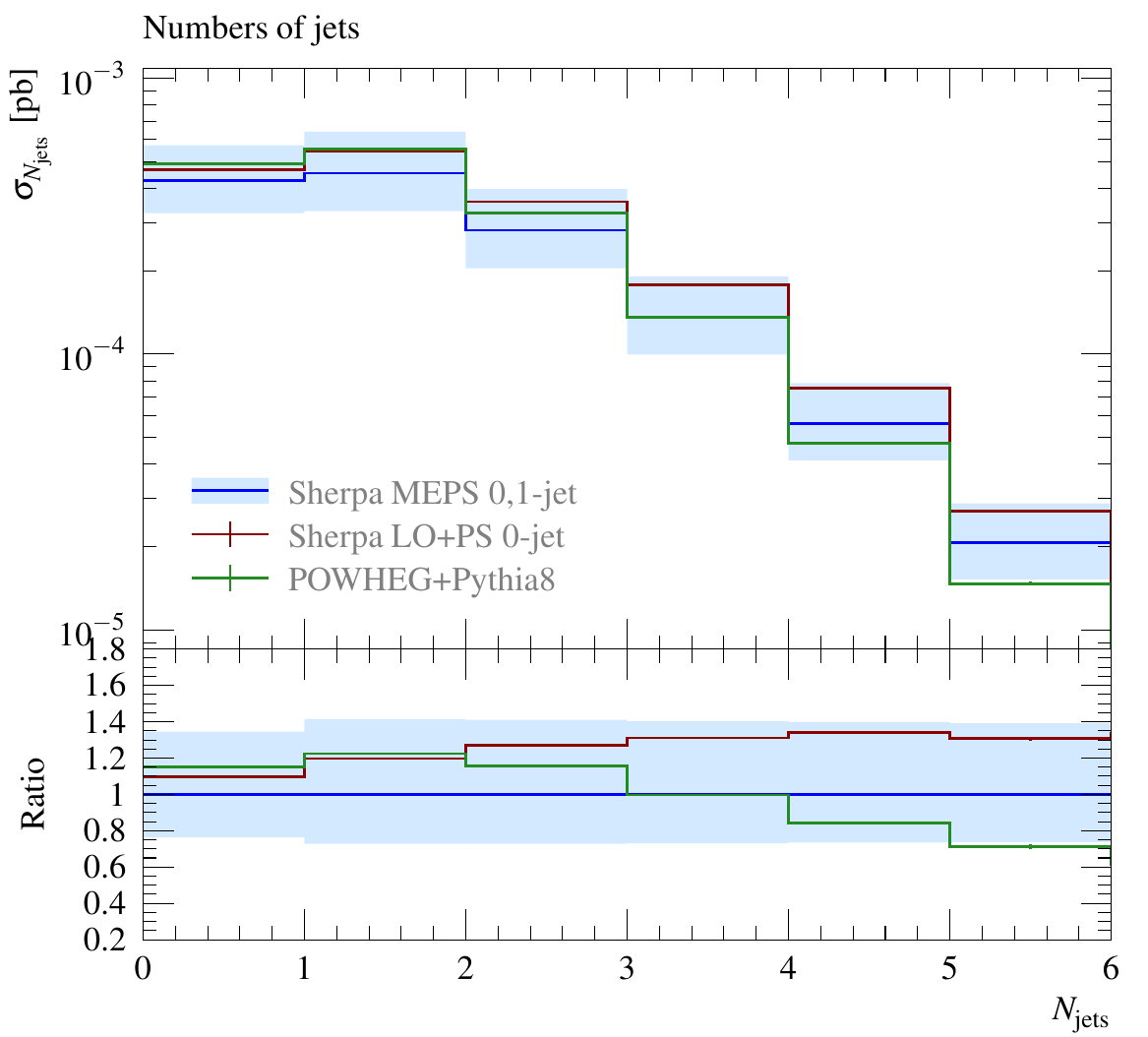}
\includegraphics[width=0.49\textwidth]{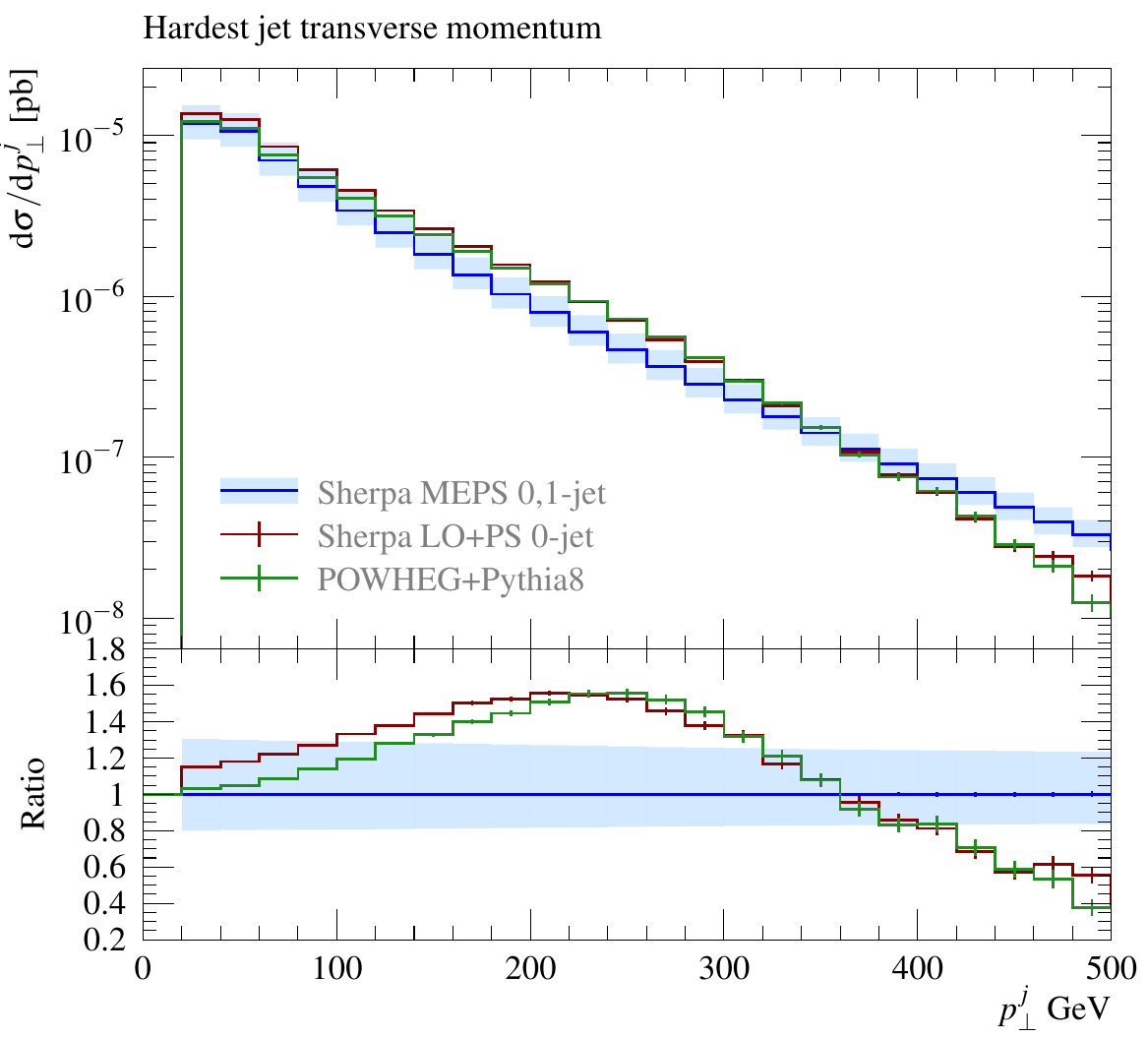}
\caption{Comparison of the 0-jet inclusive and 0,1-jets merged setups from \Sherpa and \POWHEGplus+\Pythia for the total number of hadronic jet \njet (in exclusive bins), and the transverse momentum of the leading jet \pTj. The blue error band represents the (\muR, \muF) QCD scale variations for the \Sherpa 0,1-jets sample.}
\label{fig:Higgs_ZH1j:plot3} 
\end{center}
\end{figure}

We focus on the study of a few key observables to highlight the impact of the 0,1-jets merged setup compared to the inclusive 0-jet LO+PS ones, namely:
\begin{itemize}
  \item \pTZH, the transverse momentum of the $ZH$ boson-pair,
  \item \DPhiVH, the azimuthal angular distance between the $Z$ and the Higgs boson candidates,
  \item \pTH and \pTZ, the transverse momenta of the Higgs boson and $Z$ boson candidates,
  \item \njet, the total number of jet with \qT$>25$ GeV and $|\eta|<4.5$ (in exclusive bins),
  \item \pTj, the transverse momentum of the leading jet.
\end{itemize}
Figures~\ref{fig:Higgs_ZH1j:plot1}--\ref{fig:Higgs_ZH1j:plot3} show the
comparison between the \Sherpa and \POWHEGplus+\Pythia samples for these variables. All distributions are normalized to the fiducial cross-section predicted by the respective tools and reported in Tab.~\ref{tab:Higgs_ZH1j:xs}. The uncertainty band from 7-point QCD scale variations is shown only for the \Sherpa 0,1-jets MEPS prediction. However, it is found to be very consistent for both 0-jet LO+PS samples. The QCD uncertainty is in all cases fairly flat across the differential distributions. We also note that the merging cut $Q_\text{cut}$ variations considered for the MEPS 0,1-jets setup have a negligible impact for the distributions under study, compared to the QCD scale variations, and are thus not displayed.
First, we notice that the \Sherpa and \POWHEGplus+\Pythia 0-jet LO+PS distributions show a reasonable level of agreement, with \POWHEGplus+\Pythia predicting slightly harder \qT spectra and a lower jet-multiplicity (note again that for these LO samples the emission of extra QCD radiation is only modeled by the parton-shower algorithms). Both 0-jet LO+PS predictions have very similar behavior when compared to the MEPS 0,1-jet \Sherpa setup.

In Fig.~\ref{fig:Higgs_ZH1j:plot1} we observe that the transverse momentum of the $ZH$ system \pTZH shows interesting features: up to \pTZH $\sim 400$\,GeV the 0-jet prediction is enhanced, while above this threshold the MEPS 0,1-jets sample predicts a harder \qT spectrum. Event topologies with hard QCD radiation recoiling against the $ZH$ system, which are modeled by the \Sherpa MEPS 0,1-jets matrix-element, become dominant in the high-\pTZH regime. Both \Sherpa and \POWHEGplus+\Pythia 0-jet LO+PS predictions fail to capture the harder tail of the $ZH$ \qT spectrum when compared to the MEPS 0,1-jets. 
Considering the azimuthal distance between the Higgs and $Z$ bosons \DPhiVH, we observe that the \Sherpa 0-jet LO+PS prediction seems to lead to slightly softer QCD radiation, resulting in an enhancement of event topologies where the Higgs and $Z$ bosons are produced back-to-back (\DPhiVH$\sim \pi$), while \POWHEGplus+\Pythia and \Sherpa MEPS 0,1-jets show reasonable agreement in this region. We also notice that the MEPS 0,1-jets setup does not predict a large enhancement of collinear $ZH$ topologies, which suggests that the recoil from the hard QCD emission modeled by the \Sherpa matrix-element is mainly captured by one of the bosons, with the other one remaining relatively soft.

This consideration is further supported by the Higgs and $Z$ transverse momentum distributions shown in Fig.~\ref{fig:Higgs_ZH1j:plot2}: we notice indeed that for the MEPS 0,1-jets setup the \pTH becomes harder above 350--400\,GeV, while an opposite behavior is found for the \pTZ, which is consistently softer compared to the 0-jet setups from both \Sherpa and \POWHEGplus+\Pythia. This supports the conclusion that when including the matrix-element description of extra QCD radiation, the dominant topology for \qT$ > 400$\,GeV includes a hard QCD jet mainly recoiling against the Higgs, with the production of a softer $Z$ boson. This behavior was also reported in~\cite{Hespel:2015zea}. We observe that both 0-jet LO+PS predictions do not properly model this feature.

In Fig.~\ref{fig:Higgs_ZH1j:plot3} we show the number of hadronic jets \njet, in exclusive bins, and the transverse momentum of the hardest selected jet \pTj: we notice that both 0-jet LO+PS tools are in good agreement in the 0- and 1-jet bins, with larger discrepancies for higher jet-multiplicities. We highlight that the \textit{exclusive} \njet distribution is by construction affected by the parton-shower modeling in all bins (including the 0-jet bin) for a LO+PS prediction. The \qT  of the extra QCD radiation shows a very similar behavior as the transverse momentum of the $ZH$ pair, as expected.

\subsubsection{Parton shower and matching variations}
\label{sec:Higgs_ZH1j:ps}
While being far from a robust definition of a `parton-shower uncertainty', we investigate the effect of the parton-shower setup and matching for the $gg \rightarrow ZH$ process. Relying on LO MC predictions we can expect parton-shower effects to be sizeable, since a large part of the phase-space will be populated by the shower algorithm.
In order to assess the effect of parton-shower variations for this process, we compare the alternative \POWHEGplus+\Pythia setups for wimpy and vetoed power shower introduced in Sec.~\ref{subsec:Higgs_ZH1j:PP8}, to the \Sherpa 0-jet inclusive LO sample with variations of the shower starting scale by a factor of $\sqrt{2}$ around its central value of $m_{ZH}$, introduced in Sec.~\ref{subsec:Higgs_ZH1j:SHERPA}. The \Sherpa MEPS 0,1-jets sample is not considered in this study.  We show this comparison for four observables, which capture interesting effects for this LO process: \pTZH, \pTj, \DPhiVH and \njet.
\begin{figure}[t]
\begin{center}
\includegraphics[width=0.49\textwidth]{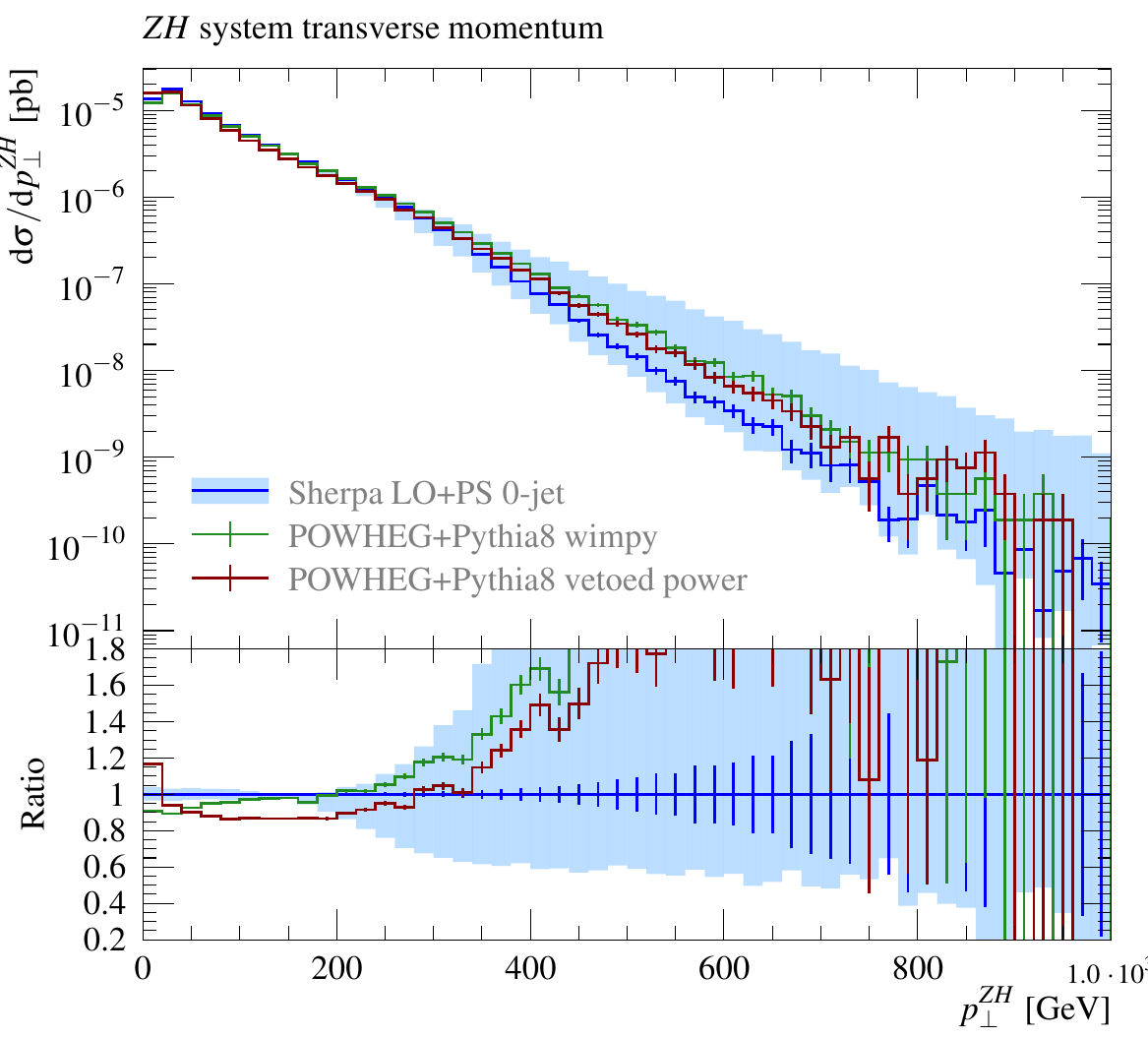}
\includegraphics[width=0.49\textwidth]{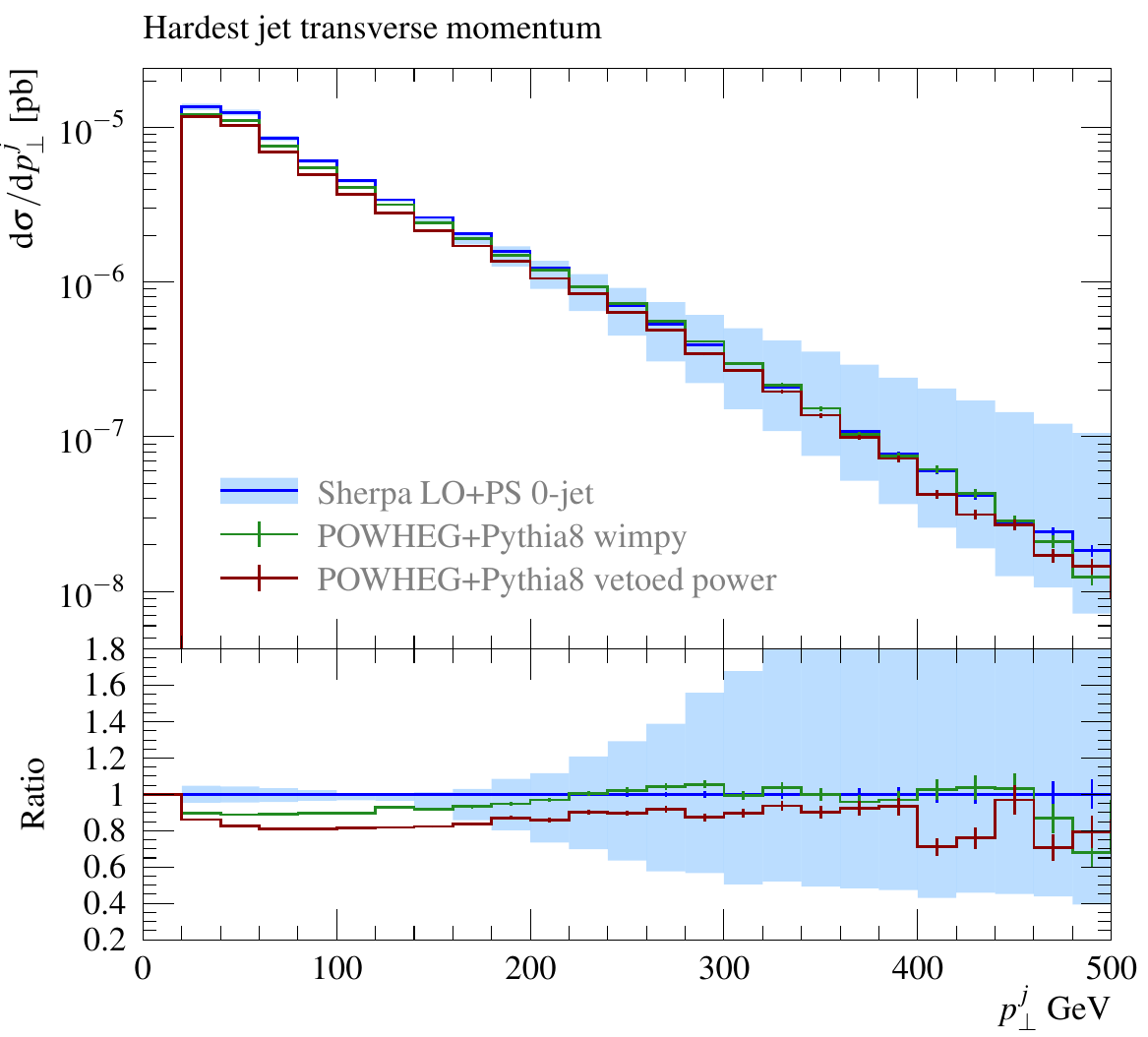}\\
\includegraphics[width=0.49\textwidth]{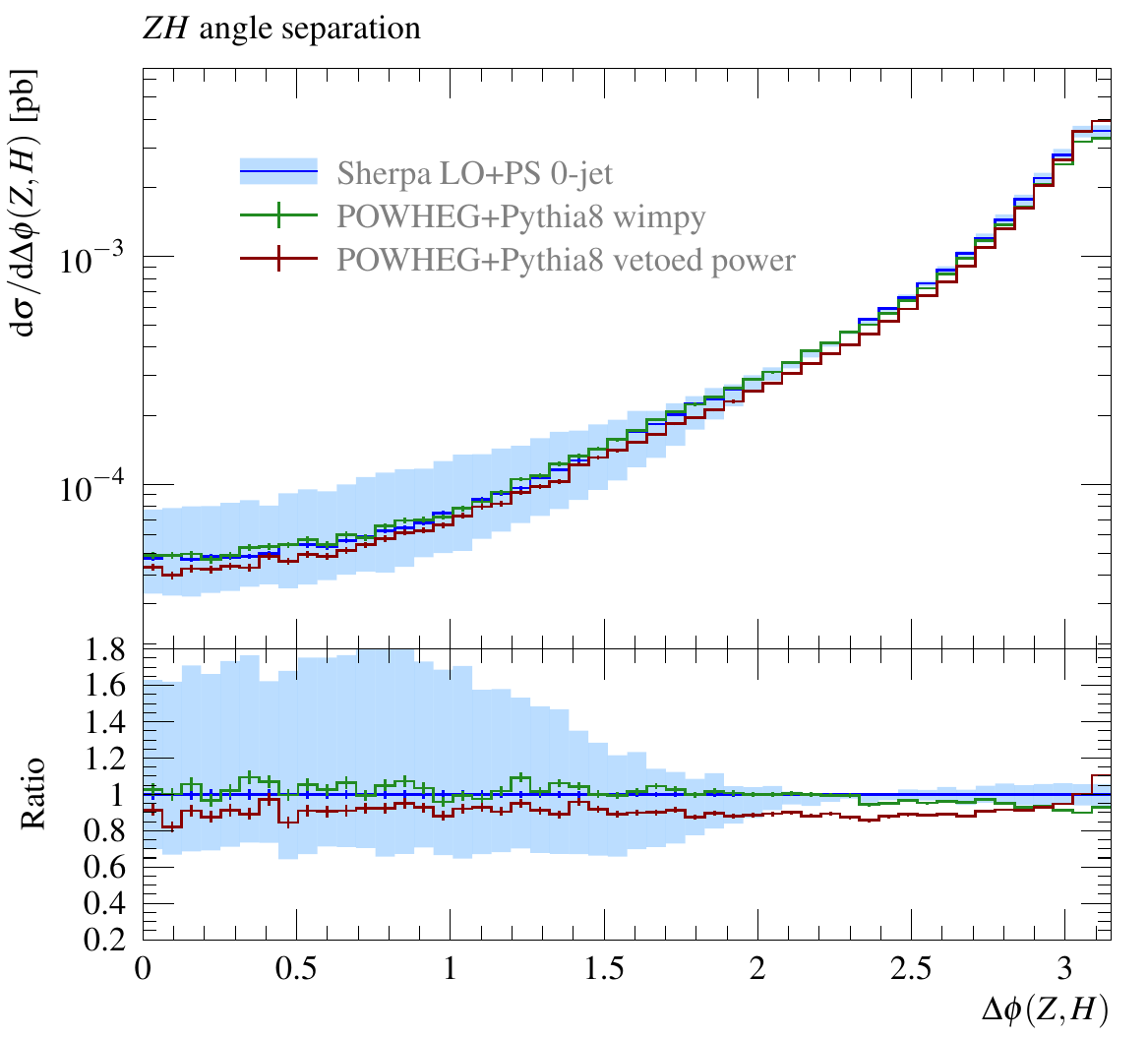}
\includegraphics[width=0.49\textwidth]{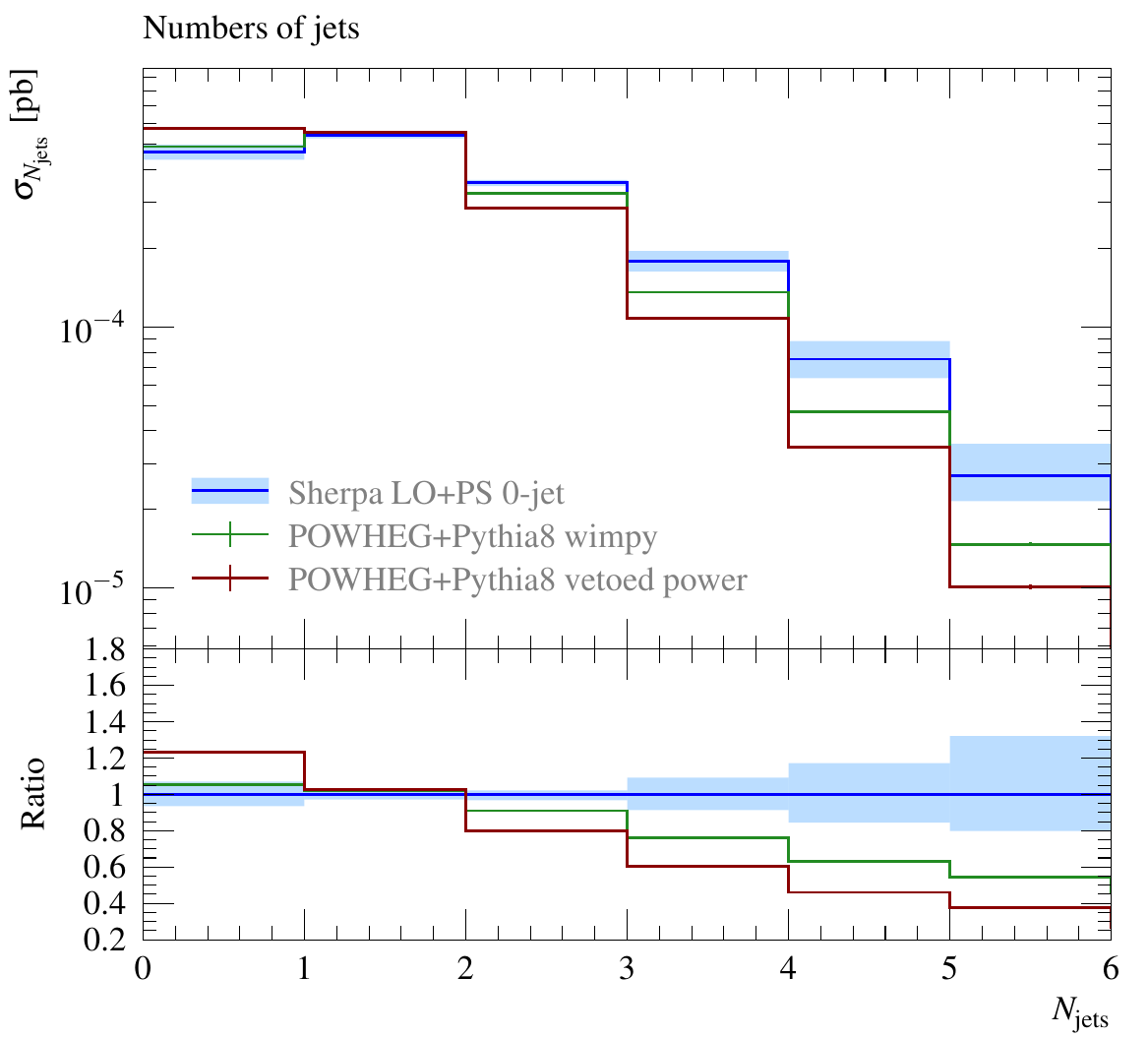}
\caption{Comparison of the 0-jet inclusive distribution for the \POWHEG+\Pythia8 wimpy and vetoed power shower setups and the \Sherpa prediction, with shower starting scale variations band shown in blue, for the transverse momentum of the $ZH$ pair \pTZH, the total number of hadronic jet \njet, and the transverse momentum of the leading jet \pTj.}
\label{fig:Higgs_ZH1j:plot4} 
\end{center}
\end{figure} 
From the transverse momentum distributions shown in Fig.~\ref{fig:Higgs_ZH1j:plot4} for the $ZH$ pair (top left) and the hardest QCD radiation (top right), we can observe how the shower starting scale variations on the \Sherpa prediction lead to a large effect when entering the regime dominated by hard QCD emission, for \qT larger than 200--300\,GeV. A similar behavior is observed for the azimuthal separation \DPhiVH (bottom left), with large shower starting scale variations away from the back-to-back $ZH$ peak.
As highlighted in Sec.~\ref{sec:Higgs_ZH1j:distributions} the 0-jet LO samples fully rely on the parton-shower in this region, and fail to reproduce the 0,1-jets MEPS prediction: the large sensitivity to the choice of shower starting scale further supports the conclusion that 0-jet LO predictions are not suited to model the $gg\rightarrow ZH$ process in this part of the phase space, and the need for a merged sample.	
We also note that the shower starting scale variation band, for \qT$> 300$--$400 $\,GeV,  becomes more important than the QCD perturbative uncertainty estimated from ($\mu_R$, $\mu_F$) scale variations (shown in Fig.~\ref{fig:Higgs_ZH1j:plot1}) which is relatively flat across the \qT spectrum.

From the comparison of the \POWHEGplus+\Pythia wimpy and  vetoed power shower algorithms we do not observe striking differences: the wimpy shower prediction is consistently slightly harder for the \qT distributions and features larger jet multiplicities. We observe that differences between the \POWHEGplus+\Pythia matching algorithms are found to be sub-dominant compared to the shower scale variation effect studied in \Sherpa. We note that the discrepancy between the \POWHEGplus+\Pythia and \Sherpa 0-jet LO \pTZH predictions is covered by the shower starting scale variation band, while this is not the case for the \njet distribution, where scale variations have a more modest effect.  
In conclusion the study of the impact of parton-shower variations for the $gg\rightarrow ZH$ process further supports the importance of a merged 0,1-jets prediction for the modeling of the high \qT regime. This observation, together with the impact of the 0,1-jets MEPS prediction on the size of shower starting scale variation, is confirmed in~\cite{Hespel:2015zea}.

\subsection{Conclusions}
\label{sec:Higgs_ZH1j:conclusions}
We present a study of the MC modeling of the $gg \rightarrow ZH$ process in the typical regions explored by the experimental analyses at the LHC, highlighting its characteristic features from the total cross-section to some of the main differential observables. We focus on the comparison between different LO+PS tools with respect to the improved merged MEPS 0,1-jets prediction from the \Sherpa generator. We observe that in the high transverse momentum regime (\qT$\gtrsim 300$\,GeV) the inclusion of $2\rightarrow 3$ matrix elements in the MEPS setup leads to a more accurate modeling: interestingly we note that the dominant event topology in this region includes a hard QCD radiation recoiling against an high-\qT Higgs, with a softer $Z$ emitted at high \DPhiVH angle. This topology is not properly modeled by the 0-jet LO+PS tools considered. The \qT asymmetry between the Higgs and $Z$ boson is a distinctive feature which might be considered for future studies to provide a better characterization of $gg \rightarrow ZH$ processes. From the study of parton-shower variations we observe a large sensitivity to the choice of shower starting scale for the 0-jet LO+PS setup, which further supports the choice of a more accurate MEPS 0,1-jets prediction, expected to strongly reduce the shower starting scale dependence.
While the study of different predictions from \POWHEGplus+\Pythia and \Sherpa provides robustness to these results, we remark that the comparison to an alternative 0,1-jets merged tool (for instance the \madgraphNLO prediction~\cite{Hespel:2015zea}) would provide more insight on the feature of the $gg \rightarrow ZH$ $2\rightarrow 3$ matrix elements, and we leave this for further studies.

\subsection{Acknowledgements}
CP acknowledges support by the CERN EP Department.
\label{sec:Higgs_ZH1j:acknowledgements}

\let\Herwig\undefined
\let\Pythia\undefined
\let\POWHEGplus\undefined
\let\POWHEG\undefined
\let\Sherpa\undefined
\let\Openloops\undefined
\let\Rivet\undefined
\let\madgraphNLO\undefined
\let\Professor\undefined
\let\eps\undefined
\let\mc\undefined
\let\mr\undefined
\let\mb\undefined
\let\tm\undefined
\let\qT\undefined
\let\pTH\undefined
\let\pTZ\undefined
\let\pTZH\undefined
\let\pTj\undefined
\let\njet\undefined
\let\DPhiVH\undefined
\let\muR\undefined
\let\muF\undefined

\newcommand{\NNLOJET}{NNLO\protect\scalebox{0.8}{JET}\xspace}

\section{A Comparative study of VBF Higgs boson production~\protect\footnote{
    A.~Buckley,
    X.~Chen,
    J.~Cruz-Martinez,
    S.~Ferrario~Ravasio,
    T.~Gehrmann,
    E.~W.~N.~Glover,
    A.~Huss,
    J.~Huston,
    C.~Oleari,
    S. Pl\"atzer,
    M.~Sch\"onherr
  }{}}

\label{sec:MC_VBF}





Vector boson fusion (VBF) is one of the crucial production channels for the Higgs boson at the LHC, and allows the determination of the Higgs boson couplings to gauge bosons. The VBF cross section is currently known experimentally to the order of 20-25\%. At high $p_T$, the VBF predicted cross section is roughly half of that from gluon-gluon fusion (ggF). Unlike gluon-gluon fusion, though, VBF production results in two jets in the final state, and both jets are quark jets. A study preformed in the context of LH17~\cite{Bellm:2019yyh} carried out a comparison of fixed-order matrix element and matrix-element-plus-parton-shower predictions for ggF Higgs boson production (along with studies of Z-boson-plus-jet and dijet production), examining the cross section dependence on jet radius. 

Here, we extend these investigations by starting a similar study of VBF Higgs boson production, a process missing in the original study. At this moment, the comparisons are yet incomplete, but we summarize the plan of the study, and present a few of the early results from the fixed order predictions. 

\subsection{Setup}
To provide theoretical guidelines for experimental analysis and simplified template cross sections (STXS), we adopt the same jet definition as in ATLAS measurements~\cite{Buhrer:2019npn,ATLAS:2019jst} and study differential observables using fiducial bins suggested by STXS stage 1.1~\cite{Berger:2019wnu}. We study on-shell Higgs boson produced from the vector boson fusion production channel and require at least two accompanying anti-$k_T$ jets each satisfying the following conditions:
\begin{center}
\begin{equation}
p_T^{\text{jet}} > 30\ \text{GeV}, \qquad  |y^{\text{jet}}|<4.4.
\end{equation}
\end{center}
We further vary the anti-$k_T$ jet radius from 0.3 to 1.0 with steps of 0.1 to study its impact on differential observables. The electroweak parameters are defined in the G$_\mu$ scheme with the gauge boson masses and widths set to:
\begin{align}
    m_{W}=80.379 \text{ GeV}, && \Gamma_{W}=2.085 \text{ GeV}. \\
    m_{Z}=91.188 \text{ GeV}, && \Gamma_{Z}=2.495 \text{ GeV}.
\end{align}
The value of $\alpha_{em}^{G_\mu}$ is 1/132.233.

The theoretical uncertainties are estimated by varying QCD renormalisation ($\mu_R$) and factorisation ($\mu_F$) scales independently by a factor of two around the central scale of $\mu=H_T^{\text{Parton}}/2$ while eliminating the two extreme combinations of $(\mu_R,\mu_F)=(\mu/2,2\mu)$ and $(\mu_R,\mu_F)=(2\mu,\mu/2)$. This is the so-called 7-point scale variation. PDF4LHC15\_30 PDFs are used, as in the LH17 study. The fixed-order results are nominally from NNLOJET, but cross sections for the ME+PS predictions will be cross-checked at fixed-order NLO. 
Comparisons of NNLOJET with SHERPA, POWHEG and HERWIG are being carried out, although no results for the latter are available at the time of these proceedings. 

\subsubsection{\NNLOJET}
The parton-level fixed-order predictions are calculated using the \NNLOJET\ package including up to NNLO QCD corrections~\cite{Cruz-Martinez:2018rod}. We use the antenna subtraction formalism~\cite{GehrmannDeRidder:2005cm,GehrmannDeRidder:2005aw,GehrmannDeRidder:2005hi,Daleo:2006xa,Daleo:2009yj,Gehrmann:2011wi,Boughezal:2010mc,GehrmannDeRidder:2012ja,Currie:2013vh} to regulate IR divergences at each stage of the fixed-order calculations and to provide fully differential predictions. The calculation is performed under the structure function approximation~\cite{Han:1992hr}, which is exact at LO and NLO, while missing non-factorisable contributions at NNLO. The \NNLOJET\ results agree with an independent earlier calculation~\cite{Cacciari:2015jma} of the NNLO QCD corrections, using the same approximations. A recent study~\cite{Liu:2019tuy} using the eikonal approximation estimates the non-factorisable contributions to be less than 2\% (with respect to LO) for differential observables. For each fixed-order predictions, we use the same PDFs (PDF4LHC15\_30) with NNLO accuracy.
\subsection{Results}
\begin{figure}[t]
\centering
\includegraphics[scale=0.5]{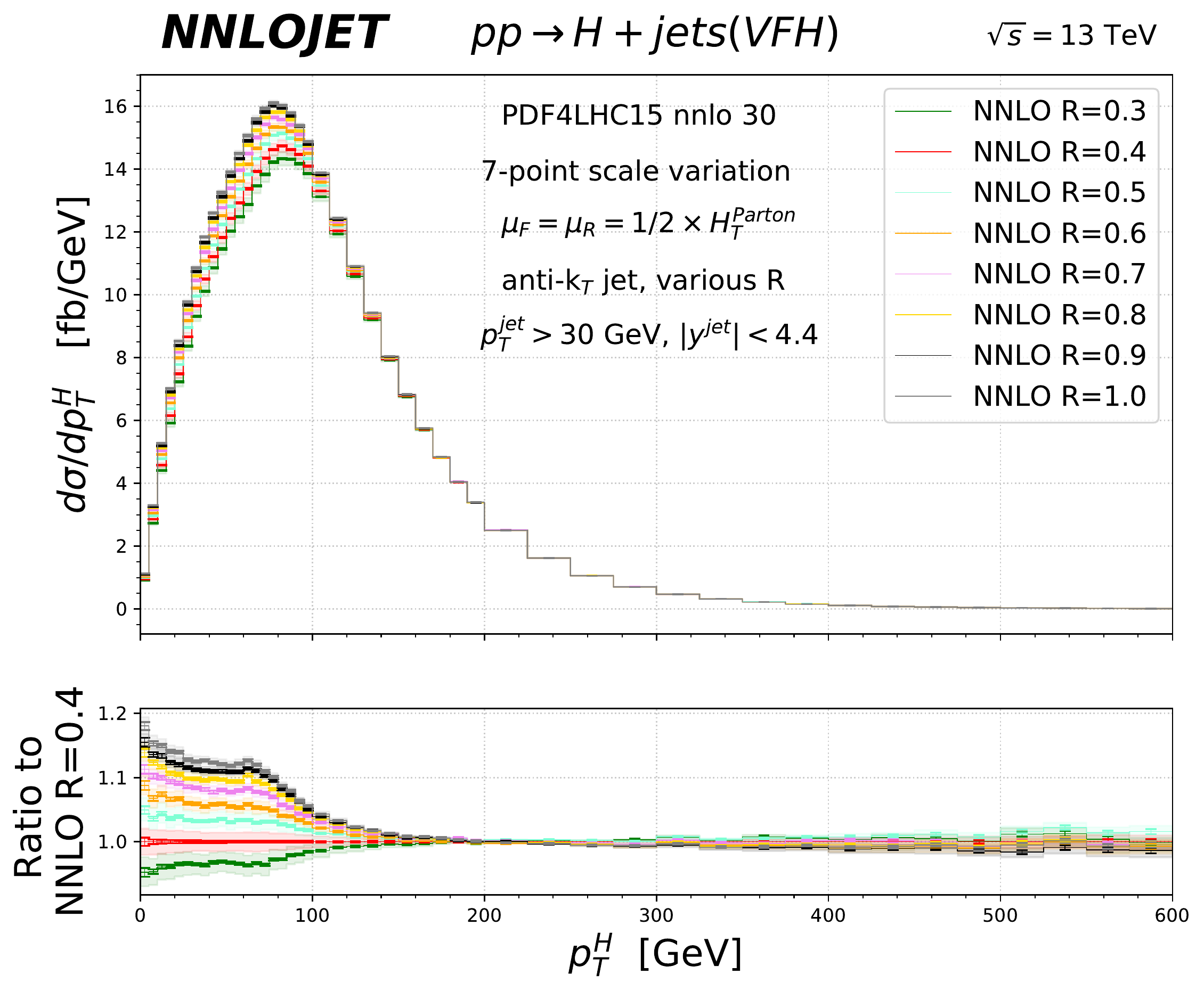}
\caption{The Higgs boson transverse momentum distribution from the VBF sub-process at NNLO as a function of jet radius. }
\label{fig:MC_VBF:Higgs_pT}
\end{figure}

As a result of two jets being present in the Born-level final state, there can be interesting R-dependent effects, even for relatively inclusive observables. For example, the Higgs boson $p_T$ distribution (from VBF production at NNLO) is shown in Figure~\ref{fig:MC_VBF:Higgs_pT} as a function of the jet radius, normalized to the result for the LHC standard jet radius of R=0.4. For $p_T^{Higgs}\ge 150$ GeV, there is little dependence of the cross section on the jet size. However, there is a sizeable variation of the cross section for lower transverse momentum values of the Higgs boson. This is due to the requirement that there be at least two jets, each with $p_T \ge 30$ GeV in the event. Thus, as the jet radius increases, combination of partons into jets becomes more likely, leading to the enhancements (or reduction for R=0.3) shown. There is a shoulder for $p_T^{Higgs}$ on the order of 70 GeV, after which all ratios converge to unity. A similar behavior is observed at NLO.  At very high $p_T$, where the Higgs boson is recoiling against the di-jet system, the cross section starts to decrease with increasing jet radius, due to the increasing probability of the two jets being reconstructed as a single jet. 

\subsection{Conclusion}

Programs involving the matching and merging of matrix elements and parton showers are most often used for comparisons to LHC data, and indeed provide a complete simulation of the final state. The highest precision, however, comes from fixed order predictions at NNLO. It is thus useful to carry out detailed comparisons for VBF Higgs boson production to understand any differences between fixed order predictions and those from ME+PS production, and any differences that might arise among ME+PS predictions. Such a study is currently in the process of being carried out. 

\subsection*{Acknowledgements}
JCM is supported by the European Research Council under the European
Union’s Horizon 2020 research and innovation Programme (grant
agreement ERC-AdG-740006).

\let\NNLOJET\undefined

\newcommand{\Herwig}{H\protect\scalebox{0.8}{ERWIG}\xspace}
\newcommand{\Pythia}{P\protect\scalebox{0.8}{YTHIA}\xspace}
\newcommand{\POWHEGplus}{P\protect\scalebox{0.8}{OWHEG}}
\newcommand{\POWHEG}{P\protect\scalebox{0.8}{OWHEG}\xspace}
\newcommand{\MCatNLO}{\protect\scalebox{0.8}{MC@NLO}\xspace}
\newcommand{\POWHEGBOX}{P\protect\scalebox{0.8}{OWHEG-BOX}\xspace}
\newcommand{\Rivet}{R\protect\scalebox{0.8}{IVET}\xspace}

\newcommand{\muR}{$\mu_\text{R}$\xspace}
\newcommand{\muF}{$\mu_\text{F}$\xspace}

\newcommand{\vNLOscales}{{\tt NLO-7-pts\xspace}}
\newcommand{\vhdamp}{{\tt hdamp-3-pts\xspace}}

\newcommand{\vISR}{{\tt isr-3-pts\xspace}}
\newcommand{\vFSR}{{\tt fsr-3-pts\xspace}}
\newcommand{\vkernels}{{\tt Pythia-AP-Kernels\xspace}}

\newcommand{\vrecoildef}{{\tt Pythia-default-recoil\xspace}}
\newcommand{\vrecoilD}{{\tt SpaceShower:dipoleRecoil=on\xspace}}
\newcommand{\vrecoilG}{{\tt TimeShower:globalRecoil=on\xspace}}
\newcommand{\vrecoilCoff}{{\tt TimeShower:recoilToColoured=off\xspace}}



\section{A study of perturbative uncertainties in top pair production at NLO+PS~\protect\footnote{
    S.~Amoroso, E.~Re }{}}

\label{sec:MC_ttbar}


\subsection{Introduction}
\label{sec:MC_ttbar:introduction}

Parton showers (PS) are at the core of Monte-Carlo event
generators. They are the algorithms that allow to turn
low-multiplicity parton-level computations into fully differential
predictions where multiple emissions off quarks and gluons are
generated according to an all-order and fully-differential resummation
of logarithms of soft/collinear origin. In this respect, therefore,
simulations based on PS algorithms are computations performed starting
from first principle QCD. Nevertheless, although progress is being
made\cite{Nagy:2007ty,Li:2016yez,Hoche:2017hno,Hoeche:2017jsi,Dasgupta:2018nvj,Bewick:2019rbu},\footnote{A new proposal for a next-to-leading logarithmic parton shower appeared while this document was finalized~\cite{Dasgupta:2020fwr}.}
the logarithmic accuracy of PS is limited, and often difficult to
assess formally.  This limited accuracy, together with the available
choices that are a priori allowed in a given parton shower algorithm
of limited logarithmic accuracy, is becoming an issue for precision
Physics at the LHC, especially if contrasted with the typical accuracy
(NLO and, at times, NNLO in QCD) with which short distance partonic
cross sections can be consistently included in modern event
generators, thanks to matching and merging methods.

Until recently, in experimental analysis at the LHC,
uncertainties related to PS have been typically estimated through a
mere comparison of two PS algorithms (typically \Pythia{}
vs. \Herwig{}). Although so far this has been acceptable, nowadays the
community agrees that such a simple approach has become a bottleneck,
and more consistent procedures need to be established. 

Estimating in a fully consistent way the PS ``uncertainty'' in
NLO+PS simulations is a very ambitious task, which goes well beyond
the aim of this study, although important steps have
been made~
\cite{Hoeche:2014lxa,Bellm:2016rhh,Bellm:2016voq,Mrenna:2016sih,Bothmann:2016nao,Cormier:2018tog,Ravasio:2018lzi,FerrarioRavasio:2019vmq}. Our original goal was to perform a study similar
to the one completed during the 2015 edition of the Les Houches
workshop (see chapter V.1 of Ref.~\cite{Badger:2016bpw}), but using
NLO+PS-accurate tools and modern frameworks to perform
``PS-reweighting'' efficiently. More precisely, we wanted to compare
results for different NLO+PS accurate generators (and possibly for
different matching methods, i.e. \MCatNLO{}-type vs. \POWHEG{}-type),
the aim being of establishing if, for the main ``variations'' of
perturbative nature available in different PS algorithms, the results
obtained with different generators are mutually compatible, at least
for observables that should only be affected by perturbative
effects. Establishing whether this is or not the case would be rather
important: it would represent a first step forward towards the main
goal, that is, the establishment of a procedure to assess the
uncertainty related to parton showers.

In this contribution, though, due to lack of time and resources, we
only limit ourselves to show the impact that different choices within
a given PS algorithm have on differential distributions, without
making any assumption on correlating among them some of these
choices.\footnote{In NLO+PS simulations, it is reasonable to expect
  that there should be a natural correlation among the variation of
  the renormalization and factorization scales in the PS evolution,
  and the variation of the same scales used to generate the hardest
  radiation (as, for instance, through the \POWHEG{} Sudakov form
  factor). In this study we don't consider any such constraints.}  We
have considered the \POWHEG{}+\Pythia{8} setup, and looked at the
production of a top pair at the LHC. We restrict our discussion to the
purely perturbative part of the event generation, i.e. we don't
include hadronization effects nor effects due to Multiple Particle
Interactions (MPI).  We have mostly focused on observables that are
expected to depend mostly on the top-pair production dynamic, and have
a minimal sensitivity to the simulation and modeling of the top-quark
decay, although we also include an observable that depends on the
modeling of the top decay.

In spite of the very limited scope of this study with respect to our
original goal, and of the fact that we are aware that not all the
variations we explored are necessarily fully consistent with an NLO+PS
simulation, we decided to document some of our findings. We present
our results in the following, in the hope that they can be used as a
starting point for more detailed and more consistent studies.

\subsection{Description of setup and parameters}

We generate partonic $t\bar t$ events in $pp$ collisions at
$\sqrt{S}=13$ TeV, using the {\tt hvq}
implementation~\cite{Frixione:2007nw} in the \POWHEGBOX{}-V2
framework~\cite{Nason:2004rx,Frixione:2007vw,Alioli:2010xd}, and
complete the NLO+PS simulation by showering these events with
\Pythia{8.301}~\cite{Sjostrand:2014zea}, switching off hadronization and MPI effects.  We set
$m_t=172.5$ GeV, as partonic distribution functions we use the set
       {\tt MMHT2014nnlo68cl}~\cite{Harland-Lang:2014zoa}, whereas in
       \Pythia{8} the set {\tt NNPDF23LO} is used, in order to be
       consistent with the {\tt Monash} tune~\cite{Skands:2014pea}.
       The central value for renormalization and factorization scales
       used in the hard scattering (i.e. in the \POWHEG{} $\bar{B}$
       function) is equal to
\begin{equation}
  \label{eq:MC_ttbar:murmuf}
  \mu_\text{F}=\mu_\text{R}=\sqrt{p_{T,t}^2 + m_t^2}.
\end{equation}

We let both top quarks decay leptonically, although we will mostly
focus on observables computed using the kinematics of top quarks
before the decay. In the following we list all the variations we
performed.

\begin{itemize}
\item In the hard scattering NLO cross section, we perform a 7-point
  variations of the renormalization and factorization scales about the
  central value of Eq.~(\ref{eq:MC_ttbar:murmuf}), i.e. we vary both
  scales independently by factors of two and one half, but omitting
  those where the two scale variation factors differ by a factor of 4.

  We label this variation as \vNLOscales{}.\\

\item We split the real emission contribution in the \POWHEG{} formula
  into a singular ($R_s$) and a finite part ($R_f$) according to the
  so-called {\tt hdamp} factor~\cite{Nason:2004rx,Alioli:2008tz}:
\begin{equation}
  \label{eq:MC_ttbar:hdamp}
  R = R_s + R_f = [ R\ h(p_{T,t\bar t}) ] + [ R\ (1- h(p_{T,t\bar t}))],
\end{equation}
where
$$
h(p_T) = \frac{{\tt hdamp}^2}{p_T^2 + {\tt hdamp}^2}\,.
$$

We pick as nominal value {\tt hdamp}=$1.5 m_t$, which seems to be the
preferred value that ATLAS and CMS have found when comparing \POWHEG{}
NLO+PS predictions against data, and we vary this value by a factor
two and a half, i.e. {\tt hdamp}=$3 m_t$ and {\tt hdamp}=$0.75
m_t$. This interval of variations is probably slightly excessive, but
finding an optimal or recommended variation is not the purpose of this
study, hence we refrain to make more ad-hoc choices.

We label this variation as \vhdamp{}.\\

\item The strength of the radiation in~\Pythia{} is governed by the
  value of the strong coupling at the evolution scale
  $\alpha_s(p_{T,{\rm evol}})$. By default, the evolution scale
  $p_{T,{\rm evol}}$ corresponds to the transverse momentum of the
  emission, both for initial- and final-state radiation (ISR and FSR,
  respectively). The exact expressions of $p_{T,{\rm evol}}$ are given
  for instance in Eq.~(15) of Ref.~\cite{Sjostrand:2014zea}. In order to show
  the dependence of our results on this perturbative aspect of the
  \Pythia{8} parton showering algorithm, we vary these renormalization
  scales by a prefactor 1/2 and 2 for all the initial and final-state
  emissions, using the ``automated variations'' procedure introduced
  in Ref.~\cite{Mrenna:2016sih}.  The parameters used to vary the
  renormalization scales in \Pythia{8}, and the values chosen, are as
  follows:
  \begin{eqnarray}
    \mbox{\tt isr:muRfac} &=& \{1/2,1,2\}\nonumber\\
    \mbox{\tt fsr:muRfac} &=& \{1/2,1,2\}\nonumber\,.
  \end{eqnarray}
  We label these variations as \vISR{} and \vFSR{}.  We also recall
  that the hardest ISR emission is not directly concerned by these
  variations, as it is generated with \POWHEG{}, and the
  renormalization and factorization scales are not varied in the
  \POWHEG{} Sudakov form factor. We don't vary the factorization scale
  used to evaluate PDFs in the PS, although this could be done in
  \Pythia{8} through the parameters \verb|isr:pdf:minus| and
  \verb|isr:pdf:plus|.\\
  
\item In recent \Pythia{8} versions it is possible to capture
  uncertainties related to variations of the non-singular parts of the
  shower kernels. We refer to Ref.~\cite{Mrenna:2016sih} for a detailed
  description of the rationale behind these variations and the details
  of the implementation. We vary the parameters {\tt isr:cNS} and {\tt
    fsr:cNS} about their default values as follows:
  \begin{eqnarray}
    \mbox{\tt isr:cNS} &=& \{-2,2\}\nonumber\\
    \mbox{\tt fsr:cNS} &=& \{-2,2\}\nonumber\,.
  \end{eqnarray}
  This yields a 5-pts variation band (nominal value plus 4
  variations), which we label as \vkernels{}.\\
  
\item From the algorithmic viewpoint, several options to determine
  how to assign the recoil of emissions are possible. We refer to the
  \Pythia{8} manual for detailed explanations, and here we only
  briefly describe the options we have tried.

  By default, in \Pythia{8} the recoil of an ISR emission is taken by
  the whole final state, whereas the full recoil of each final state
  emission is taken by one single parton, according to a dipole-style
  structure~\cite{Sjostrand:2004ef}.

  We label this recoil scheme as \vrecoildef{}.\\

  An alternative approach has recently been implemented with local
  recoils, where only one final-state parton takes the recoil of an
  emission: the existing initial-initial global recoil scheme is
  maintained for an emission off a colour line that stretches through
  the hard process, whereas the handling of initial-final dipole ends
  is changed. Here the single recoiler is picked based on the colour
  flow of the hard process. More details are given in
  Ref.~\cite{Cabouat:2017rzi}. This recoil scheme is activated through
  the setting \verb|SpaceShower:dipoleRecoil=on|.
  
  We label this recoil scheme as \vrecoilD{}.\\

  We also explore the option where the recoil for final state
  emissions is shared between all partons in the final state
  (\verb|TimeShower:globalRecoil=on|).\footnote{This is especially
    convenient for some matching algorithms, like \MCatNLO{}, where a
    full analytic knowledge of the shower radiation pattern is needed
    to avoid double-counting.}  With this alternative approach, colour
  coherence phenomena will be lost, because the radiation pattern off
  a given leg loses its correlation with colour-correlated objects.
  
  We label this recoil scheme as \vrecoilG{}.\\

  We also tried to switch off the flag
  \verb|TimeShower:recoilToColoured|, which, by default, is switched
  on in versions of \Pythia{} after v8.160. This option is expected to
  only affect how the recoil is shared when decaying coloured
  resonances are present~\cite{Brooks:2019xso}, hence we expect such
  flag to only have a visible impact on jet observables particularly
  sensitive to FSR, and, notably, those for which there's a direct or
  indirect interplay with the radiation off colored resonances.

  We label the old recoil scheme
  (\verb|TimeShower:recoilToColoured=off|) as \\~\vrecoilCoff{}.\\
\end{itemize}

\subsection{Results}

We compute different distributions at the PS level using the
\Rivet{}~\cite{Buckley:2010ar} framework. Most plots are obtained
using the ``parton level'' analysis of
Ref.\cite{Cormier:2018tog}.\footnote{We thank the authors of
  Ref.~\cite{Cormier:2018tog} for providing us with the \Rivet{}
  implementation used therein.} The jet shapes plots in
Fig.~\ref{fig:MC_ttbar:jetshapes} are obtained starting from the
analysis {\tt ATLAS\_2013\_I243871}~\cite{Aad:2013fba}, whereas the
plot showing distributions for $H_T$ (Fig.~\ref{fig:MC_ttbar:HT}) is
obtained from the {\tt MC\_JETS} \Rivet{} analysis.

\begin{figure}[t]
  \begin{center}
    \includegraphics[width=7cm]{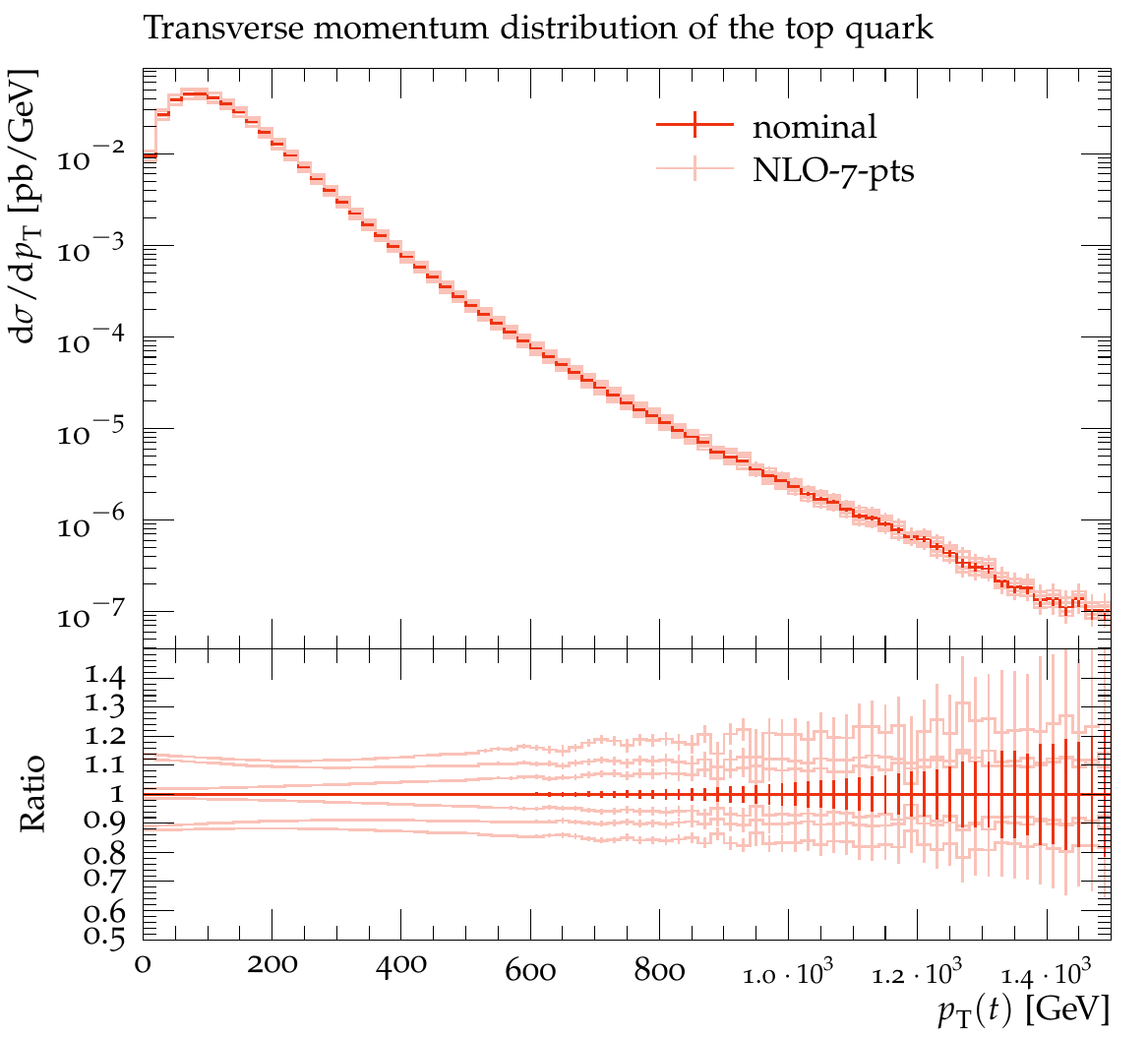}\hspace{1cm}
    \includegraphics[width=7cm]{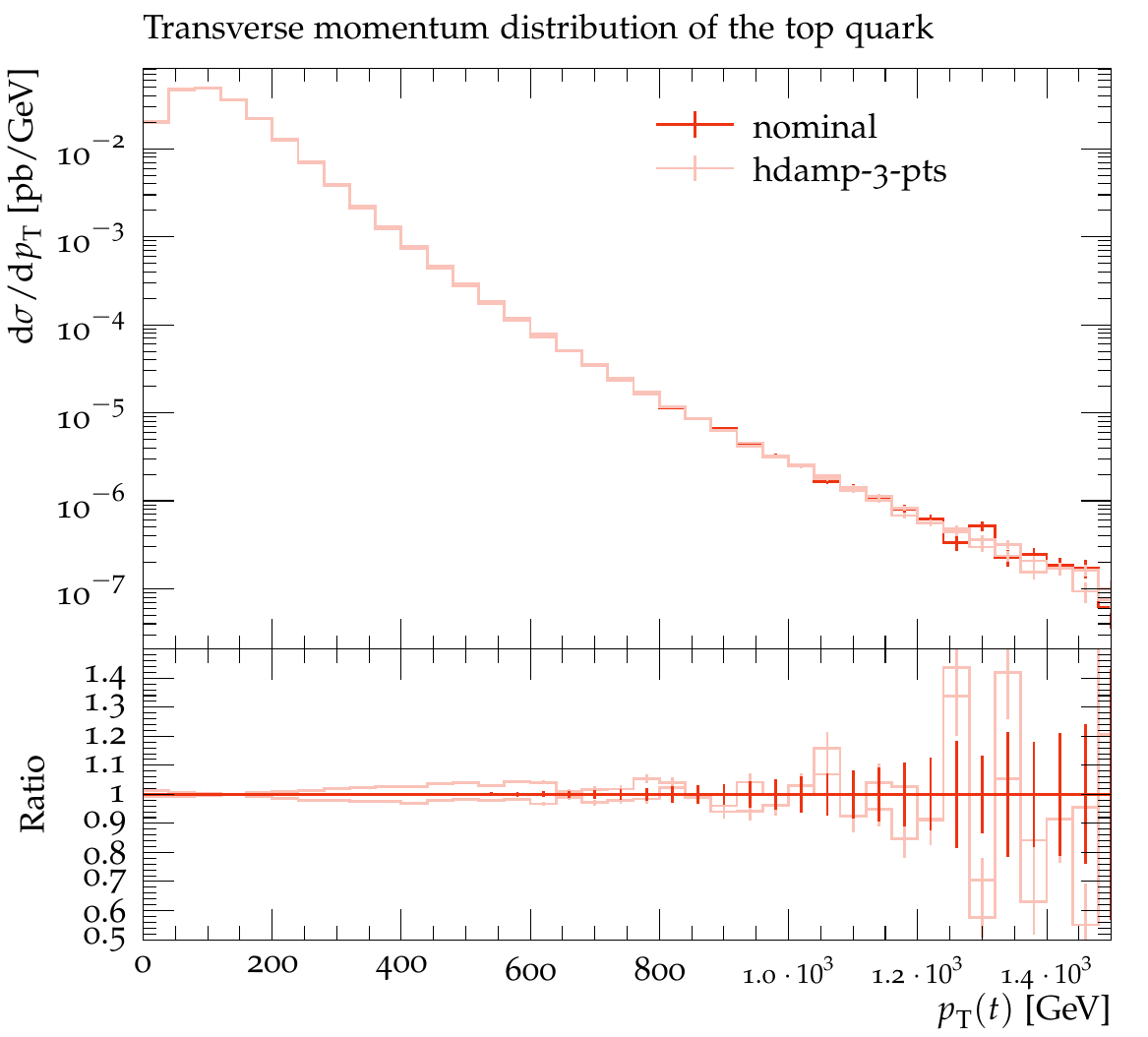}\\
    \includegraphics[width=7cm]{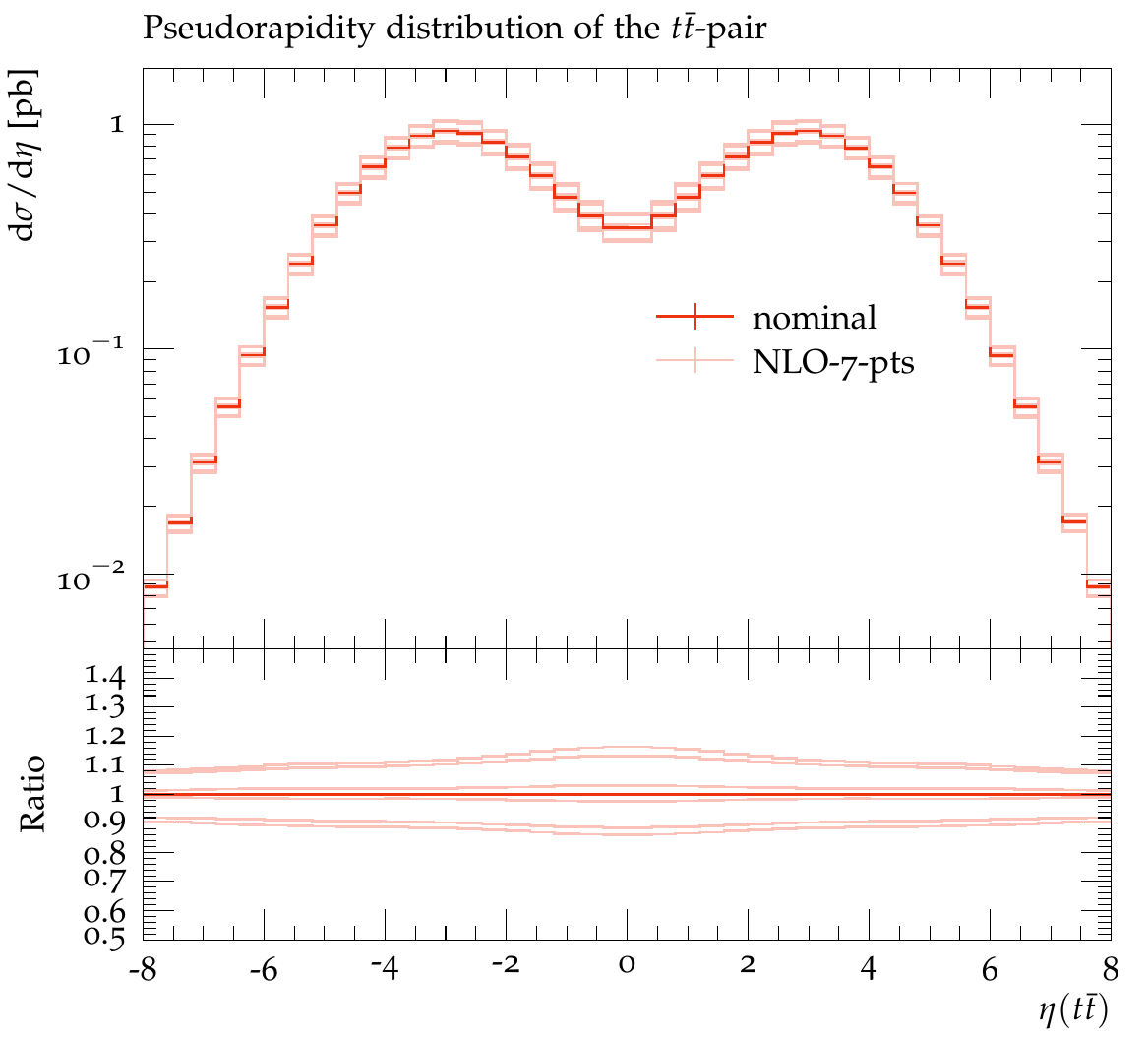}\hspace{1cm}
    \includegraphics[width=7cm]{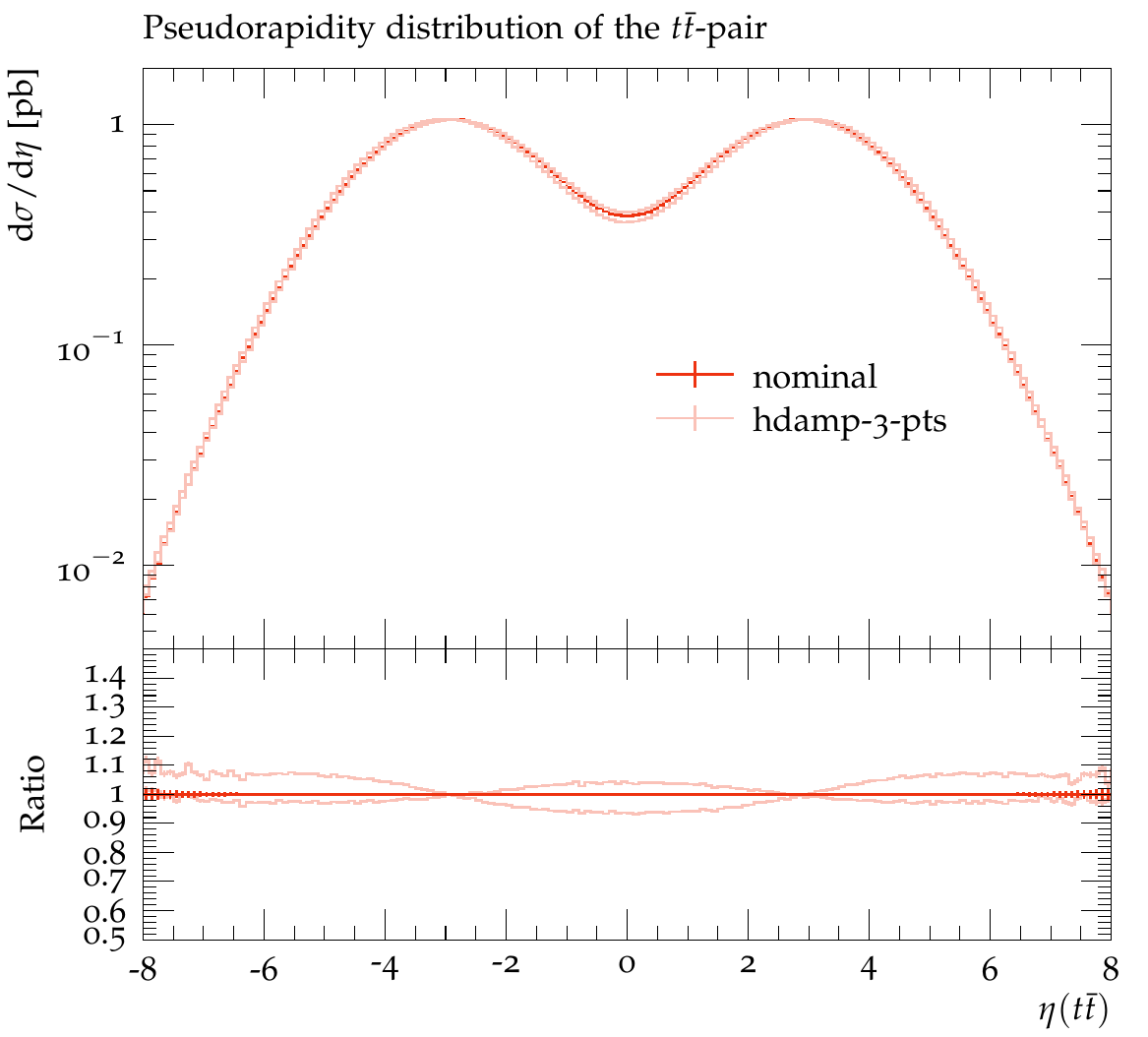}
    \caption{Effects of scale variations in the matrix element and in
      the matching. We show the transverse momentum of the top quark
      (top panels) and the the pseudorapidity of the $t\bar t$ system
      (bottom panel). In the plots on the left, we show the
      uncertainty due to the 7-pts scale variation in the NLO
      computation (\vNLOscales{}), in those on the right the
      uncertainty due to the variation of {\tt hdamp} (\vhdamp{}).}
    \label{fig:MC_ttbar:inc_hardscales}
  \end{center}
\end{figure}

In Fig.~\ref{fig:MC_ttbar:inc_hardscales} we show the effect of the scale
variations related to the computation of the hard matrix elements
(\vNLOscales{}), and of {\tt hdamp}, i.e. of the main parameter that
is currently used to assess matching uncertainties in the \POWHEG{}
scheme (\vhdamp{}). We start by focusing on two observables computed
using the kinematics of top quarks before their decay: the transverse
momentum of the top quark and the pseudorapidity of the $t\bar t$
system.

As expected, for inclusive quantities like $p_T(t)$ and
$\eta(t\bar{t})$, the uncertainty is largely dominated by the
\vNLOscales{} variations, and it amounts to an uncertainty of the
order 10\%. The effect of the \vhdamp{} variation is essentially
negligible for $p_T(t)$. It is small, but visible, on the
pseudorapidity of the $t\bar t$ system. This can be understood by
recalling that, despite $\eta(t\bar t)$ is a quantity inclusive over
radiation from initial state, different values of $\eta$ get a more
(or less) sizeable contribution from regions dominated by small or large values
of $p_T(t\bar{t})$. Large values of $|\eta|$ are dominated by the
region where $p_{T}(t\bar{t})$ is very small, and indeed we observe a
pattern similar to the one observed in the very small $p_T(t\bar{t})$
region (this is not clearly visible in the plot shown in
Fig.~\ref{fig:MC_ttbar:ptTT} because of the bin size, but we verified
it).

\begin{figure}[t]
  \begin{center}
    \includegraphics[width=7cm]{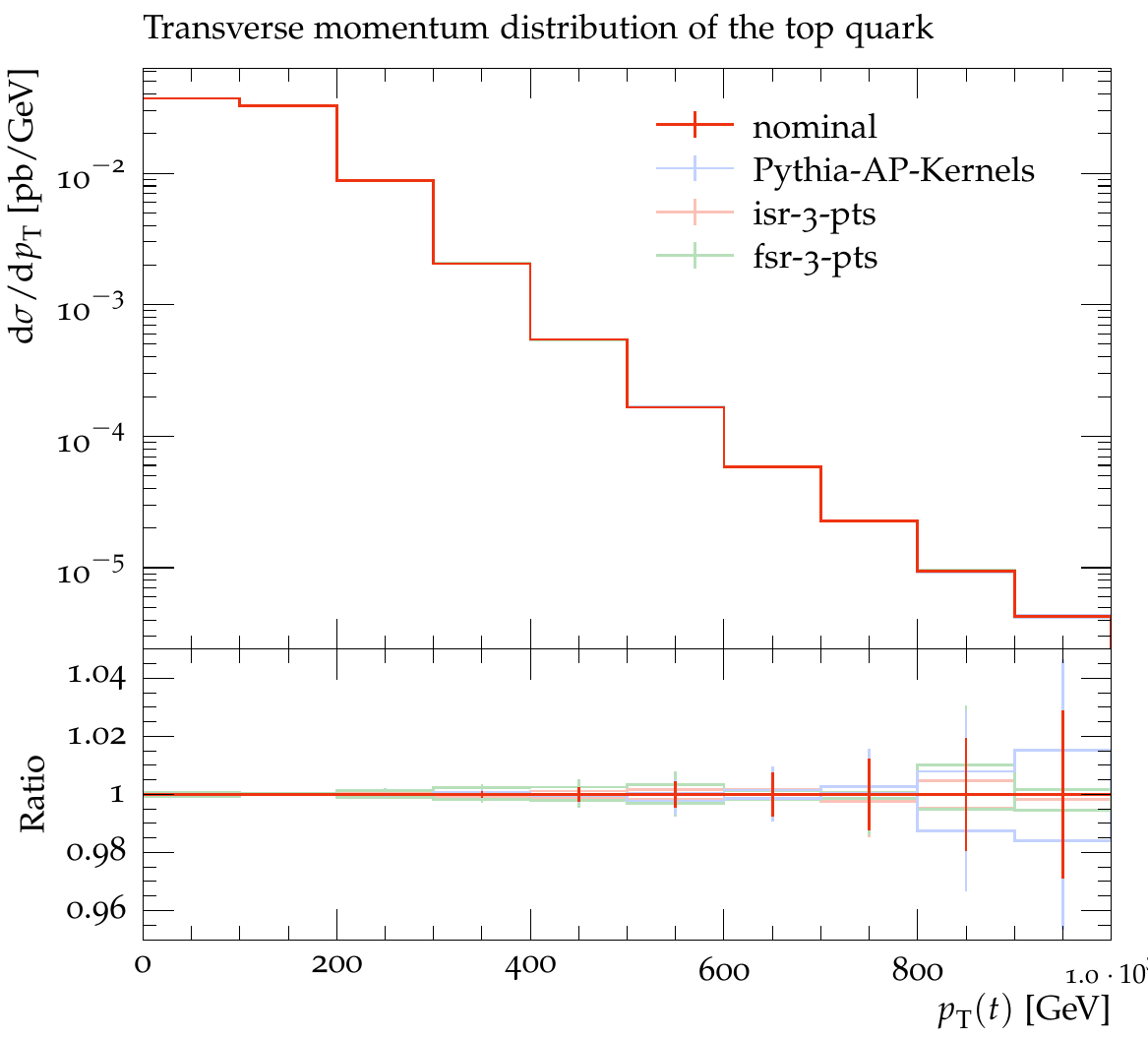}\hspace{1cm}
    \includegraphics[width=7cm]{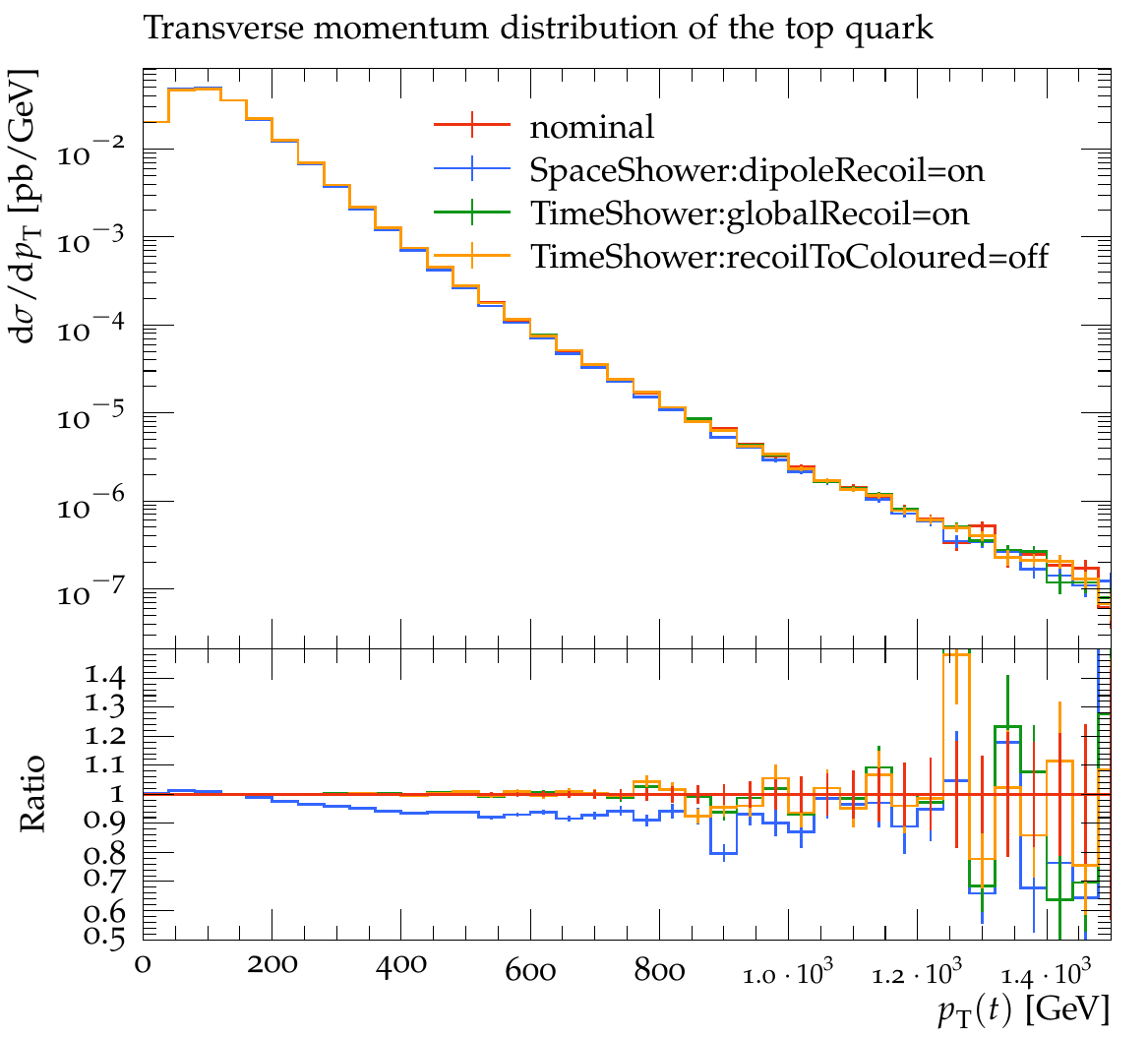}\\
    \includegraphics[width=7cm]{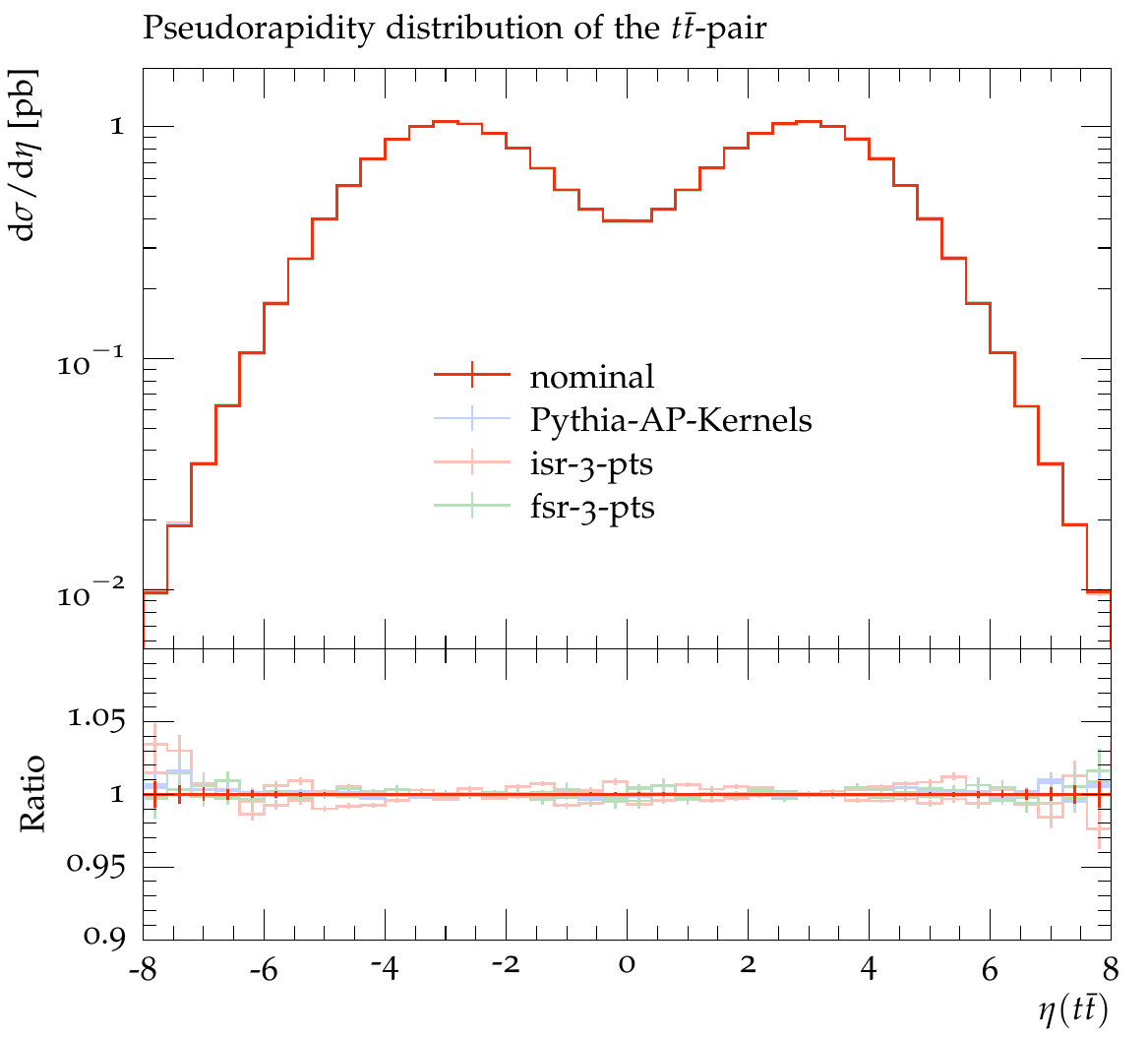}\hspace{1cm}
    \includegraphics[width=7cm]{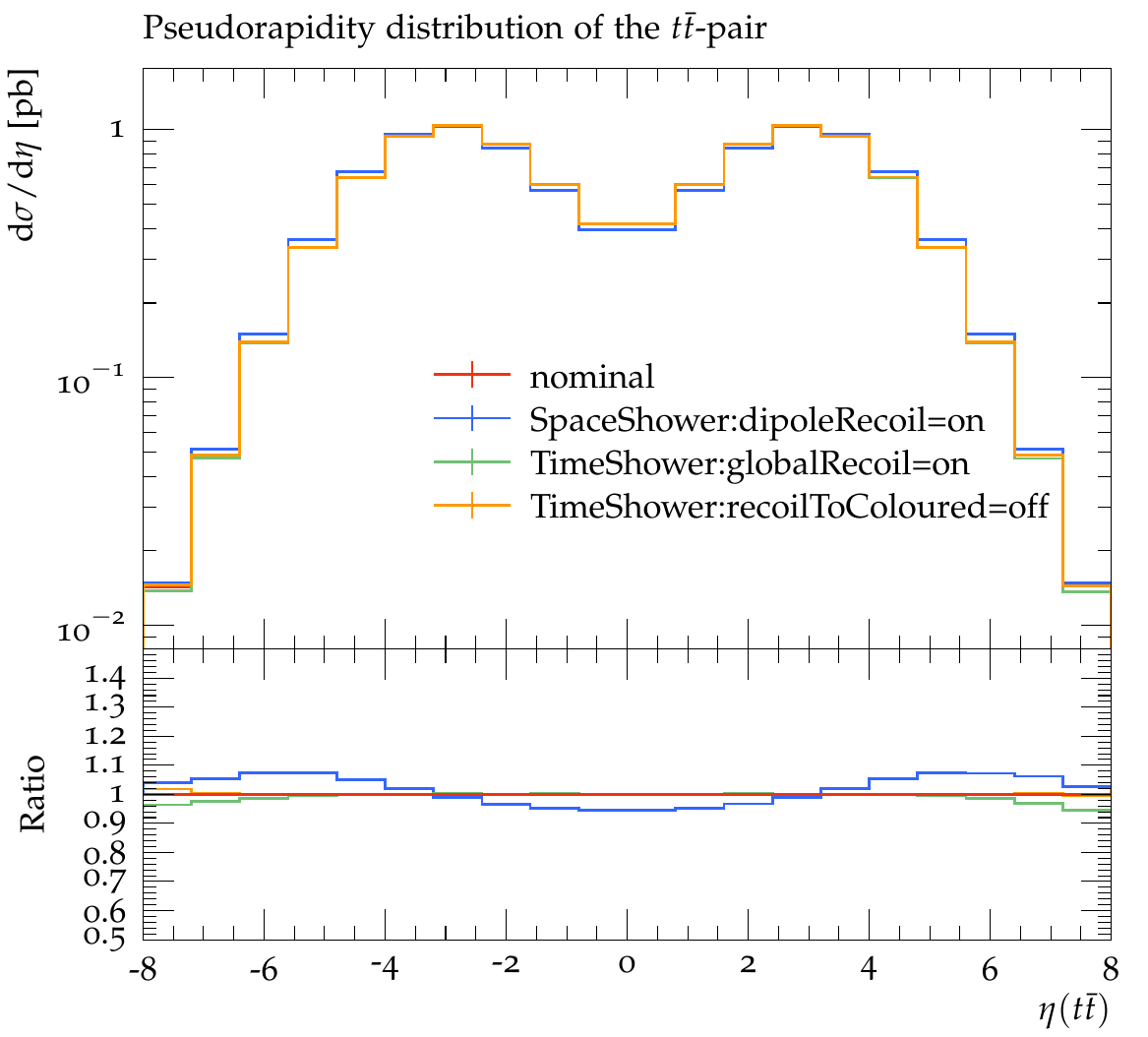}
    \caption{Effects due to choices of PS scales and finite terms of
      splitting kernels (\vISR{}, \vFSR{}, and \vkernels{}, panels on
      the left), and effects due to different recoil options (panels
      on the right). We show the same observables displayed
      in~Fig.~\ref{fig:MC_ttbar:inc_hardscales}.}
    \label{fig:MC_ttbar:inc_psvar}
  \end{center}
\end{figure}

\begin{figure}[t]
  \begin{center}
    \includegraphics[width=7cm]{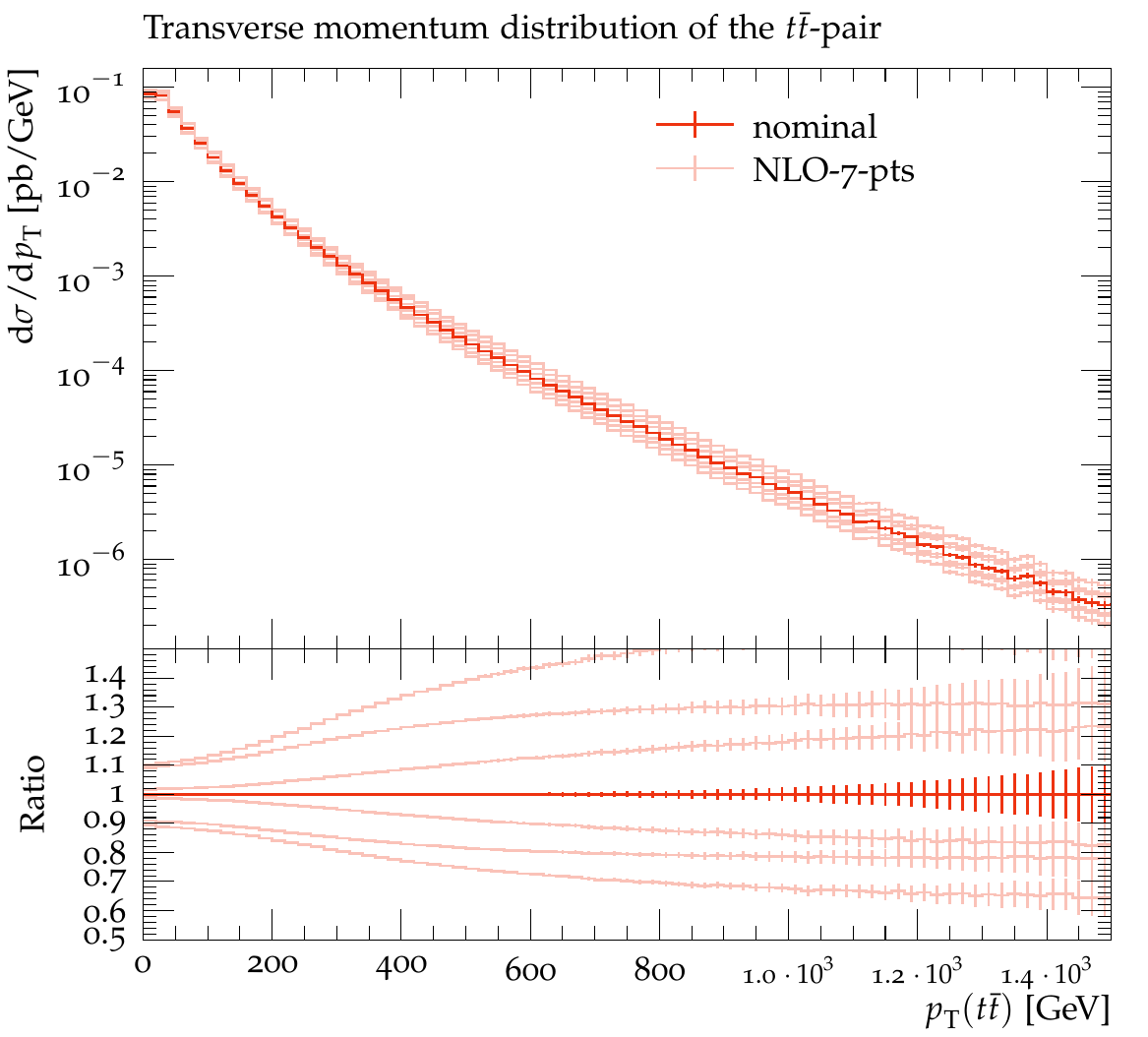}\hspace{1cm}
    \includegraphics[width=7cm]{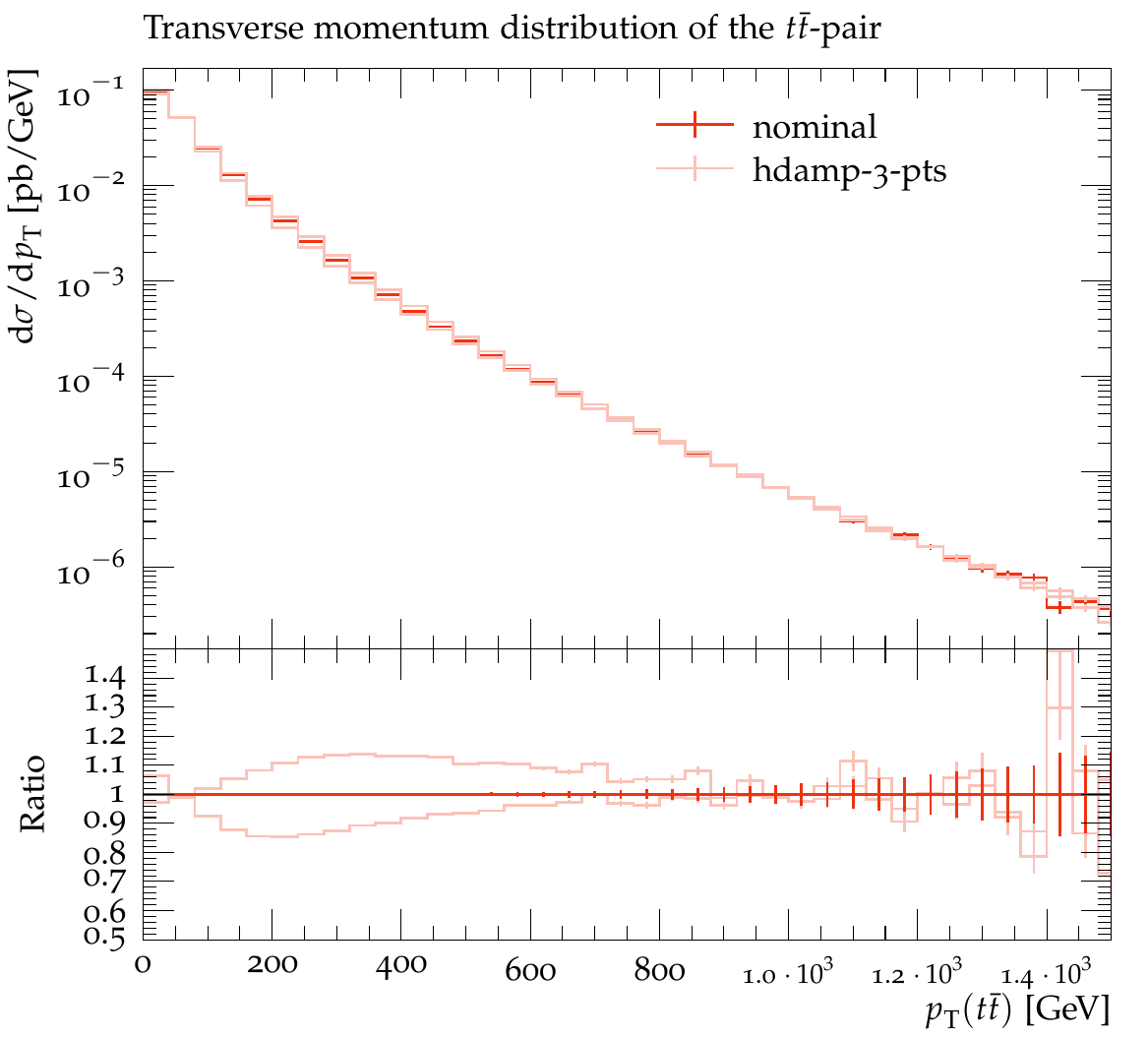}\hspace{1cm}\\
    \includegraphics[width=7cm]{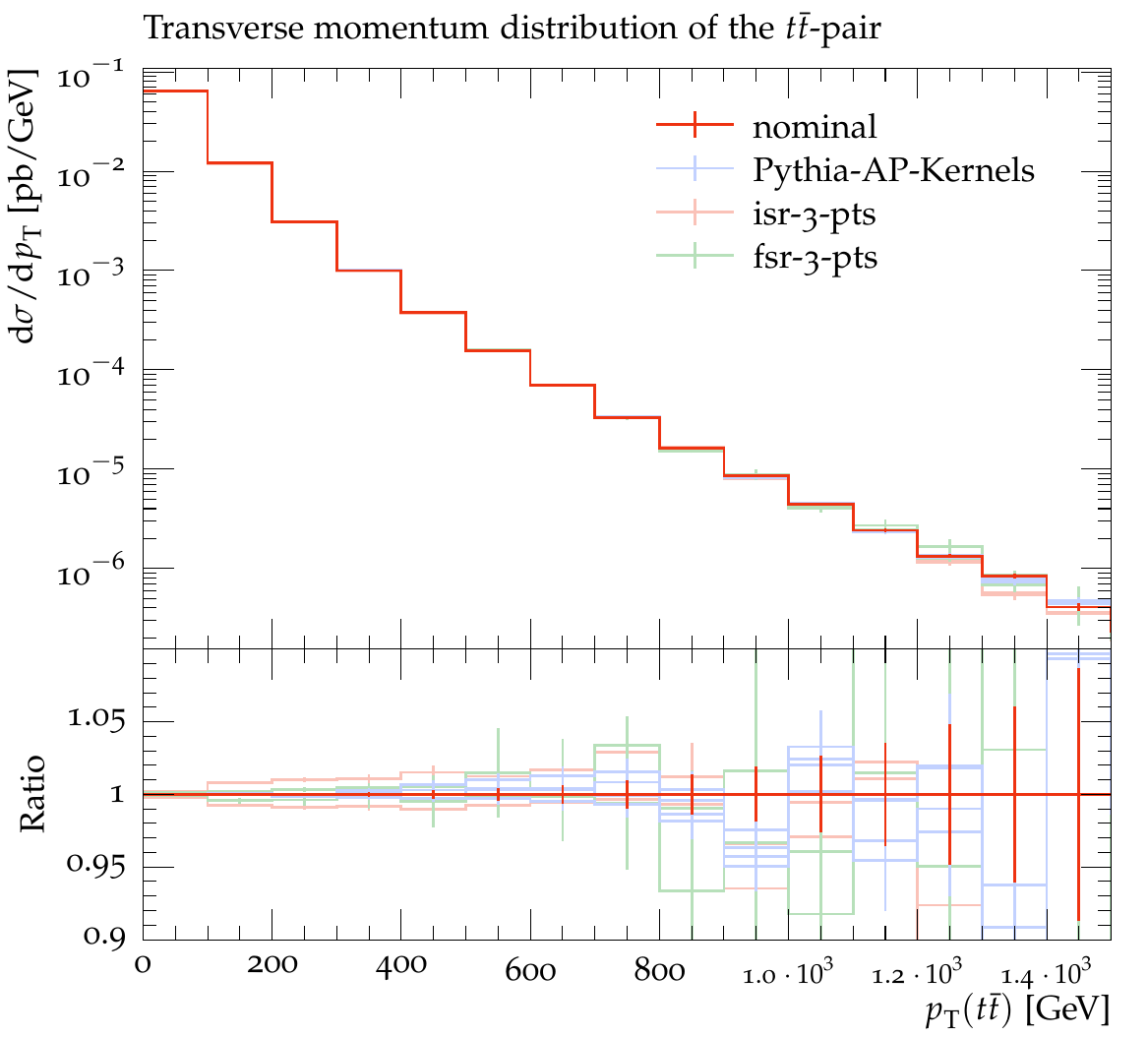}\hspace{1cm}
    \includegraphics[width=7cm]{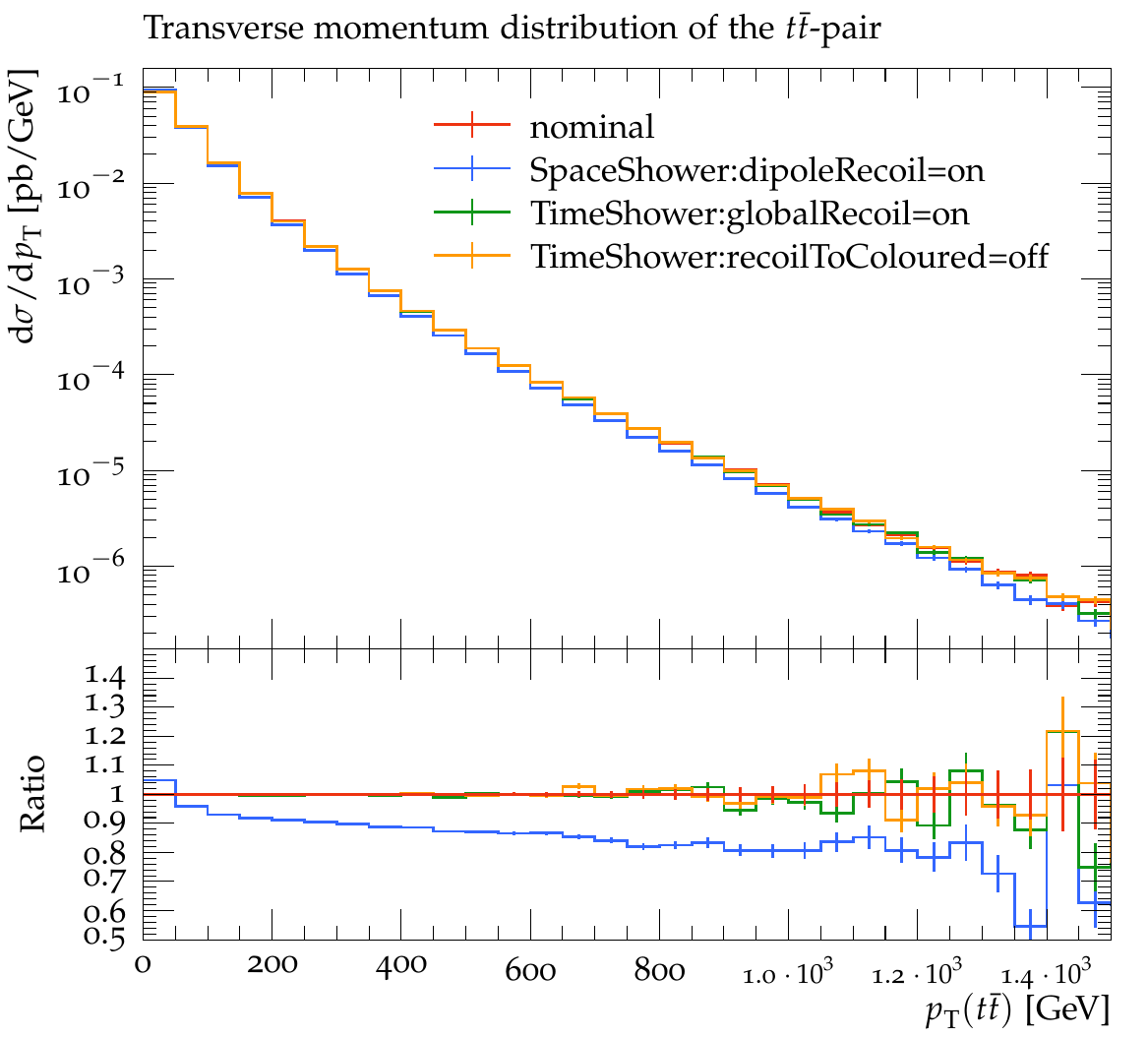}
    \caption{Effects of all the variations considered in this study on
      the transverse momentum of the $t\bar{t}$ pair. In the first row
      we show the uncertainty due to the 7 points scale variation in
      the NLO computation (left, \vNLOscales{}), and the uncertainty
      due to the variation of {\tt hdamp} (right, \vhdamp{}). In the
      second row, on the left, we show the uncertainty due to scales
      in the parton showers and the choices of the finite terms of the
      splitting kernels (\vISR{}, \vFSR{} and \vkernels{}), whereas on
      the right we show the effect of different recoil schemes.}
    \label{fig:MC_ttbar:ptTT}
  \end{center}
\end{figure}

\begin{figure}[t]
  \begin{center}
    \includegraphics[width=7cm]{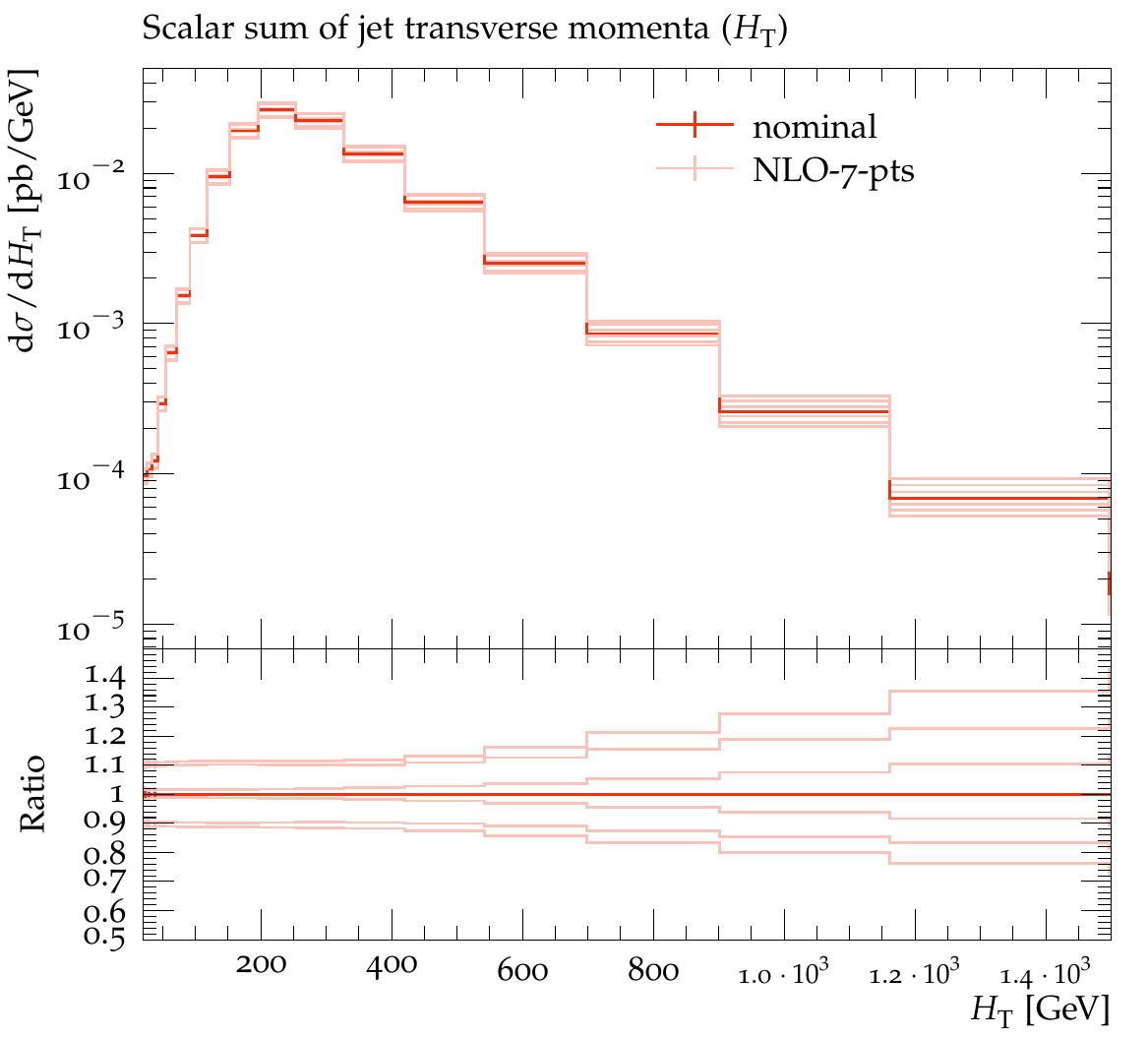}\hspace{1cm}
    \includegraphics[width=7cm]{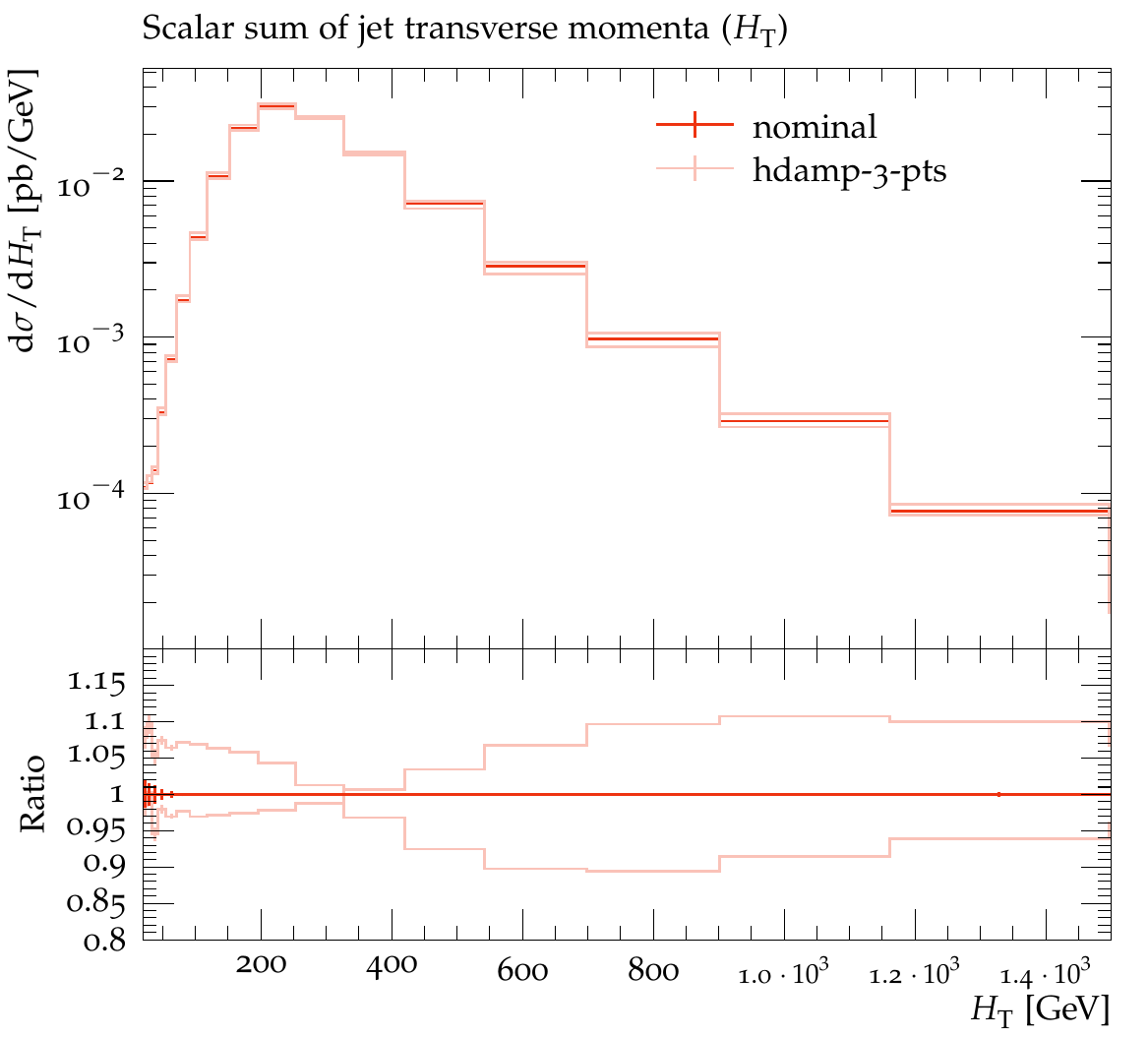}\\
    \includegraphics[width=7cm]{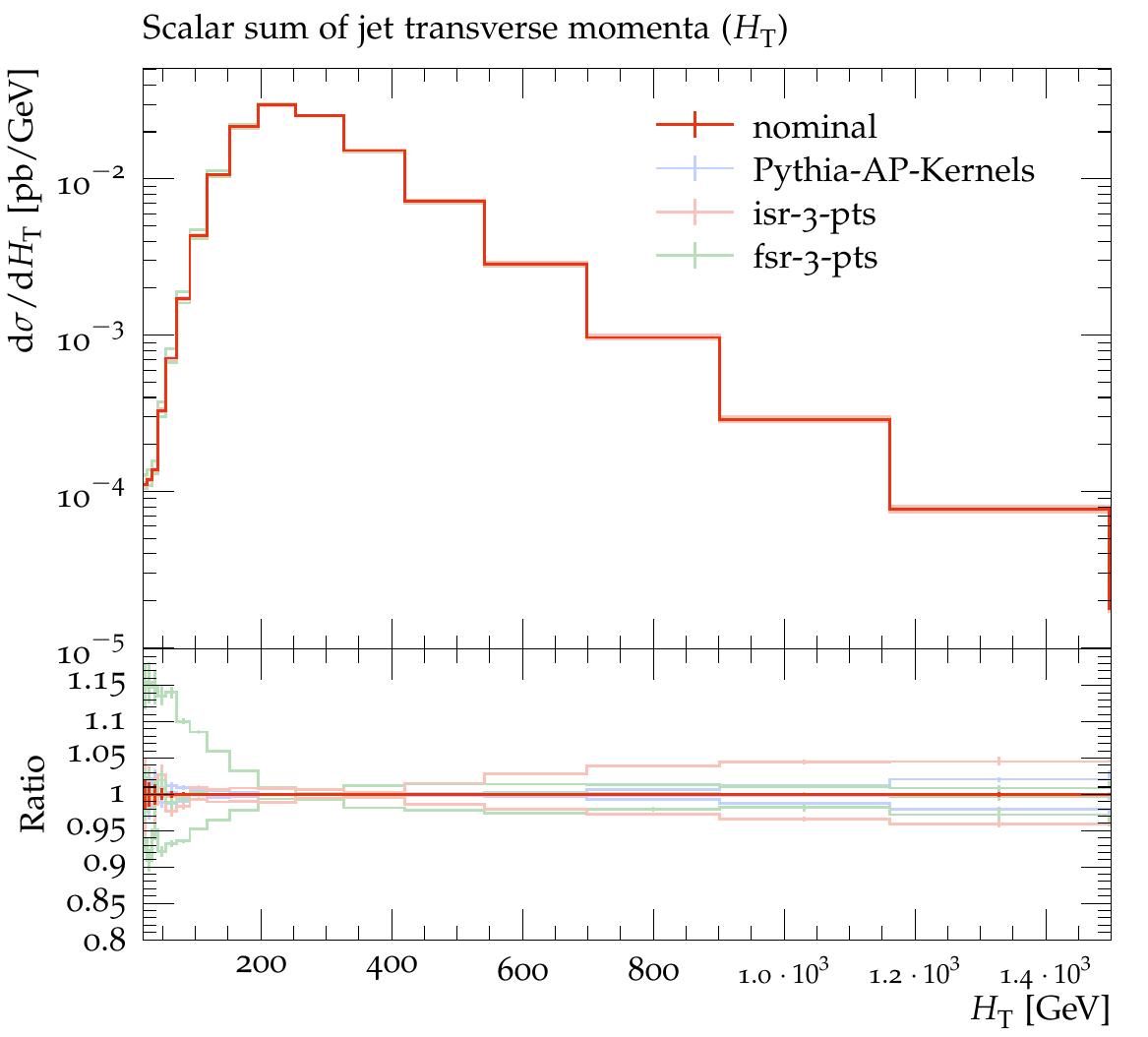}\hspace{1cm}
    \includegraphics[width=7cm]{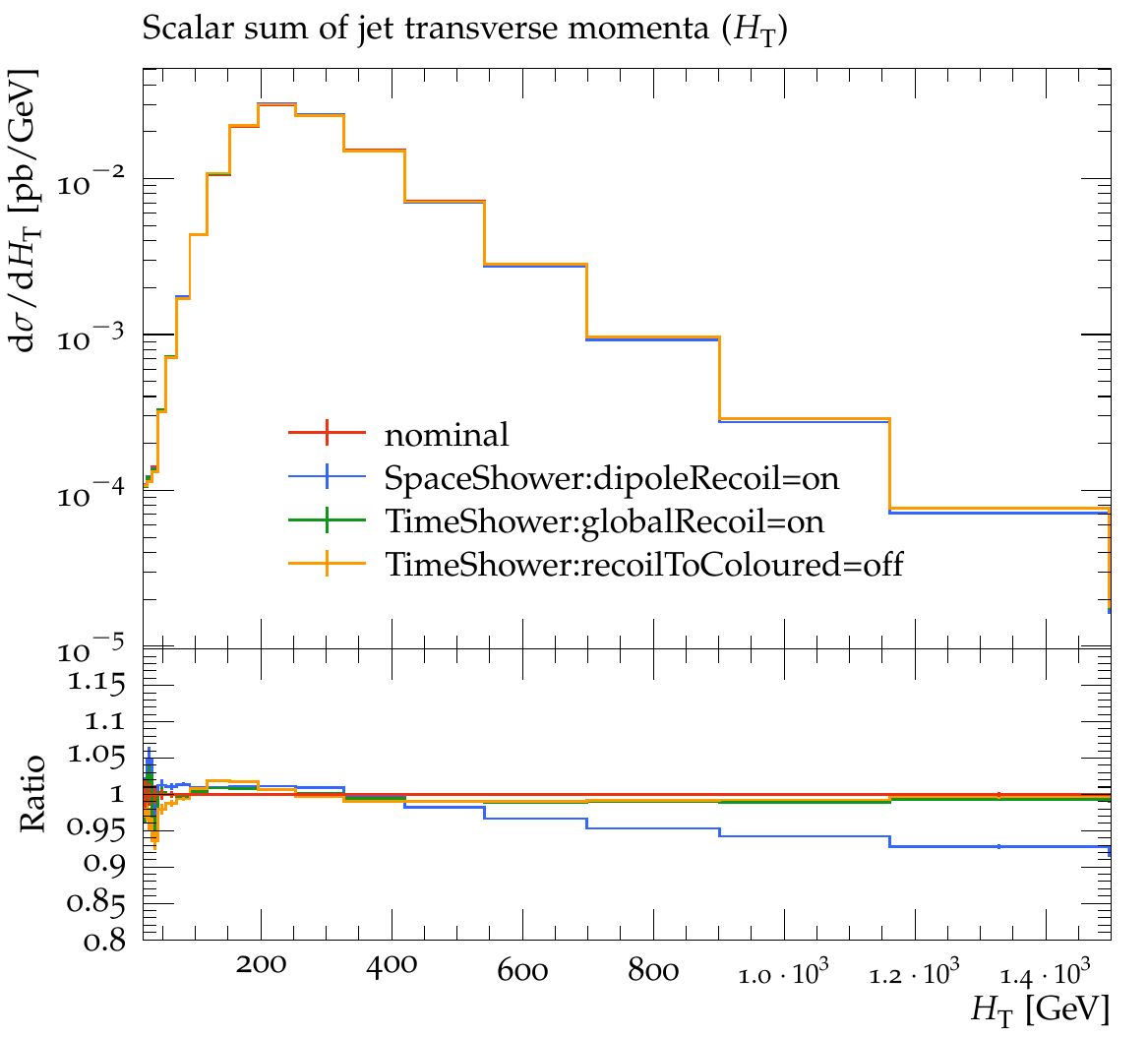}
    \caption{Same as in Fig.~\ref{fig:MC_ttbar:ptTT}, but for the
      $H_T$ observable.}
    \label{fig:MC_ttbar:HT}
  \end{center}
\end{figure}

\begin{figure}[t]
  \begin{center}
    \includegraphics[width=7cm]{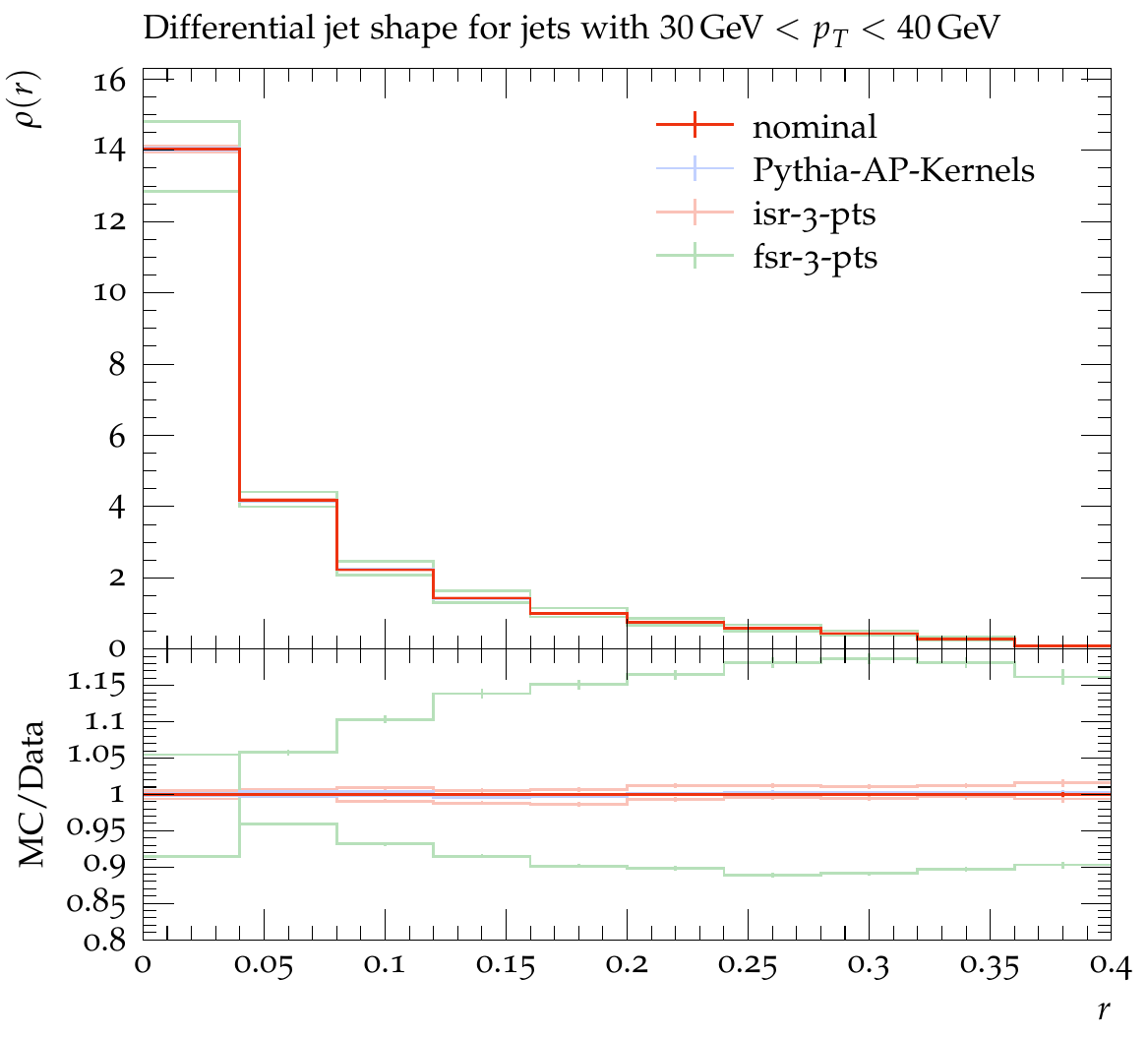}\hspace{1cm}
    \includegraphics[width=7cm]{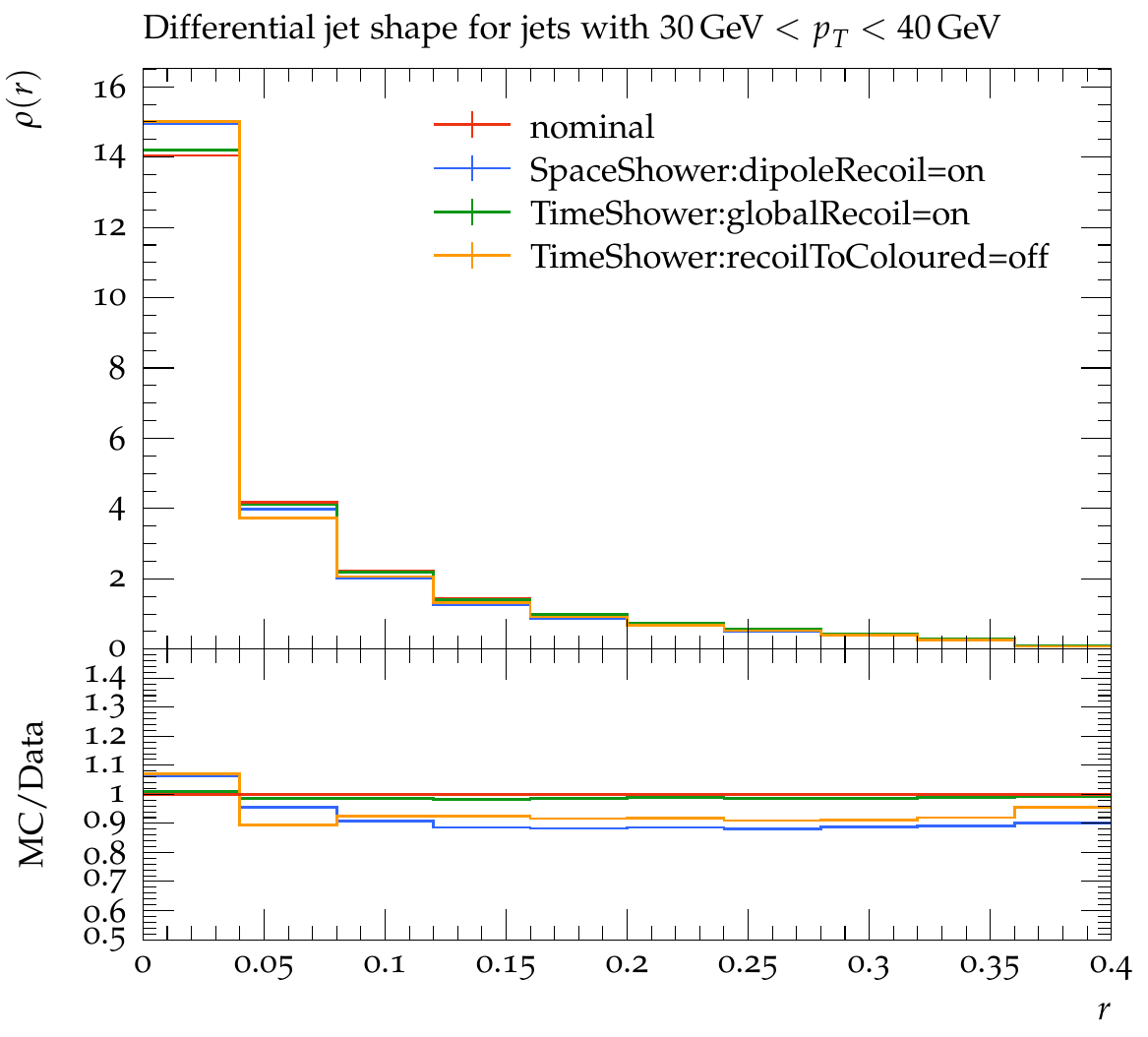}
    \caption{Results for jet shapes and their dependence on the
      variation of PS scales, on the finite terms in the splitting
      kernels, and on the different recoil schemes. See main text for
      details.}
    \label{fig:MC_ttbar:jetshapes}
  \end{center}
\end{figure}

In Fig.~\ref{fig:MC_ttbar:inc_psvar} we show, for the same
observables, the effect of the variations having to do with the
assessment of perturbative subleading effects of PS algorithms, as
well as the effects due to the algorithmic choices related to the
implementation of the recoil.
As expected, these observables are essentially unaffected by these
variations. The only visible difference is seen when the local recoil
scheme is used for radiation of initial-final state dipoles
(\vrecoilD{}). Albeit small, this difference is noticeable, and it's
probably connected to the effects seen for the $t\bar{t}$ transverse
momentum spectrum discussed below. Given the inclusive nature of these
observables, a more careful investigation of the impact of the recoil
is certainly needed. In the context of vector boson scattering
(see~\cite{Ballestrero:2018anz} and subsequent ongoing work\footnote{A
  recent update on these studies can be found here:
  \url{https://indico.cern.ch/event/826136/contributions/3560434/attachments/1926938/3190076/161019LHCHXSWGMEETING.pdf}}),
studies have shown, first, that the local dipole recoil scheme yields
sizeable differences when compared against other schemes, and later on
arguments to support this scheme over other ones were provided, at
least for VBF-type processes. Similar conclusions were reached in the
NNLO+PS study of Ref.~\cite{Monni:2019whf}.

In $t\bar t$ production, the observable that should be mostly affected
by ISR is $p_{T}(t\bar{t})$, the transverse momentum of the $t\bar
{t}$ pair. In Fig.~\ref{fig:MC_ttbar:ptTT} we show the impact of the
various variations considered in this study on this observable. As
expected, as the transverse momentum of the top-pair system is
generated by \POWHEG{} (through its Sudakov form factor as well as
through $R_f$), the effect of the variations \vNLOscales{} and
\vhdamp{} is quite visible. The variation of the scales in the hard
cross section is not flat because we split $R$ into a singular and a
regular part through {\tt hdamp}, hence at large $p_{T}(t\bar{t})$ the
uncertainty band is dominated by $R_f$, and hence it grows, reaching
the ballpark values expected (30-40\%) for a quantity whose spectrum
at LO is dominated by a matrix element which involves several powers
of the strong coupling (in the $t\bar{t}$ case, the $2\to 3$ real
corrections to $pp\to t\bar{t}$). Different choices for the scales
$\mu_\text{F}$ and $\mu_\text{R}$ in Eq.~(\ref{eq:MC_ttbar:murmuf})
might yield a slightly narrower and flatter uncertainty band.

The value of {\tt hdamp} separates smoothly the kinematic region where
the \POWHEG{} exponentiation of the full matrix element dominates and
the hard region, where a NLO+PS simulation will give a LO
prediction. Therefore we expect to observe a band that widens in the
intermediate region, and results that eventually coincide at large
values of $p_{T,t\bar{t}}$, where the perturbative uncertainty should
be dominated by scale variation in the short-distance cross
section. Because of our choice, the band indeed widens at
$p_T(t\bar{t})\sim 250$ GeV. As expected, in the tail of the
distribution all the \vhdamp{} results are in agreement.

Among the effects due to PS ``variations'' and recoils, for
$p_T(t\bar{t})$ the more sizeable effect comes from the choice of the
recoil scheme. When a local recoil for initial-final (IF) dipoles is used, the shape
of $p_{T}(t\bar{t})$ can be affected up to 10-15\% (lower panel on the
right of Fig.~\ref{fig:MC_ttbar:ptTT}). As for the top transverse
momentum, this difference calls for further studies. Since we are
performing a NLO+PS simulation, effects due to perturbative aspects of
the PS are instead expected to be subdominant for this
observable. This is what we observe in the lower panel on the left of
Fig.~\ref{fig:MC_ttbar:ptTT}: changing the prefactor of the
renormalization scale in the PS kernels gives an effect at the level
of at most 1-2\% where the plot is statistically significant. As
expected, due to the nature of this observable, this small effect is
due to the \vISR{} variation. There is essentially no effect from the
\vFSR{} variation, nor from the non-singular terms in the splitting
kernels.

In Fig.~\ref{fig:MC_ttbar:HT} we show results for the $H_T$
distribution, i.e. the scalar sum of all the transverse momenta of
jets found in the event. Jets are required to be harder than $p_T>20$
GeV, and are defined using the anti-$k_T$ algorithm with $R=0.4$ ({\tt
  MC\_JETS} analysis in \Rivet{}). In our setup all the jets passing
the cut enter the $H_T$ distribution, including those that contain a
$b$ (or $\bar{b}$) quark.

In contrast with the $p_T(t\bar{t})$ distribution, with $H_T$ we can
appreciate more clearly the effects of PS variations (PS scales and
regular terms of the splitting kernels), because the $H_T$ observable
is computed using jets, and hence it is more sensitive to radiation
generated by the PS. The dominant effects are still due to the the
scale variation in the short-distance cross section, and to the
\vhdamp{} variation (top panels of Fig.~\ref{fig:MC_ttbar:HT}). For
very small values of $H_T$, one probes the phase space region where
jets have very small transverse momenta and, not surprisingly, in this
kinematics region, even a small effect from out-of-jet radiation can
change significantly the $H_T$ value, hence we expect that variations
related to FSR emissions generated by the PS will be the dominant
ones. This is confirmed by the widening of the green band in the
bottom-left plot. Conversely, for large values of $H_T$, radiation
from initial state dominates, and indeed, among the PS variations,
\vISR{} is the more important source of uncertainty, followed by
the~\vkernels{} variation. Although more modest than for
$p_T(t\bar{t})$, effects due to change of the recoil scheme are also
visible in the tail of the distribution.

In Fig.~\ref{fig:MC_ttbar:jetshapes} we show the
differential jet shape $\rho(r)$ for jets with $30\mbox{ GeV} < p_T <
40\mbox{ GeV}$. The differential jet shape in an annulus of inner
radius $r-\Delta r/2$ and outer radius $r+\Delta r/2$ from the axis of
a given jet is defined as
\begin{equation}
  \rho(r) = \frac{1}{\Delta r} \frac{p_T(r-\Delta r/2, r+\Delta
    r/2)}{p_T(0,R)}\,.\nonumber
\end{equation}
In the analysis, $\Delta r = 0.04$ and $p_T(r_1,r_2)$ is the scalar
sum of the transverse momenta of the jet constituents with radial
distance between $r_1$ and $r_2$ with respect to the jet axis.

Jet-shapes are distributions that are expected to exhibit a noticeable
dependence on the variation of the PS scales, notably those
governing the strength of final state emissions. As for $H_T$, in our
setup all the jets passing the cuts enter into the computation of
$\rho(r)$, including the jets that contain a $b$ (or $\bar{b}$) quark.
Therefore, we expect not only the jet shape uncertainty from PS
effects to be dominated by FSR variations, but also that the
\vrecoilCoff{} option will produce visible differences. In the plots
in Fig.~\ref{fig:MC_ttbar:jetshapes} we observe the expected pattern,
i.e. no visible effect from ISR, variations (of the order of up to
10-15\%) due to FSR PS emissions, and an important shape distortion
when \vrecoilCoff{}. There is no visible impact from the variations of
the finite terms of the PS splitting kernels. This probably implies
that this observable is mostly sensitive to universal soft/collinear
effects. The fact that the dipole recoil scheme has such a significant
impact is a bit unexpected: it might simply be due to the fact that
the typical kinematics of all the radiation in the event (from
production and from decay) is affected by this recoil scheme (as can
be evinced from the global effect on $p_T(t\bar{t})$). However a more
detailed investigation is certainly needed, especially in view of the
growing evidence supporting the use of a dipole-recoil scheme, at
least for other LHC processes.

\subsection{Conclusions}
In this contribution we have tried to study the size of
perturbative uncertainties in a NLO+PS simulation of top-pair
production in hadronic collisions. We have used the
\POWHEGplus+\Pythia setup, and we have shown the impact of scale
variation in several aspects of the simulation, i.e. in the
computation of the hard matrix elements, in the \POWHEG{} matching
algorithm (i.e. the hard scale in the {\tt hdamp} factor), and in the
evaluation of the strong coupling in the parton shower. We have also
considered other sources of uncertainties due to possible choices in parton shower algorithms, namely the variation of the finite parts of the splitting
kernels, and variations of the recoil scheme. 

Our study is far from being comprehensive, and its scope is limited
with respect to the original goal, that was to compare results for
different NLO+PS accurate generators (and possibly for different
matching methods, i.e. \MCatNLO{}-type vs. \POWHEG{}-type), the aim
being of establishing if, for the main ``variations'' of perturbative
nature available in different shower algorithms, the results obtained with
different generators are mutually compatible, at least for observables
that should only be affected by perturbative effects.

Within our setup, we have found that, in most cases, each considered
variation has the expected impact on differential
distributions. Although some scale choices are different, and despite
our variation of {\tt hdamp} cannot be exactly compared to the
variation of the ``hard veto scale'', our findings are also
qualitatively similar to the {\tt NLO}$\otimes${\tt PS} results
presented in Ref.~\cite{Cormier:2018tog}. This can be considered a
first step towards the original goal outlined in the previous
paragraph.

Among our findings, we have noticed that, for the observables we
considered, the variation of the finite parts of the splitting kernels
has a negligible impact. We have also found sizeable differences in
several distributions when the local recoil scheme for initial-final
dipoles is used. These recoil effects are not typically considered in
experimental analysis, despite our study shows that their impact can
be larger than perturbative uncertainties due to parton shower scale variations.

As future theoretical developments, in addition to a comparison
against tools that have the same nominal accuracy, it would be
interesting to compare our results against dedicated but precise
computations at fixed-order and at all orders. It would also be
important to extend the study by including hadronization and MPI
effects.


\let\Herwig\undefined
\let\Pythia\undefined
\let\POWHEGplus\undefined
\let\POWHEG\undefined
\let\MCatNLO\undefined
\let\POWHEGBOX\undefined
\let\Rivet\undefined

\let\muR\undefined
\let\muF\undefined

\let\vNLOscales\undefined
\let\vhdamp\undefined

\let\vISR\undefined
\let\vFSR\undefined
\let\vkernels\undefined

\let\vrecoildef\undefined
\let\vrecoilD\undefined
\let\vrecoilG\undefined
\let\vrecoilCoff\undefined




\clearpage

\bibliography{LH19}


\end{document}